\documentclass[11pt]{yalephd}

\usepackage[
  a4paper,
  lmargin=1.5in,
  rmargin=1in,
  tmargin=1in,
  bmargin=1in,
  headsep=5pt, 
  footskip=.25in
]{geometry}
\usepackage{rotating,graphicx}	
\usepackage{lscape}
\usepackage{times}
\usepackage{float}
\usepackage{amsmath}	
\usepackage{nccmath}
\usepackage{amssymb}    
\usepackage{relsize}
\usepackage{mathtools}
\usepackage{bigints}
\usepackage{afterpage}
\usepackage[dvipsnames]{xcolor}  
\usepackage{soul}
\usepackage{xcolor}
\usepackage{subfig}
\usepackage{caption}
\usepackage{sidecap}
\usepackage{epsfig}
\usepackage{longtable}
\usepackage{xurl}
\usepackage[]{hyperref}
\def\UrlBreaks{\do\/\do-}
\usepackage{color}
\usepackage[normalem]{ulem}
\usepackage{comment}
\usepackage{bm}
\usepackage{cuted}
\usepackage{flushend}
\usepackage{cancel}
\usepackage{setspace}
\usepackage{titlesec}
\definecolor{linkColor}{rgb}{0.,0.11,0.22}
\definecolor{YaleBlue}{rgb}{0.0,0.22,0.444}

\usepackage[style=authoryear-comp,maxbibnames=2,maxcitenames=2,mincitenames=1,uniquelist=false,uniquename=false,natbib=true,backend=biber,sorting=anyt]{biblatex}

\DeclareDelimFormat{finalnamedelim}{\addspace\&\space}

\usepackage{longtable}
\usepackage[toc,page]{appendix}
\usepackage{lscape}
\usepackage[T1]{fontenc}
\usepackage{ae,aecompl}
\usepackage{enumitem}
\usepackage[bottom]{footmisc}
\usepackage{subfiles}
\usepackage[T1]{fontenc}
\usepackage{etoolbox}

\DeclareMathOperator{\sech}{sech}

\DeclareMathOperator{\erf}{erf}
\DeclareMathOperator{\sgn}{sgn}

\usepackage{macros_ub}
\graphicspath{{./}{figures/}}

\begin{document}

\newcommand{\unit}[1]{\ensuremath{\, \mathrm{#1}}}


\titleformat{\chapter}[display]
    {\fontfamily{ptm}\huge\bfseries}{\chaptertitlename\ \thechapter}{5pt}{\centering\singlespacing\huge}
\titlespacing*{\chapter}{0pt}{50pt}{30pt}

    \title{Pushing the Frontiers of Non-equilibrium Dynamics of Collisionless and Weakly Collisional Self-gravitating Systems}
	\author{Uddipan Banik}
	%
	\advisor{Prof. Frank C. van den Bosch}
	\date{May 2023}



\begin{abstract}

In the $\Lambda$CDM paradigm of cosmology, structure formation occurs via gravitational encounters and mergers between self-gravitating structures like galaxies and dark matter halos. This perturbs galaxies and halos out of equilibrium. These systems are collisionless, i.e., cannot relax within the Hubble time via two-body encounters, thereby prevailing in a state of non-equilibrium or quasi-equilibrium at best. However, such perturbed collisionless systems can relax via other mechanisms such as phase-mixing, Landau damping and violent relaxation. Phase-mixing and Landau damping take several dynamical times to achieve completion. Both these processes can be described using a linear order perturbation of the collisionless Boltzmann and Poisson equations under the assumption of a sufficiently weak perturbation. Phase-mixing is the coarse-grained destruction of a coherent response to a perturbation due to an intrinsic spread in the oscillation frequencies of the field particles. Landau damping is the fine-grained damping of the response due to energy exchanges driven by gravitational interactions between the particles, which is also known as a collective effect. Unlike the linear phenomena of phase-mixing and Landau damping, violent relaxation is fundamentally a non-linear effect and is a rapid process, achieving completion within a dynamical time. Moreover, violent relaxation is self-limiting in nature, rendering an end state that may be very different from the Maxwellian velocity distribution that ensues from two-body/collisional relaxation. While a perturbed collisionless system (subject) undergoes relaxation via the above processes, the subject response simultaneously exerts a back reaction on the perturber and slowly changes its orbital dynamics, typically draining its orbital energy and angular momentum. This phenomenon is a type of secular evolution and is known as dynamical friction. It is the key process by which the relative orbital energy of interacting galaxies and halos is dumped into their internal energies, often resulting in their merger. Gravitational encounters and dynamical friction are therefore at the basis of all structure formation in the universe.

Depending on how the timescale of perturbation ($\tau_\rmP$) compares to the oscillation periods ($\tau$) of field particles in the subject, gravitational perturbations can be impulsive ($\tau_\rmP<\tau$), resonant ($\tau_\rmP\sim\tau$) or adiabatic ($\tau_\rmP>\tau$). This dissertation investigates how gravitational encounters and collisionless relaxation occur in these three different regimes. First, we provide a general non-perturbative formalism to compute the energy change in impulsive encounters, which properly describes penetrating encounters, unlike the standard approach that only works for distant encounters. Next, we develop a comprehensive linear perturbative formalism to compute the response of a stellar disk to external perturbations. We study the cases of an infinite isothermal slab as well as a realistic disk galaxy in a non-responsive dark matter halo. The disk response phase-mixes away due to different oscillation frequencies of the stars, giving rise to local {\it phase-space spirals}. A vertically anti-symmetric (symmetric) perturbation gives rise to a bending (breathing) mode response of the disk, which triggers a one-armed (two-armed) spiral in the $z-v_z$ phase-space. Perturbations slower than the vertical oscillation period ($\tau_z$), i.e., those with $\tau_\rmP>\tau_z$, induce stronger bending modes, while faster ones trigger more pronounced breathing modes. This translates to more distant encounters with satellite galaxies causing stronger bending mode perturbations. We analyze the response of the Milky Way (MW) disk to encounters with its satellite galaxies, and find that Sagittarius (Sgr) dominates the Solar neighborhood response among all the satellites. This makes Sgr the dominant contender among the MW satellites to have triggered the Gaia phase spiral. Collisional diffusion due to the scattering of disk stars by structures like giant molecular clouds can result in a super-exponential damping of the phase spiral amplitude on a fine-grained level. The diffusion timescale in the Solar neighborhood of the MW disk turns out to be $\tau_{\rmD}^{\odot}\sim 0.6-0.7\Gyr$. This sets an approximate upper limit of $\tau_{\rmD}^{\odot}$ to the time elapsed since perturbation so that the resultant Solar neighborhood phase spiral survives collisional damping and is detectable. Only sufficiently impulsive perturbations can trigger phase spirals; adiabatic ones cannot. Near-resonant parts of the phase-space undergo gradual phase-mixing and do not develop phase spirals. It is the near-resonant response of the subject that exerts the maximum torque on the perturber, driving its orbital inspiral via dynamical friction. 

In the final chapters of this dissertation, we develop a general theory for dynamical friction on a perturber in circular orbit in a spherical host galaxy. This explains the origin of secular phenomena in $N$-body simulations of cored galaxies that are unexplained in the standard Chandrasekhar and resonance theories for dynamical friction: (i) core-stalling, the apparent cessation of dynamical friction driven infall in the core region of galaxies with a central constant density core, (ii) super-Chandrasekhar friction, an accelerated infall phase prior to core-stalling, and (iii) dynamical buoyancy, an enhancing torque that can counteract dynamical friction and push out the perturber from inside the core region. We relax the adiabatic and secular approximations adopted in the derivation of the LBK torque in the standard resonance theory, and provide a fully self-consistent perturbative formalism for dynamical friction. The LBK torque depends on the current orbital radius of the perturber, arises exclusively from {\it resonances} between the field particles and the perturber, and is always {\it retarding}. On the contrary, the self-consistent torque depends on the entire infall history of the perturber ({\it memory effect}), has a significant contribution from the {\it near-resonant orbits}, and flips sign within a certain radius in the core region, becoming {\it enhancing} instead of retarding. To overcome the limitations of linear perturbation theory near the core-stalling radius, we develop a novel, non-perturbative, orbit-based treatment of dynamical friction. Here we model dynamical friction as a circular restricted three body problem, wherein we identify the near-co-rotation resonant \horseshoen, \pacman and tadpole orbits of field particles as the dominant contributors to dynamical friction or buoyancy. Outside the core region, all these orbits exert friction. As the perturber enters the core region, it tidally disrupts the core and the inner Lagrange points undergo a bifurcation. This drastically alters the orbital topology: the friction exerting \horseshoe orbits disappear and the \pacman orbits become dominant. A shallow distribution function gradient along these \pacman orbits gives rise to an enhancing torque or dynamical buoyancy in the core region. We argue that core-stalling occurs near the radius of Lagrange point bifurcation, which marks the transition from friction to buoyancy. Bifurcation of Lagrange points and therefore core-stalling are exclusive to a galaxy with a constant density core and are absent in one with a central NFW-like cusp. We discuss some profound astrophysical implications of core-stalling and buoyancy, e.g., the potential choking of supermassive black hole (SMBH) mergers in cored galaxies, leading to a significant population of off-center, wandering SMBHs. This has implications for future detections of gravitational wave events due to SMBH mergers by Laser Interferometer Space Antenna (LISA).

\end{abstract}
	
    \maketitle
    \makecopyright
	
    \begin{center}
{\bf \large Acknowledgments}
\end{center}

\paragraph{\ul{Academic support}}

I am immensely grateful to my PhD supervisor, Prof. Frank C. van den Bosch, for providing me with invaluable advice and guidance, and inspiring me with his fantastic physical intuition and creative imagination. Successful completion of this dissertation was made possible by the ingenious ideas that were born out of our prolonged discussions on diverse topics of physics and astronomy. The pivotal role of these discussions in keeping me motivated and productive over the years cannot be overstated.

I am wholeheartedly thankful to Prof. Martin D. Weinberg for guiding me through the mathematical intricacies of various topics in galactic dynamics and kinetic theory. During the Applied Galactic Dynamics Workshop at the Center for Computational Astrophysics (CCA), Flatiron Institute, New York City, USA, I worked with him on a project on disk response and phase spirals. Stimulating and insightful discussions with him on topics of dynamics played a crucial role in my development as a dynamicist.

At the CCA Applied Galactic Dynamics Workshop, my phase spiral project heavily benefited from interactions and discussions with Prof. Kathryn V. Johnston, Jason Hunt, Adrian Price-Whelan, Elise Darragh-Ford and Prof. Elena D'Onghia. Especially, my discussions with Prof. Kathryn V. Johnston on galactic dynamics and Milky Way science were enlightening and insightful. 

I am especially grateful to Prof. Frank C. van den Bosch, Prof. Martin D. Weinberg, Prof. Kathryn V. Johnston and Prof. Elena D'Onghia for providing me with constant support regarding applications for postdoctoral research positions. I am thankful to my thesis committee members, Prof. Nikhil Padmanabhan, Prof. Marla Geha and Prof. Priyamvada Natarajan, for providing me with valuable advice on research. My occasional discussions with Prof. Greg Laughlin on various topics of astrophysical fluid dynamics were stimulating.

The valuable inputs and suggestions by the members of my research group, Dhruba Dutta-Chowdhury, Kaustav Mitra, Michael Tremmel, Nir Mandelker, Johannes Lange, Shashank Dattathri, Barry Chiang and David Hernandez, played a key role in my development as a researcher. My prolonged discussions with Kaustav Mitra and Joel Ong on various topics of physics, astronomy and mathematics were enlightening as these gave birth to new ideas, some of which contributed to this dissertation. My research also benefited from stimulating discussions with astrophysicists outside Yale including Chris Hamilton, Scott Tremaine, Simon White, Avishai Dekel, James Binney, Wyn Evans, John Magorrian, Justin Read, Andrew Pontzen, Jorge Penarrubia, Mike Petersen, Jonathan Freundlich, Seshadri Sridhar, Oleg Gnedin, Jerry Ostriker and Yuri Levin, to name a few.

And last but not least I am grateful to the Department of Astronomy at Yale University for providing me with the opportunity to blossom as a theoretical astrophysicist. The role of the dynamics group at the Center for Computational Astrophysics, Flatiron Institute, in my professional development as a dynamicist and astrophysicist cannot be overstated. My teachers and mentors at IIT Kanpur, FIITJEE and Patha Bhavan played a vital role in developing the foundations of my physics knowledge.

\paragraph{\ul{Moral support}}

This dissertation was a product of not only my efforts but also one of constant moral support from many people. My partner, Shahina, has been a source of unwavering support through the years of my PhD. Especially, when things were not going my way, her words of wisdom gave me much needed hope and confidence. She shone as a beacon of hope during dark times. 

I feel fortunate to have had amazing friends and colleagues at Yale. I shared joyous moments and unforgettable memories with Soham Jana, Aritra Ghosh, Sayoni Mitra and Mansa Srivastava. As my most treasured friends at Yale, they showed that `a friend in need is a friend indeed'. I shared many happy moments with my cohort: Aritra Ghosh, Malena Rice and Tim Miller. I cherish the stimulating conversations I engaged in with Malena Rice on several occasions.

And last but not least I am extremely thankful to my parents, Utpal Banik and Debjani Banik, for being a constant source of motivation, support and inspiration. It is after all their countless sacrifices that helped me reach where I am today. My grandparents, Lalit Mohan Banik and Arati Banik, who are no longer in this world, kindled in me a burning curiosity about the unknown and an undying thirst for knowledge at an early age. This played a crucial role in my journey as a pursuer of truth and beauty and ultimately as a theoretical physicist.

    \newpage
\thispagestyle{empty} 
\addtocounter{page}{-1} 
\begin{flushright}
\topskip0pt
\vspace*{\fill}
{\it In memory of Late Prof. Amit Dutta whose guidance kindled in me the passion and spirit of a theoretical physicist and whose teachings on Statistical Mechanics at IIT Kanpur set me on the path of investigating the dynamics of many body astrophysical systems.}
\vspace*{\fill}
\end{flushright}

    \tableofcontents %
    \listoffigures   
    \listoftables    %
    
    \mainmatter      

	\chapter{Introduction} \label{chapter: introduction}

\section{Background}

According to the $\Lambda$CDM\footnote{$\Lambda$ stands for dark energy and CDM stands for cold dark matter.} paradigm of cosmology, which is a widely accepted theory describing the formation history of our universe, structure formation proceeds in a hierarchical bottom-up fashion, i.e., smaller structures merge to form bigger ones. Galaxies and dark matter halos, which are thought to embed galaxies, undergo frequent mergers and are therefore always in a state of non-equilibrium or quasi-equilibrium at best. There exist tons of observations of gravitational encounters between galaxies that highlight this non-equilibrium or quasi-equilibrium state of self-gravitating structures in our universe: groups of interacting galaxies, galaxies in-falling in a cluster, post-merger irregular galaxies, galaxies with shells, rings, spiral arms and bars, and so on. Even our own galaxy, the Milky Way, harbours various non-equilibrium features throughout the disk as well as in the surrounding stellar halo. Many of these features have been known to exist for decades, e.g., the bar, the spiral arms and the warp, but Gaia \citep[][]{Gaia_collab.16,Gaia_collab.18a,Gaia_collab.18b} has ushered in a whole plethora of observations of non-equilibrium features in and around the Milky Way: bending and breathing waves \citep[][]{Weinberg.91,Widrow.etal.14,Banik.etal.22b}, streams of stars kicked up from the disk known as `feathers' \citep[][]{Price-Whelan.etal.15}, stellar streams \citep[][]{Malhan.etal.18}, dynamical friction wakes \citep[][]{Conroy.etal.21}, moving groups \citep[][]{Yang.Yong.etal.21}, merger remnants like the Gaia Enceladus \citep[][]{Helmi.etal.18} and the Gaia sausage \citep[][]{Belokurov.etal.18}, phase-space spirals, also known as phase spirals or snails \citep[][]{Antoja.etal.18,Bland-Hawthorn.etal.19,Hunt.etal.21,Hunt.etal.22,Gandhi.etal.22}, and the list goes on. The exquisite parallax and proper motion information from Gaia astrometry along with the radial velocity measurements from Gaia spectroscopy has provided us with an enormous amount of kinematic information about the various structures in the Milky Way galaxy. This has revolutionized the study of Milky Way's perturbation and merger history, and has ushered in a whole new era of galactic dynamics. Hence, this calls for a shift of gear in galactic dynamics research from the standard equilibrium dynamics (e.g., Jeans modeling and Schwarzschild modeling techniques to measure galaxy masses using stellar kinematics data) to the non-equilibrium dynamics of galaxies, which is what motivates this thesis.

\section{Relaxation of self-gravitating collisionless systems}

Not all structures in the universe harbour the same extent of out-of-equilibrium features. In other words, some structures are more equilibrated or `relaxed' than others. For example, there exist observations of regular-looking disk and elliptical galaxies in the field, isolated from other galaxies, that seem to have relaxed into an ordered, equilibrium state, even if they might have undergone mergers in the past. Question is: how fast can a galaxy relax after undergoing a perturbation such as an encounter or merger with another galaxy? This is an open question in the fields of galactic dynamics and galaxy formation and evolution. Galaxies are to good approximation collisionless, i.e., short range star-star interactions or collisions are scarce. The two body relaxation timescale for self-gravitating $N$-body systems \citep[][]{Binney.Tremaine.08} is given by

\begin{align}
\tau_{\rm coll} \sim \frac{N}{\ln{N}} \frac{1}{\sqrt{G\bar{\rho}}},
\end{align}
where $N$ is the number of particles and $\bar{\rho}$ is the average density of the system. For large $N$ systems like galaxies, $\tau_{\rm coll}$ exceeds the Hubble time, which is roughly the age of the universe, by many orders of magnitude. Hence, galaxies do not relax via two body encounters, and the equipartition of kinetic energy is not attained within the Hubble time. This behaviour is remarkably different from gas, where short range inter-molecular collisions are the main drivers of relaxation. Due to the collisionless nature of galaxies, the phase-space distribution function of stars in a galaxy can significantly deviate from a Maxwellian velocity distribution, something that is rapidly attained by gas molecules via collisional relaxation. Cold dark matter (CDM) halos\footnote{Dark matter is called cold if the dark matter particle was non-relativistic when it decoupled from the primordial plasma. Weakly Interacting Massive Particles (WIMPs) and QCD axions are some of the most popular particle candidates for CDM.} are also collisionless, and therefore relax via processes that are similar to those in galaxies. On the other hand, there exist weakly collisional systems such as self-interacting dark matter (SIDM) halos, where relaxation is partially driven by collisional processes. Globular clusters (GCs) and nuclear star clusters (NSCs) have a two body relaxation time smaller than the Hubble time, and are therefore more collisional than galaxies and dark matter halos.

The big questions that motivate this work are the following: 

\begin{itemize}[noitemsep]
    \item How does collisionless relaxation/equilibration occur in galaxies and dark matter halos? 
    \item How fast and how efficient in equilibration are these collisionless relaxation processes?
\end{itemize}

It has been known for some time that the primary mechanisms for collisionless relaxation include the processes of {\it phase-mixing}, {\it Landau damping} and {\it violent relaxation}. Phase-mixing is a coarse-grained damping of the perturbation in the distribution function of a collisionless system, also known as the response, that occurs due to the loss of coherence in the motion of field particles oscillating at different frequencies. Landau damping \citep[][]{Landau.46} is a fine-grained damping of the self-gravitating response of a collisionless system due to the free streaming motion of field particles exchanging energies via gravitational interactions, which is also known as a collective effect. Violent relaxation is a rapid loss of coherence in the response due to the scrambling of orbital energies driven by a time-dependent potential \citep[][]{LyndenBell.67,Sridhar.89}. While phase-mixing and Landau damping are manifest from linear perturbation theory, violent relaxation is a fundamentally non-linear phenomenon. Even after decades of research, the operating mechanism of some of these relaxation processes in self-gravitating collisionless systems is not well understood. This thesis sheds some light into the operating mechanism of some of these collisionless relaxation processes that drive the formation and evolution of self-gravitating systems like galaxies and dark matter halos, with a special emphasis on phase-mixing and the resultant spiral shaped features in the phase-space distribution of field particles known as phase-space spirals or phase spirals akin to those observed by the Gaia satellite in the disk of our Milky Way galaxy \citep[][]{Antoja.etal.18}.

\section{Different regimes of gravitational encounters}

The perturbation and relaxation of collisionless systems occurs in different regimes, based on how the perturbation timescale, $\tau_\rmP$, compares to the intrinsic oscillation periods, $\tau$, of the field particles in the subject. In case of a gravitational encounter between a subject galaxy and a perturber with a relative velocity, $v_\rmP$, and impact parameter, $b$, $\tau_\rmP$ is equal to $b/v_\rmP$. Generally, the outer part of the subject (that is located reasonably far from the perturber), where the field particles move far slower than the perturbation, i.e., $\tau\gg \tau_\rmP$, lies in the impulsive regime. Therefore, to compute energy transfer between these field particles and the perturber, one can adopt the impulse approximation: the particles gain a sudden impulsive shock due to the perturbation, which changes their velocities but not so much their positions. Examples of impulsive encounters include tidal shocks experienced by satellite galaxies or dark matter subhalos during their pericentric passage in a host galaxy/halo, or those experienced by globular clusters crossing a disk galaxy at high speed. We shall show in Chapters 4 and 5 that phase spirals due to phase-mixing of the response of a perturbed disk are only formed by sufficiently impulsive perturbations. In contrast with the outer part of the subject, the very inner part consists of fast moving particles, with $\tau\ll \tau_\rmP$, and lies in the adiabatic regime. This part of the subject is adiabatically shielded from the perturbation since any response to the perturber gets washed away due to many orbital excursions of the field particles within the perturbation timescale. It is the intermediate region of the subject, where the two timescales match up, i.e., $\tau\sim \tau_\rmP$, and resonances occur between the oscillation frequencies of the field particles and the perturbation frequency. The strongest response to the perturber therefore develops in this intermediate region where the field particle orbits are resonantly perturbed. Orbital energy exchange between the perturber and the `impulsive' and `resonant' zones of the subject ultimately drives the secular evolution of the perturber's orbit or dynamical friction.

\section{Dynamical friction}

As perturbed galaxies and halos undergo relaxation, the orbital dynamics of the perturber, which can be another satellite galaxy or dark matter subhalo or even a globular cluster or black hole, is simultaneously perturbed. This results in a secular evolution of the perturber's orbit, which occurs over a timescale much longer than the typical dynamical/orbital time. The host galaxy or halo develops a global response to the perturber. This response usually lags behind the perturber and exerts a retarding torque on it, which drains its orbital angular momentum and causes it to inspiral towards the center of the host. The gradual drainage of energy and angular momentum from the perturber to the field particles of the host, associated with the orbital inspiral of the perturber, is known as {\it dynamical friction}. It is an outcome of the back reaction of the host response on the perturber. Dynamical friction governs a vast range of astrophysical processes including (i) galaxy-galaxy mergers, (ii) galactic cannibalism, which is the inspiral and subsequent merger of galaxies in a galaxy cluster or that of satellite galaxies in a host galaxy/halo, (iii) angular momentum loss and orbital inspiral of a binary compact object (black hole, neutron star, etc.) pair due to its interaction with surrounding matter before the gravitation wave inspiral phase sets in and causes a merger. Dynamical friction plays an essential role in supermassive black hole (SMBH) mergers, which are believed to drive the formation of SMBHs found at the centers of galaxies. It is also dynamical friction that is responsible for dumping the relative orbital energy of interacting galaxies/halos into their internal degrees of freedom in the form of random kinetic energy of the field particles, thus triggering their orbital decay and merger. And since all structure formation in the non-linear scales proceeds through mergers between dark matter halos and galaxies, dynamical friction is an essential gradient of structure formation in the $\Lambda$CDM paradigm of cosmology.

There are currently two different frameworks to describe how dynamical friction operates: (i) the \citet[][]{Chandrasekhar.43} picture and (ii) the resonance picture \citep[][]{Tremaine.Weinberg.84}. The Chandrasekhar picture is the most popular picture of dynamical friction although it is highly idealistic in the sense that it describes dynamical friction as an outcome of local momentum exchanges between a massive perturber on a straight orbit and the field particles of a surrounding homogeneous medium that are also on nearly straight orbits. This description of dynamical friction is therefore fairly local. A far more sophisticated and general theory of dynamical friction is the resonance theory provided by \citet{Tremaine.Weinberg.84}, who inferred that in the realistic scenario of a massive perturber moving on a circular orbit in an inhomogeneous host galaxy, dynamical friction arises exclusively from resonances between the oscillation frequencies of the field particles and the circular frequency of the perturber. The dynamical friction torque acting on the perturber in this picture is known as the LBK torque, named after \cite{LyndenBell.Kalnajs.72} who first derived it to describe the spiral arm driven angular momentum transport in disk galaxies. The LBK/resonance picture of dynamical friction is a global picture and also more accurate than the Chandrasekhar picture, although the computation of the LBK torque is far more involved than the Chandrasekhar torque. 

Despite its obvious simplifications, predictions from the Chandrasekhar picture agree reasonably well with $N$-body simulation results. There are, however, cases where it clearly fails. For example, it cannot explain the phenomenon of {\it core-stalling}, the cessation of dynamical friction in the central constant-density core of a host galaxy or halo with cored density profile \citep[e.g.,][]{Read.etal.06c,Inoue.11,Cole.etal.12,Petts.etal.15,Petts.etal.16,DuttaChowdhury.etal.19}. On the other hand, the resonance picture explains core-stalling as an outcome of the suppression of near-co-rotation resonances in the core region of the host \citep[][]{Kaur.Sridhar.18,Kaur.Stone.22}. Still, both the Chandrasekhar and resonance pictures are incomplete. Neither of them can explain the origin of certain dynamical phenomena observed in $N$-body simulations: (i) the perturber undergoing an accelerated in-fall before stalling in the core region, which is known as super-Chandrasekhar friction \citep[][]{Read.etal.06c,Goerdt.etal.10,Zelnikov.Kuskov.16}, and (ii) the perturber often getting pushed out from inside the core region by an enhancing torque that counteracts dynamical friction, something known as {\it dynamical buoyancy} \citep[][]{Read.etal.06c,Cole.etal.12}. Thus, the standard theories of dynamical friction fail to reproduce all features of the secular evolution of massive perturbers in cored galaxies. This dissertation presents a general theory of dynamical friction that goes beyond the standard Chandrasekhar and resonance theories and explains the origin of secular phenomena in cored systems that are unexplained in the standard formalism.

\section{Dissertation outline}

This dissertation aims to explore how the perturbation and relaxation of self-gravitating collisionless systems such as galaxies and cold dark matter halos occur in three different regimes of gravitational encounters: impulsive, resonant and adiabatic. Particular emphasis is placed on the phase-mixing of the response of a perturbed galaxy/halo and the impact of the non phase-mixed near-resonant part of the response on the secular evolution of the perturber, viz., dynamical friction. This dissertation is organized as follows:

\paragraph{\ul{Chapter}~\ref{chapter: many_body_dynamics} \ul{(Review of many body dynamics)}:} This chapter provides a brief review of the standard theoretical framework adopted to study the dynamics of many body systems. First, we discuss the integrability of Hamiltonian systems, which forms the foundations of galactic dynamics. Next, we review the kinetic theory of many-body Hamiltonian systems, with a discussion of the governing equations, including the Liouville equation, BBGKY hierarchy, Balescu-Lenard equation, Boltzmann equation, collisionless Boltzmann or Vlasov equation and Fokker-Planck equation. Then we discuss the different mechanisms by which perturbed collisionless systems relax/equilibrate, and how the relaxation of self-gravitating collisionless systems contrasts with that of plasma and fluids. To this end, we discuss the advantages and shortcomings of numerical methods like N-body simulations and analytical methods like perturbation theory. Finally, we briefly review the theory of secular evolution and dynamical friction, stating the successes and failures of the standard Chandrasekhar and LBK formalisms and the potential scope for improvement. Note that this chapter is for pedagogical purposes and can be skipped if the reader is familiar with galactic dynamics and the statistical mechanics of many-body systems.

\paragraph{\ul{Chapter}~\ref{chapter: paper1} \ul{(Impulsive encounters)}:} In this chapter, we study the impulsive regime of gravitational encounters between spherical galaxies and/or CDM halos. We develop a general theory to compute the energy transfer and mass loss in an impulsive encounter between galaxies. Unlike the standard theory \citep[][]{Binney.Tremaine.87,Spitzer.58,Gnedin.etal.99} that works only for distant encounters, our theoretical framework can describe gravitational encounters for all impact parameters (including penetrating encounters) along straight as well as eccentric orbits, and yields predictions about the mass loss in galaxy-galaxy encounters that are in excellent agreement with $N$-body simulations.

\paragraph{\ul{Chapter}~\ref{chapter: paper2} \ul{(Phase-mixing in a perturbed isothermal slab)}:} In this chapter, instead of spherical galaxies, we study the relaxation of a disk galaxy modelled as a laterally homogeneous slab with a vertical isothermal profile, known as an isothermal slab. The idea is to understand some of the important features of the perturbation and relaxation of a disk galaxy in a simple setup, without resorting to the complexity of modelling a fully inhomogeneous disk. We study the phase-mixing of the response of the isothermal slab to perturbations of diverse spatio-temporal nature (e.g., encounter with a satellite galaxy), in the impulsive, resonant and adiabatic regimes. In particular, we study how the temporal nature of the perturbation dictates the dominant oscillation mode of the slab, i.e., if the slab undergoes (vertically) anti-symmetric bending mode or symmetric breathing mode oscillations. As we show, these two different modes correspond to one- and two-armed phase spirals respectively. We investigate the coarse-grained survivability of the phase-spiral, i.e., how it winds up due to vertical phase-mixing as well as how its density contrast in the phase-space damps out (in a coarse-grained sense) due to lateral mixing.

\paragraph{\ul{Chapter}~\ref{chapter: paper3} \ul{(Phase-mixing in a realistic disk galaxy)}:} In this chapter, rather than the idealized case of an isothermal slab, we perform a detailed modelling of the perturbative response of a realistic Milky Way-like disk galaxy, embedded in a dark matter halo, which for the sake of simplicity we consider to be non-responsive. We examine the nature of phase-spirals borne out of the phase-mixing of the disk response to transient spiral arm and bar perturbations as well as encounters with satellite galaxies. In particular, we, for the first time, develop a perturbative framework to compute the response of a fully inhomogeneous disk galaxy to an impacting satellite galaxy. We compare and contrast the coarse-grained survivability of phase-spirals in this case to the isothermal slab case. We also discuss the implications of collisional damping due to small scale fluctuations (e.g., scatterings of disk stars by giant molecular clouds) on the fine-grained damping of the phase spiral amplitude.

\paragraph{\ul{Chapter}~\ref{chapter: paper4} \ul{(Self-consistent perturbative treatment of dynamical friction)}:}
The previous chapters describe the perturbative response and relaxation of collisionless self-gravitating systems with different geometries. This chapter describes how the response of a perturbed host system in turn causes the secular evolution of the perturber's orbit, a process known as dynamical friction. We generalize the standard linear perturbative formalism used to compute the response of a spherical host galaxy to a point perturber on a circular orbit by perturbing the collisionless Boltzmann equation. We relax the adiabatic and secular approximations adopted in the standard theory, according to which the perturber is assumed to slowly grow and its orbit is assumed to slowly evolve. We perform a completely self-consistent treatment, i.e., compute the response using the time-evolving potential and circular frequency of an inspiraling perturber, whose radial motion is dictated by the torque computed from the response. This self-consistent perturbative formalism yields the self-consistent torque, which is a huge improvement over the standard LBK torque, since, unlike the LBK torque, it explains the origin of the different secular phenomena observed in the $N$-body simulations of cored galaxies/halos: super-Chandrasekhar dynamical friction, core-stalling and dynamical buoyancy. 

\paragraph{\ul{Chapter}~\ref{chapter: paper5} \ul{(Non-perturbative orbit-based analysis of dynamical friction)}:}
In contrast with the previous chapter which analyzes dynamical friction using linear perturbation theory, this chapter describes a novel, non-perturbative, orbit-based treatment of dynamical friction. This is motivated by the fact that non-linear perturbations in the distribution function can significantly affect the secular evolution of a massive perturber in the core region of a cored galaxy/halo, which is why linear perturbation theory is questionable in the treatment of core-stalling and dynamical buoyancy. We consider the problem of dynamical friction as a circular restricted three body problem, i.e., study the energy and angular momentum changes of field particles orbiting in the combined gravitational potential of a host galaxy and a massive perturber on a circular orbit. We identify the near-resonant orbital families that exert the strongest torque on the perturber. We find that the nature of these near-resonant orbits drastically changes as the perturber reaches a certain galactocentric radius, where the inner Lagrange points (fixed points in the co-rotating frame) undergo a bifurcation and the galaxy core is tidally disrupted by the perturber. The dynamical friction torque vanishes and the perturber stalls near this bifurcation radius, within which the torque can flip sign and become enhancing, thus exerting dynamical buoyancy.

\paragraph{\ul{Chapter}~\ref{chapter: thesis_summary} \ul{(Summary and future work)}:}
Here we summarize our findings and discuss their broader implications for galaxy formation and evolution research. We also briefly discuss some of the outstanding questions and the prospects for future investigation.

    \chapter{The Dynamics of Many body Systems}\label{chapter: many_body_dynamics}

The dynamics of many-body systems is a fascinating topic that plays out on the interface of kinetic theory, statistical mechanics and thermodynamics. How $N$-body systems form and evolve has been a matter of intense discussion and debate for a long time. Depending on the number of bodies, $N$, and the nature of interactions, the dynamics can be vastly different. A system with a very large number of particles that interact via short range forces, e.g., gas in a container, is governed by small scale collisions or two-body encounters. These collisions are ultimately responsible for equilibrating or relaxing the gas, i.e., driving it towards an equilibrium or maximum entropy state, characterized by a Maxwellian velocity distribution. This collisional equilibration happens on the two-body encounter timescale that is typically smaller for high density and low temperature systems, which therefore relax faster.

The opposite extreme of the above case is the limit of long range interactions. Self-gravitating Hamiltonian systems, including planetary systems, star-clusters, galaxies, dark matter halos, etc., are an ideal example of this kind. The behaviour of few $N$ and large $N$ self-gravitating systems is remarkably different. The simplest few $N$ self-gravitating system is a two-body system, e.g., the Sun-Earth system, where the two objects orbit each other under the central force field of gravity and move along regular conic section orbits. This system is always in perfect equilibrium (in the absence of tidal deformation of the bodies). The introduction of a third body however drastically changes the situation. The general three-body problem allows for only a few stable configurations. Most initial conditions, as proved by Poincar\'e, give rise to {\it dynamical chaos}, rendering the system non-integrable. However, the {\it hierarchical} or {\it restricted} three-body problem, where there is a strong hierarchy in the masses of the objects, can be solved perturbatively. This system allows for both regular/quasi-periodic and chaotic orbits. In fact, some orbits have commensurate oscillation frequencies, and are therefore {\it resonant}. The extent of chaos increases for more comparable masses of the three bodies. Increasing the number of objects beyond three also tends to increase the degree of chaos. In the limit of very large $N$, however, chaos is tamed and the system once again starts to harbour a larger proportion of regular orbits. Although the precise pathway to this `chaos-taming' is unclear, it is clear that this involves the restoration of symmetries in the limit of large $N$.

The presence or absence of chaos is dictated by the integrability of a Hamiltonian system (separability of Hamilton's equations of motion, which are multi-variable partial differential equations, into ordinary differential equations that can be integrated). This depends on the symmetries of the system, which in turn dictate the number of integrals of motion or conserved quantities. This decides the orbital topology, i.e., whether the system harbours regular or chaotic orbits. And this orbital topology then decides the phase-space distribution of particles. Before delving deep into the dynamics of many-body self-gravitating systems, let us first investigate under what conditions a Hamiltonian system is integrable. This will provide the foundation for understanding how many-body systems evolve and interact.

\section{Integrability of Hamiltonian systems}\label{sec:intro_H_integrability}

An autonomous (time-independent) Hamiltonian system with $n$ degrees of freedom is characterized by $n$ pairs of canonically conjugate position $(\bq)$ and momentum $(\bp)$ variables that follow Hamilton's equations of motion:

\begin{align}
\dot{q}_i = \frac{\partial H}{\partial p_i},\;\;\;\; \dot{p}_i = -\frac{\partial H}{\partial q_i},
\end{align}
where $i$ runs from $1$ to $n$, and a dot denotes a derivative with respect to time. The Hamiltonian, $H$, is given by

\begin{align}
H = \sum_{i=1}^n\frac{p^2_i}{2m} + m\,\Phi\left(\bq\right),
\label{Intro:H}
\end{align}
where $\Phi$ denotes the potential and $m$ denotes the particle mass. If the system consists of $N$ particles in $d$ dimensions, the total number of degrees of freedom is $n=Nd$. For a test particle moving in an external potential sourced by a continuous system or a discrete system with finite but large $N$, $n$ is simply equal to $d$.

A Hamiltonian system is completely integrable if and only if there exist $n$ functionally independent integrals of motion in involution, i.e., with mutually vanishing Poisson brackets. These $n$ integrals exfoliate the phase-space into $n$ hyper-surfaces, each of which confines the motion of a particle on itself; these integrals are therefore known as isolating integrals. One can define a special set of isolating integrals known as actions, denoted by $\bI=\{I_1,I_2,...,I_n\}$. An $n$-tuple of actions specifies an $n-$torus, also known as an invariant torus, along which the orbit of a particle is confined. The orbital phases are specified by $n$ angle variables, $\bw=\{\rmw_1,\rmw_2,...,\rmw_n\}$. 

Let us look at how to obtain the action-angle variables from the position-momentum variables. The canonical transformation from $(\bq,\bp)$ to canonically conjugate coordinates $(\bQ,\boldsymbol{\alpha})$, where $\boldsymbol{\alpha}$ is cyclic (conserved), is dictated by a generating function, $W(\bq,\boldsymbol{\alpha})$, known as Hamilton's principal function, such that

\begin{align}
p_i = \frac{\partial W}{\partial q_i},\;\;\;\; Q_i = \frac{\partial W}{\partial \alpha_i}.
\end{align}
$W(\bq,\boldsymbol{\alpha})$ is related to the Hamiltonian through the time-independent Hamilton-Jacobi equation,

\begin{align}
H=H\left(\bq,\bp=\frac{\partial W}{\partial \bq}\right)=H(\boldsymbol{\alpha}),
\label{Intro:Hamilton_Jacobi_eqn}
\end{align}
such that

\begin{align}
\dot{Q}_i = \frac{\partial H}{\partial \alpha_i}=\Omega_i(\boldsymbol{\alpha}), \;\;\;\; \dot{\alpha}_i = -\frac{\partial H}{\partial Q_i} = 0.
\label{Intro:Q_alpha_evolve}
\end{align}
Hence, $\alpha_i$ are constants of motion, and $Q_i=\Omega_i t + \beta_i$, where $\beta_i$ are constants. 

For a completely integrable system, $W$ can be written as a separable function of $\bq$:

\begin{align}
W(\bq,\boldsymbol{\alpha}) = \sum_{i=1}^n W_i(q_i,\boldsymbol{\alpha}),\;\;\;\; W_i = \int p_i\,\rmd q_i.
\end{align}
Hence, the different degrees of freedom get decoupled, or in other words the Hamilton-Jacobi equation~(\ref{Intro:Hamilton_Jacobi_eqn}) can be inverted to yield a quasi-periodic solution for $q_i$, i.e., $q_i=q_i(\rmw_i,\alpha_i)$, where $\rmw_i$ is the angle variable, i.e.,

\begin{align}
\rmw_i = Q_i = \Omega_i t + \beta_i.
\end{align}
The action $I_i$ corresponding to the $i^{\rm th}$ degree of freedom is defined as the area under the curve, $p_i(q_i,\alpha_i)$, i.e.,

\begin{align}
I_i(\alpha_i) = \oint p_i(q_i,\alpha_i)\, \rmd q_i. 
\end{align}
According to the Hamilton-Jacobi equation~\ref{Intro:Hamilton_Jacobi_eqn}, the Hamiltonian is expressed as a function of $\boldsymbol{\alpha}$ and therefore only the actions, i.e.,

\begin{align}
H=H(\bI).
\end{align}
From equation~\ref{Intro:Q_alpha_evolve}, it is clear that the orbital evolution of the action-angle variables is given by

\begin{align}
\dot{\rmw}_i = \Omega_i(\bI) = \frac{\partial H}{\partial I_i}, \;\;\;\; \dot{I}_i=-\frac{\partial H}{\partial \rmw_i}=0,
\end{align}
where $\Omega_i$ denotes the rate of change of the angle, $\rmw_i$, and is known as a frequency.

A completely integrable system is fully described by the $n$ action-angle variables. The orbital motion is constrained along the invariant tori, allowing only regular orbits. If there happen to exist more than $n$ isolating integrals, the system is super-integrable, in which case the motion on the invariant tori is constrained by additional integrals. The total number of isolating integrals in a $d$ dimensional spherically symmetric system, with potential $\Phi(r)$ ($r$ is the radial distance from the origin), is equal to $(2d-2)$, making it a super-integrable system for $d>2$. Let us consider the case of 3D. In this case, each orbit is confined on a plane since the angular momentum is conserved. Every orbit is specified by the magnitude of the angular momentum, $\left|\bL\right|$, its z-component, $L_z$, the radial action, $I_r$, and the longitude of the ascending node, $\rmw_\phi$, i.e., the longitude of the line where the orbital plane intersects the $x-y$ plane. The Hamiltonian, $H$, can be written as a function of $\left|\bL\right|$ and $I_r$ through the Hamilton-Jacobi equation. This system admits $4$ isolating integrals instead of $3$, which is what is required for integrability. The super-integrability of the system arises from the degeneracy of two of its frequencies and the confinement of the orbital motion on a plane, which is a consequence of spherical symmetry. 

A system with $(2n-1)$ isolating integrals is maximally super-integrable and allows only closed orbits. The Keplerian ($1/r$) and harmonic ($r^2$) potentials are special cases of spherically symmetric potentials, for which the system becomes maximally super-integrable, since it admits $(2d-1)$ isolating integrals in $d$ dimensions. In 3D, along with the Hamiltonian and the actions, the Laplace-Runge-Lenz vector serves as an additional isolating integral and specifies the orientation of an orbit. Maximal super-integrability implies that all the frequencies are degenerate, i.e., all orbits are closed.

If the number of isolating integrals of a Hamiltonian system with $2n$ degrees of freedom falls below $n$, the phase-space can no longer be globally exfoliated into invariant tori, and therefore parts of the phase-space harbour chaotic orbits. This can happen when the system lacks certain symmetries. The Hamilton-Jacobi equation is no longer separable, i.e., there exists no {\it globally} defined canonical transformation from $(\bq,\bp)$ to the action-angle variables, and the motion is no longer confined on invariant tori. For few $N$ systems, this happens whenever $N>2$. For large $N$ systems, which can be approximated by a smooth potential, the orbital motion of a particle shows chaos whenever the potential differs from the St$\ddot{\rma}$ckel family of potentials, e.g., general triaxial systems.

Partially integrable Hamiltonian systems have $m<n$ isolating integrals that are conserved globally. However, if the Hamiltonian can be expressed as a perturbative expansion about a completely integrable Hamiltonian, e.g., if a spherically symmetric system is deformed into a mildly (non-St$\ddot{\rma}$ckel) triaxial system, the system becomes near-integrable. The Kolmogorov-Arnold-M$\ddot{\rmo}$ser (KAM) theorem \citep[][]{Kolmogorov.54,Arnold.63,Moser.62} states that for small perturbations, the invariant tori of the unperturbed system, along which the frequencies are sufficiently non-resonant, are continuously deformed and therefore survive the perturbation. This means that the action integrals of the perturbed system can be expressed as analytic functions of the unperturbed actions, thereby rendering $n$ isolating integrals and making the system completely integrable {\it locally} on the surviving tori. However, the tori along which the frequencies are sufficiently close to resonances, do not survive even arbitrarily small perturbations, giving way to dynamical chaos. 

Even systems with time-dependent Hamiltonians can be deemed as near-integrable if the time-dependent part of the Hamiltonian is a small perturbation about an otherwise integrable Hamiltonian. As an example of a time-dependent near-integrable system, let us discuss the circular restricted three body problem in 3D, where we study the orbital motion of the least massive body, body $1$, under the gravitational influence of the other two bodies $2$ and $3$, that are rotating about each other in a circular orbit of frequency $\Omega_\rmP$ on a plane (with normal directed along the $z$-axis). This system conserves neither the usual Hamiltonian, $H$, nor the angular momentum, $\bL$, of body $1$, but admits another globally conserved integral, known as the Jacobi Hamiltonian, given by

\begin{align}
H_\rmJ = H - {\bf \Omega_{\rm{\bP}}} \cdot \bL.
\end{align}
If body $3$ is much less massive than body $2$, then the gravitational pull of body $3$ can be deemed as a small perturbation of the unperturbed system consisting of bodies $1$ and $2$ orbiting each other in their mutual central force field. Let the Hamiltonian of this unperturbed orbital motion of bodies $1$ and $2$ be $H_0$ and the potential due to the perturber, body $3$, be $\Phi_\rmP$. Let the actions of the unperturbed system be $I_1=I_r$, $I_2=\left|\bL\right|=L$ and $I_3=L_z$, and the corresponding unperturbed frequencies be $\Omega_1(I_1,I_2)$ and $\Omega_2(I_1,I_2)$, assuming body $2$ to be spherically symmetric, i.e., $\Omega_3=\partial H_0/\partial I_3=0$. Regions of the phase-space where the frequencies are commensurate/resonant with the perturbing frequency, $\Omega_\rmP$, i.e.,

\begin{align}
\ell_1\Omega_1(I_1,I_2)+\ell_2\Omega_2(I_1,I_2)-\ell_3\Omega_\rmP=0,
\end{align}
with $\ell_1$, $\ell_2$ and $\ell_3$ as integers, are most strongly affected by the perturbation. Surrounding each (stable) resonance, there exist a family of regular/quasi-periodic orbits trapped/librating about the resonance. For these orbits, one can define a slow action, $I_s$, and a corresponding slow angle, $w_s$, given by

\begin{align}
I_s=\frac{I_3}{\ell_3},\;\;\;\; w_s=\ell_1 w_1+\ell_2 w_2+\ell_3(w_3-\Omega_\rmP t),
\end{align}
which librate slowly about their respective values at the resonance, at a libration frequency that is much smaller than $\Omega_1$ and $\Omega_2$. The following action integrals, known as fast actions, are conserved:

\begin{align}
I_{f1}=I_1-\frac{\ell_1}{\ell_3}I_3,\;\;\;\; I_{f2}=I_2-\frac{\ell_2}{\ell_3}I_3.
\end{align}
The angles conjugate to these fast actions are fast angles, $w_{f1}=w_1$ and $w_{f2}=w_2$, that evolve with frequencies, $\Omega_1$ and $\Omega_2$, i.e., much faster than the rate at which $w_s$ and $I_s$ librate. Together with the Jacobi Hamiltonian, $H_\rmJ$, the fast actions, $I_{f1}$ and $I_{f2}$, constitute three isolating integrals, thus making the system completely integrable around the resonances. The libration zone around each resonance, also known as a resonant island, is separated from the rest of the phase-space by a `separatrix', along which the libration frequency is zero. Beyond the separatrix of a resonant island, there can be another resonant island or a chaotic island. In perturbation theory, chaos occurs whenever the libration zones of two resonances overlap, i.e., chaos is a manifestation of the {\it overlap of resonances}. The orbital dynamics of the restricted three body problem is going to be discussed in detail in chapter~\ref{chapter: paper5}.

\section{Kinetic theory of Hamiltonian systems}

Having briefly discussed the motion of a single body in a Hamiltonian system, let us now study the dynamics of a many-body Hamiltonian system as a whole. This is a sub-field of classical statistical mechanics, known as kinetic theory.

\subsection{Liouville equation}

The most general description of an $N$-body Hamiltonian system in $d$ dimensions is provided by the $N$-point distribution function (DF), $f^{(N)}(\boldsymbol{\xi}_1,\boldsymbol{\xi}_2,...,\boldsymbol{\xi}_N,t)$, where $\boldsymbol{\xi}_i=(\bq_i$, $\bp_i)$ is a $2d$-dimensional variable. The $N$-point DF is a probability density function such that the probability of $N$ particles being in the phase-space interval, $(\boldsymbol{\xi}_1+\rmd\boldsymbol{\xi}_1,\boldsymbol{\xi}_2+\rmd\boldsymbol{\xi}_2,...,\boldsymbol{\xi}_N+\rmd\boldsymbol{\xi}_N)$, is given by

\begin{align}
d^{2Nd}P = \prod_{i=1}^N \rmd^{2d} \xi_i\, f^{(N)}(\boldsymbol{\xi}_1,\boldsymbol{\xi}_2,...,\boldsymbol{\xi}_N,t).
\end{align}
The $N$-point DF is normalized such that

\begin{align}
\int \prod_{i=1}^N \rmd^{2d} \xi_i\, f^{(N)}(\boldsymbol{\xi}_1,\boldsymbol{\xi}_2,...,\boldsymbol{\xi}_N,t) = 1.
\end{align}

Due to the conservation of the differential phase-space volume, $\prod_{i=1}^N d^{2d} \xi_i$, under Hamiltonian dynamics, which is non-dissipative, the $N$-point DF follows a conservation equation, known as the Liouville equation:

\begin{align}
\frac{\rmd f^{(N)}}{\rmd t} &= \frac{\partial f^{(N)}}{\partial t} + \left[f^{(N)},H^{(N)}\right] = 0,
\label{Intro:Liouville_eq}
\end{align}
where the $N$-particle Hamiltonian, $H^{(N)}$, is given by

\begin{align}
H^{(N)} = \sum_{i=1}^N\frac{{\left|\bp_i\right|}^2}{2 m} + \sum_{i=1}^N V\left(\bq_i,t\right) + \frac{1}{2} \sum_{i=1}^N \sum_{\substack{j=1\\ j\neq i}}^N U\left(\left|\bq_i-\bq_j\right|,t\right).
\end{align}
Here $V$ is an external potential energy and $U$ is the potential energy due to pairwise inter-particle interactions. The brackets in equation~(\ref{Intro:Liouville_eq}) denote the Poisson-bracket, given by

\begin{align}
\left[Q,P\right] &= \sum_{i=1}^N \frac{\partial Q}{\partial \bq_i} \cdot \frac{\partial P}{\partial \bp_i} - \frac{\partial P}{\partial \bp_i} \cdot \frac{\partial Q}{\partial \bq_i},
\end{align}
which physically represents the net flux of probability into a phase-space element. The Liouville equation physically implies that the $N$-point DF is locally conserved. In other words, the volume occupied by a macrostate in the $2Nd$-dimensional phase-space is conserved under Hamiltonian dynamics, i.e., the flow is incompressible.

\subsection{BBGKY hierarchy}

The $N$-point DF is defined in a $2Nd$-dimensional manifold. Therefore, for large $N$ systems, solving the Liouville equation is practically an impossible task. One can however study the evolution of a lower dimensional quantity. Upon marginalizing over the phase-space coordinates of $(N-k)$ particles, one can compute the reduced $k$-point or $k$-particle DF:

\begin{align}
f^{(k)}(\boldsymbol{\xi}_1,\boldsymbol{\xi}_2,...,\boldsymbol{\xi}_k,t) = \frac{N!}{(N-k)!} \int \prod_{i=k+1}^N {\rmd}^{2d}\xi_i\, f^{(N)}(\boldsymbol{\xi}_1,\boldsymbol{\xi}_2,...,\boldsymbol{\xi}_N,t).
\end{align}
The state of an $N$-body system can be described by the $1$-particle DF, or simply DF, $f^{(1)}(\bq,\bp,t)=f(\bq,\bp,t)$, which can be obtained by marginalizing over the phase-space coordinates of $(N-1)$ particles. The number of particles in the interval, $(q_i,q_i+d q_i)$ and $(p_i,p_i+d p_i)$, where $i$ runs from $1$ to $d$, is given by

\begin{align}
\rmd^{2d} N = \rmd^d q\, \rmd^d p\, f(\bq,\bp,t).
\end{align}
Here we have dropped the subscript $1$ in the arguments of the $1$-particle DF for the sake of brevity. The number density profile of the system, $n(\bq,t)$, is equal to $\int \rmd^d p\, f(\bq,\bp,t)$, while the total number of particles, $N$, is equal to $\int \rmd^d q\, n(\bq,t) = \int \rmd^d q \int \rmd^d p\, f(\bq,\bp,t)$, which is conserved.

The evolution equation for the $k$-point DF can be obtained by integrating both sides of the Liouville equation~\ref{Intro:Liouville_eq} over the phase-space coordinates of $(N-k)$ particles, and is given by

\begin{align}
\frac{\partial f^{(k)}}{\partial t} + \left[f^{(k)},H^{(k)}\right] = \sum_{i=1}^k \int \rmd^{2d} \xi_{k+1} \frac{\partial U\left(\left|\bq_i-\bq_{k+1}\right|,t\right)}{\partial \bq_i} \cdot \frac{\partial f^{(k+1)}}{\partial \bp_i},
\end{align}
where $H^{(k)}$ is the $k$-particle Hamiltonian, given by

\begin{align}
H^{(k)} = \sum_{i=1}^k\frac{{\left|\bp_i\right|}^2}{2 m} + \sum_{i=1}^k V\left(\bq_i,t\right) + \frac{1}{2} \sum_{i=1}^k \sum_{\substack{j=1 \\ j\neq i}}^k U\left(\left|\bq_i-\bq_j\right|,t\right).
\end{align}
Note that the evolution of the $k$-point DF depends on the $(k+1)$-point DF, which leads to a hierarchy. This hierarchy of equations is known as the Bogoliubov-Born-Green-Kirkwood-Yvon hierarchy or BBGKY hierarchy. The system of equations governing the evolution of $i$-point DFs ($i$ runs from $1$ to some $k<N$) is therefore not closed. Hence, one has to truncate the BBGKY hierarchy at some order so as to obtain a closed set of equations. The philosophy behind truncation is as follows: typically, for systems in which the strength of the pairwise potential drops off fast enough with inter-particle distance, the $(k+1)$-particle DF relaxes to equilibrium much faster than the $k$-particle DF due to collisions. 

In general, the $k$-particle DF cannot be separated as a product of $k$ $1$-particle DFs, since collisions introduce correlations into the system. One can perform a Mayer cluster expansion of the $k$-particle DF, i.e., expand the $k$-particle DF in terms of the $1$-particle DF and $i$-particle correlations, with $i<k$. For example, the $2$-particle DF can be expanded as

\begin{align}
f^{(2)}(\boldsymbol{\xi}_1,\boldsymbol{\xi}_2,t) = f^{(1)}(\boldsymbol{\xi}_1,t) f^{(1)}(\boldsymbol{\xi}_2,t) + g_2(\boldsymbol{\xi}_1,\boldsymbol{\xi}_2,t),
\end{align}
where $g_2(\boldsymbol{\xi}_1,\boldsymbol{\xi}_2,t)$ is the $2$-particle correlation. Similarly, the $3$-particle DF can be expanded as

\begin{align}
f^{(3)}(\boldsymbol{\xi}_1,\boldsymbol{\xi}_2,\boldsymbol{\xi}_3,t) &= f^{(1)}(\boldsymbol{\xi}_1,t) f^{(1)}(\boldsymbol{\xi}_2,t) f^{(1)}(\boldsymbol{\xi}_3,t) \nonumber \\
& + f^{(1)}(\boldsymbol{\xi}_1,t)\, g(\boldsymbol{\xi}_2,\boldsymbol{\xi}_3,t) + f^{(1)}(\boldsymbol{\xi}_2,t)\, g(\boldsymbol{\xi}_1,\boldsymbol{\xi}_3,t) + f^{(1)}(\boldsymbol{\xi}_3,t)\, g(\boldsymbol{\xi}_1,\boldsymbol{\xi}_2,t) \nonumber\\
& + h_3(\boldsymbol{\xi}_1,\boldsymbol{\xi}_2,\boldsymbol{\xi}_3,t),
\end{align}
where $h_3(\boldsymbol{\xi}_1,\boldsymbol{\xi}_2,\boldsymbol{\xi}_3,t)$ is the $3$-particle correlation. 

\subsection{Balescu-Lenard equation}

As discussed above, the BBGKY hierarchy needs to be truncated at some order in order to obtain a closed set of equations for $k$-point DFs that can be solved. Typically, in large $N$ systems governed by short range two-body interactions, the $3$-particle correlation relaxes much faster than the $2$-particle correlation. Moreover, the probability of encounter between three or higher number of bodies is negligible compared to that of two-body encounters. This implies that the steady state $3$-particle correlation is much smaller than the steady state $2$-particle correlation, i.e., $h_3(\boldsymbol{\xi}_1,\boldsymbol{\xi}_2,\boldsymbol{\xi}_3,t)\approx 0$. This is known as the Bogoliubov ansatz, which is a valid assumption in both neutral gas and plasma (if the number of particles within a Debye length is large enough). This truncates the BBGKY hierarchy at second order. Assuming short range interactions, setting the external potential, $V$, to zero, and keeping the dominant terms, the first two equations of the BBGKY hierarchy can be manipulated to yield the following evolution equations for the $1$-point DF and the $2$-particle correlation:

\begin{align}
&\frac{\partial f^{(1)}}{\partial t} + \frac{\bp_1}{m} \cdot \frac{\partial f^{(1)}}{\partial \bq_1} -  \frac{\partial}{\partial \bq_1}U[f^{(1)},t] \cdot \frac{\partial f^{(1)}}{\partial \bp_1} = \int \rmd^{2d} \xi_2 \frac{\partial U\left(\left|\bq_1-\bq_2\right|,t\right)}{\partial \bq_1} \cdot \frac{\partial g_2}{\partial \bp_1},\nonumber \\ \\
&\frac{\partial g_2}{\partial t} + \left(\frac{\bp_1}{m}\cdot\frac{\partial}{\partial \bq_1}+\frac{\bp_2}{m}\cdot\frac{\partial}{\partial \bq_2}\right) g_2 \nonumber \\
&- \frac{\partial U\left(\left|\bq_1-\bq_2\right|,t\right)}{\partial \bq_1} \cdot \left(\frac{\partial}{\partial \bp_1}-\frac{\partial}{\partial \bp_2}\right) \left(f^{(1)}(\boldsymbol{\xi}_1,t) f^{(1)}(\boldsymbol{\xi}_2,t) +g_2(\boldsymbol{\xi}_1,\boldsymbol{\xi}_2,t)\right) \approx 0,
\label{Intro:BBGKY_f1_g2}
\end{align}
where $U[f^{(1)},t]$ is the mean field potential, given by

\begin{align}
U[f^{(1)},t] = \int \rmd^{2d}\xi_2 \, U\left(\left|\bq_1-\bq_2\right|,t\right) f^{(1)}(\bq_2,\bp_2,t).
\end{align}
One can solve the second of equations~(\ref{Intro:BBGKY_f1_g2}) for $g_2$ in terms of $f_1$ and substitute the resulting expression for $g_2$ in the first to obtain an evolution equation for the $1$-particle DF, $f^{(1)}(\boldsymbol{\xi}_1,t)=f$. In a spatially homogeneous plasma, where $f=f(\bv,t)$ with $\bv=\bp/m$, one can assume a Coulomb form for the mean field electric potential, i.e., $U=-(e^2/4\pi \epsilon_0)/\left|\bq_1-\bq_2\right|$ with $e$ the electron charge and $\epsilon_0$ the permittivity of free space. This yields the Balescu-Lenard equation governing the collisional relaxation of a plasma:

\begin{align}
\frac{\partial f}{\partial t} &= \pi\int \frac{\rmd^d k}{{\left(2\pi\right)}^d} \frac{\bk}{m} \cdot \frac{\partial}{\partial \bv} \int \rmd^d v'\, {\left|\frac{\omega^2_\rmP}{k^2\,\calD(\bk\cdot\bv,\bk)}\right|}^2 \nonumber \\
&\times \delta\left(\bk\cdot\bv-\bk\cdot\bv'\right) \frac{\bk}{m} \cdot \left(\frac{\partial}{\partial \bv} - \frac{\partial}{\partial \bv'} \right)f(\bv,t)f(\bv',t),
\end{align}
where $\omega_\rmP$ is the plasma frequency given by

\begin{align}
\omega_\rmP = \sqrt{\frac{n_e e^2}{m_e\epsilon_0}},
\label{Intro:plasma_freq}
\end{align} 
with $m_e$ the electron mass and $n_e$ the electron density. $\calD(\omega,k)$ is the {\it dielectric function} given by 

\begin{align}
\calD(\omega,k) = 1+\frac{\omega^2_\rmP}{k^2}\,\bk\cdot\int \rmd^d v\, \frac{\partial f_0/\partial \bv}{\omega - \bk \cdot \bv},
\end{align}
which physically represents charge polarization.

\subsection{Boltzmann equation}

Neutral gas is a large $N$ system governed by short range interactions. The range of interactions is in fact much smaller than the inter-particle separation. In this case, besides adopting the Bogoliubov ansatz, i.e., assuming the 3-particle correlation $h_3=0$, one can also assume that the 2-point DF, $f^{(2)}$ relaxes much faster than does the 1-point DF, $f^{(1)}$, i.e., $\partial f^{(2)}/\partial t\approx 0$. This implies that $\partial g_2/\partial t\approx 0$. Therefore, the BBGKY hierarchy is truncated at second order, which yields the following equations for $f^{(1)}$ and $f^{(2)}$:

\begin{align}
&\frac{\partial f^{(1)}}{\partial t} + \frac{\bp_1}{m} \cdot \frac{\partial f^{(1)}}{\partial \bq_1} = \int \rmd^{2d} \xi_2 \frac{\partial U\left(\left|\bq_1-\bq_2\right|,t\right)}{\partial \bq_1} \cdot \frac{\partial f^{(2)}}{\partial \bp_1},\nonumber \\
&\left(\frac{\bp_1}{m}\cdot\frac{\partial}{\partial \bq_1}+\frac{\bp_2}{m}\cdot\frac{\partial}{\partial \bq_2}\right) f^{(2)} - \frac{\partial U\left(\left|\bq_1-\bq_2\right|,t\right)}{\partial \bq_1} \cdot \left(\frac{\partial}{\partial \bp_1}-\frac{\partial}{\partial \bp_2}\right)f^{(2)} \approx 0.
\label{Intro:Boltzmann_eq_deriv_1}
\end{align}
The RHS of the first of the above equations indicates the evolution of $f^{(1)}$ due to $f^{(2)}$, i.e., due to two-body encounters or collisions. Substituting $f^{(2)}$ from the second equation in the first, one can rewrite the RHS of the first equation as

\begin{align}
C[f] &\approx \int \rmd^d p_2\, \frac{\bp_1-\bp_2}{m} \cdot\int  \rmd^d q \frac{\partial f^{(2)}}{\partial \bq},
\label{Intro:Cf1}
\end{align}
which is known as the collision functional or collision operator. Here $\bq=\bq_1-\bq_2$ is the relative position of particle 1 with respect to 2. Denoting $q_{\parallel}$ as the component of $\bq$ along the relative momentum, $\bp_1-\bp_2$ and $\bq_{\perp}$ as the vector perpendicular to it, one can integrate the above over $q_{\parallel}$ to obtain

\begin{align}
C[f] &\approx \int \rmd^d p_2\, \frac{\left|\bp_1-\bp_2\right|}{m} \nonumber \\
&\times \int  \rmd^{d-1} q_{\perp} \left[f^{(2)}(q_{\parallel}\to\infty,\bq_{\perp},\bp_1,\bp_2,t)-f^{(2)}(q_{\parallel}\to-\infty,\bq_{\perp},\bp_1,\bp_2,t)\right].
\label{Intro:Cf2}
\end{align}

At this juncture, Boltzmann adopted the assumption of molecular chaos. This entails that the momenta of any two colliding particles are uncorrelated before and after the collision, i.e.,

\begin{align}
f^{(2)}(q_{\parallel}\to-\infty,\bq_{\perp},\bp_1,\bp_2,t) &= f^{(1)}(q_{1\parallel},\bq_{1\perp},\bp_1,t) f^{(1)}(q_{2\parallel}\to \infty,\bq_{2\perp},\bp_2,t),\nonumber \\
f^{(2)}(q_{\parallel}\to\infty,\bq_{\perp},\bp_1,\bp_2,t) &= f^{(2)}(q_{\parallel}\to-\infty,\bq_{\perp},\bp'_1,\bp'_2,t) \nonumber\\
&= f^{(1)}(q_{1\parallel},\bq_{1\perp},\bp'_1,t) f^{(1)}(q_{2\parallel}\to \infty,\bq_{2\perp},\bp'_2,t),
\label{Intro:molecular_chaos}
\end{align}
where $\bp'_1$ and $\bp'_2$ are respectively the momenta of particles 1 and 2 after collision. Here we assume elastic collisions: the total linear momentum as well as the total kinetic energy is conserved, so that $\bp_1+\bp_2=\bp'_1+\bp'_2$ and ${\left|\bp_1\right|}^2+{\left|\bp_2\right|}^2={\left|\bp'_1\right|}^2+{\left|\bp'_2\right|}^2$. Even though the momenta of the two particles become correlated immediately after collision, this correlation is lost shortly afterwards due to frequent encounters with many other particles. The particle trajectories are therefore highly chaotic. The molecular chaos assumption is at the heart of the Boltzmann H theorem or the second law of thermodynamics, i.e., how chaotic motion gives rise to macroscopic irreversibility out of the microscopically reversible dynamics of molecules. 

Besides the molecular chaos assumption, we further assume that the interactions are short range, i.e., take $\bq_{1\perp}\approx \bq_{2\perp}$. Using the expression for $f^{(2)}$ in terms of $f^{(1)}$ from equations~(\ref{Intro:molecular_chaos}) in equation~(\ref{Intro:Cf2}), rewriting $\rmd^{d-1} q_{\perp}$ as $\rmd\Omega (\rmd\sigma/\rmd\Omega)$, where $\rmd\sigma/\rmd\Omega$ is the differential cross-section of interactions ($\Omega$ is the solid angle), and substituting these in equation~(\ref{Intro:Boltzmann_eq_deriv_1}), we obtain the Boltzmann equation:

\begin{align}
\frac{\rmd f}{\rmd t} &= \frac{\partial f}{\partial t} + \frac{\bp}{m} \cdot \frac{\partial f}{\partial \bq} = C[f],
\label{Intro:Boltzmann_equation}
\end{align}
where we have set $f^{(1)}=f$ and have dropped the subscript 1 from $\bp$ and $\bq$ for the sake of brevity. The collision functional or collision operator, $C[f]$, denotes the rate of change of the DF due to inter-particle collisions, and is given by

\begin{align}
C[f] &= \int \rmd^d p_2 \int \rmd\Omega\, \frac{\rmd \sigma}{\rmd \Omega}\, \frac{\left|\bp-\bp_2\right|}{m}\left[f(\bq,\bp',t)f(\bq,\bp'_2,t)-f(\bq,\bp,t)f(\bq,\bp_2,t)\right].
\label{Intro:Cf}
\end{align}
The two terms of the collision operator respectively account for the influx and outflux of particles into and out of a phase-space volume centered on $(\bq,\bp)$ due to the collisional exchange of particles with other phase-space volumes. Note that the term corresponding to the mean field potential has dropped out in the LHS of equation~(\ref{Intro:Boltzmann_equation}) due to the assumption of short range interactions.

\subsection{Collisionless Boltzmann equation}

For a collisionless system, one can neglect the two point correlation in the RHS of the first of equations~(\ref{Intro:BBGKY_f1_g2}), which therefore becomes the collisionless Boltzmann equation (CBE) or Vlasov equation. For inhomogeneous collisionless systems, this can be written as

\begin{align}
&\frac{\rmd f}{\rmd t} = \frac{\partial f}{\partial t} + \left[f,H\right] = 0\nonumber \\
&\implies \frac{\partial f}{\partial t} + \frac{\bp}{m} \cdot \frac{\partial f}{\partial \bq} - \frac{\partial}{\partial \bq}U[f,t] \cdot \frac{\partial f}{\partial \bp} = 0,
\label{Intro:Vlasov}
\end{align}
with the Hamiltonian $H$ given by

\begin{align}
H=\frac{p^2}{2m} + U[f,t],
\end{align}
and the mean field potential $U[f,t]$ is given by

\begin{align}
U[f,t] = \int \rmd^{2d}\xi_2 \, U\left(\left|\bq-\bq_2\right|,t\right) f(\bq_2,\bp_2,t).
\label{Intro:mean_field_Vlasov}
\end{align}
This physically implies that, in absence of collisions, the DF, $f$, of particles is conserved under the Hamiltonian flow. In other words, the evolution of the DF is incompressible for a collisionless system. The dynamics of each particle is governed by the mean field of all the other particles.

\subsection{Fokker-Planck equation}

The Boltzmann equation is an integro-differential equation and is therefore very difficult to solve. Although the LHS is linear in $f$, the collision operator on the RHS is non-linear, which adds to the complexity. To get over these complications, one assumes that the collisional relaxation is dominated by weak encounters. Under this assumption, $f(\bq,\bp',t)$ in the collision operator given in equation~(\ref{Intro:Cf}) can be Taylor expanded about $f(\bq,\bp,t)$, while $f(\bq,\bp'_2,t)$ is approximated as the equilibrium DF, $f_0(\bq,\bp'_2)$, which is typically Maxwellian for a collisional system. The Taylor series is then truncated at second order to yield the Fokker-Planck equation, which is nothing but the Boltzmann equation with the collision operator approximated as

\begin{align}
C[f] &\approx \frac{1}{2} \frac{\partial}{\partial p_i} \left(D_{ij} \frac{\partial f}{\partial p_j} \right) = \frac{1}{2}\frac{\partial}{\partial p_i} \left[-D_i f + \frac{\partial}{\partial p_j}\left(D_{ij} f\right) \right],
\label{Intro:Cf_Fokker_Planck}
\end{align}
where the rank $2$ diffusion tensor, $D_{ij}$, is given by

\begin{align}
D_{ij}(\bq,\bp) &= \int \rmd^d p_2 \int \rmd\Omega\, \frac{\rmd \sigma}{\rmd \Omega}\, \frac{\left|\bp-\bp_2\right|}{m} \left(p'_i-p_i\right) \left(p'_j-p_j\right) f_0(\bq,\bp_2),
\end{align}
and the rank $1$ diffusion tensor, $D_i$, can be expressed in terms of $D_{ij}$ as

\begin{align}
D_i=\frac{\partial D_{ij}}{\partial p_j}.
\end{align}
The first term in the collision operator given in equation~(\ref{Intro:Cf_Fokker_Planck}) represents the gradual drift of the peak of the DF due to collisional damping while the second term indicates the widening of the DF due to diffusion. For highly collisional systems, the steady state solution of the Fokker-Planck equation is generally a Maxwellian DF, with velocity dispersion proportional to the diagonal terms of $D_{ij}$. While both Fokker-Planck and Balescu-Lenard equations describe the collisional relaxation of many body systems, it is important to note the following fundamental difference between the two: the Fokker-Planck equation is strictly valid for weak encounters whereas the Balescu-Lenard equation takes into account the contribution from both strong and weak encounters, i.e., both large and small angle deflections. 

The Boltzmann and Fokker-Planck equations are evolution equations for the DF, which is a function of $2d+1$ variables. The high dimensionality of the problem makes it computationally expensive. Therefore, one often takes moments of the Boltzmann equation, which turn out to be the evolution equations for the zero-th moment or density, first moment or velocity, second moment or energy, and so on. The zero-th moment equation is known as the continuity equation, while the first moment equations are known as Navier-Stokes equations in case of collisional systems like fluids, and Jeans equations in case of collisionless systems. Fluids quickly relax to a Maxwellian DF and therefore possess an equation of state (EOS) that relates pressure, density and temperature, thus rendering a closed set of moment equations that can be solved. The velocity dispersion tensor in the Jeans equations generally cannot be expressed as a function of density since the DF in the end state of collisionless relaxation can substantially differ from a Maxwellian form. Thus the moment equations for collisionless systems cannot be closed. To solve them one has to impose additional constraints based on the symmetries of the problem. Therefore, while fluids can be adequately studied using moment equations, collisionless and weakly collisional systems are better studied using the collisionless Boltzmann/Vlasov and Fokker-Planck equations respectively. 

The assumption of short range interactions, which is a crucial ingredient in the derivation of the Boltzmann and Fokker-Planck equations, is valid when the impact parameter, $b$, of most collisions is much smaller than the inter-particle distance, $r_0$. This is only strictly true for collisional systems such as neutral gas, where short range van der Waals forces drive the collisions. However, in the case of self-gravitating systems that are governed by long range gravitational forces, $b\gtrsim r_0$ for many collisions. This introduces a mean field term, given by equation~(\ref{Intro:mean_field_Vlasov}), in the LHS of the Fokker-Planck equation, just as in the LHS of the Vlasov equation. Also, self-gravitating systems are inhomogeneous, and are typically characterized by quasi-periodic orbits of field particles. Hence, the dominant encounters are resonant in nature, which cannot be treated by assuming the interactions to be local. This significantly complicates the treatment of collisionality in self-gravitating systems. Recent studies have developed a Balescu-Lenard formalism for self-gravitating systems in the spirit of plasma physics calculations but in action-angle variables \citep[][]{Heyvaerts.10,Heyvaerts.etal.17,Fouvry.etal.21}. Galaxies are to very good approximation collisionless and can therefore be well described by the collisionless Boltzmann or Vlasov equation. On the other hand, the dynamics of globular clusters and nuclear star clusters is driven by collisions, especially in the central dense regions, where a Balescu-Lenard equation or a Fokker-Planck equation is more appropriate.

\subsection{Steady state}

The steady-state DF of a system is intricately related to the integrability of the Hamiltonian governing the dynamics. In steady state, the Hamiltonian is time-independent, and therefore, $f=f_0$ is also time-independent, i.e., $\partial f/\partial t=0$. Hence, the steady state solution, $f_0$, of the Vlasov equation (equation~[\ref{Intro:Vlasov}]), is given by

\begin{align}
\left[f_0,H\right] = 0.
\end{align}
This is reminiscent of the conserved quantities or action integrals, $I_i$, that commute with the Hamiltonian:

\begin{align}
\frac{\rmd I_i}{\rmd t} = \left[I_i,H\right] = 0,
\end{align}
where $i$ runs from $1$ to $m$, the total number of conserved quantities. Hence, the steady state DF, $f_0$, can be expressed as a function of the conserved quantities:

\begin{align}
f_0=f_0(I_1,I_2,...,I_m).
\end{align}
This is the statement of the strong Jeans' theorem. The higher the number of isolating integrals of a system, the more constrained is the phase-space distribution of particles, i.e., the less ergodic is the system. The DF of a maximally ergodic system is only a function of the Hamiltonian, e.g., a system with a spherically symmetric density profile that is also isotropic in velocities. Anisotropy in the velocity space on the other hand makes the DF a function of $\left|\bL\right|$ as well. Spherical symmetry only allows the DF to be a function of $H$ and $\left|\bL\right|$. But, if the density profile becomes axisymmetric with respect to the $z$-axis, then only the z-component of the angular momentum, $L_z$, and not $\bL$ as a whole is conserved. In this case, the DF can be expressed as a function of $H$ and $L_z$. Since an axisymmetric system often has a third integral of motion, $I_3$, which is approximately the $z$ action, $I_z$, its DF can be a function of $I_3$ as well. 

The strong Jeans theorem tells us that the DF of a steady state N-body system depends only on the action integrals. The exact functional form of the steady state DF however depends on the degree of collisionality. Highly collisional systems relax to a Maxwellian DF,

\begin{align}
f_0 &\propto \exp{\left[-\beta H\right]},
\end{align}
over the two body relaxation timescale, where $\beta = 1/k_\rmB T$, with $T$ the temperature of the system and $k_\rmB$ the Boltzmann constant. On the other hand, the functional form of the steady state DF of a collisionless system is not unique and generally depends on initial conditions. 

\section{Relaxation of collisionless systems}

\subsection{Vlasov-Poisson equations}

The collision operator, $C[f]$, is zero for a collisionless system, and therefore any relaxation or equilibration of the system is not collision-driven. How then does a collisionless system relax? To get to the bottom of this, one has to consider not only the evolution of the DF, which is dictated by the collisionless Boltzmann equation or Vlasov equation, but also that of the potential, which, for a self-gravitating system, is a combination of two terms: an external perturbing potential, $\Phi_\rmP$, and the self-potential, that is sourced by the evolving DF itself through the Poisson equation. Hence, the relaxation of any self-gravitating collisionless system is governed by the following Vlasov-Poisson system of equations:

\begin{align}
&\frac{\partial f}{\partial t} + \left[f,H\right] = 0, \nonumber \\
&\nabla^2 \Phi = 4\pi G\, m \int \rmd^d p \, f.
\label{Intro:CBE_Poisson}
\end{align}
Here, the (time-dependent) Hamiltonian is given by

\begin{align}
H = \sum_{i=1}^d\frac{p^2_i}{2m} + m\left[\Phi\left(\bq,t\right) + \Phi_\rmP\left(\bq,t\right)\right].
\end{align}
The Vlasov-Poisson system consists of coupled, non-linear, integro-differential equations, and is therefore extremely difficult to solve in its most general form, either analytically or numerically. Therefore they are typically solved under certain simplifying assumptions.

\paragraph{\ul{Numerical N-body simulations}}
The most commonly used numerical technique adopted to `solve' the Vlasov-Poisson system is an $N$-body simulation, where the particles sampled from some initial DF are allowed to interact gravitationally between themselves, and Hamilton's (equivalently Newton's) equations of motion are integrated to yield their positions and momenta as a function of time. This is a Lagrangian method of modelling the relaxation of an $N$-body system, where instead of evolving $f(\bq,\bp,t)$, one evolves the $\bq$ and $\bp$ of each particle. In absence of particle-particle collisions, one can use the Vlasov equation to obtain

\begin{align}
f\left(\left\{\bq(t),\bp(t)\right\}\right) = f_i\left(\left\{\bq_i,\bp_i\right\}\right).
\end{align}
Here the subscript $i$ stands for initial conditions. The final phase-space coordinates, $(\bq^j(t),\bp^j(t))$, of the $j^{\rm th}$ particle are expressed in terms of the initial ones, $(\bq^j_i,\bp^j_i)$, as follows:

\begin{align}
\bq^j(t) &= \bq^j_i + \frac{1}{m} \int_{t_i}^t\,\rmd t'\, \bp^j(t'), \nonumber \\
\bp^j(t) &= \bp^j_i - m \int_{t_i}^t\,\rmd t'\, \nabla_j \left[\Phi(\left\{\bq(t')\right\})+\Phi_\rmP(\left\{\bq(t')\right\})\right],
\label{Intro:q_p_vs_t}
\end{align}
where the self-potential, $\Phi\left(\left\{\bq(t)\right\}\right)$, obtained by integrating the Poisson equation, is given (in $3$ dimensions) as follows:

\begin{align}
\Phi(\left\{\bq(t)\right\}) = - G\, m \sum_{j=1}^N \sum_{\substack{k=1\\k<j}}^N \frac{1}{{\left|\bq_j(t)-\bq_k(t)\right|}}.
\end{align}
An $N$-body simulation solves equations~(\ref{Intro:q_p_vs_t}) using finite difference techniques, i.e., by discretizing time. Different integration algorithms are implemented, of which symplectic (Hamiltonian conserving) algorithms such as the second order leap frog integrator are some of the most widely used ones. In order to increase computational efficiency, the gravitational force on each particle is usually computed using tree algorithms such as the \cite{Barnes.Hut.86} {\tt treecode}.

Although an N-body simulation (exactly) solves the Vlasov-Poisson system, it comes with its own set of challenges. In order to preserve the collisionless nature of the system, i.e., to avoid artificial two-body relaxation, the particles have to be `softened' or represented as objects with extended density profiles instead of point objects. The optimum softening radius, $\varepsilon$, required to ensure that a system initialized in equilibrium remains so, is a function of the number of particles, $N$. Typically, the optimal $\varepsilon$ decreases with $N$. The dynamics in the very central regions of galaxies is highly susceptible to the softening protocol adopted. In fact, softening always introduces artificial cores within a few $\varepsilon$ from the center, whereas a perfectly collisionless system evolved with cosmological initial conditions tends to harbor a central cusp. With higher resolution, i.e., larger $N$, the optimal $\varepsilon$ decreases, and therefore the size of the central artificial core is also diminished. While $N$-body simulations are useful tools to study the relaxation of $N$-body systems, their finite resolution issues and computational complexity call for analytical methods that can provide valuable physical insight. One such analytical technique is perturbation theory.

\paragraph{\ul{Perturbation theory}} 

The Vlasov-Poisson system is a set of coupled non-linear equations that cannot be easily decoupled. It is therefore what is termed as a `hard problem', one that is difficult to solve in its most general form. But, if the deviations of $f$ and $\Phi$ from the equilibrium values, $f_0$ and $\Phi_0$, are small, one can analytically solve the Vlasov-Poisson system using what is known as perturbation theory. If the perturber potential, $\Phi_\rmP$, is much weaker than the unperturbed galaxy potential, $\Phi_0$, it perturbs $f$ and $\Phi$ only slightly from equilibrium. In this case, one can expand $f$ and $\Phi$ as a perturbation series:

\begin{align}
f &= f_0 + f_1 + f_2 +... \nonumber \\
\Phi &= \Phi_0 + \Phi_1 + \Phi_2 +...
\end{align}
The perturber potential, $\Phi_\rmP$, is considered as an O(1) perturbation. This perturbative expansion assumes of course that the series converges, which is the case when $f_{i+1}<f_i$ and $\Phi_{i+1}<\Phi_i$ for $i\geq 0$. For sufficiently non-linear perturbations, $f_{i+1}\sim f_i$ and $\Phi_{i+1}\sim \Phi_i$ and therefore the series can diverge, in which case one has to resort to non-perturbative techniques.

Substituting the above perturbation series in the Vlasov-Poisson system, one can sort together terms of equal order and obtain the following recursive set of equations:

\begin{align}
&\frac{\partial f_i}{\partial t} + [f_i,H_0] + [f_{i-1},\Phi_\rmP] + \sum_{j=1}^{i-1} [f_{i-j},\Phi_j] = 0, \nonumber \\
&\nabla^2 \Phi_i = 4\pi G \, m \int \rmd^d p \, f_i ,
\label{Intro:CBE_Poisson_perturb}
\end{align}
where $i\geq 1$. $f_i$ is known as the $i^{\rm th}$ order response of the system to the perturbation, while $\Phi_i$ is the potential perturbation sourced by $f_i$ and accounts for the self-gravity of the response. The unperturbed Hamiltonian is given by

\begin{align}
H_0 = \sum_{i=1}^d\frac{p^2_i}{2m} + m\,\Phi_0\left(\bq,t\right).
\end{align}
One has to solve equations~(\ref{Intro:CBE_Poisson_perturb}) order by order to obtain $f_i(\bq,\bp,t)$ and $\Phi_i(\bq,t)$ in terms of $f_0$ and $\Phi_0$. The bound orbits in an inhomogeneous, self-gravitating $N$-body system governed by an integrable Hamiltonian are characterized by action-angle variables, as discussed in section~\ref{sec:intro_H_integrability}, and hence these are the natural coordinates for solving the Vlasov equation. Both $H_0$ and $f_0$ are functions of only actions. Therefore, in terms of action-angle variables, the Poisson bracket, $[f_i,H_0]$, can be simplified as $[f_i,H_0]=\sum_j (\partial f_i/\partial \rmw_j)(\partial H_0/\partial I_j)$, while $[f_0,\Phi_1]$ and $[f_0,\Phi_\rmP]$ can be simply written as $[f_0,\Phi_1]= - \sum_j (\partial f_0/\partial I_j)(\partial \Phi_1/\partial \rmw_j)$ and $[f_0,\Phi_\rmP]= - \sum_j (\partial f_0/\partial I_j)(\partial \Phi_\rmP/\partial \rmw_j)$ respectively. Further simplification occurs if one performs a discrete Fourier transform of equations~(\ref{Intro:CBE_Poisson_perturb}) in the angle variables, since this transforms the angle derivatives to simple algebraic expressions in terms of the Fourier mode numbers and equations~(\ref{Intro:CBE_Poisson_perturb}) into evolution equations for the {\it Fourier modes} of $f_i$ and $\Phi_i$ that are much easier to integrate. There is however one complication. The Laplacian operator of the Poisson equation couples the action and angle derivatives in a non-trivial way. This complication can be overcome by implementing two techniques. Firstly, one has to adopt the Kalnajs matrix method \citep[][]{Binney.Tremaine.08,Kalnajs.77}, wherein one expands $\Phi_i$ and $\rho_i=m\int \rmd^d\, f_i$ in terms of a bi-orthogonal basis of functions that satisfy the Poisson equation with the same boundary conditions as the problem. Secondly, one has to perform a Laplace transform (in time) of the first of equations~(\ref{Intro:CBE_Poisson_perturb}), which transforms into a matrix equation in the bi-orthogonal basis. 

Implementing the full machinery of perturbation theory is not a trivial task. Therefore, perturbative analyses are usually performed under some simplifying approximations: 

\begin{itemize}
    \item Often, the self-gravity of the response is ignored, taking the Poisson equation out of the picture and rendering a bi-orthogonal basis expansion unnecessary.

    \item When $\Phi_\rmP$ is small, one may assume that the linear order response of the system dominates over the higher order ones, and can therefore solve the perturbed Vlasov equation (first of equations~[\ref{Intro:CBE_Poisson_perturb}]) at linear order to obtain $f_1$. This essentially boils down to solving a forced oscillator equation, where the stars oscillating at their natural frequencies, $\Omega_1$, $\Omega_2$,..., $\Omega_d$, are forced by an external time-varying potential, $\Phi_\rmP$. If $\Phi_\rmP$ is a periodic function of time with frequency $\Omega_\rmP$, the ${\boldsymbol {\ell}}$ mode response of stars is the strongest near resonances, i.e., when $\sum_{i=1}^d \ell_i\Omega_i - \ell_3\Omega_\rmP = 0$. The first order response oscillates with frequencies, $\ell_i\Omega_i$ ($i$ runs from $1$ to $d$), which are functions of actions, and therefore eventually phase-mixes away in the coarse-grained sense, i.e., when integrated over actions. Any persistent response of the system appears only at second or higher order.
\end{itemize}

\subsection{Mechanisms of collisionless relaxation}

Relaxation of collisionless systems, including self-gravitating $N$-body systems such as galaxies and cold dark matter halos, is fundamentally different from that of collisional systems such as cold, dense gas or plasma. Relaxation in collisional systems is primarily driven by two-body collisions, which drive the DF towards Maxwellian. This collisional diffusion manifests as viscosity in the Navier-Stokes equations (first moment equations of the Boltzmann equation) and as conductivity in the energy equation (second moment equation). Collisionless systems, however, relax through very different mechanisms that can engender a substantially non-Maxwellian DF in steady state. These are:

\begin{itemize}[noitemsep]
    \item Phase-mixing
    \item Landau damping
    \item Violent relaxation
\end{itemize}

\subsubsection{Relaxation in the linear regime: phase-mixing and Landau damping}

The origin of the above relaxation phenomena can be understood from the different terms of equations~(\ref{Intro:CBE_Poisson_perturb}). The linearized form of the perturbed CBE and Poisson equations is given by

\begin{align}
&\frac{\partial f_1}{\partial t} + [f_1,H_0] + [f_0,\Phi_\rmP] + [f_0,\Phi_1] = 0,\nonumber \\
&\nabla^2\Phi_1 = 4\pi G\, m \int \rmd^d p\,f_1.
\label{Intro:CBE_Poisson_linear_perturb}
\end{align}
We can canonically transform to the angle-action variables (\bw,\,\bI), and expand $f_1$, $\Phi_1$ and $\Phi_\rmP$ as the following Fourier series in angles:

\begin{align}
&f_1(\bw,\bI,t) = \sum_{\boldell}e^{i\boldell\cdot\bw} f_{1\boldell}(\bI,t),\nonumber \\
&\Phi_1(\bw,\bI,t) = \sum_{\boldell}e^{i\boldell\cdot\bw} \Phi_{1\boldell}(\bI,t),\;\;\;\; \Phi_\rmP(\bw,\bI,t) = \sum_{\boldell}e^{i\boldell\cdot\bw} \Phi_{\boldell}(\bI,t).
\end{align}
Substituting these in equations~(\ref{Intro:CBE_Poisson_linear_perturb}), using the mathematical machinery of bi-orthogonal basis functions detailed in \cite{Weinberg.89}, and assuming that $f_{1\boldell}(\bI,0)=0$, we obtain the following general form for $f_{1\boldell}$:

\begin{align}
f_{1\boldell}(\bI,t) &= \, i\,\boldell\cdot \frac{\partial f_0}{\partial \bI} \left[ \int_0^t \rmd\tau\,\exp{\left[-i\boldell\cdot \bf{\Omega}(\bI)\tau \right]}\, A_{\boldell}(\bI,t-\tau) \right. \nonumber \\
&+ \left. \sum_{n=1}^\infty \int_0^t \rmd\tau\,\exp{\left[\left(\gamma_n+i\omega_n\right)\tau\right]}\, B_{\boldell,n}(\bI,t-\tau) \right].
\label{Intro:linear_response}
\end{align}
In the limit of weak self-gravity of the response, $A_{\boldell}\to \Phi_{\boldell}$, and $B_{\boldell,n}$, which is a function of $\Phi_{\boldell}$, goes to zero. The first term arises from the forcing of stars oscillating at frequencies, $\bf{\Omega}$, by the perturbing potential, $\Phi_\rmP$. The oscillation frequencies are functions of actions. Therefore, stars with different actions get out of phase with each other within a few dynamical times. When integrated over a given range of actions, the first part of the response therefore undergoes phase-mixing and damps away (in a coarse-grained sense). This gives rise to spiral-shaped over- and under-densities in the phase-space distribution of particles known as phase-space spirals or phase spirals, which get more tightly wrapped over time due to phase-mixing. The topic of phase-mixing and phase spirals is going to be discussed in detail in chapters~\ref{chapter: paper2} and \ref{chapter: paper3}.

The second term of equation~(\ref{Intro:linear_response}) arises from the self-gravity of the response, and represents the coherent motion of the entire system at discrete frequencies, $\omega_n$, which exponentially damps or grows at rates, $\gamma_n$, while being forced by $\Phi_\rmP$. These discrete oscillation modes of the system are known as point modes or Landau modes. Depending on the geometry of the system and the nature of the unperturbed DF, $\gamma_n$ can be negative or positive, representing a decaying/stable or a growing/unstable point mode. Note that this damping or growth of the response occurs on a fine-grained level, unlike phase-mixing which only damps out the response on a coarse-grained level. The origin of the phenomena of collisionless damping, known as Landau damping \citep[][]{Landau.46}, and instability, can be understood as follows. In a collisional fluid, pressure and gravity act as opposing forces. In a collisionless system, the non-zero velocity dispersion plays the role of pressure and counteracts gravitational collapse. Only, unlike fluids, the velocity dispersion is not uniquely related to the density through an EOS, since collisional equilibration takes very long to occur. If the velocity dispersion, $\sigma$, is sufficiently small, then the free-streaming rate falls below the rate of gravitational collapse, and particles tend to accumulate in regions of phase-space with higher $f_1$, and consequently higher $\Phi_1$, implying the existence of unstable point modes. For large enough $\sigma$, stability depends on the nature of $f_0$. For spherical, isotropic systems, all point modes are stable for large $\sigma$ if $\partial f_0/\partial H_0<0$. This is because, for such $f_0$, more particles have energies slightly smaller than the mode energy, $E_n$, as opposed to larger than $E_n$. Hence, more particles gain energy from the mode (via mutual gravitational interactions) as opposed to losing energy to it. In other words, more particles stream from $E<E_n$ to $E>E_n$ than the other way around. As a result, the mode loses energy to the random motion of the particles and the modal response damps away. On the other hand, if $\partial f_0/\partial H_0>0$, the point modes become unstable even for large $\sigma$, since now the mode gains energy from the particles. Hence, Landau damping or instability, which can be thought of in terns of mode-particle (or wave-particle) interaction, is ultimately the outcome of a tug-of-war between free streaming and self-gravity. 

\subsubsection{Self-gravitating systems vs fluids and plasma}

It is important to discuss in some detail how the relaxation of collisionless self-gravitating systems contrasts with that of other many-body systems such as plasma and collisional systems or fluids. Short range two body encounters drive a fluid towards a {\it local thermal equilibrium} (LTE) characterized by a Maxwellian DF with roughly the same temperature {\it locally} within several mean free paths of a point. LTE, or in other words, a local equipartition of kinetic energy, is established as long as the interactions are short range, i.e., the mean free path is much shorter than the mean particle separation in a fluid. This gives rise to an equation of state (EOS) in a fluid, which directly relates its local pressure to its local density and temperature or density and entropy. The establishment of a global thermal equilibrium, i.e., constant temperature throughout a system, occurs via the random Brownian motion of particles, which is a diffusive process and takes time to equilibrate the entire system. This manifests as viscosity that tries to nullify the macroscopic velocity gradient or shear and as conductivity that tries to erase the temperature gradient. In systems governed by long range forces, e.g., self-gravitating systems and plasma, the mean free path generally exceeds the mean particle separation, i.e., such systems are collisionless. Local equipartition or LTE is not achieved in such systems, thus rendering no EOS. We shall see shortly that the relaxation of these systems occurs via collective processes rather than two body interactions.

\paragraph{\ul{Collisional systems or fluids}}

Let us first take a look at how perturbed fluids relax. There are two relevant timescales in this case: (i) the two body relaxation timescale and (ii) the timescale on which macroscopic perturbations evolve. In collisional systems like fluids, the two body relaxation timescale is much shorter than the timescale of macroscopic evolution, i.e., one can assume that LTE and therefore an EOS is well established as long as one studies the macroscopic dynamics of the system. This can be treated by simultaneously solving the moment equations of the Boltzmann equation, i.e., the continuity and Euler equations, as well as the Poisson equation, which relates the density and potential for any system governed by Coulomb forces and hence is valid for both plasma and self-gravitating fluids. In a macroscopically homogeneous self-gravitating system, one can assume small perturbations of the density $\rho(\bx,t)$, velocity $\bu(\bx,t)$ and potential $\Phi(\bx,t)$ of the fluid about their equilibrium values, and linearize the continuity, Euler and Poisson equations to obtain wave equations for the perturbations. The wave-like perturbations of a wavenumber $k$ and frequency $\omega$ follow the dispersion relation \citep[][]{Binney.Tremaine.87},

\begin{align}
\omega^2=c^2_s \left(k^2 - k^2_\rmJ\right),
\label{Intro:disp_rel_fluid_sg}
\end{align}
where $k_\rmJ=\sqrt{4\pi G \rho_0/c^2_s}$ is the Jeans wavenumber and $c_s=\sqrt{\partial p/\partial \rho}$ is the sound speed ($p=\rho \sigma^2$ is the pressure), which is proportional to the velocity dispersion, $\sigma$, along each direction. This dispersion relation is valid as long as an EOS relates the pressure and density of a fluid, which occurs whenever the two body relaxation time is much shorter than the typical sound crossing time of the system. For an isothermal EOS with constant temperature, i.e., constant $\sigma$, $c_s$ is simply equal to $\sigma$. On the other hand, for an isentropic EOS with $p\sim \rho^\gamma$, $c_s$ is equal to $\sqrt{\gamma}\sigma$. 

\begin{figure}
\centering
\subfloat{\includegraphics[width=0.7\textwidth]{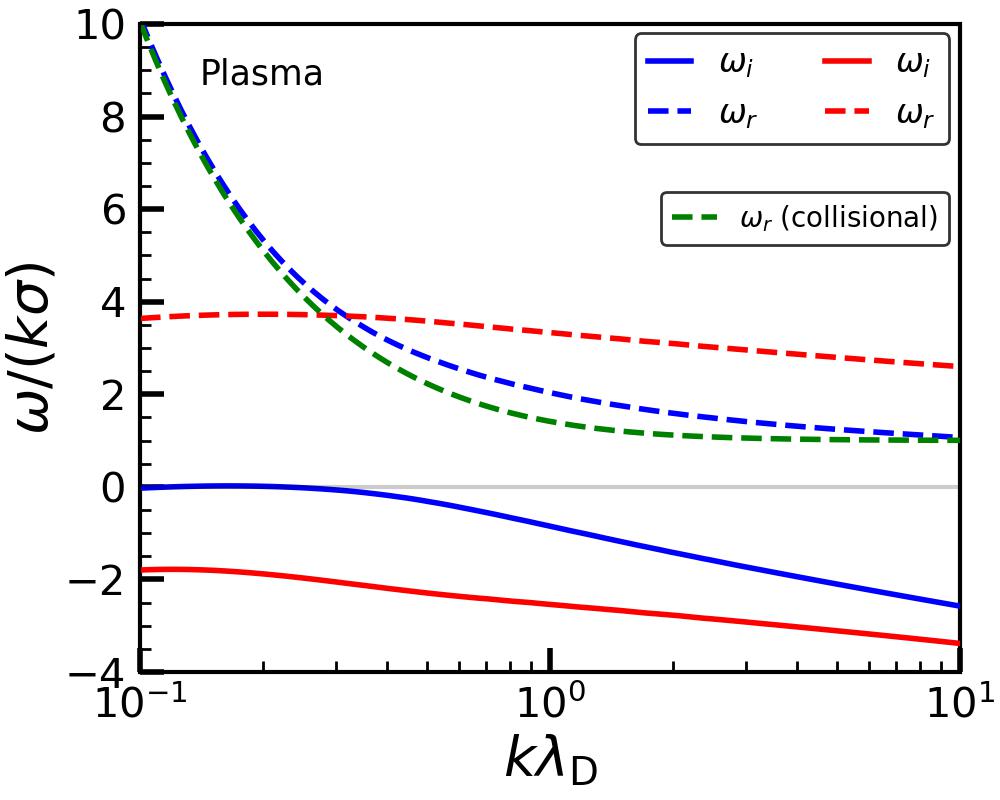}\label{plasma_modes}}\\
\subfloat{\includegraphics[width=0.7\textwidth]{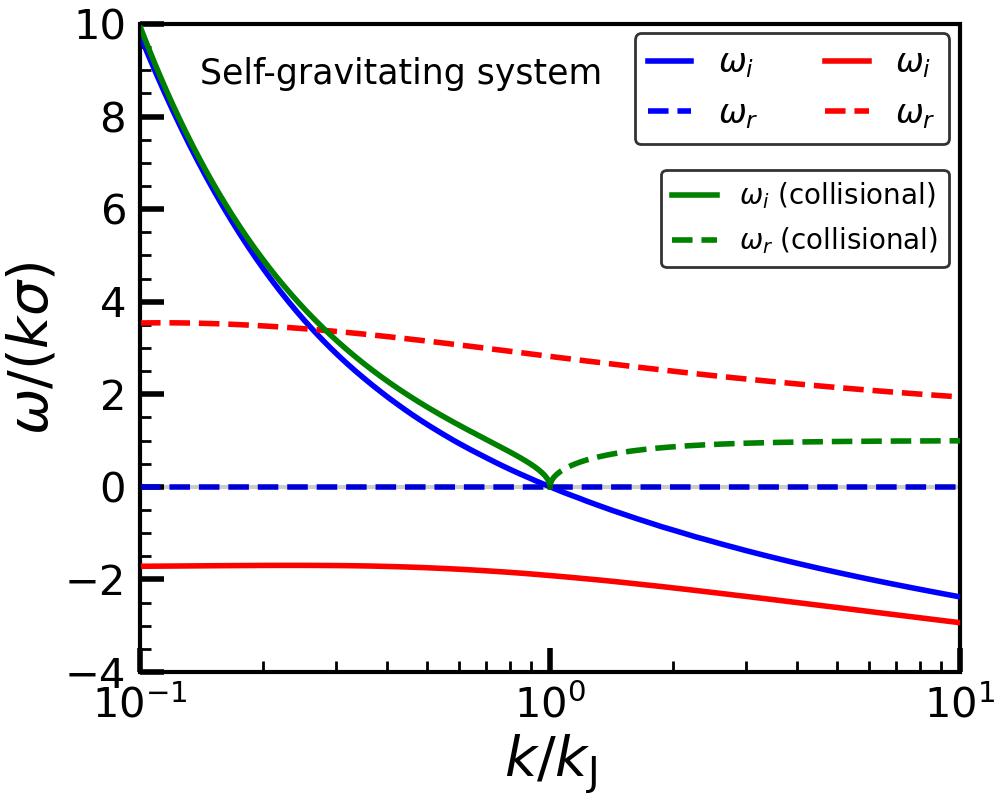}\label{self_grav_modes}}
\caption{Top panel shows the real and imaginary parts of the frequencies (in units of $k\sigma$) of perturbations in a homogeneous plasma with velocity dispersion $\sigma$ as a function of the wavenumber $k$ (in units of $1/\lambda_\rmD$). Bottom panel shows the same as a function of $k$ (in units of $k_\rmJ$) for a self-gravitating systems. Solid and dashed lines indicate the imaginary and real parts of frequencies respectively, while the blue and red colors denote two modes with the lowest damping rates. For comparison, the case of collisional systems or fluids is indicated in both cases by green lines. In the self-gravitating case, the blue mode is non-oscillatory ($\omega_r=0$) and damped for $k>k_\rmJ$ but growing for $k<k_\rmJ$. The red mode is damped and oscillating throughout. In a plasma, both modes are oscillating but damped. While the red mode behaves similarly to that in the self-gravitating case, the blue mode is very different. In the case of a plasma, it is strongly damped at small scales ($k\lambda_\rmD>1$), but very weakly damped at large scales ($k\lambda_\rmD<1$), where it becomes a long-lived oscillatory mode, known as a Langmuir mode, that oscillates at nearly the plasma frequency, $\omega_\rmP$.}
\label{fig:self_grav_plasma_resp}
\end{figure}

In the above dispersion relation, $\omega^2$ is always real. Hence, each mode of wavenumber $k$ is either oscillating ($\omega^2>0$) or growing ($\omega^2<0$). The solid and dashed lines in the top panel of Fig.~\ref{fig:self_grav_plasma_resp} respectively indicate the imaginary and real parts of the $\omega$ for a self-gravitating fluid. Note that $\omega$ is either real or imaginary. In large scales, for $k<k_\rmJ$, $\omega^2$ becomes negative, i.e., the perturbation strength either grows or decays. Of course, the growing mode soon takes over. This is known as Jeans instability. In small scales ($k>k_\rmJ$), $\omega^2>0$, and therefore the perturbations persistently oscillate and propagate like waves. These are nothing but sound waves or acoustic waves. The operating mechanism of these waves can be understood as follows. A velocity perturbation, which can be sourced gravitationally, seeds a density perturbation. This leads to a pressure perturbation due to the collisional nature of the fluid, which guarantees an EOS. Particles move towards the potential minima or low pressure and low density regions; this enhances the density there, which in turn enhances the pressure and pushes them back to their original positions. Hence, the medium undergoes alternate compressions and rarefactions that manifest as sound waves. Note that these sound waves are non-dispersive in absence of self-gravity since all modes propagate with the same velocity, $\rmd\omega/\rmd k=\omega/k = c_s$, the sound speed. In presence of self-gravity, however, the medium becomes dispersive: the group velocity, $v_\rmg=\rmd\omega/\rmd k=c^2_s/v_\rmp$ substantially differs from the phase velocity, $v_\rmp=\omega/k$, except on small scales ($k\gg k_\rmJ$).

Let us now look at the evolution of macroscopic perturbations in a plasma. Each ion in a plasma attracts the surrounding electrons, which form a polarization cloud around the ion, screening the repulsive electric force of the ion on the neighboring ions and its attractive electric force on the electrons farther away. A plasma is thus electrically neutral on a macroscopic scale. The electric field is more or less confined within the polarization cloud or the {\it Debye sphere}, whose radius, known as the {\it Debye length}, is $\lambda_\rmD=\sqrt{\epsilon_0 k_\rmB T_e/n_e e^2}$. Here $n_e$ and $T_e$ are the electron density and temperature respectively, and $\epsilon_0$ is the permittivity of free space. If there are enough particles within the Debye sphere, i.e., $\Lambda\sim n_e \lambda^3_\rmD \sim T_e^{3/2} n^{-1/2}_e \gg 1$, then the plasma becomes collisionless, and any relaxation is governed by perturbations in the mean electric field that cause a collective excitation of the plasma. On the other hand, if $\Lambda \lesssim 1$, the plasma becomes collisional since large angle scatterings occur frequently. 

The dispersion relation for oscillations in the collisional regime of a plasma, is given by an equation very similar to equation~(\ref{Intro:disp_rel_fluid_sg}):

\begin{align}
\omega^2=c^2_s k^2 + \omega^2_\rmP,
\label{Intro:disp_rel_fluid_plasma}
\end{align}
where $\omega_\rmP$ is the natural frequency of plasma oscillations on large scales, known as the plasma frequency (equation~[\ref{Intro:plasma_freq}]). Note the change of sign in the plasma dispersion relation. This arises from a change of sign on the RHS of the Poisson equation, since the Coulomb electric force can be both attractive and repulsive while the gravitational force is always attractive. Hence, we see that perturbations in in the collisional regime a plasma, i.e., on scales larger than $\lambda_\rmD$, are always stable and oscillate at the plasma frequency. These are known as {\it Langmuir oscillations} which are driven by a constant tug-of-war between the ion-electron attraction and the electron-electron repulsion. On scales smaller than $\lambda_\rmD$, the relaxation of a plasma occurs very differently, via collective collisionless processes, which is what we discuss next.

\paragraph{\ul{Plasma (collisionless regime)}}

Both self-gravitating systems and plasma are governed by long range Coulomb forces that scale as $1/r^2$ with $r$, the distance from the source. However, the relaxation of self-gravitating systems fundamentally differs from that of plasma, especially on large scales. The reason is that the electric force in a plasma can be either attractive or repulsive, leading to the {\it Debye shielding} of the electric field on large scales. But gravity is exclusively attractive in nature, implying that the gravitational field cannot be screened away.

The collisionless nature of a plasma manifests in the low density and/or high temperature limit. Let us investigate the relaxation of such a system, assuming macroscopic homogeneity of the unperturbed state for the sake of simplicity. The dynamics of a collisionless plasma is governed by the Vlasov-Poisson equations~(\ref{Intro:CBE_Poisson}). A linear order perturbative analysis of these equations, i.e., using equations~(\ref{Intro:CBE_Poisson_linear_perturb}) (replacing the gravitational potential by the Coulomb potential and neglecting the external perturber), shows that the oscillation frequencies, $\omega$, of $1$D sinusoidal perturbations in a homogeneous collisionless plasma obey the following dispersion relation as a function of wavenumber $k$ \citep[][]{Binney.Tremaine.87}:

\begin{align}
\calD(\omega,k) = 1+\frac{\omega^2_\rmP}{k}\left[{\rm P.V.}\int \frac{\partial f_0/\partial v}{\omega - k v} \rmd v - \frac{i \pi s}{k}\left.\frac{\partial f_0}{\partial v}\right|_{v=\omega/k} \right] = 0,
\label{Intro:cless_plasma_dispersion}
\end{align}
where $\omega_\rmP$ is the plasma frequency, and $s$ is defined as

\begin{align}
s&=
\begin{cases}
0,\;\;\;\; {\rm Im}(\omega)>0,\\
1,\;\;\;\; {\rm Im}(\omega)=0,\\
2,\;\;\;\; {\rm Im}(\omega)<0.
\end{cases}
\end{align}
As we shall later see, the case ${\rm Im}(\omega)>0$ arises in self-gravitating systems but not in plasma. The term within the square brackets of equation~(\ref{Intro:cless_plasma_dispersion}) represents a contour integral of the quantity, $(\partial f_0/\partial v)/(\omega-kv)$ in the complex plane of $v$. P.V. denotes the principal value of this $v$ integral, i.e., the part of the integral performed along the real axis. The second term within the square brackets arises due to the contribution from the poles at $v=\omega/k$ \citep[][]{Binney.Tremaine.87}.

Expressing $\omega$ as $\omega=\omega_r+i\omega_i$ and assuming $f_0$ to be of the Maxwellian form, i.e., $f_0=\exp{\left[-v^2/2\sigma^2\right]}/\sqrt{2\pi}\sigma$, the dispersion relation (equation~[\ref{Intro:cless_plasma_dispersion}]) is given in terms of $\omega_r$ and $\omega_i$ as the following equations:

\begin{align}
&1-\sqrt{\frac{2}{\pi}}{\left(\frac{\omega_\rmP}{k\sigma}\right)}^2 \exp{\left[-\frac{\Tilde{\omega}^2_r}{2}\right]}\, \int_{0}^\infty \rmd u\, \exp{\left[-\frac{u^2}{2}\right]} \frac{u^2-\Tilde{\omega}^2_i}{{\left(u^2+\Tilde{\omega}^2_i\right)}^2}\cosh{\left(\Tilde{\omega}_r u\right)} \nonumber\\ &= \sqrt{2\pi}\, {\left(\frac{\omega_\rmP}{k\sigma}\right)}^2 \exp{\left[-\frac{\Tilde{\omega}^2_r-\Tilde{\omega}^2_i}{2}\right]} \left[\Tilde{\omega}_i\cos{\left(\Tilde{\omega}_r\Tilde{\omega}_i\right)}-\Tilde{\omega}_r\sin{\left(\Tilde{\omega}_r\Tilde{\omega}_i\right)}\right],\nonumber\\ \nonumber\\
&\Tilde{\omega}_i\,\int_0^\infty \rmd u\, \exp{\left[-\frac{u^2}{2}\right]} \frac{u}{{\left(u^2+\Tilde{\omega}^2_i\right)}^2} \sinh{\left(\Tilde{\omega}_r u\right)}\nonumber\\ 
&= -\frac{\pi}{2} \exp{\left[\frac{\Tilde{\omega}^2_i}{2}\right]} \left[\Tilde{\omega}_r\cos{\left(\Tilde{\omega}_r\Tilde{\omega}_i\right)}+\Tilde{\omega}_i\sin{\left(\Tilde{\omega}_r\Tilde{\omega}_i\right)}\right],
\label{Intro:cless_plasma_dispersions}
\end{align}
where $\Tilde{\omega}_r=\omega_r/(k\sigma)$ and $\Tilde{\omega}_i=\omega_i/(k\sigma)$. 

A general simultaneous solution to the above equations has to be obtained numerically. For each mode of wavenumber $k$ there are multiple solutions of $\Tilde{\omega}_r$ and $\Tilde{\omega}_i$, or in other words multiple possible oscillation frequencies. Which of these is excited depends on the power spectrum of the initial perturbation. The existence of multiple frequencies for a single $k$ is a key feature of collisionless systems that differs from fluids where a given $k$ mode oscillates at a single frequency. This ultimately owes to the absence (presence) of an EOS in collisionless (collisional) systems. While both $\Tilde{\omega}_r$ and $-\Tilde{\omega}_r$ are solutions of equations~(\ref{Intro:cless_plasma_dispersions}), they only allow for negative values of $\Tilde{\omega}_i$, i.e., all modes undergo {\it Landau damping}. On large scales, i.e., for $k\lambda_\rmD \ll 1$ with $\lambda_\rmD = \sigma/\omega_\rmP$ the Debye length, the asymptotic behaviour of $\omega_r$ and $\omega_i$ is given as follows \citep[][]{Landau.46}:

\begin{align}
&\omega^2_r \approx \omega^2_\rmP + 3k^2\sigma^2\;\;\;\;\;\; = \omega^2_\rmP(1+3k^2\lambda^2_\rmD), \nonumber \\
&\omega_i \approx \frac{\pi \omega^3_\rmP}{2 k^2} \left.\frac{\partial f_0}{\partial v}\right|_{v=\omega_r/k} = -\omega_\rmP \sqrt{\frac{\pi}{8}} \frac{1}{{\left(k\lambda_\rmD\right)}^3} \exp{\left[-\frac{1}{2{\left(k\lambda_\rmD\right)}^2}\right]}.
\label{Intro:omega_ri_asymptotic}
\end{align}
In the $k\lambda_\rmD\to 0$ limit, the oscillation frequency, $\omega_\rmr$, is equal to the plasma frequency, $\omega_\rmP$, while the damping rate, $\omega_i$, rapidly goes to zero. Hence, collective effects become negligible on large scales ($k\lambda_\rmD \ll 1$), indicating that the large scale relaxation of a plasma is dominated by two body relaxation. This is because on these scales all charges are Debye shielded, making the interactions effectively two-body as in neutral fluids. On the other hand, on small scales ($k\lambda_\rmD \gg 1$), the asymptotic behaviour is given by \citep[][]{Landau.46}

\begin{align}
\omega_r &\approx \frac{\pi}{2} \frac{k\sigma}{\sqrt{\ln{\left(k\Lambda_\rmD\right)}}}, \nonumber \\
\omega_i &\approx -2k\sigma \sqrt{\ln{\left(k\lambda_\rmD\right)}}.
\end{align}
Hence, on small scales, plasma modes damp away at a rate $\sim \ln{\left(k\lambda_\rmD\right)}$ faster than the frequency at which they oscillate.

\begin{figure}
\centering
\includegraphics[width=0.8\textwidth]{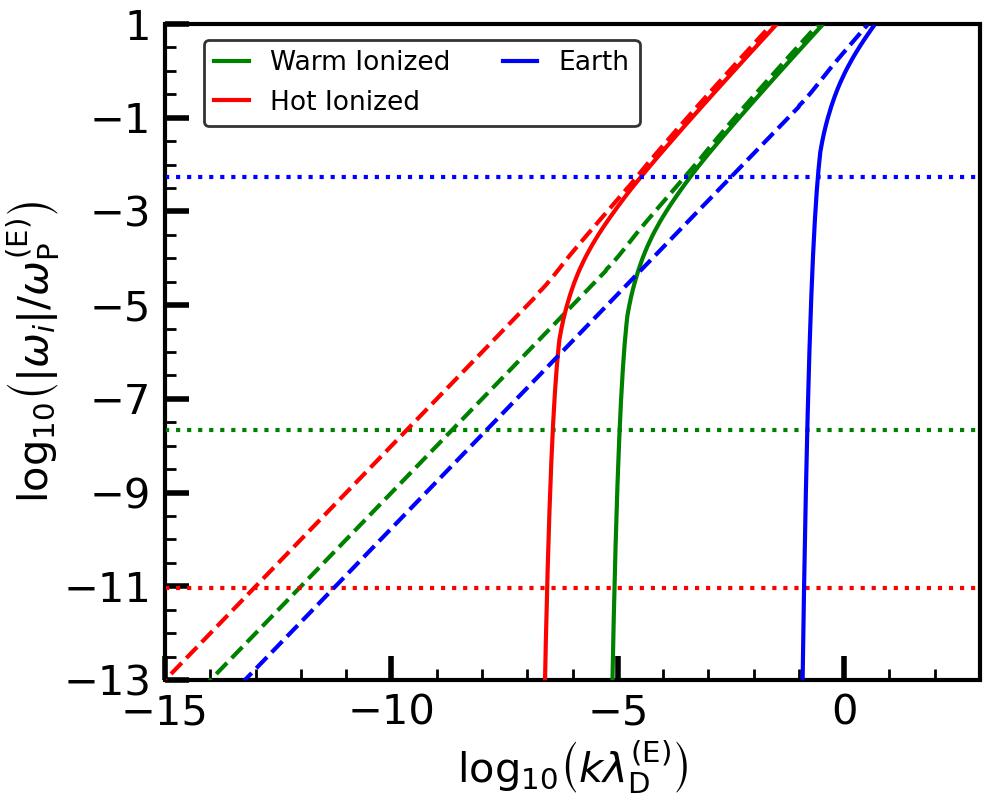}
\caption{The Landau damping rate, $\left|\omega_i\right|$ (in units of the plasma frequency in the earth's ionosphere, $\omega_\rmP^{(\rmE)}$), of the two modes shown in Fig.~\ref{fig:self_grav_plasma_resp} as a function of $k$ (in units of $1/\lambda_\rmD^{(\rmE)}$ where $\lambda_\rmD^{(\rmE)}$ is the Debye length of the earth's ionosphere). The earth's ionosphere is denoted by blue lines, and the warm and hot ionized media of the ISM are indicated by the green and red lines respectively. Solid and dashed lines indicate the modes with the lowest $\left|\omega_i\right|$ and the second lowest $\left|\omega_i\right|$ respectively. The horizontal dotted lines denote the two body relaxation frequencies, $\nu_{\rm coll}$, in the three cases. Relaxation is governed by the collective collisionless effect of Landau damping rather than collisions on scales where $\left|\omega_i\right|>\nu_{\rm coll}$ (i.e., where the solid or dashed lines are above the horizontal dotted lines), or in other words on scales smaller than $\lambda_{\rm coll}^{(i)}$, the length scale at which $\left|\omega_i\right|=\nu_{\rm coll}$. For the most weakly Landau damped mode (solid lines), $\lambda_{\rm coll}^{(i)}\sim \lambda_\rmD$, the Debye length of the medium. The extent of the collisionless relaxation regime is broader for more strongly Landau damped modes and in the ISM than in the earth's ionosphere. Astronomical structures however form at scales larger than the typical $\lambda_{\rm coll}^{(i)}$ of the ISM. This is why the relaxation of astrophysical plasma is generally governed by two-body interactions rather than Landau damping.}
\label{fig:Landau_damp_vs_coll}
\end{figure}

The top panel of Fig.~\ref{fig:self_grav_plasma_resp} plots the numerically computed values of $\Tilde{\omega}_r$ (dashed lines) and $\Tilde{\omega}_i$ (solid lines) for the two modes with the smallest damping rates. Since $\partial f_0/\partial v$ (evaluated at $v=\omega_r/k$) is negative for most realistic $f_0$ including Maxwellian, $\omega_i$ is negative for all modes, i.e., all perturbations damp away (in the linear regime) in a collisionless plasma. Both modes indicated in the figure are damped since $\omega_i$ is negative for both. As evident from equation~(\ref{Intro:omega_ri_asymptotic}) and from the dashed blue line in the top panel of Fig.~\ref{fig:self_grav_plasma_resp}, large-scale ($k\lambda_\rmD\ll 1$) modes oscillate at the plasma frequency, with $\omega_r\approx \omega_\rmP$, while at smaller scales ($k\lambda_\rmD \gg 1$), the oscillation frequency is mainly dictated by thermal pressure and the modes behave similar to acoustic modes. These plasma oscillation modes are known as {\it Langmuir modes}. These modes are very weakly damped and long-lived on large scales, i.e., $\omega_r\to 0$ as $k\lambda_\rmD\to 0$, as indicated by the solid blue line asymptoting to zero at small $k$ (also evident from the second of equations~[\ref{Intro:omega_ri_asymptotic}]). Note that the oscillation frequencies of Langmuir modes behave similarly in the collisionless and collisional (shown by the dashed green line) cases, since a plasma effectively becomes collisional on large scales. The large scale Langmuir oscillations are driven by the competition between electron-electron repulsion and ion-electron attraction. On small scales, i.e., within the Debye sphere, the oscillations are driven by the competition between {\it collective} Coulomb repulsion and attraction, since every charge simultaneously feels every other charge in the Debye sphere. Free streaming causes Landau damping of these small scale oscillations as the alternate compressions and rarefactions cannot sustain themselves due to the absence of an EOS. On the other hand, the existence of an EOS in the collisional case guarantees self-sustaining, undamped oscillations. The marked difference between large and small-scale oscillations in a plasma arises from this: the Coulomb restoring force being a long range force drives undamped large-scale oscillations while free streaming damps out the small-scale ones. This is because the ratio of the average kinetic to Coulomb potential energy of a charged particle increases as we go to smaller scales.

The relative importance of Landau damping and two body interactions as relaxation mechanism depends on the length scale of perturbation. Roughly speaking, the collective phenomenon of Landau damping drives the relaxation of a plasma on small scales. On large scales, the plasma relaxes primarily via two body encounters and its DF rapidly approaches a Maxwellian form. Instead of the Vlasov equation, the dynamics is then governed by the Boltzmann equation. The frequency of two body relaxation is given by

\begin{align}
\nu_{\rm coll} = \omega_\rmP\frac{\ln{\left(2\pi\Lambda\right)}}{2\pi\Lambda},
\label{two_body_relax_freq_plasma}
\end{align}
where $\Lambda = n_e\lambda^3_\rmD$ is roughly the number of particles within the Debye sphere. 

In Fig.~\ref{fig:Landau_damp_vs_coll} we compare the two-body relaxation frequency, $\nu_{\rm coll}$, to the Landau damping rate of a plasma, $\left|\omega_i\right|$. We plot $\left|\omega_i\right|$ (in units of the plasma frequency in the earth's ionosphere, $\omega_\rmP^{(\rmE)}\approx 10^7$rad/s), as a function of $k$ (in units of $1/\lambda_\rmD^{(\rmE)}$, with $\lambda_\rmD^{(\rmE)}\approx 0.1$ cm the Debye length in the earth's ionosphere). While the blue line denotes the $\left|\omega_i\right|$ for the earth's ionosphere ($n_e\approx 10^6\,{\rm cm}^{-3},\,T_e\approx 300\, \rmK$), the green and red lines respectively indicate that for the warm ionized component ($n_e\approx 10^{-1}\,{\rm cm}^{-3},\,T_e\approx 10^4\, \rmK$) and the hot ionized component ($n_e\approx 10^{-2}\,{\rm cm}^{-3},\,T_e\approx 10^6\, \rmK$) of the interstellar medium (ISM). The solid and dashed lines respectively indicate the $\left|\omega_i\right|$ for the two modes with the smallest $\left|\omega_i\right|$ (those shown by the blue and red lines in the top panel of Fig.~\ref{fig:self_grav_plasma_resp}). The blue, green and red horizontal dotted lines indicate the two-body relaxation frequency, $\nu_{\rm coll}$, for the earth's ionosphere, the warm ionized component and the hot ionized component of the ISM respectively. The value of $k$ at which a solid or dashed line intersects the horizontal dotted line of the same color indicates the wavenumber above which the Landau damping rate of that mode exceeds the two body relaxation frequency. Hence, above that length scale, which we denote by $\lambda_{\rm coll}^{(i)}$ for the $i^{\rm th}$ mode, the mode would damp away not due to collective effects but due to collisional diffusion. This scale is comparable to $\lambda_\rmD$ for the most weakly Landau damped mode (solid lines), but is $\sim [\Lambda/\ln{(2\pi\Lambda)}]\lambda_\rmD$ for modes with higher Landau damping rates. In case of the earth's ionosphere, $\lambda_{\rm coll}^{(1)}\approx 1$ cm while $\lambda_{\rm coll}^{(2)}\approx 3$ m. In the ISM, $\lambda_{\rm coll}^{(1)}\approx 0.1$ km for the warm component and $\approx 5$ km for the hot component. However, $\lambda_{\rm coll}^{(2)}$ is much larger: $\approx 10^3$ km and $1$ AU for the warm and hot ionized media respectively. Modes with higher Landau damping rates have higher $\lambda_{\rm coll}^{(i)}$ and therefore relax via collective effects over a larger range of scales. The above exercise suggests that the scales relevant for collisionless/collective relaxation are generally smaller than the scales at which astronomical structures form. Hence, for all practical purposes, astrophysical plasma can be assumed to primarily relax via two body collisions. The two body relaxation timescale, $\tau_{\rm coll}=1/\nu_{\rm coll}$, turns out to be maximum for the hot ionized medium, $\approx 4$ years, which is however much smaller than typical astronomical timescales. Hence, astrophysical plasma rapidly relaxes to a Maxwellian DF and can be adequately described by a one fluid model. On astronomical scales, there exists no macroscopic charge imbalance, and a plasma behaves as if it is electrically neutral. However, since ions and electrons have very different mobility, astrophysical plasma can harbour macroscopic currents that can generate macroscopic magnetic fields. Therefore, the dynamics of astrophysical plasma is well described by the equations of magnetohydrodynamics in many cases.

\paragraph{\ul{Collisionless self-gravitating systems}}

Having studied the collisionless relaxation of plasma in some detail, let us now turn our attention to that of self-gravitating systems. The contest between electron-electron repulsion plus thermal pressure on one hand and ion-electron attraction on the other is what drives Langmuir oscillations in a plasma. The key aspect of these oscillations is that they are damped, albeit very weakly, on large scales. This behaviour of collisionless plasma is exactly opposite of that of self-gravitating systems, where a form of instability, known as Jeans instability, occurs above a certain scale. The dispersion relation for homogeneous self-gravitating collisionless systems is given by equation~(\ref{Intro:cless_plasma_dispersion}) with $\omega^2_\rmP$ replaced by $-\omega^2_\rmJ=-4\pi G \rho_0$. The third term in equation~(\ref{Intro:cless_plasma_dispersion}) does not appear ($s=0$) when the imaginary part of $\omega$, $\omega_i$, is positive, i.e., when there is a growing mode. In a plasma, growing modes do not exist and therefore the third term is always present. On the contrary, self-gravitating systems allow for unstable modes, in which case the third term is zero. The existence of growing modes or instability in self-gravitating systems as opposed to plasma is a consequence of the sign change in the dispersion relation, and occurs on scales larger than the Jeans scale, $\lambda_\rmJ=1/k_\rmJ$, i.e., for

\begin{align}
k<k_\rmJ=\sqrt{\frac{4\pi G \rho_0}{\sigma^2}},
\end{align}
where $\sigma$ is the velocity dispersion. Note that $k_\rmJ$ is the self-gravitating analog of $1/\lambda_\rmD$ in plasmas. 

The dispersion relation in terms of $\omega_r$ and $\omega_i$ is now given by equations~(\ref{Intro:cless_plasma_dispersions}) with $\omega^2_\rmP$ replaced by $-\omega^2_\rmJ$, but with the RHS of both equations set to zero when $\Tilde{\omega}_i>0$. The second of equations~(\ref{Intro:cless_plasma_dispersions}) implies that $\Tilde{\omega}_r=0$ for modes with positive $\Tilde{\omega}_i$, i.e., growing modes are non-oscillatory, or in other words there are no overstable modes in a self-gravitating system. These growing modes exist only when $k<k_\rmJ$. The bottom panel of Fig.~\ref{fig:self_grav_plasma_resp} plots the numerically obtained values of $\Tilde{\omega}_r$ (dashed lines) and $\Tilde{\omega}_i$ (solid lines) for two of the modes in a self-gravitating system. Note that a non-oscillating decaying mode, indicated by blue lines, becomes a non-oscillating growing mode as $k$ falls below $k_\rmJ$. This instability occurs exponentially in the linear regime but can saturate in the non-linear regime. There are also damped oscillatory modes (red lines) for $k<k_\rmJ$, but the growing mode eventually takes over. The growth rate, $\omega_i=k\sigma\,\Tilde{\omega}_i$, of this unstable mode, can be evaluated using the first of equations~(\ref{Intro:cless_plasma_dispersions}), upon replacing $\omega^2_\rmP$ by $-\omega^2_\rmJ$ \citep[][]{Binney.Tremaine.87}:

\begin{align}
k^2=k^2_\rmJ \left[1-\sqrt{\frac{\pi}{2}}\,\Tilde{\omega}_i \exp{\left[\frac{\Tilde{\omega}^2_i}{2}\right]}\Bigg(1-{\rm erf}\left(\Tilde{\omega}_i\right)\Bigg)\right].
\end{align}

For $k>k_\rmJ$, all modes shown in the bottom panel of Fig.~\ref{fig:self_grav_plasma_resp}, are oscillating but strongly damped, with $\left|\Tilde{\omega}_i\right|\gtrsim \left|\Tilde{\omega}_r\right|$ \citep[][]{Fried.Conte.61,Ikeuchi.etal.74} (except the blue mode which is non-oscillatory but strongly damped). Hence, collisionless self-gravitating systems cannot sustain long-lived oscillations like the weakly damped Langmuir modes in plasma. Instead, small scale perturbations quickly damp away while large scale ones are vehemently unstable (in the linear regime) in a self-gravitating system. This marked contrast between plasma and self-gravitating systems (on large scales) owes to the contrast between the attractive nature of gravity and the dual nature of the electric force, which ultimately originates from the fact that masses are positive but electric charges can be either positive or negative\footnote{The sign of the RHS of the Poisson equation is negative for plasma but positive for self-gravitating systems.}.

The two body relaxation frequency, $\nu_{\rm coll}$, for a self-gravitating system with $N$ particles is given by equation~(\ref{two_body_relax_freq_plasma}) with $\omega_\rmP$ replaced by $\omega_\rmJ\sim \sqrt{G\rho_0}$ and $\Lambda$ replaced by $N$. The scale, $\lambda_{\rm coll}$, beyond which two body relaxation dominates over collisionless relaxation therefore turns out to be $\sim \lambda_\rmJ[N/\ln{N}]$. In galaxies and cold dark matter halos, $N$ is very large, which implies that $\lambda_{\rm coll}\gg \lambda_\rmJ$, and that the two body relaxation time, $\tau_{\rm coll}$, exceeds the Hubble time by many orders of magnitude. This entails that the relaxation of large $N$ self-gravitating systems like galaxies is always driven by collective, collisionless processes rather than two body interactions. Due to the long range attractive nature of gravity, collective effects are far stronger in self-gravitating systems than in plasmas, where the impact of collective excitations is shielded within the Debye sphere due to charge polarization.

\paragraph{\ul{A comparative study}}

The above differences in the relaxation of self-gravitating systems, fluids and plasma, are discussed in the context of homogeneous systems for simplicity. The primary characteristics of relaxation however remain qualitatively the same for inhomogeneous systems. The major differences in the nature of relaxation in these systems are summarized as follows. 

The relaxation of collisionless self-gravitating systems and plasmas differs from that of fluids on small scales. While perturbations undergo Landau damping on small scales in the former, the latter shows undamped oscillating modes. In fluids, each mode of a given wavenumber $k$ has a single frequency. This is remarkably different from collisionless systems which can oscillate at multiple frequencies for a given wavenumber (see Fig.~\ref{fig:self_grav_plasma_resp}). The presence (absence) of an EOS in fluids (collisionless systems) is responsible for this. Among collisionless systems, self-gravitating systems differ from plasmas on large scales. The former becomes Jeans unstable and harbors a non-oscillatory growing mode on large scales (above the Jeans scale), but the latter only shows oscillating modes. These Langmuir oscillations occur at the plasma frequency and are very weakly damped on large scales (beyond the Debye length), where the plasma essentially behaves like a collisional system that primarily relaxes via two body interactions. On small scales ($\lesssim$ the Debye length), the plasma however mainly relaxes through the collective excitation of charged particles, since every charge feels the effect of every other charge within the Debye sphere at the same time. These collective oscillations Landau damp away on scales smaller than the Debye length. Within the Jeans scale, collisionless self-gravitating systems also relax via Landau damping. The regime of collective excitations or Landau damping is however much more extended in a collisionless self-gravitating system than in a plasma, since a typical Jeans scale is much larger than a Debye length. This suggests that galaxies and cold dark matter halos primarily relax via collective effects. It is important to note that this collective relaxation involves a fine-grained damping (or growth in case of Jeans instability) of the linear order perturbation in the DF at every point in phase-space. This is because Landau damping involves a redistribution of the energies of the particles through mutual gravitational interactions. Phase-mixing on the other hand is a purely kinematic effect, i.e., involves no change in the energies, and does not damp the perturbations in the DF on a fine-grained level. Rather it involves oscillations of the (fine-grained) perturbations at different frequencies. When averaged over a small but finite sized phase-space volume, these perturbations damp away due to the intrinsic spread in the frequencies. Hence, phase-mixing damps out gradients in the DF only on a coarse-grained level. This makes it fundamentally different from Landau damping. 

Why does Landau damping occur in collisionless systems but not in fluids? The fundamental reason is this: fluids, being collisional in nature, possess an EOS that directly relates their pressure to density whereas collisionless systems do not have one that relates their velocity dispersion to density. Let us understand how the presence or absence of an EOS affects the survival of perturbations in a system. A sinusoidal perturbation in the gravitational potential seeds an in-phase density perturbation and a velocity perturbation that differs from it by a phase of $\pi/2$. Particles gain velocities and fall towards the minima of the potential well, thereby increasing the density there. In fluids, two body relaxation and the resulting EOS guarantee that a density enhancement proportionally enhances the pressure. This pushes out the particles from the potential well minima back to their original state. This is how one cycle of sinusoidal oscillations in the density, pressure and velocity perturbations operates in fluids. Even in absence of self-gravity, particles move back and forth between regions of high and low density and pressure, resulting in self-sustained, alternate compressions and rarefactions of the medium, which manifest as sound waves. On the other hand, in collisionless systems, the density enhancement in the potential minima increases the velocity dispersion via collective gravitational forces and not via two body relaxation as in fluids. Therefore, the amount of enhancement in the velocity dispersion is not proportional to that of the density enhancement, implying that the particles never go back to their original state. On small scales, the particles stream away more than self-gravity can clump them together. This damps away the small-scale density and potential perturbations if there are more particles with smaller energies, i.e., $\partial f_0/\partial E_0<0$. Particles with energies smaller than the energy of the perturbation, $E_\rmP$, gain energy from it, while those with larger energies lose energy to it. And since, in a system with $\partial f_0/\partial E_0<0$, more particles have $E_0<E_\rmP$ than $E_0>E_\rmP$, more particles gain energy from than lose energy to the perturbation. Therefore, overall, the particles outside the perturbation gain energy from those in it, causing the perturbation to Landau damp away. The opposite occurs when there are more particles with larger energies, i.e., $\partial f_0/\partial E_0>0$, in which case the perturbation grows and an instability kicks in, due to the net gain of energy by the particles in the perturbation from those outside it. On the other hand, on large scales, irrespective of whether $\partial f_0/\partial E_0$ is negative or positive, gravity being a long range force always wins over streaming, which is a local phenomenon and requires time to take effect. This is because the time, $\tau_s\sim \lambda/\sigma$, required for particles of velocity dispersion $\sigma$ to traverse the entirety of a potential well with wavelength $\lambda$, exceeds the gravitational collapse timescale, $\tau_\rmJ\sim 1/\sqrt{G \rho_0}$, where $\rho_0$ is the unperturbed density. This results in the growth of large scale perturbations, known as Jeans instability. The Jeans wavelength, $\lambda_\rmJ \sim \sigma/\sqrt{G\rho_0}$, is the wavelength of a perturbation for which $\tau_s$ becomes comparable to $\tau_\rmJ$. It is worth noting that this large-scale Jeans instability is a common feature of both collisional and collisionless self-gravitating systems since the collective effects responsible for Landau damping only manifest on small scales.

We have seen that collisionless systems do not relax via two body interactions as the mean free path significantly exceeds the mean particle separation. Rather, they equilibrate via kinematic processes like phase-mixing and collective processes like Landau damping or violent relaxation (to be discussed shortly). Phase-mixing and Landau damping are linear phenomena and typically occur over the timescale of several dynamical times. Violent relaxation on the other hand is a fundamentally non-linear phenomenon that happens quite fast, on the order of the dynamical time, but is self-limiting in nature. Till now, we have discussed how relaxation occurs in the linear regime. In the next section we discuss some prime features of relaxation in the quasi-linear and highly non-linear regimes.

\subsubsection{Non-linear response and violent relaxation}

The linear response of a self gravitating collisionless system to an external perturbation loses its coherence over time through phase-mixing and Landau damping (see equation~[\ref{Intro:linear_response}]). Therefore, according to linear perturbation theory, the system goes back to the same equilibrium state it was in before it was perturbed\footnote{Note that Landau damping does cause a lasting impact by increasing the velocity dispersion but this only shows up in second order of perturbation theory.}. In other words, the coarse-grained DF of a system is never permanently affected at linear order. However, there are cases where the system does undergo a permanent change, e.g., the encounter of a satellite galaxy with a disk galaxy. Especially, if the perturbation is strong, the post-perturbation equilibrium state of the system is different from the original one. Since the linear order response decays away in long term, any persistent change in the DF is necessarily an outcome of non-linear relaxation. While non-linear relaxation is a hard problem, useful insight can be gained by studying how relaxation occurs in the quasi-linear regime.

To understand quasi-linear relaxation, let us investigate the second order response. For simplicity, let us ignore the self-gravity of the response and only consider an external perturbing potential, $\Phi_\rmP$. At first order, the response, $f_{1\bf{0}}$, of the zero-mode, which is the only mode that survives phase-mixing, is zero. The zero-mode second order response, $f_{2\bf{0}}$, however, is non-zero. This implies that the second order response never completely phase-mixes away. Upon simultaneously solving the first and second order equations among the series of perturbed Vlasov equations given in equation~(\ref{Intro:CBE_Poisson_perturb}) in the non self-gravitating limit, one can obtain the following expression for $f_{2\bf{0}}$ \citep[][]{Carlberg.Sellwood.85}:

\begin{align}
f_{2\bf{0}} (\bI,t) = \sum_{\boldell'} \boldell' \cdot \frac{\partial}{\partial \bI} \left[ \boldell' \cdot \frac{\partial f_0}{\partial \bI} \int_0^t \rmd\tau\, \int_0^\tau \rmd\tau'\, \exp{\left[-i\boldell' \cdot \Omega \left(\tau-\tau'\right)\right]}\, \Phi_{-\boldell'}(\bI,\tau)\, \Phi_{\boldell'}(\bI,\tau') \right].
\label{Intro:second_order_response}
\end{align}

We can simplify the above expression in the special case of an impulsive perturbation, i.e., $\Phi_{\boldell'}(\bI,t)=A_{\boldell'}(\bI)\, \delta(t)$. In this impulsive limit, one can check that the steady state second order response simplifies to

\begin{align}
f_{2\bf{0}} (\bI) = \sum_{\boldell'} \boldell' \cdot \frac{\partial}{\partial \bI} \left[ \boldell' \cdot \frac{\partial f_0}{\partial \bI} {\left|A_{\boldell'}(\bI)\right|}^2 \right].
\label{Intro:second_order_response_impulsive}
\end{align}
Note that while the first order response (equation~[\ref{Intro:linear_response}]) depends on the first derivative of $f_0$ with respect to the actions, the second order response depends on both the first and second derivatives. If $f_0$ is a product of isothermal/Maxwellian distributions with velocity dispersions $\boldsymbol{\sigma}$, and the frequencies and $A_{\boldell'}$ have a much slower variation with $\bI$ than $f_0$, then equation~(\ref{Intro:second_order_response_impulsive}) can be simplified to yield

\begin{align}
f_{2\bf{0}} (\bI) \approx \sum_{\boldell'} {\left(\sum_{i=1}^d \frac{l'_i\Omega_i}{\sigma^2_i}\right)}^2\, {\left|A_{\boldell'}(\bI)\right|}^2\, f_0(\bI),
\label{Intro:second_order_response_impulsive_final}
\end{align}
where $d$ is the number of dimensions. Note that each term of the above sum is positive and therefore $f_{2\bf{0}}(\bI)$ is positive. Thus, at second order, the actions of the field particles increase due to the perturbation. This is a generic result: isothermal systems typically gain energy from an impulsive perturbation. This is why a disk galaxy usually gains kinetic energy immediately after an impulsive impact with a satellite galaxy. However, in due course, revirialization converts this kinetic energy into potential energy (due to the negative specific heat of self-gravitating systems), which puffs up and cools the disk \citep[][]{Toth.Ostriker.92}.

First and second order response theories are valid only up to the mildly non-linear or quasilinear regime. However, if the perturbation strength is too large and/or the perturbation is adiabatic or slowly varying, the problem becomes highly non-linear. In this case, one has to solve the perturbed Vlasov-Poisson system (equations~[\ref{Intro:CBE_Poisson_perturb}]) to multiple orders. Even then, the perturbation series might diverge. Hence, a rigorous analysis of highly non-linear relaxation has to be performed using non-perturbative techniques, which is beyond the scope of this dissertation. N-body simulations are very handy in this respect. They show that non-linear collisionless relaxation occurs rapidly and violently, achieving completion within a dynamical time \citep[][]{LyndenBell.67}. 
Each star experiences a time-varying gravitational potential and therefore exhibits a change in its energy. The orbital energy distribution widens and the system isotropizes. The end state of violent relaxation has been a subject of great debate for several decades. It is well known that the DF does not become Maxwellian after violent relaxation. Question is: what {\it is} the end state of violent collisionless relaxation? N-body simulations with cosmological initial conditions show that cold dark matter halos tend to have a density profile, known as the Navarro-Frenk-White (NFW) profile \citep[][]{Navarro.etal.97}, which is surprisingly insensitive to initial conditions and exhibits a universal behaviour. It is suspected that mergers and subsequent violent relaxation in the early stage of formation of a halo are the culprits behind the emergence of the NFW profile. The actual origin of this is however far from known. A steady state solution of the Vlasov-Poisson equations does not have a unique functional form, yet violent relaxation gives rise to an apparently universal profile. This indicates that the solution landscape of the Vlasov-Poisson system might have attractor states.

\section{Secular evolution and dynamical friction}

So far we have seen how a self-gravitating collisionless system (subject/host) responds to an external perturber and how this response relaxes/equilibrates over time. The orbital dynamics of the perturber is in turn affected by the gravitational effect of this response. The over- and under-densities in the perturbed host exert gravitational force and torque on the perturber, which results in an exchange of energy and angular momentum between the perturber and the field particles of the host. Typically, the perturber loses its energy and angular momentum and inspirals towards the center of the host. This process is known as dynamical friction. It involves the change in the orbital elements of the perturber due to a back reaction of the host response and is therefore a second order effect. As such, dynamical friction generally occurs over a timescale much longer than the typical dynamical time of the host or the orbital period of the perturber, and is an example of a broad class of dynamical phenomena known as secular processes.

Secular evolution occurs due to a gradual change in the mean field of a system. This slowly alters the integrals of motion and therefore the orbital dynamics of the field particles (stars or dark matter particles). Secular evolution is of utmost importance in the evolution of self-gravitating collisionless systems like galaxies and dark matter halos. These objects are always in a state of non-equilibrium since they are subject to external perturbations such as penetrating or fly-by encounters with other objects, which can be galaxies, halos, star-clusters, black holes and so on. In the last few sections we saw how the response of a system to external perturbations develops and relaxes over time. In the weak perturbation limit, phase-mixing, Landau damping and gravitational instability (also known as Jeans instability) are the only mechanisms for collisionless relaxation. Phase-mixing involves the coarse-grained destruction of a coherent response due to oscillations of field particles at different frequencies. It does not involve any changes in the orbital elements of the field particles and therefore cannot be categorized as secular evolution. Landau damping or gravitational instability on the other hand is a classic example of secular evolution in the linear regime, since it involves the self-gravitating response of the system, and steadily alters the actions of its constituent particles. Dynamical friction involves an exchange of energy and angular momentum between a host system and an external perturber, that slowly alters the orbital dynamics of both the perturber and the field particles, and is therefore an example of a secular evolution of the combined system of the host and the perturber.

A vast range of astrophysical phenomena is governed by dynamical friction. These include (i) galactic cannibalism, the orbital inspiral of galaxies towards the center of a galaxy cluster or that of satellite galaxies towards the central galaxy of a group, (ii) galaxy-galaxy mergers, (iii) formation of nuclear star cluster due to the inspiral and mergers of globular clusters, (iv) the initial phase of binary black hole mergers, etc. Structure formation in non-linear scales occurs via mergers between galaxies and dark matter halos. When two initially unbound objects gravitationally interact, each of them is distorted by the other, and the relative orbital energy is dumped into the internal energy of the field particles in each system. This is nothing but dynamical friction in action. The constant drainage of orbital energy can make the objects gravitationally bound, so that they continue to inspiral towards each other under dynamical friction until they eventually merge. Binary black holes lose their orbital energy and angular momentum to the surrounding stars, gas and dark matter through dynamical friction and inspiral towards each other before they undergo further orbital inspiral through the emission of gravitational waves and eventually merge. Hence, dynamical friction is a key ingredient of all structure formation in the universe. 

The standard picture of dynamical friction was provided by the seminal work of \cite{Chandrasekhar.43}, who considered it an outcome of {\it local} momentum exchanges between a massive perturber (of mass $M$) moving with a uniform velocity $\bv$ on a straight orbit through a homogeneous medium and surrounding field particles that are also on nearly straight orbits. Since, on an average, there exist more particles with energy lower than that of the perturber as opposed to higher than it, the perturber loses energy to the field particles. The resulting `friction' force acting on the perturber is given by the famous Chandrasekhar formula:

\begin{equation}
\bF_{\rm DF} = - \frac{4 \pi G^2 M^2}{v^2} \, \ln\Lambda \, \rho(<v) \, \frac{\bv}{v}\,,
\label{Intro:FChandra}
\end{equation}
where $\rho(<v)$ is the local density of particles with velocities less than $v$ and

\begin{align}
\Lambda = \frac{b_{\rm max}}{b_{\rm min}},
\end{align}
with $b_{\rm max}$ and $b_{\rm min}$ the maximum and minimum impact parameters for encounters between the perturber and field particles. To match the Chandrasekhar prediction with the results from $N$-body simulations of dynamical friction-driven orbital inspiral, $b_{\rm max}$ is typically taken to be the size of the host and $b_{\rm min}$ is assumed to be ${\rm max}\left[\varepsilon,b_{90}\right]$, where $\varepsilon$ is the scale radius of the perturber and $b_{90}=GM/\sigma^2$ ($\sigma$ is the local velocity dispersion of the host) is the impact parameter corresponding to $90$ degrees deflection angle for a point perturber. A better fit to simulation results is obtained for $b_{\rm max}$ is set to be $R$, the galactocentric radius of the perturber \citep[][]{Petts.etal.16,Kaur.Sridhar.18}, rather than the typical size of the host, since $R$ roughly marks the size of the perturber's region of influence.

The Chandrasekhar formalism to compute the dynamical friction force is simple to implement, but is highly idealized and thus a crude approximation to what happens in real galaxies. A stark failure of the Chandrasekhar picture is its prediction of continued dynamical friction in the core region of galaxies and halos with cored density profile while $N$-body simulations show (i) vanishing dynamical friction and stalling of the perturber at the core radius, known as core-stalling, following an accelerated infall, known as super-Chandrasekhar friction \citep[][]{Read.etal.06c,Goerdt.etal.10,Cole.etal.12} and (ii) an enhancing torque known as dynamical buoyancy that pushes out the perturber from deep inside the core region until it stalls at the stalling radius \citep[][]{Cole.etal.12}. Since the Chandrasekhar picture is a {\it local} picture that does not take into account the global curvature of the orbits of the perturber and the field particles, and therefore ignores the global host response, it is not surprising that there are cases where it fails. 

A far more general picture of dynamical friction than the standard Chandrasekhar one was provided by the seminal paper of \cite{Tremaine.Weinberg.84}, who considered the case of a perturber with potential $\Phi_\rmP$ on a circular orbit with frequency $\Omega_\rmP$ in a spherical host characterized by rosette orbits of the field particles. They discovered that if one assumes the perturber adiabatically grows over time and inspirals at a rate far slower than the orbital time of the host (which is typically the case), the dynamical friction torque is exerted only by field particles that are purely resonant with the perturber, or in other words particles with orbital frequencies exactly commensurate with the perturber's circular frequency. The resonant field particles are perturbed the most since they are `in sync' or in phase with the perturber, thereby exchanging a lot of energy and angular momentum. The torque exerted by this resonant response density on the perturber is known as the LBK torque, named after \cite{LyndenBell.Kalnajs.72} who first derived it in the context of spiral arm-driven transport of angular momentum in disk galaxies. For a spherical host with velocity isotropy, which is characterized by an unperturbed distribution function $f_0(E_0)$ ($E_0$ is the unperturbed energy), and two orbital frequencies $\Omega_1$ and $\Omega_2$, the LBK torque is expressed as

\begin{align}
\calT_{\rm LBK} &= \int \rmd \bw\, \rmd \bI\, \frac{\partial \Phi_\rmP}{\partial \phi} f_1 \nonumber\\
&= 16\pi^4 \Omega_\rmP \sum_{\ell_3=0}^\infty \sum_{\ell_1=-\infty}^\infty \sum_{\ell_2=-\infty}^\infty \ell_3^2 \int \rmd\bI\,\delta\left(\ell_1\Omega_1+\ell_2\Omega_2-\ell_3\Omega_\rmP\right)\frac{\partial f_0}{\partial E_0}\, {\left|\Phi_{\boldell}(\bI)\right|}^2.
\label{Intro:tauLBK}
\end{align}
Here $\Phi_{\boldell}$ is the $\boldell$-mode Fourier coefficient of the perturber potential, $\Phi_\rmP$. The LBK torque has several distinctive features. Firstly, the Dirac delta function of the resonant frequency, $\ell_1\Omega_1+\ell_2\Omega_2-\ell_3\Omega_\rmP$, manifests an exclusive contribution to the torque from the resonant orbits. Secondly, the LBK torque is second order in the perturber potential, just like the Chandrasekhar force (equation~[\ref{Intro:FChandra}]). And finally, all factors inside the action integral are positive definite except $\partial f_0/\partial E_0$, which is negative for all stable systems, thereby rendering the LBK torque always retarding. Hence the LBK torque only predicts dynamical friction but not buoyancy, unless $\partial f_0/\partial E_0$ becomes positive in some region of phase-space, signalling the onset of dynamical instability. On the other hand, core-stalling, which the Chandrasekhar theory fails to explain, is a natural prediction of the LBK torque in cored galaxies. As shown by \cite{Kaur.Sridhar.18}, the circular frequency of the perturber,

\begin{align}
\Omega_\rmP = \sqrt{\frac{G\left[M_\rmG(R)+M_\rmP\right]}{R^3}},
\end{align}
significantly exceeds the orbital frequencies of stars in the central core region of a cored galaxy/halo ($M_\rmP$ and $M_\rmG(R)$ are respectively the perturber mass and the enclosed mass of the host galaxy within the perturber's orbital radius $R$). This leads to a suppression of the co-rotation resonances and weakening of the torque from the surviving non-co-rotation resonances \citep[c.f.][]{Kaur.Stone.22} in the core region, causing the perturber to stall near the core radius.

Despite its obvious successes over the Chandrasekhar theory, the LBK torque does not predict several interesting phenomena in cored galaxies like super-Chandrasekhar friction and dynamical buoyancy. Moreover, the resonant theory of dynamical friction comes with its own conceptual problems. Firstly, it assumes an infinitely slow introduction of the perturber to the system (adiabatic approximation) as well as an infinitely slow radial motion due to secular evolution (secular approximation). These are unrealistic assumptions since dynamical friction is required to operate within the Hubble time in order to be astrophysically relevant. We relax these assumptions in chapter~\ref{chapter: paper4} of this thesis, where we improve upon the LBK formalism of linear perturbation theory by performing a fully self-consistent computation, i.e., taking into account the dependence of the host response on the orbital inspiral rate which is in turn dictated by the response. We find that the resulting {\it self-consistent torque} differs from the LBK torque mainly in the following aspects: 

\begin{itemize}
    \item Unlike the LBK torque, the self-consistent torque has a significant contribution from not only the pure resonances but also the {\it near-resonant orbits}.
    \item Unlike the LBK torque which is always retarding, the self-consistent torque can under certain conditions be {\it enhancing}.
\end{itemize}
 This generalization of the perturbative formalism for dynamical friction explains the origin of super-Chandrasekhar friction, dynamical buoyancy and core-stalling (balance between friction and buoyancy), something that the standard theories of dynamical friction have failed to achieve. 
 
 In chapter~\ref{chapter: paper5}, we move beyond the perturbative treatment of dynamical friction. The perturber's mass becomes comparable to the enclosed galaxy mass in the core region, implying that the perturber can no longer be deemed as a weak perturbation in this case. Moreover, as the perturber approaches the stalling radius and slows down, a large proportion of field particles gets adiabatically trapped in libration along near-resonant orbits, leading to the development of non-linear perturbations in the DF. This is where standard perturbation theory (based on the action-angle variables of the unperturbed galaxy), especially a linear order one, on which the derivation of the LBK torque and even our self-consistent torque (see chapter~\ref{chapter: paper4}) is based, becomes questionable. In fact, \cite{Tremaine.Weinberg.84} acknowledge this shortcoming of linear perturbation theory and the LBK torque, and advocate a modified version of perturbation theory using slow and fast action-angle variables \citep[][]{Lichtenberg.Lieberman.92} to compute the torque in the slow regime, i.e., when the orbital inspiral timescale exceeds the libration period of the near-resonant orbits \citep[see also][]{Chiba.Schonrich.22,Hamilton.etal.22}. However, this technique is still perturbative and works reasonably well only for near-resonant orbits but not for what are known as semi-ergodic/semi-chaotic orbits. Therefore, in chapter~\ref{chapter: paper5}, we develop a non-perturbative orbit-based treatment that addresses the contribution to dynamical friction (or buoyancy) from different orbital families. We identify the near-co-rotation-resonant \horseshoen, \pacman and tadpole orbits as the dominant contributors to dynamical friction/buoyancy. These orbits exert a retarding torque and hence dynamical friction on the perturber when it is orbiting outside the core region of a cored galaxy/halo. In the core region, however, the orbital topology drastically changes due to a bifurcation of the inner Lagrange points: the \horseshoe orbits disappear, which is synonymous to the suppression of co-rotation resonances in the core region \citep[][]{Kaur.Sridhar.18}, and the surviving \pacman orbits can under certain conditions exert an enhancing torque or dynamical buoyancy instead of friction on the perturber. 

 In the subsequent chapters of this dissertation, we present novel treatments of gravitational encounters, collisionless relaxation (via phase-mixing) and dynamical friction that go beyond the standard picture of galactic dynamics \citep[][]{Binney.Tremaine.08}.
	
%
%
%

\chapter{A Fully General, Non-Perturbative Treatment of Impulsive Heating} \label{chapter: paper1}





\begin{center}
This chapter has been published as:\\
\vspace{5pt}
\author{Uddipan Banik, Frank~C.~van den Bosch}\\
\vspace{3pt}
\textit{Monthly Notices of the Royal Astronomical Society}, Volume 502, Issue 1, p.1441-1455\\
\textit{\citep[][]{Banik.vdBosch.21b}}
\end{center}

\section{Introduction}

When an extended object, hereafter the subject, has a gravitational encounter with another massive body, hereafter the perturber, it induces a tidal distortion that causes a transfer of orbital energy to internal energy of the body (i.e., coherent bulk motion is transferred into random motion).  Gravitational encounters therefore are a means by which two unbound objects can become bound (`tidal capture'), and ultimately merge. They also cause a heating and deformation of the subject, which can result in mass loss and even a complete disruption of the subject.  Gravitational encounters thus play an important role in many areas of astrophysics, including, among others, the merging of galaxies and dark matter halos \citep[e.g.,][]{Richstone.75, Richstone.76, White.78, Makino.Hut.97, Mamon.92, Mamon.00}, the tidal stripping, heating and harassment of subhalos, satellite galaxies and globular clusters \citep[e.g.,][]{Moore.etal.96b, Gnedin.etal.99, vdBosch.etal.18a,DuttaChowdhury.etal.20}, the heating of discs \citep[][]{Ostriker.etal.72}, the formation of stellar binaries by two-body tidal capture \citep[][]{Fabian.etal.75, Press.Teukolsky.77, Lee.Ostriker.86}, and the disruption of star clusters and stellar binaries \citep[e.g.,][]{Spitzer.58, Heggie.75, Bahcall.etal.85}. Throughout this chapter, for brevity we will refer to the constituent particles of the subject as `stars'.
\begin{figure}[t!]
\centering
\includegraphics[width=1\textwidth]{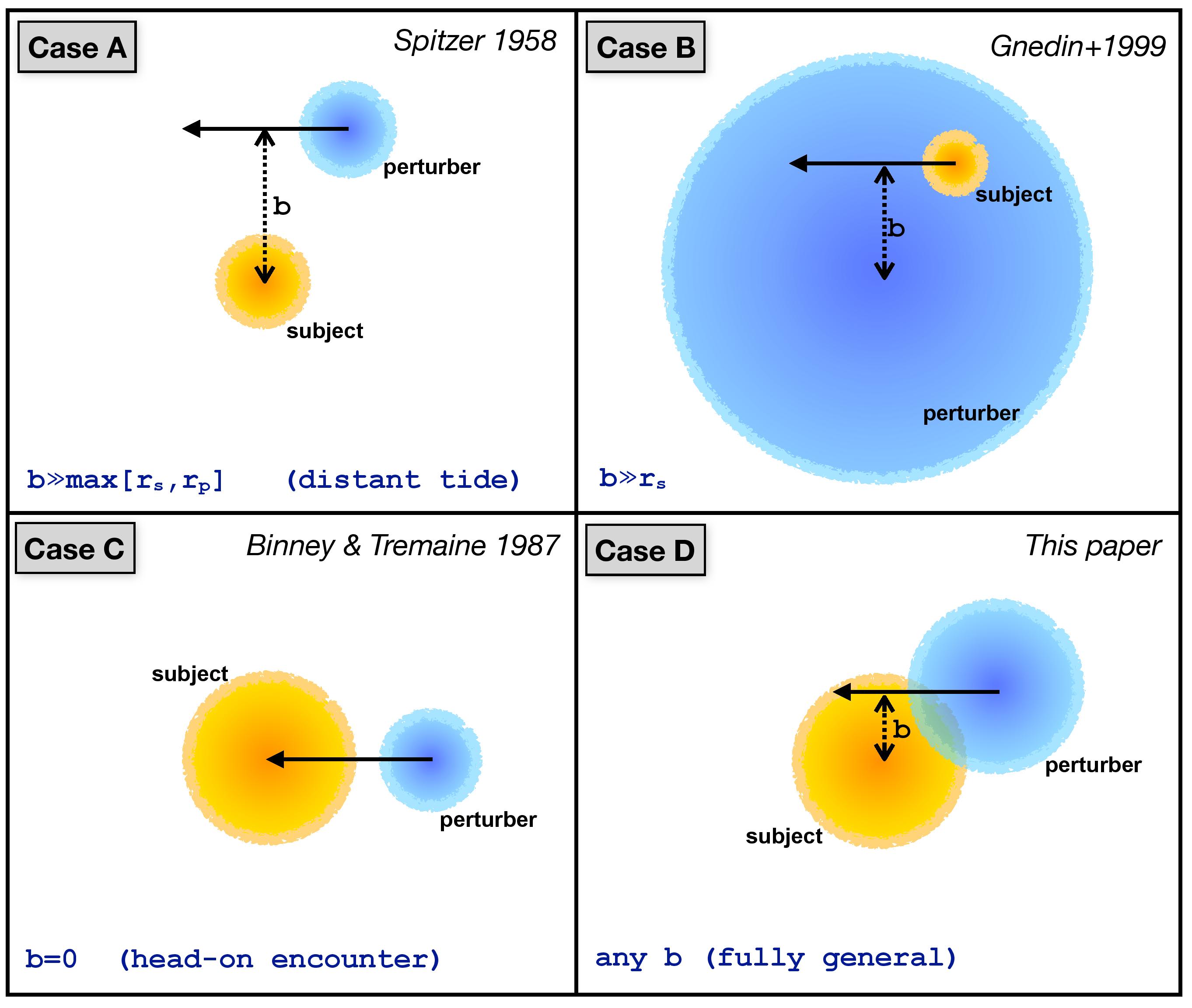}
   \caption{A pictorial comparison of impulsive encounters $(\vp\gg \sigma)$ under certain conditions for the impact parameter $b$. In the upper-right corner of each panel we cite the paper in which the impulsive energy transfer for this case was first worked out. This chapter presents the fully general case D (no constraint on $b$), as depicted in the lower right-hand panel.}
\label{fig:schematic_approximations}
\end{figure}
A fully general treatment of gravitational encounters is extremely complicated, which is why they are often studied using numerical simulations. However, in the impulsive limit, when the encounter velocity is large compared to the characteristic internal velocities of the subject, the encounter can be treated analytically. In particular, in this case, one can ignore the internal motion within the subject (i.e., ignore the displacements of the stars during the encounter), and simply compute the velocity change (the impulse) of a star using
\begin{equation}
\Delta\bv = -\int \nabla \Phi_\rmP \, \rmd t\,,
\end{equation}
where $\Phi_\rmP$ is the potential due to the perturber. And since the encounter speed, $\vp$, is high, one can further simplify matters by considering the perturber to be on a straight-line orbit with constant speed.

The impulse increases the specific kinetic energy of the subject stars by
\begin{equation}\label{dEstar}
\Delta \varepsilon = \bv \cdot \Delta \bv + {1 \over 2} (\Delta v)^2\,.
\end{equation}
Since the potential energy of the stars remains invariant during (but not after) the impulse, the increase in total {\it internal} energy of the subject is given by
\begin{equation}
\Delta E_{\rm int} = \int \rho_\rmS(\br) \Delta \varepsilon(\br) \, \rmd^3\br - {1 \over 2} M_\rmS (\Delta v_{\rm CM})^2\,.
\end{equation}
Here $M_\rmS$ and $\rho_\rmS(\br)$ are the mass and density profile of the subject and $\Delta \bv_{\rm CM}$ is the velocity impulse of the centre-of-mass of the subject.

If the encounter, in addition to being impulsive, is also distant, such that the impact parameter $b$ is much larger than the scale radii of the subject ($r_\rmS$) and the perturber ($r_\rmP$), i.e., $b \gg \max(r_\rmS, r_\rmP)$, then the internal structure of the perturber can be ignored (it can be treated as a point mass), and its potential can be expanded as a multipole series and truncated at the quadrupole term. This `distant tide approximation' (hereafter DTA, depicted as case A in Fig.~\ref{fig:schematic_approximations}) was first used by \citet[][hereafter S58]{Spitzer.58} to study the disruption of star clusters by passing interstellar clouds.  In particular, Spitzer showed that, for a spherical subject mass, $M_\rmS$, an impulsive encounter results in an internal energy increase
\begin{equation}\label{dESpitzer}
\Delta E_{\rm int} = {4 M_\rmS \over 3} \, \left({G \, M_\rmP \over \vp} \right)^2 \, {\langle r^2 \rangle \over b^4}\,,
\end{equation}
with
\begin{equation}\label{r2aver}
\langle r^2 \rangle = \frac{4 \pi}{M_\rmS} \int \rho_\rmS(r) \, r^4 \, \rmd r
\end{equation}
(see also Table~\ref{tab:comparison}). Note that $\Delta E \propto b^{-4}$, indicating that closer encounters are far more efficient in transferring energy than distant encounters.  However, as shown by \cite{Aguilar.White.85} using numerical simulations, equation~(\ref{dESpitzer}) is only accurate for relatively large impact parameters, $b \gta 10\max(r_\rmS, r_\rmP)$, for which $\Delta E_{\rm int}$ is typically extremely small (and thus less interesting).

This situation was improved upon by \citet[][hereafter GHO99]{Gnedin.etal.99}, who modified the treatment by S58 so that it can also be used in cases where $r_\rmS \ll b < r_\rmP$ (see case B in Fig.~\ref{fig:schematic_approximations}). This describes circumstances in which the subject is moving inside the perturber potential (i.e., a globular cluster moving inside a galaxy, or a satellite galaxy orbiting the halo of the Milky Way). As shown by GHO99, the resulting $\Delta E_{\rm int}$ in this case is identical to that of equation~(\ref{dESpitzer}) but multiplied by a function $\chi_{\rm st}(b)$, that depends on the detailed density profile of the perturber (see  Table~\ref{tab:comparison}).
\begin{table}
\centering
\scalebox{0.85}{\begin{tabular}{lllll}
 \hline
 Case & Impact parameter & $\Delta E_{\rm int}$\\
 (1) & (2) & (3)\\
 \hline \\
A & \begin{minipage}{2cm}\begin{equation}b\gg \max{\left(r_\rmS,r_\rmP\right)}\nonumber \end{equation}\end{minipage}    & \begin{minipage}{2cm}\begin{equation}\frac{4 M_\rmS}{3}{\left(\frac{GM_\rmP}{\vp}\right)}^2 \frac{\left<r^2\right>}{b^4},\nonumber\end{equation}\end{minipage} \\ \\
& & \begin{minipage}{2cm}\begin{align}\left<r^2\right>=\frac{4\pi}{M_\rmS}\int_0^{r_{\rm trunc}}\rmd r\, r^4\rho_\rmS(r)\nonumber\end{align}\end{minipage} &
\\ \\ \\ \\
B      & \begin{minipage}{2cm}\begin{equation}b\gg r_\rmS\nonumber \end{equation}\end{minipage}    & \begin{minipage}{2cm}\begin{equation}\frac{4 M_\rmS}{3}{\left(\frac{GM_\rmP}{\vp}\right)}^2 \left<r^2\right>\frac{\chi_{\rm st}(b)}{b^4},\nonumber \end{equation}\end{minipage}\\ \\
& & \begin{minipage}{2cm}\begin{equation}\chi_{\rm st}=\frac{1}{2}\left[{\left(3J_0-J_1-I_0\right)}^2+{\left(2I_0-I_1-3J_0+J_1\right)}^2+I^2_0\right],\nonumber\end{equation}\end{minipage}\\ \\
& & \begin{minipage}{2cm}\begin{equation}I_k(b)=\int_1^{\infty}\mu_k(b\zeta)\frac{\rmd \zeta}{\zeta^2{\left(\zeta^2-1\right)}^{1/2}},\nonumber\end{equation}\end{minipage}\\ \\
& & \begin{minipage}{2cm}\begin{equation}J_k(b)=\int_1^{\infty}\mu_k(b\zeta)\frac{\rmd \zeta}{\zeta^4{\left(\zeta^2-1\right)}^{1/2}} \;\;(k=0,1),\nonumber\end{equation}\end{minipage} \\ \\
& & \begin{minipage}{2cm}\begin{equation}\mu_0(R)=\frac{M_\rmP(R)}{M_\rmP}, \;\;\mu_1(R)=\frac{\rmd \mu_0(R)}{\rmd \ln{R}}\nonumber\end{equation}\end{minipage}\\ \\ \\ \\
C   & \begin{minipage}{2cm}\begin{equation}b=0\nonumber\end{equation}\end{minipage}     & \begin{minipage}{2cm}\begin{equation}4\pi{\left(\frac{GM_\rmP}{\vp}\right)}^2 \int_0^{r_{\rm trunc}} \frac{\rmd R}{R}I^2_0(R)\Sigma_\rmS(R),\nonumber\end{equation}\end{minipage} \\ \\
& & \begin{minipage}{2cm}\begin{equation}\Sigma_\rmS(R)=2\int_R^{r_{\rm trunc}}\rho_\rmS(r)\frac{r\,\rmd r}{\sqrt{r^2-R^2}}\nonumber\end{equation}\end{minipage} \\ \\ \\ \\
D &  \hspace{10pt} Any $b$ & \begin{minipage}{2cm}\begin{equation}2{\left(\frac{GM_\rmP}{\vp}\right)}^2 \left[\int_0^{\infty}\rmd r\,r^2\rho_\rmS(r)\calJ(r,b)-\calV(b)\right],\nonumber\end{equation}\end{minipage}\\ \\
& & \begin{minipage}{2cm}\begin{align}\calJ(r,b)=\int_0^{\pi}\rmd \theta \sin{\theta} \int_0^{2\pi}\rmd \phi\;s^2 I^2(s),\nonumber\end{align}\end{minipage}\\ \\
& & $s^2=r^2\sin^2{\theta}+b^2-2br\sin{\theta}\sin{\phi},$\\ \\
& & \begin{minipage}{2cm}\begin{equation}\calV(b)=\frac{1}{M_\rmS}{\left[\int_0^{\infty}\rmd r\, r^2 \rho_\rmS(r)\calJ_{\rm CM}(r,b)\right]}^2,\nonumber\end{equation}\end{minipage}\\ \\
& & \begin{minipage}{2cm}\begin{equation}\calJ_{\rm CM}(r,b)=\int_0^{\pi}\rmd \theta \sin{\theta} \int_0^{2\pi}\rmd \phi\; I(s)\,\left[b-r\sin{\theta}\sin{\phi}\right],\nonumber\end{equation}\end{minipage}\\ \\
& & \begin{minipage}{2cm}\begin{equation}I(s) = \int_0^{\infty} \rmd \zeta\, \frac{1}{R_\rmP}\frac{\rmd\Tilde{\Phi}_\rmP}{\rmd R_\rmP},\nonumber\end{equation}\end{minipage}\\ \\
& & \begin{minipage}{2cm}\begin{equation}\Tilde{\Phi}_\rmP = \Phi_\rmP/(GM_\rmP),\; R_\rmP=\sqrt{s^2+\zeta^2}\nonumber\end{equation}\end{minipage} \\ \\
 \hline
\end{tabular}}
\caption{Full set of expressions needed to compute $\Delta E_{\rm int}$ (considering an impulsive encounter along a straight-line orbit) for the four cases depicted in Fig.~\ref{fig:schematic_approximations}. Column [2] lists the range of impact parameters for which these expressions are accurate. Cases A, B, C and D correspond to \protect\cite{Spitzer.58}, \protect\cite{Gnedin.etal.99}, \protect\cite{vdBosch.etal.18a}, and this chapter, respectively.}
\label{tab:comparison}
\end{table}

Although this modification by GHO99 significantly extends the range of applicability of the impulse approximation, it is still based on the DTA, which requires that $b \gg r_\rmS$.  For smaller impact parameters, $\Delta E_{\rm int}$ computed using the method of GHO99 can significantly overpredict the amount of impulsive heating (see \S\ref{sec:plummer_straight_orbit}). There is one special case, though, for which $\Delta E_{\rm int}$ can be computed analytically, which is that of a head-on encounter ($b=0$; see Case C in Fig.~\ref{fig:schematic_approximations}) when both the perturber and the subject are spherical.  In that case, as shown in \cite{Binney.Tremaine.87}, the symmetry of the problem allows a simple analytical calculation of $\Delta E_{\rm int}$ (see  Table~\ref{tab:comparison}). This was used by \citet{vdBosch.etal.18a} to argue that one may approximate $\Delta E_{\rm int}(b)$ for {\it any} impact parameter, $b$, by simply setting $\Delta E_{\rm int}(b) = \min[\Delta E_{\rm dt}(b), \Delta E_0]$.  Here $\Delta E_{\rm dt}(b)$ is the $\Delta E_{\rm int}(b)$ computed using the DTA of GHO99 (case B in  Table~\ref{tab:comparison}), and $\Delta E_0$ is the $\Delta E_{\rm int}$ for a head-on encounter (case C in  Table~\ref{tab:comparison}). Although a reasonable assumption, this approach is least accurate exactly for those impact parameters ($b \sim r_\rmS$) that statistically are expected to be most relevant\footnote{For a uniform background of perturbers, the probability that an encounter has an impact parameter in the range $b$ to $b+\rmd b$ is $P(b) \rmd b \propto b \rmd b$, such that the total $\Delta E$ due to many encounters is dominated by those with $b \sim r_\rmS$.}.

Another shortcoming of using the DTA is that $\Delta E_{\rm int}$ is found to be proportional to $\langle r^2 \rangle$, the mean squared radius of the subject (see equation~[\ref{r2aver}] and Table~1). For most density profiles typically used to model galaxies, dark matter halos, or star clusters, $\langle r^2 \rangle$ diverges, unless the asymptotic radial fall-off of the density is steeper than $r^{-5}$, or the subject is physically truncated. Although in reality all subjects are indeed truncated by an external tidal field, it is common practice to truncate the density profile of the subject at some arbitrary radius rather than a physically motivated radius. And since $\langle r^2 \rangle$ depends strongly on the truncation radius adopted (see \S\ref{sec:plummer_straight_orbit}), this can introduce large uncertainties in the amount of orbital energy transferred to internal energy during the encounter.

In this chapter, we develop a fully general, non-perturbative formalism to compute the internal energy change of a subject due to an impulsive encounter. Unlike in the DTA, we do not expand the perturber potential as a multipole series, which assures that our formalism is valid for any impact parameter. For the impulse approximation to be valid, the encounter time $\tau=b/\vp$ has to be small compared to the typical orbital timescale of the subject stars. However, in the distant tide limit, when $b$ is large, the encounter time will also typically be large, rendering the impulse approximation invalid unless $\vp$ is very large. In other words, although there are cases for which the DTA and the impulse approximation are both valid, often they are mutually exclusive.  Our formalism, being applicable to all impact parameters, is not hampered by this shortcoming. Moreover, our expression for the internal energy change does not suffer from the $\langle r^2 \rangle$ divergence issue mentioned above, but instead converges, even for infinitely extended systems. This alleviates the problem of having to truncate the galaxy at an arbitrary radius.

This chapter is organized as follows. In \S\ref{sec:straight_orbit}, we present our general formalism to compute the impulse and the energy transferred in impulsive encounters along straight-line orbits. In \S\ref{sec:special} we apply our formalism to several specific perturber density profiles. In \S\ref{sec:eccentric_orbit} we further generalize the formalism to encounters along eccentric orbits, incorporating an adiabatic correction \citep[][]{Gnedin.Ostriker.99} to account for the fact that for some subject stars, those with short dynamical times, the impact of the encounter is adiabatic rather than impulsive. In \S\ref{sec:mass_loss}, as an astrophysical application of our formalism, we discuss the mass loss of \citet{Hernquist.90} spheres due to tidal shocks during mutual encounters. Finally we summarise our findings in \S\ref{sec:conclusion}.

\section{Encounters along straight-line orbits}
\label{sec:straight_orbit}

\begin{figure}[t!]
\centering
\includegraphics[width=1\textwidth]{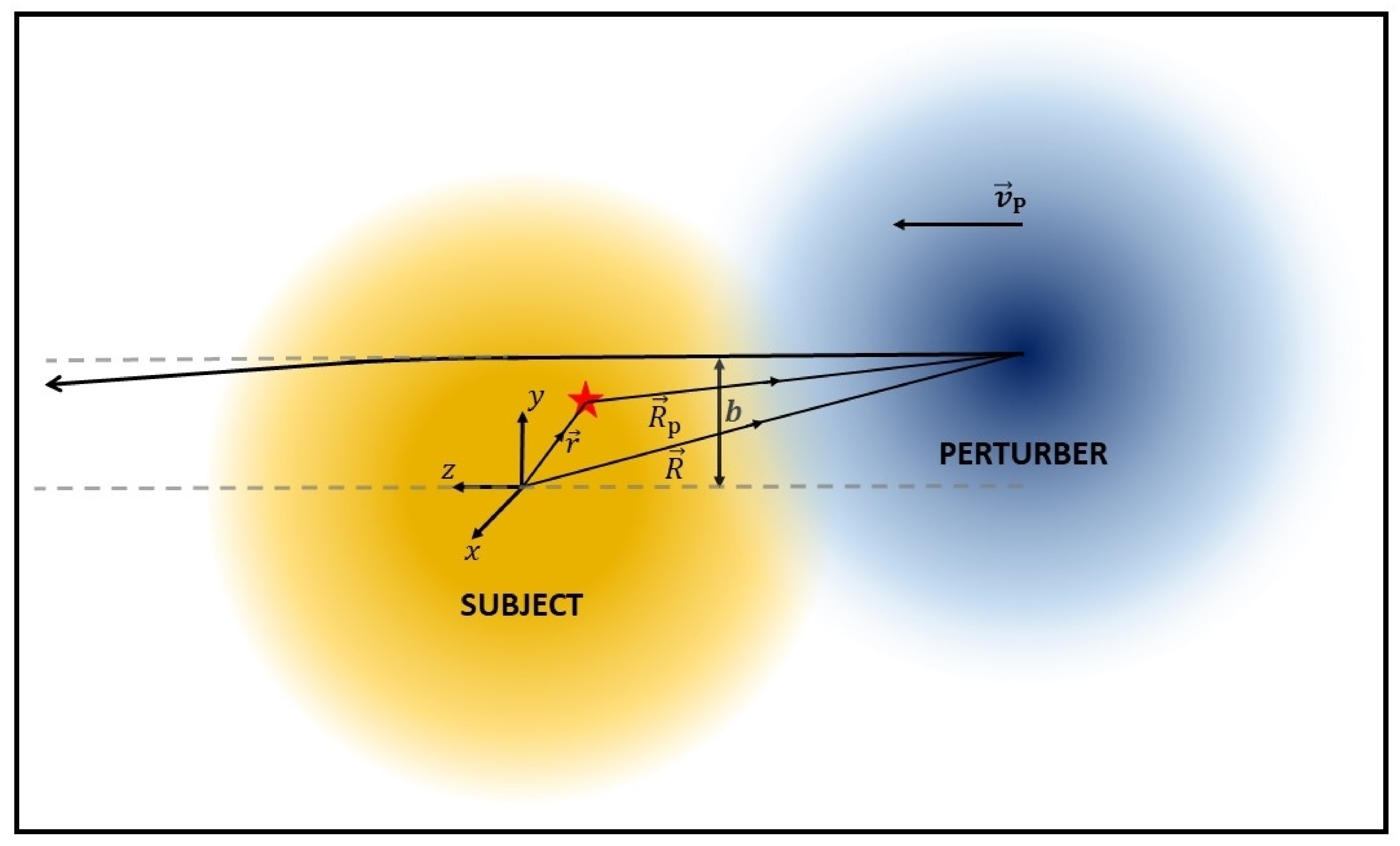}
   \caption{Illustration of the geometry of an impulsive encounter along a nearly straight orbit, specifying the coordinate axes and radial vectors used throughout this chapter.}
\label{fig:schematic_straight_orbit}
\end{figure}

Consider the gravitational encounter between two self-gravitating bodies, hereafter `galaxies'. In this section we assume that the two galaxies are mutually unbound to begin with and approach each other along a hyperbolic orbit with initial, relative velocity $\vp$ and impact parameter $b$. For sufficiently fast encounters (large $\vp$), the deflection of the galaxies from their original orbits due to their mutual gravitational interaction is small and we can approximate the orbits as a straight line. We study the impulsive heating of one of the galaxies (the subject) by the gravitational field of the other (the perturber). Throughout this chapter we always assume the perturber to be infinitely extended, while the subject is either truncated or infinitely extended. For simplicity we consider both the perturber and the subject to be spherically symmetric, with density profiles $\rho_\rmP(r)$ and $\rho_\rmS(r)$, respectively. The masses of the subject and the perturber are denoted by $M_\rmS$ and $M_\rmP$ respectively, and $r_\rmS$ and $r_\rmP$ are their scale radii. We take the centre of the unperturbed subject as the origin and define $\hat{\bz}$ to be oriented along the relative velocity $\bvp$, and $\hat{\by}$ perpendicular to $\hat{\bz}$ and directed towards the orbit of the perturber. The position vector of a star belonging to the subject is given by $\br$, that of the COM of the perturber is $\bR$ and that of the COM of the perturber with respect to the star is $\bR_\rmP=\bR-\br$ (see Fig.~\ref{fig:schematic_straight_orbit}).
\subsection{Velocity perturbation up to all orders}
\label{sec:velocity_straight_orbit}

During the encounter, the perturber exerts an external, gravitational force on each subject star. The potential due to the perturber flying by with an impact parameter $b$, on a particle located at $\br = (x,y,z)$ is a function of the distance to the particle from its center, $R_\rmP = \left|\bR-\br\right| = \sqrt{x^2+{\left(b-y\right)}^2+{\left(z-\vp t\right)}^2}$. The acceleration of the star due to the perturbing force is directed along $\widehat{\bR}_\rmP = \widehat{\bR} - \hat{\br} = \left[-x\hat{\bx}\right.$\\
$\left.+\left(b-y\right)\hat{\by}-\left(z-\vp t\right)\hat{\bz}\right]/R_\rmP$, and is equal to
\begin{align}
&\ba_\rmP = -\nabla \Phi_\rmP = \frac{1}{R_\rmP}\frac{\rmd\Phi_\rmP}{\rmd R_\rmP} \left[-x\hat{\bx}+\left(b-y\right)\hat{\by}-\left(z-\vp t\right)\hat{\bz}\right]. 
\label{force}
\end{align}
We assume that the perturber moves along a straight-line orbit from $t\rightarrow -\infty$ to $t\rightarrow \infty$. Therefore, under the perturbing force, the particle undergoes a velocity change, 
\begin{align}
&\Delta \bv = \int_{-\infty}^{\infty} \rmd t\, \ba_\rmP = \int_{-\infty}^{\infty} \rmd t\,\frac{1}{R_\rmP}\frac{\rmd\Phi_\rmP}{\rmd R_\rmP} \left[-x\hat{\bx}+\left(b-y\right)\hat{\by}-\left(z-\vp t\right)\hat{\bz}\right].
\label{deltav0}
\end{align}
The integral along $\hat{\bz}$ vanishes since the integrand is an odd function of $\left(z-\vp t\right)$. Therefore the net velocity change of the particle occurs along the $x-y$ plane and is given by 
\begin{align}
\Delta \bv = \frac{2 G M_\rmP}{\vp} I(s) \left[-x\hat{\bx} + (b-y)\hat{\by}\right],
\label{deltav1}
\end{align}
where $s^2=x^2+{\left(b-y\right)}^2$. The integral $I(s)$ is given by
\begin{align}
I(s) = \int_0^{\infty} \rmd \zeta\, \frac{1}{R_\rmP}\frac{\rmd\Tilde{\Phi}_\rmP}{\rmd R_\rmP}\,.
\label{I}
\end{align}
Here $\Tilde{\Phi}_\rmP=\Phi_\rmP/(GM_\rmP)$, $R_\rmP=\sqrt{s^2+\zeta^2}$, and $\zeta = \vp t-z$. Note that the above expression for $\Delta \bv$ is a slightly modified version of that obtained by \citet[][equation (3) of their paper]{Aguilar.White.85}. The integral $I(s)$ contains information about the impact parameter of the encounter as well as the detailed density profile of the perturber. Table.~\ref{tab:Is} lists analytical expressions for a number of different perturber potentials, including a point mass, a \cite{Plummer.11} sphere, a \cite{Hernquist.90} sphere, a NFW profile \citep[][]{Navarro.etal.97}, the Isochrone potential \citep[][]{Henon.59, Binney.14}, and a Gaussian potential. The latter is useful since realistic potentials can often be accurately represented using a multi-Gaussian expansion \citep[e.g.][]{Emsellem.etal.94, Cappellari.02}.
\begin{table*}
\centering
\scalebox{0.9}{\begin{tabular}{lllll}
 \hline
 Perturber profile & $\Phi_\rmP(r)$ & $I(s)$ \\
 (1) & (2) & (3) \\
 \hline 
 \\
Point mass      & \begin{minipage}{2cm}\begin{align}-\frac{G M_\rmP}{r}\nonumber \end{align}\end{minipage}     & \begin{minipage}{2cm}\begin{align}\frac{1}{s^2}\nonumber\end{align}\end{minipage}          
\\ \\ \\ \\
Plummer sphere      & \begin{minipage}{2cm}\begin{align}-\frac{G M_\rmP}{\sqrt{r^2+r^2_\rmP}}\nonumber\end{align}\end{minipage}  & \begin{minipage}{2cm}\begin{align}\frac{1}{s^2+r^2_\rmP}\nonumber\end{align}\end{minipage}         \\ \\ \\ \\
Hernquist sphere    & \begin{minipage}{2cm}\begin{align}-\frac{G M_\rmP}{r+r_\rmP}\nonumber\end{align}\end{minipage}  & 
\begin{minipage}{2cm}\begin{align}\frac{1}{r^2_\rmP-s^2}\left[\frac{r_\rmP}{\sqrt{r^2_\rmP-s^2}}\ln{\left(\frac{r_\rmP+\sqrt{r^2_\rmP-s^2}}{s}\right)}-1\right]\nonumber,\end{align}
\end{minipage}\\
& & $\quad \quad \quad \quad \quad \quad \quad \quad \quad \quad \quad \quad \quad \quad \quad \quad \quad \quad  s<r_\rmP\nonumber$\\ \\ \\
& & \begin{minipage}{2cm}\begin{align}\frac{1}{s^2-r^2_\rmP}\left[1-\frac{2r_\rmP}{\sqrt{s^2-r^2_\rmP}}\tan^{-1}{\sqrt{\frac{s-r_\rmP}{s+r_\rmP}}}\right],\nonumber\end{align}\end{minipage}\\
& & $\quad \quad \quad \quad \quad \quad \quad \quad \quad \quad \quad \quad \quad \quad \quad \quad \quad \quad s\geq r_\rmP\nonumber$
 \\ \\ \\ \\
 NFW profile & \begin{minipage}{2cm}\begin{align}-\frac{G M_\rmP}{r}\ln{\left(1+\frac{r}{r_\rmP}\right)}\nonumber\end{align}\end{minipage} & \begin{minipage}{2cm}\begin{align}\frac{1}{s^2}\left[\ln{\left(\frac{s}{2r_\rmP}\right)}+\frac{r_\rmP}{\sqrt{r^2_\rmP-s^2}}\ln{\left(\frac{r_\rmP+\sqrt{r^2_\rmP-s^2}}{s}\right)}\right],\nonumber\end{align}\end{minipage}\\
 & & $\quad \quad \quad \quad \quad \quad \quad \quad \quad \quad \quad \quad \quad \quad \quad \quad \quad \quad s<r_\rmP\nonumber$\\ \\ \\
& & \begin{minipage}{2cm}\begin{align}\frac{1}{s^2}\left[\ln{\left(\frac{s}{2r_\rmP}\right)}+\frac{2r_\rmP}{\sqrt{s^2-r^2_\rmP}}\tan^{-1}{\sqrt{\frac{s-r_\rmP}{s+r_\rmP}}}\right],\nonumber\end{align}\end{minipage}\\
& & $\quad \quad \quad \quad \quad \quad \quad \quad \quad \quad \quad \quad \quad \quad \quad \quad \quad \quad s\geq r_\rmP$
 \\ \\ \\ \\
Isochrone potential & \begin{minipage}{2cm}\begin{align}-\frac{G M_\rmP}{r_\rmP+\sqrt{r^2+r^2_\rmP}}\nonumber\end{align}\end{minipage} & \begin{minipage}{2cm}\begin{align}\frac{1}{s^2}-\frac{r_\rmP}{s^3}\tan^{-1}\left(\frac{s}{r_\rmP}\right)\nonumber\end{align}\end{minipage} 
\\ \\ \\ \\
Gaussian potential    &  \begin{minipage}{2cm}\begin{align}-\frac{GM_\rmP}{r_\rmP}\exp{\left[-\frac{r^2}{2r^2_\rmP}\right]}\nonumber\end{align}\end{minipage}     & \begin{minipage}{2cm}\begin{align}\frac{\sqrt{\pi}}{r^2_\rmP}\exp{\left[-\frac{s^2}{2r^2_\rmP}\right]}\nonumber\end{align}\end{minipage}         \\
 \hline
\end{tabular}}
   \caption{The $I(s)$ integral (see Eq.~\ref{I}) for different perturber profiles, where $s^2=x^2+{\left(b-y\right)}^2$ and $r^2=s^2+{\left(z-\vp t\right)}^2$. $M_\rmP$ and $r_\rmP$ are the mass and the scale radius of the perturber respectively. In case of the NFW profile, $M_\rmP=M_{\rm vir}/f(c)$ where $M_{\rm vir}$ is the virial mass and $f(c)=\ln{\left(1+c\right)}-c/(1+c)$, with $c=R_{\rm vir}/r_\rmP$ the concentration and $R_{\rm vir}$ the virial radius of the NFW perturber.}
\label{tab:Is}
\end{table*}

\subsection{Energy dissipation}
\label{sec:energy_straight_orbit}

An impulsive encounter imparts each subject star with an impulse $\Delta \bv(\br)$. During the encounter, it is assumed that the subject stars remain stagnant, such that their potential energy doesn't change. Hence, the energy change of each star is purely kinetic, and the total change in energy of the subject due to the encounter is given by
\begin{align}
\Delta E = \int \rmd^3 \br\, \rho_\rmS(\br) \, \Delta \varepsilon(\br) = \frac{1}{2}\int \rmd^3 \br\, \rho_\rmS(r) \, {\left(\Delta \bv\right)}^2.
\label{deltaE0}
\end{align}
Here we have assumed that the unperturbed subject is spherically symmetric, such that its density distribution depends only on $r =\left|\br\right|$, and $\Delta \varepsilon$ is given by equation~(\ref{dEstar}). We have assumed that the $\bv \cdot \Delta \bv$-term (see equation~[\ref{dEstar}]) in $\Delta \varepsilon$ vanishes, which is valid for any static, non-rotating, spherically symmetric subject. Plugging in the expression for $\Delta \bv$ from equation~(\ref{deltav1}), and substituting $x=r\sin{\theta}\cos{\phi}$ and $y=r\sin{\theta}\sin{\phi}$, we obtain
\begin{align}
\Delta E &= 2{\left(\frac{GM_\rmP}{\vp}\right)}^2 \int_0^{\infty} \rmd r\, r^2 \rho_\rmS(r) \calJ(r,b)\,,
\label{deltaE1}
\end{align}
where
\begin{align}
&\calJ(r,b)=\int_0^{\pi} \rmd \theta \sin{\theta} \int_0^{2\pi} \rmd\phi\,s^2 I^2(s)\,,
\label{J_1}
\end{align}
with $s^2 = x^2 + {\left(b-y\right)}^2 = r^2\sin^2{\theta} + b^2 - 2\, b\, r \sin{\theta}\sin{\phi}$. 

The above expression of $\Delta E$ includes the kinetic energy gained by the COM of the galaxy. From equation~(\ref{deltav1}), we find that the COM gains a velocity 

\begin{align}
\Delta \bv_{\rm CM} &= \frac{1}{M_\rmS}\, \int_0^\infty \rmd r\,r^2\rho_\rmS(r)\int_0^{\pi}\rmd\theta \sin{\theta}\int_0^{2\pi}\rmd \phi\, \Delta \bv \nonumber\\
&= \frac{2GM_\rmP}{\vp M_\rmS}\int_0^\infty \rmd r\,r^2\rho_\rmS(r)\calJ_{\rm CM}(r,b)\,\hat{\by}\,,
\label{deltavCM}
\end{align}
where $\calJ_{\rm CM}(r,b)$ is given by
\begin{align}
&\calJ_{\rm CM}(r,b)=\int_0^{\pi}\rmd\theta \sin{\theta} \int_0^{2\pi}\rmd \phi\, I(s)\left[b-r\sin{\theta}\sin{\phi}\right]\,.
\label{J_CM}
\end{align}
Note that $\Delta \bv_{\rm CM}$ is not the same as the velocity impulse (equation~[\ref{deltav1}]) evaluated at $\br = (0,0,0)$ since we consider perturbations up to all orders. From $\Delta \bv_{\rm CM}$, the kinetic energy gained by the COM can be obtained as follows
\begin{align}
\Delta E_{\rm CM} =\frac{1}{2}M_\rmS {\left(\Delta v_{\rm CM}\right)}^2=2{\left(\frac{GM_\rmP}{\vp}\right)}^2\calV(b),
\label{deltaECM}
\end{align}
where
\begin{align}
\calV(b)=\frac{1}{M_\rmS}{\left[\int_0^{\infty}\rmd r\, r^2 \rho_\rmS(r)\calJ_{\rm CM}(r,b)\right]}^2.
\label{V}
\end{align}

We are interested in obtaining the gain in the {\it internal} energy of the galaxy. Therefore we have to subtract the energy gained by the COM from the total energy gained, which yields the following expression for the internal energy change
\begin{align}
\Delta E_{\rm int} &= \Delta E - \Delta E_{\rm CM}=2{\left(\frac{GM_\rmP}{\vp}\right)}^2 \left[\int_0^{\infty}\rmd r\,r^2\rho_\rmS(r)\calJ(r,b)-\calV(b)\right]\,.
\label{delEint1}
\end{align}

As we show in Appendix~\ref{app:asymptote}, equation~(\ref{delEint1}) has the correct asymptotic behaviour in both the large $b$ and small $b$ limits. For large $b$ it reduces to an expression that is similar to, but also intriguingly different from the standard expression obtained using the DTA, while for $b=0$ it reduces to the expression for a head-on encounter (case C in  Table~\ref{tab:comparison}).

\section{Special cases}
\label{sec:special}

In this section we discuss two special cases of perturbers for which the expression for the impulse is analytical, and for which the expression for the internal energy change of the subject can be significantly simplified.

\subsection{Plummer perturber}
\label{sec:plummer_straight_orbit}

The first special case to be considered is that of a \cite{Plummer.11} sphere perturber, the potential and $I(s)$ of which are given in Table~\ref{tab:Is}. Substituting the latter in equation~(\ref{J_1}) and analytically computing the $\phi$ integral yields
\begin{align}
\calJ(r,b) &= \int_0^{\pi} \rmd \theta \sin{\theta} \int_0^{2\pi} \rmd\phi\,\frac{s^2}{{\left(s^2+r^2_\rmP\right)}^2} \nonumber \\
&= 4\pi\int_0^{1} \rmd \psi\,\frac{{\left(r^2-b^2-r^2\psi^2\right)}^2+r^2_\rmP\left(r^2+b^2-r^2\psi^2\right)}{{\left[{\left(r^2-b^2+r^2_\rmP-r^2\psi^2\right)}^2+4r^2_\rmP b^2\right]}^{3/2}}\,,
\label{J_plummer}
\end{align}
where $s^2 = r^2\sin^2{\theta} +b^2 - 2\,b\,r\sin{\theta}\sin{\phi}$ and $\psi=\cos{\theta}$. Similarly substituting the expression for $I(s)$ in equation~(\ref{J_CM}) yields
\begin{align}
&\calJ_{\rm CM}(r,b) = \frac{2\pi}{b} \int_0^1 \rmd \psi\,\left[1-\frac{r^2-b^2+r^2_\rmP-r^2\psi^2}{\sqrt{{\left(r^2-b^2+r^2_\rmP-r^2\psi^2\right)}^2+4r^2_\rmP b^2}}\right],
\label{J_CM_plummer}
\end{align}
which can be substituted in equation~(\ref{V}) to obtain $\calV(b)$. Both these expressions for $\calJ(r,b)$ and $\calJ_{\rm CM}(r,b)$ are easily evaluated using straightforward quadrature techniques. Finally, upon substituting $\calJ$ and $\calV$ in equation~(\ref{delEint1}), we obtain the internal energy change $\Delta E_{\rm int}$ of the subject.
\begin{figure}[t!]
\centering
\includegraphics[width=1\textwidth]{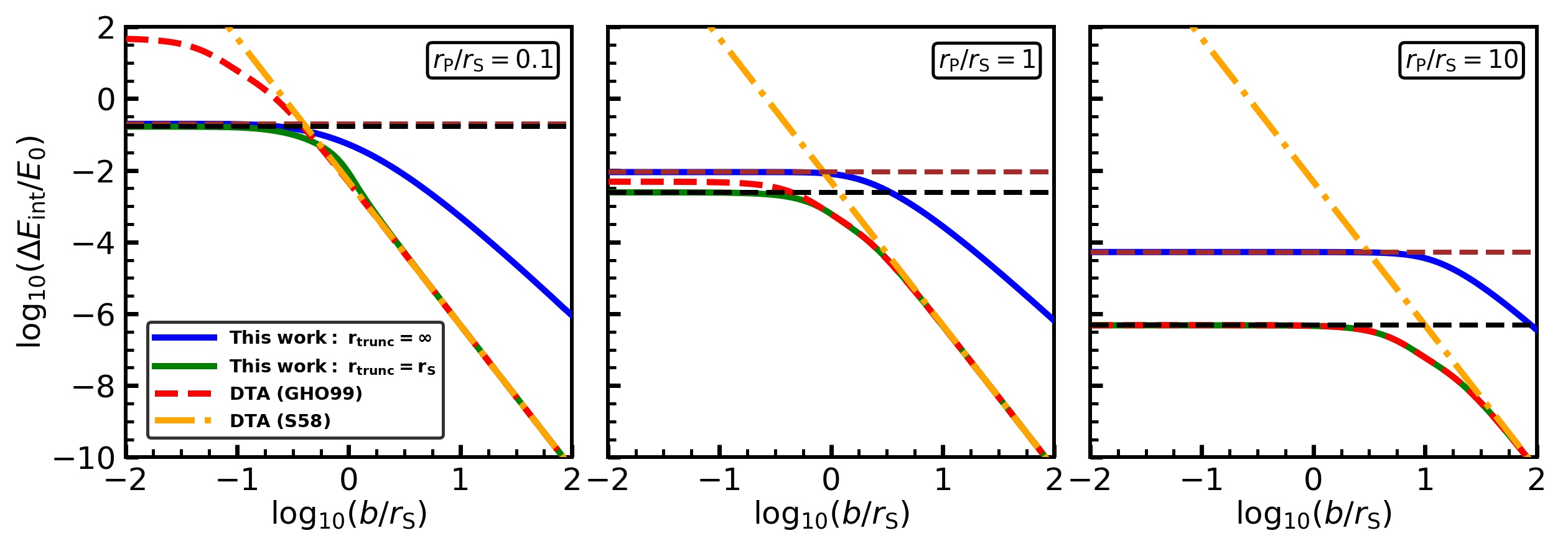}
   \caption{Impulsive heating for encounters along straight-line orbits: Each panel plots $\Delta E_{\rm int}$ in units of $E_0 = 8 \pi \, (G M_\rmP / \vp)^2 \, (M_\rmS/r^2_\rmS)$ as a function of the impact parameter $b$ in units of $r_\rmS$. Perturber and subject are modelled as Plummer and Hernquist spheres, respectively, with different panels showing results for different ratios of their characteristic radii, as indicated. The solid blue and green lines indicate $\Delta E_{\rm int}$ for infinitely extended and truncated ($r_{\rm trunc}=r_\rmS$) subjects, respectively, computed using our generalized framework (equation[~\ref{delEint1}]). The red, dashed and the orange, dot-dashed lines indicate the $\Delta E_{\rm int}$ for the truncated subject obtained using the DTA of GHO99 and S58, respectively. The brown and black dashed horizontal lines mark the head-on encounter limits for the infinite and the truncated subjects, respectively. Note that the asymptotic fall-off for the infinitely extended case (solid blue) is shallower than for the truncated case (solid green), which approaches the distant tide limit (dashed red and dot-dashed orange) for large $b$ and saturates to the head-on encounter limit for small $b$. Also note that the GHO99 approximation is in good agreement with the general result as long as the DTA is valid (i.e., $b/r_\rmS$ is large), and/or $r_\rmP$ is significantly larger than $r_\rmS$.}
\label{fig:straight_orbit}
\end{figure}
Fig.~\ref{fig:straight_orbit} plots the resulting $\Delta E_{\rm int}$, in units of $8\pi{\left(G M_\rmP/\vp\right)}^2\left(M_\rmS/r^2_\rmS\right)$, as a function of the impact parameter, $b$, for a spherical subject with a \cite{Hernquist.90} density profile. Different panels correspond to different ratios of the characteristic radii of the perturber, $r_\rmP$, and the subject, $r_\rmS$, as indicated. Solid blue lines  indicate the $\Delta E_{\rm int}$ obtained using our non-perturbative method (equation~[\ref{delEint1}]) for an infinitely extended subject, while the solid green lines show the corresponding results for a subject truncated at $r_\rmS$. For comparison, the red, dashed and orange, dot-dashed lines show the  $\Delta E_{\rm int}$ obtained using the DTA of S58 and GHO99 (cases A and B in Table~\ref{tab:comparison}), respectively, also assuming a Hernquist subject truncated at $r_\rmS$. Finally, the black and brown horizontal, dashed lines mark the values of $\Delta E_{\rm int}$ for a head-on encounter obtained using the expression of \citet{vdBosch.etal.18a} (case C in  Table~\ref{tab:comparison}) for a truncated and infinitely extended subject, respectively.

Note that $\Delta E_{\rm int}$ for the infinitely extended subject has a different asymptotic behaviour for large $b$ than the truncated case. In fact $\Delta E_{\rm int} \propto b^{-3}$ in the case of an infinitely extended Hernquist subject (when using our non-perturbative formalism), whereas $\Delta E_{\rm int} \propto b^{-4}$ for a truncated subject (see \S\ref{sec:asymptote_tidal} for more details).

For large impact parameters, our non-perturbative $\Delta E_{\rm int}$ for the truncated case (solid green line) is in excellent agreement with the DTA of S58 and GHO99, for all three values of $r_\rmP/r_\rmS$. In the limit of small $b$, though, the different treatments yield very different predictions; whereas the $\Delta E_{\rm int}$ computed using the method of S58 diverges as $b^{-4}$, the correction of GHO99 causes $\Delta E_{\rm int}$ to asymptote to a finite value as $b \rightarrow 0$, but one that is significantly larger than what is predicted for a head-on encounter (at least when $r_\rmP < r_\rmS$). We emphasize, though, that both the S58 and GHO99 formalisms are based on the DTA, and therefore not valid in this limit of small $b$. In contrast, our non-perturbative method is valid for all $b$, and nicely asymptotes to the value of a head-on encounter in the limit $b \rightarrow 0$.

It is worth pointing out that the GHO99 formalism yields results that are in excellent agreement with our fully general, non-perturbative approach when $r_\rmP/r_\rmS \gg 1$, despite the fact that it is based on the DTA. However, this is only the case when the subject is truncated at a sufficiently small radius $r_{\rm trunc}$. Recall that the DTA yields that $\Delta E_{\rm int} \propto \langle r^2 \rangle$ (see Table~\ref{tab:comparison}), which diverges unless the subject is truncated or the outer density profile of the subject has $\rmd\log\rho_\rmS/\rmd\log r < -5$. In contrast, our generalized formalism yields a finite $\Delta E_{\rm int}$, independent of the density profile of the subject.

\begin{figure}[t!]
\centering
\includegraphics[width=1\textwidth]{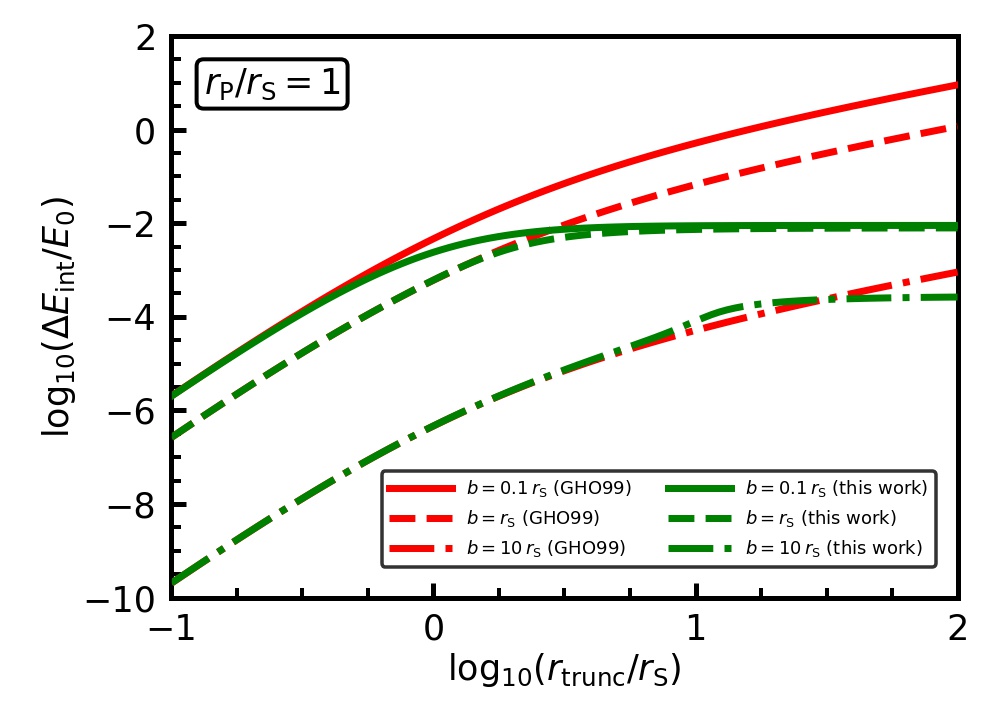}
   \caption{The increase in internal energy, $\Delta E_{\rm int}$, in units of $E_0=8\pi{\left(G M_\rmP/\vp\right)}^2\left(M_\rmS/r^2_\rmS\right)$, of a truncated Hernquist sphere due to an impulsive encounter with a Plummer sphere perturber with $r_\rmP/r_\rmS=1$ along a straight-line orbit. Results are shown as a function of the subject's truncation radius, $r_{\rm trunc}$, in units of $r_\rmS$, for three values of the impact parameter, $b/r_\rmS$, as indicated. Green and red lines correspond to the  $\Delta E_{\rm int}$ computed using our generalized framework and the DTA of GHO99, respectively.}
\label{fig:straight_orbit_rc}
\end{figure}

This is illustrated in Fig.~\ref{fig:straight_orbit_rc} which plots $\Delta E_{\rm int}$, in units of $8\pi{\left(G M_\rmP/\vp\right)}^2\left(M_\rmS/r^2_\rmS\right)$, as a function of $r_{\rm trunc}/r_\rmS$ for a Plummer perturber and a truncated Hernquist subject with $r_\rmP/r_\rmS=1$. Results are shown for three different impact parameters, as indicated. The green and red lines indicate the $\Delta E_{\rm int}$ obtained using our general formalism and that of GHO99, respectively. Note that the results of GHO99 are only in good agreement with our general formalism when the truncation radius is small and/or the impact parameter is large. 
\subsection{Point mass perturber}
\label{sec:point_mass_straight_orbit}

The next special case to discuss is that of a point mass perturber, which one can simply obtain by taking the results for a spherical Plummer perturber discussed above in the limit $r_\rmP \rightarrow 0$. In this limit the $\calJ$ integral of equation~(\ref{J_plummer}) can be computed analytically and substituted in equation~(\ref{deltaE1}) to yield 
\begin{align}
&\Delta E = 4\pi{\left(\frac{GM_\rmP}{\vp}\right)}^2 \int_0^{\infty} \rmd r\, r^2 \rho_\rmS(r) \int_0^{\pi} \rmd \theta \frac{\sin{\theta}}{\left|b^2-r^2\sin^2{\theta}\right|}\,.
\label{deltaE1_point}
\end{align}
The same applies to the $\calJ_{\rm CM}$ integral of equation~(\ref{J_CM_plummer}), which yields the following COM velocity
\begin{align}
\Delta \bv_{\rm CM} = \frac{2GM_\rmP}{\vp M_\rmS}\frac{M_{\rm enc}(b)}{b}\hat{\by}\,,
\label{deltavCM_ptmass}
\end{align}
where $M_{\rm enc}(b)$ is the galaxy mass enclosed within a cylinder of radius $b$, and is given by
\begin{align}
&M_{\rm enc}(b) = 4\pi \left[\int_0^b \rmd r\, r^2\rho_\rmS(r) + \int_b^\infty \rmd r\,r^2\rho_\rmS(r)\left(1-\sqrt{1-\frac{b^2}{r^2}}\right) \right].
\label{Menc}
\end{align}
Therefore, the kinetic energy gained by the COM in the encounter can be written as
\begin{align}
\Delta E_{\rm CM} = \frac{1}{2 M_\rmS} {\left[\frac{2GM_\rmP}{\vp}\frac{M_{\rm enc}(b)}{b}\right]}^2.
\label{deltaECM_ptmass}
\end{align}
Subtracting this from the expression for $\Delta E$ given in equation~(\ref{deltaE1_point}) and analytically computing the $\theta$ integral yields the following expression for the internal energy change
\begin{align}
\Delta E_{\rm int} &= 8\pi {\left(\frac{GM_\rmP}{\vp}\right)}^2 \int_0^{r_{\rm trunc}} \rmd r\,\rho_\rmS(r)\left[\frac{r}{\sqrt{b^2-r^2}}\tan^{-1}{\left(\frac{r}{\sqrt{b^2-r^2}}\right)}-\frac{r^2}{b^2}\right]\,.
\label{deltaEint_ptmass}
\end{align}
Here we assume the subject to be truncated at some $r_{\rm trunc}<b$, and therefore $M_{\rm enc}(b)=M_\rmS$. If $r_{\rm trunc}>b$, then the point perturber passes through the subject and imparts an infinite impulse in its neighbourhood, which ultimately leads to a divergence of $\Delta E_{\rm int}$. 

Note that the term in square brackets tends to $\frac{2}{3} (r/b)^4$ in the limit $r \ll b$. Hence, the above expression for $\Delta E_{\rm int}$ reduces to the standard distant tide expression of S58, given in equation~(\ref{dESpitzer}), as long as $b \gg r_{\rm trunc}$.
Unlike S58 though, the above expression for $\Delta E_{\rm int}$ is applicable for any $b>r_{\rm trunc}$, and is therefore a generalization of the former.

\subsection{Other perturbers}
\label{sec:other_perturbers}

The Plummer and point-mass perturbers discussed above are somewhat special in that the corresponding expression for the impulse is sufficiently straightforward that the expression for $\Delta E_{\rm int}$ (equation~[\ref{delEint1}]) simplifies considerably. For the other perturber profiles listed in Table~\ref{tab:Is}, $\Delta E_{\rm int}$ is to be computed by numerically evaluating the $\calJ$ and $\calJ_{\rm CM}$ integrals given in equations~(\ref{J_1}) and~(\ref{J_CM}), respectively. We provide a Python code, {\tt NP-impulse}\footnote{\url{https://github.com/uddipanb/NP-impulse}}, that does so, and that can be used to compute  $\Delta E_{\rm int}(b,v)$ for a variety of (spherical) perturber and subject  profiles. We emphasize that the results are in good agreement with the estimates of GHO99, which are based on the DTA, when (i) the perturber is sufficiently extended (i.e., $r_\rmP > r_\rmS$), and (ii) the subject is truncated at a radius $r_{\rm trunc} < b$. When these conditions are not met, the GHO99 formalism typically significantly overpredicts $\Delta E_{\rm int}$ at small impact parameters. Our more general formalism, on the other hand, remains valid for any $b$ and any $r_{\rm trunc}$ (including no truncation), and smoothly asymptotes to the analytical results for a head-on encounter. 

\section{Encounters along eccentric orbits}
\label{sec:eccentric_orbit}

In the previous sections we have discussed how a subject responds to a perturber that is moving along a straight-line orbit. The assumption of a straight-line orbit is only reasonable in the highly impulsive regime, when $\vp \gg \sigma$. Such situations do occur in astrophysics (i.e., two galaxies having an encounter within a cluster, or a close encounter between two globular clusters in the Milky Way). However, one also encounters cases where the encounter velocity is largely due to the subject and perturber accelerating each other (i.e., the future encounter of the Milky Way and M31), or in which the subject is orbiting within the potential of the perturber (i.e., M32 orbiting M31). In these cases, the assumption of a straight-line orbit is too simplistic.  In this section we therefore generalize the straight-line orbit formalism developed in \S\ref{sec:straight_orbit}, to the case of subjects moving on eccentric orbits within the perturber potential. Our approach is similar to that in GHO99, except that we refrain from using the DTA, i.e., we do not expand the perturber potential in multi-poles and we do not assume that  $r_\rmP \gg r_\rmS$. Rather our formalism is applicable to any sizes of the subject and the perturber. In addition, our formalism is valid for any impact parameter (which here corresponds to the pericentric distance of the eccentric orbit), whereas the formalism of GHO99 is formally only valid for $b \gg r_\rmS$. However, like GHO99, our formalism is also based on the impulse approximation, which is only valid as long as the orbit is sufficiently eccentric such that the encounter time, which is of order the timescale of pericentric passage, is shorter than the average orbital timescale of the subject stars. Since the stars towards the central part of the subject orbit much faster than those in the outskirts, the impulse approximation can break down for stars near the centre of the subject, for whom the encounter is adiabatic rather than impulsive. As discussed in \S\ref{sec:adiabatic_shielding}, we can take this `adiabatic shielding' into account using the adiabatic correction formalism introduced by \citet{Gnedin.Ostriker.99}. This correction becomes more significant for less eccentric orbits.

\subsection{Orbit characterization}
\label{sec:eccentric_orbit_character}

We assume that the perturber is much more massive than the subject ($M_\rmP \gg M_\rmS$) and therefore governs the motion of the subject. We also assume that the perturber is spherically symmetric, which implies that the orbital energy and angular momentum of the subject are conserved and that its orbit is restricted to a plane. This orbital energy and angular momentum (per unit mass) are given by
\begin{align}
E &= \frac{1}{2}\dot{R}^2+\Phi_\rmP(R)+\frac{L^2}{2R^2}, \nonumber \\
L &= R^2\dot{\theta}_\rmP,
\label{EL}
\end{align}
where $\bR$ is the position vector of the COM of the perturber with respect to that of the subject, $R= \left|\bR\right|$, and $\theta_\rmP$ is the angle on the orbital plane defined such that $\theta_\rmP = 0$ when $R$ is equal to the pericentric distance, $R_{\rm peri}$. The dots denote derivatives with respect to time. The above equations can be rearranged and integrated to obtain the following forms for $\theta_\rmP$ and $t$ as functions of $R$
\begin{align}
&\theta_\rmP(R) = \int_{1/\alpha}^{R/r_\rmP} \frac{\rmd R'}{R'^2\sqrt{2\left[\calE-\Phi'_\rmP(R')\right]/\ell^2-1/R'^2}},\nonumber \\
&t(R) = \int_{1/\alpha}^{R/r_\rmP} \frac{\rmd R'}{\ell\sqrt{2\left[\calE-\Phi'_\rmP(R')\right]/\ell^2-1/R'^2}}.
\end{align}
Here $\alpha=r_\rmP/R_{\rm peri}$, $t$ is in units of ${\left(r^3_\rmP/GM_\rmP\right)}^{1/2}$, and $\calE=E \left(r_\rmP /GM_\rmP\right)$, $\Phi'_\rmP=\Phi_\rmP \left(r_\rmP/GM_\rmP\right)$ and $\ell=L/{\left(GM_\rmP r_\rmP\right)}^{1/2}$ are dimensionless expressions for the orbital energy, perturber potential and orbital angular momentum, respectively. The resulting orbit is a rosette, with $R$ confined between a pericentric distance, $R_{\rm peri}$, and an apocentric distance, $R_{\rm apo}$. The angle between a pericenter and the subsequent apocenter is $\theta_{\rm max}$, which ranges from $\pi/2$ for the harmonic potential to $\pi$ for the Kepler potential \citep[e.g.,][]{Binney.Tremaine.87}. The orbit's eccentricity is defined as 
\begin{align}
e = \frac{R_{\rm apo}-R_{\rm peri}}{R_{\rm apo}+R_{\rm peri}},
\end{align}
which ranges from $0$ for a circular orbit to $1$ for a purely radial orbit. Here we follow GHO99 and characterize an orbit by $e$ and $\alpha = r_\rmP/R_{\rm peri}$.

\subsection{Velocity perturbation and energy dissipation}
\label{sec:velocity_energy_eccentric_orbit}

The position vector of the perturber with respect to the subject is given by $\bR=R \cos{\theta_\rmP}\hat{\by}+R\sin{\theta_\rmP}\hat{\bz}$, where we take the orbital plane to be spanned by the $\hat{\by}$ and $\hat{\bz}$ axes, with $\hat{\by}$ directed towards the pericenter. The function $R(\theta_\rmP)$ specifies the orbit of the subject in the perturber potential and is therefore a function of the orbital parameters $\alpha$ and $e$. In the same spirit as in equation~(\ref{force}), we write the acceleration due to the perturber on a subject star located at $(x,y,z)$ from its COM as
\begin{align}
&\ba_\rmP = -\nabla \Phi_\rmP = \frac{1}{R_\rmP}\frac{\rmd\Phi_\rmP}{\rmd R_\rmP} \left[-x\hat{\bx}+\left(R\cos{\theta_\rmP}-y\right)\hat{\by}+\left(R\sin{\theta_\rmP}-z\right)\hat{\bz}\right], 
\label{force_eccentric}
\end{align}
where $R_\rmP=\sqrt{x^2+{\left(R\cos{\theta_\rmP}-y\right)}^2+{\left(R\sin{\theta_\rmP}-z\right)}^2}$ is the distance of the star from the perturber. We are interested in the response of the subject during the encounter, i.e., as the perturber moves (in the reference frame of the subject) from one apocenter to another, or equivalently from $(R_{\rm apo},-\theta_{\rm max})$ to $(R_{\rm apo},\theta_{\rm max})$. During this period, $T$, the star particle undergoes a velocity perturbation $\Delta \bv$, given by
\begin{align}
\Delta \bv &= \int_{-T/2}^{T/2} \rmd t\, \ba_\rmP \nonumber \\
&= \frac{1}{L} \int_{-\theta_{\rm max}}^{\theta_{\rm max}} \rmd \theta_\rmP R^2(\theta_\rmP) \,\frac{1}{R_\rmP}\frac{\rmd\Phi_\rmP}{\rmd R_\rmP}\left[-x\hat{\bx}+\left(R\cos{\theta_\rmP}-y\right)\hat{\by}+\left(R\sin{\theta_\rmP}-z\right)\hat{\bz}\right],
\label{deltav0_eccentric}
\end{align}
where we have substituted $\theta_\rmP$ for $t$ by using the fact that $\dot{\theta}_\rmP = L/R^2$. Also, using that $L=\ell\sqrt{GM_\rmP r_\rmP}$ and $\Tilde{\Phi}_\rmP = \Phi_\rmP/(GM_\rmP)$, the above expression for $\Delta \bv$ can be more concisely written as
\begin{align}
\Delta \bv &= \sqrt{\frac{GM_\rmP}{r_\rmP}}\frac{1}{\ell(\alpha,e)}\left[-x I_1 \hat{\bx}+\left(I_2-y I_1\right) \hat{\by}+\left(I_3-z I_1\right) \hat{\bz}\right],
\label{deltav0_eccentric_conc}
\end{align}
where
\begin{align}
I_1(\br) &= \int_{-\theta_{\rm max}}^{\theta_{\rm max}} \rmd \theta_\rmP\, R^2(\theta_\rmP) \frac{1}{R_\rmP}\frac{\rmd\Tilde{\Phi}_\rmP}{\rmd R_\rmP}, \nonumber \\
I_2(\br) &= \int_{-\theta_{\rm max}}^{\theta_{\rm max}} \rmd \theta_\rmP\cos{\theta_\rmP}\, R^3(\theta_\rmP) \frac{1}{R_\rmP}\frac{\rmd\Tilde{\Phi}_\rmP}{\rmd R_\rmP}, \nonumber \\
I_3(\br) &= \int_{-\theta_{\rm max}}^{\theta_{\rm max}} \rmd \theta_\rmP\sin{\theta_\rmP}\, R^3(\theta_\rmP) \frac{1}{R_\rmP}\frac{\rmd\Tilde{\Phi}_\rmP}{\rmd R_\rmP}.
\end{align}
Note that $I_1$ has units of inverse length, while $I_2$ and $I_3$ are unitless.

Over the duration of the encounter, the COM of the subject (in the reference frame of the perturber) undergoes a velocity change 
\begin{align}
\Delta \bv_{\rm CM} & = 2 \, R_{\rm apo}\, \dot{\theta}_\rmP\vert_{\rm apo} \, \sin{\theta_{\rm max}} \, \hat{\by} = 2 \, \sqrt{\frac{GM_\rmP}{r_\rmP}} \, \alpha \, \ell(\alpha,e) \, \frac{1-e}{1+e} \, \sin{\theta_{\rm max}} \, \hat{\by}.
\end{align}
Subtracting this $\Delta \bv_{\rm CM}$ from $\Delta \bv$, we obtain the velocity perturbation $\Delta \bv_{\rm rel} = \Delta \bv - \Delta \bv_{\rm CM}$ relative to the COM of the subject, which implies a change in internal energy given by
\begin{align}
\Delta E_{\rm int} = \frac{1}{2}\int_0^\infty \rmd r\,r^2\rho_\rmS(r)\int_0^{\pi}\rmd\theta \sin{\theta}\int_0^{2\pi}\rmd \phi\, \Delta v^2_{\rm rel}.
\end{align}
Substituting the expression for $\Delta \bv$ given by equation~(\ref{deltav0_eccentric_conc}), we have that
\begin{align}
\Delta E_{\rm int} &= \frac{GM_\rmP}{2r_\rmP}\int_0^\infty \rmd r\,r^2\rho_\rmS(r)\int_0^{\pi}\rmd\theta \sin{\theta}\int_0^{2\pi}\rmd \phi\,\, \calK (\br).
\label{deltaEint_eccentric}
\end{align}
Here the function $\calK (\br)$ is given by
\begin{align}
\calK(\br) = \frac{r^2 \,I^2_1 + I'^2_2 + I^2_3 - 2\, r \,I_1  \left(I'_2\sin{\theta}\sin{\phi} + I_3\cos{\theta} \right)}{\ell^2(\alpha,e)},
\label{calK}
\end{align}
where $I'_2 = I_2 -\Delta \Tilde{v}_{\rm CM}$, with 
\begin{align}
\Delta \Tilde{v}_{\rm CM}= 2 \alpha \, \ell^2(\alpha,e) \, \frac{1-e}{1+e} \, \sin{\theta_{\rm max}}.
\end{align}

Finally, from the conservation of energy and equation~(\ref{EL}), it is straightforward to  infer that\footnote{Analytical expressions for a few specific perturber potentials are listed in Table~1 of GHO99.} 
\begin{align}
{\ell}^2(\alpha,e) = \frac{{\left(1+e\right)}^2}{2e} \, \frac{r_\rmP}{\alpha^2} \, \left[\Tilde{\Phi}_\rmP \left(\frac{r_\rmP}{\alpha} \frac{1+e}{1-e}\right) - \Tilde{\Phi}_\rmP \left(\frac{r_\rmP}{\alpha}\right) \right].
\end{align}
\begin{figure*}[t!]
\centering
\includegraphics[width=1\textwidth]{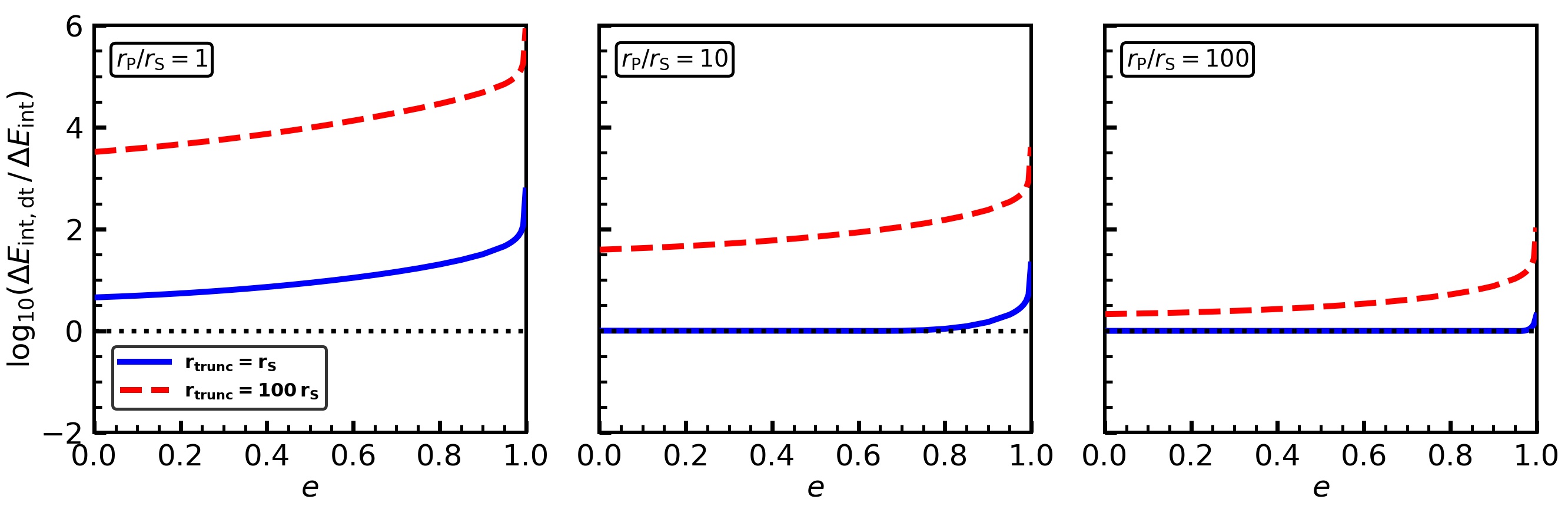}
   \caption{Impulsive heating for encounters along eccentric orbits:  Blue, solid and red, dashed lines indicate the ratio of $\Delta E_{\rm int}$ computed using the DTA of GHO99 ($\Delta E_{\rm {int,dt}}$) to that computed using our general formalism (equation~[\ref{deltaEint_eccentric_ad_corr}]) as a function of the orbital eccentricity, $e$, for cases in which the spherical Hernquist subject is truncated at $r_{\rm trunc}=r_\rmS$ and $100\,r_\rmS$, respectively. In each case, the orbital energy is $E = -0.7GM_\rmP/r_\rmP$, and the perturber is modelled as a Hernquist sphere with $M_\rmp = 1000 M_\rmS$ (here $M_\rmS$ is the subject mass enclosed within its truncation radius). Different panels correspond to different $r_\rmP/r_\rmS$, as indicated.}
\label{fig:eccentric_orbit}
\end{figure*}
\subsection{Adiabatic correction}
\label{sec:adiabatic_shielding}

The expression for the internal energy change of the subject derived in the previous section (equation~[\ref{deltaEint_eccentric}]) is based on the impulse approximation. This assumes that during the encounter the stars only respond to the perturbing force and not to the self-gravity of the subject. However, unless the encounter speed is much larger than the internal velocity dispersion of the subject, this is a poor approximation towards the center of the subject, where the dynamical time of the stars, $t_{\rm dyn}(r) \propto \left[G \rho_\rmS(r)\right]^{-1/2}$ can be comparable to, or even shorter than, the time scale of the encounter $\tau$. For such stars the encounter is not impulsive at all; in fact, if $t_{\rm dyn}(r) \ll \tau$ the stars respond to the encounter adiabatically, such that the net effect of the encounter leaves their energy and angular momentum invariant. In this section we modify the expression for $\Delta E_{\rm int}$ derived above by introducing an adiabatic correction to account for the fact that the central region of the subject may be  `adiabatically shielded' from the tidal shock.

We follow \citet{Gnedin.Ostriker.99} who, using numerical simulations and motivated by \cite{Weinberg.94a, Weinberg.94b}, find that the ratio of the actual, average energy change $\langle \Delta E \rangle(r)$ for subject stars at radius $r$ to that predicted by the impulse approximation, is given by
\begin{align}
\calA(r) = \left[ 1 + \omega^2(r) \tau^2 \right]^{-\gamma}.
\label{ad_corr}
\end{align}
Here $\tau$ is the shock duration, which is of order the timescale of pericentric passage, i.e.,
\begin{align}
\tau \sim \frac{1}{\dot{\theta}_\rmP\vert_{\rm peri}} = \sqrt{\frac{r^3_\rmP}{G M_\rmP}}\frac{1}{\alpha^2\,\ell(\alpha,e)}\,,
\end{align}
and $\omega(r) = \sigma(r)/r$ is the frequency of subject stars at radius $r$, with $\sigma(r)$ the isotropic velocity dispersion given by
\begin{align}
\sigma^2(r) = \frac{1}{\rho_\rmS(r)} \int_r^{\infty} \rmd r' \, \rho_\rmS(r') \, \frac{\rmd\Phi_\rmS}{\rmd r'}.
\end{align}

For the power-law index $\gamma$, \citet{Gnedin.Ostriker.99} find that it obeys
\begin{align}
\gamma&=
    \begin{cases}
     2.5, &\tau \lesssim t_{\rm dyn} \\
     1.5, &\tau \gtrsim 4\,t_{\rm dyn},
    \end{cases}
\end{align}
where
\begin{align}
t_{\rm dyn}=\sqrt{\frac{\pi^2 r^3_\rmh}{2GM_\rmS}}
\end{align}
is the dynamical time at the half mass radius $r_\rmh$ of the subject. In what follows we therefore adopt
\begin{align}
 \gamma = 2 - 0.5\,\rm erf \left(\frac{\tau-2.5\,t_{\rm dyn}}{0.7\,t_{\rm dyn}}\right)
\end{align}
as a smooth interpolation between the two limits. Implementing this adiabatic correction, we arrive at the following final expression for the internal energy change of the subject during its encounter with the perturber
\begin{align}
\Delta E_{\rm int} = \frac{GM_\rmP}{2r_\rmP} \int_0^\infty \rmd r \, r^2 \, \rho_\rmS(r) \, \calA(r) \, \int_0^{\pi} \rmd\theta \, \sin{\theta} \, \int_0^{2\pi} \rmd \phi\,\calK (\br)\,.
\label{deltaEint_eccentric_ad_corr}
\end{align}

We caution that the adiabatic correction formalism of \citet{Gnedin.Ostriker.99} has not been tested in the regime of small impact parameters. In addition, ongoing studies suggest that equation~(\ref{ad_corr}) may require a revision for the case of extensive tides \citep[][]{Martinez-Medina.etal.20}. Hence, until an improved and well-tested formalism for adiabatic shielding is developed, the results in this subsection have to be taken with a grain of salt. However, as long as the adiabatic correction remains a function of radius only, equation~(\ref{deltaEint_eccentric_ad_corr}) remains valid.

In Fig.~\ref{fig:eccentric_orbit}, we compare this $\Delta E_{\rm int}$ with that computed using the DTA of GHO99, which can be written in the form of equation~(\ref{deltaEint_eccentric_ad_corr}) but with $\calK(\br)$ replaced by
\begin{align}
    \calK_{\rm GHO}(\br) = \left({r \over r_\rmP} \right)^2 \, {(B_1 - B_3)^2\sin^2{\theta}\sin^2{\phi} + (B_2 - B_3)^2\cos^2{\theta} + B_3^2\sin^2{\theta}\cos^2{\phi} \over \, \ell^2(\alpha,e)},
\label{calK_GHO}
\end{align}
with $B_1$, $B_2$ and $B_3$ integrals, given by equations~(36), (37) and~(38) in GHO99, that carry information about the perturber profile and the orbit. The lines show the ratio of $\Delta E_{\rm int}$ computed using GHO99's DTA and that computed using our formalism (equations~[\ref{deltaEint_eccentric_ad_corr}] and~[\ref{calK}]) as a function of the orbital eccentricity $e$, and for an orbital energy $E=-0.7GM_\rmP/r_\rmP$.  Both the perturber and the subject are modelled as Hernquist spheres. Solid blue and dashed red lines correspond to cases in which the subject is truncated at $r_{\rm trunc} = r_\rmS$ and $100\,r_\rmS$, respectively, while different panels correspond to different ratios of $r_\rmP/r_\rmS$, as indicated.

The GHO99 results are in excellent agreement with our more general formalism when $r_{\rm trunc}=r_\rmS$ and $r_\rmP / r_\rmS \gg 1$. Note, though, that the former starts to overpredict $\Delta E_{\rm int}$ in the limit $e \to 1$. The reason is that for higher eccentricities, the pericentric distance becomes smaller and the higher-order multipoles of the perturber potential start to contribute more. Since the DTA truncates $\Phi_\rmP$ at the quadrupole, it becomes less accurate. As a consequence, the GHO99 results actually diverge in the limit $e \to 1$, while the $\Delta E_{\rm int}$ computed using our fully general formalism continues to yield finite values. The agreement between our $\Delta E_{\rm int}$ and that computed using the GHO99 formalism becomes worse for smaller $r_\rmP/r_\rmS$ and larger $r_{\rm trunc}$. When $r_\rmP/r_\rmS=1$ (left-hand panel), GHO99 overpredicts $\Delta E_{\rm int}$ by about one to two orders of magnitude when $r_{\rm trunc} = r_\rmS$, which increases to 3 to 5 orders of magnitude for  $r_{\rm trunc} = 100\,r_\rmS$. Once again, this sensitivity to $r_{\rm trunc}$ has its origin in the fact that the integral $\int_0^{r_{\rm trunc}}\rmd r\,r^4\,\rho_\rmS(r)\,\calA(r)$ diverges as $r_{\rm trunc} \to \infty$ for the Hernquist $\rho_\rmS(r)$ considered here.

\section{Mass Loss due to Tidal Shocks in Equal Mass Encounters}
\label{sec:mass_loss}

\begin{figure}
\centering
\hspace{-1mm}
\subfloat{\includegraphics[width=0.95\textwidth]{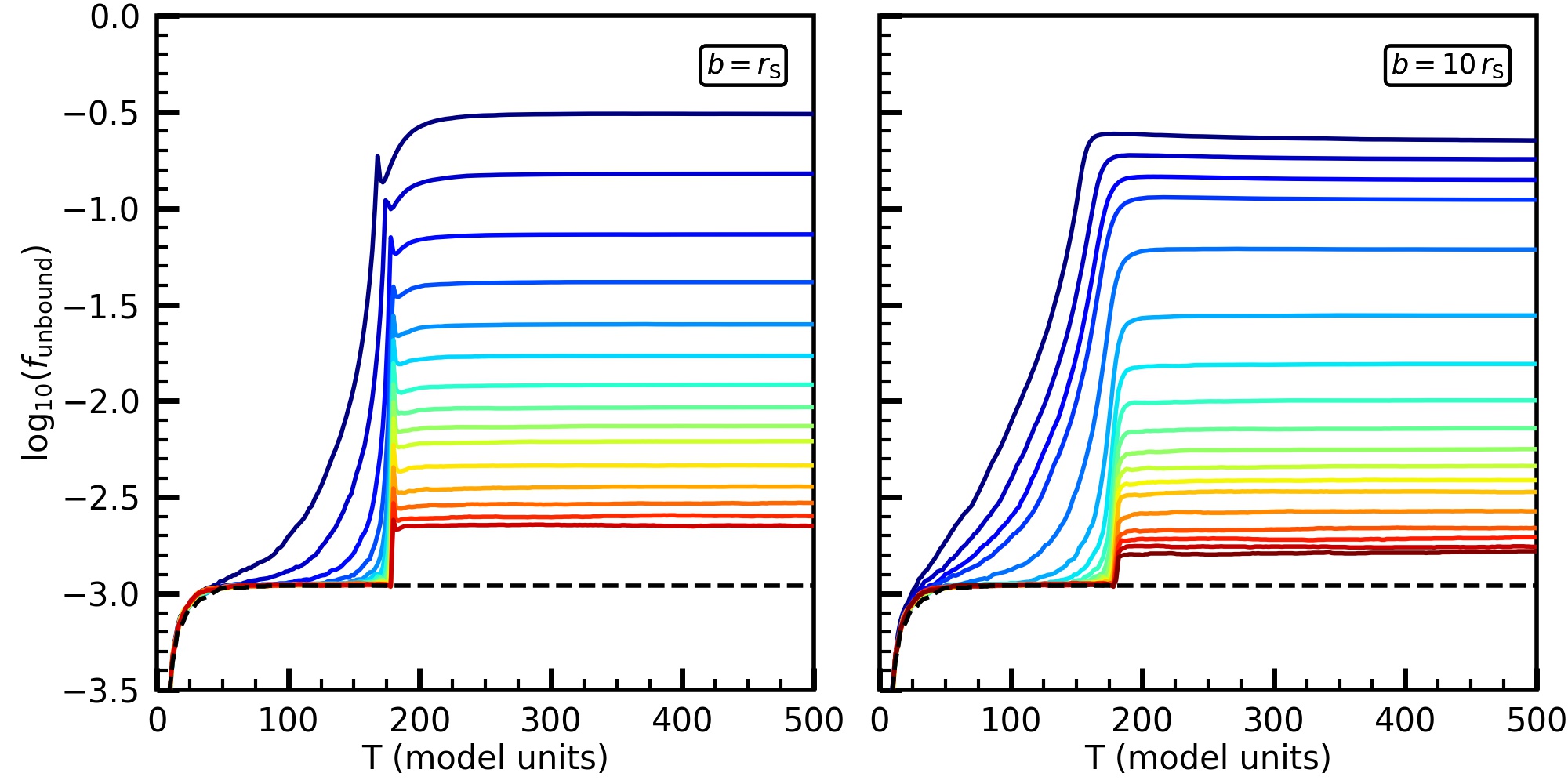}}\\
\subfloat{\includegraphics[width=0.95\textwidth]{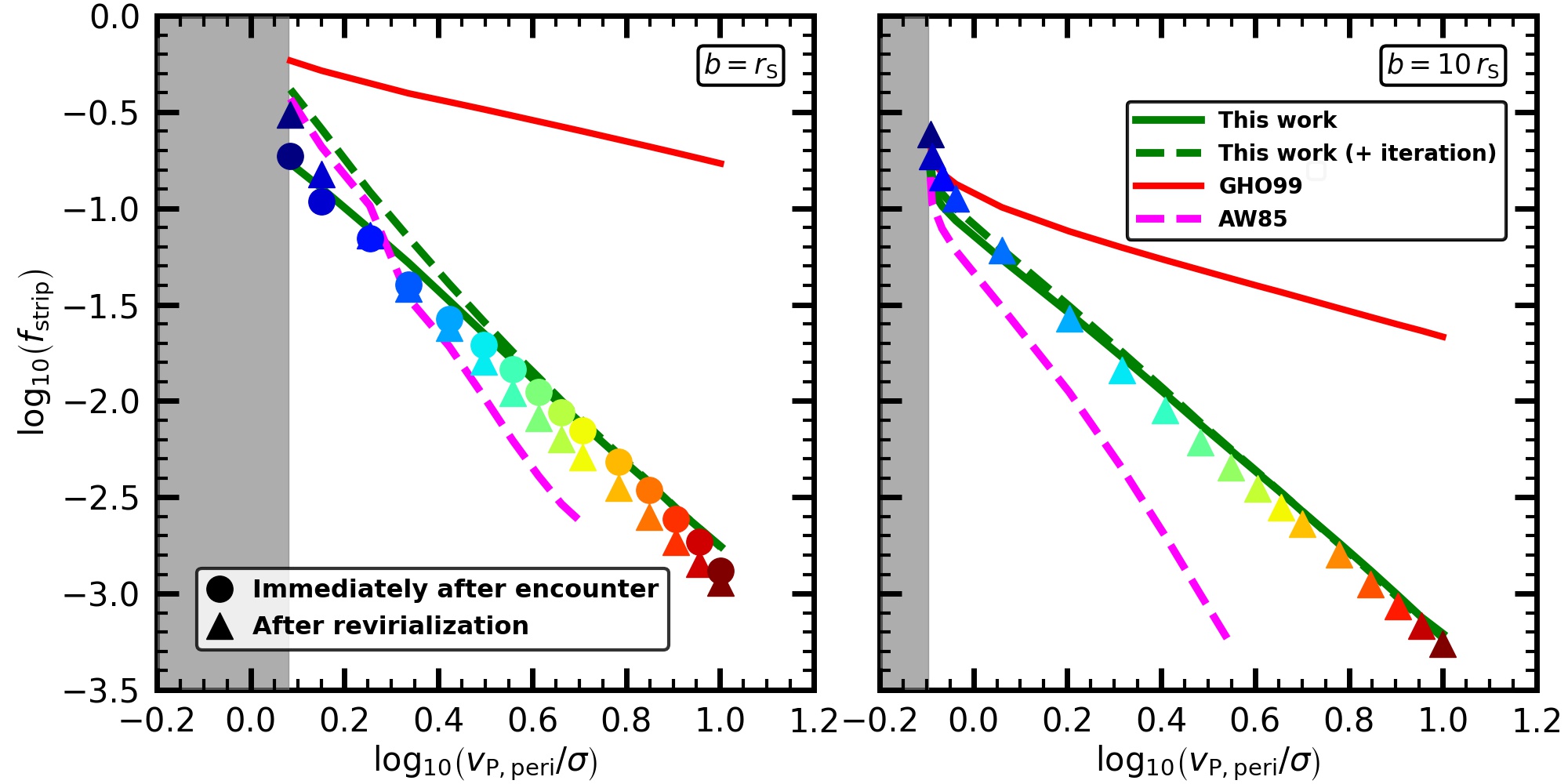}}
\caption{Comparison of numerical simulations with Monte Carlo predictions for the amount of mass loss induced by a tidal shock resulting from a penetrating encounter between two identical Hernquist spheres with impact parameter $b=r_\rmS$ (left panels) and $10\,r_\rmS$ (right panels). Upper panels show the time evolution of the unbound mass fraction, $f_{\rm unbound}$, in N-body simulations of encounters with different initial encounter velocities, $\vp$, ranging from $0.4\,\sigma$ ($0.7\,\sigma$) to $10\,\sigma$ for $b=10\,r_\rmS$ ($b=r_\rmS$), color coded from blue to red. The solid circles and triangles in the lower panels show the corresponding stripped mass fractions, $f_{\rm strip}$, as a function of $v_{\rm P,peri}/\sigma$ immediately following the encounter and after revirialization, respectively. For comparison, the solid green lines show the predictions from our general formalism for computing the impulse (equation~[\ref{deltav1}]). The solid red lines denote the predictions obtained using the DTA of GHO99. We emphasize that the DTA is not valid for penetrating encounters, and that the red lines are merely included for comparison. The green dashed lines show the predictions obtained using an iterative approach to determine the maximal subset of self-bound particles following the encounter. Finally, the dashed magenta lines show the predictions based on the fitting formula of \protect\cite{Aguilar.White.85} (AW85), which is based on a similar iterative approach, but applied to less extended objects. The grey shaded regions indicate the encounter velocities that result in tidal capture. See text for details and discussion.}
\label{fig:mass_loss}
\end{figure}

In this section we present an astrophysical application of our generalized formalism. We consider penetrating gravitational encounters between two spherical systems. In what follows we refer to them as (dark matter) haloes, though they could represent stellar systems equally well. We investigate the amount of mass loss to be expected from such encounters, and, in order to validate our formalism, compare its predictions to the results from $N$-body simulations.

\subsection{Numerical Simulations}
\label{sec:sim}

We simulate encounters between two identical, spherical \cite{Hernquist.90} haloes, whose initial density profile and potential are given by
\begin{equation}
\rho_\rmS(r) = \rho_0 \, \left({r \over r_\rmS}\right)^{-1} \, \left(1 + {r\over r_\rmS}\right)^{-3}, \;\;\;\;\;\;\;\;{\rm and}\;\;\;\;\;\;\;\;
\Phi_\rmS(r) = -\frac{G M_\rmS}{r+r_\rmS}\,,
\end{equation}
where $\rho_0=M_\rmS/2\pi r^3_\rmS$. Throughout we adopt model units in which the gravitational constant, $G$, the characteristic scale radius, $r_\rmS$, and the initial mass of each halo, $M_\rmS$, are all unity. Each halo is modelled using $N_\rmp = 10^5$ particles, whose initial phase-space coordinates are sampled from the ergodic distribution function $f = f(E)$ under the assumption that the initial haloes have isotropic velocity distributions. For practical reasons, each Hernquist sphere is truncated at $r_{\rm trunc}=1000\,r_\rmS$, which encloses 
99.8\% of $M_\rmS$. 

The haloes are initialized to approach each other with an impact parameter $b$, and an initial velocity $\vp$. We examine the cases of $b=r_\rmS$ and $10\,r_\rmS$. Initially the haloes are placed at large distance from each other, such that, depending on $\vp$, they always reach closest approach at $t \sim 200$. The simulation is continued up to $t = 500$, allowing the haloes sufficient time to re-virialize following the encounter.

The encounter is followed using the $N$-body code {\tt treecode}, written by Joshua Barnes, which uses a \cite{Barnes.Hut.86} octree to compute accelerations based on a multipole expansion up to quadrupole order, and a second order leap-frog integration scheme to solve the equations of motion. Since we use fixed time step, our integration scheme is fully symplectic. Forces between particles are softened using a simple Plummer softening. Throughout we adopt a time step of $\Delta t = 0.02$ and a softening length of $\varepsilon_{\rm soft} = 0.05$. These values ensure that the halo in isolation remains in equilibrium for well over $10 \Gyr$. For each $b$ we have run simulations for different values of $\vp/\sigma$ in the range $[0.4, 10]$. Here $\sigma=\sqrt{GM_\rmP/r_\rmS}$ is the characteristic internal velocity dispersion of the Hernquist halo.

For each of the two haloes, we measure its fraction of unbound particles $f_{\rm unbound}$ using the iterative method described in  \citet{vdBosch.etal.18b}.\footnote{Note though, that unlike in that paper, we determine the centre of mass position and velocity of each halo using all the bound particles, rather than only the 5 percent most bound particles. We find that this yields more stable results.} In the upper left (right) panel of Fig.~(\ref{fig:mass_loss}), we plot $f_{\rm unbound}$ (averaged over the two haloes) as a function of time for $b=r_\rmS$ ($10\,r_\rmS$). The solid curves correspond to different $v_\rmP$ ranging from $0.7$ to $10\,\sigma$ for $b=r_\rmS$ and from $0.4$ to $10\,\sigma$ for $b=10\,r_\rmS$, color coded from blue to red. The black dashed curve indicates the $f_{\rm unbound}$ for an isolated halo. Each halo is subject to an initial unbinding of $\sim 0.1\%$ of its mass due to numerical round-off errors in the force computation for particles in the very outskirts. The mass loss induced by the encounter occurs in two steps. First the tidal shock generated by the encounter unbinds a subset of particles. Subsequently, the system undergoes re-virialization during which the binding energies of individual particles undergo changes. This can both unbind additional particles, but also rebind particles that were deemed unbound directly following the tidal shock. Re-virialization is a more pronounced effect for more penetrating encounters, i.e., for $b=r_\rmS$. In this case, $f_{\rm unbound}$ increases steeply during the encounter and exhibits a spike before undergoing small oscillations and settling to the late time post-revirialization state. This late time value is slightly lower (higher) than the one immediately after the encounter (marked by the spike) for higher (lower) encounter velocities. For more distant encounters, i.e., the case with $b=10\,r_\rmS$, $f_{\rm unbound}$ evolves more smoothly with time as the encounter only peels off particles from the outer shells of the haloes. In this case, the late time value of $f_{\rm unbound}$ is nearly the same as that shortly after the encounter (note the absence of a temporal spike in $f_{\rm unbound}$ unlike in the $b=r_\rmS$ case).

For each simulation, we compute the unbound fraction at the end of the simulation. This is roughly $150 \, t_{\rm dyn}$ after the encounter, where $t_{\rm dyn} = \sqrt{\frac{4}{3} \pi r^3_\rms/G M_\rms}$ is a characteristic dynamical time of the Hernquist sphere. We correct for the initial unbinding of $0.1\%$ of the particles by computing the stripped fraction, $f_{\rm strip} \equiv (f_{\rm unbound}-0.001)/0.999$. The solid triangles in the bottom panels of Fig.~\ref{fig:mass_loss} plot the resulting $f_{\rm strip}$ as a function of $v_{\rm P,peri}/\sigma$, where $v_{\rm P,peri}$ is the encounter velocity at pericenter. Due to gravitational focusing $v_{\rm P,peri}$ is somewhat larger than $v_\rmP$. For the $b=r_\rmS$ case, we also compute $f_{\rm strip}$ immediately after the encounter, and thus prior to re-virialization. These values are indicated by the solid circles in the bottom left panel.

As expected, encounters of higher velocity result in less mass loss. For the strongly penetrating encounters with $b=r_\rmS$, the encounter unbinds as much as $31$\% of the mass for the smallest encounter velocity considered here, $\vp = 0.7\,\sigma$ ($v_{\rm P,peri}=1.2\,\sigma$). For smaller encounter velocities, the two haloes actually become bound, resulting in tidal capture and ultimately a merger (indicated by the grey shaded region). For $\vp = \sigma$ ($v_{\rm P,peri}=1.4\,\sigma$), the stripped mass fraction is only $\sim 15$\%.

In the case of the larger impact parameter, $b=10\,r_\rmS$, tidal capture requires somewhat lower encounter speeds ($\vp \lta 0.35\,\sigma$, i.e., $v_{\rm {P,peri}} \lta 0.8\,\sigma$), and overall the encounter is significantly less damaging than for $b=r_\rmS$, with $f_{\rm strip} \sim 6$\% for $\vp = \sigma$ ($v_{\rm P,peri}=1.15\,\sigma$). For encounter speeds a little larger than the tidal capture value, about $25\%$ of the mass is stripped. Hence, we can conclude that penetrating hyperbolic encounters between two identical Hernquist spheres can result in appreciable mass loss ($\sim 25-30$\%), but only for encounter velocities that are close to the critical velocity that results in tidal capture. For any $v_\rmP > 2\,\sigma$ ($4\,\sigma$), the stripped mass fraction is less than 5\% (1\%), even for a strongly penetrating encounter. We therefore conclude that impulsive encounters between highly concentrated, cuspy systems, such as Hernquist or NFW spheres, rarely cause a significant mass loss.

\subsection{Comparison with predictions from our formalism}

We now turn to our generalized formalism in order to predict $f_{\rm strip}$ for the different encounters simulated above. We assume that the two haloes encounter each other along a straight-line orbit with impact parameter $b_{\rm peri}$ and encounter speed $v_{\rm P,peri}$. These values are inferred from the simulation results, and differ from the initial $b$ and $v_\rmP$ due to gravitational focusing. We then compute the impulse $\Delta \bv(\br)$ for each subject star using equation~(\ref{deltav1}). In the impulsive limit ($\vp \gg \sigma$), the encounter imparts to a single subject star a specific internal energy given by
\begin{align}
\Delta \varepsilon(\br) = \bv \cdot \Delta \bv_{\rm rel} + \frac{1}{2} {\left(\Delta v_{\rm rel}\right)}^2\,,
\label{dEstarrel}
\end{align}
where $\Delta \bv_{\rm rel}(\br) = \Delta \bv(\br) - \Delta \bv_{\rm CM}$. Using our formalism for a straight-line orbit, $\Delta \bv_{\rm CM}$ is given by equation~(\ref{deltavCM}).

To compute the fraction of subject stars that become unbound, $f_{\rm strip}$, we use the Monte Carlo method of \cite{vdBosch.etal.18a} and sample the isotropic equilibrium distribution function for a Hernquist sphere with $10^6$ particles each. We then follow two different methods of calculating $f_{\rm strip}$. 

In the first method, we consider a subject star to be stripped if its $\Delta \varepsilon/|\varepsilon| > 1$, where $\varepsilon =v^2/2 +\Phi_\rmS$ is the original binding energy of the star prior to the encounter. This equates $f_{\rm strip}$ to the fraction of particles that are deemed unbound immediately following the encounter, in the COM frame of the subject\footnote{The COM velocity is computed using all particles, both bound and unbound.}. The solid green lines in the bottom panels of Fig.~\ref{fig:mass_loss} plot the $f_{\rm strip}$ thus obtained as a function of $v_{\rm P,peri}/\sigma$.

In the second method, we compute $f_{\rm strip}$ in an iterative fashion. This is motivated by \citet[][hereafter AW85]{Aguilar.White.85}, who argued that the maximal subset of self-bound particles is a better predictor of the stripped mass fraction after revirialization. In the first iteration we simply calculate the number of stars that remain bound according to the criterion of the first method. Next we compute the center of mass position and velocity from only the bound particles identified in the previous iteration, which we use to recompute the impulse $\Delta \bv_{\rm rel}(\br)$. We also recompute the potential $\Phi_\rmS$ by constructing trees comprising of only these bound particles. Now we recalculate the number of bound particles using the new $\Delta \bv_{\rm rel}(\br)$ and $\Phi_\rmS$. We perform these iterations until the bound fraction and the center of mass position and velocity of the haloes converge. The $f_{\rm strip}$ thus obtained is indicated by the dashed green lines in Fig.~\ref{fig:mass_loss}. 

Overall, both methods yield stripped mass fractions that are in good agreement with each other and with the simulation results. Only when $v_{\rm P,peri}$ is close to the critical velocity for tidal capture does the iterative method yield somewhat larger $f_{\rm strip}$ than without iteration. For the $b=r_\rmS$ case, the Monte Carlo predictions agree well with the simulation results shortly after the encounter (solid circles), while slightly overestimating (underestimating) the post-revirialization $f_{\rm strip}$ values (solid triangles) for high (low) $v_{\rm P,peri}$. This is expected since the Monte Carlo formalism based on the impulse approximation does not account for the unbinding or rebinding of material due to revirialization. We find that the iterative approach to compute the maximal subset of self-bound particles does not significantly improve this. For the $b=10\,r_\rmS$ case, revirialization has very little impact and the Monte Carlo predictions match the simulation results very well. 

We have also repeated this exercise using the initial encounter speed and impact parameter, i.e., ignoring the impact of gravitational focusing. The results (not shown) are largely indistinguishable from those shown by the green curves, except at the low velocity end ($v_{\rm P,peri}/\sigma \lta 1$) of the $b=10\,r_\rmS$ case where it underestimates $f_{\rm strip}$. In the strongly penetrating case ($b=r_\rmS$) the effect of gravitational focusing is much weaker because the impulse has a reduced dependence on the impact parameter. Hence, gravitational focusing is only important if both $v_\rmp \lta \sigma$ and $b \gta r_\rmS$. Finally, we have also investigated the impact of adiabatic correction, which we find to have a negligible impact on $f_{\rm strip}$, unless the encounter (almost) results in tidal capture. 

The magenta, dashed lines in the lower panels of Fig.~(\ref{fig:mass_loss}) show the predictions for the stripped mass fractions provided by the fitting function of AW85. These were obtained by calculating the maximal subset of self-bound particles using a similar Monte Carlo method as used here, and using the fully general, non-perturbative expression for the impulse (their equation [3], which is equivalent to our equation~[\ref{deltav1}]). Although this fitting function matches well with our iterative formalism (green dashed lines) at the low velocity end, it significantly underestimates the stripped fractions for large $v_{\rm P,peri}/\sigma$. For these encounter velocities, the impulse is small and therefore unbinds only the particles towards the outer part of the halo (those with small escape velocities). The AW85 fitting function is based on encounters between two identical, $r^{1/4}$ \cite{deVaucouleurs.1948} spheres. These have density profiles that fall-off exponentially at large radii, and are thus far less `extended' than the Hernquist spheres considered here. Hence, it should not come as a surprise that their fitting function is unable to accurately describe the outcome of our experiments.

Finally, for comparison, the red lines in the bottom panels of Fig.~\ref{fig:mass_loss} correspond to the $f_{\rm strip}$ predicted using the DTA of GHO99. Here we again use the Monte-Carlo method, but the impulse for each star is computed using equation~(10) of GHO99 for the impact parameter, $b_{\rm peri}$, and encounter velocity, $v_{\rm P,peri}$ at pericentre (i.e., accounting for gravitational focusing). Although the DTA is clearly not valid for penetrating encounters, we merely show it here to emphasize that pushing the DTA into a regime where it is not valid can result in large errors. Even for impact parameters as large as $10\,r_\rmS$, the DTA drastically overpredicts $f_{\rm strip}$, especially at the high velocity end. High velocity encounters only strip off particles from the outer shells, for which the DTA severely overestimates the impulse. This highlights the merit of the general formalism presented here, which remains valid in those parts of the parameter space where the DTA breaks down.

To summarize, despite several simplifications such as the assumption of a straight line orbit, the impulse approximation, and the neglect of re-virialization, the generalized formalism presented here can be used to make reasonably accurate predictions for the amount of mass stripped off due to gravitational encounters between collisionless systems, even if the impact parameter is small compared to the characteristic sizes of the objects involved. In particular, in agreement with AW85, we find that the impulse approximation remains reasonably accurate all the way down to encounters that almost result in tidal capture, and which are thus no longer strictly impulsive.

\section{Conclusion}
\label{sec:conclusion}

In this chapter we have developed a general, non-perturbative formalism to compute the energy transferred due to an impulsive shock. Previous studies \citep[e.g.,][]{Spitzer.58, Ostriker.etal.72, Richstone.75, Mamon.92, Mamon.00, Makino.Hut.97, Gnedin.etal.99} have all treated impulsive encounters in the distant tide limit by expanding the perturber potential as a multipole series truncated at the quadrupole term. However, this typically only yields accurate results if the impact parameter, $b$, is significantly larger than the characteristic sizes of both the subject, $r_\rmS,$ and the perturber, $r_\rmP$.  For such distant encounters, though, very little energy is transferred to the subject and such cases are therefore of limiting astrophysical interest. A noteworthy exception is the case where $r_\rmP \gg r_\rmS$, for which the formalism of GHO99, which also relies on the DTA, yields accurate results even when $b \ll r_\rmP$. However, even in this case, the formalism fails for impact parameters that are comparable to, or smaller than, the size of the subject. 

From an astrophysical perspective, the most important impulsive encounters are those for which the increase in internal energy, $\Delta E_{\rm int}$, is of order the subject's internal binding energy or larger. Such encounters can unbind large amounts of mass from the subject, or even completely destroy it. Unfortunately, this typically requires small impact parameters for which the DTA is no longer valid. In particular, when the perturber is close to the subject, the contribution of higher-order multipole moments of the perturber potential can no longer be neglected. The non-perturbative method presented here \citep[and previously in][]{Aguilar.White.85} overcomes these problems, yielding a method to accurately compute the velocity impulse on a particle due to a high-speed gravitational encounter. It can be used to reliably compute the internal energy change of a subject that is valid for any impact parameter, and any perturber profile. And although the results presented here are, for simplicity, limited to spherically symmetric perturbers, it is quite straightforward to extend it to axisymmetric, spheroidal perturbers, which is something we leave for future work.

In general, our treatment yields results that are in excellent agreement with those obtained using the DTA, but only if (i) the impact parameter  $b$ is large compared to the characteristic radii of both the subject and the perturber, and (ii) the subject is truncated at a radius $r_{\rm trunc} < b$.  If these conditions are not met, the DTA typically drastically overpredicts $\Delta E_{\rm int}$, unless one `manually' caps  $\Delta E_{\rm int}(b)$ to be no larger than the value for a head-on encounter, $\Delta E_0$ \citep[see e.g.,][]{vdBosch.etal.18a}. The $\Delta E_{\rm int}(b)$ computed using our fully general, non-perturbative formalism presented here, on the other hand, naturally asymptotes towards $\Delta E_0$ in the limit $b \to 0$. Moreover, in the DTA, a radial truncation of the subject is required in order to avoid divergence of the moment of inertia, $\langle r^2 \rangle$. Our method has the additional advantage that it does not suffer from this divergence-problem.

Although our formalism is more general than previous formalisms, it involves a more demanding numerical computation. In order to facilitate the use of our formalism, we have provided a table with the integrals $I(s)$ needed to compute the velocity impulse, $\Delta \bv(\br)$, given by equation~(\ref{deltav1}), for a variety of perturber profiles (Table~\ref{tab:Is}). In addition, we have released a public Python code, {\tt NP-impulse} (\url{https://github.com/uddipanb/NP-impulse}) that the reader can use to compute $\Delta \bv(\br)$ of a subject star as a function of impact parameter, $b$, and encounter speed, $\vp$. {\tt NP-impulse} also computes the resulting $\Delta E_{\rm int}$ for a variety of spherical subject profiles, and treats both straight-line orbits as well as eccentric orbits within the extended potential of a spherical perturber. In the latter case, {\tt NP-impulse} accounts for adiabatic shielding using the method developed in \cite{Gnedin.Ostriker.99}. We hope that this helps to promote the use of our formalism in future treatments of impulsive encounters.

As an example astrophysical application of our formalism, we have studied the mass loss experienced by a Hernquist sphere due to the tidal shock associated with an impulsive encounter with an identical object along a straight-line orbit. In general, our more general formalism agrees well with the results from numerical simulations and predicts that impulsive encounters are less disruptive, i.e., cause less mass loss, than what is predicted based on the DTA of GHO99. Encounters with $\vp/\sigma > 1$ do not cause any significant mass loss ($\lta 15$\%). For smaller encounter speeds, mass loss can be appreciable (up to $\sim 30$\%), especially for smaller impact parameters. However, for too low encounter speeds, $\vp/\sigma \lta 0.5$, the encounter results in tidal capture, and eventually a merger, something that cannot be treated using the impulse approximation. In addition, for $\vp/\sigma \lta 1$, the adiabatic correction starts to become important. Unfortunately, the adiabatic correction of \cite{Gnedin.Ostriker.99} that we have adopted in this chapter has only been properly tested for the case of disc shocking, which involves fully compressive tides. It remains to be seen whether it is equally valid for the extensive tides considered here. Ultimately, in this regime a time-dependent perturbation analysis similar to that developed in \citet{Weinberg.94b} may be required to accurately treat the impact of gravitational shocking. Hence, whereas our formalism is fully general in the truly impulsive regime, and for any impact parameter, the case of slow, non-impulsive encounters requires continued, analytical studies. A particularly interesting case to examine is the quasi-resonant tidal interaction between a disk galaxy and a satellite. This has been explored in detail by \citet{DOnghia.etal.10}, who computed the impulse on disk stars while accounting for their rotation in the disk, rather than treating them as stationary. The impulse, however, was obtained perturbatively, using the DTA, and it remains to be seen how these results change if the impulse is computed non-perturbatively, as advocated here. We intend to address this in future work.

\section*{Data availability}

The data underlying this article, including the Python code {\tt NP-impulse}, is publicly available in the GitHub Repository, at \url{https://github.com/uddipanb/NP-impulse}.



    \begin{subappendices}


\chapter*{Appendix}

\section{Asymptotic behaviour}
\label{app:asymptote}

In \S\ref{sec:straight_orbit}, we obtained the general expression for $\Delta E_{\rm int}$, which is valid for impulsive encounters with any impact parameter $b$. Here we discuss the asymptotic behaviour of $\Delta E_{\rm int}$ in both the distant tide limit (large $b$) and the head-on limit (small $b$).

\subsection{Distant encounter approximation}
\label{sec:asymptote_tidal}

In the limit of distant encounters, the impact parameter $b$ is much larger than the scale radii of the subject, $r_\rmS$, and the perturber, $r_\rmP$. In this limit, it is common to approximate the perturber as a point mass. However, as discussed above, this will yield a diverging $\Delta E_{\rm int}$ unless the subject is truncated and $b > r_{\rm trunc}$ (an assumption that is implied, but rarely mentioned). In order to avoid this issue, we instead consider a (spherical) Plummer perturber. In the limit of large $b$,  equation~(\ref{delEint1}) then reduces to an expression that is similar to, but also intriguingly different from, the standard distant tide expression first obtained by S58 by treating the perturber as a point mass, and expanding $\Phi_\rmP$ as a multipole series truncated at the quadrupole term. We also demonstrate that the asymptotic form of $\Delta E_{\rm int}$ is quite different for infinite and truncated subjects.

In the large-$b$ limit, we can assume that $r\sin{\theta} < b$, i.e., we can restrict the domains of the $\calJ$ and $\calJ_{\rm CM}$ integrals (equations~[\ref{J_plummer}] and~[\ref{J_CM_plummer}]) to the inside of a cylinder of radius $b$. The use of cylindrical coordinates is prompted by the fact that the problem is inherently cylindrical in nature: the impulse received by a subject star is independent of its distance along the direction in which the perturber is moving, but only depends on $R = r \sin\theta$ (cf. equation~[\ref{deltav0}]).  Hence, in computing the total energy change, $\Delta E$, it is important to include subject stars with small $R$ but large $z$-component, while, in the DTA, those with $R > b$ can be ignored as they receive a negligibly small impulse. Next, we Taylor expand the $\theta$-integrand in the expression for $\calJ$ about $r\sin{\theta}=0$ to obtain the following series expansion for the total energy change
\begin{align}
\Delta E &\approx 4\pi{\left(\frac{GM_\rmP}{\vp}\right)}^2 \int_0^{\infty} \rmd r\, r^2 \rho_\rmS(r) \int_0^{\pi} \rmd \theta \sin{\theta} \nonumber \\
&\times \left[\frac{1}{{\left(1+\varepsilon^2\right)}^2}\frac{1}{b^2}+\frac{1-4\varepsilon^2+\varepsilon^4}{{\left(1+\varepsilon^2\right)}^4}\frac{r^2\sin^2{\theta}}{b^4}+\frac{1-12\varepsilon^2+15\varepsilon^4-2\varepsilon^6}{{\left(1+\varepsilon^2\right)}^6}\frac{r^4\sin^4{\theta}}{b^6}+...\right]\,,
\label{deltaEint_series}
\end{align}
where $\varepsilon=r_\rmP/b$. In the large $b$ limit, the COM velocity given by equation~(\ref{deltavCM_ptmass}) reduces to
\begin{align}
\Delta \bv_{\rm CM} = \frac{2GM_\rmP}{M_\rmS\vp}\frac{\pi}{b}\int_0^{\infty} \rmd r\, r^2 \rho_\rmS(r) \int_0^{\pi} \rmd \theta \sin{\theta}\left[\frac{2}{1+\varepsilon^2}-\frac{4\varepsilon^2}{{\left(1+\varepsilon^2\right)}^3}\frac{r^2\sin^2{\theta}}{b^2}+...\right]\, \hat{\by}\,.
\end{align}
The above two integrals have to be evaluated conditional to $r\sin{\theta}<b$. Upon subtracting the COM energy, $\Delta E_{\rm CM} = \frac{1}{2} M_\rmS \, (\Delta v_{\rm CM})^2$, the first term in the $\theta$ integrand of equation~(\ref{deltaEint_series}) drops out. Integrating the remaining terms yields
\begin{align}
&\Delta E_{\rm int} \approx 4\pi {\left(\frac{GM_\rmP}{\vp}\right)}^2 \sum_{n=2}^{\infty}\calI_{n-1}\,\calC_n\left(\frac{r_\rmP}{b}\right)\frac{\calR_n(b)+\calS_n(b)}{b^{2n}}\,.
\label{tidal_general}
\end{align}
Here
\begin{align}
\calC_n(x) = \frac{P_{2n}(x)}{(1+x^2)^{2n}},
\end{align}
with $P_{2n}(x)$ a polynomial of degree $2n$. We have worked out the coefficients for $n=2$ and $3$, yielding $P_4(x) = 1 + x^4$ and $P_6(x) = 1 - 6 x^2 + 9 x^4 - 2 x^6$, and leave the coefficients for the higher-order terms as an exercise for the reader. We do point out, though, that $\calC_n(r_\rmP/b) = 1+\calO(r^2_\rmP/b^2)$ in the limit $b \gg r_\rmP$. The coefficient $\calI_n$ is given by
\begin{align}
\calI_n & = \int_{-1}^1 \rmd x\, (1-x^2)^n = 2 \sum_{m=0}^{n} \, \frac{(-1)^m}{2m+1} \, \binom{n}{m} \,,
\end{align}
while $\calR_n(b)$ and $\calS_n(b)$ are functions of $b$ given by
\begin{align}
&\calR_n(b)=\int_0^b \rmd r\,r^{2n}\rho_\rmS(r)\,, \nonumber \\
&\calS_n(b)=\int_b^{\infty} \rmd r\,r^{2n} \rho_\rmS(r) \left[1-\sqrt{1-\frac{b^2}{r^2}}\left\{1+\sum_{m=0}^{n-2}\frac{\binom{2m+1}{m}}{2^{2m+1}}{\left(\frac{b}{r}\right)}^{2m+2}\right\}\right]\,.
\end{align}
Note that $\calR_n(b)$ is the ${\left(2n-2\right)}^{\rm th}$ moment of the subject density profile, $\rho_\rmS(r)$, inside a sphere of radius $b$, while $\calS_n(b)$ is the same but for the part of the cylinder outside of the sphere. $\calR_n(b)+\calS_n(b)$ is therefore the ${\left(2n-2\right)}^{\rm th}$ moment of $\rho_\rmS(r)$ within the cylinder of radius $b$. If we truncate the series given in equation~(\ref{tidal_general}) at $n=2$, then we obtain an asymptotic form for $\Delta E_{\rm int}$ that is similar to that of the standard tidal approximation:
\begin{align}
&\Delta E_{\rm int} \approx \frac{4 M_\rmS}{3} {\left(\frac{GM_\rmP}{\vp}\right)}^2 \frac{\langle r^2\rangle_{\rm cyl}}{b^4}\,.
\label{tidal}
\end{align}
Here,
\begin{eqnarray}
\langle r^2\rangle_{\rm cyl} & = & \frac{4\pi}{M_\rmS} \left[\int_0^b \rmd r\, r^4 \rho_\rmS(r) + \int_b^{\infty} \rmd r\, r^4 \rho_\rmS(r)\left\{1-\sqrt{1-\frac{b^2}{r^2}}\left(1+\frac{b^2}{2r^2}\right)\right\} \right] \nonumber \\
&= & \langle r^2 \rangle - \frac{4\pi}{M_\rmS} \int_b^{\infty} \rmd r\, r^4 \rho_\rmS(r)\sqrt{1-\frac{b^2}{r^2}}\left(1+\frac{b^2}{2r^2}\right),
\label{moment_of_inertia}
\end{eqnarray}
which is subtly different from the moment of inertia, $\langle r^2\rangle$, that appears in the standard expression for the distant tidal limit, and which is given by equation~(\ref{r2aver}). In particular, $\langle r^2 \rangle_{\rm cyl}$ only integrates the subject mass within a cylinder truncated at the impact parameter, whereas $\langle r^2 \rangle$ integrates over the entire subject mass. As discussed above, this typically results in a divergence, unless the subject is truncated or has a density that falls of faster than $r^{-5}$ in its outskirts.

Indeed, if the subject is truncated at a truncation radius $r_{\rm trunc} < b$, then $\langle r^2 \rangle_{\rm cyl} = \langle r^2 \rangle$, and equation~(\ref{tidal}) is exactly identical to that for the `standard' impulsive encounter of S58. In addition, $\calR_n=\int_0^{r_{\rm trunc}}\rmd r\,r^{2n}\rho_\rmS(r)$, which is independent of $b$, and $\calS_n=0$. Hence, the $n^{\rm th}$-order term scales as $b^{-2n}$, and $\Delta E_{\rm int}$ is thus dominated by the quadrupole term, justifying the truncation of the series in equation~(\ref{deltaEint_series}) at $n=2$.

However, for an infinitely extended subject, or one that is truncated at $r_{\rm trunc} > b$, truncating the series at the $n=2$ quadrupole term can, in certain cases, underestimate $\Delta E_{\rm int}$ by as much as a factor of $\sim 2$. In particular, if $\rho_\rmS(r) \sim r^{-\beta}$ at large $r$, and falls off less steeply than $r^{-5}$ at small $r$, then  both $\calR_n(b)$ and $\calS_n(b)$ scale as $b^{2n+1-\beta}$, as long as $\beta < 5$. Hence, all terms in equation~(\ref{tidal_general}) scale with $b$ in the same way, and the truncation is not justified, even in the limit of large impact parameters\footnote{This is also evident from equation~(\ref{deltaEint_series}), which shows that all terms contribute equally when  $r\sin{\theta}\sim b$.}. Furthermore, in this case it is evident from equation~(\ref{tidal_general}) that $\Delta E_{\rm int}\sim b^{1-\beta}$. On the other hand, for $\beta=5$, $\calR_2$ is the dominant term and scales with $b$ as $\ln{b}$, so that $\Delta E_{\rm int}\sim \ln{b}/b^4$. For $\beta>5$, both $\calR_2$ and $\calS_2$ are the dominant terms, which add up to $\langle r^2\rangle \simeq \int_0^{\infty}\rmd r\, r^4\rho_\rmS(r)$ (which is finite in this case), such that $\Delta E_{\rm int}\sim b^{-4}$. Hence, for an infinitely extended subject with $\rho_\rmS \propto r^{-\beta}$ at large $r$ we have that
\begin{equation}
\begin{aligned}
\lim_{b\to\infty}\Delta E_{\rm int} \propto
\begin{cases}
b^{1-\beta} , & \beta < 5 \\[2pt]
b^{-4} \ln{b}, & \beta = 5 \\[2pt]
b^{-4}, & \beta > 5\,.
\end{cases}
\label{asymptote_tidal_infinite}
\end{aligned}
\end{equation}
This scaling is not only valid for an infinitely extended subject, but also for a truncated subject when the impact parameter falls in the range $\max[r_\rmS,r_\rmP] < b < r_{\rm trunc}$.

\subsection{Head-on encounter approximation}
\label{sec:asymptote_head_on}

The head-on encounter corresponds to the case of zero impact parameter (i.e., $b=0$). As long as the perturber is not a point mass, the internal energy injected into the subject is finite, and can be computed using equation~(\ref{deltaE1}) with $b=0$. Note that there is no need to subtract $\Delta E_{\rm CM}$ in this case, since it is zero. If the perturber is a Plummer sphere, the $\calJ$ integral can be computed analytically for $b=0$, which yields
\begin{align}
\Delta E_{\rm int} = 8\pi{\left(\frac{G M_\rmP}{\vp}\right)}^2 \int_0^{\infty} \rmd r\,\rho_\rmS(r)\, \calF_0(r,r_\rmP),
\label{headon}
\end{align}
where
\begin{align}
\calF_0(r,r_\rmP) = \frac{r\left(2r^2+r^2_\rmP\right)}{4{\left(r^2+r^2_\rmP\right)}^{3/2}} \ln{\left[\frac{\sqrt{r^2+r^2_\rmP}+r}{\sqrt{r^2+r^2_\rmP}-r}\right]} - \frac{r^2}{2\left(r^2+r^2_\rmP\right)}.
\end{align}
It is easily checked that $\calF_0$ has the following asymptotic behaviour in the small- and large-$r$ limits:
\begin{equation}
\begin{aligned}
\calF_0(r,r_\rmP) \sim
\begin{cases}
 \frac{2}{3} \left(\frac{r}{r_\rmP}\right)^4, & r \ll r_\rmP, \\[8pt]
 \ln{\left(\frac{2r}{r_\rmP}\right)}, & r \gg r_\rmP.
\end{cases}
\label{F0}
\end{aligned}
\end{equation}
Hence, we see that the behavior of the integrand of equation~(\ref{headon}) in the limits $r\to 0$ ($r\ll r_\rmP$) and $r\to \infty$ ($r\gg r_\rmP$), is such that $\Delta E_{\rm int}$ is finite, as long as $\rho_\rmS(r)$ scales less steeply than $r^{-5}$ at small $r$ and more steeply than $r^{-1}$ at large $r$. Both conditions are easily satisfied for any realistic astrophysical subject. Note from equation~(\ref{F0}) that, as expected, more compact perturbers (smaller $r_\rmP$) dissipate more energy and therefore cause more pronounced heating of the subject. 

Note that one obtains the same results using the expression of $\Delta E_{\rm int}$ for a head-on encounter listed under case C in Table~\ref{tab:comparison}. For a Plummer perturber, $I_0 = R^2/(R^2 + r^2_\rmp)$, which after substitution in the expression for $\Delta E_{\rm int}$, writing $R = r \sin\theta$, and solving the $\theta$-integral, yields equation~(\ref{headon}).


    \end{subappendices}

%
%
%

\chapter{A Comprehensive Perturbative Formalism for Phase-Mixing in Perturbed Disks. I. Phase spirals in an Infinite, Isothermal Slab} 
\label{chapter: paper2}

\begin{center}

This chapter has been published as:\\
\vspace{5pt}

\author{Uddipan Banik, Martin~D.~Weinberg and
   Frank~C.~van den Bosch
}
\vspace{5pt}

\textit{The Astrophysical Journal}, Volume 935, Number 2, Page 135\\

\textit{\citep[][]{Banik.etal.22b}}

\end{center}


\section{Introduction}
\label{sec:intro_2}

The relaxation or equilibration of self-gravitating systems is a ubiquitous astrophysical phenomenon that drives the formation and evolution of star-clusters, galaxies and cold dark matter halos. In quasi-equilibrium, the phase-space density of such collisionless systems can be well characterized by a distribution function (DF) which, according to the strong Jeans theorem, is a function of the conserved quantities or actions of the system. When such a system is perturbed out of equilibrium by a time-dependent gravitational perturbation, either external (e.g., encounter with another galaxy) or internal (e.g., bars or spiral arms), the original actions of the stars are modified, and the system has to re-establish a new (quasi-)equilibrium. Since disk galaxies are highly ordered, low-entropy (i.e., cold) systems, they are extremely responsive. Even small gravitational perturbations can induce oscillations in the disk, which manifest as either standing or propagating waves \citep[see][for a detailed review]{Sellwood.13}. Such oscillations consist of an initially coherent response of stars to a gravitational perturbation. This coherent response is called {\it collective} if its self-gravity is included. Over time, though, the coherence {\it dissipates}, which manifests as relaxation or equilibration and drives the system towards a new quasi-equilibrium, free of large scale oscillations. Equilibration in galactic disks is dominated by collisionless effects, including purely kinematic processes like phase-mixing (loss of coherence in the response due to different orbital frequencies of stars), and self-gravitating or collective processes like Landau damping \citep[loss of coherence due to non-dissipative damping of waves by wave-particle interactions,][]{LyndenBell.62} and violent relaxation \citep[loss of coherence due to scrambling of orbital energies in a time-varying potential,][]{LyndenBell.67}. It is noteworthy to point out that without phase-mixing neither Landau damping \citep[][]{Maoz.91} nor violent relaxation \citep[see][]{Sridhar.89} would result in equilibration. A final equilibration mechanism is chaotic mixing, the loss of coherence resulting from the exponential divergence of neighboring stars on chaotic orbits \citep[e.g.,][]{Merritt.Valluri.96,Daniel.Wyse.15,Banik.vdBosch.22}. As long as most of the phase-space is foliated with regular orbits (i.e., the Hamiltonian is near-integrable), chaotic mixing should not make a significant contribution, and phase-mixing may thus be considered the dominant equilibration mechanism.

Disk galaxies typically reveal out-of-equilibrium features due to incomplete equilibration. These may appear in the form of bars and spiral arms, which are large-scale perturbations in the radial and azimuthal directions, responsible for a slow, secular evolution of the disk. In the vertical direction, disks often reveal warps \citep[][]{Binney.92}. In the case of the Milky Way (hereafter MW) disk, which can be studied in much greater detail than any other system, recent data from astrometric and radial-velocity surveys such as SEGUE \citep[][]{Yanny.etal.09}, RAVE \citep[][]{Steinmetz.etal.06}, GALAH \citep[][]{Bland-Hawthorn.etal.19}, LAMOST \citep[][]{Cui.etal.12} and above all Gaia \citep[][]{Gaia_collab.16, Gaia_collab.18a, Gaia_collab.18b} has revealed a variety of additional vertical distortions. At large galacto-centric radii ($>10 \kpc$) this includes, among others, oscillations and corrugations \citep[][]{Xu.etal.15,Schonrich.Dehnen.18}, and streams of stars kicked up from the disk that undergo phase-mixing, sometimes referred to as `feathers' \citep[e.g.,][]{Price-Whelan.etal.15, Thomas.etal.19, Laporte.etal.22}. Similar oscillations and vertical asymmetries have also been reported in the Solar vicinity \citep[e.g.,][]{Widrow.etal.12, Williams.etal.13, Yanny.Gardner.13, Quillen.etal.18, Gaia_collab.18b, Bennett.Bovy.19, Carrillo.etal.19}. One of the most intriguing structures is the phase-space spiral discovered by \cite{Antoja.etal.18}, and studied in more detail in subsequent studies \citep[e.g.,][]{Bland-Hawthorn.etal.19,Li.Widrow.21,Li.21,Gandhi.etal.22}. Using data from Gaia DR2 \citep[][]{Gaia_collab.18a}, \cite{Antoja.etal.18} selected $\sim 900$k stars within a narrow range of galacto-centric radius and azimuthal angle centered around the Sun. When plotting the density of stars in the $(z,v_z)$-plane of vertical position, $z$, and vertical velocity, $v_z$, they noticed a faint, unexpected spiral pattern, which became more enhanced when colour-coding the $(z,v_z)$-`pixels' by the median radial or azimuthal velocities. The one-armed spiral makes 2-3 complete wraps, resembling a snail shell, and is interpreted as a signature of phase-mixing in the vertical direction following a perturbation, which \cite{Antoja.etal.18} estimate to have occurred between 300 and 900 Myr ago. More careful analyses in later studies \citep[e.g.,][etc.]{Bland-Hawthorn.etal.19,Li.21} have nailed down the interaction time to $\sim 500\Myr$ ago.

The discovery of all these oscillations in the MW disk has ushered in a new, emerging field of astrophysics, known as galactoseismology \citep[][]{Widrow.etal.12, Johnston.etal.17}. Similar to how the timbre of musical notes reveals characteristics of the instrument that produced the sound, the `ringing' of a galactic disk can (in principle) reveal its structure (both stellar disk plus dark matter halo). And similar to how the timbre can tell us whether the string of a violin was plucked (pizzicato) or bowed (arco), the ringing of a galactic disk can reveal information about the perturbation that set the disk ringing. Phase spirals are especially promising in this regard: their structure holds information about the gravitational potential in the vertical direction \citep[in particular, the vertical frequency as a function of the vertical action,][]{Antoja.etal.18} and about the type of perturbation that triggered the phase spiral \citep[e.g., bending mode vs. breathing mode, see][and Section~\ref{sec:impulsive_kick} below]{Widrow.etal.14, Darling.Widrow.19a}. In addition, by unwinding the phase spiral one can in principle determine how long ago the vertical oscillations were triggered. By studying phase spirals at multiple locations in the disk, one may even hope to use some form of triangulation to infer the direction or location from which the perturbation emerged (assuming, of course, that the phase spirals at different locations were all triggered by the same perturbation).

However promising galactoseismology may seem, many questions remain: what kind of perturbation can trigger a phase spiral? how long do phase spirals remain detectable, and what equilibration mechanism(s) causes their demise? Can we really constrain the vertical potential of the disk, or does self-gravity of the perturbation make it difficult to achieve?  What kind of constraints can we infer regarding the perturber that triggered the phase spiral? Is galactoseismology likely to be confusion limited, i.e., should we expect that each location in the disk experiences oscillations arising from multiple, independent perturbations? If so, how does this impact our ability to extract useful information? Answering these questions necessitates a deep understanding of how the MW disk, and disk galaxies in general, respond to perturbations.

To date, these questions have mainly been addressed using numerical $N$-body simulations or fairly simplified analytical approaches. In particular, numerous studies have investigated how the MW disk responds to interactions with the Sagittarius (Sgr) dwarf galaxy \citep[e.g.,][]{Gomez.etal.13, Donghia.etal.16, Laporte.etal.18, Khanna.etal.19,Hunt.etal.21}. While simulations likes these have demonstrated that the interaction with Sgr can indeed spawn phase spirals in the Solar vicinity \citep[][]{Antoja.etal.18,Binney.Schonrich.18, Darling.Widrow.19b, Laporte.etal.19, Bland-Hawthorn.etal.19, Hunt.etal.21, Bennett.etal.22}, none of them have been able to produce phase spirals that match those observed in the Gaia data. As discussed in detail in \cite{Bennett.etal.22} and \cite{Bennett.Bovy.21}, this seems to suggest that the amplitude and shape of the ``Gaia snail" cannot be produced by Sgr alone. An alternative explanation, explored by \citet{Khoperskov.etal.19}, is that the Gaia snail was created by buckling of the MW's bar. However, this explanation faces its own challenges \citep[see e.g.,][]{Laporte.etal.19, Bennett.Bovy.21}. Triggering the Gaia snail with a spiral arm \citep[][]{Faure.etal.14} is also problematic, in that it requires the spiral arms to have unusually large amplitude \citep[][]{Quillen.etal.18}. Clearly then, despite a large number of studies, pinpointing the origin of the phase spiral in the Solar vicinity still remains an unsolved problem.
 
Although simulations have the obvious advantage that they can probe the complicated response of a perturbed disk to a realistic perturbation, which often is analytically intractable, especially if the response is large (non-linear), there are also clear disadvantages. Foremost, reaching sufficient resolution to resolve the kind of fine-structure that we can observe with data sets like Gaia requires extremely large simulations with $N > 10^8-10^9$ particles \citep[][]{Weinberg.Katz.07a, Binney.Schonrich.18, Hunt.etal.21}. Although such simulations are no longer beyond our reach \citep[see e.g.,][]{Bedorf.etal.14, Fujii.etal.19, Hunt.etal.21, Peterson.etal.22}, it is clear that using such simulations to explore large areas of parameter space remains a formidable challenge. To overcome this problem, a semi-analytical approach called the {\it backward-integrating restricted N-body method} was developed originally in the context of perturbation by bars \citep[e.g.,][]{Leeuwin.etal.93,Vauterin.Dejonghe.97,Dehnen.00}, and later on used by \cite{Hunt.Bovy.18} and \cite{Hunt.etal.19} to study non-equilibrium features in the MW caused by transient spiral arms. This method is effectively a Lagrangian formalism to solve the collisionless Boltzmann equation (hereafter CBE) by integrating test particles in the perturbed potential in a restricted N-body framework, i.e., without self-consistently developing the potential perturbation from the DF perturbation. Although appropriate for studying the local kinematic distribution of particles, this approach becomes too expensive to study the global equilibration of a system. Hence, it is important to consider alternative analytical methods that can be used to investigate the global response of a disk.

In this vein, this chapter presents a rigorous, perturbative, Eulerian formalism to compute the response of a disk to perturbations. In order to gain valuable insight into the physical mechanism of phase-mixing, without resorting to the computational complexity involved in modelling a realistic disk, which we postpone to chapter~\ref{chapter: paper3} of this thesis, in this chapter we consider perturbations of an infinite slab with a vertical profile, but homogeneous in the lateral directions. Although a poor representation of a realistic galactic disk, this treatment captures most of the essential features of how disks respond to gravitational perturbations. We study the response of the slab to perturbers of various spatial and temporal scales, with a focus on the formation and dissolution of phase spirals resulting from the vertical oscillations and phase-mixing of stars.

This chapter is organized as follows. Section~\ref{sec:linear_theory_2} describes the application of perturbation theory to our infinite, isothermal slab. Section~\ref{sec:impulsive_kick} then uses these results to work out the response to an impulsive, single-mode perturbation, which nicely illustrates how phase spirals originate from vertical oscillations and how they damp out due to lateral mixing. Sections~\ref{sec:localized} and~\ref{sec:non-impulsive} generalize this to responses to localized (wave packet) and non-impulsive perturbations, respectively. In Section~\ref{sec:sat_encounter} we investigate the response to satellite encounters and examine which satellite galaxies in the halo of the MW can trigger bending and/or breathing modes strong enough to trigger phase spirals at the Solar radius (still approximating the MW disk as an infinite, isothermal slab). We summarize our findings in Section~\ref{sec:concl_2}.

\section{Linear perturbation theory for collisionless systems}
\label{sec:linear_theory_2}

\subsection{Linear perturbative formalism}

Let the unperturbed steady state distribution function (DF) of a collisionless stellar system be given by $f_0$ and the corresponding Hamiltonian be $H_0$. $f_0$ satisfies the CBE,
\begin{align}
[f_0,H_0]=0,
\end{align}
where the square brackets correspond to the Poisson bracket. Now let us introduce a small time-dependent perturbation in the potential, $\Phi_\rmP(t)$, such that the perturbed Hamiltonian becomes
\begin{align}
H=H_0+\Phi_\rmP(t)+\Phi_1(t),
\end{align}
where $\Phi_1$ is the gravitational potential sourced by the response density, $\rho_1 = \int f_1 \rmd^3\bv$, via the Poisson equation,
\begin{align}
\nabla^2\Phi_1=4\pi G\rho_1.
\end{align}
Here $f_1$ is the linear order perturbation in the DF, i.e., the linear response of the system to the perturbation in the potential. The perturbed DF can thus be written as
\begin{align}
f=f_0+f_1.
\end{align}
Assuming that the perturbations are small such that linear perturbation theory holds, the time-evolution of $f_1$ is governed by the following linearized version of the CBE
\begin{align}
\frac{\partial f_1}{\partial t}+[f_1,H_0]+[f_0,\Phi_\rmP]+[f_0,\Phi_1]=0.
\label{CBE_perturb_2}
\end{align}
In this chapter we shall neglect the self-gravity of the disk, i.e., neglect the polarization term, $[f_0,\Phi_1]$, in the lhs of the linearized CBE. We briefly discuss the impact of self-gravity in Section~\ref{sec::caveats}, leaving a more detailed analysis including self-gravity to a forthcoming publication.

\subsection{Hybrid perturbative formalism for an infinite slab}
\label{sec:slab}

We consider the simplified case of perturbations in an infinitely extended slab, uniform in $(x,y)$, but characterized by a vertical density profile $\rho(z)$. Although a rather poor approximation of a realistic galactic disk, this idealized case serves to highlight some of the main characteristics of disk response. We consider perturbations that can be described by a profile in the vertical $z$-direction and by a superposition of plane waves along the $x$-direction, such that $\Phi_\rmP$ and $f_1$ are both independent of $y$. After making a canonical transformation from the phase-space variables $(z,v_z)$ to the corresponding action angle variables $(I_z,w_z)$, Equation~(\ref{CBE_perturb_2}) becomes
\begin{align}
\frac{\partial f_1}{\partial t}+\frac{\partial H_0}{\partial I_z}\frac{\partial f_1}{\partial w_z}+\frac{\partial H_0}{\partial v_x}\frac{\partial f_1}{\partial x}-\frac{\partial \Phi_\rmP}{\partial w_z}\frac{\partial f_0}{\partial I_z}-\frac{\partial \Phi_\rmP}{\partial x}\frac{\partial f_0}{\partial v_x}=0.
\label{CBE_perturb1}
\end{align}
The unperturbed Hamiltonian $H_0$ can be written as
\begin{align}
H_0 = \frac{v^2_x+v^2_y}{2} + \frac{v^2_z}{2} + \Phi_z(z),
\end{align}
where $v_x$, $v_y$ and $v_z$ are the unperturbed velocities of stars along $x$, $y$ and $z$ respectively, and $\Phi_z(z)$ is the unperturbed potential that dictates the oscillatory vertical motion of the stars.
We expand $\Phi_\rmP$ and $f_1$ as Fourier series that are discrete along $z$ but continuous along $x$:
\begin{align}
\Phi_\rmP(z,x,t)&=\sum_{n=-\infty}^{\infty}\int \rmd k\, \exp{\left[i (n w_z + k x)\right]}\, \Phi_{nk}(I_z,t),\nonumber \\
f_1(z,v_z,x,v_x,v_y,t)&=\sum_{n=-\infty}^{\infty}\int \rmd k\, \exp{\left[i (n w_z + k x)\right]}\, f_{1nk}(I_z,v_x,v_y,t).
\end{align}
Here $z$ can be expressed as the following implicit function of $w_z$ and $I_z$,
\begin{align}
w_z = \Omega_z \int_0^{z} \frac{\rmd z'}{\sqrt{2\left[E_z(I_z)-\Phi_z(z')\right]}},
\label{z_wz_Iz}
\end{align}
where $\Omega_z=\Omega_z(I_z)$ is the vertical frequency of stars with vertical action $I_z$, given in equation~(\ref{omzvx}) below.

Here and throughout this chapter we express any dependence on the continuous wave number $k$ with an index rather than an argument, i.e., $\Phi_{nk}(I_z,t)$ rather than $\Phi_n(k,I_z,t)$. This implies that any function that carries $k$ as an index is in Fourier space.

\begin{figure*}[h!]
  \centering
  \includegraphics[width=0.95\textwidth]{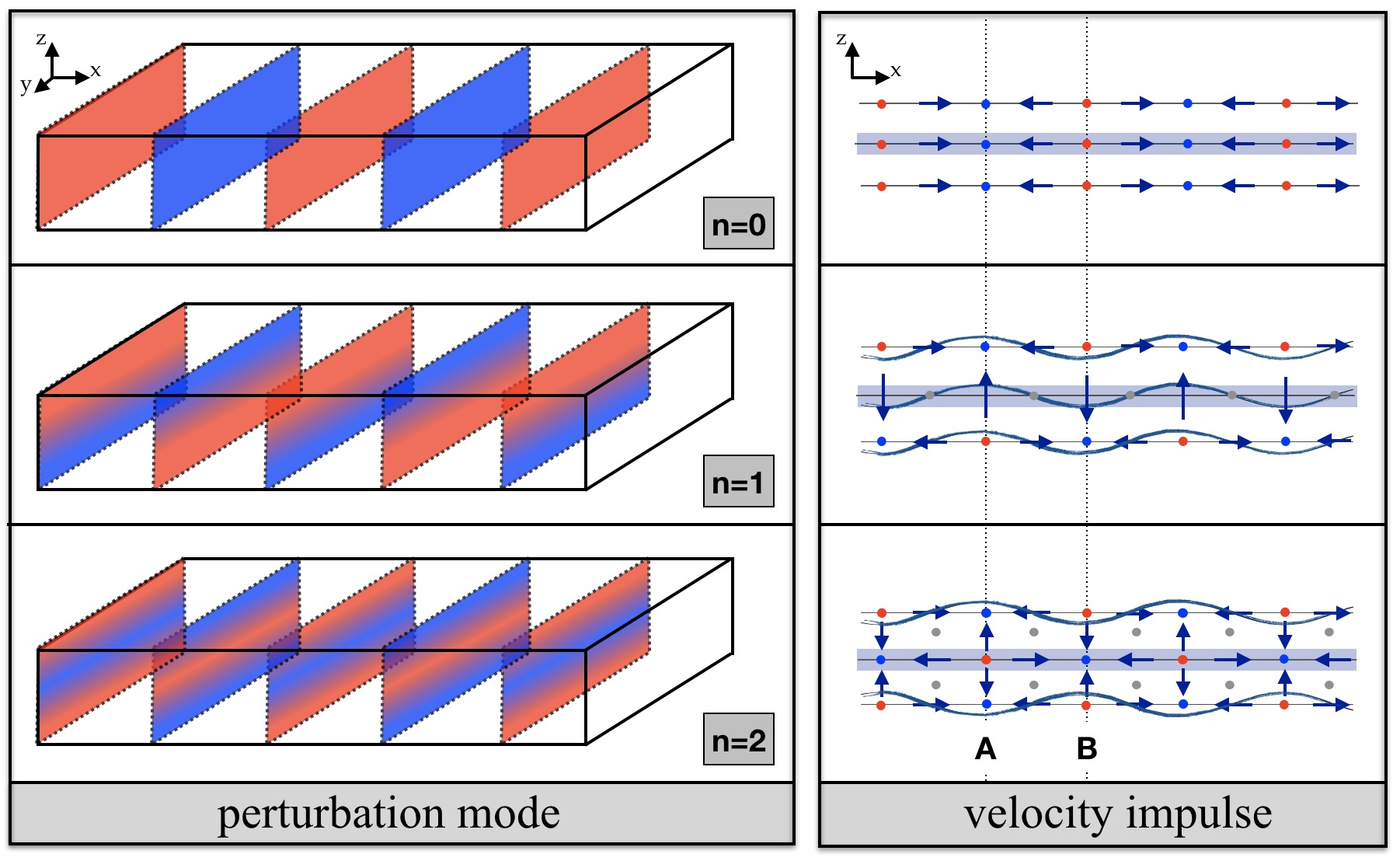}
  \caption{Illustration of the $n=0$, $n=1$ and $n=2$ plane-wave perturbation modes in a laterally uniform and vertically isothermal slab (left-hand panel) and the velocity impulses corresponding to these modes (right-hand panel) in the case of an instantaneous/impulsive perturbation. In the left-hand panel, the rectangular box indicates a random section of the slab, centered on the slab's midplane ($z=0$), while red and blue colors indicate positive and negative $\Phi_\rmP$. For clarity, this color coding is only shown at the extrema (peaks and troughs) of the mode, which has a wave-vector that is pointing in the $x$-direction. The right-hand panel shows an edge-on view of the slab, with arrows indicating the local direction of the velocity impulse caused by the instantaneous perturbation $\Phi_\rmP$, and dots marking locations in the disk where the velocity impulse is zero. Whereas the $n=0$ mode corresponds to a longitudinal perturbation, both $n=1$ and $n=2$ correspond to transverse perturbations; the former is a bending mode, while the latter is a breathing mode (note though that both these modes also cause velocity impulses in the lateral directions). Finally, `A' and `B' mark two specific locations in the slab to which we refer in the text and in Figs.~\ref{fig:impulsive_slab_n1} and~\ref{fig:impulsive_slab_n2}.}
  \label{fig:modes}
\end{figure*}

We express the perturber potential and the DF perturbation or response as linear superpositions of Fourier modes. Since we do not take into account the self-gravity of the response itself, i.e., do not self-consistently solve the Poisson equation along with the CBE, these are not dynamical or normal modes of the system. In other words, the oscillation frequencies of the Fourier modes are just the unperturbed frequencies, $\Omega_z$, and do not follow a dispersion relation as in the self-gravitating case. To aid the visualization of the various Fourier modes, Fig.~\ref{fig:modes} illustrates what the $n=0$, $n=1$ and $n=2$ modes for one particular value of the wavenumber $k$ look like. The figure also indicates the direction of the velocity impulses resulting from an instantaneous perturbation of each mode.

Substitution of the above expressions in equation~(\ref{CBE_perturb1}) yields the following evolution equation for $f_{1nk}$
\begin{align}
\frac{\partial f_{1nk}}{\partial t}+i(n\Omega_z+k v_x)f_{1nk}=i\left(n\frac{\partial f_0}{\partial I_z}+k\frac{\partial f_0}{\partial v_x}\right)\Phi_{nk},
\label{f1nk_de_2}
\end{align}
where we have used that
\begin{align}
\Omega_z = \frac{\partial H_0}{\partial I_z}\,, \;\;\; v_x=\frac{\partial H_0}{\partial v_x}.
\label{omzvx}
\end{align}
The above first order differential equation in time is easily solved using the Green's function technique. With the initial condition, $f_{1nk}(t_\rmi)=0$, we obtain the following integral form for $f_{1nk}$ for a given time dependence of the perturber potential,
\begin{align}
f_{1nk}(I_z,v_x,v_y,t)&=i\left(n\frac{\partial f_0}{\partial I_z}+k\frac{\partial f_0}{\partial v_x}\right)\int_{t_i}^{t}\rmd\tau \exp{\left[-i(n\Omega_z+k v_x)(t-\tau)\right]}\, \Phi_{nk}(I_z,\tau).
\label{f1nk_slabsol}
\end{align}
This solution is analogous to the particular solution for a forced oscillator with natural frequencies, $n\Omega_z$ and $k v_x$, which is being forced by an external perturber potential, $\Phi_{nk}$. The time-dependence of this external perturbation ultimately dictates the temporal evolution of the perturbation in the DF, $f_{1nk}$. A {\it net} response requires gradients in the (unperturbed) DF with respect to the actions and/or velocities. Similar solutions for the response of perturbed, collisionless systems have been derived in a number of previous studies \citep[e.g.,][]{LyndenBell.Kalnajs.72, Tremaine.Weinberg.84, Carlberg.Sellwood.85, Weinberg.89, Weinberg.91, Weinberg.04, Kaur.Sridhar.18, Banik.vdBosch.21a, Kaur.Stone.22, Chiba.Schonrich.22}, often in the context of phenomena like angular momentum transport, radial migration or dynamical friction.

\subsection{Perturbation in an isothermal slab}
\label{sec:iso_slab}

The infinite slab has a non-uniform (uniform) density profile along the vertical (horizontal) direction. Therefore the unperturbed motion of the stars is only vertically bounded by a potential but is unbounded horizontally. This implies that the unperturbed DF, $f_0$, involves a potential $\Phi_z$ only along $z$. For simplicity, we assume it to be isothermal but with different velocity dispersions in the vertical direction, $\sigma_z$, and the in-plane directions, $\sigma_x = \sigma_y \equiv \sigma$, i.e.,
\begin{align}
f_0(v_x,v_y,E_z)= \frac{\rho_c}{{(2\pi)}^{3/2}\sigma_z\,\sigma^2} \, \exp\left[-\frac{E_z}{\sigma^2_z}\right] \, \exp\left[-\frac{v^2_x+v^2_y}{2\sigma^2}\right],
\label{f_iso}
\end{align}
where 
\begin{align}
E_z = \frac{1}{2} v^2_z +\Phi_z(z)
\label{E_z}
\end{align}
is the energy involving the $z$-motion. The corresponding density and potential profiles in the vertical direction are given by
\begin{align}
\rho_z(z) = \rho_c \, {\sech}^2(z/h_z),\;\;\;\;\;\;\;\;\;\;\;\Phi_z(z) = 2 \sigma^2_z \,\ln\left[\cosh(z/h_z)\right],
\end{align}
where $h_z$ is the vertical scale height \citep[][]{Spitzer.42,Camm.50}. The vertical action, $I_z$, can be obtained from the unperturbed Hamiltonian, $E_z$, as follows
\begin{align}
I_z = \frac{1}{2\pi} \oint v_z \, \rmd z= \frac{2}{\pi} \int_0^{z_{\rm max}} \sqrt{2[E_z - \Phi_z(z)]} \, \rmd z,
\end{align}
where $\Phi_z(z_{\rm max})=E_z$, i.e., $z_{\rm max} = h_z \cosh^{-1}\left(\exp{\left[E_z/2\sigma^2_z\right]}\right)$. The time period of vertical oscillation is given by
\begin{align}
T_z = \oint \frac{\rmd z}{v_z} = 4\int_0^{z_{\rm max}} \frac{\rmd z} {\sqrt{2\left[E_z-\Phi_z(z)\right]}},
\label{T_z_2}
\end{align}
and the vertical frequency is $\Omega_z = 2\pi/T_z$. Throughout this chapter, to compute the perturbative response of the slab, we shall use typical MW parameter values, i.e., $h_z=0.4$ kpc, $\sigma_z=23$ km/s, and $\sigma=1.5\,\sigma_z=35$ km/s \citep[][]{McMillan.11}.

Substituting the above form for $f_0$ (Equation~[\ref{f_iso}]) in Equation~(\ref{f1nk_slabsol}) and using that $\Omega_z = \Omega_z(I_z) = \partial E_z/\partial I_z$ yields the following closed integral form for $f_{1nk}$:
\begin{align}
f_{1nk}(I_z,v_x,v_y,t) &= -i\left(\frac{n\Omega_z}{\sigma^2_z} + \frac{k v_x}{\sigma^2}\right) \, f_0(v_x,v_y,E_z) \nonumber \\
&\times \int_{t_i}^{t}\rmd \tau \, \exp{\left[-i(n\Omega_z+k v_x)(t-\tau)\right]} \, \Phi_{nk}(I_z,\tau).
\label{f1nk_isosol}
\end{align}

\subsection{Perturber potential}
\label{sec:per_pot}

The slab response depends on the spatio-temporal nature of the perturber. In this chapter we consider two different functional forms of the perturber potential described below.

\subsubsection{Separable potential}
\label{sec:sep_per_pot}

In order to capture the essential physics of perturbative collisionless dynamics without much computational complexity, we shall consider the following separable form for the perturber potential:
\begin{align}
\Phi_\rmP(z,x,t) = \Phi_\rmN\, \calZ(z) \calX(x) \calT(t),
\label{Phip_sep}
\end{align}
where $\Phi_\rmN$ has the units of potential, and $\calZ$, $\calX$ and $\calT$ are dimensionless functions of $z$, $x$ and $t$ respectively that specify the spatio-temporal profile of $\Phi_\rmP$. Thus, the Fourier transform of $\Phi_\rmP$ can also be written in the following separable form,
\begin{align}
\Phi_{nk}(I_z,t) = \Phi_\rmN\, \calZ_n(I_z) \calX_k\, \calT(t).
\label{Phip_sep_fourier}
\end{align}
Here $\calZ_n(I_z)$ is the $n^{\rm th}$ Fourier coefficient in the discrete Fourier series expansion of $\calZ(z)$ in the vertical angle, $w_z$, given by
\begin{align}
\calZ_n(I_z) = \frac{1}{2\pi} \int_0^{2\pi} \rmd w_z \, \calZ(z) \, \exp{\left[-in w_z\right]},
\label{Phip_sep_fourier_z}
\end{align}
where we have used the implicit expression for $z$ in terms of $w_z$ and $I_z$ given in equation~(\ref{z_wz_Iz}). $\calX_k$ is the Fourier transform of $\calX(x)$, given by
\begin{align}
\calX_k = \frac{1}{2\pi} \int_{-\infty}^{\infty} \rmd x\,  \calX(x) \, \exp{\left[-ikx\right]}.
\label{Phip_sep_fourier_x}
\end{align}
In the following sections, we investigate the slab response to perturbers with various functional forms for $\calX(x)$ and $\calT(t)$, while keeping the form for $\calZ(z)$ arbitrary. We start in Section~\ref{sec:impulsive_kick} with an impulsive ($\calT(t)=\delta(t)$) single-mode ($\calX(x) = \exp[ikx]$) perturbation, followed in Section~\ref{sec:localized} by a perturbation that is temporally impulsive but spatially localized ($\calX(x)=\exp{\left[-x^2/\Delta^2_x\right]}$). In Section~\ref{sec:non-impulsive} we consider the same spatially localized perturbation, but this time temporally extended ($\calT(t)=\exp{\left[-\omega^2_0 t^2\right]}$).

\subsubsection{Satellite galaxy along straight orbit}
\label{sec:sat_per_pot}

As a practical astrophysical application of our perturbative formalism, we also study the response of an isothermal slab to a satellite galaxy or DM subhalo undergoing an impact along a straight orbit with a uniform velocity $\vp$ at an angle $\thetap$ (with respect to the disk normal). We model the impacting satellite as a point perturber, whose potential is given by

\begin{align}
\Phi_\rmP(z,x,t) = -\frac{GM_\rmP}{\sqrt{{\left(z-\vp\cos{\thetap} t\right)}^2+{\left(x-\vp\sin{\thetap} t\right)}^2}}.
\label{Phip_sat}
\end{align}
In this case the spatial and temporal parts are coupled and thus the slab response needs to be evaluated by performing the $\tau$ integral before the $w_z$ and $x$ integrals (to find $\Phi_{nk}$), as shown in Appendix~\ref{App:sat_disk_resp}.

\section{Response to an Impulsive Perturbation}
\label{sec:impulsive_kick}

In order to gain some insight into the perturbative response of the slab, we start by solving equation~(\ref{f1nk_isosol}) for an instantaneous impulse at $t=0$; i.e., $\calT(t) = \delta(t)$. Here the normalization factor $\Phi_\rmN$ has the units of potential times time. With the initial time $t_i<0$, the integral over $\tau$ yields $\exp{\left[-i(n\Omega_z+k v_x)t\right]}$. Further integrating $f_{1nk}$ over $v_x$ and $v_y$ and summing over all $n$ modes, yields the following form for any given $k$ mode of the perturbed DF for a given action $I_z$ and angle $w_z$:
\begin{align}
f_{1k}(I_z,w_z,t)&=\sum_{n=-\infty}^{\infty}\exp{\left[in w_z\right]}\int_{-\infty}^{\infty}\rmd v_y\int_{-\infty}^{\infty}\rmd v_x\, f_{1nk}(I_z,v_x,v_y,t)\nonumber \\
&=A_{\rm norm}\, D_k(t)\, R_k(I_z,w_z,t),
\label{f1_delta}
\end{align}
where 

\begin{align}
A_{\rm norm}=\frac{\rho_c}{\sqrt{2\pi}\sigma_z} \exp{\left[-E_z/\sigma^2_z\right]}
\end{align}
is a normalization factor reflecting the vertical structure of the unperturbed disk,

\begin{align}
D_k(t)=\exp{\left[-\frac{k^2\sigma^2t^2}{2}\right]}
\end{align}
is a damping term that describes the temporally Gaussian decay of the response by lateral mixing, and

\begin{align}
R_k(I_z,w_z,t)=-\Phi_\rmN\calX_k\sum_{n=-\infty}^{\infty}\calZ_n(I_z)\left(k^2t+i\frac{n\Omega_z}{\sigma^2_z}\right)\exp{\left[i n \left(w_z - \Omega_z\,t\right)\right]}
\label{Rk_impulse}
\end{align}
is a (linear) response function that includes vertical phase-mixing.

Equation~(\ref{f1_delta}) is the basic `building block' for computing the response of our infinite isothermal slab to a perturbation mode $k$ in the impulsive limit. Using the canonical transformation from $(w_z,I_z)$ to $(z,v_z)$, i.e., using equations~(\ref{z_wz_Iz}) and~(\ref{E_z}), we can transform $f_{1k}(I_z,w_z,t)$ to $f_{1k}(v_z,z,t)$. Upon multiplying this by $\exp{\left[ikx\right]}$ and integrating over $k$, and then integrating further over $v_z$ at a fixed $z$, one obtains the response density as a function of both time and position:
\begin{align}
\rho_1(z,x,t) &=-\frac{\rho_c \Phi_\rmN}{\sqrt{2\pi}\sigma_z} \sum_{n=-\infty}^{\infty} \int_0^{\Tilde{I}_z} \rmd I_z\, \frac{\Omega_z}{\sqrt{2\left[E_z-\Phi_z(z)\right]}} \exp{\left[-E_z/\sigma^2_z\right]} \exp{\left[i n \left(\Tilde{w}_z - \Omega_z\,t\right)\right]} \nonumber \\
&\times \calZ_n(I_z) \int \rmd k\,\exp{\left[ikx\right]}\,  \exp{\left[-\frac{k^2\sigma^2t^2}{2}\right]}\left(k^2t+i\frac{n\Omega_z}{\sigma^2_z}\right)\calX_k,
\end{align}
where $\Tilde{I}_z$ is the solution of $E_z(I_z)=\Phi_z(z)$, and $\Tilde{w}_z$ is the solution for $w_z(z,I_z)$ from equation~(\ref{z_wz_Iz}).

\begin{figure*}[t!]
  \centering
  \includegraphics[width=0.95\textwidth]{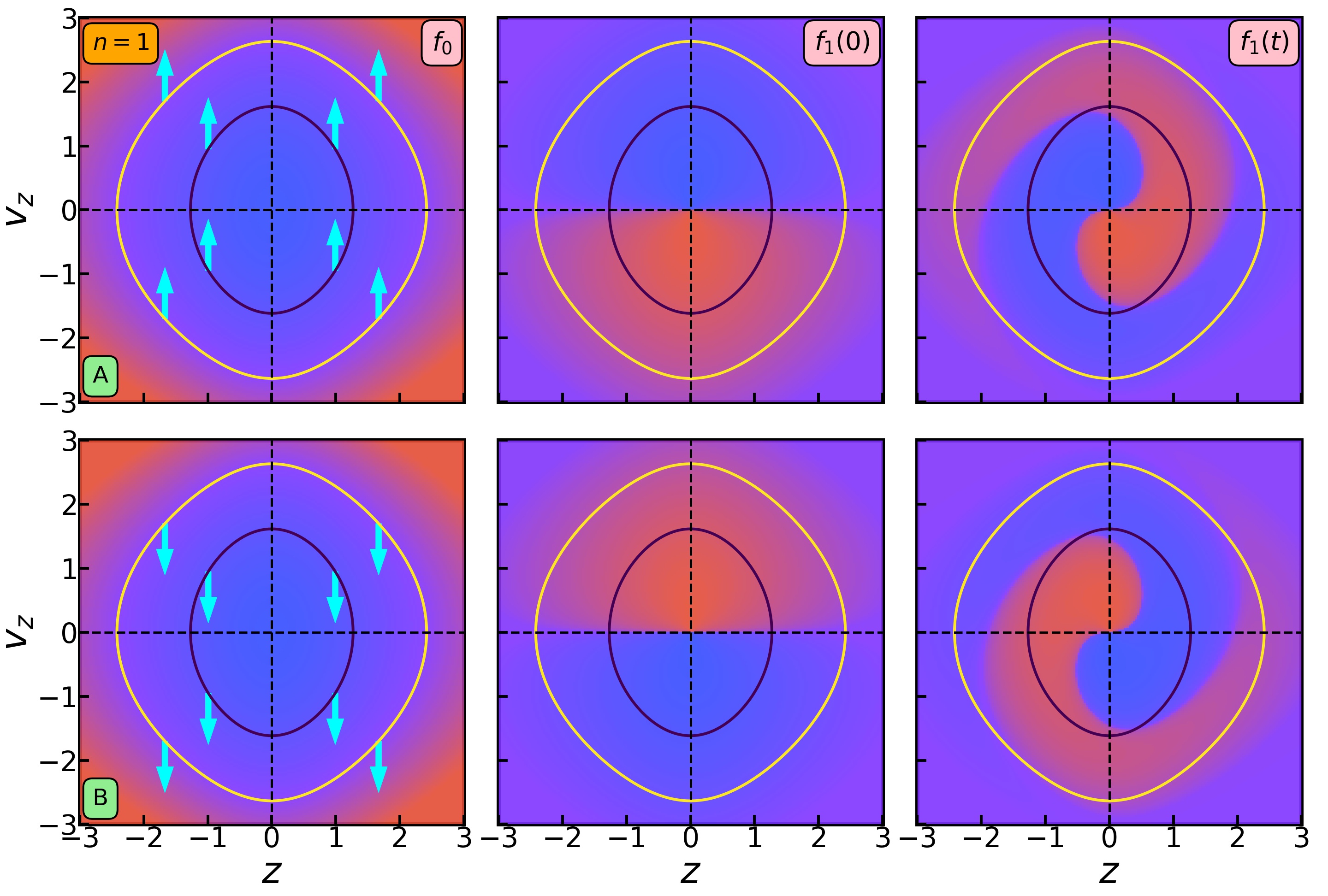}
  \caption{The formation of a one-armed phase spiral due to an impulsive $n=1$ bending-mode perturbation. The color-coding in the left-hand panels shows the unperturbed distribution function $f_0(z,v_z)$ (equation~[\ref{f_iso}]) in the isothermal slab at neighboring locations A (top) and B (bottom), separated by a lateral distance of $\pi/k$, with blue (red) indicating a higher (lower) phase-space density. Locations A and B coincide with extrema in the perturbation mode as depicted in Fig.~\ref{fig:modes}. The black and yellow contours indicate the phase-space trajectories for two random values of $E_z$ (or, equivalently, $I_z$). The cyan arrows indicate the velocity impulses resulting from the instantaneous perturbation at different locations in phase-space. Note that, in the case of the $n=1$ mode considered here, at the extrema A and B all velocity impulses $\Delta v_z$ are positive and negative, respectively (cf. Fig~\ref{fig:modes}). The middle panels indicate the response $f_1$ immediately following the instantaneous response (at $t=0$), with blue (red) indicating a positive (negative) response density. Finally, the right-hand panels show the response after some time $t$, computed using equation~(\ref{f1_delta}). Note how the response at A reveals a one-armed phase spiral that is exactly opposite of that at location B, i.e., they exactly cancel each other. Hence, lateral mixing causes damping of the phase spiral amplitude.}
  \label{fig:impulsive_slab_n1}
\end{figure*}

In order to gain insight into the slab response for a particular $I_z$ and $w_z$, let us start by analyzing equation~(\ref{f1_delta}) for the $n=0$ mode, an in-plane density wave, for which the perturbation causes an in-plane velocity impulse as depicted in Fig.~\ref{fig:modes}. The response is a standing, longitudinal oscillation in density. The response function for this mode is $R_k(I_z,w_z,t) = \Phi_\rmN \calZ_0(I_z)\calX_k\, k^2 t$, indicating that the amplitude of oscillation initially grows linearly with time.  However, this growth is inhibited by the Gaussian damping function $D_k(t) = \exp[-k^2 \sigma^2 t^2/2]$, which describes lateral mixing due to the non-zero velocity dispersion of stars in the $k$-direction. The Gaussian form of this temporal damping term owes its origin to the assumed Gaussian/Maxwellian form of the unperturbed velocity distribution along the plane. Hence, following the perturbation, the $n=0$ mode starts to grow linearly with time, but then rapidly damps away; the response loses its coherence due to mixing in the direction of the wave-vector. In the cold slab limit $(\sigma\to 0)$, without any lateral streaming motion to damp it out, the response will grow linearly in time until it eventually becomes non-linear. This is because in an infinite, laterally homogeneous slab there is no restoring force in the lateral directions, causing the stars to stream uninhibited towards (away from) the minima (maxima) of $\Phi_\rmP$ due to the initial velocity impulse induced. This leads to over- and under-density spikes which cannot be treated using linear theory. Hence, Equation~(\ref{f1_delta}) can only adequately describe the response to an instantaneous $n=0$ mode at early times, or if the damping time $\tau_\rmD \sim (k \sigma)^{-1}$ is shorter than the time-scale of formation of density spikes. The latter is roughly the time needed to cross one quarter of the perturbation's wavelength with the velocity impulse triggered at the zeroes of $\Phi_\rmP$. Therefore, in order for linear theory to be valid, we require that $\sigma > (2/\pi) \, |\Delta v|_{\rm max}$, where $|\Delta v|_{\rm max}=k\, \Phi_\rmN \calZ_0(I_z)\calX_k$. Moreover, upon including self-gravity, it can be found that the $n=0$ mode is linearly stable only for $k>k_\rmJ\approx \sqrt{4\pi G \rho_c}/\sigma$ \citep[][]{Binney.Tremaine.08}, or in other words $\lambda<\lambda_\rmJ\approx \sigma\sqrt{\pi/G \rho_c}$, where $k_\rmJ$ and $\lambda_\rmJ=2\pi/k_\rmJ$ refer to the Jeans wave-number and Jeans wavelength respectively. In the $\sigma\to 0$ limit, the Jeans wave-length, $\lambda_\rmJ\to 0$, and thus the $n=0$ mode becomes globally unstable. Hence, the condition of Jeans stability requires an additional constraint on $\sigma$: $\sigma>\sqrt{4\pi G\rho_c}/k$. The validity of linear perturbation theory thus requires that for each $k$,

\begin{align}
\sigma > \max{\left[\frac{\sqrt{4\pi G \rho_c}}{k},\frac{2k}{\pi}\Phi_\rmN \calZ_0(I_z)\calX_k\right]}.
\end{align}

\begin{figure*}[t!]
  \centering
  \includegraphics[width=0.95\textwidth]{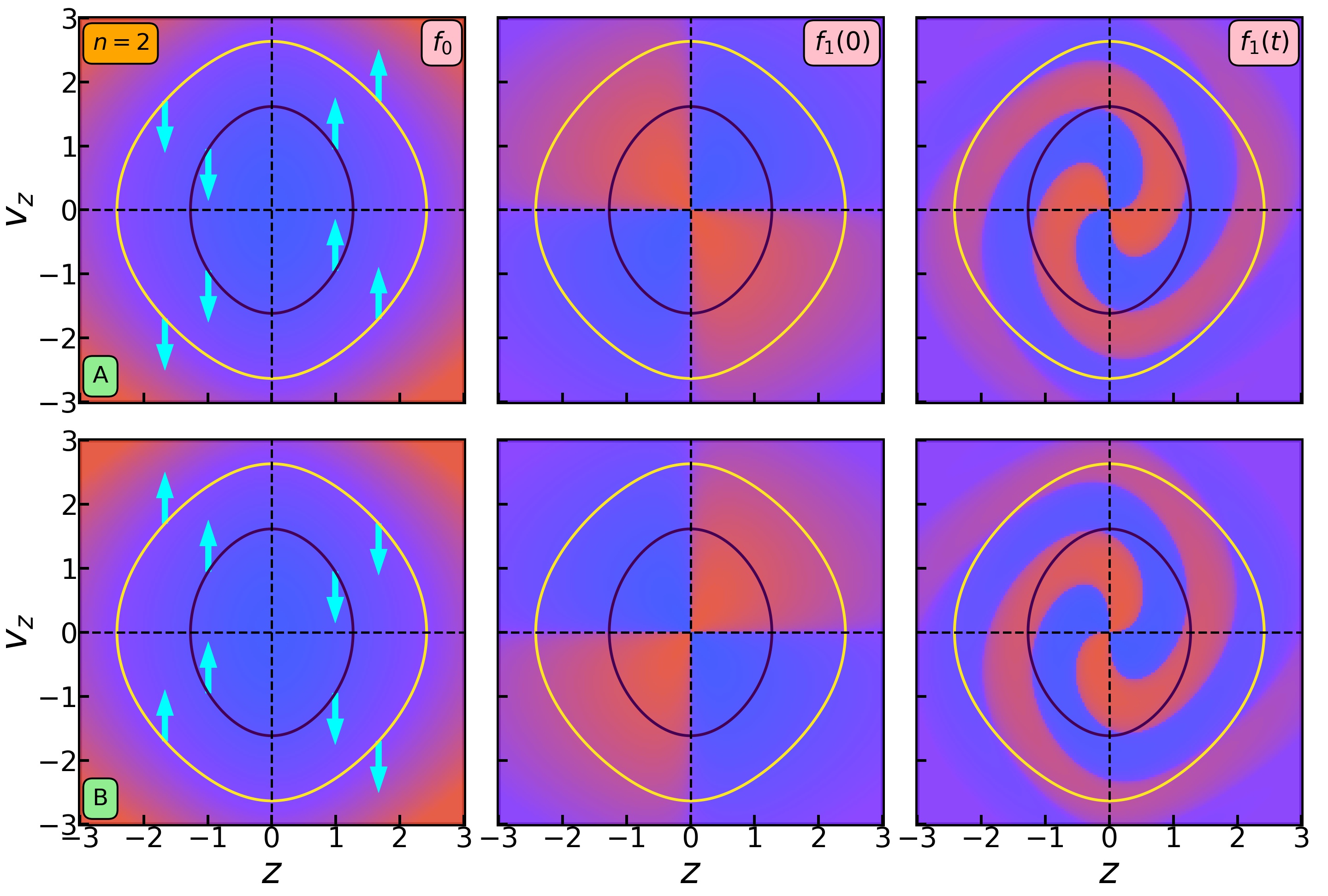}
  \caption{Same as Fig.~\ref{fig:impulsive_slab_n1}, except for a pure $n=2$ breathing mode. Note how in this case the velocity impulses above and below the mid-plane are of opposite sign (cyan arrows in left-hand panels). As a consequence, the response density immediately following the perturbation has a quadrupole signature (middle panels), which ultimately gives rise to two-armed phase spirals (right-hand panels). Note how once again, the phase spirals at A and B are each other's additive inverse.}
  \label{fig:impulsive_slab_n2}
\end{figure*}

For $n=1$, the perturbation is a standing, transverse wave on the slab, formally known as the bending wave. The perturbation induces velocity impulses in the direction perpendicular to the slab, as indicated in  Fig.~\ref{fig:modes}. At the locations marked A and B, separated by a lateral distance of $\pi/k$, these velocity impulses point in the positive and negative $z$-directions, respectively. The top panels of Fig.~\ref{fig:impulsive_slab_n1} illustrate the impact this has at location A. The left-hand panels indicate the velocity impulses (cyan arrows) in the $(z,v_z)$-plane. Prior to the perturbation, due to the vertical restoring force from the slab, each star executes a periodic oscillation in this plane. The black and yellow contours indicate the corresponding phase-space trajectories for two values of $I_z$, while the heat-map indicates phase-space density (bluer colors indicate higher density). The top-middle panel shows that immediately following the impulse, the phase-space density is boosted (reduced) where $v_z>0$ ($v_z<0$), resulting in a dipole pattern in the phase-space distribution of stars. After the impulse, the stars continue to execute periodic motion in the $(z,v_z)$-plane, but starting from their new position (corresponding to a modified action $I_z$). The angular frequency of this periodic motion is $\Omega_z$, which is a function of the (modified) action, and hence, stars with different actions oscillate in the $(z,v_z)$-plane at different frequencies. As a consequence, the perturbed phase-space density shown in the middle panels is wound-up into a {\it phase spiral} of over- and under-densities as depicted in the right-most panels of Fig.~\ref{fig:impulsive_slab_n1}. The bottom panels of Fig.~\ref{fig:impulsive_slab_n1} show what happens following the impulsive perturbation at location B. Since the velocity impulses are now reversed in direction, the phase spiral that emerges is exactly the opposite of that at location A.

The creation of phase spirals is an outcome of phase-mixing in the $z$-direction and is described by the oscillatory factor, $\exp[i\,n(w_z -\Omega_z t)]$, that is part of the response function $R_k(I_z,w_z,t)$. It consists of two terms: a term that scales as $k^2 t$, which describes the lateral streaming motion of stars due to the non-zero velocity impulses in the lateral directions (see Fig.~\ref{fig:modes}), and a term that scales as $n\Omega_z/\sigma^2_z$ which purely describes the vertical oscillations. As in the case of the $n=0$ mode, the lack of a restoring force in the lateral directions\footnote{If accounting for self-gravity of the response density, there will be non-zero forces in the lateral direction, but these will promote growth rather than act as a restoring force. This ultimately leads to exponential growth (according to linear theory) of unstable modes and Landau damping of stable modes, which occurs exponentially, i.e., more slowly than the Gaussian lateral mixing in the absence of self-gravity.} causes the perturbation to grow linearly with time in the absence of lateral streaming (for a cold disk with $\sigma \approx 0$). Meanwhile, the phase spirals continue to wind-up, which implies that the vertical bending loses its coherence. Over time, phase-mixing in the vertical direction will ensure that the disk regains mirror-symmetry with respect to the midplane, but with a scale-height, $h_z$, that would be a periodic function of $x$, with a wavelength equal to $\pi /k$ (i.e., half the wave-length of the original perturbation).
 
However, all this ignores lateral mixing due to the unconstrained motion with non-zero velocity dispersion in the $x$ direction. Stars that received an impulse $\Delta v_z>0$ create phase spirals that are exactly the inverse of those created by neighboring stars for which the impulse was negative. Thus lateral mixing between neighboring points on the slab causes a damping of the phase spiral amplitude at any location, a process that is captured by the damping function $D_k(t)$. The lateral mixing timescale is $\tau_\rmD \sim 1/k\sigma$, indicating, as expected, that small scale perturbations (larger $k$) mix faster, and that mixing is more efficient for larger velocity dispersion in the lateral direction. After a few mixing time-scales, the slab will once again be completely homogeneous (laterally), with a scale-height $h_z$ that is independent of location. In addition, the density of stars in the $(z,v_z)$-plane will once again be perfectly symmetric without any trace of a phase spiral. The slab has completely equilibrated, and the only impact that remains of the impulsive perturbation is that the new scale-height is somewhat larger than it was originally, i.e., the impulsive perturbation has injected energy into the disk, which causes it to puff-up in the vertical direction. Hence, the final outcome is as envisioned in the impulsive-heating scenario discussed in the seminal study of \citet{Toth.Ostriker.92}. This persistent effect in the vertical density profile is however only captured in perturbation theory at second order \citep[e.g.,][]{Carlberg.Sellwood.85}; to first order the perturbation simply phase-mixes away in the impulsive limit considered here.

For $n=2$, the perturbation triggers a breathing mode, as depicted in Fig.~\ref{fig:modes}, i.e., at a given location A on the slab, the velocity impulses for this mode are positive (negative) for positive (negative) $z$. As evident from Fig.~\ref{fig:impulsive_slab_n2}, this leads to a quadrupole pattern for the initial perturbed phase-space distribution of stars, which becomes a two-armed phase spiral over time, as opposed to the one-armed phase spiral resulting from the $n=1$ mode. This reveals an important lesson: the structure of a phase spiral depends, among others, on which perturbation mode(s) are triggered. The phase spirals in regions A and B are each other's additive inverse. Hence, once again lateral mixing will cause damping of the phase spiral's amplitude, as described by the damping function $D_k(t)$. \cite{Hunt.etal.21} have shown using N-body simulations that two-armed phase spirals can indeed arise from breathing mode oscillations and that both bending and breathing modes can be excited at different locations on the MW disk by satellite-induced perturbations such as the passage of Sagittarius (see section~\ref{sec::MW_satellites} for detailed discussion).

To summarize, we see that, in case of our infinite slab, equilibration after an impulsive perturbation is driven by a combination of phase-mixing in the vertical direction and free-streaming damping in the horizontal direction. While the former gives rise to phase spirals in the $(\sqrt{I_z} \cos w_z, \sqrt{I_z} \sin w_z)$ or equivalently the $(z,v_z)$ plane, the latter causes them to damp away by lateral mixing. Due to vertical phase-mixing the phase spiral will continue to wrap itself up into a more and more tightly wound pattern, until its structure can no longer be discerned observationally due to finite-$N$ noise \citep[][]{Beraldo_e_Silva.etal.19a,Beraldo_e_Silva.etal.19b} and measurement errors in the actions and angles of individual stars (this is an example of coarse-grain mixing). Hence, even without lateral mixing phase spirals are only detectable for a finite duration.

\section{Response to a localized perturbation}
\label{sec:localized}

In the previous section we investigated the slab response to an external disturbance with a single wavenumber $k$. Realistic perturbations are however localized in space and thus consist of many wavenumbers. In this section we shall look into what happens when the slab is hit by an impulsive perturbation that is spatially localized. 

For simplicity, we assume that the external perturber behaves as a Gaussian packet with half-width $\Delta_x$ along the $x$ direction, i.e., $\Phi_\rmP$ is given by equation~(\ref{Phip_sep}) with
\begin{align}
\calX(x) = \exp{\left[-x^2/2\Delta_x^2\right]}.
\end{align}
The $\calZ(z)$ term in equation~(\ref{Phip_sep}) denotes the vertical structure of the perturber potential, which is part of what dictates the relative strength of bending and breathing mode oscillations. We shall see in the next section, though, that the relative strength of the modes is mostly dictated by the form of $\calT(t)$. For simplicity, we only consider localization along the $x$ and $z$-directions; along the $y$-direction the perturbation is assumed to extend out to infinity. We emphasize, though, that this assumption does not impact the essential physics of phase-mixing and lateral mixing discussed below.

The Fourier transform of the perturber potential, $\Phi_{nk}$, is given by equation~(\ref{Phip_sep_fourier}), with
\begin{align}
\calX_k = \frac{\Delta_x}{\sqrt{2\pi}}\, \exp{\left[-k^2\Delta_x^2/2\right]}.
\label{Phink_gaussian}
\end{align}
Upon substituting the above expression for $\calX_k$ in equation~(\ref{f1_delta}) we obtain the response for a single $k$ mode, $f_{1k}$. After multiplying this by $\exp{\left[ikx\right]}$, integrating over all $k$ and summing over all $n$ modes, we obtain the following final form for the slab response density in the case of a (laterally) Gaussian perturber:
\begin{align}
f_1(I_z,w_z,x,t) &= \sum_{n=-\infty}^{\infty} \exp{\left[i n w_z\right]} \int_{-\infty}^{\infty} \rmd k\, \exp{\left[ikx\right]}\, f_{1k}(I_z,w_z,t) \nonumber \\
&=A_{\rm norm}\, D(x,t)\, R(I_z,w_z,x,t),
\label{f1_local_impulsive}
\end{align}
where 
\begin{align}
A_{\rm norm}=\frac{\rho_c}{\sqrt{2\pi}\sigma_z} \exp{\left[-E_z/\sigma^2_z\right]}
\end{align}
is the same normalization factor as in equation~(\ref{f1_delta}),
\begin{align}
D(x,t)=\frac{\Delta_x}{\sqrt{\Delta_x^2+\sigma^2 t^2}} \exp{\left[-\frac{x^2}{2\left(\Delta_x^2+\sigma^2 t^2\right)}\right]}
\end{align}
is a factor that captures the decay of the response by lateral mixing, and
\begin{align}
R(I_z,w_z,x,t)&=-\Phi_\rmN \sum_{n=-\infty}^{\infty}\calZ_n(I_z) \nonumber \\
&\times \left[\frac{t}{\Delta_x^2+\sigma^2 t^2}\left(1-\frac{x^2}{\Delta_x^2+\sigma^2 t^2}\right)+i\frac{n\Omega_z}{\sigma^2_z}\right]\exp{\left[i n \left(w_z - \Omega_z\,t\right)\right]}
\label{R_local_impulsive}
\end{align}
with $\calZ_n(I_z)$ given by equation~(\ref{Phip_sep_fourier_z}), corresponds to the remaining part of the response that includes vertical phase-mixing.

The above expression (equation~[\ref{f1_local_impulsive}]) for the slab response to a localized disturbance has several important features. Firstly, the profile of the slab response is nearly Gaussian in $x$ since we assumed a Gaussian form (along $x$) for the perturber potential. Secondly, the $D(x,t)$ factor describes the decay of the response amplitude and widening of the response profile due to mixing by lateral streaming. The mixing in this case occurs as a power law in time rather than like a Gaussian as for a single $k$ mode (see equation~[\ref{f1_delta}]), since the power spectrum of the Gaussian perturber is dominated by small $k$ which mix very slowly, at a timescale, $\tau_\rmD\sim 1/k\sigma$. Thirdly, the $R$ factor captures two important effects: (i) a transient response reflecting an initial linear growth due to the perturber-induced velocity impulse, followed by a subsequent decay by lateral mixing, and (ii) vertical oscillations of stars (for $n\neq 0$) at different frequencies resulting in phase-mixing over time and the formation of phase spirals as described in detail in Section~\ref{sec:impulsive_kick}. The $n=0$ modes, i.e., perturbations confined to the slab, damp out faster than the non-zero $n$ modes that manifest the vertical oscillations of stars. Since the perturber was introduced impulsively by means of a Dirac delta function in time, the higher order oscillations are stronger for the same value of $\calZ_n(I_z)$ as the corresponding changes in the vertical actions have larger amplitude. Typically, for $n\geq 2$, $\calZ_n(I_z)$ gets smaller with larger $n$; hence the $n=2$ breathing mode turns out to be the dominant mode of oscillation for impulsive disturbances. The response characteristics however change as we move to non-impulsive or more temporally extended perturbers in the next section.

It takes time for the local response to propagate along the slab by lateral streaming. Initially the perturber's gravity draws in stars towards the center of impact, $x=0$. Thus, immediately following the impulse, the region near the center of impact has a larger concentration of stars, which laterally stream outwards due to non-zero velocity dispersion. This leads to a damping of the response amplitude at small $x$ and growth at large $x$, or equivalently damping and widening of the response profile, which occurs at the rate,
\begin{align}
\calD_x(t) = \frac{\rmd}{\rmd t} \sqrt{\Delta^2_x + \sigma^2 t^2} = \frac{\sigma^2 t}{\sqrt{\Delta^2_x + \sigma^2 t^2}}.
\end{align}
This rate of outward streaming of slab material is initially equal to
\begin{align}
\lim_{t\to 0} \calD_x(t) = \frac{\sigma^2 t}{\Delta_x},
\end{align}
but at later times asymptotes to a constant value,
\begin{align}
\lim_{t\to \infty} \calD_x(t) = \sigma.
\end{align}

To summarize, the response to a spatially localized perturbation can be understood in the context of that to a single mode plane wave perturbation discussed in the previous section, as follows. In both cases, the response involves vertical oscillations that phase-mix away, thus giving rise to phase spirals. However, whereas the plane wave response maintains its sinusoidal profile in the lateral direction with an overall Gaussian decay of the amplitude due to lateral mixing, the response profile in the case of localized perturbation changes its shape and undergoes both decay and widening. This is because in the latter case the response is a linear superposition of responses to many plane wave perturbations with different $k$, each decaying in amplitude over a time-scale, $\tau_\rmD\sim 1/k\sigma$, due to lateral mixing by free-streaming. Since the spatially Gaussian profile considered here has a Gaussian power spectrum and thus more power on large scales (small $k$) that mix more slowly, the combined response from all $k$ modes undergoes much slower lateral mixing (as a power law) than that from a single $k$ mode. The typical timescale of coarse-grained survival (against free-streaming damping) of the phase spiral in this case turns out to be $\sim (f_{\rm max}/f_{\rm res})\,\Delta_x/\sigma$. Here $f_{\rm max}$ is the maximum amplitude of the phase spiral, which is attained at $t=0$, and $f_{\rm res}$ is the resolution limit. The power law nature of free-streaming damping implies that the response to spatially and temporally localized perturbations (e.g., encounters with satellite galaxies) can be sustained in the disk for a long time.

\section{Response to a non-impulsive perturbation}
\label{sec:non-impulsive}

Thus far we have only considered impulsive perturbations of our slab, with $\calT(t)=\delta(t)$. However, a realistic disturbance would not only have a spatial structure, the effects of which we studied in the previous section, but also be extended in time. In this section we investigate the effect of non-impulsive or temporally extended disturbances on the slab oscillations. In particular, we broaden the Dirac delta pulse from the previous section into a Gaussian in time, i.e., $\Phi_\rmP$ is given by equation~(\ref{Phip_sep}) with $\calT(t) = \frac{1}{\sqrt{\pi}}\,\exp{\left[-\omega^2_0 t^2\right]}$, where $\omega_0$ is the pulse frequency. We define the pulse-width or pulse-time as $\tau_\rmP=\sqrt{2}/\omega_0$. We also assume that the pulse is localized and follows a Gaussian profile in $x$ as in the previous section, i.e., $\calX(x)=\exp{\left[-x^2/2\Delta^2_x\right]}$. As before, $\calZ(z)$ in equation~(\ref{Phip_sep}) denotes some generic vertical profile. The (spatial) Fourier transform of this potential, $\Phi_{nk}$, is provided in equation~(\ref{Phip_sep_fourier}) with $\calX_k$ given by equation~(\ref{Phink_gaussian}) and $\calZ_n$ given by equation~(\ref{Phip_sep_fourier_z}). We can substitute this in equation~(\ref{f1nk_slabsol}) and perform the integration over $\tau$ and $v_x$ to obtain the following expression for the response for a single $k$ mode,
\begin{align}
&f_{1k}(I_z,w_z,t)=A_{\rm norm}\, D_k(t)\,R_k(I_z,w_z,t),
\label{f1k_gaussian}
\end{align}
where 
\begin{align}
A_{\rm norm}=\frac{\rho_c}{\sqrt{2\pi}\sigma_z} \exp{\left[-E_z/\sigma^2_z\right]}
\end{align}
is the same normalization factor as in equation~(\ref{f1_delta}),

\begin{align}
D_k(t) = \frac{\calQ^3}{2\omega_0}\,  \exp{\left[-\calQ^2\frac{k^2\sigma^2t^2}{2}\right]}
\end{align}
is a factor that describes the damping of the response due to lateral mixing, and

\begin{align}
R_{k}(I_z,w_z,t) &=-\Phi_\rmN\calX_k \sum_{n=-\infty}^{\infty} \calZ_n(I_z) \nonumber \\
& \times \left\{\,
S_{nk} \, \Upsilon_{nk}(t) \, \left(k^2 t + i \frac{n\Omega_z}{\sigma^2_z}\right) \exp{\left[i\, n(w_z-\calQ\,\Omega_z t)\right]} - \calG_{nk}(w_z,t)\right\},
\label{Rk_gaussian}
\end{align}
with $\calZ_n(I_z)$ given by equation~(\ref{Phip_sep_fourier_z}), includes the vertical phase-mixing of the response. Here $\calQ$ is a factor that depends on the pulse frequency, $\omega_0$, and the wavenumber $k$, and is given by
\begin{align}
\calQ=\calQ(\omega_0,k\sigma)=\frac{\omega_0}{\sqrt{\omega^2_0+\frac{k^2\sigma^2}{2}}}.
\end{align}
The mode-strength,
\begin{align}
S_{nk} = \exp{\left[-\frac{1}{\omega^2_0+\frac{k^2\sigma^2}{2}}\frac{n^2\Omega^2_z}{4}\right]}
\label{mode_strength_gaussian}
\end{align}
is a function that indicates the strength of each $n$ mode,
\begin{align}
\Upsilon_{nk}(t) &=  1+\erf\left\{\calQ \left(\omega_0 t-i\frac{n\Omega_z}{2\omega_0}\right)\right\}
\label{growth_gaussian}
\end{align}
describes the temporal build-up of the response and the decay of transient oscillations, and
\begin{align}
\calG_{nk}(w_z,t) = \frac{k^2}{\sqrt{\pi}\,\omega_0\calQ} \exp{\left[-\calQ^2\omega^2_0 t^2\right]} \exp{\left[in w_z\right]}
\label{transient_gaussian}
\end{align}
is another rapidly decaying transient feature. In the $\omega_0 \to \infty$ limit, both $\Upsilon_{n}(t)$ and the mode strength $S_{nk}$ become unity, and $\calG_{nk}(w_z,t) \to 0$, such that we recover the response to impulsive perturbations given in equation~(\ref{f1_delta}) as required.

It is interesting to contrast this response to an extended pulse to that in the impulsive limit. First of all, the damping factor, $D_k(t)$, which still owes its origin to lateral mixing due to non-zero velocity dispersion, now depends not only on $k$ and $\sigma$ but also on the pulse frequency $\omega_0$. The damping time is given by
\begin{align}
\tau_{\rmD} = \frac{1}{k\sigma} \sqrt{1+\frac{k^2\sigma^2}{2\omega^2_0}},
\end{align}
which scales as $\sim 1/k\sigma$ in the impulsive/short pulse ($\omega^2_0 \gg k^2\sigma^2/2$) limit indicating that the response mixes away laterally with small scale perturbations mixing faster. In the adiabatic/long pulse ($\omega^2_0 \ll k^2\sigma^2/2$) limit, though, $\tau_\rmD \to 1/\sqrt{2}\omega_0$, i.e., the damping of the response follows the temporal behaviour of the perturbing pulse itself, independent of $k$.

The mode-strength reveals several important trends: it exponentially damps away with $n^2$, implying that the lower order modes are much stronger for perturbations that are slower \citep[see also][]{Widrow.etal.14} and/or have larger wavelength (smaller $k$). Therefore the $n=1$ bending modes dominate over the $n=2$ breathing modes for a sufficiently slow pulse. Note, though, that if the pulse is too slow ($\omega_0 \to 0$) the mode strength is super-exponentially suppressed, especially at large scales (small $k$), or if the slab has a small lateral velocity dispersion, $\sigma$, compared to that along the vertical direction, $\sigma_z$ (recall that $\Omega_z\sim \sigma_z/h_z$). This is a classic signature of adiabatic shielding of the slab due to the averaging out of the net response to zero by many oscillations of stars within the (very long) perturbation timescale \citep[cf.][]{Weinberg.94a,Weinberg.94b,Gnedin.Ostriker.99}.

\begin{figure*}
  \centering
  \includegraphics[width=0.95\textwidth]{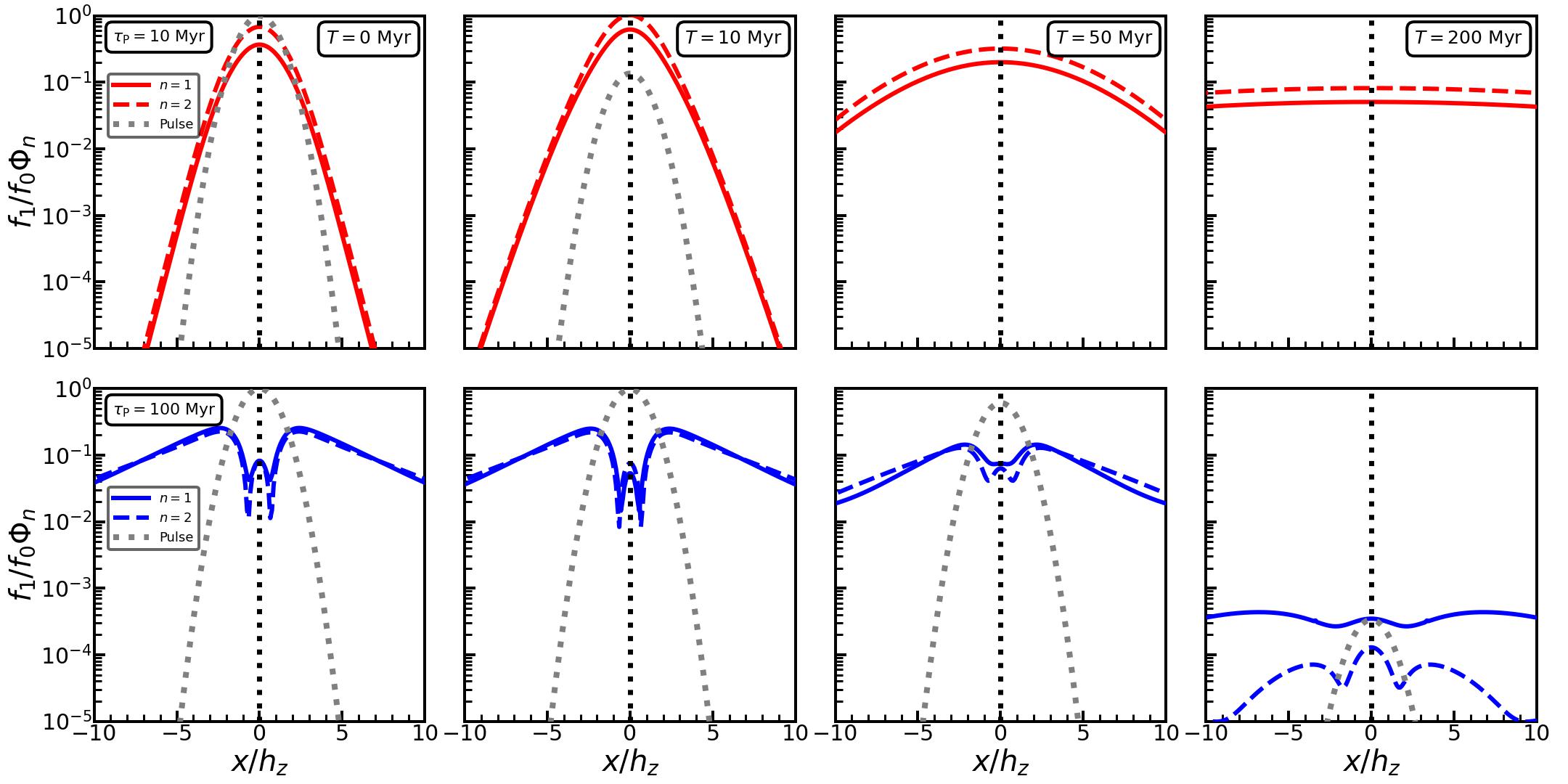}
  \caption{Amplitude of the slab response to a Gaussian (in both $x$ and $t$) packet of half-width $\Delta_x=h_z$ as a function of $x$ for different times since the maximum pulse-strength. The two rows indicate two different pulse times, as indicated. We adopt our fiducial MW parameters (see Section~\ref{sec:iso_slab}) and take $I_z=0.5\, h_z\sigma_z$. Solid (dashed) lines show $n=1$ ($n=2$) bending (breathing) modes, while the grey-dotted lines show the perturbing pulse, $\calT(t)\calX(x)$. The response density initially grows and then damps away due to lateral mixing. In the short pulse limit, the response density is Gaussian in $x$, which damps out and widens like a power law in time. The response in the longer pulse behaves like a sinusoid at small $x$ (see Appendix~\ref{App:ad_lim_resp}) and its intensity shows a transient growth followed by exponential damping before it falls off as a power law. The bending (breathing) mode eventually dominates in the slow (fast) pulse limit.}
  \label{fig:gaussian_slab_x}
\end{figure*}

Finally, if the perturbation is not impulsive the frequency with which the slab stars oscillate in the vertical direction is modified with respect to their natural frequency according to
\begin{align}
\Omega_z \to \frac{\omega^2_0}{\omega^2_0+\frac{k^2\sigma^2}{2}} \Omega_z,
\end{align}
which goes to $\Omega_z$ in the impulsive limit, as expected. For slower pulses however, the vertical motion of the stars couples to the lateral motion \citep[see also][]{Binney.Schonrich.18}, resulting in a reduced oscillation frequency, especially for smaller wavelengths (larger $k$). In the extremely slow/adiabatic limit, $\Omega_z \to 0$, signalling a lack of vertical phase-mixing. This is easy to understand; a forced oscillator remains in phase with the perturber if the frequency of the latter is much lower than the natural frequency. In fact, in the adiabatic limit, the response only consists of resonant stars, for which $n\Omega_z+k v_x=0$ (see Appendix~\ref{App:ad_lim_resp}), and thus no phase spiral emerges.

\begin{figure*}
  \centering
  \includegraphics[width=1\textwidth]{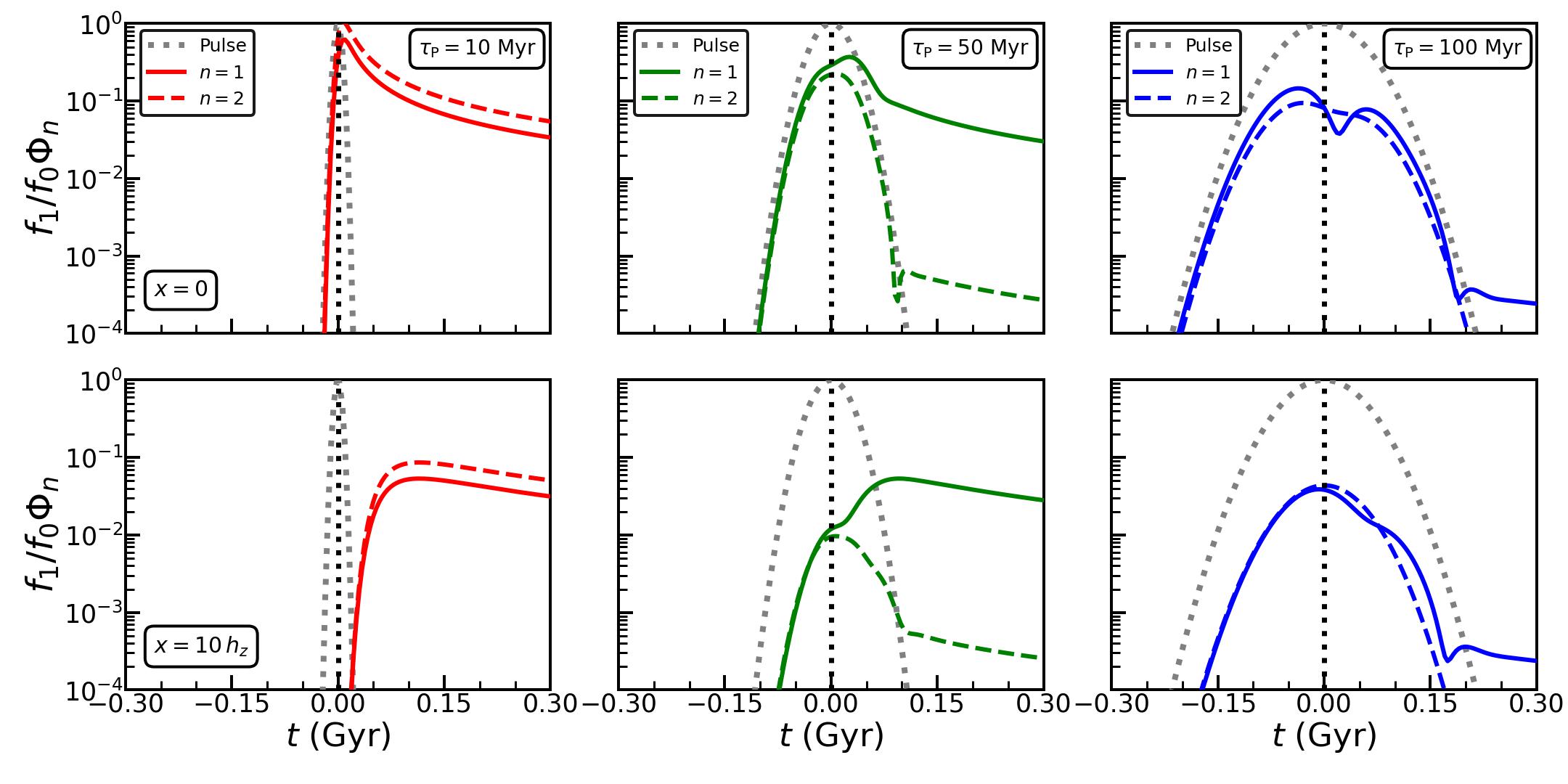}
  \caption{Amplitude of the slab response to a Gaussian perturbation (in both $x$ and $t$) at two locations in the slab: at the location of impact, $x = 0$, shown in the top panels, and at a distance $x = 10 h_z$ away, shown in the bottom panels. As in Fig.~\ref{fig:gaussian_slab_x}, the spatial Gaussian wave-packet, $\calX(x)$, has a half-width of $\Delta_x = h_z$.  Different columns correspond to different values of the Gaussian pulse-widths, $\tau_\rmP$, as indicated. The grey-dotted line in each panel shows the perturbing pulse $\calT(t)$ at $x=0$, while solid and dashed lines show responses for the $n=1$ (bending) and $n=2$ (breathing) modes. The response to shorter pulses shows a transient growth followed by a power law fall-off with time. Response to longer pulses initially grows and then damps away as a Gaussian before finally transitioning to a power law fall-off. For longer pulses, the bending modes dominate in the long run, while for shorter pulses, the breathing modes are stronger.}
  \label{fig:gaussian_slab_t}
\end{figure*}

The above response corresponds to a temporally Gaussian pulse for a fixed wavenumber $k$. To get the full response to a localized perturber, we substitute the expression for $\calX_k$ given in equation~(\ref{Phink_gaussian}), in the $k$-response of equation~(\ref{f1k_gaussian}), multiply it by $\exp{\left[ikx\right]}$ and integrate over all $k$. The resultant response is an oscillating function of $w_z$ and has a profile along $x$ which varies with time. For the short pulse/impulsive case, we recover the expression given in equation~(\ref{f1_local_impulsive}). In Fig.~\ref{fig:gaussian_slab_x} we plot the amplitude (relative to the unperturbed DF) of this oscillating response (normalized by the $z$ Fourier component of the perturber potential, $\calZ_n$) as a function of $x$. The columns correspond to four different times since the time of maximum pulse strength, and the rows correspond to two different pulse-times, as indicated. The solid and dashed lines indicate the bending ($n=1$) and breathing ($n=2$) modes, respectively. The short pulse response shown in the upper panels has a Gaussian profile centered on the point of impact at $x=0$ with the initial width very similar to that of the $\Phi_\rmP$ profile (see equations~[\ref{f1_local_impulsive}]-[\ref{R_local_impulsive}]). Over time, this response profile gets weaker and wider like a power law, as the unconstrained lateral motion of the stars causes an outward streaming, and thus decay, of the response. The long pulse response in the lower panels has a different, more extended profile than in the short pulse case; it exhibits some ripples along $x$ besides having an overall smooth behaviour (see Appendix~\ref{App:ad_lim_resp} for the response derived in the adiabatic limit). As time goes on, the response decays away and widens out due to lateral mixing. Unlike the short pulse case, here the response initially decays like $\sim \exp{\left[-\omega^2_0 t^2\right]}$ over a timescale of the pulse-time, $\tau_\rmP=\sqrt{2}/\omega_0$, before attaining a power law decay at large time.

The temporal behaviour of the response becomes even clearer in Fig.~\ref{fig:gaussian_slab_t}, where we plot the amplitude of the response as a function of time at two different positions on the slab (different rows), and for three different pulse-times (different columns). As before, the solid and dashed lines indicate the $n=1$ and $n=2$ modes, respectively. Initially the slab response grows nearly hand in hand with the perturbing pulse. This is captured by the $\Upsilon_{nk}(t)$ term (equation~[\ref{growth_gaussian}]) in the expression for $R_k(I_z,w_z,t)$, which scales as $\exp{\left[-\calQ^2 \omega^2_0 t^2\right]}$ at small $t$, but asymptotes to a constant value of $2$ at late times. As the perturber strength falls off, the response decays as a Gaussian for each $k$, as described by the damping factor, $D_k(t) \propto \exp[-{\calQ}^2k^2\sigma^2 t^2/2]$. The combined response from all $k$ however decays at a different rate. For the shortest pulse, for which the response asymptotes to that given by equation~(\ref{f1_local_impulsive}), the damping factor, $D(x,t) \propto 1/t$ at late times. In the intermediate and long pulse cases, the response initially tends to follow the same $\sim \exp{\left[-\omega^2_0 t^2\right]}$ decay as the perturbing pulse, before finally transitioning to a power law fall-off, which typically occurs as $\sim 1/t$, just as in the short pulse case. Importantly, this transition sets in later for longer lasting pulses, such that the late-time response for slower perturbations is drastically suppressed with respect to faster perturbations. From the bottom panels, it is evident that the region ($x=10 h_z$) farther away from the center of impact responds later, with a time-lag of $\Delta t= 10\, h_z/\sigma$ (timescale of lateral streaming), which is $\sim 115$ Myr for the typical MW parameter values adopted here. The breathing mode is the dominant mode in the short pulse case ($\tau_\rmP = 10 \Myr$) while in both the intermediate ($\tau_\rmP = 50 \Myr$) and long ($\tau_\rmP = 100 \Myr$) pulse scenarios the bending mode eventually dominates. Note, though, that if the pulse becomes too long, the long-term response is adiabatically suppressed. Hence, there is only a narrow window in pulse-widths for which bending modes dominate and cause a detectable response. In the next section we examine whether any of the MW satellites have encounters with the disk over timescales that fall in this regime.

The response formalism for localized, non-impulsive perturbations developed so far can be used to model the response to transient bars and spiral arms. Encounters with such features can cause transient vertical perturbations in the potential over timescales comparable to the vertical oscillation periods of stars, thereby creating phase spirals. We discuss this in detail in chapter~\ref{chapter: paper3} for realistic disk galaxies.

\section{Encounters with satellite galaxies}
\label{sec:sat_encounter}

In all cases considered above we have made the simplifying assumption that the perturber potential is separable, i.e., can be written in the form of equation~(\ref{Phip_sep}). However a realistic perturber is seldom of such simple form. For example, the potential due to an impacting satellite galaxy or DM subhalo (approximated as a point perturber) cannot be written in separable form, thereby making the analysis significantly more challenging. In this section, as an astrophysical application of the perturbative formalism developed in this chapter, we compute the response of the infinite slab to a satellite encounter. We relegate the far more involved computation of the response of a realistic disk to an impacting satellite to chapter~\ref{chapter: paper3}.

As shown in Appendix~\ref{App:sat_disk_resp}, the $n\neq 0$ response to a satellite impacting the slab with a uniform velocity $\vp$ along a straight orbit at an angle $\thetap$, at a distance $x$ from the point of impact, can be approximated as
\begin{align}
f_1(I_z,w_z,x,t) &= \frac{\rho_c}{\sqrt{2\pi}\sigma_z} \exp{\left[-E_z/\sigma^2_z\right]} \nonumber \\
& \times i\frac{2G\Mp}{\vp} \sum_{n=-\infty}^{\infty} \frac{n\Omega_z}{\sigma^2_z}\, \Psi_n(x,I_z)\, \exp{\left[i\,\frac{n\Omega_z \sin{\thetap}}{\vp}x\right]} \exp{\left[i n\left(w_z-\Omega_z t\right)\right]},
\label{f1_sat}
\end{align}
where
\begin{align}
\Psi_n(x,I_z)&= \frac{1}{2\pi} \int_0^{2\pi} \rmd w_z\, \exp{\left[-i n \left(w_z - \frac{\Omega_z \cos{\thetap} z}{\vp}\right)\right]} \nonumber\\
&\times K_0\left[\,\left|\frac{n\Omega_z \left(x\cos{\thetap}-z\sin{\thetap}\right)}{\vp}\right|\,\right],
\label{Psi_n}
\end{align}
with $K_0$ the zero-th order modified Bessel function of the second kind. This expression for the response is only valid far away from the point of impact ($x\gtrsim \sigma t$), such that the response can be approximated as a plane wave along $x$, and at late times, after the perturber has moved far enough away from the disk, i.e., for $t \gg (x\sin{\thetap}+z\cos{\thetap})/\vp$).

There are several salient features of this response that deserve special attention. The strength of the response is dictated by the $K_0$ function whose argument depends on $\Omega_z \cos{\thetap}\, x/\vp$ (for small $I_z$), which is basically the ratio of the encounter timescale,
\begin{align}
\tau_{\rm enc} = \frac{x\cos{\thetap}}{\vp},
\label{tau_enc}
\end{align}
and the vertical dynamical time of the stars,
\begin{align}
\tau_z = \frac{1}{\Omega_z} \sim \frac{h_z}{\sigma_z}.
\label{tau_z}
\end{align}
From the asymptotic limits of $K_0$ it follows that the response scales as a power law ($\sim \vp^{-1}$) in the impulsive ($\tau_{\rm enc}\ll \tau_z$) limit and as $\sim\exp{\left[-\left|n\Omega_z\cos{\thetap}\right|x/\vp\right]}$ in the adiabatic ($\tau_{\rm enc}\gg \tau_z$) limit. The response peaks roughly at the maximum of the $K_0$ function, which occurs when the encounter timescale is comparable to the vertical dynamical time of the stars, i.e., when $\tau_{\rm enc}\approx 0.6\,\tau_z$, or in other words the `resonance' condition,
\begin{align}
\frac{x\cos{\thetap}}{\vp} \approx \frac{0.6}{\Omega_z},
\end{align}
is satisfied. For encounters faster than this, the response is suppressed like a power law, while for slower encounters it is exponentially suppressed. The $\vp^{-1}$ scaling of the response in the impulsive limit is a well known feature of impulsive perturbations \citep[e.g.,][]{Spitzer.58, Aguilar.White.85, Weinberg.94a, Weinberg.94b, Gnedin.etal.99, Banik.vdBosch.21b}, and the exponential suppression is a telltale signature of adiabatic shielding\footnote{While the adiabatic response in one degree-of-freedom cases, e.g., the vertical phase spiral in the isothermal slab, is exponentially suppressed, that in multiple degree-of-freedom systems such as inhomogeneous disks is usually not because of resonances \citep[][]{Weinberg.94a,Weinberg.94b}.}, similar to the adiabatic suppression of the mode-strength factor in the response to slow Gaussian pulses discussed in section~\ref{sec:non-impulsive}.

While the response is heavily damped for very slow encounters, something interesting happens in the mildly slow regime, $\tau_{\rm enc}=x\cos{\thetap}/\vp \gtrsim \tau_z$. In this regime, the ratio of the $n=2$ breathing to the $n=1$ bending mode response scales as
\begin{align}
f_{21} \equiv {f_{1,n=2} \over f_{1,n=1}} \sim \sqrt{2}\,\exp{\left[-\frac{\Omega_z\cos{\thetap}\,x}{\vp}\right]}.
\label{f21}
\end{align}
Thus the bending mode response exponentially dominates over that of the breathing mode for slower (smaller $\vp$), more distant (large $x$), and more perpendicular ($\thetap\approx 0$) encounters. The bending mode is also more pronounced for stars with larger $\Omega_z$ or equivalently smaller $I_z$. On the other hand, for encounters with $\tau_{\rm enc}=x\cos{\thetap}/\vp < \tau_z$, the breathing modes dominate.

Finally, the slab response to the impacting satellite, given in equation~(\ref{f1_sat}), consists of oscillating functions of time, lateral distance $x$, and the vertical oscillation amplitude, $\sqrt{2I_z/\nu}$ (see equations~[\ref{Psi_n_app_epi}] and [\ref{Phin_sat_app}]). This implies that the satellite not only induces temporal oscillations, which give rise to phase-mixing and thus phase spirals due to different oscillation frequencies of the stars (see section~\ref{sec:impulsive_kick}), but also spatial corrugations. These vertical and lateral waveforms have wavenumbers given by

\begin{align}
k_z = \frac{n\Omega_z\cos{\thetap}}{\vp},\;\;\;\;\;\;\;{\rm and}\;\;\;\;\;\;\;\;
k_x = \frac{n\Omega_z\sin{\thetap}}{\vp},
\end{align}
respectively. Thus, perpendicular impacts induce only vertical corrugations while planar ones excite waves only laterally. An inclined encounter, on the other hand, spawns corrugations along both directions. Both wavelengths get longer with decreasing mode order, increasing impact velocity, and decreasing vertical frequencies, i.e., increasing actions.

\begin{figure*}[t!]
    \centering
    \subfloat{\includegraphics[width=0.352\textwidth]{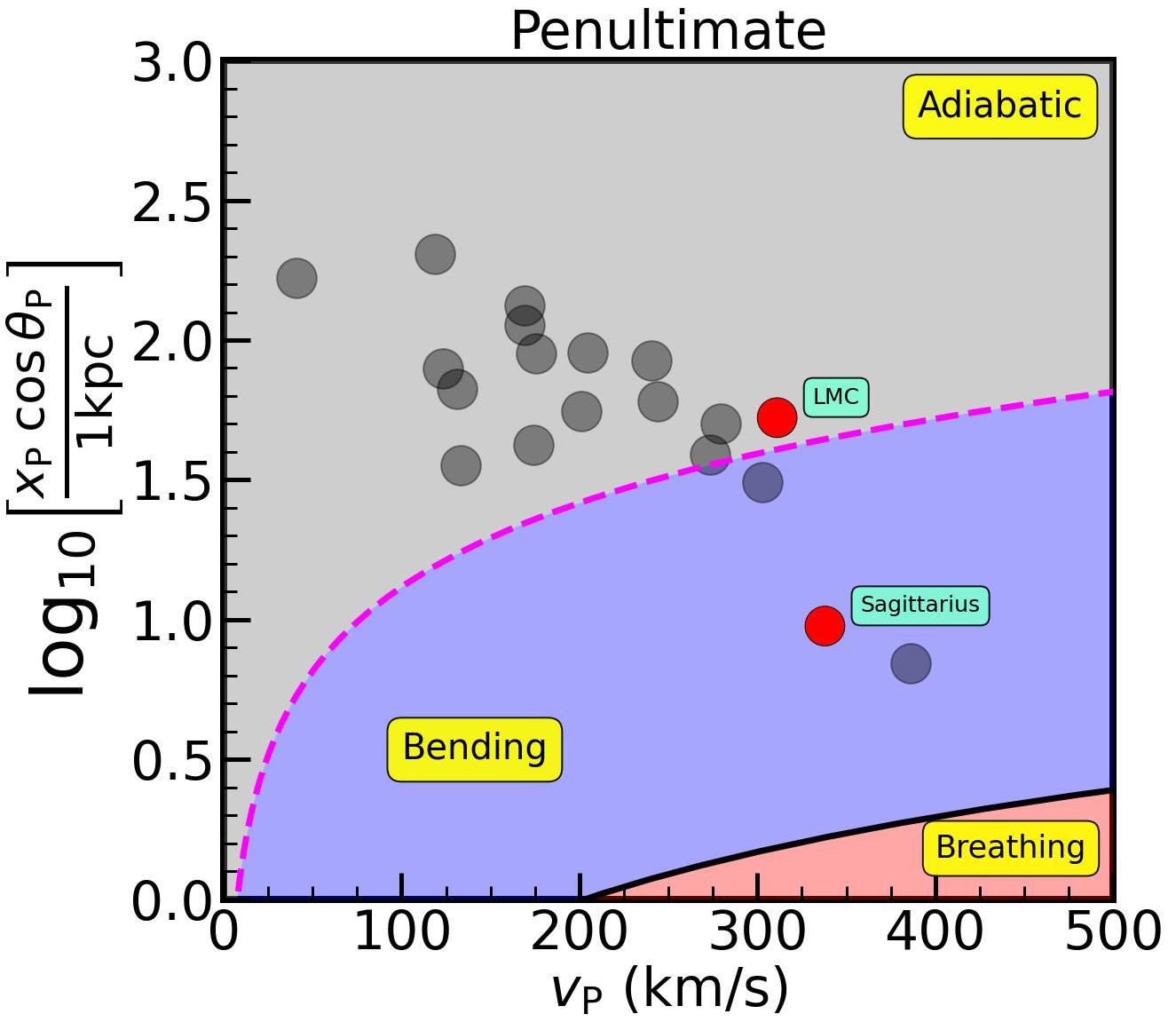}
    \label{disk_resp_n1_0.5pi_1}}
    \subfloat{\includegraphics[width=0.63\textwidth]{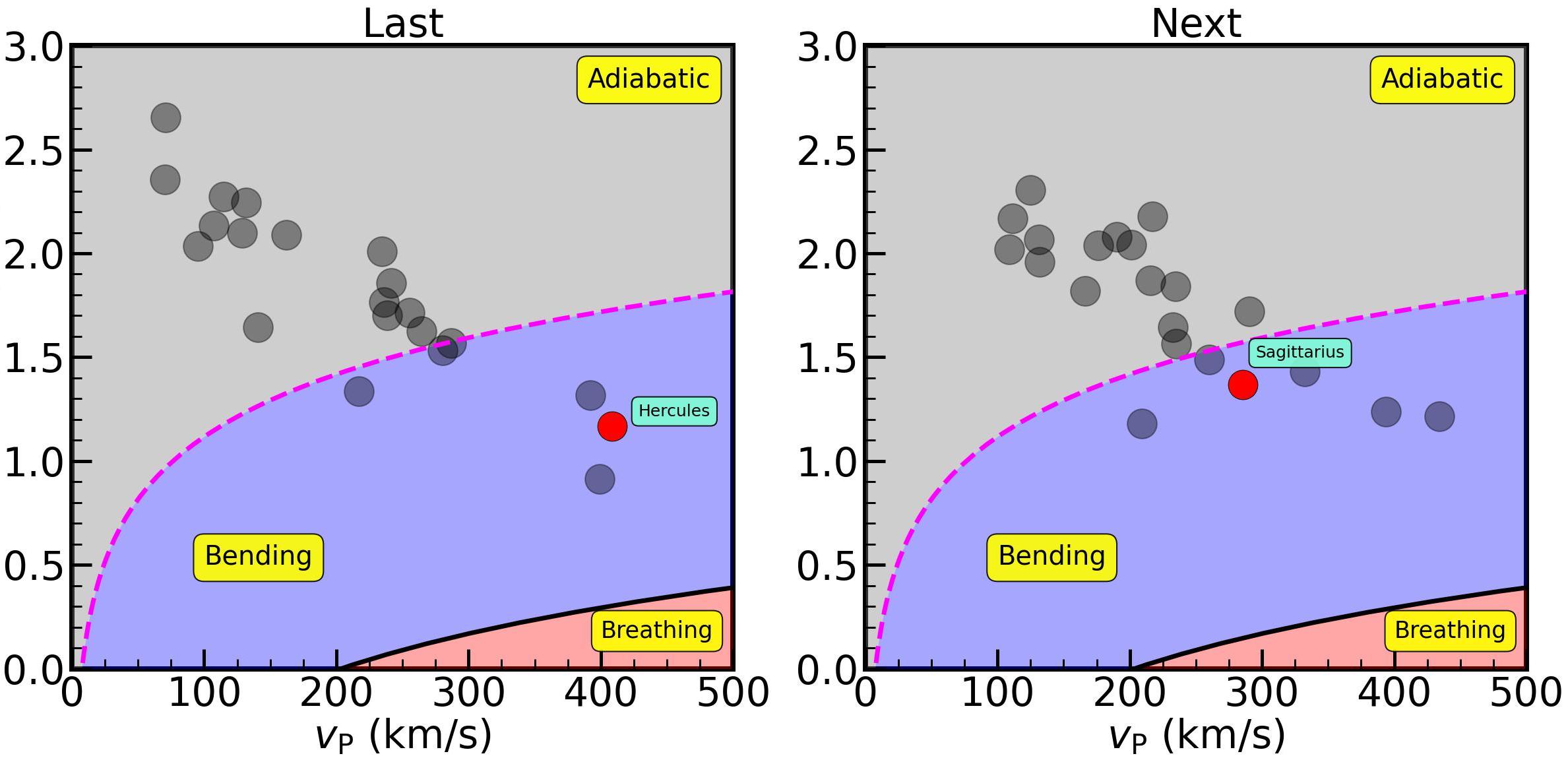}
    \label{disk_resp_n1_0.5pi_2}}
    \caption{Regions in the space of impact parameter, $\xp\cos{\thetap}$, and velocity, $\vp$, of a satellite galaxy, corresponding to bending (blue) and breathing (red) mode responses in the Solar neighborhood. Response is adiabatically suppressed in the grey region. The circles in the left, middle and right panels indicate the values of $\xp\cos{\thetap}$ and $\vp$ for several MW satellites during their penultimate, last and next disk crossings respectively. The satellites that induce a relative bending mode response, $f_{1,n=1}/f_0\gtrsim 10^{-4}$, for $I_z=h_z\sigma_z$ in the Solar neighborhood, are indicated by red circles, while the others are denoted in grey. All the MW satellites lie outside the breathing region and thus preferentially excite bending modes in the vicinity of the Sun.}
  \label{fig:satellite_constraints1}
\end{figure*}

\subsection{Impact of satellite galaxies on the Milky Way disk}
\label{sec::MW_satellites}

The MW halo harbors many satellite galaxies. Some of these are quite massive, with DM halo mass comparable to the disk mass, and either underwent or are about to undergo an encounter with the MW disk within a few hundred Myr from the present day. Hence we expect at least some of them to perturb the disk significantly. Here we use existing data on MW satellites to obtain a rough estimate of the disk response to their encounters with the MW stellar disk.

Our formalism provides physical insight into the trends and scalings of the disk response as a function of impact parameters and velocities of the MW satellites. We emphasize upfront, though, that the precise numerical estimates of the responses are to be taken with a grain of salt. These estimates only serve as a crude, order-of-magnitude attempt to compare the relative disk responses to different satellite galaxies. As discussed in more detail in section~\ref{sec::caveats}, these estimates are subject to a number of oversimplifications and caveats. First of all, the MW disk is modelled as an isothermal slab, and we only consider the {\it direct} impact of the satellites. We ignore indirect effects due to the self-gravity of the response. Our approach also ignores the presence of a dark matter halo, which can impact the disk response in several ways (see section~\ref{sec::caveats}). Because of all these shortcomings, we caution against using the following response estimates for comparison with actual data and/or detailed numerical simulations.

\begin{figure*}
  \centering
  \includegraphics[width=1\textwidth]{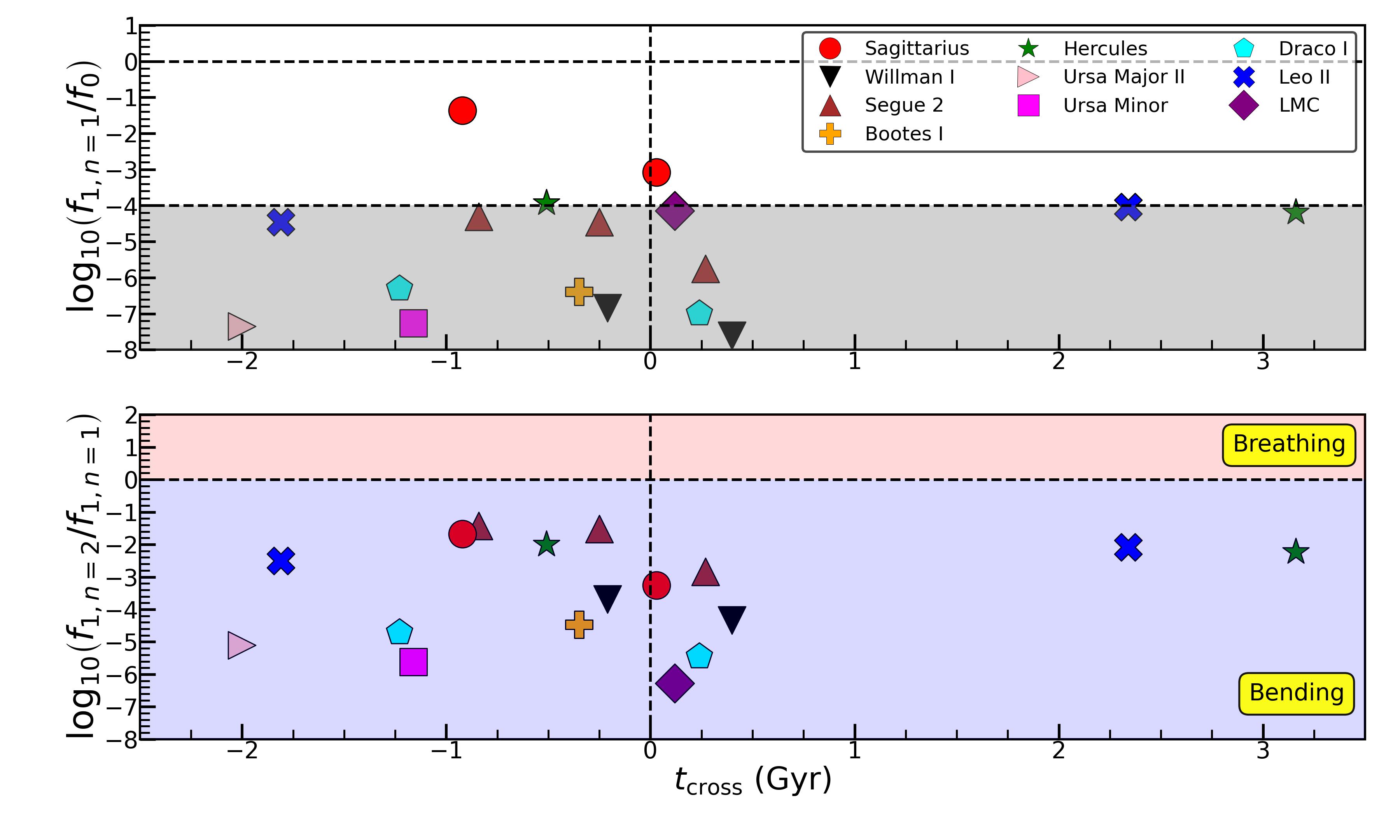}
  \caption{Bending mode strength, $f_{1,n=1}/f_0$ (upper panel), and the corresponding breathing vs bending ratio, $f_{1,n=2}/f_{1,n=1}$ (lower panel), in the Solar neighborhood for the MW satellites, as a function of the disk crossing time, $t_{\rm cross}$, in Gyr, where $t_{\rm cross}=0$ marks today. The previous two and the next impacts are shown. Here we consider $I_z=h_z \sigma_z$, with fiducial MW parameters. In the upper panel, the region with bending mode response, $f_{1,n=1}/f_0<10^{-4}$, has been grey-scaled, indicating that the response from the satellites in this region is far too adiabatic and weak. Note that the response is dominated by that due to Sgr, followed by Hercules, Leo II, Segue 2 and the Large Magellanic Cloud (LMC). Also note that the previous two and next impacts of all the satellites shown here excite bending modes in the Solar neighborhood.}
  \label{fig:satellite_constraints2}
\end{figure*}

We consider the MW satellites with parallax and proper motion measurements from Gaia DR2 \citep[][]{Gaia_collab_sat.18b} and the corresponding galactocentric coordinates and velocities computed and documented by \cite{Riley.etal.19} \citep[table A.2, see also][]{Li.etal.20} and \cite{Vasiliev.Belokurov.20}. Of these, we only consider the satellites with known dynamical mass estimates \citep[][]{Simon.Geha.07,Bekki.Stanimirovic.09,Lokas.09,Erkal.etal.19}. Adopting the initial conditions for galactocentric positions ($R,z,\phi$) and velocities ($v_R,v_z,v_\phi$) as the median values quoted by \cite{Riley.etal.19} and \cite{Vasiliev.Belokurov.20}, we simulate the orbits of the galaxies in the combined gravitational potential of the MW halo, disk and bulge, which are respectively modelled by a spherical NFW \citep[][]{Navarro.etal.97} profile (virial mass $M_h=9.78\times 10^{11}\Msun$, scale radius $r_h=16$ kpc, and concentration $c=15.3$), a Miyamoto-Nagai \citep[][]{Miyamoto.Nagai.75} profile (mass $M_d=9.5\times 10^{10} \Msun$, scale radius $a=4$ kpc, and scale-height $b=0.3$ kpc), and a spherical \cite{Hernquist.90} profile (mass $M_b=6.5\times 10^9\Msun$ and scale radius $r_b=0.6$ kpc)\footnote{Our MW potential is similar to {\tt GALPY MWPOTENTIAL2014} \citep[][]{Bovy.15} except for the power-law bulge which has been replaced by an equivalent Hernquist bulge.}. The total mass of our fiducial MW model is thus $1.08\times 10^{12}\Msun$. We evolve the positions and velocities of the satellites both forwards and backwards in time from the present day, using a second order leap-frog integrator. For simplicity, we ignore the effect of dynamical friction\footnote{Dynamical friction might play an important role in the orbital evolution of massive satellites like the Large Magellanic Cloud (LMC) and Sgr, pushing their orbital radius farther out in the past.}. From each individual orbit, we note the time, $t_{\rm cross}$, when the satellite crosses the disk (i.e., crosses $z=0$), and record the corresponding distance, $\xp$, from the Sun, which we integrate backwards/forwards in time using a purely circular orbit up to $t_{\rm cross}$. We also record the velocity, $\vp=\sqrt{v^2_R+v^2_z+v^2_\phi}$, and the angle of impact with respect to the disk normal, $\thetap=\cos^{-1}{(v_z/\vp)}$. Finally, we compute the disk response to the satellite encounter using equation~(\ref{f1_sat}). Results are summarized in Table~\ref{tab:MW_sat_resp_2} and Figs.~\ref{fig:satellite_constraints1} and~\ref{fig:satellite_constraints2}.

%

%
%

In Fig.~\ref{fig:satellite_constraints1}, we plot the impact parameter, $\xp\cos{\thetap}$ (with respect to the Sun), as a function of the encounter velocity, $\vp$, of the satellites, for the penultimate (left-hand panel), last (middle panel), and next (right-hand panel) disk crossings. The red (grey) symbols denote the satellites that induce a strong (weak) amplitude of the bending mode response, $f_{1,n=1}/f_0$, for $I_z=h_z\sigma_z = 9.2 \kpc\kms$. As shown in Appendix~\ref{App:detect_crit}, we consider $f_{1,n=1}/f_0=\delta=10^{-4}$ as a rough estimate for the minimum detectable relative response, i.e., the boundary between strong and weak responses to satellite passage. The solid black line indicates the boundary between bending and breathing modes, i.e., where the breathing-to-bending ratio, $f_{21}$ (equation~[\ref{f21}]), is equal to unity. Hence, the blue and red shaded regions indicate where the response is dominated by bending and breathing modes, respectively. The magenta, dashed line roughly denotes the boundary between a strong bending response (blue shaded region) and a response that is adiabatically suppressed (grey shaded region). The latter is defined by the condition $\exp{\left[-\Omega_z \xp \cos{\thetap}/\vp\right]} < \delta = 10^{-4}$.

In Fig.~\ref{fig:satellite_constraints2}, we plot the amplitude of the bending mode response, $f_{1,n=1}/f_0$ (upper panel), and the breathing-to-bending ratio, $f_{21}=f_{1,n=2}/f_{1,n=1}$ (lower panel), in the Solar neighborhood, as a function of the time $t_{\rm cross}$ (in Gyr) when the satellite crosses the plane of the disk, assuming the fiducial MW parameters. Negative and positive $t_{\rm cross}$ correspond to disk crossings in the past and future, respectively, and we once again consider stars with $I_z = h_z \sigma_z = 9.2 \kpc\kms$.

Both Fig.~\ref{fig:satellite_constraints1} and the lower panel of Fig.~\ref{fig:satellite_constraints2} make it clear that {\it all} the disk crossings considered here preferentially excite bending rather than breathing modes in the Solar neighborhood. As shown in Section~\ref{sec:impulsive_kick} these trigger one-armed phase spirals in the Solar neighborhood, in qualitative agreement with the MW snail observed in the Gaia data. However, as is evident from the upper panel of Fig.~\ref{fig:satellite_constraints2}, most satellites only trigger a minuscule response in the disk, with  $f_{1,n=1}/f_0 < \delta = 10^{-4}$, either because the satellite has too low mass, or because the encounter, from the perspective of the Sun, is too slow such that the local response is adiabatically suppressed.  The strongest response by far is triggered by encounters with Sgr, for which the bending mode response, $f_{1,n=1}/f_0$, is at least $1-2$ orders of magnitude larger than that for any other satellite. Based on our orbit-integration, it had its penultimate disk crossing, which closely coincides with its last pericentric passage, about $900\Myr$ ago, triggering a strong response of $f_{1,n=1}/f_0 \sim 0.04$ in the Solar neighborhood. The last disk crossing, which nearly corresponds to the last apocentric passage, occurred about $300\Myr$ ago, triggering a very weak (adiabatically suppressed) response. Sgr is currently near its pericenter and will undergo the next disk crossing in about $30\Myr$, which we estimate to only trigger a moderately strong response with $f_{1,n=1}/f_0\sim 0.001$. We caution, though, that in addition to the caveats listed above and in Section~\ref{sec::caveats} these estimates ignore dynamical friction and are sensitive to the MW potential and the current phase-space coordinates of the satellites. We have checked that a heavier MW model with a total mass of $1.5\times 10^{12}\Msun$ does not change the relative amplitudes of the satellite responses significantly, but brings most of the disk crossing times closer to the present day since the satellites are more bound in a heavier MW. For example, the previous pericentric and apocentric passages of Sgr occur at $\sim 600$ and $200\Myr$ ago in the heavier case. The only satellite apart from Sgr that triggers a response $f_{1,n=1}/f_0 > \delta = 10^{-4}$ is Hercules, whose disk crossing $\sim 500\Myr$ ago caused a bending-mode response, $f_{1,n=1}/f_0 = 1.2\times 10^{-4}$. Segue 2 induces a response that is marginally below the detection threshold. Disk crossings of LMC and Leo II trigger responses that are comparable in strength to that of Hercules, but the crossing times are too far in the past or future for them to be considered as candidates for triggering the Gaia snail. All in all, it is clear then that Sgr is by far the most likely candidate among the MW satellite galaxies considered here to have triggered the one-armed phase spiral in the Solar neighborhood discovered in Gaia DR2 by \citet{Antoja.etal.18}.

\renewcommand{\arraystretch}{1.4}
\begin{sidewaystable}
\centering
\tabcolsep=0.2 cm
\scalebox{0.9}{\begin{tabular}{c|c|cc|cc|cc}
 \hline
MW satellite & Mass & $f_{1,n=1}/f_0$ & $t_{\rm cross}$ & $f_{1,n=1}/f_0$ & $t_{\rm cross}$ & $f_{1,n=1}/f_0$ & $t_{\rm cross}$ \\
name & $(\Msun)$ & & $(\Gyr)$ & & $(\Gyr)$ & & $(\Gyr)$ \\
 & & Penultimate & Penultimate & Last & Last & Next & Next \\
 (1) & (2) & (3) & (4) & (5) & (6) & (7) & (8) \\
 \hline
 Sagittarius & $10^9$ & $4.3\times 10^{-2}$ & $-0.92$ & $1.4\times 10^{-10}$ & $-0.3$ & $8.3\times 10^{-4}$ & $0.03$ \\
 Hercules & $7.1\times 10^6$ & -- & $-3.57$ & $1.2\times 10^{-4}$ & $-0.51$ & $6.4\times 10^{-5}$ & $3.16$ \\
 Leo II & $8.2\times 10^6$ & -- & $-3.61$ & $3.5\times 10^{-5}$ & $-1.81$ & $9.3\times 10^{-5}$ & $2.34$ \\
 Segue 2 & $5.5\times 10^5$ & $5\times 10^{-5}$ & $-0.84$ & $3.4\times 10^{-5}$ & $-0.25$ & $1.8\times 10^{-6}$ & $0.27$ \\
 LMC & $1.4\times 10^{11}$ & $1.4\times 10^{-4}$ & $-6.97$ & -- & $-2.37$ & $7.2\times 10^{-5}$ & $0.12$ \\
 SMC & $6.5\times 10^9$ & $3.6\times 10^{-8}$ & $-3.22$ & -- & $-1.39$ & $1.2\times 10^{-9}$ & $0.22$ \\
 Draco I & $2.2\times 10^7$ & -- & $-2.43$ & $5\times 10^{-7}$ & $-1.23$ & $1\times 10^{-7}$ & $0.24$ \\
 Bootes I & $10^7$ & -- & $-1.65$ & $4.1\times 10^{-7}$ & $-0.35$ & -- & $0.87$ \\
 Willman I & $4\times 10^5$ & -- & $-0.63$ & $1.4\times 10^{-7}$ & $-0.21$ & $2.5\times 10^{-8}$ & $0.4$ \\
 Ursa Minor & $2\times 10^7$ & -- & $-2.26$ & $5.5\times 10^{-8}$ & $-1.16$ & $8.6\times 10^{-9}$ & $0.29$ \\
 Ursa Major II & $4.9\times 10^6$ & $4.5\times 10^{-8}$ & $-2$ & $6.2\times 10^{-10}$ & $-0.1$ & -- & $0.9$ \\
 Coma Berenices I & $1.2\times 10^6$ & $7\times 10^{-10}$ & $-2.47$ & -- & $-0.25$ & -- & $0.69$ \\
 Sculptor & $3.1\times 10^7$ & -- & $-2.7$ & $2\times 10^{-10}$ & $-0.46$ & -- & $1.47$\\
 \hline
\end{tabular}}
\caption{MW disk response ($l=m=0$) to satellites for stars with $I_z=h_z\sigma_z$ in the Solar neighborhood. Column (1) indicates the name of the MW satellite and Column (2) indicates its dynamical mass estimate from literature \citep[][]{Simon.Geha.07,Bekki.Stanimirovic.09,Lokas.09,Erkal.etal.19,Vasiliev.Belokurov.20}. We assume $10^9\Msun$ for the Sagittarius mass; note that there is a discrepancy between its measured mass of $\sim 4\times 10^8\Msun$ \citep[][]{Vasiliev.Belokurov.20} and the required mass of $10^9-10^{10}\Msun$ for observable phase spiral signatures in N-body simulations \citep[see for example][]{Bennett.etal.22}. Columns (3) and (4) respectively indicate the bending mode response assuming fiducial MW parameters and the crossing time for the penultimate disk-crossing. Columns (5) and (6) show the same for the last disk-crossing, while columns (7) and (8) indicate it for the next one. Only the satellites that trigger a bending mode response, $f_{1,n=1}/f_0\geq 10^{-10}$, in at least one of the three cases are shown. The responses smaller than $10^{-10}$ are considered far too adiabatic and negligible and are marked by dashes.}
\label{tab:MW_sat_resp_2}
\end{sidewaystable}

We emphasize that the results shown in Figs.~\ref{fig:satellite_constraints1} and~\ref{fig:satellite_constraints2} correspond to stars with a vertical action $I_z=h_z\sigma_z=9.2\, \kpc\kms$. As mentioned above, the strength of the response depends on the ratio of the encounter time scale, $\tau_{\rm enc}$ (equation~[\ref{tau_enc}]) and the vertical oscillation period of stars in the Solar neighborhood, $\tau_z$ (equation~[\ref{tau_z}]). The latter is longer for stars with larger vertical action, and from the perspective of such stars the encounter is more impulsive, resulting in a stronger response. Since the response does not scale linearly with $\tau_{\rm enc}/\tau_z$, the relative response strength of different satellites depends somewhat on the vertical action. We have verified that for $I_z/(h_z\sigma_z) < 3$, which is roughly the range covered by the Gaia phase spiral, the direct response from the encounter with Sgr remains larger than that of any other satellite considered here by at least $1-2$ orders of magnitude. However, for stars with larger actions (larger vertical excursions), the LMC can dominate the response. In particular, for stars with $I_z/(h_z\sigma_z) \gtrsim 6.5$ ($z_{\rm max}\gtrsim 4\, h_z$), which make up the thick disk, the LMC is expected to trigger a stronger response than Sgr during its upcoming disk crossing.

To summarize, our analysis suggests that the MW satellites during their most recent and forthcoming disk crossings preferentially excite bending modes in the Solar neighborhood. This is because satellite encounters are fairly distant from the Sun and thus the encounter time exceeds the vertical oscillation time of the stars. However, as previously discussed in section~\ref{sec:sat_encounter} and as evident from the N-body simulation of MW-Sgr encounter by \cite{Hunt.etal.21} (especially the earlier disk passages of Sgr), a satellite passage can trigger breathing modes closer to the point of impact, where the encounter is more impulsive. Since almost all the MW satellites undergo their disk-crossings at $R\gg 8\kpc$, future observations of the outskirts of the disk might reveal breathing instead of bending mode oscillations if they are excited by any of the satellites considered here.

\subsection{Caveats}
\label{sec::caveats}

The above calculation of the response of the MW disk to perturbations is subject to a number of oversimplifications and caveats discussed below.

The MW disk is modelled as an isothermal slab, which lacks the axisymmetric density profile and velocity structure that characterize a realistic disk. In particular, whereas the lateral motion in our slab is uninhibited, the in-plane motion in a realistic disk consists of an azimuthal rotation combined with a radial epicyclic motion. Among others, this will have important implications for the global disk response and the rate at which phase spirals damp out due to lateral mixing. In chapter~\ref{chapter: paper3} we apply our perturbative formalism to a realistic self-gravitating disk galaxy with a pseudo-isothermal distribution function \citep[][]{Binney.10}, and consider both external perturbations (encounters with satellites) and internal perturbations (bars and spiral arms).
    
All responses calculated in this chapter only account for the direct response to a perturbing potential. In general, though, the response also has an indirect component that arises from the fact that neighboring regions in the disk interact with each other gravitationally. This self-gravity of the response, which we have ignored, triggers long-lived normal mode oscillations of the slab that are not accounted for in our treatment. Several simulation-based studies have argued that including self-gravity is important for a realistic treatment of phase spirals \citep[e.g.,][]{Darling.Widrow.19a, Bennett.Bovy.21}. Using the Kalnajs matrix method \citep[][]{Kalnajs.77, Binney.Tremaine.08}, we have made some initial attempts to include the self-gravity of the response in our perturbative analysis, along the lines of \citet{Weinberg.91}. Our preliminary analysis shows that the self-gravitating response is a linear superposition of two terms: (i) a continuum of modes given in equation~(\ref{f1nk_slabsol}), dressed by self-gravity, that undergo phase-mixing and give rise to the phase spiral, and (ii) a discrete set of modes called point modes or normal modes \citep[c.f.][]{Mathur.90,Weinberg.91} that follow a dispersion relation. The continuum response can be amplified by self-gravity when the continuum frequencies, $n\Omega_z+k v_x$, are close to the point mode frequencies, $\nu$. Depending on the value of $k$, the normal modes can be either stable or unstable. \cite{Araki.85} find that in an isothermal slab the bending normal mode undergoes fire hose instability below a certain critical wavelength if $\sigma_z/\sigma \lesssim 0.3$ while the breathing normal mode becomes unstable above the Jeans scale. In the regime of stability, the normal modes are undamped oscillation modes in absence of lateral streaming \citep[][]{Mathur.90} but are Landau damped otherwise \citep[][]{Weinberg.91}. For an isothermal slab with typical MW-like parameter values, the point modes are strongly damped since their damping timescale (inverse of the imaginary part of $\nu$) is of order their oscillation period (inverse of the real part of $\nu$), which turns out to be of order the vertical dynamical time, $h_z/\sigma_z$. Moreover, the normal mode oscillations are coherent oscillations of the entire system, independent of the vertical actions of the stars, and are decoupled from the phase spiral in linear theory since the full response is a linear superposition of the two. Based on the above arguments, we conclude that self-gravity has little impact on the evolution of phase spirals in the isothermal slab, at least in the linear regime. We emphasize that \citet{Darling.Widrow.19a}, who found their phase spirals to be significantly affected by the inclusion of self-gravity, assumed a perturber-induced velocity impulse with magnitude comparable to the local velocity dispersion in the Solar neighborhood; hence their results are likely to have been impacted by non-linear effects. Moreover, the self-gravitating response of an inhomogeneous disk embedded in a dark matter halo, as in the simulations of \citet{Darling.Widrow.19a}, can be substantially different from that of the isothermal slab. We intend to include a formal treatment of self-gravity along the lines of \citet{Weinberg.91} in future work.
    
The disk of our MW is believed to be embedded in an extensive dark matter halo, something we have not taken into account. The presence of such a halo has several effects. First of all, the satellite not only perturbs the disk, but also the halo. In particular, it induces both a local wake and a global modal response\footnote{The torque from the local as well as global halo response is responsible for dynamical friction acting on the satellite.} \citep[e.g.,][]{Weinberg.89, Tamfal.etal.21}. The former typically trails the satellite galaxy, and boosts its effective mass by about a factor of two \cite[][]{Binney.Tremaine.08}, which might boost the (direct) disk response by about the same factor. The global halo response is typically dominated by a strong $l=1$ dipolar mode followed by an $l=2$ quadrupolar mode \citep[][]{Tamfal.etal.21}, which might have a significant impact on the disk. The presence of a halo also modifies the total potential. At large disk radii and vertical heights, the halo dominates the potential and will therefore significantly modify the actions and frequencies of the stars, and consequently the shape of the phase spirals. Finally, since the disk experiences the gravitational force of the halo, a (sufficiently massive) satellite galaxy can excite normal mode oscillations of the disk in the halo \citep[see for example][]{Hunt.etal.21}. We intend to incorporate some of these effects of the MW halo in chapter~\ref{chapter: paper3}.

\section{Conclusion}
\label{sec:concl_2}

In this chapter we have used linear perturbation theory to compute the response of an infinite, isothermal slab to various kinds of external perturbations with diverse spatio-temporal characteristics. Although a poor description of a realistic disk galaxy, the infinite, isothermal slab model captures the essential physics of perturbative response and collisionless equilibration via phase-mixing in the disk, and thus serves as a simple yet insightful case for investigation.

We use a hybrid (action-angle variables in the vertical direction and position momentum variables in the lateral direction) linear perturbative formalism to perturb and linearize the collisionless Boltzmann equation and compute the response in the distribution function of the disk to a gravitational perturbation. We have considered external perturbations of increasing complexity, ranging from an instantaneous (laterally) plane-wave perturbation (Section~\ref{sec:impulsive_kick}), an instantaneous localized perturbation, represented as a wave-packet (Section~\ref{sec:localized}), a non-impulsive, temporally extended, localized perturbation (Section~\ref{sec:non-impulsive}), and ultimately an encounter with a satellite galaxy moving along a straight-line orbit (Section~\ref{sec:sat_encounter}). This multi-tiered approach is ideal for developing the necessary insight into the complicated response that is expected from a realistic disk galaxy exposed to a realistic perturbation. We summarize our conclusions below.

\begin{itemize}
    \item The two primary Fourier modes of slab oscillation are the $n=1$ bending mode and the $n=2$ breathing mode, which correspond to anti-symmetric and symmetric oscillations about the mid-plane, respectively. For a sufficiently impulsive perturbation, the dominant mode is the breathing mode, which initially causes a quadrupolar distortion in the $(z,v_z)$ phase-space, that evolves into a two-armed phase spiral as the stars with different vertical actions oscillate with different vertical frequencies. If the perturbation is temporally more extended (less impulsive), the dominant mode is the bending mode. This causes a dipolar distortion in $(z,v_z)$ phase-space that evolves into a one-armed phase spiral \citep[see also][]{Hunt.etal.21, Widrow.etal.14}. Due to vertical phase-mixing, the phase spiral wraps up tighter and tighter until it becomes indistinguishable from an equilibrium distribution in the coarse-grained sense.
    
    \item Besides vertical phase-mixing the survivability of the phase spiral is also dictated by the lateral streaming motion of stars. The initial lateral velocity impulse towards the minima of $\Phi_\rmP$ tends to linearly boost the contrast of the phase spiral. This is however quickly taken over by lateral streaming (with velocity dispersion $\sigma$), which causes mixing between the over- and under-densities, and damps out the phase spiral amplitude. For an impulsive, laterally sinusoidal perturbation, the disk response is also sinusoidal and damps out like a Gaussian (due to the Maxwellian/Gaussian distribution of the unconstrained lateral velocities) over a timescale of $\tau_\rmD \sim 1/k\sigma$, i.e., small scale perturbations damp out faster, as expected.
    
    \item Lateral mixing operates differently for a spatially localized perturbation which can be expressed as a superposition of many plane waves. The response to each of them damps out like a Gaussian (if the perturber is impulsive). Since the power spectrum of a spatially localized perturber with a lateral Gaussian profile is dominated by its largest scales (small $k$) that mix and damp out slower, the net response from all $k$ damps away as $\sim t^{-1}$ (the response profile spreads out as $\sim t$), much slower than the Gaussian damping in case of a sinusoidal perturber.
    
    \item The disk response to a non-impulsive perturbation is substantially different from that to an impulsive one. If the temporal strength of the perturber follows a Gaussian pulse with pulse frequency, $\omega_0$ (e.g., a transient bar or spiral arm), the response grows and decays following the temporal profile of the pulse before eventually attaining a $\sim 1/t$ power law fall-off. The response peaks when the pulse frequency, $\omega_0$, is comparable to the vertical oscillation frequency, $\Omega_z$. The response to more impulsive perturbations ($\omega_0 \gg \Omega_z$) is suppressed as $\sim 1/\omega_0$, whereas much slower ($\omega_0 \ll \Omega_z$) perturbations trigger a super-exponentially ($\sim \exp{\left[-n^2\Omega^2_z/4\omega^2_0\right]}$ at small $k$) suppressed response. In this adiabatic limit, the stars tend to remain in phase with the perturber, oscillating at frequencies much smaller than $\Omega_z$, which inhibits the formation of a phase spiral.
    
    \item The timescale of perturbation dictates the excitability of different modes, with slower (faster) pulses triggering stronger bending (breathing) modes. An encounter with a satellite galaxy that hits the disk with a uniform velocity $\vp$ and an angle $\thetap$ with respect to the normal at a distance $\xp$ away from an observer in the disk, perturbs the potential at an observer's location with a characteristic time scale $\tau_{\rm enc} \sim \xp\cos{\thetap}/\vp$. If $\tau_{\rm enc}$ is long (short) compared to the typical vertical oscillation time, $\tau_z \sim h_z / \sigma_z$, at the observer's location, the dominant perturbation mode experienced is a bending (breathing) mode. Thus, bending modes are preferentially excited not only by low velocity encounters, but also by more distant and more perpendicular ones. Since the velocities of all MW satellites are much larger than $\sigma_z$, the decisive factor for bending {\it vs.} breathing modes is the distance from the point of impact. This is in qualitative agreement with the results from $N$-body simulations of the MW-Sgr encounter performed by \cite{Hunt.etal.21}, which show more pronounced bending (breathing) modes further from (closer to) the location where Sgr impacts the disk. Moreover, for a given encounter, stars with larger actions undergo stronger breathing mode oscillations since they oscillate slower.
    
    \item Besides phase spirals satellite encounters also induce spatial corrugations in the disk response, with vertical and lateral wave-numbers given by $k_z=n\Omega_z\cos{\thetap}/\vp$ and $k_x=n\Omega_z\sin{\thetap}/\vp$, respectively.
\end{itemize}

As an astrophysical application of our formalism, we have investigated the direct response of the MW disk (approximated as an isothermal slab) to several of the satellite galaxies in the halo for which dynamical mass estimates and galactocentric phase-space coordinates from Gaia parallax and proper motion measurements are available. We integrate the orbits of these satellites in the MW potential and note the impact velocity, $\vp$, angle of impact, $\thetap$, with respect to the normal, and the impact distance from the Solar neighborhood, $\xp$, during their penultimate, last and next disk crossings. We use these parameters to compute the direct response to the MW satellites and find that all of them excite bending modes and thus one-armed phase spirals in the Solar neighborhood, similar to that discovered in the Gaia data by \citet{Antoja.etal.18}. In the Solar vicinity, the largest direct response, by far, is due to the encounter with Sgr. The direct responses triggered by other satellites, most notably Hercules and the LMC, are at least $1-2$ orders of magnitude smaller. Hence, we conclude that, if the Gaia phase spiral was triggered by an encounter with a MW satellite, the strongest contender is Sgr. Although Sgr has been considered as the agent responsible for the Gaia phase spiral and other local asymmetries and corrugations, several studies have pointed out that it cannot be the sole cause of all these perturbations \citep[see e.g.,][]{Bennett.etal.22, Bennett.Bovy.21}. Our work argues, though, that the direct response in the Solar neighborhood from the other MW satellites, including the LMC, is not significant enough, at least in the range of actions covered by the Gaia snail. Of course, as discussed in section~\ref{sec::caveats}, the indirect response from the DM halo of the MW might play an important role especially for the more massive satellites such as Sgr and the LMC. Moreover the global response of a realistic disk will be different from that of the isothermal slab model considered here. We investigate the realistic disk response in chapter~\ref{chapter: paper3} and leave a sophisticated analysis incorporating self-gravity and halo response for future work. It remains to be seen whether a combination of Sgr plus other (internal) perturbations due to for example spiral arms \citep[][]{Faure.etal.14} or the (buckling) bar \citep[e.g.,][]{Khoperskov.etal.19} can explain the fine-structure in the Solar neighborhood, or whether perhaps a solution requires modifying the detailed MW potential. It is imperative, though, to investigate the structure of phase spirals at other locations in the MW disk, in particular whether they are one-armed or two-armed. This would help to constrain both the time-scale and location of the perturbation responsible for the various out-of-equilibrium features uncovered in the disk of our MW.


    \begin{subappendices}


\chapter*{Appendix}\label{chapter: paper2_app}

\section{Adiabatic limit of slab response}
\label{App:ad_lim_resp}

In the adiabatic/slow limit, the slab response can be computed by taking the $\omega_0\to 0$ limit and performing the $\tau$ integral in equation~(\ref{f1nk_isosol}) to obtain

\begin{align}
f_{1nk} = -i \pi\, \Phi_\rmN \calZ_n(I_z) \calX_k \left(\frac{n\Omega_z}{\sigma^2_z}+\frac{k v_x}{\sigma^2}\right) f_0(I_z,v_x,v_y)\, \delta(n\Omega_z+k v_x).
\end{align}
The Dirac delta function implies that only the resonant stars, i.e., those for which $n\Omega_z + k v_x=0$, contribute to the response in this slow limit. Substituting the expression for $f_0$ from equation~(\ref{f_iso}) in the above equation, integrating over $v_x$ and then summing over $n$, we obtain

\begin{align}
f_{1k} = -i\pi\,\Phi_\rmN \frac{\calX_k}{\left|k\right|} \sum_{n=-\infty}^{\infty} \calZ_n(I_z) \exp{\left[-\frac{n^2\Omega^2_z}{2 k^2\sigma^2}\right]} n\Omega_z \left(\frac{1}{\sigma^2_z}-\frac{1}{\sigma^2}\right) \exp{\left[i n w_z\right]}.
\end{align}
Substituting the Gaussian form for $\calX_k$ given in equation~(\ref{Phink_gaussian}) in the above expression, multiplying it by $\exp{\left[i k x\right]}$ and integrating over all $k$, we obtain the following final expression for the slab response in the slow limit:

\begin{align}
f_1(I_z,w_z,x) = -i \pi\,\Phi_\rmN \frac{\calX_k}{\left|k\right|} \sum_{n=-\infty}^{\infty} \calZ_n(I_z) \calJ_n(x)\, n\Omega_z \left(\frac{1}{\sigma^2_z}-\frac{1}{\sigma^2}\right) \exp{\left[i n w_z\right]},
\end{align}
where

\begin{align}
\calJ_n(x) = \int_{-\infty}^{\infty} \rmd k\, \frac{\exp{\left[i k x\right]}}{\left|k\right|} \exp{\left[-k^2\Delta^2_x/2\right]} \exp{\left[-\frac{n^2\Omega^2_z}{2 k^2\sigma^2}\right]}.
\end{align}
The above integral can be approximately evaluated in the small and large $x$ limits by the saddle point method to obtain the following asymptotic behaviour of $\calJ_n(x)$:

\begin{align}
\calJ_n(x) &\sim 
\begin{cases}
\sqrt{\pi \sigma/2 \left|n\right| \Omega_z \Delta_x}\, \exp{\left[-\left|n\right|\Omega_z \Delta_x/\sigma\right]}\, \cos{\left(\sqrt{\frac{\left|n\right|\Omega_z}{\sigma \Delta_x}}x\right)}, & \text{small\;} x,\nonumber \\
\sqrt{2\pi}\,\frac{\Delta_x}{x} \exp{\left[-x^2/2\Delta^2_x\right]}, & \text{large\;} x.
\end{cases}
\end{align}
\\

\section{Slab response to satellite encounters}
\label{App:sat_disk_resp}

The perturbing potential, $\Phi_\rmP$, at $(x,z)$ due to a satellite galaxy impacting the disk along a straight orbit with uniform velocity $\vp$ at an angle $\thetap$ with respect to the normal is given by equation~(\ref{Phip_sat}). Computing the Fourier transform, $\Phi_{nk}$, of $\Phi_\rmP$, and substituting this in equation~(\ref{f1nk_isosol}) yields

\begin{align}
f_{1nk}(I_z,v_x,v_y,t) &=i\frac{G \Mp}{\vp} f_0(v_x,v_y,E_z) \left(\frac{n\Omega_z}{\sigma^2_z}+\frac{k v_x}{\sigma^2}\right) \exp{\left[-i\left(n\Omega_z+k v_x\right) t\right]}\, \calF_{nk}(t),
\label{f1nk_sat_1}
\end{align}
where

\begin{align}
\calF_{nk}(t) &= \frac{1}{{\left(2\pi\right)}^2} \int_0^{2\pi}\rmd w'_z \exp{\left[-i n w'_z\right]} \int_{-\infty}^{\infty} \rmd x' \exp{\left[-i k x'\right]} \nonumber \\
&\times \int_{-\infty}^{t} \rmd \tau\, \frac{\exp{\left[i\left(n\Omega_z+k v_x\right) \tau\right]}}{\sqrt{{\left(\tau-\frac{z'\cos{\thetap}+x'\sin{\thetap}}{\vp}\right)}^2+\frac{{\left(x'\cos{\thetap}-z'\sin{\thetap}\right)}^2}{\vp^2}}}.
\end{align}
The $\tau$ integral can be computed in the large $t$ limit to yield

\begin{align}
\calF_{nk}(t\to \infty) &= \frac{1}{2\pi^2} \int_0^{2\pi}\rmd w'_z \exp{\left[-i n w'_z\right]} \int_{-\infty}^{\infty} \rmd x' \exp{\left[-i k x'\right]} \nonumber \\
& \times \exp{\left[i \frac{\left(n\Omega_z+k v_x\right)\cos{\thetap} z'}{\vp} \right]} \exp{\left[i \frac{\left(n\Omega_z+k v_x\right)\sin{\thetap} x'}{\vp} \right]} \nonumber \\
& \times K_0\left[\left(n\Omega_z+k v_x\right)\frac{\left(x'\cos{\thetap}-z'\sin{\thetap}\right)}{\vp}\right],
\end{align}
where $K_0$ denotes the zero-th order modified Bessel function of the second kind. Recalling that the unperturbed DF is isothermal, given by equation~(\ref{f_iso}), we integrate equation~(\ref{f1nk_sat_1}) over $v_x$ and $v_y$ to obtain

\begin{align}
&\int_{-\infty}^{\infty}\rmd v_y\int_{-\infty}^{\infty}\rmd v_x\, f_{1nk}(I_z,v_x,v_y,t) \approx \frac{\rho_c}{\sqrt{2\pi}\sigma_z} \exp{\left[-E_z/\sigma^2_z\right]} \frac{G\Mp}{\vp} \nonumber \\
&\times \frac{1}{2\pi^2} \int_0^{2\pi}\rmd w'_z \exp{\left[-i n w'_z\right]} \exp{\left[i \frac{n\Omega_z\cos{\thetap} z'}{\vp} \right]} \nonumber\\
&\times \int_{-\infty}^{\infty} \rmd x' \exp{\left[-i k x'\right]} \exp{\left[i \frac{n\Omega_z\sin{\thetap} x'}{\vp} \right]} \nonumber \\
&\times \exp{\left[-\frac{1}{2}k^2\sigma^2 {\left(t-\frac{\calS}{\vp}\right)}^2\right]} \left[k^2 \left(t-\frac{\calS}{\vp}\right)+i\frac{n\Omega_z}{\sigma^2_z}\right]  \nonumber \\
&\times K_0\left[\left(n\Omega_z- i k^2 \sigma^2 \left(t-\calS/\vp\right)\right)\frac{\left(x'\cos{\thetap}-z'\sin{\thetap}\right)}{\vp}\right],
\label{f1nk_sat_2}
\end{align}
where we have defined

\begin{align}
\calS=z'\cos{\thetap}+x'\sin{\thetap}.
\end{align}
Multiplying equation~(\ref{f1nk_sat_2}) by $\exp{\left[ikx\right]}$ and integrating over $k$ yields

\begin{align}
&\int_{-\infty}^{\infty} \rmd k\,\exp{\left[i k x\right]}\int_{-\infty}^{\infty}\rmd v_y\int_{-\infty}^{\infty}\rmd v_x\, f_{1nk}(I_z,v_x,v_y,t) \approx \frac{\rho_c}{\sqrt{2\pi}\sigma_z} \exp{\left[-E_z/\sigma^2_z\right]} \frac{G\Mp}{\vp} \nonumber \\ 
&\times \frac{1}{2\pi^2} \int_0^{2\pi}\rmd w'_z \exp{\left[-i n w'_z\right]} \exp{\left[i \frac{n\Omega_z\cos{\thetap} z'}{\vp} \right]} \nonumber \\
&\times \sqrt{2\pi} \int_{-\infty}^{\infty} \rmd \Delta x\, \frac{1}{\sigma t'} \exp{\left[-\frac{1}{2}\frac{{(\Delta x)}^2}{\sigma^2 t'^2}\right]} \left[\frac{1}{\sigma^2 t'}\left(1+\frac{{(\Delta x)}^2}{\sigma^2 t'^2}\right)+i\frac{n\Omega_z}{\sigma^2_z}\right] \nonumber \\
&\times \exp{\left[i \frac{n\Omega_z\sin{\thetap} x'}{\vp} \right]} K_0\left[\left(n\Omega_z+ i \frac{{(\Delta x)}^2}{\sigma^2 t'^3}\right)\frac{\left(x'\cos{\thetap}-z'\sin{\thetap}\right)}{\vp}\right],
\end{align}
where $\Delta x = x-x'$, and $t'=t-\calS/\vp$. In the large time limit, using the identity that 

\begin{align}
\lim_{t'\to \infty} \exp{\left[-{(\Delta x)}^2/2\sigma^2 t'^2\right]}\Big/\sigma t'=\sqrt{2\pi}\delta (\Delta x),
\end{align}
the integration over $\Delta x$ is simplified. Upon performing this integral, multiplying the result by $\exp{\left[i n w_z\right]}$ and summing over all $n$, we obtain the following response:

\begin{align}
f_1(I_z,w_z,x,t) &\approx \frac{\rho_c}{\sqrt{2\pi}\sigma_z} \exp{\left[-E_z/\sigma^2_z\right]}\times \frac{2G\Mp}{\vp} \nonumber \\
&\times \sum_{n=-\infty}^{\infty} \left[\frac{1}{\sigma^2 t}+i\frac{n\Omega_z}{\sigma^2_z}\right]\, \Psi_n(x,I_z)\, \exp{\left[i\,\frac{n\Omega_z \sin{\thetap}}{\vp}x\right]} \exp{\left[i n\left(w_z-\Omega_z t\right)\right]},
\label{f1_sat_app}
\end{align}
where
\begin{align}
\Psi_n(x,I_z)&= \frac{1}{2\pi} \int_0^{2\pi} \rmd w_z\, \exp{\left[-i n \left(w_z - \frac{\Omega_z \cos{\thetap} z}{\vp}\right)\right]} \nonumber \\
&\times K_0\left[\,\left|\frac{n\Omega_z \left(x\cos{\thetap}-z\sin{\thetap}\right)}{\vp}\right|\,\right].
\label{Psi_n_app}
\end{align}

The above expression for $\Psi_n$ can be simplified by evaluating the $w_z$ integral under the epicyclic approximation (small $I_z$ limit), to yield the following approximate form,
\begin{align}
\Psi_n(x,I_z) &\approx K_0\left(\frac{\left|n\Omega_z \cos{\thetap}\right|}{\vp}x\right) \Phi_n^{(0)}(I_z) - i\frac{n\Omega_z \sin{\thetap}}{\vp} K'_0\left(\frac{\left|n\Omega_z \cos{\thetap}\right|}{\vp}x\right) \Phi_n^{(1)}(I_z)\nonumber \\
&- \frac{1}{2} {\left(\frac{n\Omega_z \sin{\thetap}}{\vp}\right)}^2 K''_0\left(\frac{\left|n\Omega_z \cos{\thetap}\right|}{\vp}x\right) \Phi_n^{(2)}(I_z)+...\,.
\label{Psi_n_app_epi}
\end{align}
Here each prime denotes a derivative with respect to the argument of the function. $\Phi_n^{(j)}(I_z)$, for $j=0,1,2,...$, is given by
\begin{align}
\Phi_n^{(j)}(I_z) &= \frac{1}{2\pi} \int_0^{2\pi}\rmd w_z\, z^j\, \exp{\left[-i n \left(w_z - \frac{\Omega_z \cos{\thetap} z}{\vp}\right)\right]} \nonumber \\
&\approx {\left(\frac{2I_z}{\nu}\right)}^{j/2} J_{n,j}\left(\frac{n\Omega_z \cos{\thetap}}{\vp}\sqrt{\frac{2I_z}{\nu}}\right).
\label{Phin_sat_app}
\end{align}
Here the implicit relation between $z$, $w_z$ and $I_z$ given in equation~(\ref{z_wz_Iz}), which yields $z=\sqrt{2 I_z/\nu}\, \sin{w_z}$ for small $I_z$, has been used. $J_{n,j}$ denotes the $j^{\rm th}$ derivative of the $n^{\rm th}$ order Bessel function of the first kind, and $\nu=\sqrt{2}\,\sigma_z/h_z$ is the vertical epicyclic frequency. In equation~(\ref{f1_sat_app}), well after the encounter (large $t$), the term, $1/\sigma^2 t$, can be neglected relative to $i n\Omega_z/\sigma^2_z$ for $n\neq 0$, thus yielding the expression for the disk response to satellite encounters given in equation~(\ref{f1_sat}).

\section{Detectability criterion for the phase spiral}
\label{App:detect_crit}

The demarcation between strong and weak amplitudes of a phase spiral is dictated by the minimum detectable relative response, $\delta$, which can be determined in the following way. Let there be a phase spiral that we want to detect with a total number, $N_*$, of stars by binning the phase-space distribution in the $\sqrt{I_z}\cos{w_z}-\sqrt{I_z}\sin{w_z}$ plane. Let us define the unperturbed DF, $f_0$, and the normalized unperturbed DF, $\bar{f}_0$, such that

\begin{align}
N_* = \iint f_0\, \rmd I_z\, \rmd w_z,\;\;\;\; \bar{f}_0 = \frac{f_0}{N_*}.
\end{align}
The perturber introduces a perturbation in the (normalized) DF, $\bar{f}_1$, which manifests as a spiral feature in the phase-space due to phase-mixing. To recover $\bar{f}_1$ we bin the data in $I_z$ and $w_z$, such that the perturbation in the number of stars in each bin ($\Delta I_z,\Delta w_z$) is given by

\begin{align}
N(\Delta I_z,\Delta w_z) = N_* \bar{f}_1 \Delta I_z \Delta w_z.
\end{align}
The optimum binning strategy can be determined as follows. The phase spiral is a periodic feature in both $I_z$ and $w_z$. Therefore, to pull out the periodicity in $I_z$, we need to sample with a frequency exceeding the Nyquist frequency, i.e., the bin size, $\Delta I_z$, should be less than $I_{z,\rm max}/N_{\rm wind}$, where $I_{z,\rm max}$ is the maximum $I_z$ in the sample and $N_{\rm wind}$ is the number of winds of the spiral. Moreover, $\Delta I_z$ is required to exceed the Gaia measurement error so that the error is dominated by Poisson noise, i.e., we require $\Delta I_z/I_z > \Delta_{\rm Gaia} \sim 10^{-2}$ \citep[see][for parallax and radial velocity errors, the two dominant sources of measurement errors in Gaia]{Luri.etal.18,Katz.etal.19}. Within each $I_z$ bin, the data is further divided into $N_a$ azimuthal bins, each of size $\Delta w_z=2\pi/N_a$. For optimum sampling in $w_z$, $N_a$ should be greater than $2n$ (for spiral mode $n$) and less than $2\pi/\Delta_{\rm Gaia}$. After binning the data as discussed above, a reliable detection of the phase spiral can be made with a given signal to noise ratio, $S/N$, when the perturbation in the number of stars in each bin,

\begin{align}
N(\Delta I_z,\Delta w_z) = N_* \times \frac{\bar{f}_1}{\bar{f}_0} \times \frac{2\pi \bar{f}_0(I_z) \Delta I_z}{N_a} \geq {\left(S/N\right)}^2.
\end{align}
Here we have assumed that the error in recovering the spiral feature is dominated by Poisson noise. This yields the following estimate for the minimum detectable relative response for an isothermal slab,

\begin{align}
\frac{\bar{f}_1}{\bar{f}_0} \geq \delta = 3.6\times 10^{-4} \times {\left(\frac{S/N}{3}\right)}^2 \left(\frac{10^6}{N_*}\right) \left(\frac{N_a}{10}\right) \left(\frac{0.1}{\Delta I_z/I_z}\right) \frac{h_z\sigma_z}{I_z}\, \exp{\left[\frac{E_z(I_z)}{\sigma^2_z}\right]}.
\end{align}
Provided that there are about a million stars in the Gaia data of the Solar neighborhood \citep[][]{Antoja.etal.18}, we consider $\delta=10^{-4}$ to be a rough estimate for the minimum detectable relative response.

\label{lastpage}
    \end{subappendices}
    
    \chapter{A Comprehensive Perturbative Formalism for Phase-Mixing in Perturbed Disks. 
II. Phase spirals in an Inhomogeneous Disk Galaxy with a Non-responsive Dark Matter Halo} 
\label{chapter: paper3}

\begin{center}

This chapter has been published as:\\
\vspace{5pt}

\author{Uddipan Banik,  Frank~C.~van den Bosch and Martin~D.~Weinberg}

\vspace{5pt}

\textit{The Astrophysical Journal}, Volume 952, Number 1, Page 65\\

\textit{\citep[][]{Banik.etal.22a}}

\end{center}


\section{Introduction}
\label{sec:intro_3}

Disk galaxies are characterized by large-scale ordered motion and are therefore highly responsive to perturbations. Following a time-dependent gravitational perturbation, the actions of the disk stars are modified. This in turn causes a perturbation in the distribution function (DF) of the disk known as the response. Over time the response decays away as the system `relaxes' towards a new quasi-equilibrium via collisionless processes that include kinematic processes like phase-mixing (loss of coherence in the response due to different oscillation frequencies of stars) and secular/self-gravitating/collective processes like Landau damping \citep[loss of coherence due to wave-particle interactions,][]{LyndenBell.62}. As pointed out by \cite{Sridhar.89} and \cite{Maoz.91}, phase-mixing is the key ingredient of all collisionless relaxation and re-equilibration. 

The timescale of collisionless equilibration is typically longer than the orbital periods of stars. Therefore disk galaxies usually harbour prolonged features of incomplete equilibration following a perturbation, e.g., bars, spiral arms, warps and other asymmetries. An intriguing example is the one-armed phase-space spiral, or phase spiral for short, discovered in the Gaia DR2 data \citep[][]{Gaia_collab.18a} by \cite{Antoja.etal.18} and discussed in more detail in subsequent studies \citep[e.g.,][]{Bland-Hawthorn.etal.19, Laporte.etal.19, Li.Widrow.21, Li.21, Gandhi.etal.22}. \cite{Antoja.etal.18} plotted the density of stars in the Solar neighborhood in the $(z,v_z)$-plane of vertical position, $z$, and vertical velocity, $v_z$, and noticed a faint spiral pattern which became more pronounced when colour-coding the $(z,v_z)$-`pixels' by the median radial or azimuthal velocities. The one-armed spiral shows 2-3 complete wraps like a snail shell, and is interpreted as an indication of vertical phase-mixing following a perturbation that is anti-symmetric about the midplane (bending mode) and occurred $\sim 500\Myr$ ago. More recently, \cite{Hunt.etal.22} used the more extensive Gaia DR3 data to study the distributions of stars in $z-v_z$ space at different locations in the MW disk. They found that unlike the one-armed phase spiral or bending mode at the Solar radius, the inner disk shows a two-armed phase spiral that corresponds to a breathing mode or symmetric perturbation about the midplane. They inferred that while the one-armed spiral in the Solar neighborhood might have been caused by the impact of a satellite galaxy such as the Sagittarius dwarf, the two-armed spiral in the inner disk could not have been induced by the same since almost all satellite impacts are far too slow/adiabatic from the perspective of the inner disk. Rather, they suggested that the two-armed phase spiral might haven been triggered by a transient spiral arm or bar.

The phase spiral holds information about the perturbative history and gravitational potential of the disk and can therefore serve as an essential tool for galacto-seismology \citep[][]{Widrow.etal.14, Johnston.etal.17}. For a given potential, the winding of the spiral is an indication of the time elapsed since the perturbation occurred with older spirals revealing more wraps. A one-armed (two-armed) phase spiral corresponds to a bending (breathing) mode. Which mode dominates, in turn, depends on the time-scale of the perturbation, with temporally shorter (longer) perturbations (e.g., a fast or slow encounter with a satellite) predominantly triggering breathing (bending) modes \citep[][]{Widrow.etal.14,Banik.etal.22b}.

In addition to depending on the nature of the perturbation, the phase spiral also encodes information about the oscillation frequencies of stars and thus the detailed potential. In particular, the shape of the spiral depends on how the vertical frequencies, $\Omega_z$, vary as a function of the vertical action, $I_z$, which in turn depends on the underlying potential. In chapter~\ref{chapter: paper2} \citep[][]{Banik.etal.22b} we showed that the amplitude of the phase spiral can damp away due to lateral mixing, with a damping rate that depends on both the spatio-temporal nature of the perturbation and the frequency structure of the galaxy. This damping, though, only affects the response in the coarse-grained sense, i.e., upon marginalization of the response over the lateral degrees of freedom (the action-angle variables). Damping at the fine-grained level requires collisional diffusion, such as that arising from  gravitational scattering of stars against giant molecular clouds (GMCs), or dark matter (DM) substructure \citep[][]{Tremaine.etal.22}.

Chapter~\ref{chapter: paper2} addresses the problem of inferring the nature of the perturbation from the amplitude and structure of the phase spiral using a model of an infinite, isothermal slab for the unperturbed disk. This simple, yet insightful, model provides us with essential physical understanding of the perturbative response of disks without the complexity of modelling a realistic, inhomogeneous disk. However it suffers from certain glaring caveats: (i) lateral uniformity leading to an incorrect global structure of the response in the lateral direction, (ii) Maxwellian distribution of velocities in the lateral direction that overpredicts lateral mixing and thereby the rate at which the amplitude of the phase spiral damps out, (iii) absence of a DM halo and (iv) absence of self-gravity of the response. In this chapter we relax the first three assumptions. We consider an inhomogeneous disk characterized by a realistic DF similar to the pseudo-isothermal DF \citep[][]{Binney.10}, that properly captures the orbital dynamics of the disk stars in 3D. In addition, we consider the effect of an underlying DM halo which for the sake of simplicity we consider to be non-responsive. This ambient DM halo alters the potential and thus the frequencies of stars, which can in turn affect the shape of the phase spiral and its coarse-grained survival. We also consider the impact of small-scale collisionality on the fine-grained survival of the phase spiral. Since in this chapter we are primarily interested in the phase-mixing of the disk response that gives rise to phase spirals, we ignore self-gravity of the response which to linear order spawns coherent point mode oscillations of the disk \citep[for treatments of the self-gravitating response of isothermal slabs, see][]{Mathur.90,Weinberg.91}. 

This chapter is organized as follows. Section~\ref{sec:linear_theory_3} describes the standard linear perturbation theory for collisionless systems and its application to a realistic disk galaxy embedded in a DM halo that is exposed to a general perturbation. Sections~\ref{sec:disk_resp_spiral} and \ref{sec:disk_resp_sat} are concerned with computing the disk response for different perturber models. In Section~\ref{sec:disk_resp_spiral} we compute the disk response and phase spirals induced by bars and spiral arms. We also discuss the impact of collisional diffusion on the fine-grained survivability of the phase spiral. In Section~\ref{sec:disk_resp_sat} we compute the response to encounters with satellite galaxies. Section~\ref{sec:pot_const} describes how phase spirals can be used to constrain the galactic potential. We summarize our findings in Section~\ref{sec:concl_3}.


\section{Linear perturbation theory for galaxies}
\label{sec:linear_theory_3}

\subsection{Linear perturbative formalism}

A galaxy, to very good approximation, is devoid of star-star collisions. However, there are other potential sources of collisions such as scatterings due to gravitational interactions of stars with giant molecular clouds (GMCs) or DM substructure. The dynamics of stars in such a system is governed by the Boltzmann equation:
\begin{align}
\frac{\partial f}{\partial t}+[f,H]=C[f],
\label{CBE_master}
\end{align}
where $f$ denotes the DF, $H$ denotes the Hamiltonian, square brackets denote the Poisson bracket, and $C[f]$ denotes the collision operator due to small-scale fluctuations, which can be approximated by a Fokker-Planck operator \citep[see Appendix~A of][]{Tremaine.etal.22}:
\begin{align}
C[f] = \frac{1}{2} \frac{\partial}{\partial \xi_i} \left(D_{ij} \frac{\partial f}{\partial \xi_j} \right),
\label{coll_op}
\end{align}
where $\boldsymbol{\xi}=\left(\bq,\bp\right)$ with $\bq$ and $\bp$ denoting the canonically conjugate position and momentum variables, and $D_{ij}$ denotes the diffusion coefficient tensor.

Let the unperturbed steady state Hamiltonian of the galaxy be $H_0$ and the corresponding DF be given by $f_0$, which satisfies the unperturbed Fokker-Planck equation (FPE),
\begin{align}
[f_0,H_0]=C[f_0].
\end{align}
In presence of a small time-dependent perturbation in the potential, $\Phi_\rmP(t)$, the perturbed Hamiltonian can be written as
\begin{align}
H=H_0+\Phi_\rmP(t)+\Phi_1(t),
\end{align}
where $\Phi_1$ is the gravitational potential related to the response density, $\rho_1 = \int f_1 \rmd^3\bv$,
via the Poisson equation,
\begin{align}
\nabla^2\Phi_1=4\pi G\rho_1.
\end{align}
The perturbed DF can be written as
\begin{align}
f=f_0+f_1,
\end{align}
where $f_1$ is the linear order perturbation in the DF. In the weak perturbation limit where linear perturbation theory holds, the time-evolution of $f_1$ is dictated by the following linearized form of the FPE:
\begin{align}
\frac{\partial f_1}{\partial t}+[f_1,H_0]+[f_0,\Phi_\rmP]+[f_0,\Phi_1]=C[f_1].
\label{CBE_perturb_3}
\end{align}
Throughout this chapter we neglect the self-gravity of the disk response, which implies that we set the polarization term, $[f_0,\Phi_1]=0$. The implications of including self-gravity are discussed in chapter~\ref{chapter: paper2}.

\subsection{Response of a Galactic Disk to a realistic perturbation}
\label{sec:galdisk}

The dynamics of a realistic disk galaxy like the Milky Way (MW) is quasi-periodic, i.e., can be characterized by oscillations in the azimuthal, radial and vertical directions. In close proximity to the mid-plane and under radial epicyclic approximation, the Hamilton-Jacobi equation becomes separable, implying that all stars confined within a few vertical scale heights from the mid-plane of the disk are on regular, quasi-periodic orbits that are characterized by a radial action, $I_R$, an azimuthal action $I_\phi$, and a vertical action $I_z$. Hence, the motion of each star is characterized by three frequencies:
\begin{align}
\Omega_R = \frac{\partial H_0}{\partial I_R}\,,\;\;\; 
\Omega_\phi = \frac{\partial H_0}{\partial I_\phi}\,,\;\;\; 
\Omega_z = \frac{\partial H_0}{\partial I_z}\,.
\label{freqs}
\end{align}
This quasi-periodic nature of the orbits near the mid-plane is approximately preserved even in the presence of a (non-triaxial) DM halo since this preserves the axi-symmetry of the potential. Typically, as discussed in section~\ref{sec:pot_const}, the presence of a halo increases the oscillation frequencies of the disk stars.

In terms of these canonical conjugate action-angle variables, using equation~(\ref{freqs}), the linearized form of the FPE given in Equation~(\ref{CBE_perturb_3}) becomes
\begin{align}
&\frac{\partial f_1}{\partial t}+\Omega_z\frac{\partial f_1}{\partial w_z}+\Omega_R\frac{\partial f_1}{\partial w_R}+\Omega_\phi\frac{\partial f_1}{\partial w_\phi}-\frac{\partial \Phi_\rmP}{\partial w_z}\frac{\partial f_0}{\partial I_z}-\frac{\partial \Phi_\rmP}{\partial w_R}\frac{\partial f_0}{\partial I_R}-\frac{\partial \Phi_\rmP}{\partial w_\phi}\frac{\partial f_0}{\partial I_\phi} \nonumber \\
&= D^{(z)}_I\frac{\partial}{\partial I_z}\left(I_z\frac{\partial f_1}{\partial I_z}\right) + \frac{D^{(z)}_I}{4 I_z}\frac{\partial^2 f_1}{\partial w^2_z}+\frac{D^{(R)}_I}{4 I_R}\frac{\partial^2 f_1}{\partial w^2_R}.
\label{CBE_perturb_gen}
\end{align}
Here we have performed several simplifications of the Fokker-Planck operator (see Appendix~\ref{sec:FPE_pert_sol} for details). Firstly, the diffusion coefficients are computed using the epicyclic approximation, i.e., small $I_z$ and $I_R$, since only such stars are significantly affected by collisional scattering. In addition, following \cite{Tremaine.etal.22}, we consider $D^{(z)}_I$ and $D^{(R)}_I$ to be nearly constant, something that is implied by the age-velocity dispersion relation of the MW disk stars. Secondly, the $I_R$ diffusion of the response $f_1$ is negligible since the frequencies do not depend on $I_R$ under the radial epicyclic approximation (and only mildly depend on $I_R$ without it) and therefore the response does not develop $I_R$ gradients. Thirdly, following \cite{Binney.Lacey.88}, we have neglected diffusion in $I_\phi$ and $w_\phi$ since the terms involving $D_{\phi\phi}$, $D_{r\phi}$ and $D_{\phi z}$ are smaller than the $I_z$ and $I_R$ diffusion terms by factors of at least $\sigma_R/v_c$ or $\sigma_z/v_c$, which are typically much smaller than unity ($\sigma_R$ and $\sigma_z$ are radial and vertical velocity dispersions respectively, and $v_c$ is the circular velocity along $\phi$). We have retained the $w_z$ and $w_R$ diffusion terms for the sake of completeness, but as we point out later, the diffusion in angles typically occurs over much longer timescales than that in actions and hence is comparatively less important.

Since the stars move along quasi-periodic orbits characterized by actions and angles, we can expand the perturbations, $\Phi_\rmP$ and $f_1$, as discrete Fourier series in the angles as follows
\begin{align}
\Phi_\rmP\left(\bw,\bI,t\right)&=\sum_{n=-\infty}^{\infty} \sum_{\ell=-\infty}^{\infty} \sum_{m=-\infty}^{\infty} \exp{\left[i (n w_z + \ell w_R + m w_\phi)\right]}\, \Phi_{n\ell m}\left(\bI,t\right),\nonumber \\
f_1\left(\bw,\bI,t\right)&=\sum_{n=-\infty}^{\infty} \sum_{\ell=-\infty}^{\infty} \sum_{m=-\infty}^{\infty} \exp{\left[i (n w_z + \ell w_R + m w_\phi)\right]}\, f_{1,n\ell m}(\bI,t),
\label{fourier_series_gen}
\end{align}
where $\bw=(w_z,w_R,w_\phi)$ and $\bI=(I_z,I_R,I_\phi)$. Substituting these Fourier expansions in equation~(\ref{CBE_perturb_gen}) yields the following differential equation for the evolution 
of $f_{1,n\ell m}$:
\begin{align}
\frac{\partial f_{1,n\ell m}}{\partial t}+i(n\Omega_z+\ell\Omega_R+m\Omega_\phi)f_{1,n\ell m}&=i\left(n\frac{\partial f_0}{\partial I_z}+\ell\frac{\partial f_0}{\partial I_R} + m\frac{\partial f_0}{\partial I_\phi}\right)\Phi_{n\ell m} \nonumber \\
&+ D^{(z)}_I\frac{\partial}{\partial I_z}\left(I_z\frac{\partial f_{1,n\ell m}}{\partial I_z}\right) \nonumber \\ 
&- \left[\frac{n^2 D^{(z)}_I}{4 I_z} + \frac{\ell^2 D^{(R)}_I}{4 I_R}\right] f_{1,n\ell m}.
\label{f1nk_de_3}
\end{align}
This can be solved using the Green's function technique, with the initial condition, $f_{1,n\ell m}(t_\rmi)=0$, to yield the following closed integral form for $f_{1,n\ell m}$:
\begin{align}
f_{1,n\ell m}(\bI,t)&=i\left(n\frac{\partial f_0}{\partial I_z}+\ell\frac{\partial f_0}{\partial I_R}+m\frac{\partial f_0}{\partial I_\phi}\right) \calI_{n\ell m}(\bI,t).
\label{f1nk_gensol}
\end{align}
Here, for $D^{(z)}_I \ll \sigma^2_z$ ($\sigma_z$ is the vertical velocity dispersion), which is typically the case, $\calI_{n\ell m}(\bI,t)$ can be approximately expressed as

\begin{align}
\calI_{n\ell m}(\bI,t)&\approx \int_{t_\rmi}^{t}\rmd \tau\, \calG_{n\ell m}(\bI,t-\tau)\, \Phi_{n\ell m}(\bI,\tau).
\end{align}
Here $\calG_{n\ell m}(t-\tau)$ is the Green's function (see Appendix~\ref{sec:FPE_pert_sol} for detailed derivation), given by
\begin{align}
\calG_{n\ell m}(\bI,t-\tau) &\approx \exp{\left[-i(n\Omega_z+\ell\Omega_R+m\Omega_\phi)(t-\tau)\right]} \nonumber \\
&\times \exp{\left[-\left(\frac{n^2 D^{(z)}_I}{4 I_z}+\frac{\ell^2 D^{(R)}_I}{4 I_R}\right) \left(t-\tau\right)\right]}\, \exp{\left[-\frac{{\left(n\Omega_{z1}\right)}^2 D^{(z)}_I I_z}{3}{\left(t-\tau\right)}^3\right]},
\end{align}
where $\Omega_{z1}=\partial \Omega_z/ \partial I_z$. The sinusoidal factor represents the oscillations of stars at their natural frequencies which vary with actions, leading to the formation of phase spirals (see section~\ref{sec:spiral_cless} for details). The first exponential damping factor indicates the damping of the response due to diffusion in angles while the second damping factor manifests the damping of the $I_z$ gradients of the response by diffusion in $I_z$. As discussed in section~\ref{sec:spiral_c}, the diffusion in actions is much more efficient than that in angles.

Each $(n,\ell,m)$ Fourier coefficient of the response acts as a forced damped oscillator with three different natural frequencies, $n\Omega_z$, $\ell\Omega_R$ and $m\Omega_\phi$, which is being driven by an external time-dependent perturber potential, $\Phi_{n\ell m}$, and damped due to collisional diffusion. A similar expression, albeit without allowing for collisionality, for the DF perturbation has been derived by \cite{Carlberg.Sellwood.85} in the context of spiral arm induced perturbations and radial migrations in the galactic disk, and by other previous studies \citep[e.g.,][]{LyndenBell.Kalnajs.72, Tremaine.Weinberg.84, Carlberg.Sellwood.85, Weinberg.89, Weinberg.91, Weinberg.04, Kaur.Sridhar.18, Banik.vdBosch.21a, Kaur.Stone.22} in the context of dynamical friction in spherical systems. To obtain the final expression for $f_{1,n\ell m}$, we need to specify the spatio-temporal behavior of the perturber potential, $\Phi_\rmP$, as well as the DF, $f_0$, of the unperturbed galaxy, which is addressed below.

\subsection{The unperturbed galaxy}
\label{sec:disk_model_3}

Under the radial epicyclic approximation (small $I_R$), the unperturbed DF, $f_0$, for a rotating MW-like disk galaxy can be well approximated as a pseudo-isothermal DF, i.e., written as a nearly isothermal separable function of the azimuthal, radial and vertical actions. Following \cite{Binney.10}, we write
\begin{align}
f_0 \approx \frac{\sqrt{2}}{\pi^{3/2} \, \sigma_z h_z} {\left(\frac{\Omega_\phi \Sigma}{\kappa\, \sigma^2_R}\right)}_{\Rc} \, \exp{\left[-\frac{\kappa I_R}{\sigma^2_R}\right]} \, \exp{\left[-\frac{E_z(I_z)}{\sigma^2_z}\right]} \, \Theta(L_z)\,.
\label{DF_MW}
\end{align}
The vertical structure of this disk is isothermal, while the radial profile is pseudo-isothermal\footnote{In the limit of small $I_R$, the radial energy $E_R$ can be approximated as $\kappa I_R$. In this case, the isothermal form of the unperturbed DF, $\exp{\left[-E_R/\sigma^2_R\right]}$, reduces to $\exp{\left[-\kappa I_R/\sigma^2_R\right]}$, which is known as a pseudo-isothermal distribution.}. Here $\Sigma = \Sigma(R) = \int_{-\infty}^\infty \rmd z\, \rho(R,z)$ is the surface density of the disk, $L_z$ is the $z$-component of the angular momentum, which is equal to $I_\phi$, $\Rc = \Rc(L_z)$ is the guiding radius, $\Omega_\phi$ is the circular frequency, and $\kappa = \kappa(\Rc) = \lim_{I_R \to 0}{\Omega_R}$ is the radial epicyclic frequency \citep[][]{Binney.Tremaine.87}. $\Theta(x)$ is the Heaviside step function. Thus we assume that the entire galaxy is composed of prograde stars with $L_z>0$. 

The density profile, $\rho(R,z)$, of the disk corresponding to the above DF is the product of a radially exponential profile with scale radius $h_r$ and a vertically isothermal ($\sech^2$) profile with scale height $h_z$ (equation~[\ref{disk_exp_iso_rho_app}]).
As shown by \cite{Smith.etal.15}, this density profile is accurately approximated by a sum of three \cite{Miyamoto.Nagai.75} disks\footnote{the 3MN profile as implemented in the {\tt Gala Python package} \citep[][]{gala, gala_code_adrian}.}, which has a simple, analytical form for the associated potential. Throughout, we therefore use this 3MN approximation for our disk since this drastically simplifies the computation of orbital frequencies. For the purpose of computing the disk response, we assume typical MW like parameters for the various quantities, i.e., $\Rsun=8$ kpc, disk mass $M_\rmd=5\times 10^{10}\Msun$, $h_R=2.2 \kpc$, $\sigma_R(\Rsun) = \sigma_{R,\odot} = 35\, {\rm km}/\rms$, $h_z=0.4 \kpc$ and $\sigma_z(\Rsun) = \sigma_{z,\odot} = \sqrt{2\pi G h_z \Sigma(\Rsun)} = 23\, {\rm km}/\rms$ \citep[][]{McMillan.11, Bovy.Rix.13}. For this set of parameters, the Toomre $Q=\sigma_R \kappa /(3.36 G \Sigma)$ for the stellar disk turns out to be $2.26$ in the Solar neighborhood, indicating that the disk is gravitationally stable. The isothermal vertical distribution of disk stars adopted here ignores the potential influence of gas near the midplane, which may increase the shear, $d\Omega_z/d I_z$, for small $I_z$. This may affect the shape of the phase spiral, making it more tightly wound, but should not substantially impact its amplitude.

The disk is assumed to be embedded in an extended DM halo characterized by a spherical NFW \citep[][]{Navarro.etal.97} density profile, with virial mass $\Mvir$, concentration $c$, scale radius $r_s$ and the corresponding potential $\Phi_\rmh$ given by equation~(\ref{Phi0_halo_app}). For the NFW DM halo, we adopt $\Mvir=9.78\times 10^{11}\Msun$, $r_s=16$ kpc, and $c=15.3$ \citep[][]{Bovy.15}.

The combined potential experienced by the disk stars is simply the sum of disk and halo potentials, i.e.,
\begin{align}
\Phi_0(R,z)=\Phi_\rmd(R,z)+\Phi_\rmh(R,z).
\end{align}
The total energy of a disk star under the radial epicyclic approximation is $E= L^2_z/2 R^2_c + \Phi_0(\Rc,0) + \kappa I_R + E_z$, where the vertical part of the energy is given by $E_z=v^2_z/2+\Phi_z(\Rc,z)$, with $\Rc(L_z)$ the guiding radius given by $L^2_z/\Rc^3=\partial \Phi_0/\partial R|_{R=\Rc}$. The vertical potential, $\Phi_z(\Rc,z)$, is given by
\begin{align}
\Phi_z(\Rc,z)=\Phi_0(\Rc,z)-\Phi_0(\Rc,0).
\end{align}
The vertical action, $I_z$, can be obtained from $E_z$ as follows
\begin{align}
I_z = \frac{1}{2\pi} \oint v_z \, \rmd z= \frac{2}{\pi} \int_0^{z_{\rm max}} \sqrt{2[E_z - \Phi_z(\Rc,z)]} \, \rmd z\,,
\end{align}
where $\Phi_z(\Rc,z_{\rm max})=E_z$. This implicit equation can be inverted to obtain $E_z(\Rc,I_z)$. The time period of vertical oscillation can then be obtained using
\begin{align}
T_z(\Rc,I_z) = \oint \frac{\rmd z}{v_z} = 4\int_0^{z_{\rm max}} \frac{\rmd z} {\sqrt{2\left[E_z(\Rc,I_z)-\Phi_z(\Rc,z)\right]}},
\label{T_z_3}
\end{align}
which yields the vertical frequency, $\Omega_z(\Rc,I_z) = 2\pi/T_z(\Rc,I_z)$.

Substituting the expression for $f_0$ given by Equation~(\ref{DF_MW}) in Equation~(\ref{f1nk_gensol}), we obtain the following integral form for $f_{1,n\ell m}$,
\begin{align}
f_{1,n\ell m}(\bI,t) & \approx -\frac{2i}{\pi\sigma^2_R}\, \frac{1}{\sqrt{2\pi}h_z\sigma_z} \exp{\left[-\frac{\kappa I_R}{\sigma^2_R}\right]} \exp{\left[-\frac{E_z(I_z)}{\sigma^2_z}\right]} \nonumber \\
&\times \left[\left\{\left(\frac{n\Omega_z}{\sigma^2_z} + \frac{\ell\kappa}{\sigma^2_R}\right) {\left(\frac{\Omega_\phi \Sigma}{\kappa}\right)} - m\frac{\rmd}{\rmd L_z}\left(\frac{\Omega_\phi \Sigma}{\kappa}\right)\right\} \Theta(L_z) - m\frac{\Omega_\phi \Sigma}{\kappa} \delta(L_z)\right] \nonumber\\
&\times \calI_{n\ell m}(\bI,t).
\label{f1nk_gensol_f0}
\end{align}
As we shall see, the first order disk response expressed above phase mixes away and gives rise to phase spirals due to oscillations of stars with different frequencies except when they are resonant with the frequency of the perturber. However this `direct' response of the disk does not include certain effects. First of all, we ignore the self-gravity of the response. As discussed in chapter~\ref{chapter: paper2}, to linear order self-gravity gives rise to point mode oscillations of the disk that are decoupled from the phase-mixing component of the response which is what we are interested in.  Secondly, for the sake of simplicity, we consider the ambient DM halo to be non-responsive and therefore ignore the indirect effect of the halo response on disk oscillations. We leave the inclusion of these two effects in the computation of the disk response for future work.

The spatio-temporal nature of the perturbing potential dictates the disk response. In this chapter we explore two different types of perturbation to which realistic disc galaxies can be exposed, and which are thus of general astrophysical interest. The first is an in-plane spiral/bar perturbation with a vertical structure, either formed as a consequence of secular evolution, or triggered by an external perturbation. We consider both short-lived (transient) and persistent spirals. The second type of perturbation that we consider is that due to an encounter with a massive object, e.g., a satellite galaxy or DM subhalo.

\section{Disk response to spiral arms and bars}
\label{sec:disk_resp_spiral}

We model the potential of a spiral arm perturbation as one with a vertical profile and a sinusoidal variation along radial and azimuthal directions,
\begin{align}
\Phi_\rmP(R,\phi,z) &= -\frac{2\pi G \Sigma_\rmP}{k_R}\, \left[\alpha\,\calM_\rmo(t)\,\calF_\rmo(z) + \calM_\rme(t)\,\calF_\rme(z)\right] \nonumber \\
&\times \sum_{m_\phi=0,2} \sin{\left[k_R R + m_\phi \left(\phi-\Omega_\rmP t\right)\right]}\,.
\label{Phip_spiral}
\end{align}
Here $\Omega_\rmP$ is the pattern speed and $k_R$ is the horizontal wave number of the spiral perturbation. The long wavelength limit, $k_R\to 0$, corresponds to a bar. We consider the in-plane part of $\Phi_\rmP$ to be a combination of an axisymmetric ($m_\phi=0$) and a 2-armed spiral mode ($m_\phi=2$), and the vertical part to be a combination of anti-symmetric/odd and symmetric/even perturbations respectively denoted by $\calF_\rmo$ and $\calF_\rme$, that are modulated by growth functions, $\calM_\rmo(t)$ and $\calM_\rme(t)$, capturing the growth and/or decay of the spiral strength over time. The ratio of the maximum strengths of the anti-symmetric and symmetric parts of the perturbation is $\alpha$. We consider the following two functional forms for $\calM_j(t)$ (where the subscript $j=\rmo$ or $\rme$):
\begin{align}
\calM_j(t) &= 
\begin{cases}
\frac{1}{\sqrt{\pi}}\exp{\left[-\omega^2_j t^2\right]}, & \text{Transient spiral/bar} \\
\exp{\left[\gamma_j t\right]} + \left(1-\exp{\left[\gamma_j t\right]}\right)\,\Theta(t), & \text{Persistent spiral/bar.}
\end{cases}
\label{modulation}
\end{align}
The first option describes a transient spiral/bar that grows and decays like a Gaussian pulse with a characteristic life-time $\tau_{\rmP j} \sim 1/\omega_j$ \citep[][]{Banik.etal.22b}. The second form describes a persistent spiral perturbation that grows exponentially on a timescale $\tau_{\rmG j} \sim 1/\gamma_j$ and then saturates to a constant amplitude. We shall see shortly that these two kinds of spiral perturbations perturb the disk in very different ways.

The vertical part of the perturbation consists of an anti-symmetric function, $\calF_\rmo(z)$, and a symmetric function, $\calF_\rme(z)$, which, for the sake of simplicity, we take to be the following trigonometric functions: 
\begin{align}
\calF_\rmo(z) &= \sin{\left(k_z^{(\rmo)} z\right)}, \nonumber \\
\calF_\rme(z) &= \cos{\left(k_z^{(\rme)} z\right)}.
\end{align}
Here $k^{(\rmo)}_z$ and $k^{(\rme)}_z$ denote the vertical wave-numbers of the anti-symmetric and symmetric perturbations, respectively. Since the above functions form a complete Fourier basis in $z$, any (vertical) perturber profile can be expressed as a linear superposition of $\calF_\rmo$ and $\calF_\rme$. The disk response involves the Fourier coefficients of the perturbing potential, $\Phi_{n\ell m}$, which can be obtained by taking the Fourier transform of $\Phi_\rmP$ given in Equation~(\ref{Phip_spiral}) with respect to the angles, $w_R$, $w_\phi$ and $w_z$, as detailed in Appendix~\ref{App:fourier_spiral}.

\begin{figure}[H]
\centering
\subfloat{\includegraphics[width=0.49\textwidth]{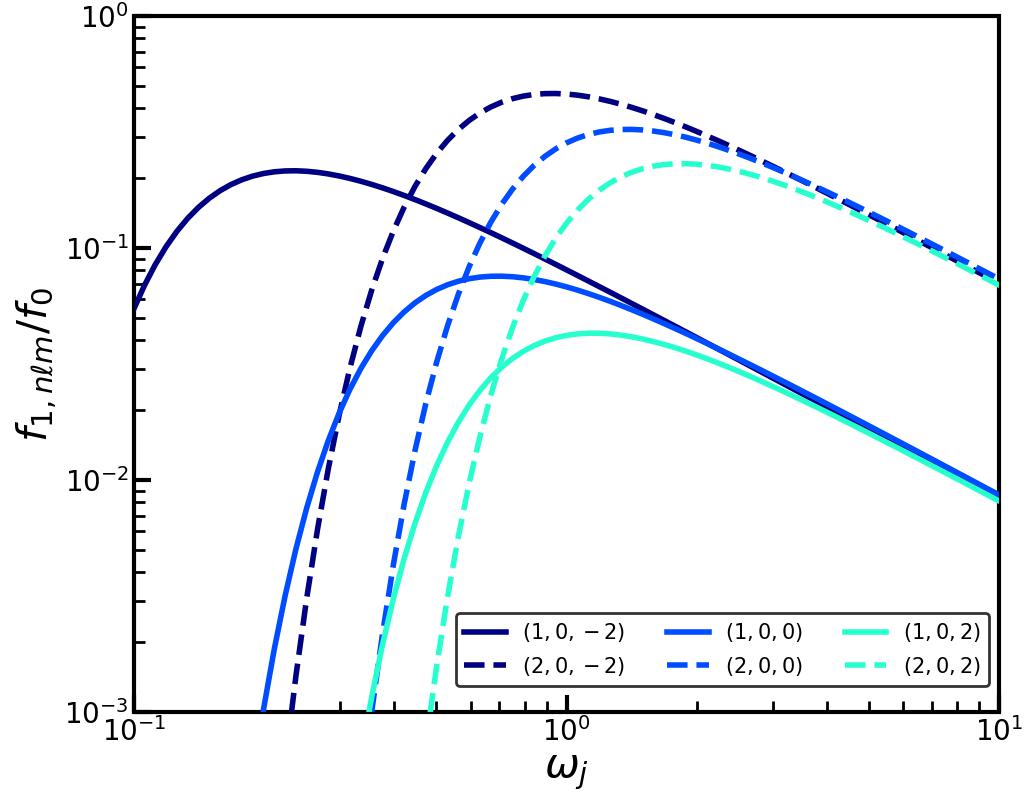} \label{fig:f1_vs_omega_gauss}}
\subfloat{\includegraphics[width=0.49\textwidth]{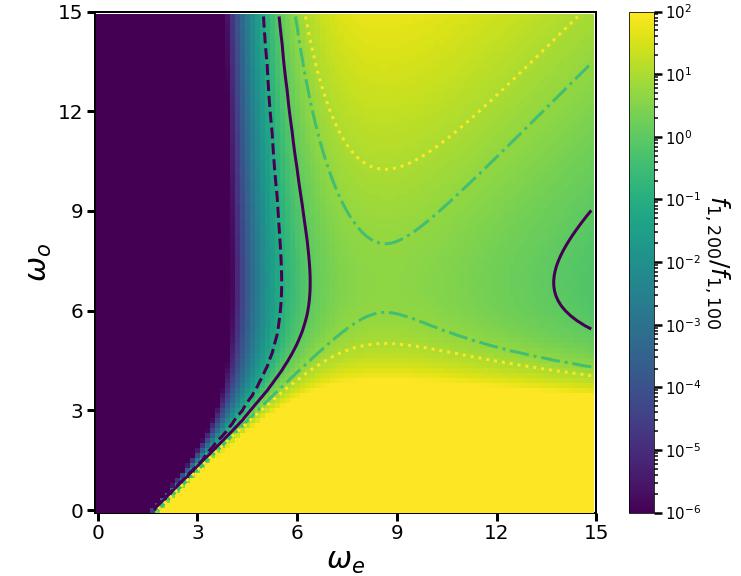} \label{fig:BB_gauss}}
\caption{MW disk response to transient bars/2-armed spirals with Gaussian temporal modulation in absence of collisional diffusion: Left panel shows the steady state ($t\to \infty$) amplitude of the disk response, $f_{1,n\ell m}/f_0$, in the Solar neighborhood, computed using equations~(\ref{trans_spiral_Resp}) and (\ref{Pnlm_trans_spiral}) in presence of an ambient DM halo, as a function of the pulse frequency, $\omega_j$, where the subscript $j=\rmo$ and $\rme$ for vertically anti-symmetric (odd $n$) and symmetric (even $n$) perturbations. Solid (dashed) lines indicate the $n=1$ bending ($n=2$ breathing) modes and different colors denote $(\ell,m)=(0,-2)$, $(0,0)$ and $(0,2)$ respectively. We consider $I_z = I_{z,\odot} \equiv h_z\sigma_{z,\odot}$ and marginalize the response over $I_R$. Note that the response peaks at intermediate values of $\omega_j$, which is different for different modes, and is suppressed like a power law in the impulsive (large $\omega_j$) limit and super-exponentially in the adiabatic (small $\omega_j$) limit. Right panel shows the breathing-to-bending ratio, $f_{1,200}/f_{1,100}$, as a function of $\omega_\rme$ and $\omega_\rmo$, the pulse frequencies of the bending and breathing mode perturbations respectively. The dashed, solid, dot-dashed and dotted contours correspond to breathing-to-bending ratios of $0.1,1,5$ and $10$ respectively. The breathing-to-bending ratio rises and falls with increasing $\omega_\rme$ at fixed $\omega_\rmo$, while the reverse occurs with increasing $\omega_\rmo$ at fixed $\omega_\rme$, leading to a saddle point at $(\omega_\rme,\omega_\rmo)\approx (9,7)$.}
\label{fig:trans_spiral}
\end{figure}

\subsection{Computing the disk response}

The expression for the disk response to bars or spiral arms can be obtained by substituting the Fourier coefficient of the perturber potential given in Equation~(\ref{spiral_fourier_app}) in Equation~(\ref{f1nk_gensol_f0}) and performing the $\tau$ integration with the initial time, $t_\rmi\to -\infty$. This yields the modal response, $f_{1,n\ell m}$ (Equation~[\ref{f1nk_gensol_f0}]), with $\calI_{n\ell m}(\bI,t)$ given by
\begin{align}
\calI_{n\ell m}(\bI,t) &= \alpha\,\Psi^{(\rmo)}_{n\ell m}(\bI)\, \calP^{(\rmo)}_{n\ell m}(\bI,t) + \Psi^{(\rme)}_{n\ell m}(\bI)\, \calP^{(\rme)}_{n\ell m}(\bI,t),
\label{trans_spiral_Resp}
\end{align}
where $\Psi^{(\rmo)}_{n\ell m}$ and $\Psi^{(\rme)}_{n\ell m}$ respectively denote the time-independent parts of the odd and even terms in the expression for $\Phi_{n\ell m}$, and
\begin{align}
\calP^{(j)}_{n\ell m}(\bI,t) &= \exp{\left[-i\,m\Omega_\rmP\,t\right]} \int_0^{\infty} \rmd \tau \exp{\left[-i\,\Omega_{\rm res}\,\tau\right]}\, \exp{\left[-\left(\frac{n^2 D^{(z)}_I}{4 I_z}+\frac{\ell^2 D^{(R)}_I}{4 I_R}\right) \tau\right]}\nonumber \\
&\times \exp{\left[-\frac{{\left(n\Omega_{z1}\right)}^2 D^{(z)}_I I_z}{3}{\tau}^3\right]} \, \calM_j(t-\tau),
\label{Pnlm_spiral}
\end{align}
which characterizes the temporal evolution of the response. Here the subscript $j=\rmo$ or $\rme$, and the resonance frequency, $\Omega_{\rm res}$, is given by
\begin{align}
\Omega_{\rm res} = n\Omega_z + \ell\kappa + m(\Omega_\phi-\Omega_\rmP).
\end{align}

\subsubsection{Collisionless limit}\label{sec:spiral_cless}

First we examine the response in the limit of zero diffusion, i.e., $D^{(z)}_I=0$, where each star acts as a forced oscillator. 

\paragraph{\ul{Transient spiral arms and bars}}

First we consider the case of transient spiral arm or bar perturbations that grow and decay in strength over time, i.e., the temporal modulation $\calM_j(t)$ is given by the first of equations~(\ref{modulation}). In this case,
\begin{align}
\calP_{n\ell m}^{(j)}(\bI,t) &= 
\frac{1}{2\,\omega_j}\,\exp{\left[-\frac{\Omega^2_{\rm res}}{4\omega^2_j}\right]} \left[1+\erf{\left(\omega_j t - i\frac{\Omega_{\rm res}}{2\omega_j}\right)}\right] \exp{\left[-i(n\Omega_z+\ell\kappa+m\Omega_\phi)t\right]}\nonumber \\
&\xrightarrow{t\to\infty} \frac{1}{\omega_j}\,\exp{\left[-\frac{\Omega^2_{\rm res}}{4\omega^2_j}\right]}\, \exp{\left[-i(n\Omega_z+\ell\kappa+m\Omega_\phi)t\right]}.
\label{Pnlm_trans_spiral}
\end{align}
The error function describes the growth and transient oscillations of the response amplitude; over time the transients die away, and in the limit $t\to\infty$ the response saturates to a constant amplitude (in the absence of collisional diffusion).

The left-hand panel of Fig.~\ref{fig:trans_spiral} plots the amplitude of the steady state disk response to transient spiral/bar perturbations, relative to the unperturbed DF, as a function of the modulation/pulse frequency, $\omega_j$ ($j=\rmo$ and $\rme$ for bending and breathing modes respectively), for different modes indicated in different colors. Solid and dashed lines correspond to the $n=1$ bending modes and the $n=2$ breathing modes, respectively. We adopt $\Sigma_\rmP=5.5\Msun \pc^{-2}$, $\Omega_\rmP=12 \kms\kpc^{-1}$, $k^{(\rmo)}_z=k^{(\rme)}_z=1\kpc$, $k_R=10\kpc$, and $I_z = I_{z,\odot} \equiv h_z\sigma_{z,\odot}=9.2\kms$, and marginalize the response over $I_R$. We set $\alpha=1$, implying equal maximum strengths for the bending and breathing modes. As evident from this figure, and also from equation~(\ref{Pnlm_trans_spiral}), the long-term strength of the disk response (after the initial transients have died out like $e^{-\omega^2_j t^2}$) scales as $\sim 1/\omega_j$ in the impulsive (large $\omega_j$) limit, but is super-exponentially suppressed ($\sim \exp\left[-\Omega^2_{\rm res}/4 \omega^2_j\right]$) in the adiabatic (small $\omega_j$) limit away from resonances, i.e., for $\Omega_{\rm res}\neq 0$. The adiabatic suppression scales differently with $\omega_j$ for other functional forms of $\calM_j(t)$, e.g., for $\calM_j(t)=1/\sqrt{1+\omega^2_j t^2}$ the response strength is exponentially suppressed $(\sim \exp[-\Omega_{\rm res}/\omega_j])$. The response of resonant modes ($\Omega_{\rm res}=0$) however does not undergo adiabatic suppression and scales as $\sim 1/\omega_j$ throughout, becoming non-linear in the adiabatic regime. Since there are many resonance modes, the cumulative response in the adiabatic limit of {\it all} modes combined is only suppressed as a power-law, rather than an exponential, in $\omega_j$ \citep[][]{Weinberg.94a,Weinberg.94b}.

The sinusoidal factor, $\exp{\left[-i(n\Omega_z+\ell\kappa+m\Omega_\phi)t\right]}$, in $\calP_{n\ell m}^{(j)}$ describes the oscillations of stars with three different frequencies, $\Omega_z$, $\kappa$ and $\Omega_\phi$, along the vertical, radial and azimuthal directions, respectively. Due to the dependence of these frequencies on the actions, that of $\Omega_z$ on $I_z$ and of $\kappa$ and $\Omega_\phi$ on $I_\phi=L_z$, the response integrated over actions eventually phase mixes away. This manifests as phase spirals in the $I_z\cos{w_z}-I_z\sin{w_z}$ and $I_\phi\cos{\phi}-I_\phi\sin{\phi}$ phase-spaces, which are proxies for the $z-v_z$ and $\phi-\dot{\phi}$ phase-spaces, respectively. As is evident from equation~(\ref{Pnlm_spiral}), $\calP_{n\ell m}^{(j)} \sim \exp{\left[-i m \Omega_\rmP t\right]}$ in the adiabatic limit ($\omega_j\to 0$); hence, in this limit the  sinusoidal factor, $\exp{\left[-i(n\Omega_z+\ell\kappa+m\Omega_\phi)t\right]}$ is absent from the response, which implies that phase spirals only occur for sufficiently impulsive perturbations. As shown in chapter~\ref{chapter: paper2}, $n=1$ bending modes involve a dipolar perturbation in the vertical phase-space ($I_z\cos{w_z}-I_z\sin{w_z}$) distribution immediately after the perturbing pulse reaches its maximum strength. This dipolar distortion is subsequently wound up into a one-armed phase spiral since $\Omega_z$ is a function of $I_z$. Breathing modes, on the other hand, involve an initial quadrupolar perturbation in the phase-space distribution which is subsequently wrapped up into a two-armed phase spiral. Since $\Omega_z$, $\Omega_\phi$ and $\Omega_R$ all depend on $L_z$, the amplitude of the $I_z\cos{w_z}-I_z\sin{w_z}$ phase spiral damps out over time due to mixing between stars with different $L_z$. The modal response, $f_{1,n\ell m}$, when marginalized in a narrow bin of size $\Delta L_z$ around $L_z$, damps out as follows:
\begin{align}
\left<f_{1,n\ell m}\right>(\bI,t) &= \frac{1}{\Delta L_z} \int_{L_z-\Delta L_z/2}^{L_z+\Delta L_z/2} \rmd L_z\, f_{1,n\ell m} (\bI,t) \nonumber \\ \nonumber \\
&\approx \dfrac{\;\;\;\sin{\left[\left(\dfrac{\partial}{\partial L_z}\left(n\Omega_z+\ell\kappa+m\Omega_\phi\right)\right)\dfrac{\Delta L_z}{2} t\right]}\;\;\;} {\;\;\;\;\;\left(\dfrac{\partial}{\partial L_z}\left(n\Omega_z+\ell\kappa+m\Omega_\phi\right)\right)\dfrac{\Delta L_z}{2} t}\, f_{1,n\ell m}(\bI,t).
\label{lateral_mixing_disk}
\end{align}
Since the frequencies vary with $L_z$, marginalizing over $L_z$ mixes phase spirals that differ slightly in phases, giving way to a $\sim 1/t$ damping accompanied by a beat-like modulation with a characteristic lateral mixing timescale,

\begin{align}
\tau_\rmD^{(\rm{LM})} &= \dfrac{1} {\left(\dfrac{\partial}{\partial L_z}\left(n\Omega_z+\ell\kappa+m\Omega_\phi\right)\right)\dfrac{\Delta L_z}{2}}.
\end{align}
This explains why the density-contrast of the Gaia phase spiral is enhanced upon color-coding by $v_\phi$ or, equivalently, $L_z$ \citep[][]{Antoja.etal.18, Bland-Hawthorn.etal.19}. Radial phase-mixing is also present, but is typically much weaker because none of the frequencies depend on $I_R$ under the radial epicyclic approximation and only mildly depend on $I_R$ without it. Hence, due to ordered motion, the phase spiral amplitude in a realistic disk galaxy damps out at a much slower rate, as $\sim 1/t$ (in absence of collisional diffusion), than the lateral mixing damping in the isothermal slab case considered in chapter~\ref{chapter: paper2}, which arises from the unconstrained lateral velocities of the stars and exhibits a Gaussian temporal behavior.

It is worth emphasizing that not all frequencies undergo phase-mixing. In fact the resonant frequencies, for which
\begin{align}
\Omega_{\rm res}=n\Omega_z+\ell\kappa+m(\Omega_\phi-\Omega_\rmP)=0,
\end{align}
do not phase mix away. Hence, parts of the phase-space closer to a resonance take longer to phase-mix away. Moreover, as manifest from the adiabatic suppression factor, $\exp[-\Omega^2_{\rm res}/4\omega^2_j]$, the near-resonant modes with $\Omega_{\rm res} \ll 2\omega_j$ have much larger amplitude than those with $\Omega_{\rm res} \gg 2\omega_j$ that are far from resonance. Therefore the long-term disk response consists of stars in (near) resonance with the perturbing bar or spiral arm. Most of the strong resonances are confined to the disk-plane, including the co-rotation resonance $(n,\ell,m)=(0,0,m)$, the Lindblad resonances $(0,\pm 1, \pm 2)$, the ultraharmonic resonances $(0,\pm 1,\pm 4)$, and so on. For thin disks with $h_z \ll h_R$ , the vertical degrees of freedom are generally not in resonance with the radial or azimuthal ones since $\Omega_z$ is much larger than $\Omega_\phi$ or $\kappa$. Hence the vertical oscillation modes ($n\neq 0$) such as the $n=1$ bending or $n=2$ breathing modes undergo phase-mixing and give rise to phase spirals. However, if the disk has significant thickness, then the vertical degrees of freedom can be in resonance with the horizontal ones, e.g., banana orbits ($\Omega_z=2\Omega_r$) in barred disks.

The excitability of the bending and breathing modes is dictated by the perturbation timescale, or more precisely by the ratio of the pulse frequency, $\omega_j$, and the resonant frequency, $\Omega_{\rm res}$. The right panel of Fig.~\ref{fig:trans_spiral} shows the breathing-to-bending ratio, $f_{1,200}/f_{1,100}$, as a function of $\omega_\rme$ and $\omega_\rmo$, with blue (yellow) shades indicating low (high) values. In general, the breathing-to-bending ratio rises steeply and falls gradually with $\omega_\rme$ at fixed $\omega_\rmo$ while the trend is reversed as a function of $\omega_\rmo$ at fixed $\omega_\rme$, resulting in a saddle point at $(\omega_\rme,\omega_\rmo)\approx (9,7)$. This owes to the super-exponential suppression in the adiabatic ($\omega_j \ll \Omega_{\rm res}$) limit and the power-law suppression in the impulsive ($\omega_j \gg \Omega_{\rm res}$) limit. Along the $\omega_\rmo=\omega_\rme$ line, the bending (breathing) modes dominate in the adiabatic (impulsive) limit, as evident from the left panel of Fig.~\ref{fig:trans_spiral}. All this suggests that bending modes dominate over breathing modes when (i) the anti-symmetric perturbation is more impulsive, i.e., evolves faster than the symmetric one, or (ii) both symmetric and anti-symmetric perturbations occur over comparable timescales but slower than the stellar vertical oscillation period.

\paragraph{\ul{Persistent spiral arms and bars}}

Next we consider perturbations caused by a persistent spiral arm or bar that grows exponentially until it saturates at a constant strength. The corresponding temporal modulation $\calM_j(t)$ is given by the second of equations~(\ref{modulation}). In this case, as shown by equation (19) of \cite{Banik.vdBosch.21a},
\begin{align}
\calP_{n\ell m}^{(j)}(\bI,t) &= \frac{\exp{\left[\gamma_j t\right]} \exp{\left[-i m \Omega_\rmP t\right]}}{\gamma_j+i\Omega_{\rm res}}\left[1-\Theta(t)\right] \nonumber \\
&+i\left[\frac{\gamma_j \exp{\left[-i(n\Omega_z+\ell\kappa+m\Omega_\phi)t\right]}}{\Omega_{\rm res}(\gamma_j+i\Omega_{\rm res})}-\frac{\exp{\left[-i m \Omega_\rmP t\right]}}{\Omega_{\rm res}}\right]\Theta(t).
\end{align}
Up to $t=0$ when the perturber amplitude stops growing, the response from all modes oscillates with the pattern speed $\Omega_\rmP$ and grows hand in hand with the perturber. Subsequently, as the perturbation attains a steady strength, the disk response undergoes temporary phase-mixing due to the oscillations of stars at different frequencies, giving rise to phase spirals. These transients, however, are quickly taken over by long term oscillations driven at the forcing frequency $\Omega_\rmP$. 

For a slowly growing spiral/bar, i.e., in the `adiabatic growth' limit ($\gamma \to 0$), the entire disk oscillates at the driving frequency, $\Omega_\rmP$, i.e.,
\begin{align}
\calP_{n\ell m}^{(j)}(\bI,t) &\xrightarrow{\gamma_j\to 0} \exp{\left[-im\Omega_\rmP t\right]} \left[\pi\delta(\Omega_{\rm res})-\frac{i}{\Omega_{\rm res}}\right].
\end{align}
This has two major implications. First of all, since all stars, both resonant and non-resonant, are driven at the pattern speed of the perturbing spiral/bar, transient phase-mixing does not occur and thus no phase spiral arises. Secondly, the response is dominated by the resonances, $\Omega_{\rm res}=0$. In fact the resonant response diverges, reflecting the failure of (standard) linear perturbation theory near resonances. The adiabatic invariance of actions is partially broken near these resonances, causing the stars to get trapped in librating near-resonant orbits. A proper treatment of the near-resonant response can be performed by working with `slow' and `fast' action-angle variables \citep[][]{Tremaine.Weinberg.84, Lichtenberg.Lieberman.92, Chiba.Schonrich.22, Banik.vdBosch.22, Hamilton.etal.22}, which are uniquely defined for each resonance as linear combinations of the original action-angle variables. The fast actions remain nearly invariant while the fast angles oscillate with periods comparable to the unperturbed orbital periods of stars. The slow action-angle variables, on the other hand, undergo large amplitude oscillations about their resonance values over a libration timescale that is typically much longer than the orbital periods. For example, at co-rotation resonance ($n=\ell=0$), angular momentum behaves as the slow action while the radial and vertical actions behave as the fast ones.

Based on the above discussion, we infer that phase spirals can only be excited in the galactic disk by transient spiral/bar perturbations whose amplitude changes over a timescale comparable to the vertical oscillation periods of stars. Persistent spirals or bars rotating with a fixed pattern speed cannot give rise to phase spirals. Rather they trigger stellar oscillations at the pattern speed itself, which manifests in phase-space as a steadily rotating dipole or quadrupole depending on whether the $n=1$ or $2$ mode dominates the response. Thus, a phase spiral is necessarily always triggered by a transient perturbation.

\subsubsection{Impact of collisions on the disk response}\label{sec:spiral_c}

\begin{figure*}
  \centering
  \includegraphics[width=0.9\textwidth]{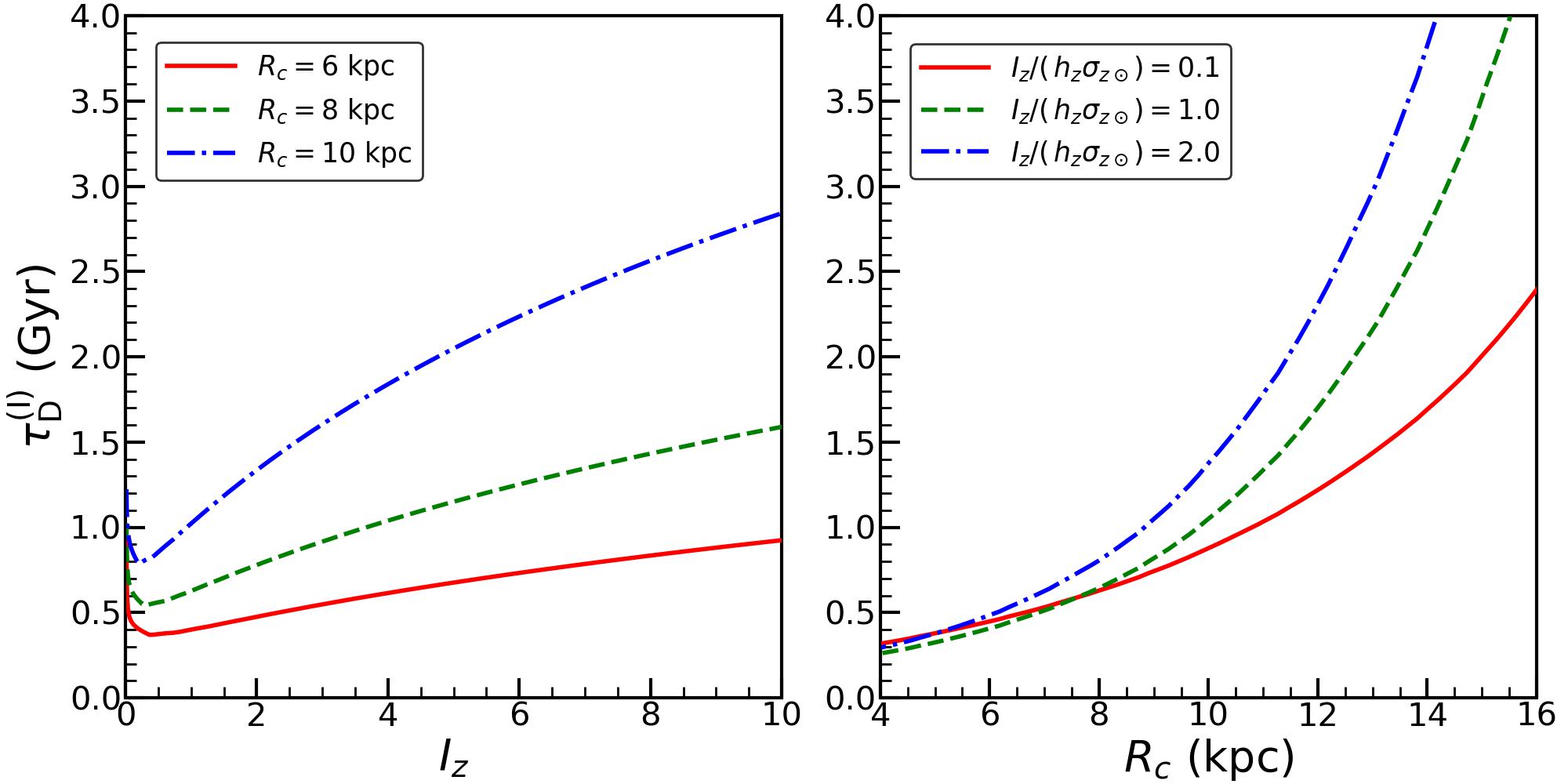}
  \caption{Timescale at which the disk response damps away due to collisional diffusion, i.e., small-scale scatterings of stars with structures like GMCs, is plotted as a function of $I_z$ ($R_c$) for three different values of $R_c$ ($I_z$) as indicated, in the left (right) panel. Typically, collisional diffusion occurs faster for smaller $I_z$ and smaller $R_c$.}
  \label{fig:diffusion_timescale}
\end{figure*}

In the above section we discussed the characteristics of the disk response in the absence of collisions. However, in a real galaxy like the MW disk, small-scale collisionality can potentially damp away any coherent response to a perturbation. Collisional diffusion arises not from star-star collisions, which is typically negligible, but from gravitational scattering with other objects, such as GMCs, DM substructure, etc. As discussed in Section~\ref{sec:galdisk}, the impact of collisional diffusion is mainly captured by the diffusion coefficients $D^{(z)}_I$ and $D^{(R)}_I$. Following \cite{Tremaine.etal.22} we assume that the disk stars have gained their mean vertical and radial actions over the age of the disk, $T_{\rm disk}=10\Gyr$, due to collisional heating, which implies that $D_a=\left<I_a\right>/T_{\rm disk}$ where $a$ is either $z$ or $R$ and $\left<I_a\right>=\int \rmd I_a \, I_a \, f_0 \, /\int \rmd I_a \, f_0$.

\begin{figure*}
\centering
\subfloat{\includegraphics[width=1\textwidth]{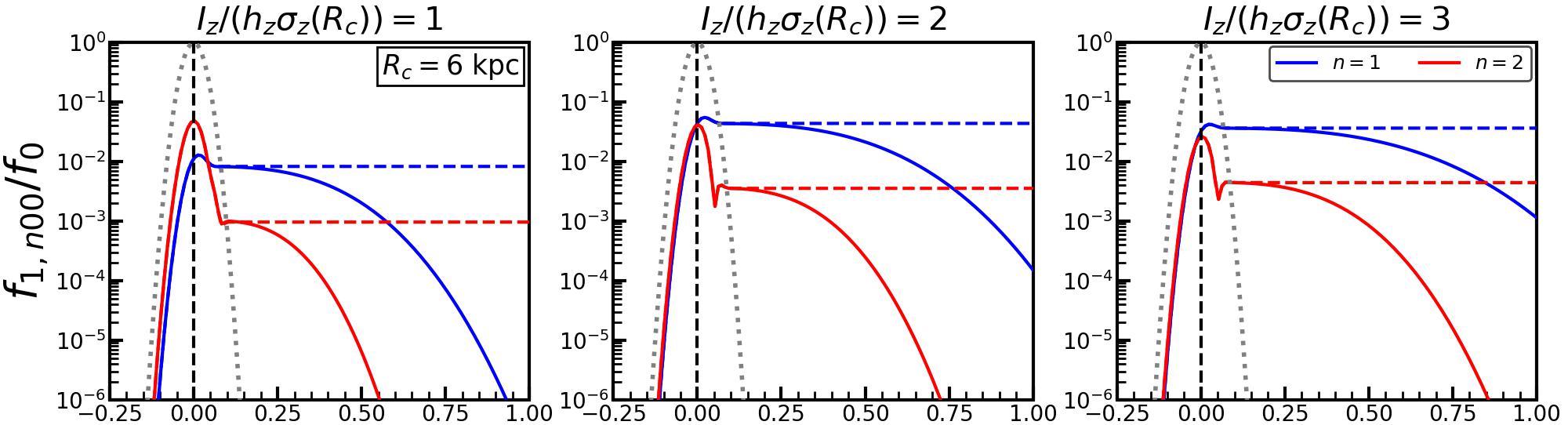}
  \label{fig:f1_vs_t_1}}\\
\subfloat{\includegraphics[width=1\textwidth]{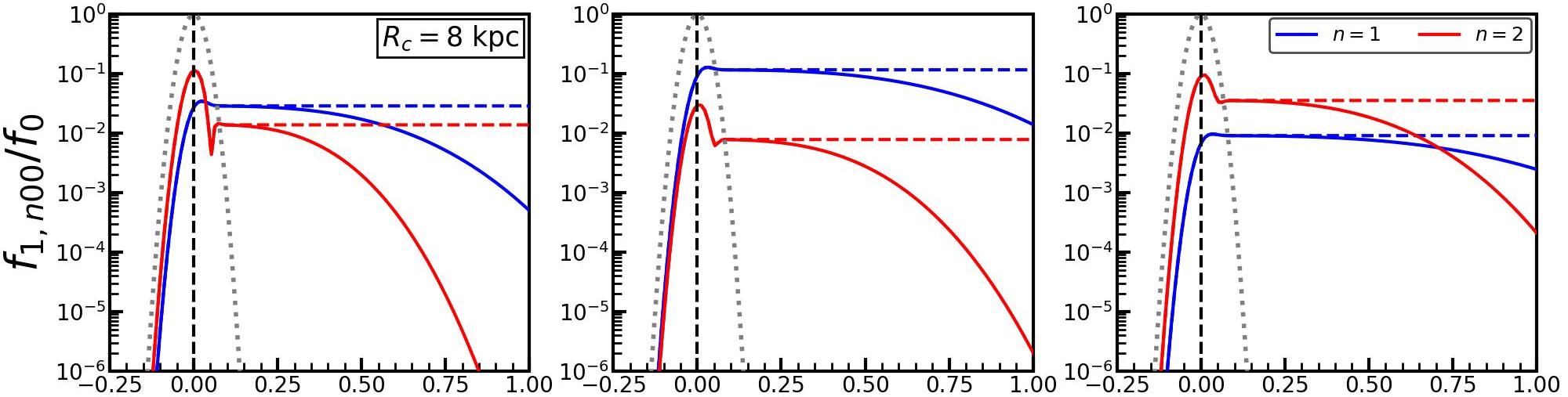}
  \label{fig:f1_vs_t_2}}\\
\subfloat{\includegraphics[width=1\textwidth]{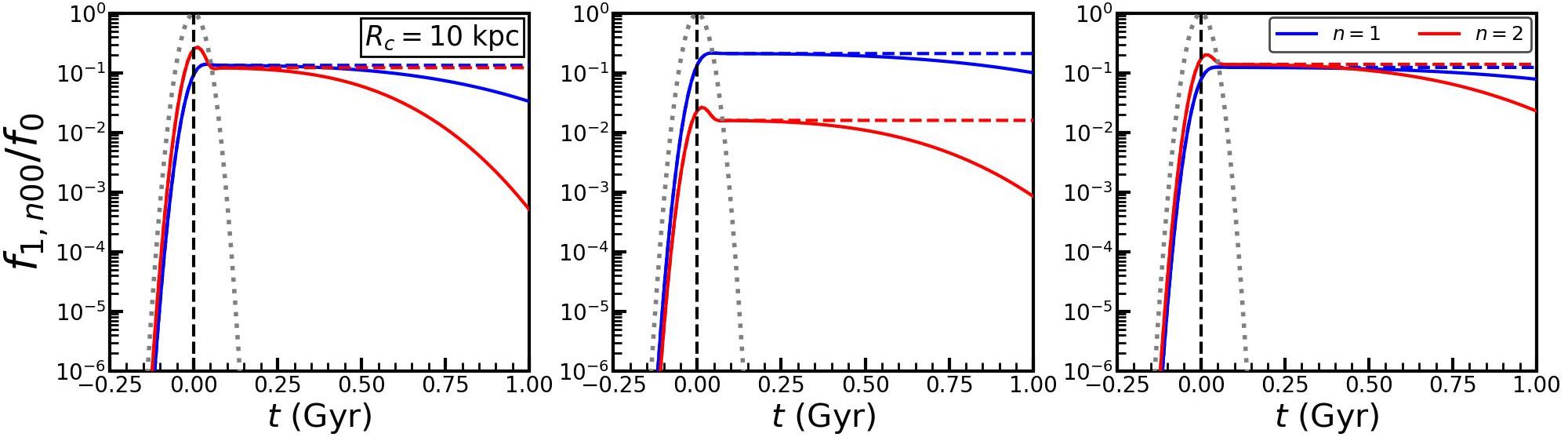}
  \label{fig:f1_vs_t_3}}
\caption{MW disk response to transient bars/2-armed spirals with Gaussian temporal modulation of pulse frequency, $\omega_\rmo=\omega_\rme=0.5\, \sigma_{z,\odot}/h_z$: the amplitude of the disk response, $f_{1n00}/f_0$, is plotted as a function of time. The rows and columns respectively denote different values of $R_c$ and $I_z$ as indicated. Blue and red lines indicate the $n=1$ and $2$ modes, while the solid and dashed lines respectively denote the cases with and without collisional diffusion (due to interactions of stars with structures like GMCs). The disk response initially rises and falls hand in hand with the perturbing pulse (indicated by the grey dotted line), before saturating to a steady state in the collisionless case and undergoing super-exponential damping in the collisional case. Note that collisional damping is faster for smaller $I_z$, smaller $R_c$ and larger $n$ modes.}
\label{fig:trans_spiral_damp}
\end{figure*}

For a transient bar/spiral with pulse frequency $\omega_j$, $\calP^{(j)}_{n\ell m}$ is given by equation~(\ref{Pnlm_spiral}). In the impulsive limit ($\omega_j \to \infty$), we have that $\calM_j(t-\tau) \to \omega_j \delta(t-\tau)$. Upon absorbing $\omega_j$ in the prefactor, the expression for $\calP^{(j)}_{n\ell m}$ then simplifies to
\begin{align}
\calP^{(j)}_{n\ell m} (\bI,t) &\approx \Theta(t)\, \exp{\left[-i\,(n\Omega_z+\ell\kappa+m\Omega_\phi)\,t\right]} \nonumber \\
&\times \exp{\left[-\left(\frac{n^2 D^{(z)}_I}{4 I_z}+\frac{\ell^2 D^{(R)}_I}{4 I_R}\right)t\right]} \, \exp{\left[-\frac{{\left(n\Omega_{z1}\right)}^2 D^{(z)}_I I_z}{3}t^3\right]}.
\end{align}
This demonstrates that, in the impulsive limit, the disk response instantaneously grows and spawns phase spirals whose amplitude decays due to collisional diffusion, manifest in the exponential damping terms. The first and second exponential factors, respectively, characterize the diffusion in vertical angle and action, which occur over the following timescales:
\begin{align}
\tau_\rmD^{(\rmw)} &= {\left[\frac{n^2 D^{(z)}_I}{4 I_z}+\frac{\ell^2 D^{(R)}_I}{4 I_R}\right]}^{-1}, \;\;\;\; \tau_\rmD^{(\rmI)} = {\left[\frac{3}{{\left(n\Omega_{z1}\right)}^2 D^{(z)}_I I_z}\right]}^{1/3}.
\label{diffusion_timescale}
\end{align}
Of these, the timescale for the diffusion in angles, $\tau_\rmD^{(\rmw)}$, typically exceeds that for the diffusion in actions, $\tau_\rmD^{(\rmI)}$, by at least an order of magnitude, implying that angle diffusion is negligible. Hence collisional diffusion mainly causes the abatement of action gradients in the phase-space structure of the response (arising from the action dependence of the frequencies, i.e., $\Omega_{z1}\neq 0$). The left (right) panel of Fig.~\ref{fig:diffusion_timescale} plots the diffusion timescale, $\tau_\rmD^{(\rmI)}$, as a function of $I_z$ ($\Rc$) for three different values of $\Rc$ ($I_z$) as indicated. Note that $\tau_\rmD^{(\rmI)}$ diverges in the small $I_z$ limit, attains a minimum around $I_z\sim 0.2-0.5\, h_z\sigma_{z,\odot}$, and increases as $I_z^\beta$ with $\beta<1$ at large $I_z$. As a function of $R_\rmc$, $\tau_\rmD^{(\rmI)}$ shows an approximately exponential rise. This owes to the fact that $\left<D^{(z)}_I\right> \sim h_z \, \sigma_z(R_\rmc) / T_{\rm disk} \sim \exp{\left[-R_\rmc/2h_R\right]} / T_{\rm disk}$ for the 3MN profile adopted for the MW disk. At $R_\rmc=\Rsun=8\kpc$, $\tau_\rmD^{(I)}\sim 0.6-0.7\Gyr$, in agreement with \cite{Tremaine.etal.22}. Hence, we see that collisional diffusion in action space is fairly efficient, and thus that phase spirals are short-lived features.

Fig.~\ref{fig:trans_spiral_damp} plots the amplitude of the disk response (for $I_R=0$) to a transient spiral of pulse frequency, $\omega_\rmo=\omega_\rme=0.5 \sigma_{z,\odot}/h_z$, computed using equations~(\ref{f1nk_gensol_f0}), (\ref{trans_spiral_Resp}) and (\ref{Pnlm_spiral}), as a function of time. Dashed and solid lines show the results with and without collisional diffusion, respectively. The rows correspond to different values of $\Rc$ while the columns denote different values of $I_z/(h_z\sigma_{z,\odot})$ as indicated. The blue and red lines denote the response for the $(n,\ell,m)=(1,0,0)$ and $(2,0,0)$ modes, respectively, and the dotted grey line represents the Gaussian pulse strength. The response for both bending and breathing modes initially grows hand in hand with the perturbing pulse. Following the point of maximum pulse strength, the response follows the decaying pulse strength before saturating to the steady state amplitude given in equation~(\ref{Pnlm_trans_spiral}) in the collisionless limit. In the presence of collisional diffusion, however, the response continues to damp out as $\sim \exp{[-(t/\tau_\rmD^{(\rmI)})^3]}$ after temporarily saturating at the collisionless steady state. Note that the collisional damping is faster for smaller $\Rc$ and smaller $I_z$. In addition, $n=2$ breathing modes damp out faster than the $n=1$ bending modes due to the $n^{-2/3}$ dependence of $\tau_\rmD^{(\rmI)}$.

To summarize, we have shown that phase spirals can be triggered by impulsive perturbations resulting from transient spiral arms or bars, but are subject to super-exponential damping due to collisional diffusion that is likely to be dominated by scattering against GMCs. This collisional damping is more efficient in the inner disk, for stars with smaller $I_z$, and for modes of larger $n$.

\section{Disk response to satellite encounter}
\label{sec:disk_resp_sat}

In addition to the spiral arm/bar perturbations considered above, we also consider disk perturbations triggered by encounters with a satellite galaxy. For the sake of brevity, we only compute the disk response in the collisionless limit. In the case of impulsive encounters, the impact of collisional diffusion is simply expressed by multiplying the collisionless response by the collisional damping factor $\exp[-(t/\tau_\rmD^{(\rmI)})^3]$, with $\tau_\rmD^{(\rmI)}$ given by equation~(\ref{diffusion_timescale}).

\begin{SCfigure}
  \centering
  \includegraphics[width=0.5\textwidth]{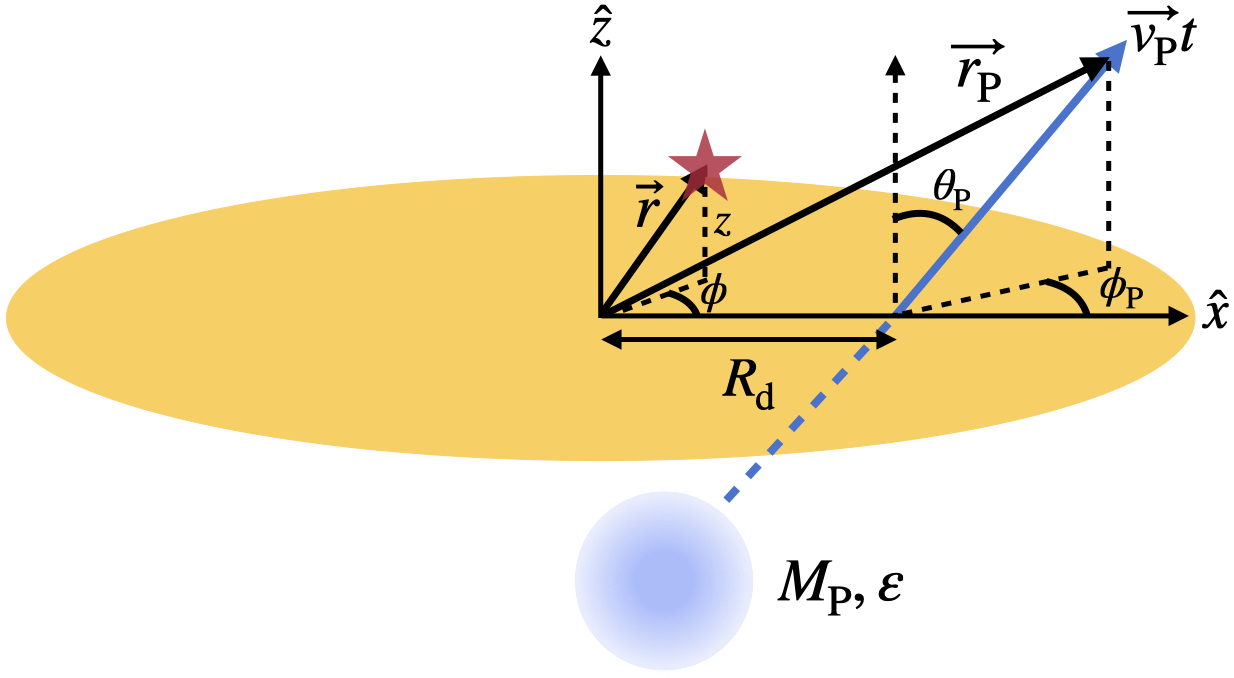}
  \caption{Illustration of the geometry of a satellite galaxy with mass $M_\rmP$ impacting a disk galaxy with uniform velocity $\vp$ along a straight line. The impact occurs at a galactocentric distance $\rd$. The orientation of $\bvp$ is specified by $\thetap$, the angle between $\bvp$ and the $z$-axis, and $\phip$, the angle between the projection of $\bvp$ on the mid-plane and the $x$-axis.}
  \label{fig:sat_enc_orient}
\end{SCfigure}

For simplicity, we assume that the satellite is moving with uniform velocity $\vp$ along a straight line, impacting the disk at a galactocentric distance $\rd$ with an arbitrary orientation, specified by the angles, $\thetap$ and $\phip$, which are respectively defined as the angles between $\bvp$ and the $z$-axis, and between the projection of $\bvp$ on the mid-plane and the $x$-axis (see Fig.~\ref{fig:sat_enc_orient}). Thus the position vector of the satellite with respect to the galactic center can be written as
\begin{align}
\brp = (\rd + \vp\sin{\thetap}\cos{\phip}\,t)\,\hat{\bx} + \vp\sin{\thetap}\sin{\phip}\,t\,\hat{\by} + \vp\cos{\thetap}\,t\,\hat{\bz}\,,
\label{rp_sat}
\end{align}
while that of a star is given by
\begin{align}
\br = R(\cos{\phi}\,\hat{\bx} + \sin{\phi}\,\hat{\by}) + z\,\hat{\bz}\,.
\label{r_sat}
\end{align}
We consider the satellite to be a Plummer sphere of mass $M_\rmP$ and size $\varepsilon$, such that its gravitational potential at location $\br$ is given by
\begin{align}
\Phi_\rmP &= G M_\rmP \left[-\frac{1}{\sqrt{{\left|\br-\brp\right|}^2+\varepsilon^2}} + \frac{\br \cdot \brp}{{\left(\brp^2+\varepsilon^2\right)}^{3/2}}\right]\,.
\label{sat_pot}
\end{align}
Here the first term is the `direct' term and the second is the `indirect' term that accounts for the reflex motion of the disk and the fact that the disk center is accelerated by the satellite and is thus non-inertial. Typically, the first one dominates over the second.

In order to compute the disk response to this external perturbation, we need to compute its Fourier coefficients, which is challenging. Rather, we first evaluate the $\tau$-integral in Equation~(\ref{f1nk_gensol_f0}), setting $t_\rmi\to -\infty$, and then compute the Fourier transform of the result, as worked out in Appendix~\ref{App:sat_disk_Resp}. For $I_R\approx 0$ (this is justified since we adopt the radial epicyclic approximation in this chapter), this yields a modal response, $f_{1,n\ell m}$ (Equation~[\ref{f1nk_gensol_f0}]), with $\calI_{n\ell m}(\bI,t)$ given by

\begin{align}
&\calI_{n\ell m}(\bI,t)\approx-\frac{2G M_\rmP}{\vp} \exp{\left[-i\Omega t\right]} \times \exp{\left[-i\frac{\Omega\sin{\thetap}\cos{\phip}}{\vp} \rd\right]} \nonumber \\ 
&\times \frac{1}{{\left(2\pi\right)}^2} \int_0^{2\pi} \rmd w_z \exp{\left[-in w_z\right]} \exp{\left[i\frac{\Omega \cos{\thetap}}{\vp} z\right]} \nonumber \\
&\times \int_0^{2\pi} \rmd \phi\,\exp{\left[-im\phi\right]} \exp{\left[i\frac{\Omega \sin{\thetap} \cos{\left(\phi-\phip\right)}}{\vp} \Rc\right]} \nonumber \\ 
&\times K_{0i}\left(\frac{\Omega\sqrt{\calR^2_\rmc+\varepsilon^2}}{\vp},\frac{\vp t-\calS_\rmc}{\sqrt{\calR^2_\rmc+\varepsilon^2}}\right),
\label{sat_gen}
\end{align}
where $\Omega$ is given by
\begin{align}
\Omega = n\Omega_z + \ell\kappa + m\Omega_\phi.
\end{align}
Here $\calR_\rmc=\calR(\Rc)$ and $\calS_\rmc=\calS(\Rc)$ with $\calR$ and $\calS$ given by equation~(\ref{sat_RS_app}). $K_{0i}$ is given by equation~(\ref{K0i_app}), which asymptotes to the modified Bessel function of the second kind, $K_0\left(\left|\Omega\right|\sqrt{\calR^2_\rmc+\varepsilon^2}/\vp\right)$, in the large time limit. A more precise expression for $\calI_{n\ell m}$ that is valid for higher values of $I_R$ is given by equation~(\ref{sat_gen_IR_app}) of Appendix~\ref{App:sat_disk_Resp}.

The expression for $\calI_{n\ell m}$ given in equation~(\ref{sat_gen}) exhibits several key features of the disk response to satellite encounters. The $\exp{\left[-i\Omega t\right]}$ factor encodes the phase-mixing of the response due to oscillations at different frequencies, giving rise to phase spirals. The $\exp{\left[i\left(\Omega \cos{\thetap}/\vp\right)z\right]}$ and $\exp{\left[i\left(\Omega \sin{\thetap} \cos{(\phi-\phip)}/\vp\right) \Rc\right]}$ factors respectively indicate that the satellite induces wave-like perturbations in the disk with two characteristic wave-numbers: the vertical wave-number, $k_z\approx \Omega\cos{\thetap}/\vp$ and the horizontal wave-number, $k_R\approx \Omega\sin{\thetap}/\vp$. Therefore, the disk response will be vertically (horizontally) stratified in case of a perpendicular (planar) impact of the satellite. As shown in Appendix~\ref{App:special}, the response to a satellite having a face-on, perpendicular, impulsive encounter through the center of the disk can be obtained from an asymptotic analysis of the general response to satellite encounters, given by equations~(\ref{f1nk_gensol_f0}) and (\ref{sat_gen}).

\subsection{Asymptotic behaviour of the response}

It is instructive to study the two extreme cases of encounter speed, the impulsive limit (large $\vp$) and the adiabatic limit (small $\vp$). Using the asymptotic form of the $K_0$ Bessel function that appears in equation~(\ref{sat_gen}), we obtain the following approximate asymptotic behaviour of $f_{1,n\ell m}$ at large time:
\begin{align}
f_{1,n\ell m} &\sim \frac{G M_\rmP}{\vp} \exp{\left[-i\Omega t\right]} \times
\begin{cases} 1, & \vp\to \infty \\ \\
\sqrt{\vp/\Omega b}\, \exp{\left[-\Omega b/\vp\right]}, & \vp \to 0,
\end{cases}
\label{sat_asymptote}
\end{align}
where $b$ is the impact parameter of the encounter, defined as the perpendicular distance of the nearest star on the mid-plane from the satellite's (straight) orbit, and expressed as
\begin{align}
b = \left|\rd - \Rc \right| \, \sqrt{1-\sin^2{\thetap}\cos^2{\phip}}.
\label{impact_parameter}
\end{align}
It is clear from these limits that the disk response is most pronounced for intermediate velocities, $\vp\sim\Omega b$. For impulsive encounters, the response is suppressed as a power law in $\vp$, whereas in the adiabatic limit the response is exponentially suppressed, except at resonances, $\Omega=n\Omega_z+\ell\kappa+m\Omega_\phi=0$. In this limit, far from the resonances, the perturbation timescale, $b/\vp$, is much larger than $\Omega^{-1}$, and the net response is washed away due to many oscillations during the perturbation (i.e., the actions are adiabatically invariant), a phenomenon known as adiabatic shielding \citep[][]{Weinberg.94a,Weinberg.94b,Gnedin.Ostriker.99}.

\renewcommand{\arraystretch}{1.4}
\begin{sidewaystable}
\centering
\tabcolsep=0.2 cm
\scalebox{0.88}{\begin{tabular}{c|c|cc|cc|cc}
 \hline
MW satellite & Mass & $f_{1,n=1}/f_0$ & $t_{\rm cross}$ & $f_{1,n=1}/f_0$ & $t_{\rm cross}$ & $f_{1,n=1}/f_0$ & $t_{\rm cross}$ \\
name & $(\Msun)$ & & $(\Gyr)$ & & $(\Gyr)$ & & $(\Gyr)$ \\
 & & Penultimate & Penultimate & Last & Last & Next & Next \\
 (1) & (2) & (3) & (4) & (5) & (6) & (7) & (8) \\
 \hline
 Sagittarius & $10^9$ & $2.7\times 10^{-1}$ & $-1.01$ & $4.9\times 10^{-8}$ & $-0.35$ & $1.3\times 10^{-1}$ & $0.03$ \\
 Hercules & $7.1\times 10^6$ & $8.4\times 10^{-8}$ & $-3.78$ & $2.4\times 10^{-3}$ & $-0.5$ & $2.4\times 10^{-3}$ & $3.18$ \\
 Leo II & $8.2\times 10^6$ & -- & $-3.86$ & $1.6\times 10^{-3}$ & $-1.78$ & $3.2\times 10^{-3}$ & $2.31$ \\
 Segue 2 & $5.5\times 10^5$ & $6.2\times 10^{-4}$ & $-0.84$ & $8.5\times 10^{-4}$ & $-0.25$ & $6.3\times 10^{-5}$ & $0.28$ \\
 LMC & $1.4\times 10^{11}$ & $5.1\times 10^{-2}$ & $-7.63$ & -- & $-2.67$ & $2.3\times 10^{-2}$ & $0.11$ \\
 SMC & $6.5\times 10^9$ & $2.8\times 10^{-5}$ & $-3.32$ & -- & $-1.44$ & $8.7\times 10^{-6}$ & $0.22$ \\
 Draco I & $2.2\times 10^7$ & -- & $-2.46$ & $9.9\times 10^{-5}$ & $-1.24$ & $7.1\times 10^{-6}$ & $0.24$ \\
 Bootes I & $10^7$ & $1.8\times 10^{-7}$ & $-1.67$ & $3.7\times 10^{-5}$ & $-0.35$ & -- & $0.88$ \\
 Willman I & $4\times 10^5$ & $1.6\times 10^{-8}$ & $-0.66$ & $1.2\times 10^{-6}$ & $-0.21$ & $9.3\times 10^{-6}$ & $0.41$ \\
 Ursa Minor & $2\times 10^7$ & -- & $-2.28$ & $1.7\times 10^{-5}$ & $-1.17$ & $2.6\times 10^{-6}$ & $0.29$ \\
 Ursa Major II & $4.9\times 10^6$ & $5.8\times 10^{-6}$ & $-2.12$ & $2.5\times 10^{-6}$ & $-0.09$ & -- & $0.97$ \\
 Coma Berenices I & $1.2\times 10^6$ & $9.2\times 10^{-7}$ & $-2.58$ & $3.7\times 10^{-8}$ & $-0.25$ & -- & $0.71$ \\
 Sculptor & $3.1\times 10^7$ & -- & $-2.74$ & $3.4\times 10^{-8}$ & $-0.46$ & -- & $1.48$\\
 \hline
\end{tabular}}
\caption{Steady state response of the MW disk to encounters with satellites in the collisionless limit, for the $(n,\ell,m)=(1,0,0)$ and $(2,0,0)$ modes and for stars with $I_z = h_z\sigma_{z,\odot}$ in the Solar neighborhood. We have marginalized the response over $I_R$. Columns (1) and (2) list the name and dynamical mass of each satellite. The latter is taken from the literature \citep[][]{Simon.Geha.07,Bekki.Stanimirovic.09,Lokas.09,Erkal.etal.19,Vasiliev.Belokurov.20}, except for Sagittarius for which we adopt a mass of $10^9\Msun$. Note that there is a discrepancy between its estimated mass of $\sim 4\times 10^8\Msun$ \citep[][]{Vasiliev.Belokurov.20} and the mass required ($10^9-10^{10}\Msun$) to produce detectable phase spiral signatures in N-body simulations \citep[see for example][]{Bennett.etal.22}. Columns (3) and (4) respectively denote the bending mode response assuming our fiducial MW parameters and the penultimate disk-crossing time. Columns (5) and (6) indicate the same for the last disk-crossing, while columns (7) and (8) show it for the next one. Only satellites that induce a bending mode response, $f_{1,n=1}/f_0\geq 10^{-8}$, in at least one of the three cases are shown. Any response weaker than $10^{-8}$ is considered negligible and is indicated with a horizontal dash.}
\label{tab:MW_sat_resp_3}
\end{sidewaystable}

\subsection{Response of the MW disk to satellites}

The MW halo harbors several fairly massive satellite galaxies that repeatedly perturb the MW disk. Here we use existing data on the phase-space coordinates of those MW satellites to compute the disk response of satellite encounters that occurred in the past few hundred Myr, which are those for which we may expect phase spirals that were triggered to have survived to the present day.

\begin{figure*}
  \centering
  \includegraphics[width=1\textwidth]{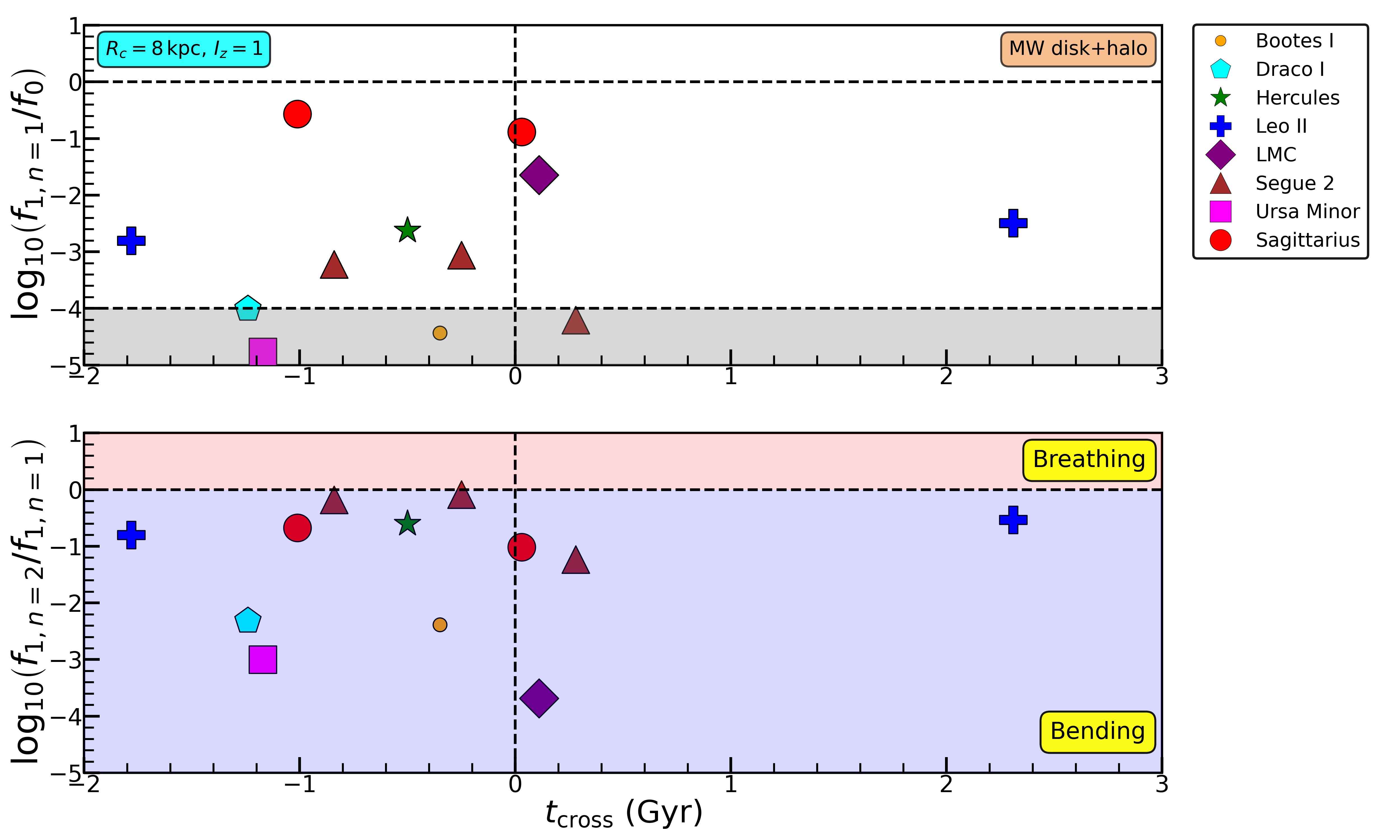}
  \caption{Steady state MW disk response to satellite encounter in the collisionless limit: bending mode strength, $f_{1,n=1}/f_0$ (upper panel), and the corresponding breathing vs bending ratio, $f_{1,n=2}/f_{1,n=1}$ (lower panel) for the $(\ell,m)=(0,0)$ modes, in the Solar neighborhood for the MW satellites, as a function of the disk crossing time, $t_{\rm cross}$, in Gyr, where $t_{\rm cross}=0$ marks today. The previous two and the next impacts are shown. Here we consider $I_z=h_z \sigma_{z,\odot}$, with fiducial MW parameters, and marginalize over $I_R$. The effect of the (non-responsive) ambient DM halo on the stellar frequencies is taken into account. The estimates of $t_{\rm cross}$ are very sensitive to the detailed potential of the MW system, while the response estimates are fairly robust (see text for details). In the upper panel, the region with bending mode response, $f_{1,n=1}/f_0<10^{-4}$, has been grey-scaled, indicating that the response from the satellites in this region is far too weak and adiabatic to be detected by Gaia. Note that the response is dominated by that due to Sgr, followed by Hercules, Leo II, Segue 2 and the Large Magellanic Cloud (LMC). Also note that the previous two and next impacts of all the satellites excite bending modes in the Solar neighborhood.}
  \label{fig:MW_sat_Resp}
\end{figure*}

To compute the disk response to the MW satellites, we proceed as follows. As in chapter~\ref{chapter: paper2}, we adopt the galactocentric coordinates and velocities computed and documented by \cite{Riley.etal.19} \citep[table A.2, see also][]{Li.etal.20} and \cite{Vasiliev.Belokurov.20} as initial conditions for the MW satellites. We then simulate their orbits in the combined gravitational potential of the MW halo, disk plus bulge (modelled as a spherical \cite{Hernquist.90} profile with mass $M_b=6.5\times 10^9\Msun$ and scale radius $r_b=0.6$ kpc) using a second order leap-frog integrator. For each individual orbit, we record the times, $t_{\rm cross}$, and the galactocentric radii, $\rd$, corresponding to disk crossings. We also register the corresponding impact velocities, $\vp=\sqrt{v^2_z+v^2_R+v^2_\phi}$, and the angles of impact, $\thetap=\cos^{-1}{(v_z/\vp)}$ and $\phip=\tan^{-1}{(v_\phi/v_R)}$. We substitute these quantities in equation~(\ref{sat_gen_IR_app}) and compute the disk response (integrated over $I_R$) following the satellite encounter, using equations~(\ref{f1nk_gensol_f0}) and (\ref{sat_gen}). Results are summarized in Table~\ref{tab:MW_sat_resp_3}. Fig.~\ref{fig:MW_sat_Resp} plots the amplitude of the Solar neighborhood (for which $\Rc(L_z)=\Rsun=8\kpc$) bending mode response, $f_{1,n=1}/f_0$ (top panel), and breathing-to-bending ratio, $f_{1,n=2}/f_{1,n=1}$ (bottom panel), as a function of $t_{\rm cross}$. Here we only show the responses for $(\ell,m)=(0,0)$ modes, and consider stars with $I_z = I_{z,\odot} = 9.2\kpc\kms$.

It is noteworthy that the responses in the realistic MW disk computed here are $\sim 1-2$ orders of magnitude larger than those evaluated for the isothermal slab model shown in Fig.~7 of chapter~\ref{chapter: paper2}. This owes to the reduced damping of the phase spiral amplitude due to lateral mixing, which is more pronounced in the isothermal slab with unconstrained lateral velocities than in the realistic disk with constrained, ordered motion. From the lower panel of Fig.~\ref{fig:MW_sat_Resp} it is evident that, as in the isothermal slab case, almost all satellites trigger a bending mode response in the Solar neighborhood, resulting in a one-armed phase spiral in qualitative agreement with the Gaia snail. However, as is evident from the upper panel, only five of the satellites trigger a detectable response in the disk, with $f_{1,n=1}/f_0 > \delta_{\rm min} \equiv 10^{-4}$ (see Appendix~C of chapter~\ref{chapter: paper2_app} for a derivation of this approximate detectability criterion for Gaia). The response to encounters with the other satellites is weak either because they have too low mass or because the encounter with respect to the Sun is too slow and adiabatically suppressed. Sgr excites the strongest response by far; its bending mode response, $f_{1,n=1}/f_0$, is at least $1-2$ orders of magnitude above that for any other satellite. Its penultimate disk crossing, about the same time as its last pericentric passage $\sim 1\Gyr$ ago, triggered a strong response of $f_{1,n=1}/f_0 \sim 0.3$ in the Solar neighborhood. For comparison, the response from its last disk crossing, which nearly coincides with its last apocentric passage about $350\Myr$ ago, triggered a very weak, adiabatically suppressed response ($\sim 5\times 10^{-8}$) that falls below the lower limit of Fig.~\ref{fig:MW_sat_Resp}. Its next disk crossing in about $30\Myr$ is estimated to trigger a strong response with $f_{1,n=1}/f_0\sim 0.1$. Besides Sgr, the satellites that excite a detectable response, $f_{1,n=1}/f_0 > \delta_{\rm min}$ are Hercules, Segue 2, Leo II and the LMC. The imminent crossing of LMC is estimated to trigger $f_{1,n=1}/f_0 \sim 2\times10^{-2}$, which is an order of magnitude below Sgr. Only for $I_z/I_{z,\odot} \gtrsim 4.5$ ($z_{\rm max}\gtrsim 3.4 h_z$), the LMC response dominates over Sgr. This exercise therefore suggests that Sgr is the leading contender, among the MW satellites considered here, for triggering the Gaia snail in the Solar neighborhood, in agreement with several previous studies \citep[][]{Antoja.etal.18,Binney.Schonrich.18, Laporte.etal.18, Laporte.etal.19, Darling.Widrow.19b, Bland-Hawthorn.etal.19, Hunt.etal.21, Bland-Hawthorn.Garcia.21, Bennett.etal.22}.

\begin{figure}[H]
\centering
\subfloat{\includegraphics[width=0.48\textwidth]{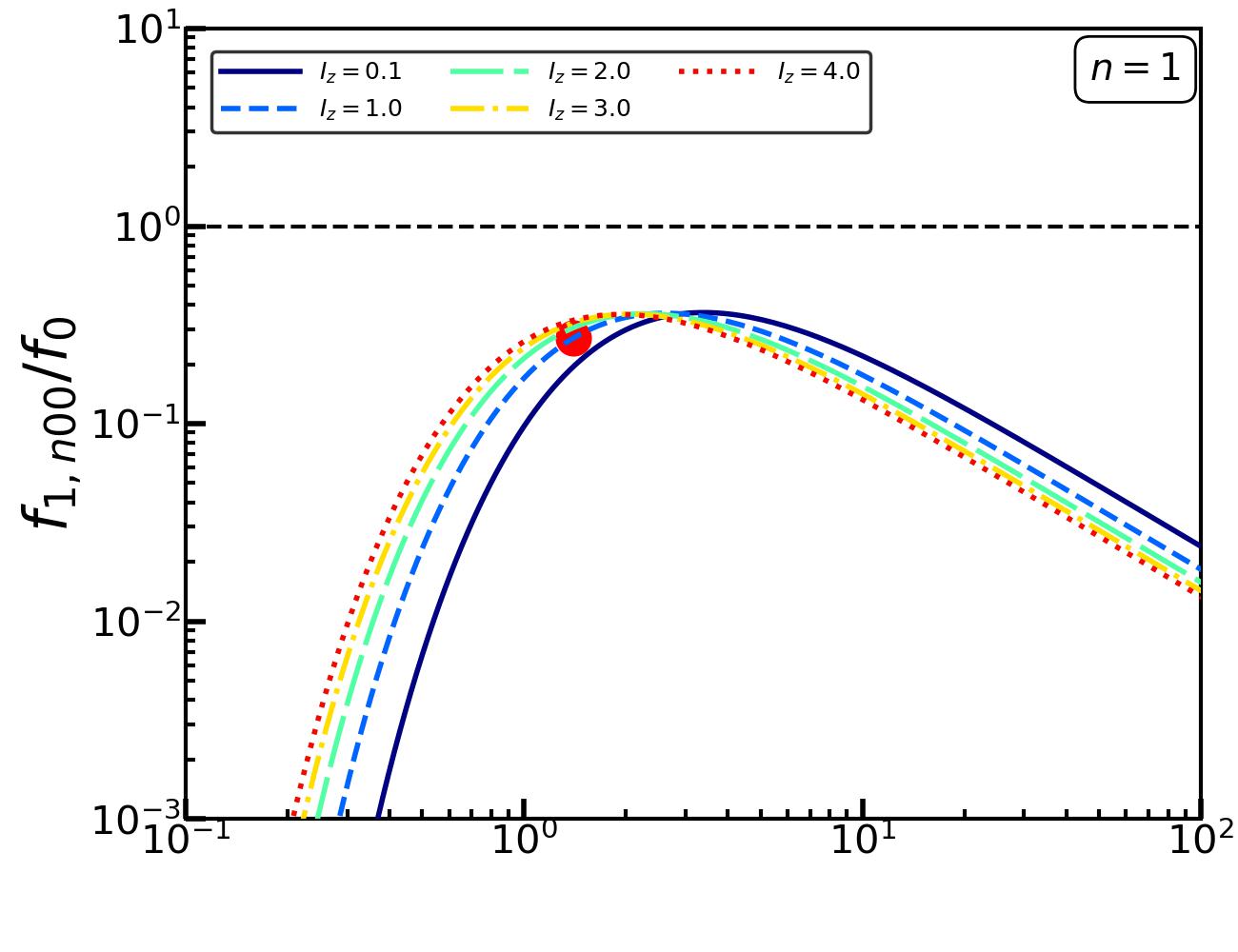}
    \label{disk_Resp_n1_0.5pi_Iz}}
\subfloat{\includegraphics[width=0.48\textwidth]{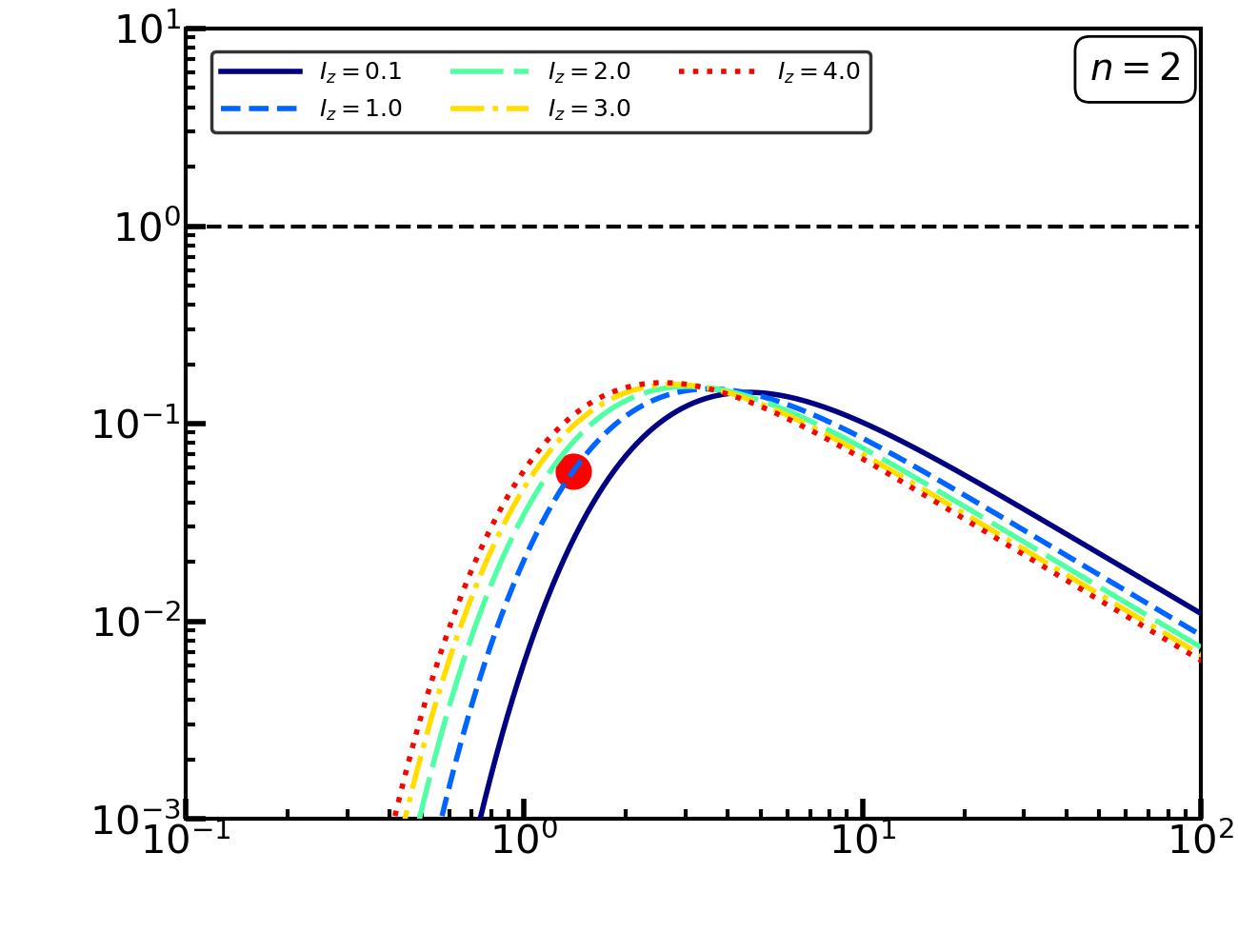}
    \label{disk_Resp_n2_0.5pi_Iz}}\\
\subfloat{\includegraphics[width=0.48\textwidth]{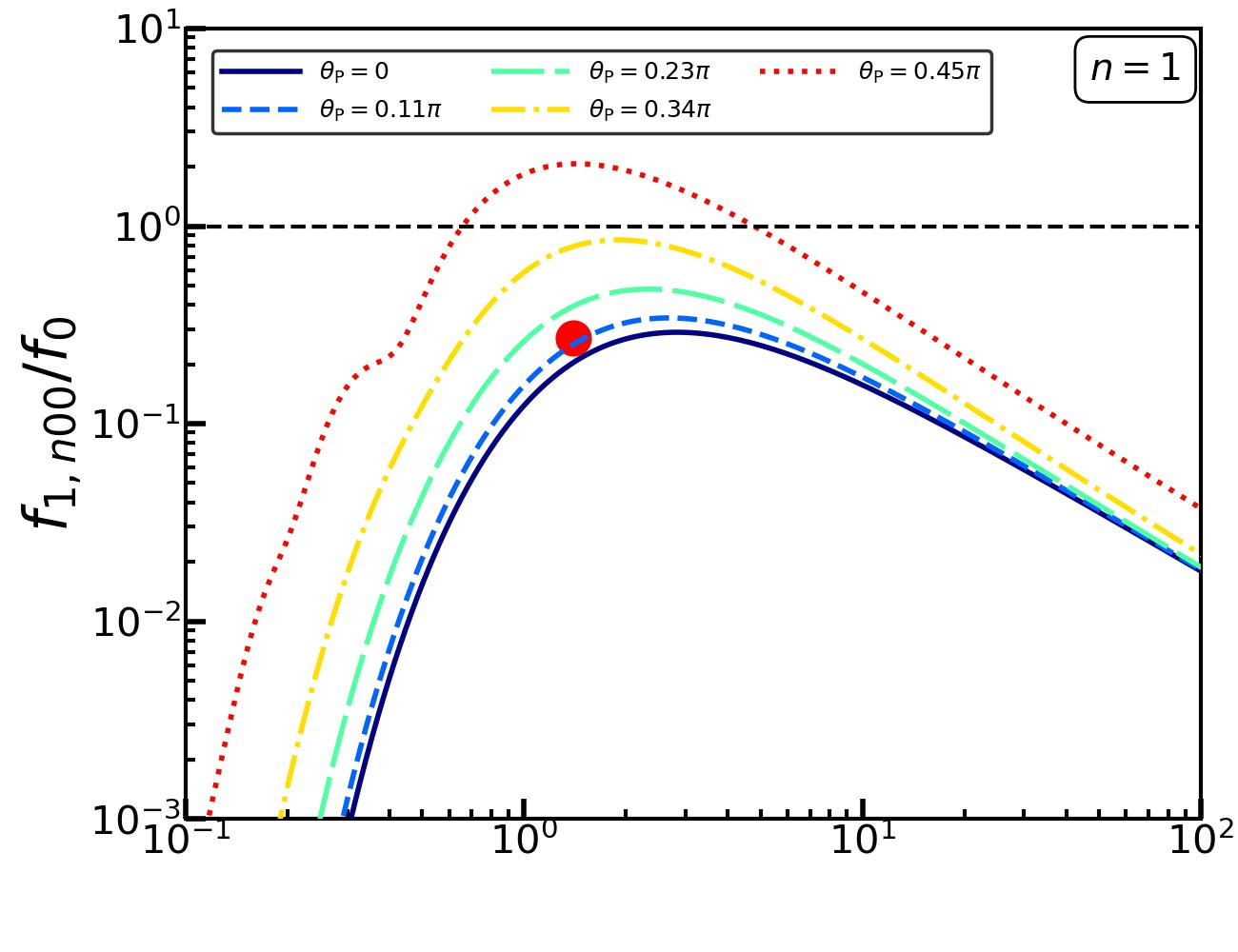}
    \label{disk_Resp_n1_0.5pi_thetap}}
\subfloat{\includegraphics[width=0.48\textwidth]{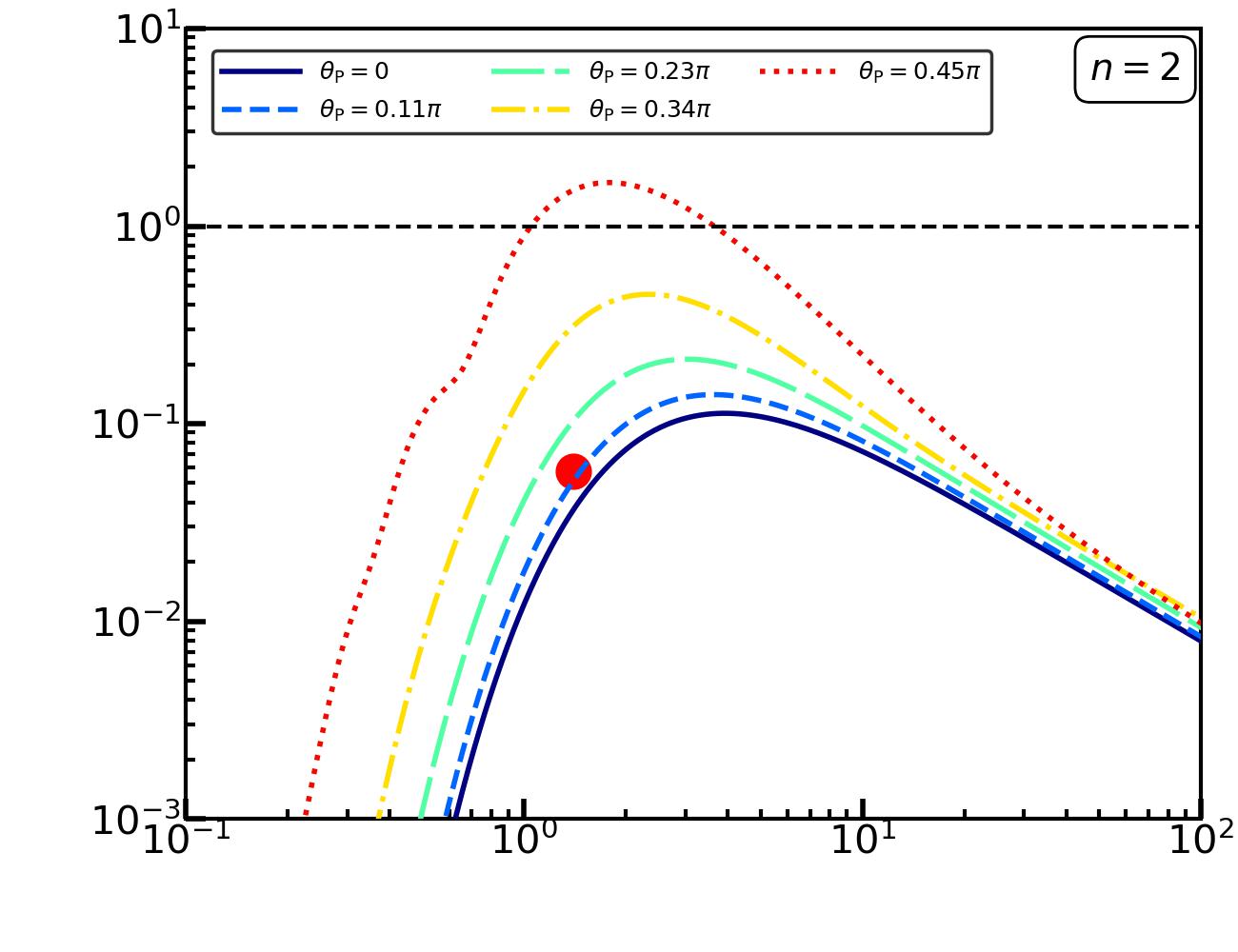}
    \label{disk_Resp_n2_0.5pi_thetap}}\\
\subfloat{\includegraphics[width=0.48\textwidth]{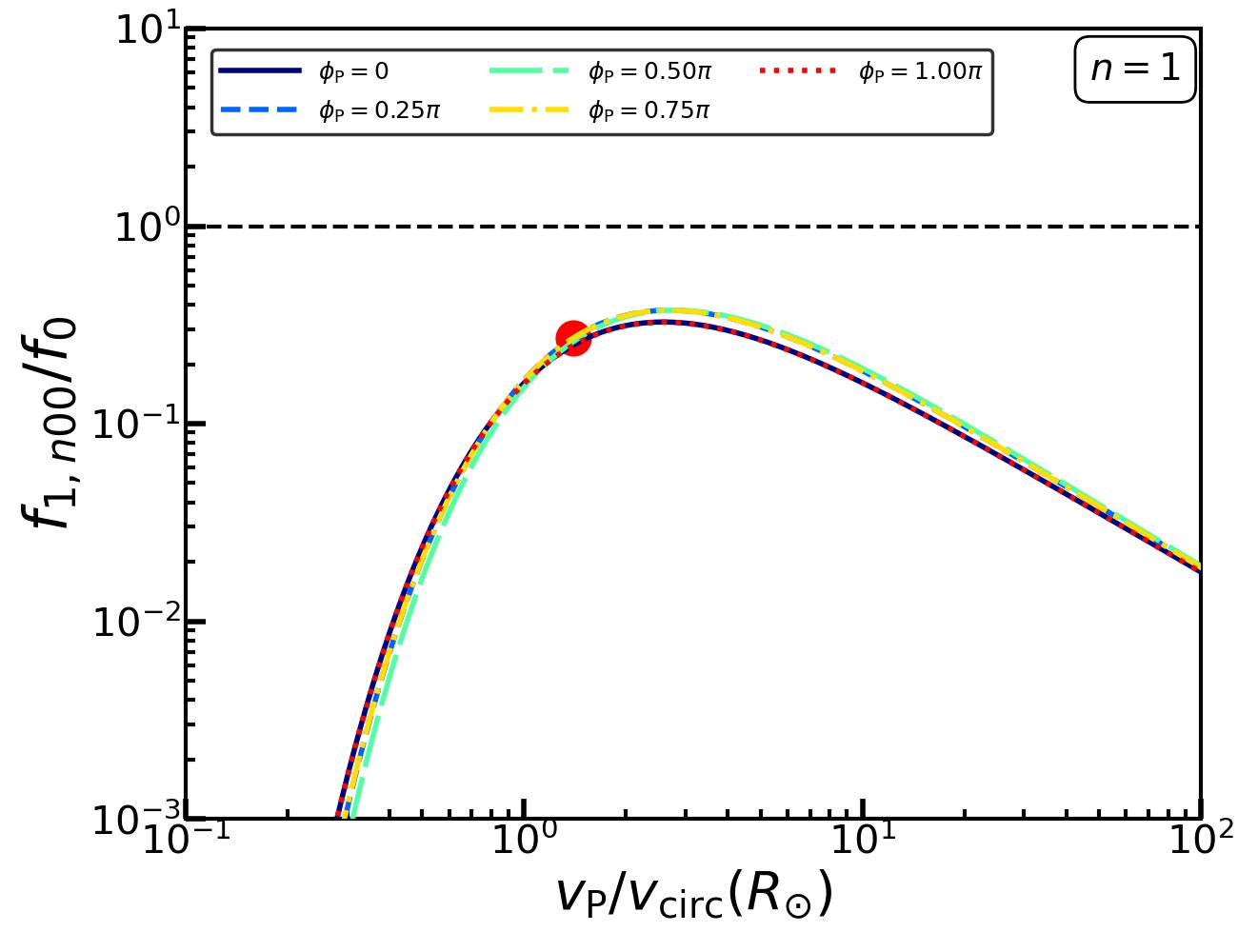}
    \label{disk_Resp_n1_0.25pi_phip}}
\subfloat{\includegraphics[width=0.48\textwidth]{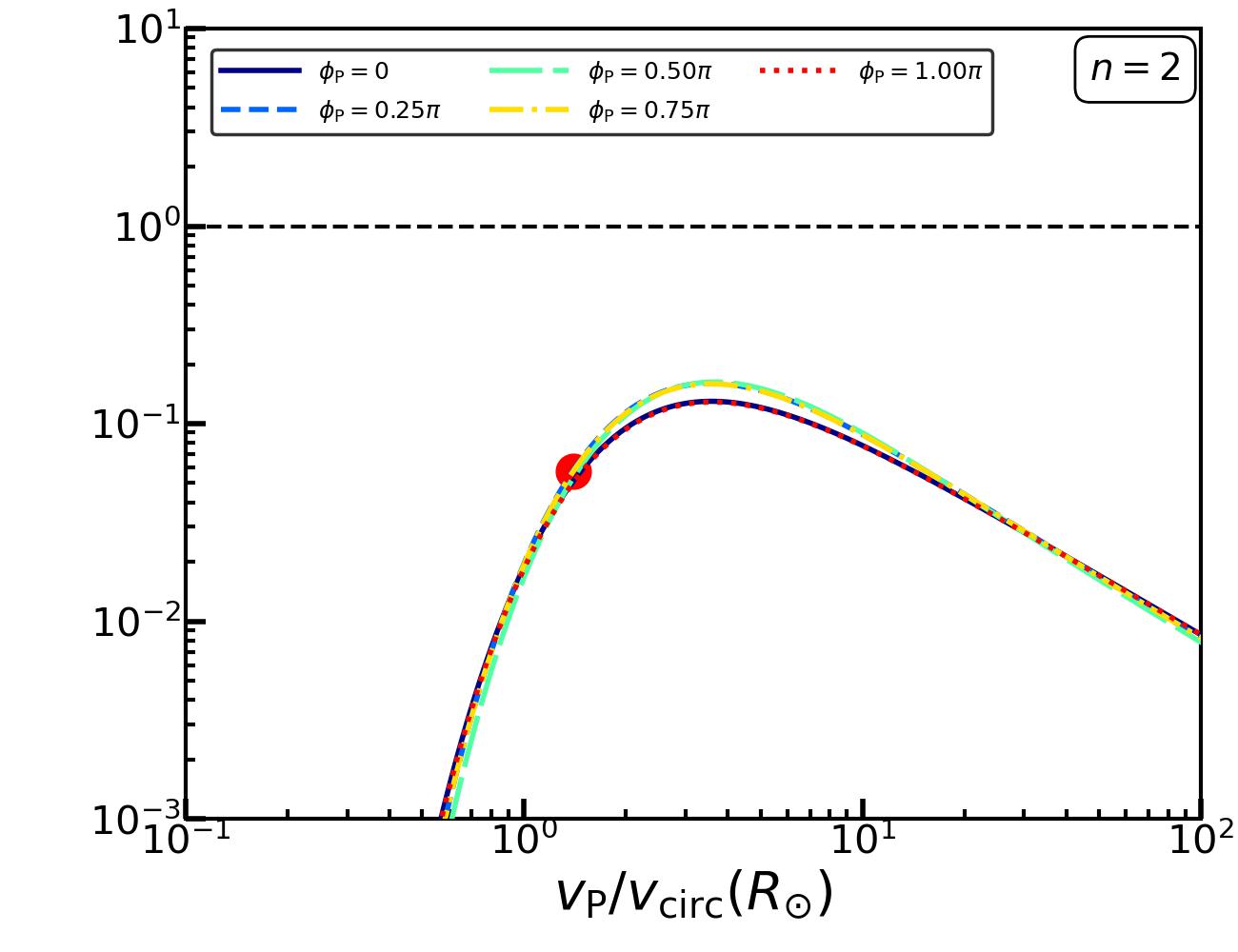}
    \label{disk_Resp_n2_0.25pi_phip}}
\caption{Steady state MW disk response to satellite encounter in the collisionless limit: each panel shows the behaviour of the disk response amplitude, $f_{1,n00}/f_0$ (evaluated using equations~[\ref{f1nk_gensol_f0}] and [\ref{sat_gen_IR_app}]) and marginalized over $I_R$), as a function of the impact velocity, $\vp$, in the Solar neighborhood, i.e., $\Rc = \Rsun = 8\kpc$, in presence of an ambient DM halo. The left and right columns respectively indicate the response for the $n=1$ bending and $n=2$ breathing modes. The top, middle and bottom rows show the same for different values of $I_z$ (in units of $I_{z,\odot}$), $\thetap$ and $\phip$ respectively as indicated, with the fiducial parameters corresponding to $I_{z,\odot}$ and the parameters for Sgr impact, the response amplitude for which is indicated by the red circle. Note that the response is suppressed as $\vp^{-1}$ in the impulsive (large $\vp$) limit but exponentially suppressed in the adiabatic (small $\vp$) regime, and peaks at an intermediate velocity, $\vp\sim 2-3\, v_{\rm circ}(\Rsun)$ (which is very similar to the encounter speed of Sgr). The peak of the response shifts to smaller $\vp$ for larger $I_z$, since $\Omega_z$ decreases with $I_z$. The response depends only very weakly on $\phip$ but is quite sensitive to $\thetap$; more planar encounters, i.e., increasing $\thetap$ triggers stronger responses.}
\label{fig:disk_Resp_vp}
\end{figure}

We caution that the above estimates of the disk response are obtained under the assumption of nearly straight line orbits of the satellites. This assumption is well-justified as long as the satellite orbit is sufficiently eccentric. However, realistic orbits with low eccentricities can trigger a substantially different disk response. Also, there is a large uncertainty in Sgr's orbit (as well as that of the other satellites) due to the uncertainty in its observed phase-space coordinates and the MW parameters. These can explain the apparent discrepancy between our results and those of \cite{Bennett.etal.22}: they find in their $N$-body simulations that Sgr triggers a weaker response in the Solar neighborhood than what we compute in this chapter. This probably owes to the fact that Sgr's orbit induces an adiabatic perturbation in the Solar neighborhood in their simulations. To integrate the satellite orbits for our analysis, we pick the median values of their observed phase-space coordinates quoted by \cite{Riley.etal.19} and \cite{Vasiliev.Belokurov.20} as the initial conditions. Other values within the margin of error may lead to substantially different orbits, implying different $R_\rmd$, $\vp$, $\thetap$ and $\phip$ and therefore different responses. Moreover, some of these orbits can be mildly eccentric, for which the straight line orbit approximation adopted in this chapter is not well-justified. A perturbative analysis of the disk response to satellites along general orbits is beyond the scope of this dissertation and deserves future investigation. We also ignore the effect of dynamical friction on the satellite orbits and thus on the resulting disk response. Moreover, the disk crossing times are sensitive to the satellite orbits and therefore to the detailed MW potential and the current phase-space coordinates of the satellites. For example, a heavier MW model with a total mass of $1.5\times 10^{12}\Msun$ leaves the relative amplitudes of the satellite responses (in the collisionless limit) nearly unchanged, but makes the satellites more bound, bringing most of the disk crossing times closer to the present day. In particular, the last disk passage of Sgr that triggers a significant response now occurs $\sim 600\Myr$ ago (as opposed to $1 \Gyr$ ago in the fiducial case) which is closer to the winding time of $\sim 500\Myr$ inferred from the phase spiral observed in the Solar neighborhood \citep[][]{Bland-Hawthorn.etal.19}.

In this section, we have computed the responses in the collisionless limit. In reality, collisional diffusion due to interactions of stars with GMCs, etc. would damp away the response super-exponentially over a timescale that is $\sim 0.6-0.7\Gyr$ in the Solar neighborhood (see section~\ref{sec:spiral_c}). This would almost completely wash away the response to any satellite encounter that occurred $\gtrsim 1\Gyr$ ago. For example, the present day response to the last pericentric passage of Leo II that occurred $\sim 1.8\,\Gyr$ ago would be completely erased. If the last disk crossing of Sgr that induced a strong response occurred $\sim 1\Gyr$ ago as in the fiducial MW model, the response would have been damped out by $\sim 2$ orders of magnitude by today, deeming Sgr unlikely to be the agent behind the Gaia snail. However, as discussed above, the disk crossing times are sensitive to the satellite orbits. The heavier MW model with a total mass of $1.5\times 10^{12}\Msun$ implies a Sgr crossing time of $\sim 0.6\Gyr$ instead of $1\Gyr$. In this case the response would only have been damped by a factor of $\sim 0.4$. Therefore, the collisionality argument suggests that if the Gaia snail was indeed triggered by Sgr, the impact causing it must have happened within $\sim 0.6-0.7\Gyr$ from the present day.

\subsection{Exploring parameter space}

\begin{figure*}[t!]
\centering
\subfloat{\includegraphics[width=0.49\textwidth]{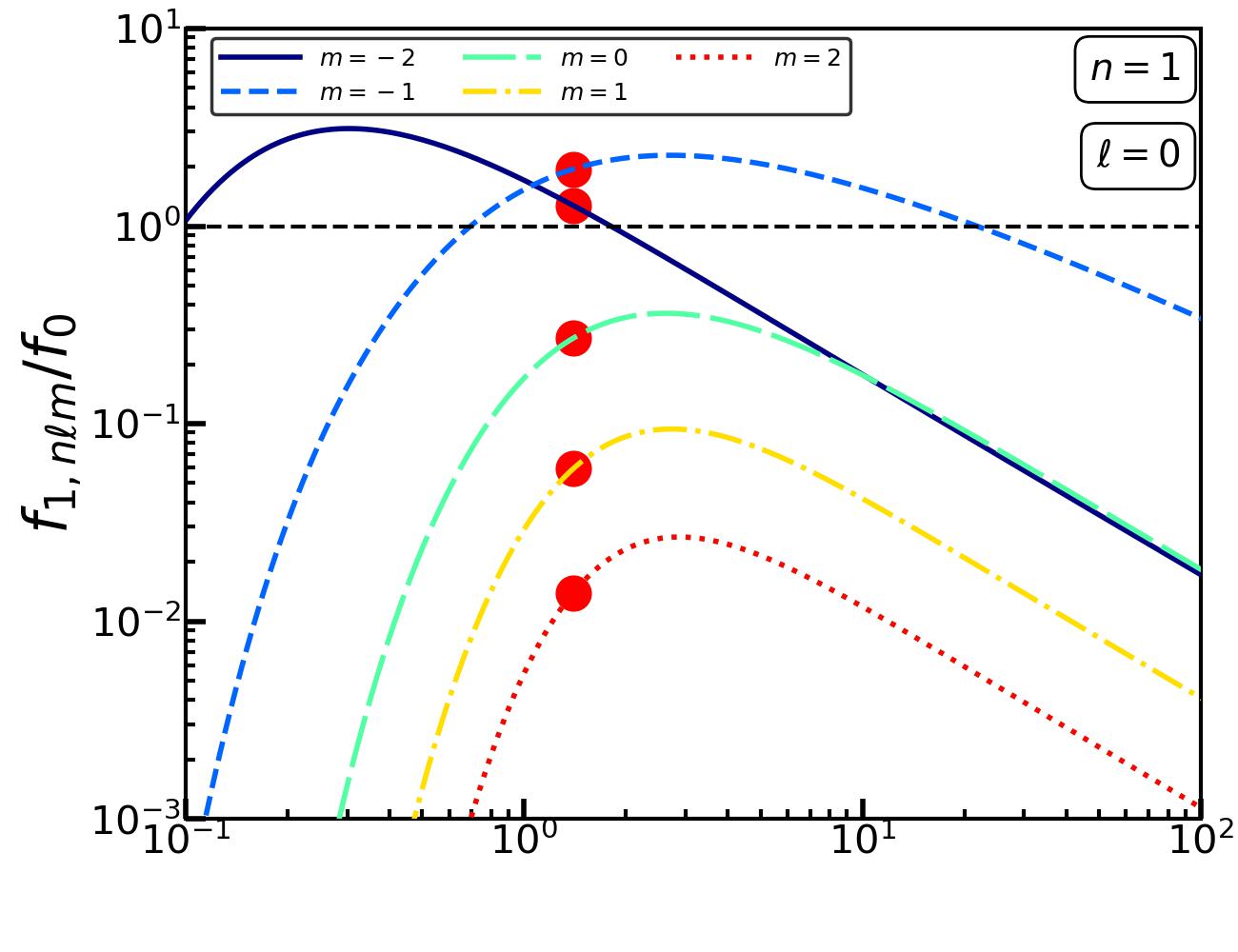}
  \label{disk_Resp_n1_l0}}
\subfloat{\includegraphics[width=0.49\textwidth]{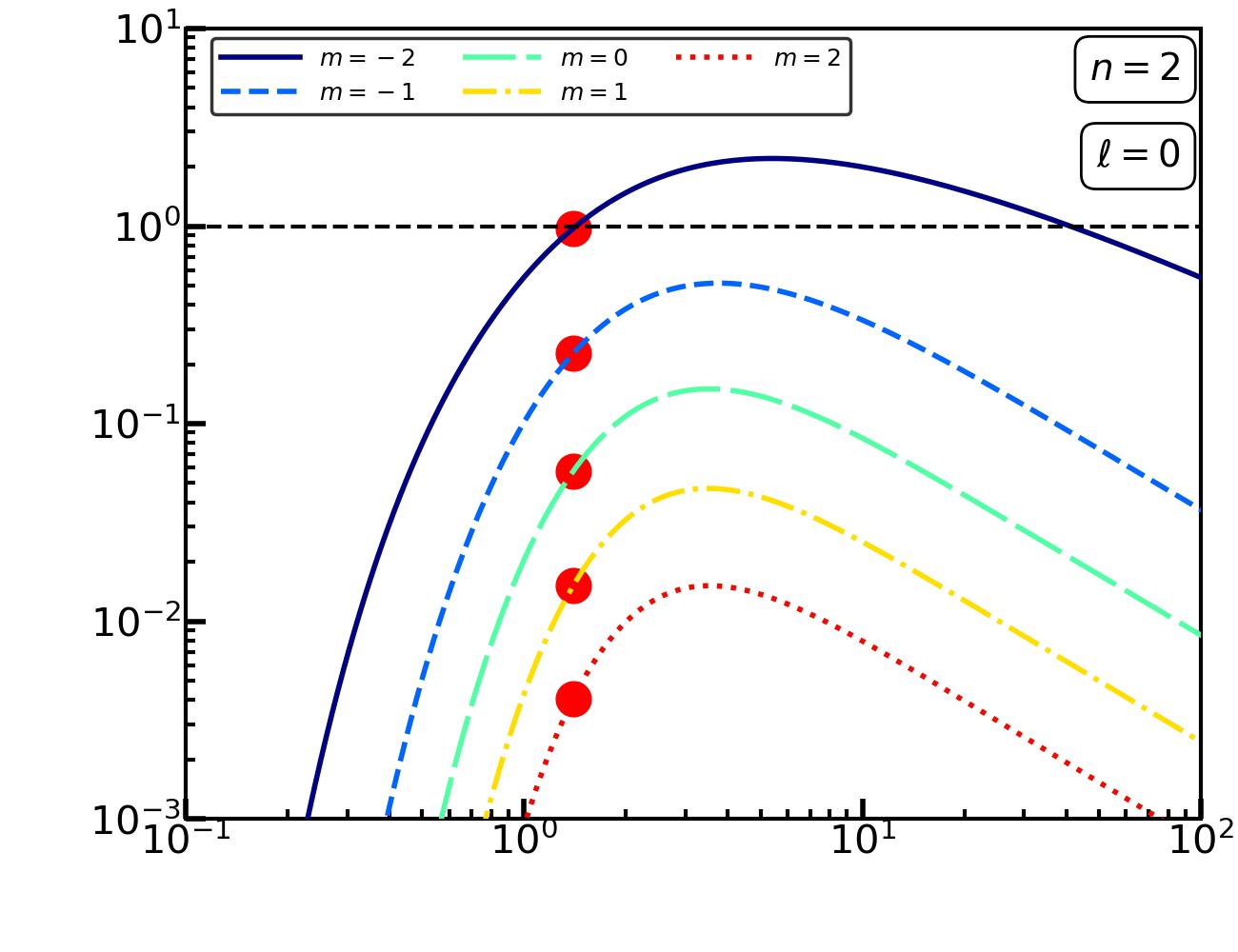}
  \label{disk_Resp_n2_l0}}\\
\subfloat{\includegraphics[width=0.49\textwidth]{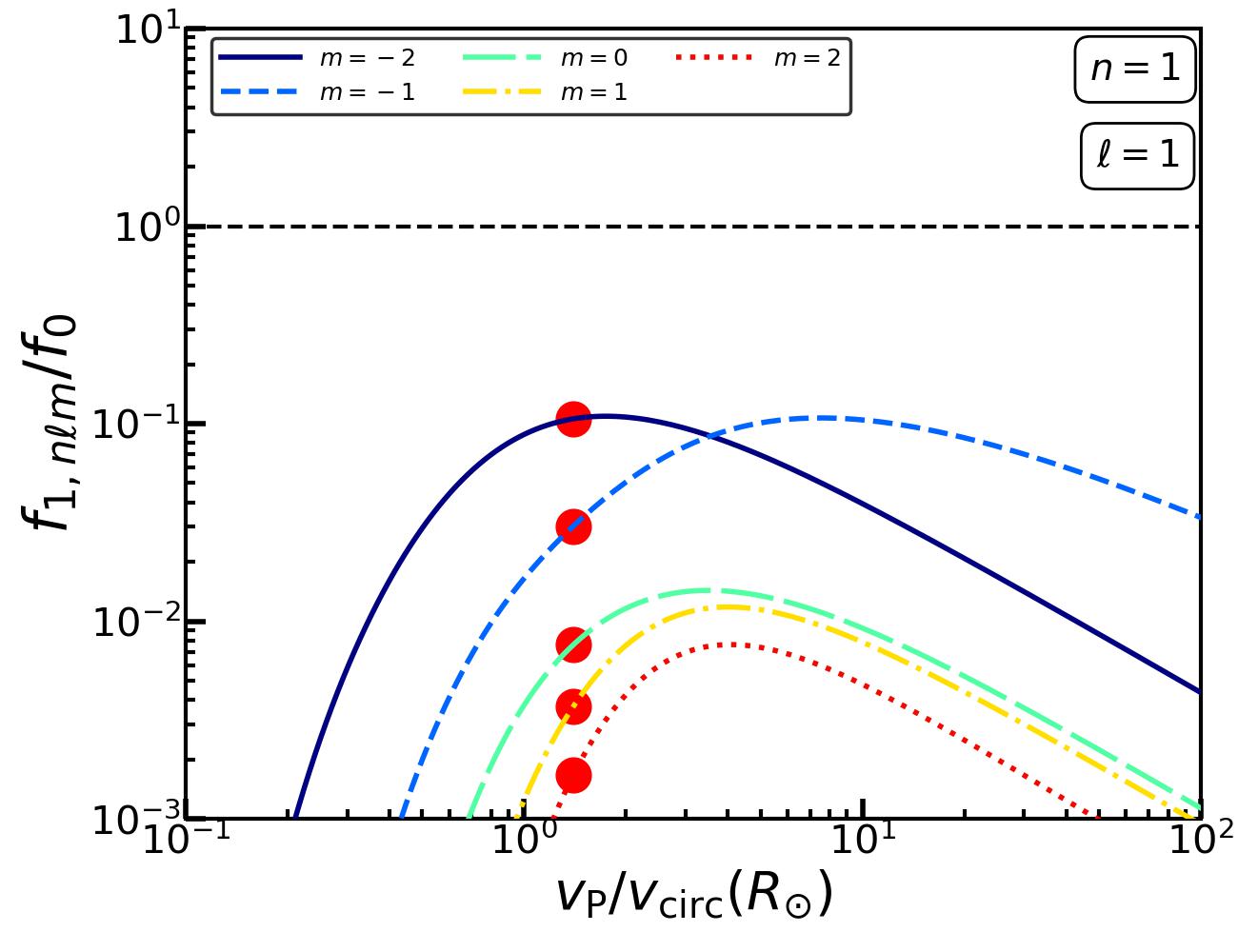}
  \label{disk_Resp_n1_l1}}
\subfloat{\includegraphics[width=0.49\textwidth]{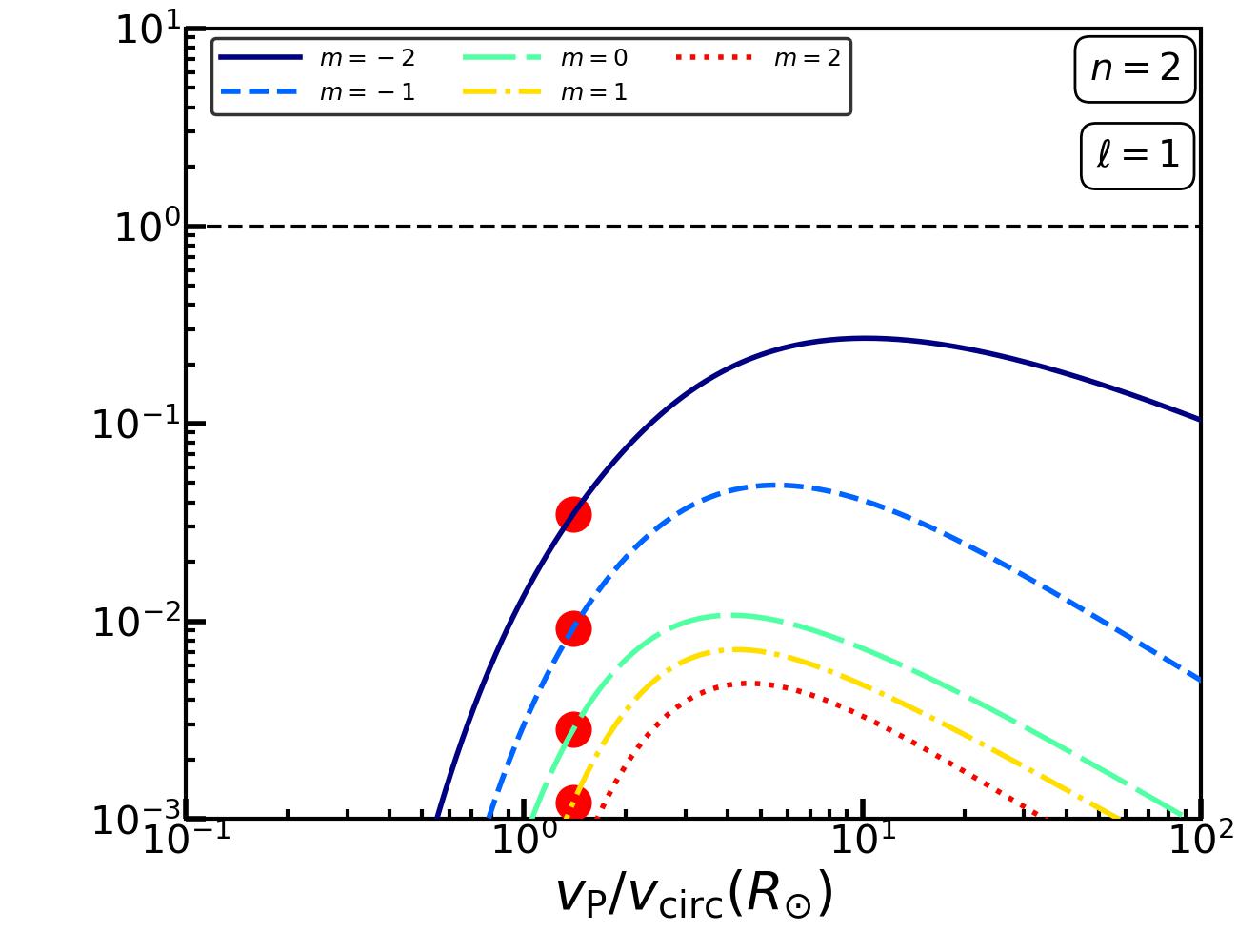}
  \label{disk_Resp_n2_l1}}
\caption{Steady state MW disk response to satellite encounter in the collisionless limit: each panel shows the behaviour of the disk response amplitude, $f_{1,n\ell m}/f_0$ (marginalized over $I_R$), as a function of the impact velocity, $\vp$, in the Solar neighborhood, in presence of an ambient DM halo. Different lines correspond to different $m$ modes as indicated. The top and bottom rows show the response for $\ell=0$ and $1$ while the left and right columns indicate it for the $n=1$ bending and $n=2$ breathing modes. The fiducial parameters correspond to $I_z=I_{z,\odot}$ and the parameters for Sgr impact, the response amplitudes for which are indicated by the red circles in each panel. The response is dominated by the $(n,\ell,m)=(1,0,-2)$ mode or the two-armed warp at small $\vp$ and the $(2,0,-2)$ mode or the two-armed spiral at large $\vp$. Typically, the $m=-2$ and $-1$ responses dominate over $m=0,1$ and $2$, while the $\ell=0$ response is more pronounced than $\ell=1$.}
\label{fig:disk_Resp_lm}
\end{figure*}

\begin{figure*}[t!]
\centering
\subfloat{\includegraphics[width=0.49\textwidth]{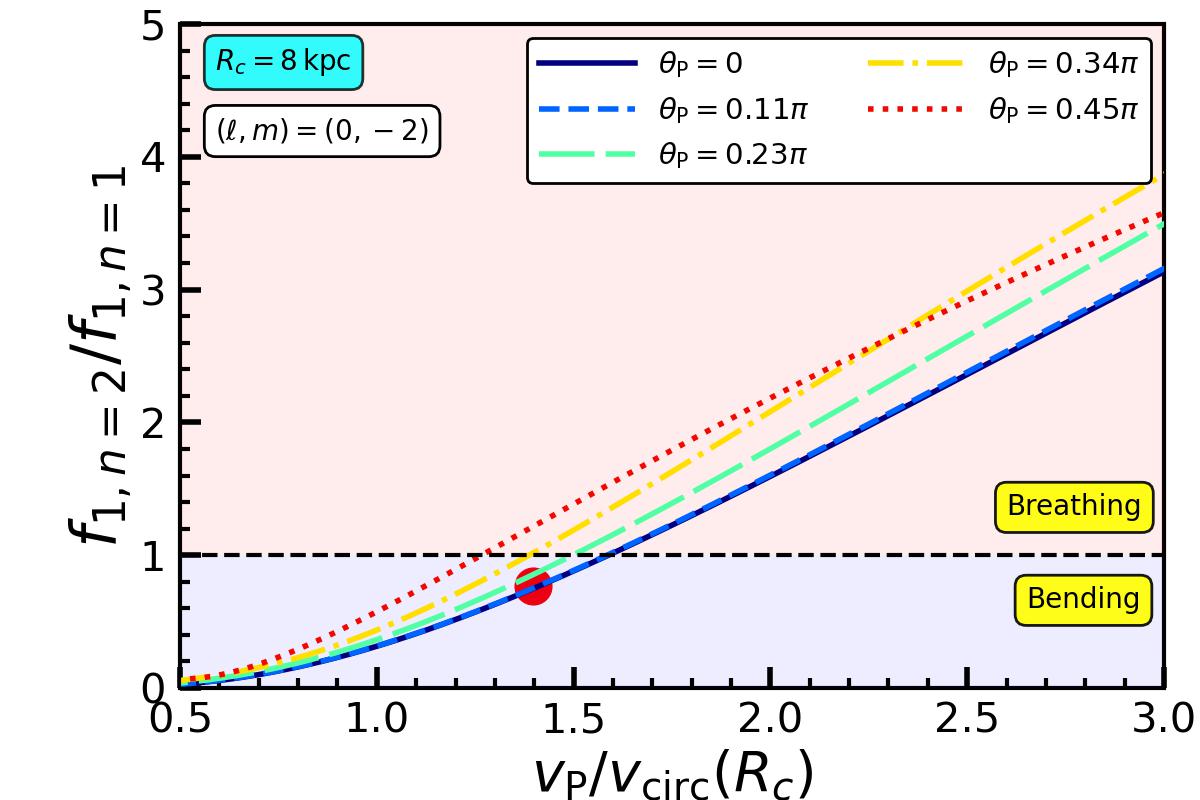}
  \label{disk_Resp_Ratio_thetap}}
\subfloat{\includegraphics[width=0.49\textwidth]{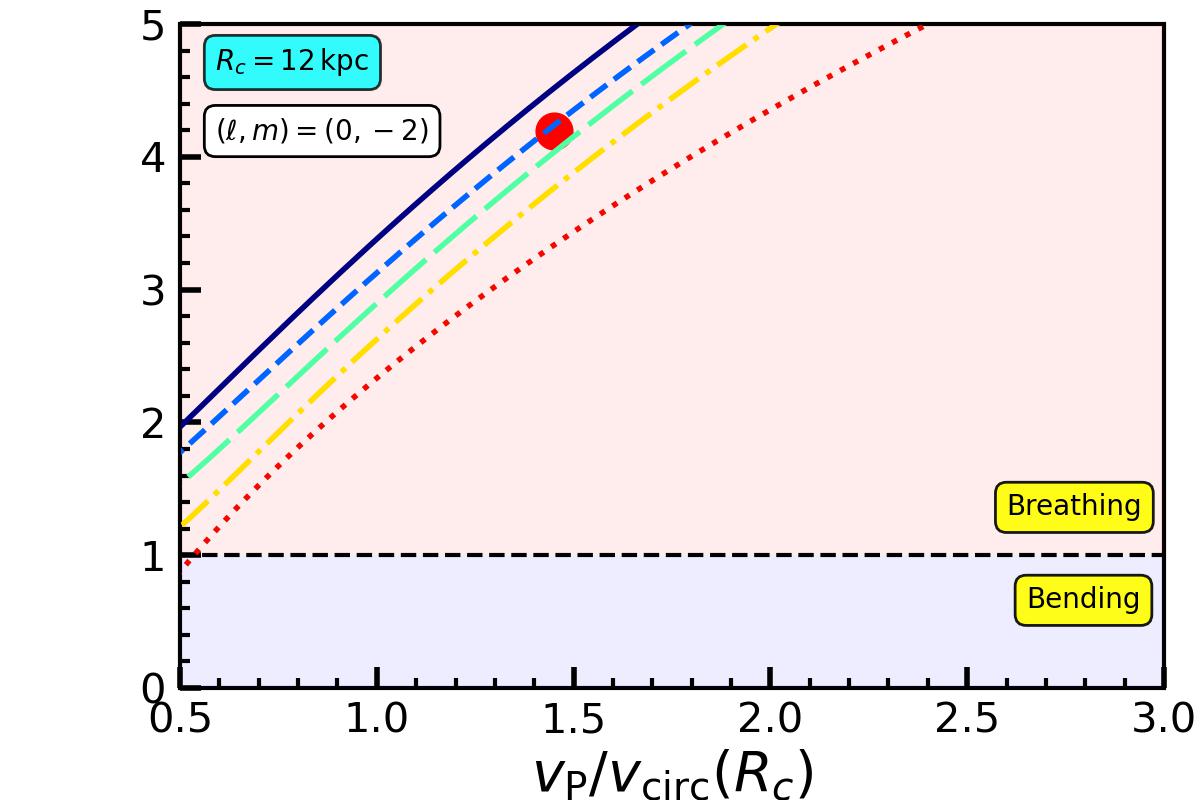}
  \label{disk_Resp_Ratio_phip}}
\caption{MW disk response to satellite encounter: breathing-to-bending ratio or the relative strength of the $n=2$ and $n=1$ modes of disk response to a Sgr-like impact is plotted as a function of the impact velocity, $\vp$, at $\Rc=\Rsun=8\kpc$ and $\Rc=1.5\Rsun=12\kpc$ shown in the left and right columns respectively, for the $(\ell,m)=(0,-2)$ mode which typically dominates the response. Different lines correspond to different values of $\thetap$ as indicated. We consider $I_z=I_{z,\odot}$ and the fiducial parameters to correspond to those for Sgr encounter, for which the breathing-to-bending ratio is denoted by the red circle. Bending modes dominate over breathing modes at small $\vp$ and vice versa at large $\vp$. Breathing modes are relatively more pronounced than bending modes in the outer disk, closer to the Sgr impact radius, $\rd=17\kpc$. More planar (perpendicular) encounters trigger larger breathing-to-bending ratios farther away from (closer to) the impact radius.}
\label{fig:disk_Resp_Ratio}
\end{figure*}

Having computed the MW disk response to its satellites, we now investigate the sensitivity of the response to the various encounter parameters. In Fig.~\ref{fig:disk_Resp_vp} we plot the amplitude of the Solar neighborhood response, $f_{1,n\ell m}/f_0$ (marginalized over $I_R$), as a function of the impact velocity, $\vp$ (in units of the circular velocity at $\Rc=\Rsun$), for the $(n,\ell,m)=(1,0,0)$ bending and $(n,\ell,m)=(2,0,0)$ breathing modes, shown in the left and right columns respectively. The top, middle and bottom rows show the results for varying $I_z$, $\thetap$ and $\phip$ respectively, assuming the fiducial parameters to be those for Sgr (mass $M_\rmP=10^9 \Msun$, scale radius $\varepsilon=1.6$ kpc) during its penultimate disk crossing (most relevant for the Gaia snail), i.e., impact radius $\rd=17\kpc$, impact velocity $\vp=340$ km/s, and angles of impact, $\thetap={21}^\circ$ and $\phip={150}^\circ$. In Fig.~\ref{fig:disk_Resp_lm} we plot the bending and breathing mode response amplitudes (in the Solar neighborhood) as a function of $\vp$ for different $(\ell,m)$ modes, with the fiducial parameters again corresponding to Sgr. The left and right columns respectively indicate the $n=1$ bending and $n=2$ breathing modes, while the top and bottom rows correspond to $\ell=1$ and $\ell=2$ respectively. The different lines in each panel denote the responses for $m=-2,-1,0,1$ and $2$. Fig.~\ref{fig:disk_Resp_Ratio} shows the ratio of the bending and breathing response amplitudes as a function of $\vp$ for the dominant mode $(\ell,m)=(0,-2)$. Different lines indicate breathing-to-bending ratios for different values of $\thetap$, while the left and right columns respectively indicate the ratios observed at $\Rc=8$ and $12\kpc$.

From Figs.~\ref{fig:disk_Resp_vp} and \ref{fig:disk_Resp_lm} it is evident that, as shown in equation~(\ref{sat_asymptote}), the disk response is suppressed like a power law ($\sim v^{-1}_\rmP$) in the high velocity/impulsive limit and exponentially ($\sim \exp{\left[-\Omega b/\vp\right]}$) suppressed in the low velocity/adiabatic limit. The response is the strongest for intermediate velocities, $\vp\sim 2-3\, v_{\rm circ}(\Rsun)$, where the time periods of the vertical, radial and azimuthal oscillations of the stars are nearly commensurate with the encounter timescale, $\sqrt{b^2+\varepsilon^2}/\vp$. The $v^{-1}_\rmP$ and $K_{0i}$ factors in equation~(\ref{sat_gen}) conspire to provide the near-resonance condition for maximum response,

\begin{align}\label{rescond}
n\Omega_z + \ell\kappa + m\Omega_\phi \approx \frac{0.6\,\vp}{\sqrt{b^2 + \varepsilon^2}},
\end{align}
where $b$ is the impact parameter of the encounter, given by equation~(\ref{impact_parameter}). From the top panels of Fig.~\ref{fig:disk_Resp_vp}, it is clear that the peak response shifts to smaller $\vp$ with increasing $I_z$. This is easy to understand from the fact that the corresponding vertical frequency, $\Omega_z$, decreases with increasing $I_z$, making the encounter more impulsive for larger actions. The middle and bottom panels show that the response depends strongly on the polar angle of the encounter, $\thetap$, but very mildly on the azimuthal angle, $\phip$. Moreover, the middle panels indicate that more planar encounters (larger $\thetap$) induce stronger responses. 

The in-plane structure of the disk response depends on the relative contribution of the different $(\ell,m)$ modes. From Fig.~\ref{fig:disk_Resp_lm} it is evident that a typical Sgr-like encounter predominantly excites $(\ell,m)=(0,-1)$ and $(\ell,m)=(0,-2)$ in the Solar neighborhood. The dominant mode for slower encounters is $(n,\ell,m)=(1,0,-2)$ while that for faster ones is $(n,\ell,m)=(2,0,-2)$. Since $f_{1,n\ell m}/f_0\gtrsim 1$ in these cases, the response to the impact by Sgr is in fact non-linear in the Solar neighborhood. Either way, a satellite encounter is typically found to excite strong $m=-2$ modes, i.e., $2$-armed warps ($n=1$) and spirals ($n=2$). This is due to a quadrupolar tidal distortion of the disk by the satellite, which manifests as a stretching of the disk in the direction of the impact and a compression perpendicular to it.

Fig.~\ref{fig:disk_Resp_Ratio} elucidates that the bending mode response dominates for slower encounters, i.e., smaller $\vp$, and at guiding radii far from the impact radius, $\rd$. More planar impacts trigger larger breathing-to-bending ratios farther away from the impact radius while this trend reverses closer to it. This is because more planar encounters cause more vertically symmetric perturbations farther away from the impact radius. The predominance of bending modes for low $\vp$ encounters while that of breathing modes for high $\vp$ ones has been observed by \cite{Widrow.etal.14} and \cite{Hunt.etal.21} in their N-body simulations of satellite-disk encounters. As demonstrated by \cite{Widrow.etal.14}, slower encounters provide energy to the stars near one of the vertical turning points while drain energy from those near the other turning point, thereby driving bending wave perturbations that are asymmetric about the mid-plane. On the other hand, fast satellite passages are impulsive and impart energy to the stars near both the turning points, thus triggering symmetric breathing waves. 

The predominance of breathing (bending) modes closer to (farther away from) the impact radius is qualitatively similar to the observation by \cite{Hunt.etal.21} in their simulations of MW-Sgr encounter that the outer part of the MW disk which is closer to the impact radius shows a preponderance of two-armed phase spirals or breathing modes. This can be understood within the framework of our formalism by noting that the impact parameter, $b$, and therefore the encounter timescale $\sim \sqrt{b^2+\varepsilon^2}/\vp$ decreases with increasing proximity to the point of impact; hence the impact is faster than the vertical oscillations of stars near the point of impact, driving stronger breathing mode perturbations. However, contrary to these predictions for the MW-Sgr encounter, \cite{Hunt.etal.22}, using Gaia DR3 data, revealed two-armed phase spirals, and therefore breathing modes, in the inner disk ($\Rc \sim 6-7\kpc$). Our analysis suggests that none of the MW satellites could have caused this. Using N-body simulations of an isolated MW system, \cite{Hunt.etal.22} suggested that a transient spiral arm or bar could be a potential trigger for breathing modes in the inner disk. However, such a transient perturbation would have to be sufficiently impulsive, i.e., occur over a timescale that is comparable to or smaller than the vertical oscillation timescale in the inner disk (see section~\ref{sec:spiral_cless}), in order to produce two-armed phase spirals with density contrast as strong as in the data. Such short timescales are unlikely to arise from the secular evolution of the disk alone and may instead require forcing of the inner disk by perturbations in the MW halo. Another possible trigger of this feature is the recent passage of dark satellite(s) through the inner disk. The true origin of this feature is however unclear. Hence, we conclude that the presence of two-armed phase spirals in the inner disk is rather unexpected, and that its origin poses an intriguing conundrum.

\section{Phase spirals and the Galactic potential}
\label{sec:pot_const}

Thus far we mainly focused on how the nature of the perturbation dictates the vertical (i.e., bending and breathing modes) as well as the in-plane (various $(\ell,m)$ modes) structure of the disk response. However, the detailed structure, in particular the winding, of the phase spiral not only depends on the triggering agent but also holds crucial information about the underlying potential in which the stars move, and can thus be used to constrain the potential of the combined disk plus halo system \citep[see also][]{Widmark.etal.22a,Widmark.etal.22b}.

The winding of the vertical phase spiral can be characterized by the pitch-angle, $\phi_\rmI$, along the ridge of maximum density. It is defined as the angle between the azimuthal direction and the tangent to the line of constant density \citep[][]{Binney.Tremaine.87}. It is related to the local dependence of the vertical frequency on the vertical action according to:
\begin{align}
\phi_\rmI = \cot^{-1}{\left[\left|I_z \frac{\rmd \Omega_z }{\rmd I_z}\right| t\right]} = \cot^{-1}{\left[\left|\frac{\rmd \Omega_z }{\rmd \ln{I_z}}\right| t\right]}.
\label{pitch_angle}
\end{align}
Following a perturbation, the pitch angle increases with time, asymptoting towards zero, as the spiral winds up as a consequence of the ongoing phase-mixing. Based on the above expression for $\phi_\rmI$, we can define the following timescale of phase-mixing,
\begin{align}
\tau_\phi = \left|\frac{\rmd \ln{I_z}}{\rmd \Omega_z}\right|\,.
\label{phase_mix_timescale}
\end{align}
This timescale, which determines the rate of winding of the spiral, is a function of both the guiding radius, $\Rc$, and the action, $I_z$, and is ultimately dictated by the (unperturbed) potential of the disk+halo system, which sets $\rmd \Omega_z/\rmd I_z$. Hence, the detailed shape of the phase spiral at a given location in the disk is sensitive to the local, relative strengths of the disk and halo, thereby opening up interesting avenues for constraining the detailed potential of the MW by examining phase spirals throughout the disk.

\begin{figure}
\centering
\subfloat{\includegraphics[width=0.33\textwidth]{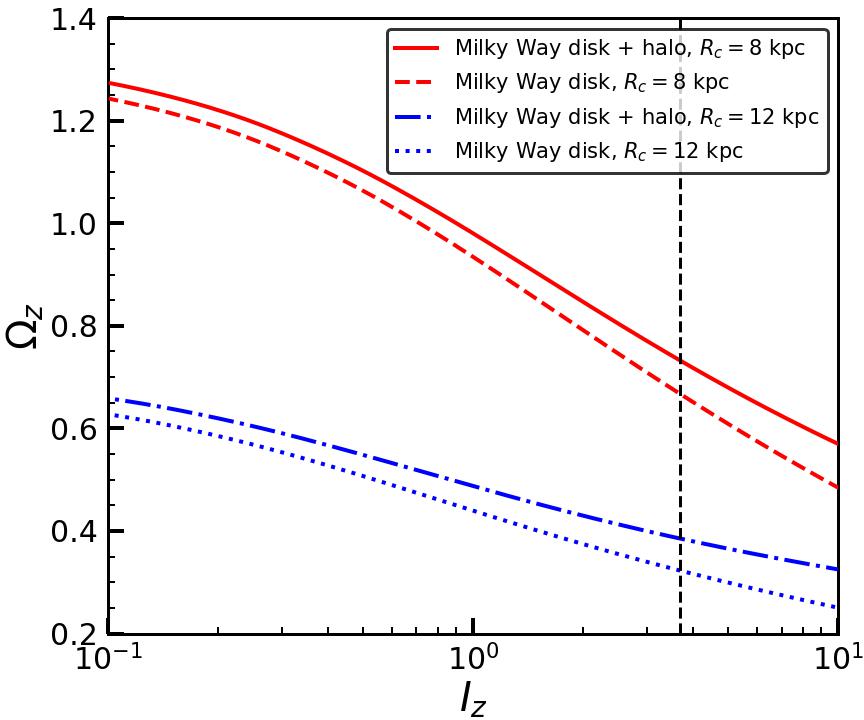}
\label{freq_Iz}}
\subfloat{\includegraphics[width=0.33\textwidth]{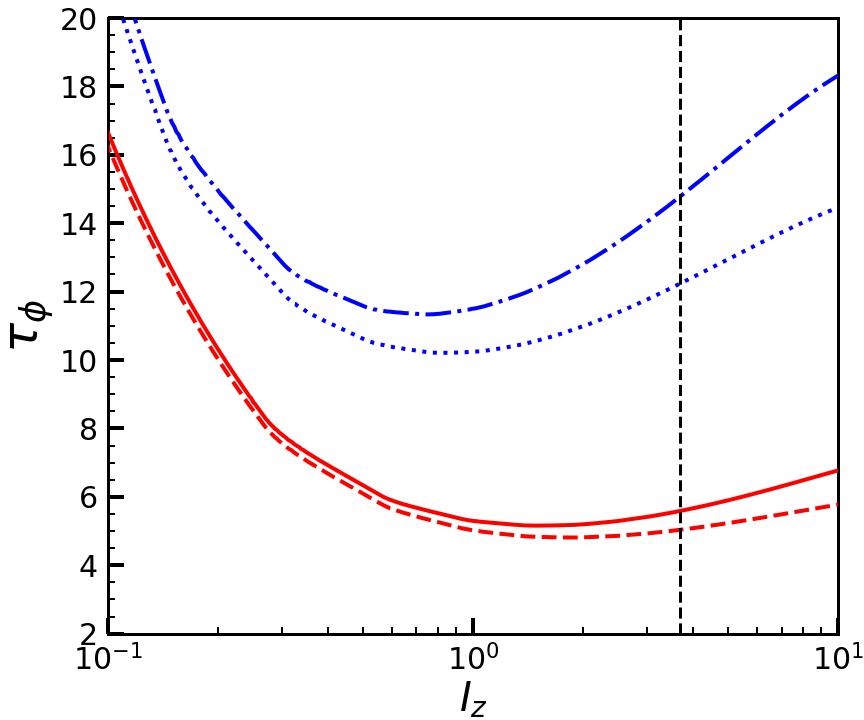}
\label{tauphi_Iz}}
\subfloat{\includegraphics[width=0.33\textwidth]{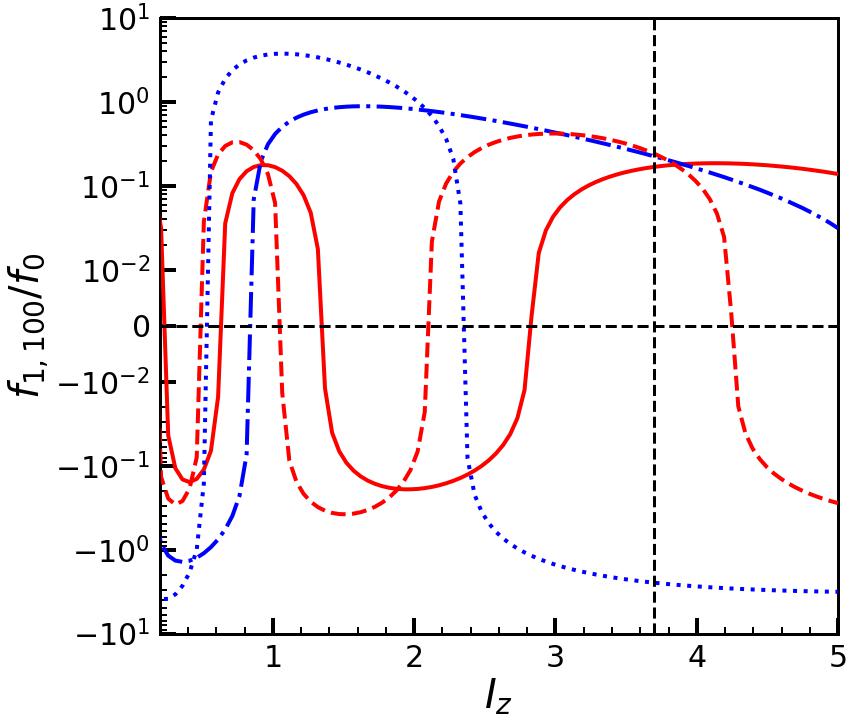}
\label{f1_Iz}}
\caption{Impact of DM halo on vertical phase-mixing: the panels from left to right respectively indicate the vertical frequency, $\Omega_z$ (units of $\sigma_{z,\odot}/h_z$), the vertical phase-mixing timescale, $\tau_\phi$ (given by equation~[\ref{phase_mix_timescale}]), and the $w_z=0$ cuts of the phase spirals shown in Fig.~\ref{fig:snailn1} as a function of the vertical action, $I_z$ (units of $h_z\sigma_{z,\odot}$). The solid and dashed red lines denote the cases with and without a halo for $\Rc=\Rsun=8\kpc$ while the dot-dashed and dotted blue lines show the same for $\Rc=12\kpc$. The vertical dashed line indicates roughly the maximum $I_z$ for which a phase spiral is discernible in the Gaia data. Note that phase-mixing occurs the fastest for $I_z\sim1$ and that the inner disk phase mixes faster than the outer disk. Also note that the presence of a DM halo increases $\Omega_z$ as well as $\tau_\phi$, leading to slower phase-mixing and therefore slower wrapping of the phase spiral. This effect is more pronounced in the outer disk.}
\label{fig:phase_mix}
\end{figure}
\begin{figure*}
\subfloat{\includegraphics[width=0.5\textwidth]{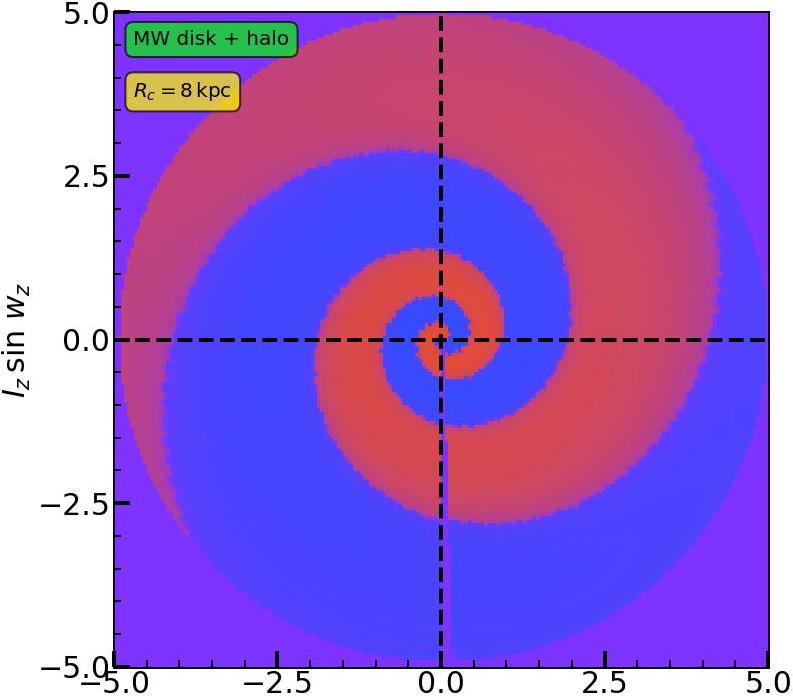}
    \label{snailn1a_1}}
\subfloat{\includegraphics[width=0.475\textwidth]{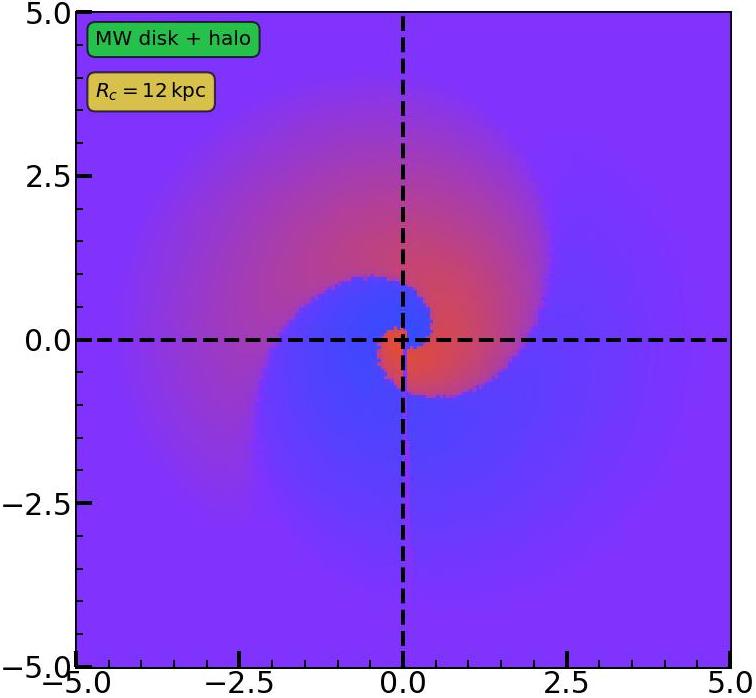}
    \label{snailn1b_1}}\\
\subfloat{\includegraphics[width=0.5\textwidth]{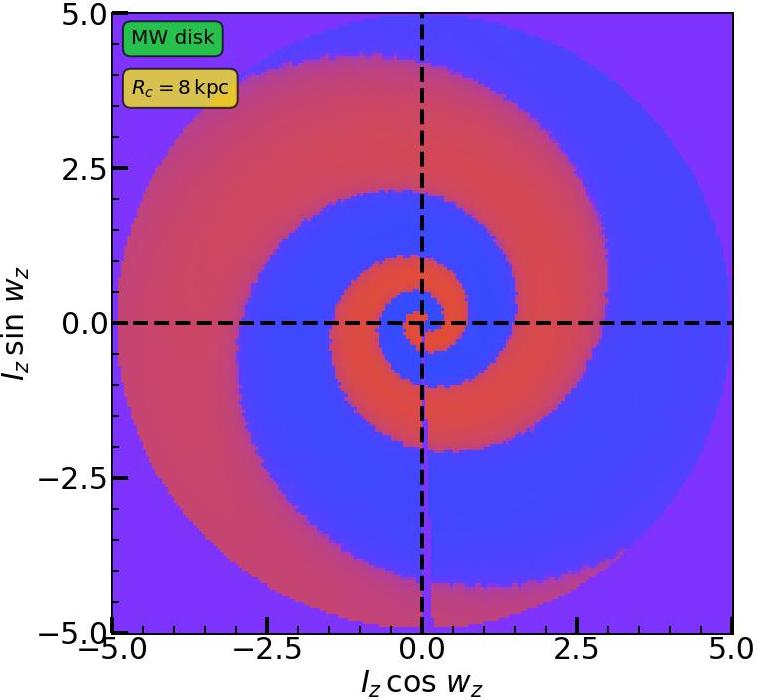}
    \label{snailn1a_2}}
\subfloat{\includegraphics[width=0.476\textwidth]{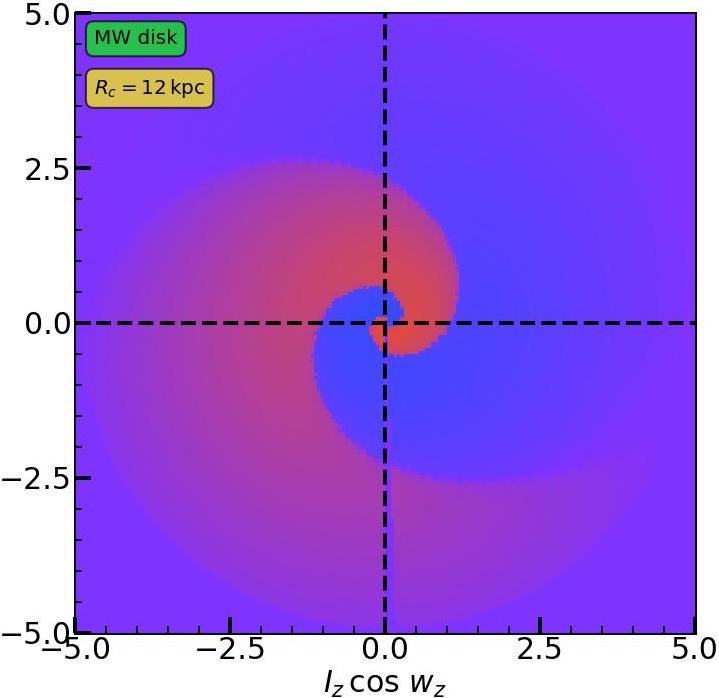}
    \label{snailn1b_2}}
\caption{Vertical phase-mixing: one-armed phase spiral corresponding to $n=1$ bending mode excited by the encounter with Sgr for MW disk+halo and MW disk models (columns) at $\Rc=8\kpc$ and $12\kpc$ (rows). The presence of DM halo slows down the rate of phase-mixing, leading to more loosely wrapped phase spirals. Phase-mixing occurs more rapidly in the inner disk than in the outer disk.}
\label{fig:snailn1}
\end{figure*}

The left panel of Fig.~\ref{fig:phase_mix} plots the vertical frequency, $\Omega_z$, as a function of the logarithm of the action, $I_z$, for the MW potential with and without the halo and at guiding radii, $\Rc=8$ (red) and $12\kpc$ (blue). The middle panel shows the behaviour of the corresponding phase-mixing timescale, $\tau_\phi$, as a function of $I_z$. Fig.~\ref{fig:snailn1} shows the $(n,\ell,m)=(1,0,0)$ phase spirals $400\Myr$ after the penultimate disk crossing of Sagittarius, color coded by the MW disk response, $f_{1,100}$, with blue (red) indicating higher (lower) phase-space density. Results for the same four cases are shown as indicated. Finally, the right panel of Fig.~\ref{fig:phase_mix} shows the $w_z=0$ cuts of the normalized response, $f_{1,100}/f_0$, as a function of $I_z$, for the four different phase spirals. The vertical frequency, $\Omega_z$, is a decreasing function of $\ln{I_z}$ in all cases, indicating that stars with larger actions (i.e., larger vertical excursion amplitudes) oscillate slower. Note that $\left|\rmd\Omega_z/\rmd \ln{I_z}\right|$ is an increasing (decreasing) function of $I_z$ at small (large) $I_z$, reaching a maximum at intermediate $I_z$. Consequently, the phase-mixing timescale, $\tau_\phi$, which is the inverse of $\left|\rmd\Omega_z/\rmd \ln{I_z}\right|$, attains its minimum at $I_z/(h_z\sigma_z)\sim 1$. Thus phase-mixing occurs the fastest at intermediate actions and slows down at larger actions, causing the spiral to become more loosely wound (smaller pitch angle) farther away from its origin. 

The rate of phase-mixing is different in the four different cases. Closer to the galactic center where the potential is deeper and steeper, stars have a larger range of $\Omega_z$, or in other words $\Omega_z$ falls off more steeply with $\ln{I_z}$ in the inner disk than in the outskirts. This leads to faster phase-mixing and therefore a much more tightly wound phase spiral in the inner disk (left panels of Fig.~\ref{fig:snailn1}) as opposed to the outer disk (right panels). The difference in the phase-mixing rates is also manifest in the $w_z=0$ response shown in the right panel of Fig.~\ref{fig:phase_mix}; note the longer oscillation wavelengths of the blue lines (outer disk) as opposed to the red lines (inner disk). Hence, in agreement with expectations, the inner part of the disk equilibrates much faster than the outer part. 

The presence of a DM halo deepens the potential well and thus boosts the oscillation frequencies. But the halo also steepens the potential such that the range of frequencies is reduced, i.e., $\Omega_z$ falls off more mildly with $\ln{I_z}$ than in the disk only case. This leads to slower phase-mixing and therefore more loosely wound phase spirals in the presence of the halo (upper panels of Fig.~\ref{fig:snailn1}) than in its absence (lower panels), the effect being more pronounced in the outer (right panels) than in the inner (left panels) disk. Equivalently, the $w_z=0$ response in the right panel of Fig.~\ref{fig:phase_mix} shows longer wavelength wiggles in presence of the halo.

The above sensitivity of the phase-mixing timescale to the detailed galaxy potential implies that one can use phase spirals to constrain it. One can unwind the observed phase spiral by adopting a form for the galactic potential. Only for the correct potential will the spiral be properly unwound, i.e., the pitch-angle, $\phi_\rmI$, go to zero for all $I_z$ (modulo measurement errors) at the same time, $t_0$, in the past. This $t_0$ then corresponds to the time elapsed since the maximum strength of the perturbation that triggered the phase spiral. However, this method to constrain the total potential (disk plus dark matter) of the MW relies on the assumption of a single, impulsive perturbation as the trigger. In reality, the phase spirals may have been impacted by multiple, overlapping perturbations and/or by large-scale temporal fluctuations in the overall potential, which would severely hamper this technique \citep[][]{Tremaine.etal.22}. We intend to investigate the promise of phase spirals as probes of the galactic potential for different kinds of perturbation in future work.

\section{Conclusion}
\label{sec:concl_3}

In this chapter, we have developed a linear perturbative formalism to analyze the response of a realistic disk galaxy (characterized by a pseudo-isothermal DF) embedded in an ambient spherical DM halo (modelled by an NFW profile) to perturbations of diverse spatiotemporal nature: bars, spiral arms, and encounters with satellite galaxies. Adopting the radial epicyclic approximation, we perturb the FPE up to linear order (in action-angle space) in presence of a perturbing potential, $\Phi_\rmP$, to compute the post-perturbation linear response in the DF, $f_1$. Without self-gravity to reinforce the response, the oscillations in the response phase mix away due to an intrinsic spread in the frequencies of stars, giving rise to spiral features in the phase-space distribution known as phase spirals. Depending on the timescale of $\Phi_\rmP$, different modes of disk oscillation, corresponding to different phase spiral structures, are excited. We summarize our conclusions as follows:

\begin{itemize}
    \item Following an impulsive perturbation, the $(n,\ell,m)$ mode of the disk response consists of stars oscillating with frequencies, $n\Omega_z$, $\ell\Omega_r\approx \ell\kappa$ and $m\Omega_\phi$, along vertical, radial and azimuthal directions respectively. Since the frequencies depend on the actions, primarily on the vertical action $I_z$ and the angular momentum $L_z$, the response phase mixes away, spawning phase spirals. The dominant modes of vertical oscillation are the anti-symmetric bending ($n=1$) and symmetric breathing ($n=2$) modes, which induce initial dipolar and quadrupolar perturbations in the $z-v_z$ or $I_z\cos{w_z}-I_z\sin{w_z}$ phase-space. Over time these features are phase-wrapped into one- and two-armed phase spirals, respectively, due to the variation of $\Omega_z$ with $I_z$.
    
    \item Since $\Omega_z$ and $\Omega_\phi$ both depend on $L_z$, the amplitude of the $I_z\cos{w_z}-I_z\sin{w_z}$ phase spiral damps away over time, typically as $\sim 1/t$ (equation~\ref{lateral_mixing_disk}), at a coarse-grained level, i.e., upon marginalization over $L_z$. Therefore, in a realistic disk with ordered motion, lateral mixing causes phase spirals to damp out much slower than in the isothermal slab with unconstrained lateral velocities discussed in chapter~\ref{chapter: paper2}, where it occurs like a Gaussian in time.
    
    \item Collisional diffusion due to scatterings of stars by GMCs, DM substructure, etc. damps away the disk response to a perturbation, and therefore the phase spiral amplitude, at a fine-grained level. Typically, the diffusion in actions is much more efficient than that in angles. The action gradients of the response, which predominantly arise from the action dependence of the oscillation frequencies, are erased by collisional diffusion, causing a super-exponential damping of the response over a timescale, $\tau_\rmD^{(\rmI)}$, which is $\sim 0.6-0.7\Gyr$ in the Solar neighborhood. The diffusion timescale is shorter in the inner disk, for stars with smaller $I_z$, and for higher-$n$ modes.
    
    \item The response to a bar or spiral arm with a fixed pattern speed, $\Omega_\rmP$, is dominated by the near-resonant stars ($\Omega_{\rm res}=n\Omega_z+\ell\kappa+m(\Omega_\phi-\Omega_\rmP) \approx 0$), especially in the adiabatic regime (slowly evolving perturber amplitude). Moreover, phase-mixing occurs gradually in the near-resonant parts of phase-space. Most of the strong resonances are confined to the disk-plane, such as the co-rotation ($n=\ell=0$) and Lindblad ($n=0,\ell=\pm 1,m=\pm 2$) resonances. For a {\it transient} bar or spiral arm whose amplitude varies over time as $\sim \exp{\left[-\omega^2_0 t^2\right]}$, the response is maximal when $\omega_0 \sim \Omega_{\rm res}$. In the impulsive limit ($\omega_0 \gg \Omega_{\rm res}$), the response is power-law suppressed, while in the adiabatic limit ($\omega_0 \ll \Omega_{\rm res}$) it is suppressed (super)-exponentially. 
    
    \item For a thin disk, since $\Omega_z$ is very different from $\Omega_\phi$ and $\kappa$, the vertical modes ($n\neq 0$) are generally not resonant with the radial and azimuthal ones and thus undergo phase-mixing. The strength of a vertical mode primarily depends on the nature of the perturbing potential, most importantly its timescale. Slower pulses trigger mainly bending $(n=1)$ modes, while faster pulses excite more pronounced breathing $(n=2)$ modes. Therefore, a transient bar or spiral arm with amplitude $\sim \exp{\left[-\omega^2_0 t^2\right]}$ triggers a bending (breathing) mode when the pulse-frequency, $\omega_0$, is smaller (larger) than $\Omega_z$. The response to very slow perturbations ($\omega_0 \ll \Omega_z$) is however heavily suppressed (adiabatic shielding).
    
    \item For a persistent bar or spiral arm with a fixed pattern speed, $\Omega_\rmP$, that grows and saturates over time, the response initially develops a phase spiral. However, this transient response is quickly taken over by coherent oscillations at the driving frequency, $\Omega_\rmP$, which manifest in phase-space as a steadily rotating dipole (quadrupole) for the bending (breathing) mode. Therefore, a transient (pulse-like) perturbation, such as a bar or spiral arm whose amplitude varies over a timescale comparable to the vertical oscillation period, $T_z\sim h_z/\sigma_z$, is essential for the formation of a phase spiral in $z$-$v_z$ space.
    
    \item The above analysis suggests that if the recently discovered two-armed Gaia phase spiral (breathing mode) in the inner disk of the MW was indeed induced by a spiral arm/bar as suggested by \citet{Hunt.etal.22} using N-body simulations, the spiral arm/bar was probably a transient one with a predominantly symmetric vertical profile whose amplitude varied over a timescale comparable to the vertical oscillation period. However, it remains to be seen whether such a rapid excitation and decay of a spiral arm/bar perturbation is realistic.

    \item We have computed the response of the MW disk, embedded in an extended DM halo, to disk-crossing perturbations by several of its satellite galaxies. We find that the response in the Solar neighborhood is dominated by the perturbations due to Sgr, followed by those due to the LMC, Hercules and Leo II. This implies that, if the Gaia snail near the Solar radius was indeed triggered by a MW satellite (which is still subject to debate), Sgr is the leading contender \citep[see also][]{Banik.etal.22b}. However, if that is the case, then the impact (disk crossing) must have happened within the last $\sim 0.6-0.7\Gyr$ in order for the response to have survived damping due to collisional diffusion.
    
    \item The amplitude of the response (at a fixed guiding radius $\Rc$) to satellite encounters for all modes scales as $\vp^{-1}$ in the impulsive (large $\vp$) limit, but is exponentially suppressed in the adiabatic (small $\vp$) limit, a phenomenon known as adiabatic shielding. The resonant modes with $n\Omega_z+\ell\kappa+m\Omega_\phi=0$ are not suppressed but rather become non-linear in the adiabatic regime. The peak response of a mode (with $n\Omega_z+\ell\kappa+m\Omega_\phi \neq 0$) is achieved at intermediate velocities for which the encounter timescale is commensurate with the oscillation periods of the stars, i.e., the near-resonance condition given by equation~(\ref{rescond}) is satisfied.

    \item The response of a disk to an encounter with a satellite galaxy depends primarily on three parameters:  (i) impact velocity $\vp$, (ii) polar angle of impact $\thetap$, and (iii) position on the disk relative to the point of impact where the satellite crossed the disk. Slower (faster) encounters excite predominantly $n=1$ bending ($n=2$ breathing) modes. More planar encounters (those with larger $\thetap$) typically result in larger breathing-to-bending ratios farther away from the impact radius while this trend gets reversed closer to it. In general, breathing modes dominate over bending modes closer to the point of impact, in agreement with $N$-body simulations of the MW-Sgr encounter \citep[][]{Hunt.etal.21}. Since the impact velocities of the MW satellites are all fairly similar to the local circular velocity, the decisive factor for breathing {\it vs.} bending modes is not so much the impact velocity, but rather the distance from the point of impact.
    
    \item The $(n,m)=(1,-2)$ and $(-1,2)$ modes generally dominate the response for slower satellite encounters, e.g., that of Sgr with respect to the Solar neighborhood, due to the tidal distortion of the disk by the satellite. The in-plane spatial structure of the disk response therefore generally resembles a two-armed warp ($n=1$) or spiral ($n=2$).

    \item The presence of an extended DM halo causes phase-mixing to occur slower, and modifies the structural appearance of the phase spirals (i.e., the pitch angle as function of vertical action). Hence, provided that the phase spiral was triggered by a single, impulsive perturbation, the detailed shape of the phase spiral can in principle be used to constrain not only the time elapsed since the perturbation \citep[][]{Darragh-Ford.etal.23} but also the total (disk$+$halo) potential. If the phase spiral has been triggered and/or impacted by multiple, overlapping perturbations the situation is less clear. In future work we intend to investigate the constraining power of phase spirals for different kinds of perturbation.
\end{itemize}

This chapter focused on the analysis of the phase-mixing component (phase spirals) of the `direct' disk response to various perturbations such as bars, spiral arms and satellite galaxies. However this leaves out some other potentially important features of the disk response. First of all, we considered the ambient DM halo to be non-responsive. In reality, the DM halo will also be perturbed, for example by an impacting satellite, and this halo response, which can be enhanced by self-gravity, can indirectly perturb the disk. A preliminary, perturbative analysis based on the $N$-body simulation of the MW-Sgr encounter by \cite{Hunt.etal.21} suggests that the indirect disk response to the halo perturbations is comparable, but sub-dominant, to the direct response to Sgr. However, a more detailed analysis is warranted, which we leave for future work. Secondly, we have neglected the self-gravity of the disk response. As discussed in chapter~\ref{chapter: paper2}, the dominant effect of self-gravity is to cause coherent point mode oscillations \citep[][]{Mathur.90, Weinberg.91} of the disk, which in linear theory are decoupled from the phase-mixing component of the response. The point modes manifest as coherently rotating features in the phase-space, which can make the phase spiral look less wound than in the non self-gravitating case \citep[][]{Darling.Widrow.19a, Widrow.23}, before the point modes Landau damp away. Usually, there is one dominant point mode for a given set of actions. In linear theory, one can always transform to the momentarily comoving reference frame of the dominant point mode, or in other words subtract away the point mode response, to obtain the pure phase-mixing contribution to the phase spiral. Besides triggering coherent point mode oscillations that make the phase spiral less wound, self-gravity can also enhance the amplitude of the phase-mixing component of the response, thus increasing the density contrast of the phase spiral. Recent work by \cite{Dootson.Magorrian.22} has shed some light on the self-gravitating response of razor-thin disks to bar perturbations, while that by \cite{Widrow.23} has investigated the impact of self-gravity on vertical phase spirals in a shearing box for impulsive perturbations. However, a more generic theoretical description of the self-gravitating response of inhomogeneous, thick disks to general perturbations (bars, spiral arms, satellite galaxies, etc) is still lacking. We hope to include the effects of self-gravity on disk perturbations in future work.




    \begin{subappendices}


\chapter*{Appendix}

\section{Perturbative solution of the Fokker-Planck equation}
\label{sec:FPE_pert_sol}

To solve the FPE in the action-angle space, the collision operator, $C[f]$, given in equation~(\ref{coll_op}) first needs to be expressed in terms of the action-angle variables. For this purpose, we follow \cite{Tremaine.etal.22} and invoke the epicyclic approximation in both vertical and radial directions. This is somewhat justified by the fact that stars with smaller $I_z$ and $I_R$ are more strongly scattered by structures like giant molecular clouds, since these are concentrated towards the midplane of the MW disk. This implies that $z\approx\sqrt{2I_z/\nu}\sin{w_z}$, $p_z\approx\sqrt{2\nu I_z}\cos{w_z}$, $R \approx R_c(L_z)+\sqrt{2I_R/\kappa}\sin{w_z}$, $p_R\approx\sqrt{2\kappa I_R}\cos{w_R}$, where $\nu$ and $\kappa$ are respectively the vertical and radial epicyclic frequencies. This also implies that $D_{zz}=D^{(z)}_I/\nu$, $D_{p_z p_z}=\nu D^{(z)}_I$, $D_{RR}=D^{(R)}_I/\kappa$ and $D_{p_R p_R}=\kappa D^{(R)}_I$, with $D^{(z)}_I$ and $D^{(R)}_I$ the first order diffusion coefficients in $I_z$ and $I_R$ respectively. Substituting these in equation~(\ref{coll_op}), the expression for 

\begin{align}
C[f] &= \frac{1}{2} \frac{\partial}{\partial z} \left(D_{zz} \frac{\partial f}{\partial z} \right) + \frac{1}{2} \frac{\partial}{\partial p_z} \left(D_{p_z p_z} \frac{\partial f}{\partial p_z} \right) \nonumber \\
&+ \frac{1}{2} \frac{\partial}{\partial R} \left(D_{RR} \frac{\partial f}{\partial R} \right) + \frac{1}{2} \frac{\partial}{\partial p_R} \left(D_{p_R p_R} \frac{\partial f}{\partial p_R} \right)
\label{coll_op_app_1}
\end{align}
simplifies to

\begin{align}
C[f] &\approx \frac{\partial}{\partial I_z} \left(D^{(z)}_I I_z \frac{\partial f}{\partial I_z}\right) + \frac{1}{4 I_z} \frac{\partial}{\partial w_z}\left( D^{(z)}_I \frac{\partial f}{\partial w_z}\right) \nonumber \\
&+ \frac{\partial}{\partial I_R} \left(D^{(R)}_I I_R \frac{\partial f}{\partial I_R}\right) + \frac{1}{4 I_R} \frac{\partial}{\partial w_R}\left( D^{(R)}_I \frac{\partial f}{\partial w_R}\right).
\label{coll_op_app_2}
\end{align}
We have made several approximations here. Firstly, we have adopted the epicyclic approximation. Secondly, we have neglected the $z-p_z$ and $R-p_R$ cross-terms for simplicity. Thirdly, following \cite{Binney.Lacey.88}, we have neglected diffusion in $\phi$ and $p_\phi$ since the terms involving $D_{\phi\phi}$, $D_{r\phi}$ and $D_{\phi z}$ are smaller than the $I_z$ and $I_R$ diffusion terms by factors of at least $\sigma_R/v_c$ or $\sigma_z/v_c$, which are typically much smaller than unity ($\sigma_R$ and $\sigma_z$ are radial and vertical velocity dispersions respectively, and $v_c$ is the circular velocity along $\phi$). These approximations together imply the form for the collision operator given in equation~(\ref{coll_op_app_2}). A close inspection of this tells us that the diffusion coefficients in the action-angle variables satisfy the relations: $D^{(z)}_{II}=2 D^{(z)}_I I_z$, $D^{(z)}_{ww}=D^{(z)}_I/(2 I_z)$, $D^{(R)}_{II}=2 D^{(R)}_I I_R$, and $D^{(R)}_{ww}=D^{(R)}_I/(2 I_R)$. These diffusion coefficients approximately preserve the form of the unperturbed DF (equation~[\ref{DF_MW}]) of the MW disk \citep[][]{Binney.Lacey.88}.

From the age-velocity dispersion relation of the MW disk stars, we deem $D^{(z)}_I$ and $D^{(R)}_I$ as constants and approximate them as $D^{(z)}_I=\left<I_z\right>/T_{\rm disk}$ and $D^{(R)}_I=\left<I_R\right>/T_{\rm disk}$, with $T_{\rm disk}$ the age of the disk \citep[][]{Tremaine.etal.22}. Here, $\left<I_a\right>=\int \rmd I_a \, I_a \, f_0 \, /\int \rmd I_a \, f_0$, where $a$ is either $z$ or $R$. This yields

\begin{align}
C[f] &\approx D^{(z)}_I \frac{\partial}{\partial I_z} \left(I_z \frac{\partial f}{\partial I_z}\right) + \frac{D^{(z)}_I}{4 I_z} \frac{\partial^2 f}{\partial w_z^2} \nonumber \\
&+ D^{(R)}_I \frac{\partial}{\partial I_R} \left(I_R \frac{\partial f}{\partial I_R}\right) + \frac{D^{(R)}_I}{4 I_R} \frac{\partial^2 f}{\partial w_R^2}.
\end{align}
Substituting this form of the collision operator in equation~(\ref{CBE_perturb_3}) and neglecting the $I_R$ diffusion term, since the response does not develop strong $I_R$ gradients due to the mild dependence of the stellar frequencies on $I_R$, we obtain the evolution equation for the response, $f_1$, given in equation~(\ref{CBE_perturb_gen}). Using the Fourier series expansions of $f_1$ and $\Phi_\rmP$ given in equations~(\ref{fourier_series_gen}) in equation~(\ref{CBE_perturb_gen}), the evolution equation for the Fourier mode of $f_1$ (equation~[\ref{f1nk_de_3}]) can be obtained and rearranged to yield the following inhomogeneous differential equation for $f_{1,n\ell m}$:

\begin{align}
&\frac{\partial f_{1,n\ell m}}{\partial t}+i(n\Omega_z+\ell\Omega_R+m\Omega_\phi)f_{1,n\ell m} - D^{(z)}_I\frac{\partial}{\partial I_z}\left(I_z\frac{\partial f_{1,n\ell m}}{\partial I_z}\right) \nonumber \\
& + \left[\frac{n^2 D^{(z)}_I}{4 I_z} + \frac{\ell^2 D^{(R)}_I}{4 I_R}\right] f_{1,n\ell m} = i\left(n\frac{\partial f_0}{\partial I_z}+\ell\frac{\partial f_0}{\partial I_R} + m\frac{\partial f_0}{\partial I_\phi}\right)\Phi_{n\ell m}(I_z,I_R,I_\phi).
\label{f1nk_de_app1}
\end{align}
If $D^{(z)}_I$ is much smaller than $\sigma^2_z$, which is typically the case, then one can assume the $I_z$ diffusion term to be small, i.e., the effect of diffusion to be localized in $I_z$. More specifically, one can look for a solution in the neighborhood of $I_z=I_{z0}$, i.e., for $I_z=I_{z0}+\Delta I_z$, where $\Delta I_z \ll I_{z0}$ \citep[][]{Tremaine.etal.22}. Thus, equation~(\ref{f1nk_de_app1}) can be rewritten as

\begin{align}
&\frac{\partial f_{1,n\ell m}}{\partial t}+i\left[\left(n\Omega_{z0}+\ell\Omega_R+m\Omega_\phi\right)+n\Omega_{z1}\Delta I_z\right]f_{1,n\ell m} \nonumber \\
&- D^{(z)}_I I_{z0}\frac{\partial^2 f_{1,n\ell m}}{\partial {\left(\Delta I_z\right)}^2} + \left[\frac{n^2 D^{(z)}_I}{4 I_{z0}} + \frac{\ell^2 D^{(R)}_I}{4 I_R}\right] f_{1,n\ell m} \nonumber \\
&= i\left(n\frac{\partial f_0}{\partial I_{z0}}+\ell\frac{\partial f_0}{\partial I_R} + m\frac{\partial f_0}{\partial I_\phi}\right)\Phi_{n\ell m}(I_{z0},I_R,I_\phi).
\label{f1nk_de_app2}
\end{align}
The above inhomogeneous differential equation can be solved using the Green's function technique. With the initial condition, $f_{1,n\ell m}(t=t_i)=0$, and in the limit of $\Delta I_z \to 0$, one can obtain the following solution for $f_{1,n\ell m}$:

\begin{align}
f_{1,n\ell m}(I_{z0},I_R,I_\phi,t)&=i\left(n\frac{\partial f_0}{\partial I_{z0}}+\ell\frac{\partial f_0}{\partial I_R}+m\frac{\partial f_0}{\partial I_\phi}\right) \nonumber \\
&\times \int_{t_\rmi}^{t}\rmd \tau\, \calG_{n\ell m}(I_{z0},I_R,I_\phi,t-\tau)\, \Phi_{n\ell m}(I_{z0},I_R,I_\phi,\tau),
\end{align}
where $\calG_{n\ell m}(\bI,t-\tau)$ is the Green's function, i.e., the solution of the homogeneous equation,

\begin{align}
&\frac{\partial \calG_{n\ell m}}{\partial t}+i\left[\left(n\Omega_{z0}+\ell\Omega_R+m\Omega_\phi\right)+n\Omega_{z1}\Delta I_z\right]\calG_{n\ell m} - D^{(z)}_I I_{z0}\frac{\partial^2 \calG_{n\ell m}}{\partial {\left(\Delta I_z\right)}^2} \nonumber \\
&+ \left[\frac{n^2 D^{(z)}_I}{4 I_{z0}} + \frac{\ell^2 D^{(R)}_I}{4 I_R}\right] \calG_{n\ell m} = 0.
\label{Greens_func_sol1}
\end{align}

The Green's function can be computed as a function of $\Delta I_z$, and later evaluated in the $\Delta I_z \to 0$. The computation proceeds as follows. First, we expand $\calG_{n\ell m}$ as a continuous Fourier series:

\begin{align}
\calG_{n\ell m}(I_{z0}+\Delta I_z,I_R,I_\phi,t)=\int_{-\infty}^\infty \rmd \omega \exp{\left[i\omega t\right]}\, g_{n\ell m}(I_{z0}+\Delta I_z,I_R,I_\phi,\omega).
\end{align}
This reduces equation~(\ref{Greens_func_sol1}) to the following differential equation for $g_{n\ell m}$:

\begin{align}
&i\left[\left(\omega+n\Omega_{z0}+\ell\Omega_R+m\Omega_\phi\right)+n\Omega_{z1}\Delta I_z\right]\, g_{n\ell m} - D^{(z)}_I I_{z0}\frac{\partial^2 g_{n\ell m}}{\partial {\left(\Delta I_z\right)}^2} \nonumber\\
&+ \left[\frac{n^2 D^{(z)}_I}{4 I_{z0}} + \frac{\ell^2 D^{(R)}_I}{4 I_R}\right] g_{n\ell m} = 0,
\end{align}
where $\Omega_{z1}=\partial \Omega_z/\partial I_z$ evaluated at $I_z=I_{z0}$. The above equation can be written more concisely as

\begin{align}
\frac{\partial^2 g_{n\ell m}}{\partial x^2} + k x\, g_{n\ell m} = 0,
\label{Airy_eqn}
\end{align}
where

\begin{align}
x &= i\left(\Delta I_z + \frac{\omega+n\Omega_{z0}+\ell\Omega_R+m\Omega_\phi}{n\Omega_{z1}}\right) + \frac{D^{(z)}_I I_{z0}}{n\Omega_{z1}}\left(\frac{n^2 D^{(z)}_I}{4 I_{z0}}+\frac{\ell^2 D^{(R)}_I}{4 I_R}\right),\nonumber \\
k &= \frac{n\Omega_{z1}}{D^{(z)}_I I_{z0}}.
\label{xk}
\end{align}
The differential equation~(\ref{Airy_eqn}) is known as the Airy equation, which has the general solution,

\begin{align}
g_{n\ell m}(x) = c_1 A_i\left({(-k)}^{1/3} x\right) + c_2 B_i\left({(-k)}^{1/3} x\right),
\end{align}
with $c_1$ and $c_2$ constants, and $A_i$ and $B_i$ the Airy functions. The long time behaviour of $\calG_{n\ell m}$ is captured by the $A_i$ part of $g_{n\ell m}$. $A_i(z)$ is given by the following integral form:

\begin{align}
A_i(z) = \frac{1}{2\pi} \int_{-\infty}^\infty \rmd s \, \exp{\left[i s^3/3\right]} \exp{\left[i s z\right]}.
\end{align}
Upon substituting $x$ and $k$ from equations~(\ref{xk}) in $g_{n\ell m}\approx A_i\left({(-k)}^{1/3} x\right)$, and performing the inverse Fourier transform of $g_{n\ell m}$, we obtain

\begin{align}
\calG_{n\ell m}(I_{z0}+\Delta I_z,I_R,I_\phi,t) \approx \int_{-\infty}^\infty \rmd s \,\, \delta\left(t+is\frac{{\left(-k\right)}^{1/3}}{n\Omega_{z1}}\right) \exp{\left[i s^3/3\right]} \exp{\left[i s {\left(-k\right)}^{1/3} x'\right]},
\end{align}
where

\begin{align}
x' &= i\left(\Delta I_z + \frac{n\Omega_{z0}+\ell\Omega_R+m\Omega_\phi}{n\Omega_{z1}}\right) + \frac{D^{(z)}_I I_{z0}}{n\Omega_{z1}}\left(\frac{n^2 D^{(z)}_I}{4 I_{z0}}+\frac{\ell^2 D^{(R)}_I}{4 I_R}\right).
\label{x'}
\end{align}
Integrating over $s$, we obtain the following form for $\calG_{n\ell m}$:

\begin{align}
\calG_{n\ell m}(I_{z0}+\Delta I_z,I_R,I_\phi,t) &\approx \exp{\left[-i(n(\Omega_{z0}+\Omega_{z1}\Delta I_z)+\ell\Omega_R+m\Omega_\phi)t\right]} \nonumber \\
&\times \exp{\left[-\left(\frac{n^2 D^{(z)}_I}{4 I_{z0}}+\frac{\ell^2 D^{(R)}_I}{4 I_R}\right) t\right]}\, \exp{\left[-\frac{{\left(n\Omega_{z1}\right)}^2 D^{(z)}_I I_{z0}}{3}{t}^3\right]},
\end{align}
which, in the limit $\Delta I_z \to 0$, reduces to

\begin{align}
\calG_{n\ell m}(I_{z0},I_R,I_\phi,t) &\approx \exp{\left[-i(n\Omega_{z0}+\ell\Omega_R+m\Omega_\phi)t\right]} \nonumber \\
&\times \exp{\left[-\left(\frac{n^2 D^{(z)}_I}{4 I_{z0}}+\frac{\ell^2 D^{(R)}_I}{4 I_R}\right) t\right]}\, \exp{\left[-\frac{{\left(n\Omega_{z1}\right)}^2 D^{(z)}_I I_{z0}}{3}{t}^3\right]}.
\end{align}

\section{The unperturbed galaxy}
\label{sec:disk_model_app_3}

Under the radial epicyclic approximation (small $I_R$), the unperturbed DF, $f_0$, for a rotating MW-like disk galaxy can be well approximated as a pseudo-isothermal DF, i.e., written as a nearly isothermal separable function of the azimuthal, radial and vertical actions. Following \cite{Binney.10}, we write
\begin{align}
f_0 = \frac{1}{\pi} {\left(\frac{\Omega_\phi \Sigma}{\kappa\, \sigma^2_R}\right)}_{R_\rmc} \left(1+\tanh{\frac{L_z}{L_0}}\right)\times \exp{\left[-\frac{\kappa I_R}{\sigma^2_R}\right]} \times \frac{1}{\sqrt{2\pi}h_z\sigma_z} \exp{\left[-\frac{E_z(I_z)}{\sigma^2_z}\right]}\,.
\end{align}
The vertical structure of this disk is isothermal, while the radial profile is pseudo-isothermal. Here $\Sigma=\Sigma(R)$ is the surface density of the disk, $L_z$ is the $z$-component of the angular momentum, which is equal to $I_\phi$, $R_\rmc = R_\rmc(L_z)$ is the guiding radius, $\Omega_\phi$ is the circular frequency, and $\kappa = \kappa(\Rc) = \lim_{I_R \to 0}{\Omega_R}$ is the radial epicyclic frequency \citep[][]{Binney.Tremaine.87}. If $L_0$ is sufficiently small, then we can further approximate the above form for $f_0$ as
\begin{align}
f_0 \approx \frac{\sqrt{2}}{\pi^{3/2} \, \sigma_z h_z} {\left(\frac{\Omega_\phi \Sigma}{\kappa\, \sigma^2_R}\right)}_{\Rc} \, \exp{\left[-\frac{\kappa I_R}{\sigma^2_R}\right]} \, \exp{\left[-\frac{E_z(I_z)}{\sigma^2_z}\right]} \, \Theta(L_z)\,,
\label{DF_MW_app}
\end{align}
where $\Theta(x)$ is the Heaviside step function. Thus we assume that the entire galaxy is composed of prograde stars with $L_z>0$. 

The corresponding density profile can be written as a product of an exponential radial profile and an isothermal ($\sech^2$) vertical profile, i.e.,
\begin{align}
\rho(R,z) = \rho_\rmc \, \exp{\left[-\frac{R}{h_R}\right]} \, \sech^2{\left(\frac{z}{h_z}\right)},
\label{disk_exp_iso_rho_app}
\end{align}
where $h_R$ and $h_z$ are the radial and vertical scale heights, respectively. Throughout we adopt the thin disk limit, i.e., $h_z \ll h_R$. The surface density profile is given by
\begin{align}
\Sigma(R) = \int_{-\infty}^{\infty} \rmd z\, \rho(R,z) = \Sigma_\rmc \exp{\left[-\frac{R}{h_R}\right]},
\label{disk_exp_iso_Sigma_app}
\end{align}
where $\Sigma_\rmc=\rho_\rmc h_z$ is the central surface density of the disk. We assume a radially varying vertical velocity dispersion, $\sigma_z$, satisfying $\sigma^2_z(R) = 2\pi G h_z \Sigma(R)$ \citep[][]{Binney.Tremaine.08}. We assume a similar profile for $\sigma^2_R$ such that the ratio, $\sigma_R/\sigma_z$ is constant throughout the disk \citep[][]{Binney.10} and equal to the value at the Solar vicinity.

Throughout chapter~\ref{chapter: paper3}, for the ease of computation of the frequencies (because of a simple analytic form of the potential), we approximate the above density profile by a combination of three \cite{Miyamoto.Nagai.75} disk profiles \citep[][]{Smith.etal.15}, i.e., the 3MN profile as implemented in the {\tt Gala} Python package \citep[][]{gala, gala_code_adrian}. The corresponding disk potential is given by
\begin{align}
\Phi_\rmd(R,z) = -\sum_{i=1}^{3} \frac{G M_i}{\sqrt{R^2+{\left(a_i+\sqrt{z^2+b^2_i}\right)}^2}},
\label{Phi0_disk_app}
\end{align}
where $M_i$, $a_i$ and $b_i$, with $i=1,2,3$, are the mass, scale radius and scale height corresponding to each of the MN profiles.

The MW disk is believed to be embedded in a much more extended DM halo, which we model using a spherical NFW \citep[][]{Navarro.etal.97} profile with potential

\begin{align}
\Phi_\rmh(R,z)=-\frac{G \Mvir}{\Rvir} \, \frac{c}{f(c)} \,
\frac{\ln(1+r/r_\rms)}{r/r_\rms}.
\label{Phi0_halo_app}
\end{align}
Here $\Mvir$ is the virial mass of the halo, $r_s$ is the scale radius, $c=R_{\rm vir}/r_s$ is the concentration ($R_{\rm vir}$ is the virial radius), and $f(c)=\ln{\left(1+c\right)}-c/(1+c)$. The combined potential experienced by the disk stars is thus given by
\begin{align}
\Phi_0(R,z)=\Phi_\rmd(R,z)+\Phi_\rmh(R,z).
\end{align}

\section{Fourier coefficients of spiral arm or bar perturbing potential}
\label{App:fourier_spiral}

An essential ingredient of the disk response to spiral arm or bar perturbations is the Fourier component of the perturber potential, $\Phi_{n\ell m}$. This can be computed as follows:

\begin{align}
\Phi_{n\ell m}(\bI,t) &= \frac{1}{{\left(2\pi\right)}^3} \int_0^{2\pi} \rmd w_z \int_0^{2\pi} \rmd w_R \int_0^{2\pi} \rmd w_\phi\, \exp{\left[-i(n w_z + \ell w_R + m w_\phi)\right]}\, \Phi_\rmP \left(\br,t\right).
\end{align}
To evaluate this first we need to calculate $\br=(z,R,\phi)$ as a function of $\left(\bw,\bI\right)=(w_z,w_\phi,w_R,I_z,I_\phi,I_R)$ where $I_\phi=L_z$, the angular momentum. Under the epicyclic approximation, $R$ can be expressed as a sum of the guiding radius and an oscillating epicyclic term, i.e.,

\begin{align}
R\approx \Rc(L_z)+\sqrt{\frac{2 I_R}{\kappa}}\sin{w_R}\,,
\label{R_epi_app}
\end{align}
and the azimuthal angle, $w_\phi$, is given by

\begin{align}
w_\phi \approx \phi - \frac{2\,\Omega_\phi}{\Rc \kappa} \sqrt{\frac{2 I_R}{\kappa}}\cos{\theta_R}\,.
\end{align}

The vertical distance $z$ from the mid-plane is related to $\Rc(L_z)$ and $\left(w_z,I_z\right)$, according to
\begin{align}
w_z=\Omega_z(\Rc,I_z)\int_0^z \frac{\rmd z'}{\sqrt{2\left[E_z(\Rc,I_z)-\Phi_z(\Rc,z')\right]}},
\label{z_wz_Iz_app}
\end{align}
where $\Omega_z(\Rc,I_z)=2\pi/T_z(\Rc,I_z)$, with $T_z(\Rc,I_z)$ given by Equation~(\ref{T_z_3}). The above equation can be numerically inverted to obtain $z(\Rc,w_z,I_z)$. 

Upon substituting the above expressions for $R$, $\phi$ and $z$ in terms of $(\bw,\bI)$ in the expression for $\Phi_\rmP$ given in equation~(\ref{Phip_spiral}), we obtain

\begin{align}
\Phi_{n\ell m} \left(\bI,t\right) &= -\frac{2\pi G\Sigma_\rmP}{k_R}\left(\sum_{m_\phi=0,2,-2}\delta_{m,m_\phi}\right) \frac{\sgn(m)\exp{\left[i\,{\rm sgn}(m) k_R \Rc(I_\phi)\right]}}{2i}\nonumber \\
&\times \exp{\left[i\, \ell \tan^{-1}{\frac{2 m \Omega_\phi}{\Rc \kappa}\sqrt{\frac{2 I_R}{\kappa}}}\right]} \, J_\ell\left(\sqrt{k_R^2+{\left(\frac{2 m \Omega_\phi}{\Rc \kappa}\right)}^2} \sqrt{\frac{2 I_R}{\kappa}}\right) \nonumber \\
&\times\left[\alpha\,\calM_\rmo(t)\Phi_n^{(\rmo)}(I_z)+\calM_\rme(t)\Phi_n^{(\rme)}(I_z)\right] \exp{\left[-i m \Omega_\rmP t\right]}\,,
\label{spiral_fourier_app}
\end{align}
where $J_\ell$ is the $\ell^{\rm th}$ order Bessel function of the first kind,
\begin{align}
\sgn(m) &=
\begin{cases}
1, & m\geq 0, \\
-1, & m<0,
\end{cases}
\end{align}
and $\Phi_n^{(\rmo)}(I_z)$ and $\Phi_n^{(\rme)}(I_z)$ are given by
\begin{align}
\Phi_n^{(\rmo)}(I_z) &= \frac{1}{2\pi} \int_0^{2\pi} \rmd w_z \sin{n w_z}\, \calF_\rmo\left(z,k_z^{(\rmo)}\right),\nonumber \\
\Phi_n^{(\rme)}(I_z) &= \frac{1}{2\pi} \int_0^{2\pi} \rmd w_z \cos{n w_z}\, \calF_\rme\left(z,k_z^{(\rme)}\right).
\label{wz_int_spiral_app}
\end{align}
In deriving equation~(\ref{spiral_fourier_app}) we have used the Hansen-Bessel formula which provides the following integral representation for Bessel functions of the first kind,
\begin{align}
\int_0^{2\pi} \rmd x \, \exp{\left[-i \ell x\right]} \, \exp{\left[i \alpha \sin{x}\right]} = 2\pi J_\ell\left(\alpha\right),
\label{Bessel_int_id}
\end{align}
and the identity for expansion in products of Bessel functions given in equation~(8.530.2) of \cite{Gradshteyn.Ryzhik.65} (see also section~6.1 of \cite{Binney.Lacey.88}). We have also used the identity,
\begin{align}
\int_0^{2\pi}\rmd \phi\, \exp{\left[-i m\phi\right]} = 2\pi\,\delta_{m,0}\,.
\end{align}

\section{Perturbation by encounter with satellite galaxy}

\subsection{Computation of the disk response}
\label{App:sat_disk_Resp}

To evaluate the disk response to satellite encounters using equation~(\ref{f1nk_gensol_f0}) we first evaluate the $\tau$ integral (with $t_\rmi\to -\infty$) of the satellite potential given in equation~(\ref{sat_pot}) and then compute the Fourier transform of the result. This yields the expression for the response in equation~(\ref{f1nk_gensol_f0}) with
\begin{align}
&\calI_{n\ell m}(\bI,t)=\exp{\left[-i\Omega t\right]}\int_{-\infty}^{t}\rmd \tau\, \exp{\left[i\Omega \tau\right]}\, \Phi_{n\ell m}\left(\bI,\tau\right) \nonumber\\
&=\frac{\exp{\left[-i\Omega t\right]}}{{\left(2\pi\right)}^3} \int_0^{2\pi}\rmd w_z \exp{\left[-inw_z\right]}\int_0^{2\pi}\rmd w_R \exp{\left[-i\ell w_R\right]}\int_0^{2\pi}\rmd w_\phi \exp{\left[-imw_\phi\right]} \nonumber \\
&\times \int_{-\infty}^{t}\rmd \tau\, \exp{\left[i\Omega \tau\right]}\, \Phi_\rmP(z,R,\phi,\tau),
\label{Inlm_FT_app}
\end{align}
where
\begin{align}
\Omega = n\Omega_z + \ell\Omega_R + m\Omega_\phi.
\end{align}
We perform the inner $\tau$ integral of $\Phi_\rmP$ to obtain
\begin{align}
&\int_{-\infty}^{t}\rmd \tau\, \exp{\left[i\Omega \tau\right]}\, \Phi_\rmP(z,R,\phi,\tau) = -\frac{G M_\rmP}{\vp}\exp{\left[i\frac{\Omega\calS}{\vp}\right]} \int_{-\infty}^{t-\calS/\vp}\rmd\tau \frac{\exp{\left[i\Omega \tau\right]}}{\sqrt{\tau^2+{\left(\calR^2+\varepsilon^2\right)}/{v^2_\rmP}}}\nonumber \\
&=-\frac{G M_\rmP}{\vp}\exp{\left[i\frac{\Omega\calS}{\vp}\right]} \int_{-\infty}^{\left(\vp t-\calS\right)/\sqrt{\calR^2+\varepsilon^2}}\rmd x \frac{\exp{\left[i\left(\Omega\sqrt{\calR^2+\varepsilon^2}/\vp\right) x\right]}}{\sqrt{x^2+1}} \nonumber \\
&=-\frac{2\, G M_\rmP}{\vp}\exp{\left[i\frac{\Omega\calS}{\vp}\right]} K_{0i}\left(\frac{\Omega\sqrt{\calR^2+\varepsilon^2}}{\vp},\frac{\vp t-\calS}{\sqrt{\calR^2+\varepsilon^2}}\right).
\label{K0i_app}
\end{align}
Here $K_{0i}$ is defined as
\begin{align}
K_{0i}(\alpha,\beta) = \frac{1}{2}\int_{-\infty}^\beta\rmd x\, \frac{\exp{\left[i\alpha x\right]}}{\sqrt{x^2+1}},
\end{align}
which asymptotes to the zero-th order modified Bessel function of the second kind, $K_0\left(\left|\alpha\right|\right)$, in the limit $\beta \to \infty$. $\calR$ and $\calS$ are respectively the perpendicular and parallel projections along the direction of ${\bf \vp}$ of the vector connecting the point of observation, $(z,R,\phi)$, with the point of impact, and are given by
\begin{align}
\calR^2&= {\left[R\sin{(\phi-\phip)}+\rd\sin{\phip}\right]}^2 + {\left[(R\cos{(\phi-\phip)}-\rd\cos{\phip})\cos{\thetap}-z\sin{\thetap}\right]}^2\nonumber \\
\calS &= (R\cos{(\phi-\phip)}-\rd\cos{\phip})\sin{\thetap}+z\cos{\thetap}.
\label{sat_RS_app}
\end{align}
In deriving equation~(\ref{K0i_app}), we have only considered the direct term in the expression for $\Phi_\rmP$ given in equation~(\ref{sat_pot}); the indirect term turns out to be sub-dominant.

In the large time limit, i.e., $t\gg \calS/\vp$, $K_{0i}$ asymptotes to $K_0\left(\left|\Omega\right|\sqrt{\calR^2+\varepsilon^2}/\vp\right)$. We substitute $\phi\approx w_\phi$ and the expressions for $R$ and $z$ in terms of $(\bw,\bI)$ given in equations~(\ref{R_epi_app}) and (\ref{z_wz_Iz_app}) in the above expressions for $\calR$ and $\calS$. Further substituting the resultant $\tau$ integral from equation~(\ref{K0i_app}) in equation~(\ref{Inlm_FT_app}), adopting the small $I_R$ limit and performing the $w_R$ integral, we obtain 
\begin{align}
&\calI_{n\ell m}(\bI,t)\approx-\frac{2G M_\rmP}{\vp} \exp{\left[-i\Omega t\right]} \times \exp{\left[-i\frac{\Omega\sin{\thetap}\cos{\phip}}{\vp} \rd\right]} \nonumber \\ 
&\times \exp{\left[i\, \ell \tan^{-1}{\frac{2 m \Omega_\phi}{\Rc \kappa}\sqrt{\frac{2 I_R}{\kappa}}}\right]} \times \frac{1}{{\left(2\pi\right)}^2} \int_0^{2\pi} \rmd w_z \exp{\left[-in w_z\right]} \exp{\left[i\frac{\Omega \cos{\thetap}}{\vp} z\right]} \nonumber \\
&\times \int_0^{2\pi} \rmd \phi\,\exp{\left[-im\phi\right]}\, \exp{\left[i\frac{\Omega \sin{\thetap} \cos{\left(\phi-\phip\right)}}{\vp} \Rc\right]} \nonumber \\
&\times J_\ell\left(\sqrt{{\left(\frac{\Omega \sin{\thetap}}{\vp}\right)}^2\cos^2{\left(\phi-\phip\right)} + {\left(\frac{2 m \Omega_\phi}{\Rc \kappa}\right)}^2} \sqrt{\frac{2 I_R}{\kappa}}\right) \nonumber\\
&\times K_{0i}\left(\frac{\Omega\sqrt{\calR^2_\rmc+\varepsilon^2}}{\vp},\frac{\vp t-\calS_\rmc}{\sqrt{\calR^2_\rmc+\varepsilon^2}}\right),
\label{sat_gen_app}
\end{align}
where $\calR_\rmc=\calR(R=\Rc)$ and $\calS_\rmc=\calS(R=\Rc)$. Here we have used the integral representation of Bessel functions of the first kind given in equation~(\ref{Bessel_int_id}) and the identity given in equation~(8.530.2) of \cite{Gradshteyn.Ryzhik.65}.

The expression for $\calI_{n\ell m}$ given in equation~(\ref{sat_gen_app}) consists of the leading order expansion in $\sqrt{2 I_R/\kappa}$. A more precise expression that is accurate up to second order in $\sqrt{2 I_R/\kappa}$ is given, in the large time limit, as
\begin{align}
&\calI_{n\ell m}(\bI,t)\approx-\frac{2G M_\rmP}{\vp} \exp{\left[-i\Omega t\right]} \times \exp{\left[-i\frac{\Omega\sin{\thetap}\cos{\phip}}{\vp} \rd\right]} \nonumber \\ 
&\times \frac{1}{{\left(2\pi\right)}^2} \int_0^{2\pi} \rmd w_z \exp{\left[-in w_z\right]} \exp{\left[i\frac{\Omega \cos{\thetap}}{\vp} z\right]} \nonumber \\ &\int_0^{2\pi} \rmd \phi\,\exp{\left[-im\phi\right]}\, \exp{\left[i\frac{\Omega \sin{\thetap} \cos{\left(\phi-\phip\right)}}{\vp} \Rc\right]} \nonumber \\
&\times \exp{\left[i\, \ell \tan^{-1}{\frac{2 m \Omega_\phi}{\Rc \kappa}\sqrt{\frac{2 I_R}{\kappa}}}\right]} \, \left[\zeta^{(0)} J_\ell\left(\chi\right) - i\zeta^{(1)} J'_\ell\left(\chi\right) -\frac{1}{2}\zeta^{(2)} J''_\ell\left(\chi\right) \right],
\label{sat_gen_IR_app}
\end{align}
where
\begin{align}
\chi = \sqrt{{\left(\frac{\Omega \sin{\thetap}}{\vp}\right)}^2\cos^2{\left(\phi-\phip\right)} + {\left(\frac{2 m \Omega_\phi}{\Rc \kappa}\right)}^2} \sqrt{\frac{2 I_R}{\kappa}},
\end{align}
and
\begin{align}
\zeta^{(0)} &= K_0\left(\eta\right), \nonumber \\
\zeta^{(1)} &= \sqrt{\frac{2 I_R}{\kappa}}\, \frac{\partial \calR_\rmc}{\partial \Rc} \frac{\calR_\rmc}{\sqrt{\calR^2_\rmc+\varepsilon^2}} \frac{\left|\Omega\right|}{\vp} K'_0\left(\eta\right), \nonumber \\
\zeta^{(2)} &= \frac{2 I_R}{\kappa}\, \left[{\left(\frac{\partial \calR_\rmc}{\partial \Rc}\right)}^2\frac{\calR^2_\rmc}{\calR^2_\rmc+\varepsilon^2}\frac{\Omega^2}{v^2_\rmP}K''_0(\eta)\right. \nonumber \\
&\left.+ \left\{\frac{\partial^2 \calR_\rmc}{\partial R^2_c}\frac{\calR_\rmc}{\sqrt{\calR^2_\rmc+\varepsilon^2}}+{\left(\frac{\partial \calR_\rmc}{\partial \Rc}\right)}^2\frac{\varepsilon^2}{{\left(\calR^2_\rmc+\varepsilon^2\right)}^{3/2}}\right\} \frac{\left|\Omega\right|}{\vp}K'_0(\eta)\right],
\end{align}
with
\begin{align}
\eta &= \frac{\left|\Omega\right|\sqrt{\calR^2_\rmc+\varepsilon^2}}{\vp}.
\label{eta_app}
\end{align}
Here each prime denotes a single derivative of the function with respect to its argument.

\subsection{Special case: disk response for face-on impulsive encounters}
\label{App:special}

The disk response in the general case, expressed by equation~(\ref{sat_gen}), depends on several encounter parameters: $\rd$, $\thetap$, $\phip$, and is complicated to evaluate. Therefore, as a sanity check, here we compute the response as well as corresponding energy change for the special case of a satellite undergoing an impulsive, perpendicular passage through the center of the disk. 

As shown in \cite{vdBosch.etal.18a} \citep[see also][]{Banik.vdBosch.21b}, the total energy change due to a head-on encounter of velocity $v_\rmP$ with a Plummer sphere of mass $M_\rmP$ and size $\varepsilon$ is given by
\begin{align}
 \Delta E = 4 \pi \left({G M_\rmp \over v_\rmP}\right)^2 \int_0^{\infty} I^2_0(R) \Sigma(R) {\rmd R \over R},
\end{align}
where
\begin{align}
 I_0(R) = \int_1^{\infty} {M_\rmP(\zeta R) \over M_\rmp} \, {\rmd \zeta \over \zeta^2 (\zeta^2 - 1)^{1/2}}.
\end{align}
Using that the enclosed mass profile of a Plummer sphere is given by $M_\rmP(R) = M_\rmP R^3 (R^2 + \varepsilon^2)^{-3/2}$, we have that $I_0(R)=R^2/(R^2+\varepsilon^2)$, which yields
\begin{align}
 \Delta E = 4 \pi \left({G M_\rmp \over v_\rmP}\right)^2 \int_0^{\infty} \Sigma(R) {R^3 \rmd R \over (R^2 + \varepsilon^2)^2}.
\label{deltaE_impulse}
\end{align}

Now we compute the disk response to the face-on satellite encounter using equations~(\ref{f1nk_gensol_f0}) and (\ref{sat_gen_IR_app}-\ref{eta_app}). For a perpendicular face-on impact through the center of the disk we have $\rd=0$ and $\thetap=0$, implying that $\calR_\rmc$ becomes $\Rc$. The corresponding response is greatly simplified. In the large time and small $I_R$ limit, it is given by equation~(\ref{f1nk_gensol_f0}) with

\begin{align}
\calI_{n\ell m}(\bI,t) &\approx -\frac{2GM_\rmP}{\vp} \exp{\left[-i\Omega t\right]}\, \delta_{m,0}\times \frac{1}{2\pi}\int_0^{2\pi} \rmd w_z \exp{\left[-in w_z\right]} \exp{\left[i\frac{\Omega z}{\vp}\right]} \nonumber \\
&\times \frac{1}{2\pi} \int_0^{2\pi} \rmd w_R\, \exp{\left[-i\ell w_R\right]}\, K_0\left[\frac{\left|\Omega\right|}{\vp}\sqrt{\varepsilon^2+{\left(\Rc+\sqrt{\frac{2I_R}{\kappa}}\sin{w_R}\right)}^2}\right],
\end{align}
where the $\phi$ integral only leaves contribution from the axisymmetric $m=0$ mode. The $w_R$ integrand can be expanded as a Taylor series and the $w_R$ integral can be performed to yield the following leading order expression for $\calI_{n\ell m}$:
\begin{align}
\calI_{n\ell m}(\bI,t) &\approx i\frac{GM_\rmP}{\vp} \exp{\left[-i\Omega t\right]}\, \delta_{m,0}\,\left(\delta_{\ell,1}-\delta_{\ell,-1}\right) \times \frac{1}{2\pi}\int_0^{2\pi} \rmd w_z \exp{\left[-in w_z\right]} \exp{\left[i\frac{\Omega z}{\vp}\right]} \nonumber \\
& \times \sqrt{\frac{2 I_R}{\kappa}} \frac{\Rc}{\sqrt{\varepsilon^2+R^2_c}}\, \frac{\left|\Omega\right|}{\vp} K'_0\left[\frac{\left|\Omega\right|}{\vp}\sqrt{\varepsilon^2+R^2_c}\right].
\end{align}
In the impulsive limit, $\vp\to \infty$, this becomes
\begin{align}
\calI_{n\ell m}(\bI,t) &\approx i\,\delta_{n,0} \delta_{m,0} \left(\delta_{\ell,1}-\delta_{\ell,-1}\right) \frac{GM_\rmP}{\vp} \sqrt{\frac{2 I_R}{\kappa}} \frac{\Rc}{\varepsilon^2+R^2_c} \exp{\left[-i\, \ell\kappa\, t\right]}\,,
\end{align}
which can be substituted in equation~(\ref{f1nk_gensol_f0}) to yield
\begin{align}
&f_{1,n\ell m}\left(\bI,t\right) = f_0(\bI) \times \delta_{n,0} \delta_{m,0} \left(\delta_{\ell,1}-\delta_{\ell,-1}\right) \frac{G M_\rmP}{\vp} \frac{\ell\kappa}{\sigma^2_R} \sqrt{\frac{2 I_R}{\kappa}} \frac{\Rc}{\varepsilon^2+R^2_c} \exp{\left[-i\, \ell\kappa\, t\right]}\,,
\label{f1nlm_sat_app}
\end{align}
with $f_0$ given by equation~(\ref{DF_MW}). Hence, the response is given by
\begin{align}
f_1\left(\bw,\bI,t\right)&=\sum_{n=-\infty}^{\infty} \sum_{\ell=-\infty}^{\infty} \sum_{m=-\infty}^{\infty} \exp{\left[i (n w_z + \ell w_R + m w_\phi)\right]}\, f_{1,n\ell m}(\bI,t) \nonumber \\ &= f_0(\bI) \times \frac{2 G M_\rmP}{\vp} \frac{\sqrt{2 \kappa I_R}}{\sigma^2_R} \frac{\Rc}{\varepsilon^2+R^2_c} \cos{\left(w_R-\kappa t\right)}\,,
\label{f1_sat_spl}
\end{align}
which shows that the satellite passage introduces a relative overdensity, $f_1\left(\bw,\bI,t\right)/f_0(\bI)$, that scales as $\sim \Rc/\left(\varepsilon^2+R^2_c\right)$, which increases from zero at the center, peaks at $\Rc=\varepsilon$, and asymptotes to zero again at large $\Rc$. The $\cos(w_R - \kappa t)$-term describes the radial epicyclic oscillations in the response.

To compute the energy change due to the impact, we note that $\rmd E/\rmd t = \partial E/\partial \bI \cdot \rmd \bI/\rmd t$, where $\partial E/\partial \bI = {\bf\Omega}=(\Omega_z,\Omega_R,\Omega_\phi)$ and $\rmd \bI/\rmd t = \partial \Phi_\rmP/\partial \bw$ from Hamilton's equations of motion. Thus the total phase-averaged energy injected per unit phase-space can be obtained as follows:
\begin{align}
\left<\Delta E \left(\bI\right)\right> &= \frac{1}{{\left(2\pi\right)}^3} \int \rmd \bw \int_{-\infty}^{\infty} \rmd t \frac{\rmd E}{\rmd t} f_1(\bI,t) = \frac{1}{{\left(2\pi\right)}^3} \int \rmd \bw \int_{-\infty}^{\infty} \rmd t\; {\bf\Omega} \cdot \frac{\partial \Phi_\rmP}{\partial \bw} f_1(\bI,t).
\end{align}
We can substitute the Fourier series expansions of $\Phi_\rmP$ and $f_1$ given in equations~(\ref{fourier_series_gen}) in the above expression and integrate over $\bw$ to obtain \citep[][]{Weinberg.94a,Weinberg.94b}
\begin{align}
\left<\Delta E \left(\bI\right)\right> &= i\sum_{n\ell m} \left(n\Omega_z+\ell\kappa+m\Omega_\phi\right) \int_{-\infty}^{\infty} \rmd t\, \Phi^*_{n\ell m}(\bI,t) f_{1,n\ell m}(\bI,t).
\label{delE_sat_app}
\end{align}
We can now substitute the form of $\Phi_\rmP$ for a Plummer perturber given in equation~(\ref{sat_pot}), with $\brp$ and $\br$ given by equations~(\ref{rp_sat}) and (\ref{r_sat}). The time integral can thus be written as
\begin{align}
&\int_{-\infty}^{\infty} \rmd t\, \Phi^*_{n\ell m}(\bI,t) f_{1,n\ell m}(\bI,t) \nonumber \\
&= - \frac{1}{{\left(2\pi\right)}^3} \int_0^{2\pi}\rmd w_z \exp{\left[inw_z\right]}\int_0^{2\pi}\rmd w_R \exp{\left[i\ell w_R\right]}\int_0^{2\pi}\rmd w_\phi \exp{\left[imw_\phi\right]} \nonumber \\
&\times \int_{-\infty}^{\infty} \rmd t \frac{G M_\rmP}{\sqrt{{\left(\vp t - z\right)}^2+R^2+\varepsilon^2}} f_{1,n\ell m}(\bI,t).
\end{align}
Using equations~(\ref{R_epi_app}) and (\ref{z_wz_Iz_app}) to express $R$ and $z$ in terms of $(\bw,\bI)$, and substituting the form for $f_{1,n\ell m}(\bI,t)$ from equation~(\ref{f1nlm_sat_app}), we can perform the above integrals over $\bw$ and $t$. Substituting the result in equation~(\ref{delE_sat_app}) we obtain
\begin{align}
\left<\Delta E \left(\bI\right)\right> &= {\left(\frac{G M_\rmP}{\vp}\right)}^2 f_0(\bI)\, \frac{2\kappa I_R}{\sigma^2_R}\, \frac{R^2_c}{{\left(\varepsilon^2+R^2_c\right)}^2}.
\end{align}

The total energy, $\Delta E_{\rm tot}$, imparted into the disk by the impulsive satellite passage can be computed by integrating the above expression over $\bI$ and $\bw$ (which simply introduces a factor of ${\left(2\pi\right)}^3$ since $\left<\Delta E \left(\bI\right)\right>$ is already phase-averaged), using equation~(\ref{DF_MW}) and transforming from $L_z$ to $\Rc$ using the Jacobian $\rmd L_z/\rmd \Rc = \Rc \kappa^2/2 \Omega_\phi$. This yields
\begin{align}
\Delta E_{\rm tot} = 4\pi{\left(\frac{GM_\rmP}{\vp}\right)}^2 \int_0^{\infty}\rmd \Rc\, \Rc\, \Sigma(\Rc) \frac{\Rc^2}{{\left(\varepsilon^2+\Rc^2\right)}^2}.
\end{align}
This is indeed the expression for $\Delta E_{\rm tot}$ derived under the impulse approximation given by equation~(\ref{deltaE_impulse}).

%
    \end{subappendices}
    
%
%
%

\chapter{A Self-Consistent, Time-Dependent Treatment of Dynamical Friction:\\ New Insights regarding Core Stalling and Dynamical Buoyancy} 
\label{chapter: paper4}

\begin{center}

This chapter has been published as:

\vspace*{5pt}

\author{Uddipan Banik,
   Frank~C.~van den Bosch
}

\vspace*{5pt}

\textit{The Astrophysical Journal}, Volume 912, Number 1, Page 43\\

\textit{\citep[][]{Banik.vdBosch.21a}}\\

\end{center}


\section{Introduction}
\label{sec:intro_4}

Dynamical friction is an important ingredient of hierarchical structure formation. It is the dynamical process by which galaxies merge, and by which globular clusters and black holes sink to the centers of their host systems where they can form bulges and binary black holes, respectively. In his seminal 1943 paper, Chandrasekhar showed that dynamical friction arises from the transfer of energy and momentum from the subject to the individual particles that make up the host system traversed by the subject. In particular, \cite{Chandrasekhar.43} considered a subject mass $M$ moving on a straight orbit through a uniform and isotropic sea of background (or `field') particles of mass $m \ll M$. When a field particle encounters the subject, it experiences velocity changes $\Delta v_{\perp}$ and $\Delta v_{\parallel}$ in the directions perpendicular and parallel to the direction of the relative velocity. Chandrasekhar summed the velocity changes from the encounters between the subject mass and all field particles, treating them as independent two-body encounters, and showed that the net result is a frictional force acting on $M$ given by
\begin{equation}\label{aDF}
\bF_{\rm DF} = - \frac{4 \pi G^2 M^2}{v^2} \, \ln\Lambda \, \rho(<v) \, \frac{\bv}{v}\,,
\end{equation}
\citep[see e.g.,][]{MBW10}. Here $\bv$ is the velocity of the subject mass, $\rho(<v)$ is the density of field particles with a speed less than $v = \vert \bv \vert$, and $\ln\Lambda = \ln[b_{\rm max}/b_{\rm min}]$ is the Coulomb logarithm, with $b_{\rm max}$ and $b_{\rm min}$ the maximum and minimum impact parameters of the encounters contributing to the drag.

Equation~(\ref{aDF}) is used routinely in astrophysics, even though it formally only applies to a uniform, isotropic background of field particles. While numerous studies have shown that it gives a reasonably accurate description of the orbital decay rates in galaxies and dark matter halos \citep[e.g.,][]{Lin.Tremaine.83, Cora.etal.97, vdBosch.etal.99, Hashimoto.etal.03, Boylan-Kolchin.etal.08, Jiang.etal.08}, there are also cases in which it clearly fails. For instance, according to equation~(\ref{aDF}) the drag force is proportional to the local density. Hence, a subject mass orbiting outside of a galaxy or halo of finite extent should experience no drag. This is inconsistent with numerical experiments, which show that even in such cases the subject loses orbital angular momentum \citep[][]{Lin.Tremaine.83}. Another example where the standard treatment of dynamical friction fails is `core-stalling', the cessation of dynamical friction in the central constant-density core of a halo or galaxy \citep[e.g.,][]{Read.etal.06c, Inoue.11, Cole.etal.12, Petts.etal.15, Petts.etal.16, DuttaChowdhury.etal.19}.

Since the seminal work by Chandrasekhar, dynamical friction has been studied using a variety of different techniques and aproaches. This includes the Fokker-Planck method, in which dynamical friction arises from the momentum exchange described by the first-order diffusion coefficient \citep[][]{Rosenbluth.etal.57, Binney.Tremaine.08}, stochastic approaches based on the fluctuation-dissipation theorem, in which dynamical friction arises from a correlation between the perturber's velocity vector and the stochastic force it experiences from the field particles \citep[][]{Bekenstein.Maoz.92, Maoz.93, Nelson.Tremaine.99, Fouvry.Bar-Or.18}, and a wide variety of methods that treat dynamical friction as a drag force arising from a `wake', or `polarization cloud' developing behind the perturber \citep[][]{Marochnik.68, Kalnajs.71, Mulder.83, Weinberg.89, Colpi.Pallavicini.98}. An excellent in-depth account of how all these methods relate to a generalized Landau equation derived from a truncation of the BBGKY hierarchy issued from the Liouville equation can be found in \citet{Chavanis.13}.

A shortcoming of many, though not all, of these methods is that they only treat dynamical friction as a local phenomenon and/or that they have only been worked out for perturbers moving through homogeneous density distributions. The first study to overcome this, and to treat dynamical friction in a more realistic, non-uniform density distribution, is that by \citet[][hereafter TW84]{Tremaine.Weinberg.84}. Using a perturbative approach, in which the subject mass, $M,$ is treated as a small, time-dependent perturbation on a circular orbit in a spherical system, they show that dynamical friction is entirely due to resonant orbits that give rise to a net retarding torque on the perturbing subject mass\footnote{Throughout this chapter we will use `subject' and `perturber' without distinction.}. In particular, by integrating the torque exerted by a single resonant field particle, multiplied by the (unperturbed) velocity distribution function along the orbits perturbed to second order and summing over all resonances, they obtain a torque that is second order in the perturber's mass (i.e., proportional to $M^2$). This torque is known as the LBK torque, after \citet{LyndenBell.Kalnajs.72} who first derived it in their treatment of angular momentum transport due to spiral structure in disk galaxies. Note that this resonance picture of dynamical friction gives a natural explanation for the non-zero drag experienced by a subject mass orbiting outside of the galaxy, as it simply arises from the net torque due to resonant interactions with stars inside of the galaxy. 

A slightly different perturbative approach was recently taken by \citet[][hereafter KS18]{Kaur.Sridhar.18}; rather than perturbing the resonance orbits, they use the collisionless Boltzmann equation to compute the perturbed distribution function of field particles, which they integrate along the unperturbed resonant orbits. This once again yields a net torque that is second order in $M,$ and consistent with the LBK torque obtained by TW84. Interestingly, KS18 then proceed to show that when the perturber enters the core region of a galaxy the number of low-order resonances (which dominate the torque) is suppressed and the strength of the remaining resonances is weakened. Hence, core stalling has a natural explanation in terms of the LBK torque. However, it fails to explain two related phenomena that have been identified in numerical simulations. The $N$-body simulations by \cite{Cole.etal.12} manifest `dynamical buoyancy' in that perturbers initially placed near the center of a cored galaxy are found to be `pushed out'. Others have reported that when a perturber approaches a core, it first experiences a phase of strongly enhanced `super-Chandrasekhar' dynamical friction, followed by a `kick-back' effect in which the pertuber is pushed out again \citep[][]{Goerdt.etal.10, Read.etal.06c, Zelnikov.Kuskov.16}. These simulation results have thus far eluded a proper explanation, and appear inconsistent with the notion that dynamical friction arises from the LBK torque which is always retarding (at least in a spherical, isotropic system, see Section~\ref{sec:torque}).

The notion that dynamical friction arises solely from resonant interactions can be traced back to the assumption that it is a secular process. This \textit{secular approximation} implies that the actions of the perturber change only very slowly, on a time scale much longer than the dynamical time. In addition, it is assumed that the perturber is introduced to the system on a long time scale (i.e., the past evolution was also secular). We shall refer to this as the \textit{adiabatic approximation}. Both the secular and adiabatic approximations underlie the treatments of TW84 and KS18, which are based on Hamiltonian perturbation, as well as all other treatments that have inferred that dynamical friction arises exclusively from resonances. This includes treatments in action-angle space that use kinetic theory and/or the fluctuation-dissipation theorem \citep[e.g.,][]{Chavanis.12, Chavanis.13, Heyvaerts.etal.17, Fouvry.Bar-Or.18}.

It is important, though, to realize that the secular and adiabatic approximations are really only justified if the dynamical friction time, defined as the timescale on which the perturber sinks to the center of its host, is much longer than the dynamical time. Such cases, though, are of limiting astrophysical interest. If instead we focus on systems for which the dynamical friction time is (significantly) shorter than the Hubble time, we are unavoidably in a regime for which the secular and adiabatic approximations may no longer be justified. In this chapter we examine the impact of relaxing both these approximations. We do this by properly accounting for the past orbital evolution of the perturber in a self-consistent way. The resulting `self-consistent' torque differs from the standard LBK torque in two important ways. First of all, the self-consistent torque makes it explicitly clear that the dynamical friction torque arises from both resonant and near-resonant orbits. Secondly, while the exact resonances always exert a retarding torque, the near-resonant orbits can exert both retarding and enhancing torques. As long as the orbital decay rate is slow, the self-consistent torque can be written as the sum of two terms: (i) an `instantaneous' torque, which is the torque experienced by a perturber introduced abruptly to the host galaxy, and (ii) a `memory' torque, which depends on the entire orbital history of the perturber. The instantaneous torque builds up slowly, and then starts to oscillate in amplitude. Over time these `transient' oscillations damp out due to phase-mixing, after which the instantaneous torque reduces to the LBK torque due to the pure resonant orbits. The memory torque, which always has a non-zero contribution from both resonant and near-resonant orbits, starts out sub-dominant, but becomes the dominant contributor to the total torque when the perturber approaches the core region of a galaxy. When this happens the orbital decay enters a phase of accelerated, super-Chandrasekhar infall, which ceases after the perturber crosses a critical radius at which the torque flips sign. Inside of this radius the torque is enhancing, giving rise to `dynamical buoyancy'. Hence, we argue that core stalling occurs at or near this critical radius, as a manifestation of a delicate balance between dynamical friction outside and buoyancy within. 

This chapter is organized as follows. In Section~\ref{sec:standard_perturbation} we first relax the adiabatic approximation. We use Hamiltonian perturbation theory to derive an expression for the `generalized LBK torque', and discuss how it differs from the standard LBK torque using the analogy of a forced, damped oscillator. In Section~\ref{sec:general_perturbation} we subsequently also relax the secular approximation and derive an expression for the `self-consistent torque', which self-consistently accounts for the orbital evolution (decay) of the perturber. We use this torque in Section~\ref{sec:core_stalling} to discuss the orbital decay of a perturber in a cored background galaxy, providing new insight regarding core stalling, dynamical buoyancy, and super-Chandrasekhar dynamical friction. We summarize our results in Section~\ref{sec:concl_4}.
 
\section{Hamiltonian Perturbation Theory and the generalized LBK Torque}
\label{sec:standard_perturbation}

Throughout this chapter we follow TW84 and KS18, and consider a rigid perturber\footnote{Throughout we ignore potential mass loss of the perturber due to the tidal field of the host.} of mass $M_\rmP$ moving on a circular orbit in a spherical host potential (hereafter the `galaxy') with mass profile $M_\rmG(R)$. The host is made up of a large number of `stars', or `field particles', of mass $m$, and we have that $m \ll M_\rmP \ll M_\rmG$.

\subsection{Hamiltonian Dynamics in the Co-Rotating Frame}
\label{sec:Hamilton_4}

Since the total, perturbed gravitational potential, and hence the Hamiltonian for each field particle, is time-variable, energy is not a conserved quantity. And due to the lack of spherical symmetry, neither is angular momentum. However, as is well known \citep[see e.g.,][]{Binney.Tremaine.08}, the Jacobi Hamiltonian
\begin{equation}
H_{\rm J} = E -{\bf \Omega_{\rm{\bP}}} \cdot \bL
\label{Ej_4}
\end{equation}
is a conserved quantity (if we ignore time evolution of ${\bf \Omega_{\rm{\bP}}}$). $E$ and $\bL$ are, respectively, the perturbed energy and angular momentum of the field particle in the non-rotating, inertial frame, given by
\begin{align}\label{Eperturbed_4}
E& =E_0+\Phi'_\rmP=\frac{1}{2}{\dot{\br}}^2 + \Phi_\rmG\left(\br\right) + \Phi'_\rmP\left(\br\right),\\
\bL &= \br \times \bf{\dot{r}}\,.
\end{align}
Here $\br$ is the position vector of the field particle with respect to the galactic center and $\dot{\br}$ is the velocity of the field particle in the inertial frame. The angular frequency of the galaxy-perturber system is given by ${\bf \Omega_{\rm{\bP}}} = (0,0,\Omega_\rmP)$, where
\begin{equation}
\Omega_\rmP = \sqrt{\frac{G\,[M_\rmG(R) + M_\rmP]}{R^3}}\,,
\end{equation}
with $R$ the galacto-centric radius of the perturber. $E_0$ is the unperturbed energy, i.e., the part of the Hamiltonian without the perturber potential, $\Phi_\rmG$ is the gravitational potential due to the galaxy, and $\Phi'_\rmP$ is the perturber potential, which consists of both direct and indirect terms and is given by
\begin{align}
\Phi'_\rmP& = \Phi_\rmP + G M_\rmP\frac{\br\cdot\bR}{R^3}\,,
\end{align}
with $\Phi_\rmP = -G M_\rmP/\vert\br-\bR\vert$ for a point perturber. The first term is the direct term, while the second term is the indirect term which accounts for the fact that the galaxy center (the origin), is rotating about the common COM with the perturber. 

In reality the perturbation in the potential also includes a gravitational `polarization' term which arises from the perturbation in the stellar distribution function induced by the perturber (also known as the `wake'). That term manifests the collective effects due to the self-gravity of the stars and significantly complicates the analysis. As first shown by \cite{Weinberg.89}, using Hamiltonian perturbation theory, and more recently by \cite{Chavanis.12}, \cite{Heyvaerts.etal.17} and \cite{Fouvry.Bar-Or.18} using the fluctuation-dissipation theorem, the collective effects primarily `dress' the perturber potential $\Phi'_\rmP$ by introducing a prefactor which is the gravitational equivalent of the dielectric function in plasma physics. In particular, as nicely summarized in \cite{Fouvry.Bar-Or.18}, taking collective effects into account in the stochastic picture yields an inhomogeneous Lenard-Balescu equation in which the diffusion coefficients involve the dressed potential\footnote{Upon using the bare potential instead (i.e., ignoring collective effects due to self-gravity of the field particles), these diffusion coefficients reduce to those of the inhomogeneous Landau equation, which in turn implies a dynamical friction force consistent with the LBK torque.}. Given the formidable challenge in computing the dressed potentials, and given that the impact of collective effects is likely less important than for plasmas \citep[see][for detailed discussion]{Chavanis.13} we follow TW84 and KS18, and neglect the effects of self-gravity for the sake of simplicity.

\bigskip

\subsection{Perturbation Analysis}
\label{sec:perturbation_analysis}

The dynamics of the unperturbed system in the co-rotating frame is governed by the Jacobi Hamiltonian $H_{0\rmJ}=E_0-{\bf \Omega_{\rm{\bP}}} \cdot \bL$, which is a conserved quantity and therefore commutes with the unperturbed distribution function $f_0$, i.e.,
\begin{align}
\left[f_0,H_{0\rmJ}\right]=0.
\end{align}
where $[A,B]$ denotes the Poisson bracket of $A$ and $B$. This is nothing but the steady state form of the collisionless Boltzmann equation (hereafter CBE) for the unperturbed galaxy in the co-rotating frame. When we introduce a perturber, the system is no longer in equilibrium and its dynamics is governed by the perturbed Hamiltonian. The perturber potential $\Phi'_\rmP$ gives rise to a perturbation in the distribution function $f_1$, which in turn exerts a torque on the perturber. This is ultimately responsible for dynamical friction\footnote{This is the key idea behind linear response theory.}. In what follows, we shall, in the spirit of KS18, perturb the CBE up to linear order to obtain the expression for $f_1$ and use it to compute the torque on the perturber.

The perturbation in the distribution function can be computed by perturbing the collisionless Boltzmann equation 
\begin{align}
\frac{\partial f}{\partial t}+[f,H_\rmJ]=0.
\label{CBE}
\end{align}
Up to linear order, we write $f$ and $H_\rmJ$ as the following perturbative series
\begin{align}
&f=f_0+f_1, \nonumber \\
&H_\rmJ=H_{0\rmJ}+\Phi_1^{\rm ext}\,. 
\label{series}
\end{align}
We follow TW84 and KS18 and let the external perturbation grow as $\Phi_1^{\rm ext}(R,t) = g(t) \, \Phi'_\rmP(R)$, where the growth function
\begin{equation}
\begin{aligned}
g(t) = 
\begin{dcases}
\rme^{\gamma t}, & t<0 \\
1,\,             & t \geq 0\,,
\label{Phi1ext}
\end{dcases}
\end{aligned}
\end{equation}
and $\gamma>0$. This indicates that the perturber grows its mass exponentially from $t\to -\infty$ to $t=0$ on a characteristic time-scale $\tau_{\rm grow} \equiv 1/\gamma$, while remaining at a fixed host-centric radius $R$. We do not consider this realistic, but before we consider an alternative we first aim to clarify the implications of this assumption. Note also that we have neglected the implicit time dependence of $\Phi_1^{\rm ext}$ and $\Omega_\rmP$ through $R(t)$. This constitutes the secular approximation that $R$ changes on a time scale much longer than the dynamical time.

Substituting the series expansions given in equation~(\ref{series}) in the CBE of equation~(\ref{CBE}), we obtain the following evolution equation for $f_1$ up to linear order
\begin{align}
&\frac{\partial f_1}{\partial t} + [f_1,H_{0\rmJ}] + [f_0,\Phi_1^{\rm ext}] = 0\,.
\label{CBEn}
\end{align}
In general, one can obtain the solution for $f_1$ once $f_0$ is known. The unperturbed distribution function $f_0$ is a solution of the unperturbed CBE and therefore, by the Jeans Theorem, is a function of the conserved quantities of the dynamical system, which in case of a spherical galaxy correspond to the three actions $I_1$, $I_2$ and $I_3$. These consist of the radial action $I_\rmr$, the total angular momentum $L$, and the $z$ component of the angular momentum, $L_\rmz$, or linear combinations thereof. Throughout, we consider the $z$-axis to coincide with the normal to the orbital plane of the perturber. In order to simplify the dynamics, we make a canonical transformation from $(\br,\bp)$ phase-space to $(\bw,\bI)$ action-angle space spanned by the action vector $\bI=\{I_1,I_2,I_3\}$ and the corresponding angle vector $\bw=\{w_1,w_2,w_3\}$. Recall that $f_0$ and $H_{0\rmJ}$ are both functions of only $I_1, I_2$ and $I_3$ and do not depend on the angles, while $f_1$ and $\Phi'_\rmP$ are functions of both actions and angles. Therefore, in action-angle space the Poisson brackets in the above equations become
\begin{align}
&[f_1,H_{0\rmJ}]=\frac{\partial f_1}{\partial w_k} \frac{\partial H_{0\rmJ}}{\partial I_k}, \\
&[f_0,\Phi'_\rmP]=-\frac{\partial f_0}{\partial I_i} \frac{\partial \Phi'_\rmP}{\partial w_i}\,.
\end{align}
Here and throughout, the Einstein summation convention is implied, and indices $k$ and $i$ run from 1 to 3 and 1 to 2, respectively. In action-angle space, $[f_1,H_{0\rmJ}]$ reduces to 
\begin{align}
[f_1,H_{0\rmJ}]=\Omega_k \frac{\partial f_1}{\partial w_k}\,,
\end{align}
where the frequencies $\Omega_k$ are given by
\begin{align}
&\Omega_1 = \frac{\partial H_{0\rmJ}}{\partial I_1} = \frac{\partial E_0}{\partial I_1}\,, \nonumber \\
&\Omega_2 = \frac{\partial H_{0\rmJ}}{\partial I_2} = \frac{\partial E_0}{\partial I_2}\,, \nonumber \\
&\Omega_3 = \frac{\partial H_{0\rmJ}}{\partial I_3} = -\Omega_\rmP\,.
\end{align}
Here $I_3=L_\rmz$, and $I_1$ and $I_2$ are linear combinations of $I_\rmr$ and $L$, respectively. Since $\Phi_\rmG$ is a spherically symmetric {potential}, $E_0$ is a function of $I_\rmr$ and $L$ only, or in other words of only $I_1$ and $I_2$. And since $f_0$ is a function of $E_0$ and $L$ only, it has a similar dependence on the actions, i.e., $\partial f_0/\partial I_3 = 0$.

Following KS18 we expand $f_1$ and $\Phi'_\rmP$ as a Fourier series in $\bw$ using
\begin{align}
&f_1(\bw,\bI,t)=\sum_{\boldell} \hat{f}_{1,\boldell}(\bI,t)\, \rme^{i\bw\cdot\boldell}, \nonumber \\
&\Phi'_\rmP(\bw,\bI)=\sum_{\boldell} \hat{\Phi}'_{\boldell}(\bI)\, \rme^{i\bw\cdot\boldell},
\label{fourier_series}
\end{align}
where the summation is over all integer triplets $\boldell = (\ell_1, \ell_2, \ell_3)$. Note that, since both $f_1$ and $\Phi'_\rmP$ are real, we have that $\hat{f}_{1,-\boldell} = \hat{f}^{\ast}_{1,\boldell}$ and $\hat{\Phi}_{-\boldell} = \hat{\Phi}^{\ast}_{\boldell}$, where $A^{\ast}$ indicates the complex conjugate of $A$. Substituting the Fourier series in equation~(\ref{CBEn}) yields the following evolution equation for $\hat{f}_{1,\boldell}$
\begin{equation}
\frac{\partial \hat{f}_{1,\boldell}}{\partial t}+i\ell_k\Omega_k \hat{f}_{1,\boldell} = g(t) \, i\ell_i \, \frac{\partial f_0}{\partial I_i} \, \hat{\Phi}'_{\boldell}\,.
\label{f1l_evolve}
\end{equation}

At this point in their analysis KS18 assume that the perturbation in the distribution function evolves in a similar way as the external perturber, i.e., $f_1 \propto \rme^{\gamma' t}$ with $\gamma' = \gamma$. Under this assumption, the above differential equation becomes a simple algebraic equation that can be solved for $\hat{f}_{1,\boldell}$. KS18 thus assume that the response density builds up on the same time scale as that on which the perturber is introduced. This is a fair assumption as long as $\gamma$ is sufficiently small, such that there is sufficient time for the host to respond. However, if dynamical friction is very efficient, then $\gamma'$ can be different from $\gamma$. In fact, in general the perturbation does not have a single growth rate. Different parts of the phase-space respond to the perturber at different rates. Therefore, in what follows we will not make any a priori assumption about the growth rate of the response density due to the perturber. Rather, we solve the differential equation for $\hat{f}_{1,\boldell}$ using the Green's function technique with the initial condition that $\hat{f}_{1,\boldell}(t\to -\infty)=0$. We find the Green's function to be $e^{-i\ell_k\Omega_k \left(t-\tau\right)}$, which can be used to obtain the following particular solution for $\hat{f}_{1,\boldell}$
\begin{equation}
\begin{aligned}
\hat{f}_{1,\boldell}(\bI,t) & = i\ell_i \, \frac{\partial f_0}{\partial I_i} \, \hat{\Phi}'_{\boldell}(\bI) \, \rme^{-i\ell_k\Omega_k t}
\begin{dcases}
\int_{-\infty}^{t} \rmd\tau \, \rme^{(\gamma + i\ell_k\Omega_k)\tau}, & t<0, \\ \\
\int_{-\infty}^{0} \rmd\tau \, \rme^{(\gamma + i\ell_k\Omega_k)\tau} + \int_{0}^{t} \rmd\tau \, \rme^{i\ell_k\Omega_k\tau}, & t\geq 0.
\end{dcases}
\end{aligned}
\end{equation}
This can be integrated to yield
\begin{align}
\hat{f}_{1,\boldell}(\bI,t) & =  i\ell_i \, \frac{\partial f_0}{\partial I_i} \, \hat{\Phi}'_{\boldell}(\bI)
\begin{dcases}
\frac{ \rme^{\gamma t}}{\gamma + i\ell_k\Omega_k}, & t<0 \\ \\
\left[\frac{ \rme^{-i\ell_k\Omega_k t}}{\gamma + i\ell_k\Omega_k} + \frac{1 - \rme^{-i\ell_k\Omega_k t}}{i\ell_k\Omega_k}\right], & t \geq 0.
\end{dcases}
\label{f1l}
\end{align}
Note that the solution for $t<0$ is identical to that obtained by KS18.

\medskip

\subsection{The Generalized LBK Torque}
\label{sec:torque}

As shown in KS18, the torque on the perturber by a field particle is given by $\partial \Phi_{1,\rm ext}/\partial \phi$. Hence, the total torque on the perturber can be computed by weighting $\partial \Phi_{1, \rm ext}/\partial \phi$ by the perturbed distribution function and then integrating over all of phase-space as follows
\begin{align}
\calT &= \int \rmd \br \int \rmd \bp \,\frac{\partial \Phi_1^{\rm ext}}{\partial \phi} \left(f_0+f_1\right) = \int \rmd \br \int \rmd \bp \,\frac{\partial \Phi_1^{\rm ext}}{\partial \phi} f_1.
\end{align}
Note that, since $f_0$ is independent of $\phi$ and $\oint \rmd \phi\, \left(\partial \Phi_1^{\rm ext}/\partial \phi\right)=0$, the leading order contribution to the torque comes from $f_1$. And since both $\Phi_1^{\rm ext}$ and $f_1$ are first order in $M_\rmP$, the torque itself is second-order in the mass of the perturber.

To evaluate the torque we follow KS18 and note that $\partial \Phi_1^{\rm ext}/\partial \phi = -[p_\phi,\Phi_1^{\rm ext}] = -[L_\rmz,\Phi_1^{\rm ext}]$. And since the Poisson bracket is invariant under canonical transformation, we thus have that $\partial \Phi_1^{\rm ext}/\partial \phi = -[I_3,\Phi_1^{\rm ext}] = \partial \Phi_1^{\rm ext}/\partial w_3$. Moreover the volume element $\rmd \br\, \rmd \bp$ is also invariant under canonical transformation and becomes $\rmd \bw\, \rmd \bI$. Therefore, using equation~(\ref{Phi1ext}), the torque can be written in action-angle space as
\begin{align}
\calT & = g(t) \int \rmd \bw \int \rmd \bI\, \frac{\partial \Phi'_\rmP}{\partial w_3} f_1\,,
\end{align}
After substituting the Fourier expansions of $\Phi'_\rmP$ and $f_1$, and using the reality condition, i.e. $\hat{f}_{1,-\boldell'}(\bI,t)=\hat{f}^{\ast}_{1,\boldell'}(\bI,t)$, we obtain that
\begin{align}
\calT & = g(t)  \sum_{\boldell} \sum_{\boldell'} i \ell_3 \, \int \rmd \bI\, \hat{\Phi}'_{\boldell}(\bI)  \hat{f}^{\ast}_{1,\boldell'}(\bI,t) \int \rmd \bw\,e^{i\left(\boldell-\boldell'\right)\cdot \bw},
\label{torque_intermediate}
\end{align}
which can be integrated over $\bw$ using the following identity for the Dirac delta function,
\begin{align}
\delta^3(\bx) = {1 \over {\left(2\pi\right)}^3} \int \rmd \bw\,e^{i\bw\cdot \bx},
\end{align}
and summed over the $\boldell'$ indices to yield
\begin{align}
\calT & = (2\pi)^3 \, g(t)  \sum_{\boldell} i \ell_3 \, \int \rmd \bI\, \hat{\Phi}'_{\boldell}(\bI)  \hat{f}^{\ast}_{1,\boldell}(\bI,t)\,.
\label{torque}
\end{align}
Substituting $\hat{f}_{1,\boldell}(\bI)$ given by equation~(\ref{f1l}) in the above expression, the second order torque can be written as
\begin{align}
\calT_2 = \Tgen = 16\pi^3 \reducedsum \ell_3 \int \rmd \bI\, \calJ(\ell_k\Omega_k,t)\, \ell_i\frac{\partial f_0}{\partial I_i}\, {\left|\hat{\Phi}'_{\boldell}(\bI)\right|}^2\,,
\label{Torque_gen}
\end{align}
where 
\begin{align}
\reducedsum = \sum_{\ell_1=-\infty}^{\infty} \, \sum_{\ell_2=-\infty}^{\infty} \, \sum_{\ell_3=1}^{\infty}\,, 
\end{align}
is the `reduced' summation over the positive-$\ell_3$ hemisphere of $(\ell_1,\ell_2,\ell_3)$-space, which arises from applying the mirror symmetry operation $\boldell\to -\boldell$ and retaining the symmetric part of the integrand. The function $\calJ(\ell_k\Omega_k,t)$ is given by
\begin{equation}
\begin{aligned}
\calJ(\ell_k\Omega_k, t) & = \frac{\gamma}{\gamma^2+{\left(\ell_k \Omega_k\right)}^2} 
\begin{dcases}
\rme^{2\gamma t}, & t < 0 \\ 
\cos{\ell_k\Omega_k t} + \gamma \, \frac{\sin{\ell_k\Omega_k t}}{\ell_k\Omega_k}, & t \geq 0.
\end{dcases}
\label{J_4}
\end{aligned}
\end{equation}
Note that this torque contains the contribution from all orbits: resonant, near-resonant and non-resonant. We therefore refer to this as the generalized LBK torque, which is based on the secular, but not the adiabatic approximation (i.e., we did not take the limit $\gamma \to 0$). The amplitude of the torque scales as the Lorentzian-like function $\gamma / [\gamma^2 + (\ell_k \Omega_k)^2]$, which peaks at the resonances, where the commensurability condition of the frequencies,
\begin{equation}
\ell_k \Omega_k = \ell_1 \Omega_1(I_1,I_2) + \ell_2 \Omega_2(I_1,I_2) - \ell_3 \Omega_\rmP = 0\,
\label{resonance_condition}
\end{equation} 
is satisfied. The width of the Lorentzian is proportional to $\gamma$, and determines the relative contributions to the torque from resonant and near-resonant orbits, with larger values of $\gamma$ (i.e., smaller $\tau_{\rm grow}$) resulting in a more dominant contribution from the near-resonant orbits.

An important feature of the generalized LBK torque is that it can be either retarding ($\calT < 0$) or enhancing ($\calT > 0$), depending on the sign of $\calJ$. Using that $\ell_i\partial f_0/\partial I_i = \ell_i\Omega_i \partial f_0/\partial E_0 < 0$ for stable distribution functions \citep[e.g.,][]{Doremus.Feix.Baumann.71, Binney.Tremaine.08}, we see that a coherent retarding (enhancing) torque corresponds to $\calJ > 0$ ($\calJ < 0$). To get some insight, we start by examining the generalized LBK torque in the limits of both small and large $\gamma$, which correspond to adiabatic growth and instantaneous introduction of the perturber, respectively.

\subsubsection{Adiabatic Growth of Perturber Potential}
\label{sec:adiabatic}

Adiabatic growth of the perturber potential implies that the perturber has to be introduced on a time scale that is long compared to all other relevant dynamical times in the problem. Only then can the distribution function $f_0(I_1,I_2,I_3)$, expressed according to the Jeans theorem as a function of its actions, remain perfectly invariant. The longest time scales of relevance are the libration times $T_{\rm lib} \sim 1/\ell_k\Omega_k$, which become infinitely long for orbits that satisfy the commensurability condition. Hence, assuring strict adiabatic invariance requires that we take the limit $\gamma \to 0$. In this limit, $\calJ(\ell_k\Omega_k,t)$ converges according to
\begin{align}
\lim_{\gamma \to 0}\calJ(\ell_k\Omega_k,t) = \pi\delta(\ell_k\Omega_k),\;\;\;\;\;-\infty<t<\infty,
\label{J_ad}
\end{align}
where we have used that the Lorentzian function becomes a Dirac delta function in the limit of vanishing width. Substituting this expression for $\calJ$ in equation~(\ref{Torque_gen}) yields the standard LBK torque
\begin{align}
\calT_{2} = \Tlbk &= 16\pi^4 \reducedsum \ell_3 \int \rmd \bI\,\delta\left(\ell_k\Omega_k\right) \left(\ell_1\Omega_1+\ell_2\Omega_2\right)\frac{\partial f_0}{\partial E_0}\, {\left|\hat{\Phi}'_{\boldell}(\bI)\right|}^2 \nonumber \\
&= 16\pi^4 \Omega_\rmP \reducedsum \ell_3^2 \int \rmd \bI\,\delta\left(\ell_k\Omega_k\right)\frac{\partial f_0}{\partial E_0}\, {\left|\hat{\Phi}'_{\boldell}(\bI)\right|}^2,
\label{LBK}
\end{align}
which has a non-zero contribution from only the exact resonances. And since $\partial f_0/\partial E_0<0$ for a stable distribution function, we see that $\Tlbk < 0$ for all resonances. In other words, the standard second-order LBK torque is always retarding in nature. 

This makes it clear that the LBK torque is ultimately an outcome of taking the adiabatic limit ($\gamma \to 0$). This should not come as a surprise: in the limit where the perturber takes infinitely long to present itself, the only contribution to a net torque that is not phase mixed away (see Section~\ref{sec:analogy} and Fig.~\ref{fig:genLBK}) is that from orbits in perfect resonance with the perturber.  However, to what extent the LBK torque is relevant for dynamical friction ultimately depends on the time scale on which phase-mixing removes the transient contributions. The rate at which an orbit and the perturber get out of phase depends on the incommensurability between their frequencies. If large, phase-mixing is fast, and the contribution to the total torque vanishes rapidly. However, phase-mixing the contribution from the near-resonant orbits can easily take many dynamical times. And since a typical galaxy or dark matter halo is only a few dynamical times old (at least in its outskirts), we are not a priori justified in assuming that dynamical friction is dominated by the LBK torque.

\subsubsection{Instantaneous Introduction of the Perturber}
\label{sec:instantaneous}

Idealized numerical simulations that examine dynamical friction  \citep[e.g.,][]{Lin.Tremaine.83, White.83, Cora.etal.97, vdBosch.etal.99, Jiang.Binney.00, Hashimoto.etal.03, Boylan-Kolchin.etal.08, Inoue.09, Inoue.11, Tamfal.etal.21} typically do not adiabatically grow the perturber potential over time, but rather introduce it instantaneously to the host system. We can use our expression for the generalized torque to examine such a scenario. Instantaneous introduction of the perturber corresponds to taking $\gamma \to \infty$, for which 
\begin{align}
\lim_{\gamma \to \infty}\calJ(\ell_k\Omega_k,t) = 
\begin{dcases}
0, & t<0, \nonumber \\ \\
\frac{\sin{\ell_k\Omega_k t}}{\ell_k\Omega_k}, & t\geq 0.
\end{dcases}
\end{align}
This yields the following expression for the torque 
\begin{align}
\calT_{2} &= \Tinst \equiv 16\pi^3 \reducedsum \ell_3 \int \rmd \bI\,\frac{\sin{\ell_k\Omega_k t}}{\ell_k\Omega_k} \left(\ell_1\Omega_1+\ell_2\Omega_2\right)\frac{\partial f_0}{\partial E_0}\, {\left|\hat{\Phi}'_{\boldell}(\bI)\right|}^2.
\label{Torque_inst}
\end{align}
We shall hereafter refer to this as the instantaneous torque. At small $t$, i.e., shortly after the instantaneous introduction of the perturber, we have that $\Tinst \propto t$, indicating that the second-order torque builds up linearly with time. Note that the contribution to the torque from the non-resonance orbits can be either retarding or enhancing. In particular, for modes that are sufficiently far away from resonance, $\sin{\ell_k\Omega_k t}$ can become negative a short time after the instantaneous introduction of the perturber, which can result in an enhancing torque. In addition, since $\ell_1 \Omega_1 + \ell_2 \Omega_2$ can be either positive or negative, it is even possible (at least in principle) for $\Tinst$ to be enhancing when $\sin{\ell_k\Omega_k t}$ is positive.

Finally, note that in the $t \to \infty$ limit, $\sin{(\ell_k\Omega_k  t)} / \ell_k\Omega_k \to \pi \, \delta(\ell_k\Omega_k)$, and we recover the familiar LBK formula for the torque, with a non-zero contribution only from the exact resonances. Hence, in accordance with the analogy of the forced, damped oscillator discussed in Section~\ref{sec:analogy}, following the instantaneous introduction of the perturber, the torque initially builds up linearly with time, then undergoes oscillations (corresponding to transients) that slowly phase mix away, after which only the LBK torque due to the perfect resonances remains. 

\begin{figure*}[t!]
\centering
\includegraphics[width=0.9\textwidth]{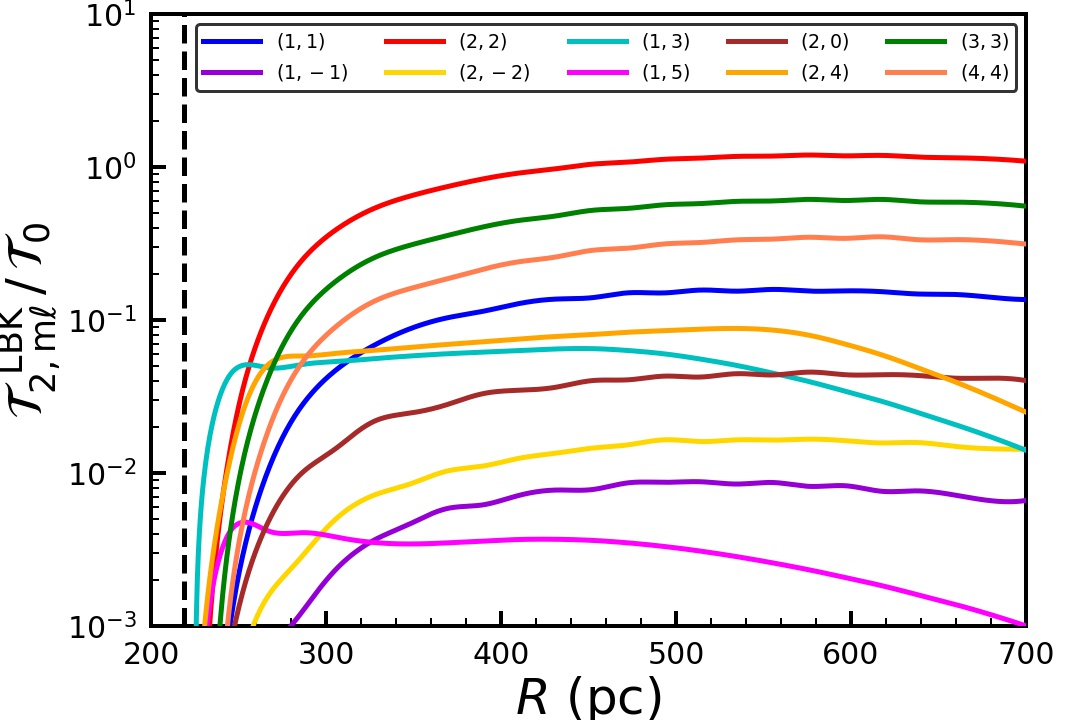}
\caption{The LBK torque on a point mass perturber of mass $M_\rmP$ on a circular orbit in a spherical isochrone potential of mass $M_\rmG = 8000 M_\rmP$, in units of $\calT_0 \equiv G M^2_\rmP/b$, as a function of the galacto-centric radius of the perturber, $R$. Different curves show the contribution due to the ten $(m,\ell) = (m,\ell,m)$ resonance orbits (modes) that dominate the total LBK torque, as indicated. Note how all the $m=|\ell|$ modes contribute a torque with a similar $R$-dependence, and that the LBK torque dies out as the perturber approaches the `filtering radius' $R_\ast= 0.22\,b = 220\pc$, indicated by the black vertical, dashed line. As discussed in KS18, this decline of the (LBK) torque as the perturber approaches the central core region is responsible for the phenomenon of core stalling (but see section~(\ref{sec:core_stalling}) for a somewhat different explanation).}
\label{fig:LBKiso}
\end{figure*}

\subsection{Dynamical Friction in an Isochrone Sphere}
\label{sec:iso}

In order to illustrate how the instantaneous torque differs from the LBK torque, we compare the two for the case of a point mass perturber on a circular orbit in a spherical, isotropic isochrone galaxy \citep[][]{Henon.59}. This configuration was also considered by KS18 and has the advantage that (i) it is a fairly realistic representation of a galaxy, (ii) many of its physical quantities can be computed analytically, and (iii) it has a central constant density core, which allows us to examine core stalling. 

The gravitational potential of Henon's isochrone sphere with mass $M_\rmG$ and scale radius $b$ is given by
\begin{align}\label{IsoPot}
\Phi_\rmG = -\frac{GM_\rmG}{b+\sqrt{b^2+r^2}}\,,
\end{align}
and its corresponding density profile, 
\begin{align}\label{IsoDens}
\rho_\rmG(r) &= \frac{M_\rmG}{4\pi} \left[\frac{3\left(b^2+r^2\right)\left(b+\sqrt{b^2+r^2}\right)-r^2\left(b+3\sqrt{b^2+r^2}\right)}{{\left(b+\sqrt{b^2+r^2}\right)}^3{\left(b^2+r^2\right)}^{3/2}}\right]
\end{align}
falls off as $r^{-4}$ at large $r$, and asymptotes to a constant core value of $3M_\rmG/16\pi b^3$ as $r \rightarrow 0$. Following KS18, we adopt $M_\rmG = 1.6\times 10^9 \Msun$ and $b=1\kpc$. These parameters were chosen by KS18 such that the isochrone sphere has the same core radius and central density as the \citet{Burkert.95} sphere used in the high-resolution $N$-body simulation of \citet{Inoue.11}. Following both  KS18 and \citet{Inoue.11}, we adopt a point mass perturber of mass $M_\rmP = 2\times 10^5 \Msun$ (corresponding to a mass ratio $M_\rmP/M_\rmG = 1.25\times 10^{-4}$), which we consider to be on a circular orbit. In what follows we shall refer to this set-up as our fiducial example.

As detailed in Appendix~\ref{app:model} and KS18, the commensurability condition for this system can be written as
\begin{align}
\ell_k\Omega_k = n\,\Omega_w + \ell\,\Omega_g - m\,\Omega_\rmP 
\end{align}
where, following KS18, we have used $\ell_1 = n$, $\ell_2=\ell$ and $\ell_3 =m$. The frequencies $\Omega_w$ and $\Omega_g$ are related to the radial and angular frequencies in the orbital plane, as described in Appendix~\ref{app:model}. Although the total (generalized) LBK torque is the sum over all $(n,\ell,m)$, KS18 have shown that the torque is dominated by the co-rotation resonances, which have $m=n$. In addition, the torque is typically stronger for lower order modes $(|\ell| \lsim 3m)$. In what follows we therefore restrict ourselves to the $(m,\ell) = (m,\ell,m)$ modes with dominant LBK torque. 

Fig.~\ref{fig:LBKiso} plots the LBK torque (computed using equation~[\ref{Torque_LBK_isochrone}] as detailed in  Appendix~\ref{app:model}), as a function of the galacto-centric distance of the perturber, $R$, for the 10 dominant $(m,\ell)$ modes. Note how the LBK torque is dominated by that due to the $(m,\ell) = (2,2)$ resonance, and that the LBK torque dies out as the perturber approaches the `filtering radius' $R_\ast = 0.22\, b = 220 \pc$, marked by the black vertical dashed line. As detailed in KS18, at this radius, the circular frequency $\Omega_\rmP$ equals $\Omega_b = 0.5\sqrt{G M_\rmG/b^3}$, roughly the circular frequency of stars in a central core of the isochrone sphere. As a result, the phase-space contributing to the resonances shrinks, strongly suppressing the contribution of the dominant, lower order modes to the total torque.

\begin{figure*}
\centering
\includegraphics[width=1\textwidth]{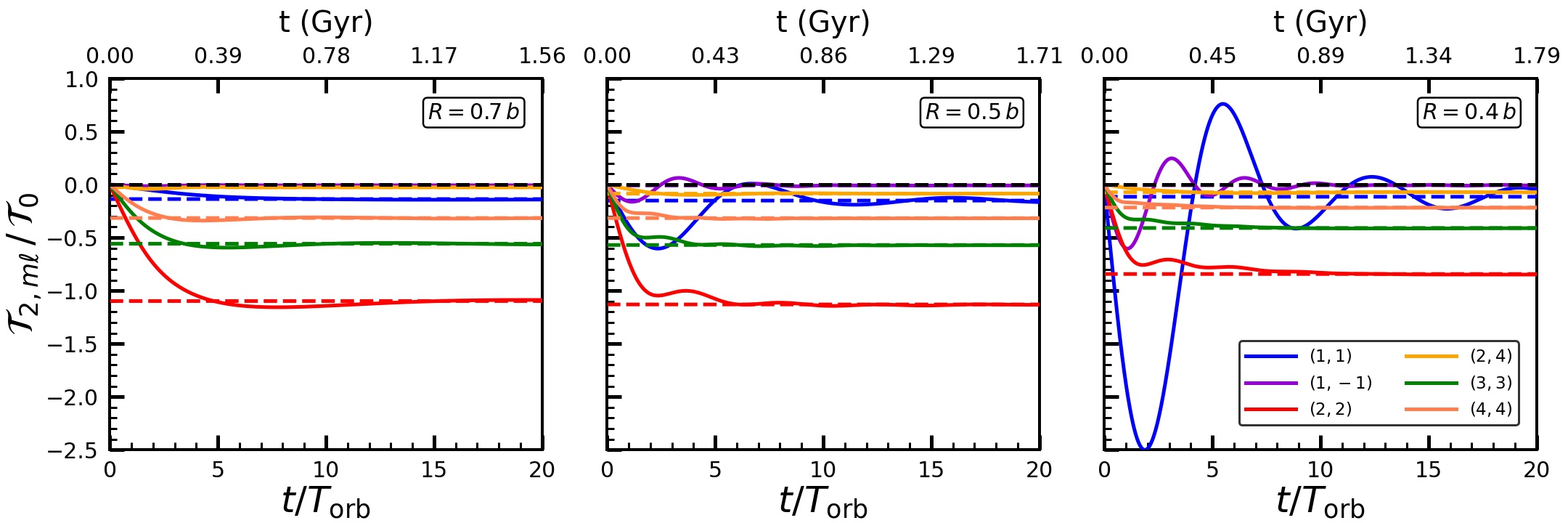}
\caption{The instantaneous, generalized LBK torque (assuming an isochrone model for the galaxy and point perturber with $M_\rmP/M_\rmG=1.25\times 10^{-4}$), in units of $\calT_0=G M^2_\rmP/b$ as a function of $t/T_{\rm orb}$ ($T_{\rm orb}=2\pi/\Omega_\rmP$ is the orbital period of the perturber) when the perturber is introduced at $R=0.7\, b$ (left panel), $0.5\, b$ (middle panel) and $0.4\,b$ (right panel). The solid lines show the instantaneous torque for six of the dominant $(m,\ell)$ modes as indicated. The dashed lines show the corresponding LBK torque. Note that the instantaneous torque converges to the LBK torque as $t\to\infty$ as all but the perfectly resonant orbits get phase-mixed away.}
\label{fig:genLBK}
\end{figure*}

The solid lines in Fig.~\ref{fig:genLBK} plot the instantaneous torque of equation~(\ref{Torque_inst_isochrone}) as a function of time in units of $T_{\rm orb} = 2\pi/\Omega_\rmP$ for six modes that dominate the total torque either at early and/or late times. Results are shown for three different radii of introduction of the perturber, $R=0.7\, b$ (left panel), $0.5\, b$ (middle panel) and $0.4\, b$ (right panel). For comparison, we also plot the corresponding LBK torque as horizontal dashed lines. All modes initially show a coherent, retarding torque, causing the torque to build-up linearly with time, before undergoing oscillations about the corresponding LBK value that slowly damp away with time. This is a classic example of phase-mixing in collisionless systems where all but the purely resonant responses in the distribution function are damped out. Note, though, that the transients from some modes take many orbital times to phase-mix away, especially at smaller radii (closer to the core). The reason is that the core region with a nearly constant density has a much narrower dynamic range in orbital frequencies, resulting in a more coherent response. Interestingly, the transient oscillations can even contribute a positive, enhancing torque at times. Therefore, while the late time dynamics is governed by the LBK torque from the perfectly resonant orbits, the transient behaviour following the introduction of the perturber is driven by the near-resonant orbits. 

\bigskip

\subsection{Analogy: the Forced, Damped Oscillator}
\label{sec:analogy}

It is insightful to compare the galaxy-plus-perturber system to a forced, damped oscillator. As is well known, the general solution of a sinusoidally forced, damped oscillator is a sum of a transient solution that depends on the initial conditions and a steady state solution that is independent of the initial conditions. With time, the transients damp away, causing the response to settle towards the steady state solution, which is a sinusoidal oscillation with a frequency equal to that of the driver, and an amplitude that depends on the driving amplitude, the driving frequency, the eigenfrequency of the (undamped) oscillator, and the damping ratio.

This is similar to how a galaxy responds to a perturber, $M_\rmP$, on a (circular) orbit with frequency $\Omega_\rmP$, which can be regarded as the forcing frequency. The galaxy, in turn, acts as a damped oscillator. In fact, the galaxy is an ensemble of many individual oscillators (the individual orbits), each with three frequencies (corresponding to the three actions). Although each of these orbital frequencies can be considered to correspond to an individual {\it undamped} oscillator, the collective response of all orbits acts as if damped. The source of damping is phase-mixing; initially, when the perturber is introduced (i.e., the forcing commences), all near-resonant orbits respond in phase, and the galaxy response is dominated by transient behavior. Due to phase-mixing, though, the responses of different orbits get out of phase, causing the transient behavior to die out. The only orbits that never get out of phase with the forcing are the resonant orbits, which due to the commensurability of their orbital frequencies with $\Omega_\rmP$, remain in phase, thereby resisting phase-mixing. As a consequence, the `steady-state response' of the galaxy is the LBK torque due to the resonance orbits.

This analogy shows that taking the limit $\gamma \to 0$ corresponds to `skipping' the transient behavior, assuming that phase-mixing has caused the response of all non-resonant orbits to die out. However, it is clear from the analogy that one can ignore the transients only after a sufficiently long time. If the perturber inspirals rapidly (i.e., $\dot{\Omega}_\rmP$ is large), then the forcing with frequency $\Omega_\rmP$ may not last sufficiently long for phase-mixing to nullify the net response of all non-resonant orbits. In particular, the near-resonant orbits, whose response takes the longest to phase mix away, are expected to make a significant, if not dominant, contribution. The generalized LBK torque includes the transient response due to non-resonant orbits and presents a proper description of how dynamical friction builds up in idealized numerical simulations that introduce the perturber instantaneously. However, it is obtained by only relaxing the adiabatic approximation while still relying on the secular approximation, i.e. the perturber undergoes a slow inspiral under dynamical friction. In the following section we relax this assumption and develop a fully self-consistent treatment for dynamical friction, that includes the `memory effect' due to the entire past orbital history of the perturber.

\bigskip

\section{Self-consistent computation of the torque}
\label{sec:general_perturbation}

The previous section has given some useful insight as to how transients that result from an instantaneous introduction of the perturber ($\gamma \to\infty$) phase mix away, ultimately giving rise to the LBK torque that one obtains in the adiabatic limit ($\gamma \to 0$). However, that entire analysis is based on the generalized LBK torque (equations~[\ref{Torque_gen}]--[\ref{J_4}]), which assumes that the perturber grows its mass exponentially, on a time scale  $\tau_{\rm grow} = 1/\gamma$, while continuing to orbit at a fixed radius $R$. This is not realistic. The proper way to introduce the perturber is to self-consistently account for its past trajectory $R(t)$, from $t=0$ to the present, which is what we tackle in this section.

The main shortcoming with the derivation of the generalized LBK torque,
or with that of the standard LBK torque in TW84 and KS18, is that it ignores the time dependence of $\Phi'_\rmP$ and $\Omega_\rmP$ in solving the evolution equation for $\hat{f}_{1,\boldell}$ (equation~[\ref{f1l_evolve}]). In this section we are going to relax this assumption of secular evolution and compute $\hat{f}_{1,\boldell}$ and ultimately the torque in a fully self-consistent way. Here we assume that the perturber starts out at $t=0$ at some large initial radius, $R_0$, and slowly makes its way inwards following a circular orbit with time-dependent radius $R(t)$. Therefore, both the external perturbation and its circular frequency depend implicitly on time according to
\begin{align}
\Phi_1^{\rm ext} = \Phi'_\rmP(R(t)),\;\;\;\;\;\;\;\;\;\;
\Omega_\rmP = \Omega_\rmP (R(t))\,.
\end{align}
In what follows, for brevity, we simply write these dependencies as $\Phi'_\rmP(t)$ and $\Omega_\rmP(t)$ and consider it understood that the time-dependence enters implicitly via the perturber's orbit, $R(t)$.

As in \S\ref{sec:standard_perturbation}, we can perturb the CBE up to linear order and expand $f_1$ and $\Phi'_\rmP$ as the Fourier series of equation~(\ref{fourier_series}) to obtain the following evolution equation for $\hat{f}_{1,\boldell}$
\begin{align}
\frac{\partial \hat{f}_{1,\boldell}}{\partial t} + i\left[\ell_i\Omega_i - \ell_3\Omega_\rmP(t)\right] \hat{f}_{1,\boldell} = i\ell_i\frac{\partial f_0}{\partial I_i}\hat{\Phi}'_{\boldell}(\bI,t),
\end{align}
where, as before, the index $i$ runs from 1 to 2. The above equation can be solved using the Green's function with the initial condition that $\hat{f}_{1,\boldell}(\bI,0)=0$ to yield the following form for $\hat{f}_{1,\boldell}(\bI,t)$
\begin{align}
\hat{f}_{1,\boldell}(\bI,t) = i\ell_i\frac{\partial f_0}{\partial I_i}\int_{0}^t \rmd \tau e^{-i\zeta(\tau)} \, \hat{\Phi}'_{\boldell}(\bI,t-\tau),
\end{align}
where
\begin{align}
\zeta(\tau) = \ell_i \Omega_i \tau - \ell_3 \int_0^{\tau} \Omega_\rmP(t') \rmd t'.
\end{align}
We substitute this expression in equation~(\ref{torque}) without the $g(t)$ factor to obtain the following form for the self-consistent torque
\begin{equation}
\calT_{2} = \Tsc = 16\pi^3 \reducedsum \ell_3 \int \rmd \bI\, \ell_i\frac{\partial f_0}{\partial I_i}\, \left[\calJ_{1\boldell}(\bI,t)+\calJ_{2\boldell}(\bI,t)\right]\,.
\label{Torque_SC}
\end{equation}
Here $\calJ_{1\boldell}(\bI,t)$ and $\calJ_{2\boldell}(\bI,t)$ are given by
\begin{align}
\calJ_{1\boldell}(\bI,t) = \rm{Re}\left[ \hat{\Phi}^{'\ast}_{\boldell}(\bI,t) \int_{0}^t \rmd \tau \cos\zeta(\tau) \, \hat{\Phi}'_{\boldell}(\bI,t-\tau)\right],\nonumber \\
\calJ_{2\boldell}(\bI,t) = \rm{Im}\left[ \hat{\Phi}^{'\ast}_{\boldell}(\bI,t) \int_{0}^t \rmd \tau \sin\zeta(\tau) \, \hat{\Phi}'_{\boldell}(\bI,t-\tau)\right],
\label{calJsc}
\end{align}
where $\rm{Re}(z)$ and $\rm{Im}(z)$ are the real and imaginary parts of $z$, respectively. As we shall see shortly, $\calJ_{1\boldell}$ is generally the dominant term and $\calJ_{2\boldell}$ is sub-dominant. 

Equation~(\ref{Torque_SC}) is the most general form for the dynamical friction torque in the framework of linear perturbation theory in the absence of collective effects. This self-consistent torque differs from the instantaneous, generalized LBK torque of equation~(\ref{Torque_inst}) in two important ways. First of all, it  modifies the resonances by introducing a time-dependence to the circular frequency $\Omega_\rmP$. Mathematically, this implies that the argument $\ell_k\Omega_k \tau$ of the sinusoidal function in the instantaneous torque is replaced by $\ell_i \Omega_i \tau - \ell_3 \int_0^{\tau} \Omega_\rmP(t') \rmd t'$. Secondly, the self-consistent torque properly accounts for the fact that the perturber potential $\Phi'_\rmP$ evolves as the perturber falls in. This implies that the ${|\hat{\Phi}'_{\boldell}(R(t))|}^2$ term in the instantaneous torque is replaced by a convolution term, i.e.
\begin{align}
\frac{\sin{\ell_k\Omega_k t}}{\ell_k\Omega_k} {\left|\hat{\Phi}'_{\boldell}(\bI,t)\right|}^2 \to \rm{Re} \left[\hat{\Phi}^{'\ast}_{\boldell}(\bI,t)\int_{0}^t \rmd \tau  \cos\zeta(\tau) \, \hat{\Phi}'_{\boldell}(\bI,t-\tau)\right].
\end{align}
In the secular limit where the temporal evolution is very slow, such that $\hat{\Phi}'_{\boldell}(t-\tau) \approx \hat{\Phi}'_{\boldell}(t)$ and $\Omega_\rmP(t') \approx \Omega_\rmP(t)$, we have that
\begin{align}
\calJ_{1\boldell}(\bI,t) \approx  {\left|\hat{\Phi}'_{\boldell}(\bI,t)\right|}^2 \int_0^t \rmd\tau \cos{\ell_k\Omega_k \tau} = {\left|\hat{\Phi}'_{\boldell}(\bI,t)\right|}^2 \, \frac{\sin{\ell_k\Omega_k t}}{\ell_k\Omega_k},
\end{align}
while $\calJ_{2\boldell}\approx 0$. Substituting this in equation~(\ref{Torque_SC}), one recovers the expression for the instantaneous torque of equation~(\ref{Torque_inst}), as required.

While the instantaneous torque only depends on the current time $t$, the self-consistent torque takes into account the entire infall history of the perturber, thereby introducing temporal correlation into the system. This is reminiscent of how in the linear response theory of \cite{Colpi.Pallavicini.98} the overdensity along the trail marked by the perturber exerts a retarding torque on it, i.e., dynamical friction originates from a memory effect involving the stars along the path of the perturber. The self-consistent torque properly accounts for this memory effect, which is ignored in both the instantaneous torque and the LBK torque. While \cite{Colpi.Pallavicini.98} compute the torque for an impulsive, straight orbit of the perturber through a homogeneous medium, our self-consistent torque (equation~[\ref{Torque_SC}]) describes dynamical friction for the more realistic case of a circular orbit in an inhomogeneous background. We also emphasize that even the inhomogeneous Lenard-Balescu equation derived by \cite{Heyvaerts.10}, \cite{Chavanis.12} and \cite{Fouvry.Bar-Or.18}, which is considered the most complete kinetic theory for gravitational systems to date, accounting for both inhomogeneity and collective effects, ignores the memory effect by assuming, in the computation of the diffusion coefficients, that the motion of the perturber is given by the mean-field limit with time-invariant actions (the secular approximation).

\bigskip

\subsection{Orbital Decay}
\label{sec:orbdec}

Under the assumption that the evolution of $R(t)$ is governed by the second-order dynamical friction torque, we have that
\begin{align}
\frac{\rmd R}{\rmd t} = \frac{\rmd R}{\rmd L_\rmP} \, \frac{\rmd L_\rmP}{\rmd t} = \left(M_\rmP\frac{\rmd}{\rmd R}\left[R^2\Omega_\rmP(R)\right]\right)^{-1} \calT_2,
\label{dRdt_torque}
\end{align}
where $L_\rmP=M_\rmP R^2\Omega_\rmP$ is the angular momentum of the perturber. And since the torque $\calT_2$ itself depends on time both explicitly and implicitly (through $R(t)$), we have that the evolution of $R$ is governed by the following integro-differential equation
\begin{align}
&\frac{\rmd R}{\rmd t} = 16\pi^3 \left(M_\rmP\frac{\rmd}{\rmd R}\left[R^2\Omega_\rmP(R)\right]\right)^{-1} \reducedsum \ell_3 \int \rmd \bI\, \ell_i\frac{\partial f_0}{\partial I_i}\, \left[\calJ_{1\boldell}(\bI,t)+\calJ_{2\boldell}(\bI,t)\right].
\label{dRdt}
\end{align}

Solving this equation is rather challenging. However, we can obtain some powerful insight by expanding $R(t)$ using a Taylor series expansion. As long as the rate of infall,  $\rmd R/\rmd t$, varies sufficiently slowly (i.e., $\rmd^2 R/\rmd t^2$ is small), we have that, to good approximation, $R(t-\tau) \approx R(t) - \tau\,\rmd R/\rmd t$. Since $\rmd R/\rmd t \sim M_\rmP/M_\rmG$ and is therefore typically small, $\hat{\Phi}'_{\boldell}(t-\tau)$ can be expanded as a Taylor series and truncated at the leading order to obtain
\begin{align}
\hat{\Phi}'_{\boldell}(t-\tau) \approx \hat{\Phi}'_{\boldell}\big( R - \tau\,\rmd R/\rmd t\big) \approx \hat{\Phi}'_{\boldell}(t) - \frac{\rmd \hat{\Phi}'_{\boldell}}{\rmd R} \, \frac{\rmd R}{\rmd t} \, \tau\,,
\end{align}
where we remind the reader that the time dependence of $\hat{\Phi}'_{\boldell}$ only enters through $R(t)$. Next we note that $\zeta(\tau) = \tau\left[\ell_1 \Omega_1 + \ell_2 \Omega_2 - \ell_3 \overline{\Omega}_\rmP(\tau)\right]$, where
\begin{align}
\overline{\Omega}_\rmP(\tau) = \frac{1}{\tau} \int_0^{\tau} \Omega_\rmP(t') \rmd t'
\label{barOmegaP}
\end{align}
is the time-averaged value of the circular frequency of the perturber. If we now make the assumption that $\overline{\Omega}_\rmP \simeq \Omega_\rmP$, i.e., we neglect the temporal evolution of $\Omega_\rmP$\footnote{This is a reasonable approximation for a low-mass perturber on a circular orbit inside or close to a constant density core, which is the case of interest here.}, then $\zeta(\tau) \to \ell_k\Omega_k\tau$. We thus have that $\calJ_{2\boldell}\approx 0$, and 
\begin{align}
\calJ_{1\boldell}(\bI,t) &\approx  {\left|\hat{\Phi}'_{\boldell}(\bI,t)\right|}^2 \int_0^t \rmd\tau \cos{\ell_k\Omega_k \tau} - \frac{1}{2} \frac{\rmd \vert\hat{\Phi}'_{\boldell}(\bI,t)\vert^2}{\rmd R}\, \frac{\rmd R}{\rmd t} \int_0^t \rmd\tau\,\tau\cos{\ell_k\Omega_k \tau} \nonumber \\
\nonumber \\
&= {\left|\hat{\Phi}'_{\boldell}(\bI,t)\right|}^2\,\frac{\sin{\ell_k\Omega_k t}}{\ell_k\Omega_k} - \frac{1}{2} \frac{\rmd \vert\hat{\Phi}'_{\boldell}(\bI,t)\vert^2}{\rmd R} \, \frac{\rmd R}{\rmd t}\left(t\frac{\sin{\ell_k\Omega_k t}}{\ell_k\Omega_k}-\frac{1-\cos{\ell_k\Omega_k t}}{{\left(\ell_k\Omega_k\right)}^2}\right).
\label{J1l}
\end{align}
Substituting the above expression for $\calJ_{1\boldell}$ in equation~(\ref{Torque_SC}), we can write the second-order self-consistent torque as
\begin{equation}
\calT_{2} = \Tinst + \Tmem,
\label{Torque_SC_fast}
\end{equation}
where
\begin{equation}
\Tmem = -8 \pi^3 \frac{\rmd R}{\rmd t} \reducedsum \ell_3 \int \rmd \bI\, \ell_i \frac{\partial f_0}{\partial I_i}\, \frac{\rmd \vert\hat{\Phi}'_{\boldell}(\bI,t)\vert^2}{\rmd R} \,\left(t\frac{\sin{\ell_k\Omega_k t}}{\ell_k\Omega_k} - \frac{1-\cos{\ell_k\Omega_k t}}{{\left(\ell_k\Omega_k\right)}^2}\right)\,.
\label{Torque_mem}
\end{equation}
Hence, the torque is the sum of the instantaneous torque given by equation~(\ref{Torque_inst}), and a leading order correction term due to the inward radial motion of the perturber. In what follows, we refer to this second term as the memory term.

We can substitute the above expression for the torque (equation~[\ref{Torque_SC_fast}]) in equation~(\ref{dRdt_torque}) to obtain the following evolution equation for $R$,
\begin{align}
\dfrac{\rmd R}{\rmd t} = \dfrac{\Tinst}{\rmd L_\rmP/\rmd R - \Pmem}
\label{dRdt_selfconst}
\end{align}
where $\Pmem \equiv \Tmem/(\rmd R/\rmd t)$ is a momentum term associated with the orbital decay (i.e., the `sinking') of the perturber, and 
\begin{align}
\frac{\rmd L_\rmP}{\rmd R} = \frac{L_\rmP}{R} \, \left[2 + \frac{\rmd\ln\Omega_\rmP}{\rmd\ln R} \right]
\end{align}
is related to the momentum of the perturber in the absence of orbital decay.

At small $t$, $\Pmem$ is small compared to $\rmd L_\rmP/\rmd R$ and the infall is driven by the instantaneous torque, which is subject to transients that slowly die out due to phase-mixing. As time goes on, and orbital decay becomes significant, $\Pmem$ which we find to be typically positive, becomes more and more important, causing the denominator to become smaller. This in turn enhances the orbital decay rate. Hence, the memory term of the self-consistent torque has a destabilizing effect on the orbital decay. This is similar to the destabilizing `dynamical feedback' discussed in TW84. As we will see below, this becomes particularly important when the perturber approaches a central constant density core.

\begin{figure}[t!]
\centering
\includegraphics[width=0.8\textwidth]{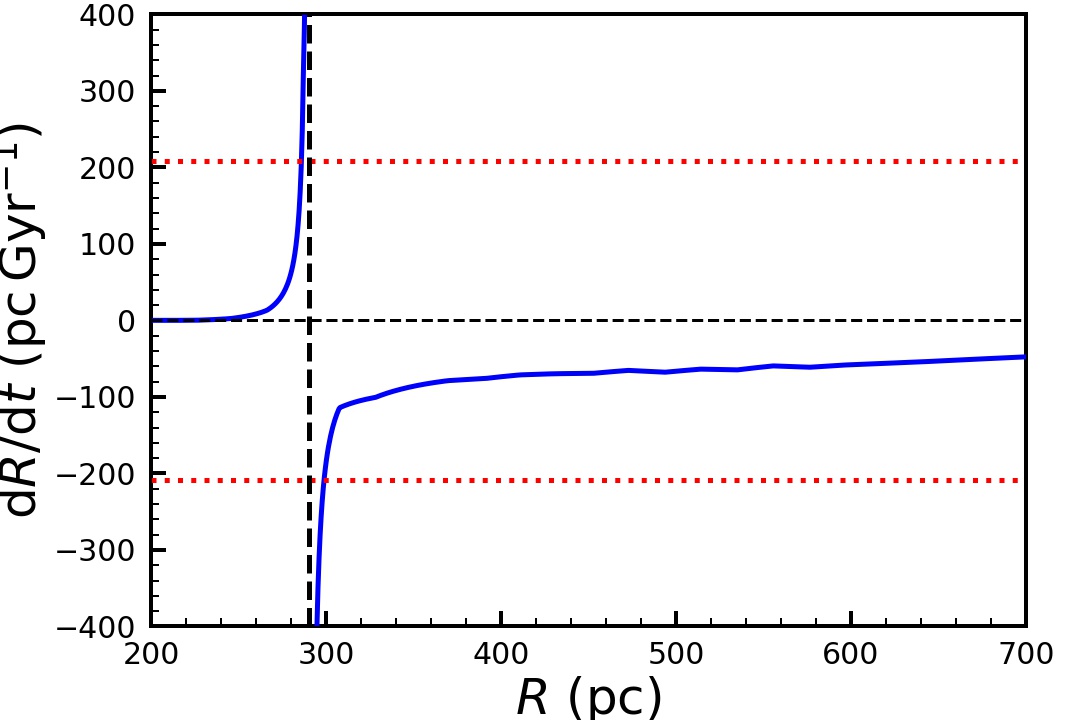}
\caption{The orbital decay rate, $\rmd R/\rmd t$, for our fiducial isochrone plus point-mass perturber system as a function of radius $R$ (equation~\ref{dRdt_selfconst}) for the asymptotic (large $t$) value of  the self-consistent torque exerted by the 10 dominant $(m,l)$ modes shown in Fig.~\ref{fig:LBKiso}. Note that under the approximation of a linear order truncation in $\calJ_{1\boldell}$ as assumed in deriving equation~(\ref{dRdt_selfconst}), $\rmd R/\rmd t \to \pm \infty$ as $R\to R_{\rm crit}=290$ pc (marked by the black vertical dashed line) from left or right. In order to avoid this singular behavior when calculating the orbital decay, we implement a maximum cut-off for $\left|\rmd R/\rmd t\right|$, indicated by the red, dotted lines.}
\label{fig:dRdt_vs_R}
\end{figure}

Note that equation~(\ref{dRdt_selfconst}) indicates the potential presence of a singularity at a critical radius, $\Rcrit$, where $\rmd L_\rmP/\rmd R = \Pmem$. Whether such a radius exists or not depends on both the galaxy potential and the mass of the perturber\footnote{In absence of the perturber, for a cored profile, $\rmd L_\rmP/\rmd R \sim R$ for small $R$. $\Pmem$ on the other hand, typically increases with decreasing $R$. Therefore, for small enough $R$, $\Pmem$ will overtake $\rmd L_\rmP/\rmd R$ and $\rmd R/\rmd t$ will flip sign.}. In the limit $R \downarrow \Rcrit$, the orbital decay rate $\rmd R/\rmd t \to -\infty$. Inside of $\Rcrit$, the denominator flips sign and $\rmd R/\rmd t \to +\infty$ as $R \uparrow \Rcrit$. Fig.~\ref{fig:dRdt_vs_R} demonstrates this by plotting $\rmd R/\rmd t$ as a function of radius $R$ for our fiducial isochrone galaxy plus point mass perturber (see Section~\ref{sec:iso} for details). Here we have assumed the asymptotic (large time) forms for both the instantaneous torque (which equates to the LBK torque) and for the `memory torque', $\Tmem$. The fact that $\rmd R/\rmd t$ flips sign when crossing $\Rcrit$ suggests that this radius must act as an attractor for the dynamical evolution of the perturber, and we therefore associate $\Rcrit$ with the `core-stalling' radius.

\bigskip

\section{Super-Chandrasekhar dynamical friction, Dynamical buoyancy and core-stalling}
\label{sec:core_stalling}

For our fiducial example of a point mass perturber of mass $M_\rmP = 2\times 10^5 \Msun$ on a circular orbit in a spherical isochrone galaxy of mass  $M_\rmG = 1.6\times 10^9 \Msun$ and scale radius $b=1\kpc$ (see Section~\ref{sec:iso}), we use equation~(\ref{dRdt_selfconst}) and a fourth order Runge-Kutta integrator to evolve the radius $R(t)$ of the perturber.  As before, we only consider the contribution to the torque from the ten dominant $(m,\ell)$ modes shown in Fig.~\ref{fig:LBKiso}. We have verified that this sampling of only the dominant modes does not significantly impact the results; in fact, we obtain virtually identical results if we were to only use the eight most dominant modes. In order to avoid problems with the integrator close to the singularity at $\Rcrit \simeq 0.29 b = 290\pc$, we implement a maximum cut-off in $\vert \rmd R/\rmd t\vert$ (shown by dotted, red lines in Fig.~\ref{fig:dRdt_vs_R}).

\begin{figure}[t!]
\centering
\includegraphics[width=0.9\textwidth]{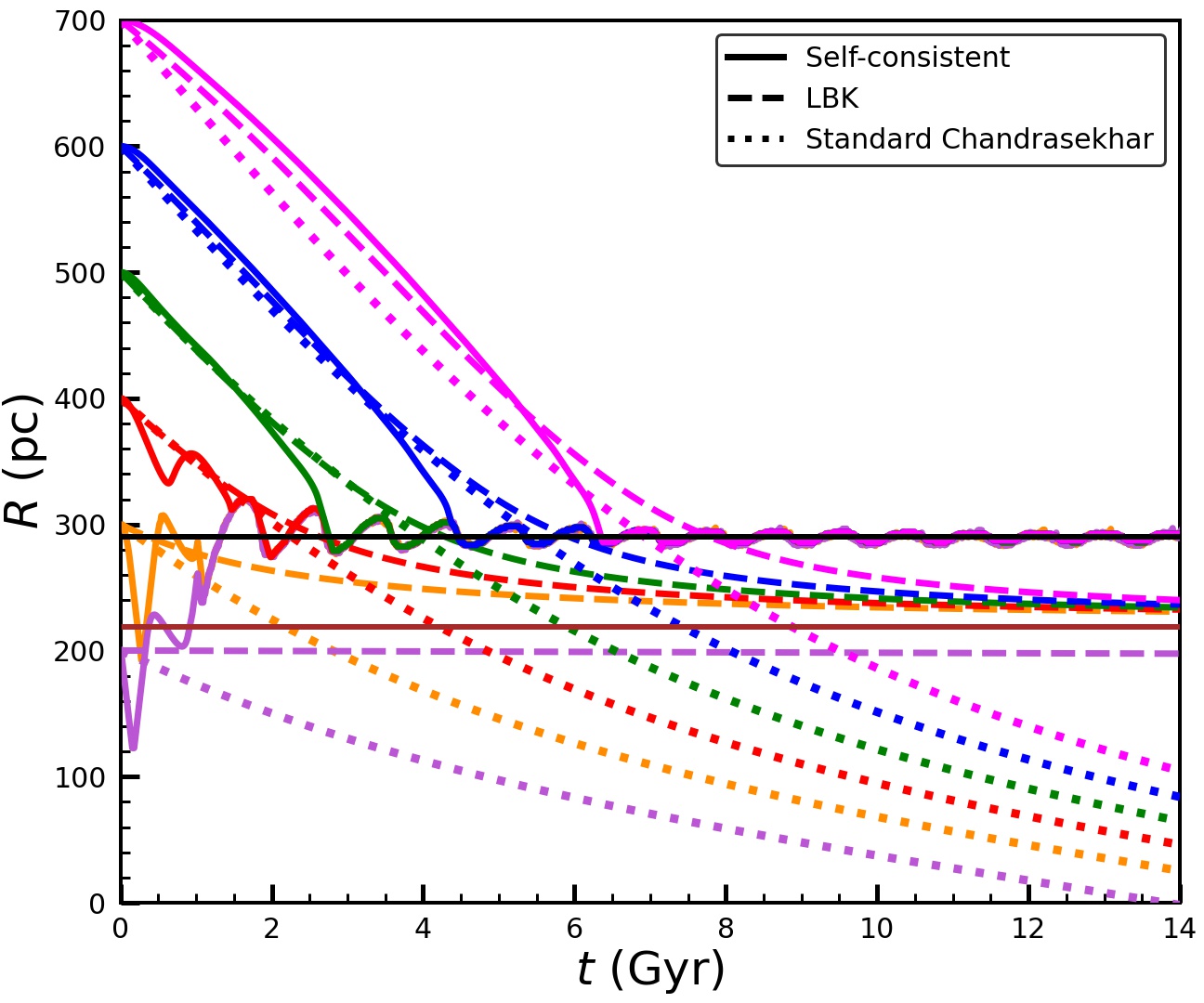}
\caption{The orbital decay of a point mass perturber in our fiducial isochrone sphere. Solid and dashed lines show the results obtained using the self-consistent and LBK torques, respectively, computed using the 10 dominant $(m,\ell)$ modes shown in Fig.~\ref{fig:LBKiso}. The dotted curves show the results obtained using the standard Chandrasekhar formalism, as described in the text. Different colors correspond to different initial radii $R_0 = 700\pc, 600\pc,..., 200\pc$. The horizontal black line indicates the critical radius, $R_{\rm crit}$, where the perturber stalls its infall in our self-consistent formalism. Note the transients at early times when $R_0 \sim R_{\rm crit}$, and the super-Chandrasekhar decay shortly before stalling. For comparison, based on the LBK torque stalling happens at the somewhat smaller filtering radius, $R_\ast$ (horizontal, brown line), defined in KS18 as the radius where $\Omega_\rmP(R)=\Omega_b$. Note that no stalling is expected with the standard Chandrasekhar formalism.}
\label{fig:R_vs_t}
\end{figure}

\begin{figure*}
\centering
\includegraphics[width=1\textwidth]{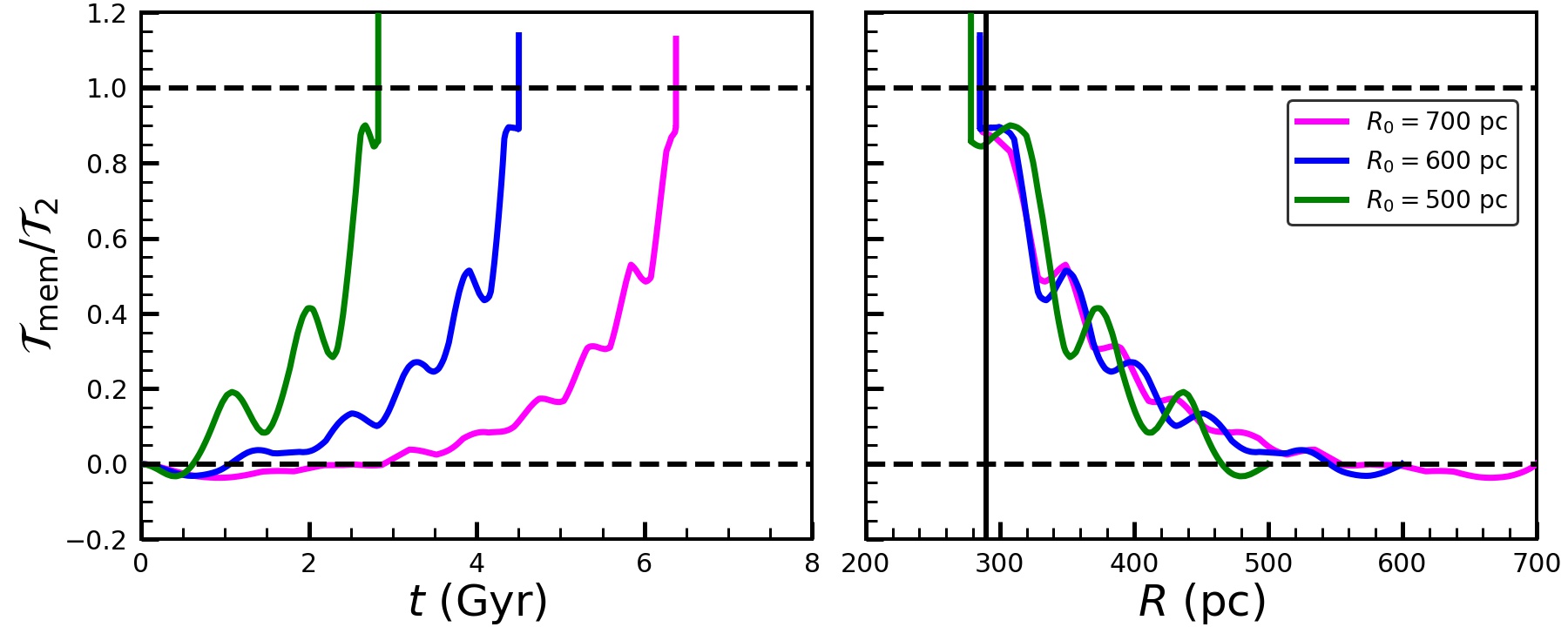}
\caption{The memory torque $\calT_{\rm mem}$ normalized by the total torque $\calT_2$ and computed using the $10$ dominant $(m,\ell)$ modes shown in Fig.~\ref{fig:LBKiso}, for the orbital decay of a point perturber in our fiducial isochrone sphere. Left (right) panel plots  $\calT_{\rm mem}/\calT_2$ vs $t$ ($R$) for three different initial radii $R_0$ as indicated. Note that the memory torque is initially retarding and sub-dominant but gradually gains strength, while undergoing oscillations, until it dominates (causing the accelerated Super-Chandrasekhar infall) near the critical radius $\Rcrit$ (marked by the vertical black line in the right-hand panel), where it flips sign, making the total torque enhancing (dynamical buoyancy).}
\label{fig:Torque_mem}
\end{figure*}

The solid lines in Fig.~\ref{fig:R_vs_t} plot the resulting orbital decay tracks, $R(t)$, obtained for 6 different initial radii, $R_0 = 700\pc, 600\pc,..., 200\pc$, which bracket $\Rcrit$. For comparison, we also show for each case the orbital decay track obtained using the LBK torque (dashed lines), and the standard Chandrasekhar formalism (dotted lines), which are obtained by solving equation~(\ref{dRdt_torque}) with $\calT_2 = \Tlbk$ and $\calT_2 = \bR \times \bF_{\rm DF}$, respectively. Here $\bF_{\rm DF}$ is given by equation~(\ref{aDF}) where we follow KS18 by setting $\ln{\Lambda}=\ln(R/a)$, with $a=10\pc$ the assumed scale radius for the perturber,\footnote{Since $a \ll b$, we are justified to treat the perturber as a point mass in the computation of the torque.} and properly compute $\rho(<v)$ from the isotropic distribution function of the isochrone sphere (equation~[\ref{f0}]).

When the initial radius of introduction, $R_0$, is large (compared to $R_{\rm crit}$), the orbital decay is characterized by four distinct phases of infall:
\begin{itemize}
    \item Phase I: Following the introduction into the system, the perturber falls in at a slightly slower rate than what is predicted by the LBK torque alone. This is because it takes time for the torque to build up and saturate to the asymptotic LBK value.
    
    \item Phase II: Once the transients have died out, and the torque has reached the steady LBK value, the infall rate of the perturber matches that predicted by the LBK torque.
    
    \item Phase III: As it approaches $\Rcrit$, the perturber starts to fall in at an accelerated pace, much faster than predicted by either the LBK torque or the standard Chandrasekhar formalism. This enhancement of the torque occurs only in the core region of the galaxy and is known as super-Chandrasekhar dynamical friction \citep[see e.g.,][]{Read.etal.06c, Goerdt.etal.10, Zelnikov.Kuskov.16}.
    
     \item Phase IV: Finally, the perturber reaches the stalling radius $R_{\rm st} \simeq \Rcrit$ about which it oscillates under the action of dynamical friction (retarding torque) outside and buoyancy (enhancing torque) within.
\end{itemize}

When the initial radius $R_0$ is close to $R_{\rm crit}$ the transients due to the self-consistent torque become more pronounced, in agreement with Fig.~\ref{fig:genLBK}. Introducing the perturber inside of the critical radius (i.e., $R_0 < R_{\rm crit}$) results in it being pushed out to $R_{\rm crit}$ (following initial transient oscillations).  Following \citet{Cole.etal.12} we refer to this as `dynamical buoyancy'.

This complicated behavior is in excellent, qualitative agreement with numerical $N$-body simulations, which have revealed how perturbers, upon approaching a central core, undergo accelerated super-Chandrasekhar friction, followed by a `kick-back' effect in which the perturber is pushed out again before it ultimately settles (stalls) at some radius, typically close to the core radius (see Section~\ref{sec:intro_4} for references).  

None of this is predicted by the standard Chandrasekhar formalism, according to which the perturber continues to sink all the way towards the center, albeit at a rate that becomes smaller towards the core. The latter owes to the fact that both $\rho(<v)$ and the Coulomb logarithm $\ln{\Lambda}=\ln(R/a)$ decrease with decreasing $R$. However, the resulting decline in the Chandrasekhar torque is insufficient to result in stalling. 

The LBK torque is more successful, in that it clearly predicts core stalling. As discussed in detail in KS18, the stalling is expected to occur at or near the `filtering radius', $R_\ast$, defined as the radius where the circular frequency of the perturber is equal to that of stars in the central core region \citep[see also][]{Read.etal.06c}. KS18 showed that the lower order modes, which otherwise exert a strong torque on the perturber, drop out of resonance, causing a significant reduction in the amplitude of the LBK torque (see also Fig.~\ref{fig:LBKiso}). This suppression of resonances arises from the fact that just outside of $R_\ast$, the circular frequency of the perturber is just a little bit lower than that of the core stars. Indeed, as shown by the dashed curves in Fig.~\ref{fig:R_vs_t}, based on the LBK torque one predicts that the infall stalls just outside of $R_\ast$ (indicated by the horizontal, brown line).  Note, though, that the LBK torque does neither predict a super-Chandrasekhar phase, nor dynamical buoyancy (the LBK torque is always retarding). In fact, introducing a perturber at $R_0 = 200 \pc < R_{\rm crit}$, the formalism based on the LBK torque predicts that it remains at that radius (see purple, dashed line in Fig.~\ref{fig:R_vs_t}),

Our formalism based on the self-consistent torque predicts a much richer dynamics, including dynamical buoyancy and super-Chandrasekhar infall. It also implies an explanation for core-stalling that is intriguingly different from that based on the LBK torque. Rather than resulting from a diminishing of the dynamical friction torque, core-stalling is an outcome of a balance between friction and buoyancy. All of this owes to the memory torque, which becomes dominant over the instantaneous torque close to $R_{\rm crit}$, and which causes the total torque to flip sign upon crossing $R_{\rm crit}$. This is illustrated in Fig.~\ref{fig:Torque_mem}, which plots $\calT_{\rm mem}/\calT_2$ as a function of time (left-hand panel) and radius (right-hand panel), respectively, for three different values of the initial radius $R_0$, as indicated. Initially, the torque is dominated by the instantaneous term, $\calT_{\rm inst}$, given by equation~(\ref{Torque_inst}). The memory torque slowly gains strength, while undergoing oscillations, and starts dominating when the perturber approaches $R_{\rm st}$ (indicated by the vertical black line in the right-hand panel). At this radius, $\calT_{\rm mem}$ (and thus also the total torque) flips sign and becomes enhancing, giving rise to dynamical buoyancy when $R < R_{\rm st}$, even though the instantaneous torque remains retarding.

\medskip

\subsection{Caveats and Outstanding Issues}
\label{sec:caveats}

Despite its success in reproducing previously unexplained aspects of dynamical friction observed in numerical simulations, in particular super-Chandrasekhar infall and dynamical buoyancy, the treatment of orbital decay based on the self-consistent torque presented above is subject to a few caveats.

First of all, we have ignored the time-evolution of $\Omega_\rmP$ (i.e., we assumed that $\Omega_\rmP = \overline{\Omega}_\rmP$). Although this is likely to be a reasonable approximation close to a constant density core, where $\Omega_\rmP(R)$ is nearly independent of radius, it remains to be seen how a proper treatment with a non-zero $\rmd\Omega_\rmP/\rmd t$ impacts the orbital decay. Unfortunately, since the temporal evolution of $\Omega_\rmP$, as quantified by $\overline{\Omega}_\rmP$ (equation~[\ref{barOmegaP}]), enters as an argument of the cosine and sine in the expressions for $\calJ_{1\boldell}(\bI,t)$ and $\calJ_{2\boldell}(\bI,t)$, respectively, numerically evaluating the corresponding integrals is non-trivial. 

Secondly, in deriving the expression for $\rmd R/\rmd t$ (equation~[\ref{dRdt_selfconst}]) we expanded $R(t)$ as a Taylor series that we truncated at first order. This is only valid as long as $\rmd^2 R/\rmd t^2$ is sufficiently small. Unfortunately, this is likely to be violated during the super-Chandrasekhar phase, when the rate of infall rapidly accelerates. This caveat, which is also responsible for the singular behavior at $\Rcrit$, can be overcome by using a higher-order truncation of the Taylor series, or by trying to directly solve the integro-differential equation~(\ref{dRdt}). We leave this as an exercise for future investigations. 

Finally, the entire formalism is based on perturbation theory, and therefore hinges on the assumption that the perturbation parameter $\vert \Phi_1^{\rm ext}/H_{0\rmJ}\vert$ is small. This assumption becomes questionable whenever the galaxy mass enclosed by the perturber, $M_\rmG(R)$, becomes comparable to the perturber mass. Unfortunately, in numerical simulations core stalling often happens at a radius at which $M_\rmG(R) \sim M_\rmP$ \citep[][]{Petts.etal.15, Petts.etal.16, DuttaChowdhury.etal.19}. In addition, when the perturber stalls at a fixed radius, the resonances no longer sweep by the stars fast enough to prevent non-linear perturbations from developing (i.e., one is no longer in what TW84 refer to as the `fast regime'). These non-linearities can even reverse the gradient of the distribution function near the resonances and contribute to an enhancing torque (which may counteract the retarding torque from the `fast' resonances and stall the infall) if the stars remain near-resonant for long enough ($\rmd\Omega_\rmP/\rmd t$ is slow enough), as is the case near the stalling radius. \cite{Sellwood.2006} finds such an effect in $N$-body simulations of a rotating bar-like perturbation in a spherical galaxy.

All of this suggests that a proper treatment of core stalling may not be possible with perturbation theory. In chapter~\ref{chapter: paper5} \citep[][]{Banik.vdBosch.22}, we therefore examine dynamical friction, and core-stalling in particular, using a non-perturbative, orbit-based approach. This reveals a family of (perturbed) orbits that exert a coherent, enhancing torque, thus contributing to dynamical buoyancy. When the perturber approaches the central core region, the nature of the near-resonant orbits changes, due to a bifurcation of the inner Lagrange points, causing buoyancy to become dominant over friction. Hence, the non-perturbative, orbit-based approach lends support to our conclusion that central core regions manifest dynamical buoyancy, something that was first noticed in the numerical simulations by \citet{Cole.etal.12}.

\bigskip

\section{Conclusion}
\label{sec:concl_4}

Various approaches to describe dynamical friction in inhomogeneous systems have shown that it ultimately arises from a torque that has a non-zero contribution only from stars in resonance with the perturber. Ultimately, this notion that only the resonances contribute to the torque has its origin in the assumption that the orbital decay rate of the perturber is (secular approximation) and always has been (adiabatic approximation) very slow compared to the dynamical time of the host\footnote{This is similar to Bogoliubov's ansatz in plasma physics that the two-point correlation function relaxes much faster than the one-point distribution function.}. In the Hamiltonian perturbation theory of TW84 and KS18 the adiabatic approximation is enforced by multiplying the perturber potential by $\rme^{\gamma t}$ and taking the limit $\gamma \to 0$, while the secular approximation enters when the assumption is made that the orbital radius and circular frequency of the perturber are time-invariant over a dynamical time. This, in turn, implies that the equations of motion of both the perturber and the field particles are predominantly determined by the mean field, and thus characterized by slowly varying actions. Note that the same assumptions also underlie other approaches to dynamical friction in an inhomogeneous background, such as that based on the generalized Landau equation \citep[e.g.,][]{Chavanis.13} or the stochastic approach in action-angle space based on the fluctuation dissipation theorem \citep[e.g.,][]{Fouvry.Bar-Or.18}.

The secular and adiabatic approximations are justified when the mass of the perturber is sufficiently small. In that case, the dynamical friction time is much longer than the dynamical time. However, dynamical friction is mainly of astrophysical interest if the friction time is shorter than the Hubble time, which typically implies a perturber mass $M_\rmP$ in excess of 1-10 percent of the host mass. For such massive perturbers the dynamical friction time is no longer well separated from the dynamical time, and the secular and adiabatic assumptions are no longer justified. This breakdown is especially acute in the case of super-Chandrasekhar dynamical friction observed in numerical simulations when a massive perturber approaches a constant density core.

In this chapter we have examined implications of relaxing the adiabatic and secular assumptions. Using Hamiltonian perturbation theory similar to KS18, but without taking the limit $\gamma \to 0$ and without the assumption that the response density builds up on the same time scale as that on which the perturber is introduced, we first relaxed the adiabatic approximation and derived an expression for the generalized LBK torque (equation~[\ref{Torque_gen}]). This differs from the standard LBK torque in that it depends on the growth rate $\gamma$ and has contribution from all orbits, resonant and non-resonant. Taking the adiabatic limit $\gamma\to 0$, i.e., assuming an extremely slow growth of the perturber potential, we recovered the LBK torque with a non-zero contribution only from the pure resonances. The opposite limit, $\gamma\to \infty$, corresponding to an instantaneous introduction of the perturber as typically done in idealized numerical simulations, leads to a time-dependent torque with a non-zero contribution from the near-resonant orbits along with the purely resonant ones. This `instantaneous' torque builds up linearly with time before undergoing oscillations (`transients') about the LBK value. Over time these oscillations damp out, and in the long-term the generalized torque reduces to the LBK torque. This behavior is analogous to how a forced, damped oscillator undergoes transients before settling to a steady-state solution in which the frequency of the response matches the driving frequency. The main difference is that here the damping is due to phase-mixing of the responses from the individual orbits (each with its own frequencies), rather than due to some dissipative processes. The time-scale of relevance here is the time-scale on which the transients damp away, which is proportional to the dynamic range in orbital frequencies of the field particles that make up the host. Typically this range is sufficiently large and phase-mixing is very efficient, causing the generalized LBK torque to quickly transition to the LBK torque. This justifies the standard treatments of dynamical friction, in which only the resonances contribute, even when the adiabatic approximation is not necessarily justified. However, there is one important exception, which is the case when the perturber is introduced close to a central constant-density core. Here the dynamic range of frequencies is drastically suppressed, causing large transient oscillations that can take many orbital periods of the perturber to phase mix away (see Fig.~\ref{fig:genLBK}).

Although the generalized LBK torque gives useful insight as to how transients that result from a non-adiabatic introduction of the perturber phase mix away, it is still based on the unphysical ansatz that the perturber grows its mass exponentially over time, on a characteristic time $\tau_{\rm grow} = 1/\gamma$, while remaining at a fixed host-centric radius, $R$. This time invariance of $R$ is a manifestation of the secular approximation. In order to improve on this, we next computed the torque in a self-consistent manner, in which we retained the information about the time dependence of the potential and circular frequency of the perturber throughout the entire evolution of the perturber's orbital radius, $R(t)$ (we relaxed the secular approximation). This self-consistent torque differs from the (generalized) LBK torque in that the instantaneous circular frequency of the perturber is replaced by its time-averaged value, and that it includes a convolution term that embodies the temporal correlation of the perturber potential. As a consequence, the self-consistent torque always has a non-zero contribution from the near-resonant orbits, and depends on the entire infall history, $R(t)$, which in turn is dictated by the torque itself. A proper description of dynamical friction thus requires solving an integro-differential equation for $R(t)$ (equation~[\ref{dRdt}]).

While solving this equation in full generality is highly non-trivial, we obtained some valuable insight by Taylor expanding $R(t)$ and truncating it at first order. This is valid as long as  $\rmd^2 R/\rmd t^2$ is sufficiently small, i.e., the rate of infall, $\rmd R/\rmd t$ varies slowly. If, in addition, we assume that the time-dependence of the perturber's frequency is small, which is a valid assumption at or near a central core region, we can write the self-consistent torque as a sum of two terms, the instantaneous torque, which depends on $R(t)$, and a memory torque, which is proportional to $\rmd R/\rmd t$ besides having an $R$ dependence. We used this simplified form of the self-consistent torque to evolve the radius $R(t)$ of a point mass perturber in an isochrone galaxy (which has a central core). We found that the infall of the perturber occurs in four subsequent phases: (i) sub-LBK infall during the initial (linear) build-up of the torque, (ii) infall at the LBK rate as the instantaneous torque asymptotes to the LBK torque, (iii) accelerated super-Chandrasekhar infall due to a destabilizing effect of the memory torque, and (iv) kick-back of the perturber from within a critical radius $\Rcrit$ due to buoyant effects followed by stalling at that radius. The instantaneous torque dominates the early phase of the infall while the memory torque becomes dominant near the critical radius. It is responsible for the super-Chandrasekhar infall and flips sign at $\Rcrit$, causing the total torque to become enhancing for $R<\Rcrit$. When the perturber is introduced inside of $\Rcrit$, it is consequently pushed out (dynamical buoyancy) to $\Rcrit$ by this enhancing memory torque.

These phenomena of super-Chandrasekhar infall followed by kick-back and core-stalling, as well as dynamical buoyancy inside central core regions, have been observed in numerous $N$-body simulations \citep[][]{Read.etal.06c, Goerdt.etal.10, Inoue.11, Cole.etal.12, DuttaChowdhury.etal.19}, but have thus far eluded a proper explanation \citep[but see][for some phenomenological explanations]{Read.etal.06c, Petts.etal.15, Petts.etal.16, Zelnikov.Kuskov.16}. Although KS18 had shown that the LBK torque strongly diminishes as one approaches a core, which they advocated as an explanation for core stalling, they were unable to explain either super-Chandrasekhar infall or dynamical buoyancy. Based on our results, we argue that core-stalling is ultimately a consequence of a subtle balance between dynamical friction (retarding torque) and buoyancy (enhancing torque), which is preceded by a phase of super-Chandrasekhar friction caused by the destabilizing effect of the memory torque that depends on the past infall history.

Finally, while wrapping up the paper \citep[][]{Banik.vdBosch.21a}, on which this chapter is based, we became aware of an unpublished study by M. Weinberg \citep[][]{Weinberg.04}, in which they also point out the problematic nature of the `time-asymptotic limit' (i.e., taking $\gamma \to 0$) used to derive the LBK torque. Using Hamiltonian perturbation theory similar to what is presented here, but using Laplace transforms rather than Green's functions to solve for the response, they obtain a time-dependent torque (equation~[14] in their paper) that is identical to our self-consistent torque of equation~(\ref{Torque_SC}), except that it doesn't explicitly account for a time-dependence of the perturber frequency. They then proceed to examine how the time-dependent torque differs from the LBK torque for the examples of a slowing bar and a decaying satellite. In the latter case, rather than calculating the orbital decay of the satellite self-consistently, as done here, they first compute the orbital decay $R(t)$ using the local Chandrasekhar formula, which is then substituted in the expression for the time-dependent torque. In agreement with our results, they show that massive perturbers, which decay rapidly, are significantly impacted by transients that are not accounted for in the LBK torque.





    \begin{subappendices}


\chapter*{Appendix}

\section{The isochrone sphere}
\label{app:model}

All specific examples presented in chapter~\ref{chapter: paper4} correspond to a point mass perturber, with mass $M_\rmP$, moving on a circular orbit in an isotropic isochrone sphere, whose potential and density are given by equations~(\ref{IsoPot}) and~(\ref{IsoDens}), respectively. In addition, the distribution function of the (unperturbed) isotropic isochrone sphere of mass $M_\rmG$ is given by
\begin{align}
f_0(\varepsilon) &= \frac{M_\rmG}{\sqrt{2}{(2\pi)}^3 {(G M_\rmG b)}^{3/2}} \nonumber \\
&\times \frac{\sqrt{\varepsilon}}{{\left[2(1-\varepsilon)\right]}^4} \left[27-66\varepsilon+320\varepsilon^2 - 240\varepsilon^3 + 64\varepsilon^4 + 3(16\varepsilon^2 + 28\varepsilon - 9) \frac{\arcsin \sqrt{\varepsilon}}{\sqrt{\varepsilon(1-\varepsilon)}} \right],
\label{f0}
\end{align}
\citep[e.g.,][]{Binney.Tremaine.08}, where $\varepsilon = -E_0\, b/G M_\rmG$ and covers the range $0<\varepsilon\leq 1/2$. 

In absence of the perturber, the orbits of the field particles are characterized by four isolating integrals of motion: the energy $H_0$ and the three actions $(I_r,L,L_z)$. Following KS18 we make a canonical transformation from $(I_r,L,L_z)$ to $(I,L,L_z)$ where $I$ is given by $2 I_r + L$, with $0\leq L \leq I$ and $-L\leq L_z\leq L$. In terms of $I$ and $L$, the orbital energy per unit mass is given by
\begin{align}
E_0(I,L) = -\frac{2{\left(G M_\rmG\right)}^2}{{\left[I+\sqrt{I^2_\rmb+L^2}\right]}^2}\,,
\label{E0}
\end{align}
where $I_b \equiv 2\sqrt{G M_\rmG b}$. While $E_0$ is conserved in the inertial frame, in the co-rotating perturbed frame the conserved quantity is the Jacobi Hamiltonian, given by
\begin{align}
H_{\rmJ 0}(I,L,L_z) = E_0(I,L) - \Omega_\rmP L_z\,.
\label{HJ0}
\end{align}

Corresponding to the actions are the conjugate angles $(w,g,h)$, whose corresponding frequencies are given by
\begin{align}
&\Omega_w(I,L) = \frac{\partial H_{\rmJ 0}}{\partial I} = \frac{\Omega_r}{2} = \frac{4{\left(G M_\rmG\right)}^2}{{\left[I+\sqrt{I^2_\rmb+L^2}\right]}^3}, \nonumber \\
&\Omega_g(I,L) = \frac{\partial H_{\rmJ 0}}{\partial L} = \Omega_\psi-\frac{\Omega_r}{2} = \frac{L}{\sqrt{I^2_\rmb+L^2}} \Omega_w(I,L), \nonumber \\
&\Omega_h(I,L) = \frac{\partial H_{\rmJ 0}}{\partial L_z} = -\Omega_\rmP.
\label{frequencies}
\end{align}
Here $\Omega_g$ is the frequency of periapse precession. $\Omega_r$ and $\Omega_\psi$ are the radial and angular frequencies in the orbital plane, which can be expressed in terms of the actions $I$ and $L$ as
\begin{align}
&\Omega_r (I,L) = \frac{8{\left(G M_\rmG\right)}^2}{{\left[I+\sqrt{I^2_\rmb+L^2}\right]}^3}, \nonumber \\
&\Omega_\psi (I,L) = \frac{\Omega_r}{2} \left(1+\frac{L}{\sqrt{I^2_\rmb+L^2}}\right)\,.
\end{align}
The fact that all these (unperturbed) frequencies can be expressed as simple algebraic functions is what makes the isochrone potential ideal for an analytical exploration of core-stalling. 

Following KS18, we ignore the torque from the stars outside of the core, which allows us to truncate the integration over $I$ at a maximum value $I_{\rm max} \ll I_b$. We follow KS18 and adopt $I_{\rm max}=0.1\,I_b$. Under this approximation the expressions for the frequencies can be simplified as follows
\begin{align}
&\Omega_w \approx \Omega_b\left(1-3\frac{I}{I_b}\right), \nonumber \\
&\Omega_g \approx \Omega_b \frac{L}{I_b},
\label{Omega_wg}
\end{align}
where $\Omega_b = 0.5\sqrt{G M_\rmG/b^3}$ is the central frequency of the galaxy.

Substituting the above expressions for the frequencies, we have the following expression for the resonance angle
\begin{align}
\ell_k\Omega_k = n\,\Omega_w + \ell\,\Omega_g - m\,\Omega_\rmP = s\,\Omega_b - m\,\Omega_\rmP\,,
\end{align}
where 
\begin{align}
s = s(I,L) \equiv \left[n \left(1 - 3\frac{I}{I_b}\right) + \ell\frac{L}{I_b}\right]\,.
\end{align}
KS18 find that the co-rotation resonances with $n=m$ exert much stronger torque than the ones with $n\neq m$; therefore we only study co-rotation modes in chapter~\ref{chapter: paper4}. 

Substituting the above expressions in equation~(\ref{Torque_inst}), we arrive at the following form for the instantaneous torque for the $(m,\ell,m)$ mode
\begin{align}
\calT_{2,m\ell} &= 16\pi^3 m \Omega_b \int_0^{I_{\rm max}} \rmd I \int_0^I \rmd L \,\dfrac{\sin{\left[s\,\Omega_b-m\,\Omega_\rmP\right] t}}{s\,\Omega_b-m\,\Omega_\rmP} \,s(I,L) \,\frac{\partial f_0}{\partial E_0}\, P_{m\ell m}(I,L),
\label{Torque_inst_isochrone}
\end{align}
where $P_{m\ell m}(I,L)$ is given by
\begin{align}
P_{m\ell m}(I,L)=\int_{-L}^{L} \rmd L_z {\left|\hat{\Phi}'_{m\ell m}(I,L,L_z)\right|}^2.
\end{align}
We compute the Fourier coefficients $\hat{\Phi}'_{m\ell m}(I,L,L_\rmz)$ using the analytical expressions given in Appendix A of KS18. 

The corresponding LBK torque is given by
\begin{align}
\calT_{2,m\ell}^{\rm LBK} &= 16\pi^4 m^2\, \Omega_\rmP \int_0^{I_{\rm max}} \rmd I \int_0^I \rmd L \,\delta\left[s\,\Omega_b - m\,\Omega_\rmP\right] \frac{\partial f_0}{\partial E_0}\, P_{m\ell m}(I,L).
\label{Torque_LBK_isochrone}
\end{align}

    \end{subappendices}

%
%
%

\chapter{Dynamical Friction, Buoyancy and Core-Stalling -- A Non-perturbative Orbit-based Analysis} 
\label{chapter: paper5}

\begin{center}

Majority of this chapter has been published as:

\vspace*{5pt}

\author{Uddipan Banik, Frank~C.~van den Bosch

\vspace*{5pt}

}

\textit{The Astrophysical Journal}, Volume 926, Number 2, Page 215 \\
\textit{\citep[][]{Banik.vdBosch.22}}

\end{center}


\section{Introduction}
\label{sec:intro_5}

Dynamical friction is an important relaxation mechanism in gravitational $N$-body systems like galaxies and clusters. Massive objects such as black holes, globular clusters and dark matter subhaloes lose energy and angular momentum to the field particles and sink to the centers of their host systems, driving the system towards equipartition. \cite{Chandrasekhar.43} was the first to derive an expression for the dynamical friction force on a massive object (hereafter the `perturber') travelling through a homogeneous medium on a straight orbit, by summing the velocity changes from independent two body encounters with the field particles. Despite its obvious over-simplifications, applying the formula for Chandrasekhar's friction force using the {\it local} density and velocity distribution of the particles in an {\it inhomogeneous} body, such as a halo or galaxy, yields results that are in fair agreement with numerical simulations \citep[][]{Lin.Tremaine.83, Cora.etal.97, vdBosch.etal.99, Hashimoto.etal.03, Boylan-Kolchin.etal.08, Jiang.etal.08}. However, this `local approximation' fails to account for the cessation of dynamical friction in the central regions of halos or galaxies with a constant-density core. This so-called core-stalling has been observed in $N$-body simulations \citep[e.g.,][]{Read.etal.06c, Inoue.11,  Petts.etal.15, Petts.etal.16, DuttaChowdhury.etal.19} but is still not properly understood. In addition, the simulations also show that prior to stalling the object often experiences a short phase of enhanced `super-Chandrasekhar friction', followed by a `kick-back' effect in which it is pushed out before it settles at the `core-stalling radius' \citep[][]{Goerdt.etal.10, Read.etal.06c, Zelnikov.Kuskov.16}. In fact,  \citet{Cole.etal.12} have shown that massive objects initially placed near the center of a cored galaxy experience a `dynamical buoyancy' that pushes them out towards this stalling radius. This complicated phenomenology cannot be explained using Chandrasekhar's treatment of dynamical friction, which instead predicts that the orbits of massive objects continue to decay inside a central core region, albeit at a reduced rate \citep[e.g.,][]{Hernandez.Gilmore.98, Banik.vdBosch.21a}. 

Dynamical buoyancy can have important astrophysical implications in cored galaxies, where it can either push out massive objects such as nuclear star clusters and supermassive black holes from the central regions, or stall their in-fall (core-stalling) by counteracting the effect of dynamical friction. The latter has been invoked by \cite{Goerdt.etal.10} and \cite{Cole.etal.12} to explain the survival of the globular clusters in the Fornax dwarf galaxy, hinting at the possibility of a central dark matter core.

Given that Chandrasekhar's expression for the dynamical friction force is based on the highly idealized assumption of straight orbits in a uniform, isotropic background, it should not come as a surprise that there are circumstances under which it fails. \citet[hereafter TW84]{Tremaine.Weinberg.84} generalized the description of dynamical friction to a more realistic system of an inhomogeneous spherical galaxy with a small, time-dependent perturbation (bar or satellite). Using Hamiltonian perturbation theory to perturb the actions of the field particles (or `stars') up to second order in the perturbation parameter, they infer that dynamical friction arises from a net retarding torque on the perturber from stars along purely resonant orbits (whose orbital frequencies are commensurable with the circular frequency of the perturber). This torque, known as the LBK torque, was first derived by \cite{LyndenBell.Kalnajs.72} in the context of angular momentum transport driven by spiral arms in disk galaxies. \citet[hereafter KS18]{Kaur.Sridhar.18} showed that for a cored \cite{Henon.59} Isochrone galaxy the LBK torque vanishes at a certain radius in the core due to the suppression in the number of contributing resonances and reduction of the strength of the torque from the surviving resonances, causing the perturber to stall. However their treatment does not explain the origin of super-Chandrasekhar dynamical friction or dynamical buoyancy.

In \citet[hereafter BB21]{Banik.vdBosch.21a}, we showed that an exclusive contribution from resonances between the perturber and the field particles to the LBK torque, as obtained by TW84 and KS18, is ultimately a consequence of two key assumptions, the adiabatic (slow growth of the perturber) and secular (slow in-fall under dynamical friction) approximations, which effectively boil down to ignoring the effect of friction-driven in-fall in the computation of the torque. In BB21 we relaxed these two assumptions and properly accounted for the time dependence of the location and circular frequency of the perturber (due to its radial in-fall motion) to compute the response density and the corresponding self-consistent torque, $\calT_{\rm SC}$. This differs from the standard LBK torque in two key aspects: (i) it has a significant contribution from near-resonant orbits, and (ii) it not only depends on the instantaneous orbital radius of the perturber, $R(t)$, but on its entire in-fall history by involving a temporal correlation of the perturber potential. We showed that super-Chandrasekhar dynamical friction, dynamical buoyancy and core-stalling can all be explained as consequences of this ``memory effect".

Although this self-consistent formalism is more general than the standard LBK formalism and offers predictions related to core stalling that qualitatively match those from numerical simulations, it suffers from a few caveats. First of all, in order to avoid having to solve the complicated integro-differential equation for the self-consistent evolution of $R(t)$, BB21 assume the in-fall rate, $\rmd R/\rmd t$, to be slowly varying over time. This allows $\calT_{\rm SC}$ to be written as the sum of an instantaneous torque, $\calT_{\rm inst}$, that depends on time $t$ and the orbital radius $R(t)$, and a memory torque, $\calT_{\rm mem}$, that is proportional to $\rmd R/\rmd t$. The latter becomes dominant in the core region and acts as a source of destabilizing feedback, giving rise to an accelerated super-Chandrasekhar in-fall outside a critical radius, $\Rcrit$. Inside $\Rcrit$, the memory torque flips sign and becomes enhancing, i.e., exerts dynamical buoyancy. The perturber is thus found to stall at $\Rcrit$ due to a balance between friction outside and buoyancy within, i.e., $\Rcrit$ acts as an attractor. However, the critical behaviour near this radius ($\rmd R/\rmd t \to \pm \infty$ as $R\to \Rcrit$ instead of approaching zero as is typical for a stable attractor) is an artefact of the assumption of a near-constant $\rmd R/\rmd t$, which becomes questionable close to $\Rcrit$ as the perturber undergoes an accelerated in-fall before stalling at this radius. This critical behaviour can be smoothed out by solving the integro-differential equation for $R(t)$ in its full generality, which is however a non-trivial exercise. 

The second caveat of the self-consistent formalism (and of previous studies like TW84 and KS18) is related to the concept of resonances in linear perturbation theory. In this perturbative picture, dynamical friction is driven by resonances between the unperturbed frequencies of the stars and the perturber. But these resonances themselves drastically change (`perturb') the actions and frequencies of the resonant stars, questioning the very assumption of a weak perturbation. TW84 address this philosophical issue by introducing the concept of `sweeping through the resonances', i.e., linear perturbation theory only holds in the `fast' regime, where the circular frequency of the perturber changes rapidly under dynamical friction such that the stars fall out of resonance before their actions can change significantly and give rise to non-linear perturbations in the distribution function. However, in a cored galaxy, as the perturber slows down upon approaching the stalling radius, stars no longer sweep fast enough through the resonances. Therefore, perturbation theory, especially a linear order one, becomes questionable in this `slow' regime. 

The final caveat relates to the fact that linear perturbation theory assumes a weak perturbing potential, i.e., the mass of the perturber, $M_\rmP$, is much smaller than the galaxy mass enclosed within $R$, $M_\rmG(R)$. Numerical simulations, though, have shown that near the stalling radius $M_\rmG(R)$ is actually comparable to $M_\rmP$ \citep[][]{Petts.etal.15, Petts.etal.16, DuttaChowdhury.etal.19}, indicating that the torque is likely to have an appreciable contribution from non-linear perturbations in the distribution function.

Simply put, then, linear perturbation theory is inadequate to describe the dynamics related to core stalling. In order to overcome this conceptual problem, in this chapter we develop a {\it non-perturbative} formalism to investigate how dynamical friction operates in the `slow' regime, i.e., near the core stalling radius. We adopt a circular restricted three body framework and integrate the orbits of massless field particles in the combined potential of a host galaxy and a massive perturber (to arbitrary order) moving along a circular orbit. We find that the dominant contribution to the torque comes from a family of \textit{near-co-rotation-resonant} orbits that slowly drift (librate) around the Lagrange points in the co-rotating frame. The nature of these orbits is found to change drastically as one approaches the core region of a galaxy. This causes a transition from a state in which the majority of orbits cause a retarding torque on the perturber (`dynamical friction'), to one in which the torque becomes predominantly enhancing (`dynamical buoyancy'). This transition is associated with a bifurcation in the Lagrange points that occurs whenever the perturber reaches a  characteristic radius, $R_{\rm bif}$, which we associate with the core stalling radius.

This chapter is organized as follows. In Section~\ref{sec:concept} we first conceptualize, without resorting to mathematics, how dynamical friction on a massive perturber arises from a net torque exerted by particles on near-co-rotation-resonant orbits. We then introduce, in Section~\ref{sec:threebody}, the restricted three-body framework used throughout this chapter. In Section~\ref{sec:orbits} we introduce the various orbital families that arise in the presence of a massive perturber, and briefly discuss how they contribute to dynamical friction. In Section~\ref{sec:resonances} we describe a non-perturbative method to compute the integrated energy and angular momentum transfer from individual orbits, and show that certain orbital families in a cored galaxy can give rise to a positive, enhancing torque (dynamical buoyancy) in the core region, the origin of which we examine in Section~\ref{sec:core}. We summarize our findings in Section~\ref{sec:concl_5}.

\begin{figure*}
\includegraphics[width=1\textwidth]{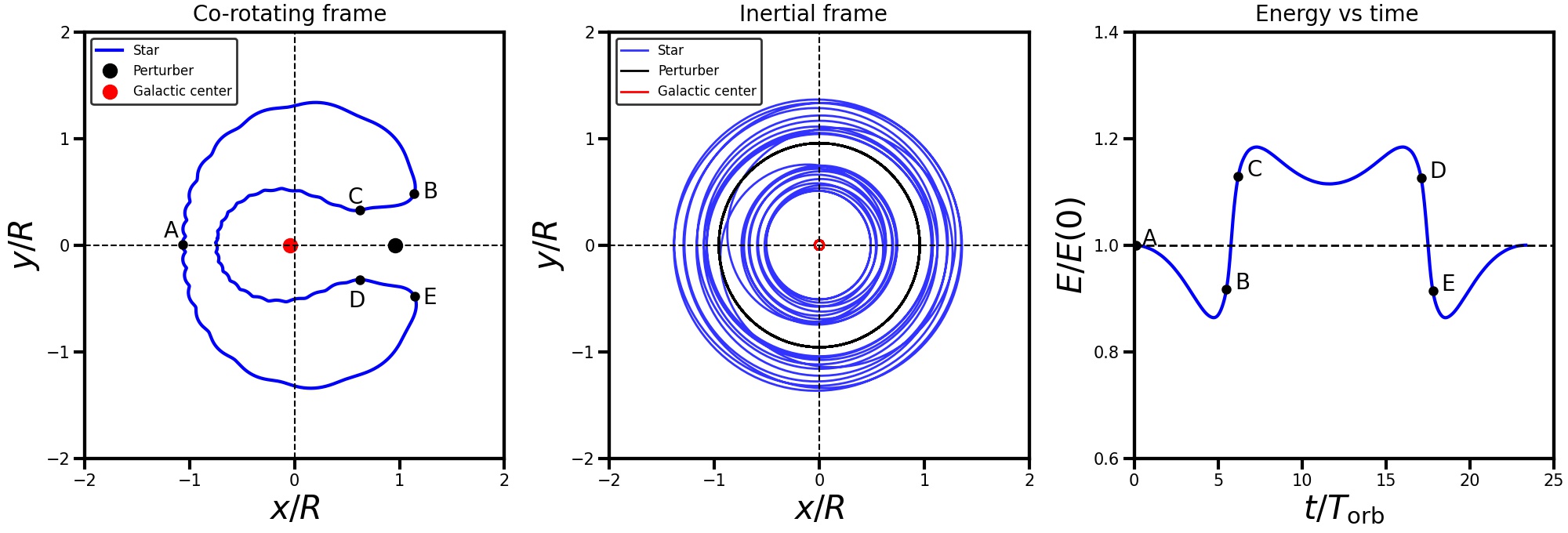}
  \caption{\small Example of a NCRR \horseshoe orbit. The left-hand panel shows the orbit in the co-rotating frame, in which the perturber (indicated by a thick, solid black dot) is at rest at $(x,y) = (R,0)$. The red dot marks the center of the galaxy, while the letters A,B,..,E mark specific points along the orbit. The middle panel shows the same orbit, but now in the inertial frame. Note how the orbit librates back and forth between regions inside and outside of the perturber. The right-hand panel depicts how a field particle moving along this \horseshoe orbit changes its orbital energy with time. Because of the near-co-rotation resonance nature of this orbit, it takes many orbital periods of the perturber, $T_{\rm orb}$, to complete one \horseshoe (in this case, the libration time $T_{\rm lib} \sim 24\, T_{\rm orb}$). The largest energy changes occur when the field particle moves from outside of the perturber (outer section) to inside (inner section), and vice-versa, which corresponds to the transitions from B to C and from D to E, respectively.}
  \label{fig:horseshoe}
\end{figure*}

\section{Conceptualizing Dynamical Friction}
\label{sec:concept}

The non-perturbative framework adopted here gives an alternative, complementary view of dynamical friction, which is subtly different from the standard resonance picture presented in TW84 and KS18. In this section we conceptualize this alternative view using the example of a single orbit. Without going into any mathematical detail, which is relegated to Sections~\ref{sec:threebody}-\ref{sec:core}, the goal is to illustrate, in a pictorial view, how dynamical friction arises. This serves to underscore the complicated, higher-order nature of dynamical friction, and to hopefully clarify the more technical treatment that follows.

As we do throughout this chapter, we consider a massive body, the perturber, orbiting a large system (hereafter the galaxy) consisting of a large number, $N$, of `field' particles or stars. Throughout, we  simplify the picture by assuming that both perturber and galaxy are spherically symmetric, and that the perturber is on a planar, circular orbit within the galaxy at a galacto-centric radius $R$. We assume that the mass of the field particles, $m$, is negligible compared to that of either the perturber, $M_\rmP$, or the galaxy, $M_\rmG$. In addition, we ignore the radial motion of the perturber due to dynamical friction/buoyancy, since we are interested in the dynamics near the stalling radius. Hence, we can treat our dynamical system as a circular restricted three body problem, which dramatically simplifies the dynamics since the gravitational potential is now static in the frame co-rotating with the perturber. Here, and throughout this section, we assume an isotropic \cite{Plummer.11} galaxy and a point mass perturber with a mass that is 0.4 percent of the galaxy mass on a circular orbit at half the scale radius of the galaxy.
 
As we discuss in Section~\ref{sec:orbits}, one can distinguish a number of different orbital families in the co-rotating frame. Here we focus on one example; the \horseshoe orbit, which, as we will show, is one of the key actors in our dynamical friction narrative. Fig.~\ref{fig:horseshoe} shows an example of a \horseshoe orbit, both in the co-rotating frame (left-hand panel), in which it takes on a shape to which it owes its name, and in the inertial frame (middle panel). A field particle on this orbit is in near-co-rotation resonance (hereafter NCRR) with the perturber in that the azimuthal frequency, $\Omega_\phi$, with which it circulates the center of the unperturbed galaxy is very similar to that of the perturber's circular orbit, $\Omega_\rmP$. Since we assume that the perturber orbits in the anti-clockwise direction, all orbits in the co-rotating frame will have a net clockwise drift motion around their center of circulation. The NCRR orbits librate about the Lagrange points and are therefore often called `trapped' orbits \citep[e.g.,][]{Barbanis.76, Sellwood.Binney.02, Daniel.Wyse.15, Contopoulos.73, Contopoulos.79, Goldreich.Tremaine.82}. However, since many of these orbits are not strictly trapped, in that they often undergo separatrix crossings (see Section~\ref{sec:orbfam} below), we consider the nomenclature NCRR more explicit.

Let us assume that the field particle starts out at position A (indicated in the left-hand panel of Fig.~\ref{fig:horseshoe}) on the \horseshoe orbit. Since it is farther away from the center-of-mass than the perturber, it circulates slower. Slowly, with an angular speed of roughly $\Omega_\rmP - \Omega_\phi$, the perturber catches up with the field particle, coming closer and closer. In the co-rotating frame, this corresponds to the field particle travelling upwards, clockwise, along its orbit. As it slowly librates from A ($t=0$) to B, its energy and angular momentum increase (note the gradual decrease in $E/E(0)$ from A to B in the right-hand panel of Fig.~\ref{fig:horseshoe}). When it reaches point B, the perturber exerts an inward accelerating force, pulling the particle onto the inner, more bound arc of the orbit. As the particle moves from B to C, it crosses co-rotation resonance; its orbital energy decreases steeply and its azimuthal frequency, $\Omega_\phi$, now becomes larger than $\Omega_\rmP$. Note that, since the Hamiltonian of our perturbed system is time-variable, energy is not a conserved quantity (and neither is angular momentum nor $\Omega_\phi$). However, the total energy of the system is conserved, and the energy that the field particle loses as it transits from B to C is transferred to the perturber, which will move (very slightly) outward; this is the opposite of dynamical friction, to which we refer as dynamical buoyancy. 

Once the field particle arrives at C, the particle now circulates {\it faster} than the perturber, and it starts to drift farther and farther ahead of the perturber (in the co-rotating frame). It circulates around the center of the galaxy (as we will see below, it has to go all the way around the center because of the potential barrier associated with an unstable Lagrange point, or saddle, in between the perturber and the center), and ultimately makes its way to point D, where the perturber exerts an outward pulling force, which puts the particle back on the outer arc of its orbit. This time, the perturber gives energy to the field particle, thus experiencing dynamical friction. Once at point E, the particle starts to lag behind the perturber again, until it drifts back to (close to) its original position A. 

In the restricted three-body problem considered here, the Jacobi energy, unlike the orbital energy, is a conserved quantity (see Section~\ref{sec:threebody}). This ensures that the energy gain experienced by the perturber at $B \rightarrow C$ balances the energy loss experienced at $D \rightarrow E$. In other words, the {\it net} effect on the perturber of a field particle along this NCRR orbit is zero. 

\begin{figure*}[t!]
  \centering
  \includegraphics[width=1\textwidth,height=0.27\textwidth]{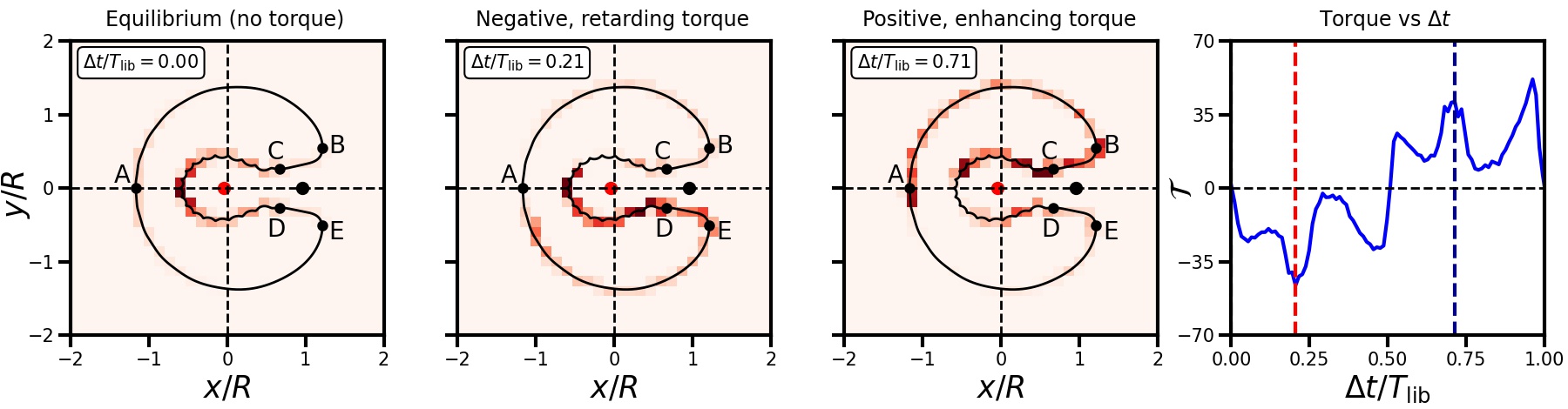}
  \caption{Illustration of the origin of torque on the perturber from a NCRR orbit. The heat maps show the distribution of field particles in the co-rotating frame along a \horseshoe orbit as in Fig.~\ref{fig:horseshoe}, with darker colors indicating a larger number density. The rightmost panel shows the evolution of the torque (as a function of time in units of $T_{\rm lib}$, the libration time or the time taken for $2\pi$ circulation in the co-rotating frame) as the field particles move along the orbit. At $\Delta t=0$ (first panel), the unperturbed density distribution of field particles is spherically symmetric, and there is no net torque on the perturber. However, some time later (second panel, corresponding to $\Delta t$ marked by the red dashed line in the right-most panel), the particles have shifted along the orbit, resulting in an enhanced density of field particles lagging behind the perturber, giving rise to a retarding torque. If the perturber would remain on its original orbit, then some time later (many orbital periods since the drift/libration time along the \horseshoe is long) the particles would have drifted to the location depicted in the third panel (at $\Delta t$ marked by the blue dashed line in the rightmost panel), exerting an enhancing torque exactly opposite to that depicted in the second panel. When integrating over the entire libration period, the net torque is therefore zero. Dynamical friction arises only because the initial torque is retarding, after which the perturber moves in, and the near-resonant frequencies change (i.e., one never makes it to the point shown in the third panel).}
  \label{fig:wake}
\end{figure*}

So how, then, does dynamical friction arise? The two key ingredients that give rise to net dynamical friction are the long libration (or `drift') time of these NCRR orbits, and the non-uniform density distribution of field particles as a function of orbital phase. The libration time, $T_{\rm lib}$, is the time in which the field particle completes a full \horseshoe (i.e., from $A \rightarrow B \rightarrow C \rightarrow D \rightarrow E \rightarrow A$). Because the orbit is in near-co-rotation resonance, this is much longer than the orbital periods of the perturber or the field particle. The non-uniform distribution of particles along the orbit can be understood as follows: in the limit of large $N$, there are many field particles that are on the same (or at least on a very similar) orbit. All these particles have different orbital phases, though. Consider the unperturbed galaxy, which is assumed to be in equilibrium and characterized by a distribution function $f_0(\bx,\bv)$. This unperturbed distribution function determines how many field particles are mapped onto each phase of each orbit once the perturber is introduced (here, for the sake of simplicity, we assume that the perturber is introduced instantaneously). Typically, since the density increases towards the center, the number density of particles on the inner arc of the \horseshoe ($C-D$) is larger than along the outer arc ($E-A-B$). This is depicted in the left-most panel of Fig.~\ref{fig:wake}, where darker colors indicate a larger number density of field particles. These have been computed using the (isotropic) distribution function of our (unperturbed) Plummer sphere, under the assumption that this captures the distribution of field particles along this orbit at time $t=0$, when the perturber is introduced. Some time $\Delta t < T_{\rm lib}$ later, all the particles have drifted along the \horseshoen, and the phase-dependent number density distribution now looks similar to that in the second panel: because of the initial non-uniformity in orbital phases, there are now more particles along the $D \rightarrow E$ part of the orbit than along the $B \rightarrow C$ part; there are more energy gainers than energy losers, causing a net energy loss of the perturber. Or, in terms of angular momentum, the overdensity of field particles trailing the perturber, exerts a torque that reduces the perturber's orbital angular momentum (note the negative, retarding torque at this time, marked by the red dashed line in the rightmost panel that shows the evolution of the torque exerted by the particles). Hence, during this phase of the evolution, the perturber experiences (net) dynamical friction from the field particles associated with this \horseshoe orbit.
0
If the perturber would remain at its current orbital radius (i.e., if we temporarily ignore the consequences of dynamical friction), then the phase of the overdensity of particles along the \horseshoe orbit would continue to drift around, ultimately making its way to points $B$ and $C$ (depicted in the third panel of Fig.~\ref{fig:wake}), where it would exert a positive, enhancing torque/ buoyancy on the perturber (marked by the blue dashed line in the rightmost panel) which nullifies the initial dynamical friction on the perturber\footnote{The alternating phases of retarding and enhancing torques from the NCRR orbits are responsible for oscillations in the pattern speed of a galactic bar in the slow regime of dynamical friction, as noted by \cite{Chiba.Schonrich.22}.}. However, because of the long drift time, the time between this net friction and equal, but opposite, net buoyancy is very long ($\sim 10\, T_{\rm orb}$ for the specific \horseshoe orbit shown in Fig.~\ref{fig:wake}). 
During this time, the initial net friction from many NCRR orbits will have caused the perturber to move inward, to a more bound orbit. This changes its orbital frequency, $\Omega_\rmP$, such that, by the time the overdensity {\it would} have reached point $B$, the system has changed sufficiently that new field particles have now entered near-co-rotation resonance with the perturber and those associated with our original \horseshoe orbit have fallen out of resonance. Dynamical friction is therefore a secular process; the field particles drain energy from the perturber, causing it to in-fall, which in turn changes the orbital frequencies, facilitating further energy transfer. This process of `sweeping through the resonances' by the perturber is crucial for dynamical friction to operate, as emphasized in great detail in TW84. 

\subsection{The Role of Resonances}

In the perturbative framework of TW84 and KS18, dynamical friction arises from the LBK torque which only has a non-zero contribution from pure resonances, i.e., orbits that obey a commensurability condition between the (circular) frequency of the perturber, $\Omega_\rmP$, and the frequencies of the field particles in the {\it unperturbed} potential. Even the more general, self-consistent torque introduced by BB21, is formulated in terms of these frequencies. 

In the non-perturbative framework adopted in this chapter, in which we consider fully perturbed orbits\footnote{To clarify the paradoxical use of `perturbed orbits in a `non-perturbative framework'; perturbative is used to mean `as pertaining to perturbation theory', whereas perturbed means `impacted by the in-falling, perturbing mass'.} in the galaxy$+$perturber potential to arbitrary order, the frequencies of the individual field particles vary with time due to energy and angular momentum exchanges with the perturber; the original actions of the unperturbed galaxy are no longer conserved, and neither are the frequencies associated with the corresponding angles \citep[][]{Tremaine.Weinberg.84,Fouvry.Bar-Or.18}. Hence, a field particle will not satisfy a commensurability condition throughout its orbital evolution but rather will find itself `trapped', librating around resonance(s) with the perturber. In fact, this is what happens when the field particle along the \horseshoe orbit in Fig.~\ref{fig:horseshoe} moves from B to C and from D to E; it's azimuthal frequency, $\Omega_\phi$, is swept back and forth through a near-co-rotation resonance with the circular frequency of the perturber,  $\Omega_\rmP$. This same principle also underlies the physics of radial migration in disks due to interactions with transient spirals \citep[e.g.,][]{Carlberg.Sellwood.85, Sellwood.Binney.02, Daniel.Wyse.15}. Dynamical friction arises from an imbalance between the number of field particles that `sweep up' versus `sweep down' in frequency space, and this imbalance itself arises from gradients in the distribution function.

\section{The Restricted Three Body Problem}
\label{sec:threebody}

We treat dynamical friction as a restricted three-body problem, in which the mass of the field particles is negligible compared to that of the galaxy and the perturber. Throughout, we assume that both galaxy and perturber are spherically symmetric, and that the perturber is moving along a circular orbit of galacto-centric radius $R$ within the galaxy. In this setting the gravitational potential is static (in the absence of dynamical friction) in the co-rotating frame, which greatly simplifies the analysis that follows. As the perturber only feels the gravitational field of the galaxy mass enclosed within a sphere of radius $R$ centered on the galactic center, denoted by $M_\rmG(R)$, we follow \cite{Inoue.11} and KS18 in assuming that $M_\rmP$ and $M_\rmG(R)$ rotate about their common center of mass (hereafter COM).

\begin{figure}[t!]
\centering
\includegraphics[width=0.65\textwidth,height=0.65\textwidth]{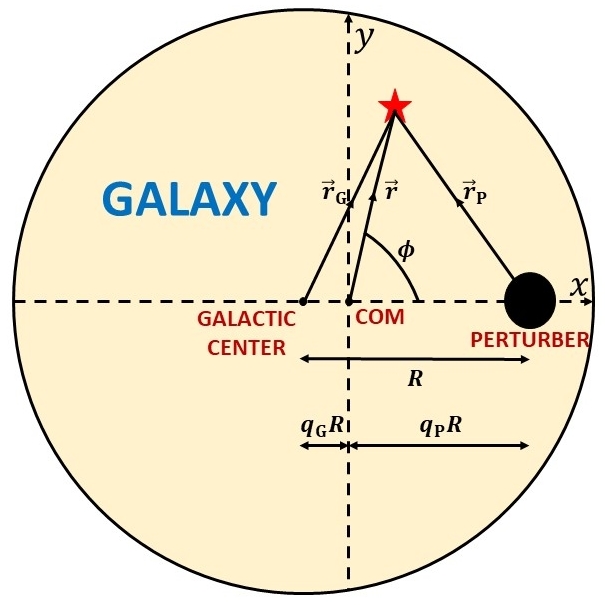}
\caption{\small Schematic of a massive perturber on a circular orbit in a spherically symmetric galaxy. The co-rotating $(x,y)$-frame is centered on the COM with the $x$ axis pointing in the direction of the perturber.}
\label{fig:schematic}
\end{figure}

\subsection{Models for the Galaxy and the Perturber}
\label{sec:models}

The geometry of our dynamical model is illustrated in Fig.~\ref{fig:schematic}. It depicts the galaxy (large, shaded circle), the perturber (solid black dot), and the COM in the co-rotating $(x,y)$-frame that we will adopt throughout. For convenience, we define the following mass ratios: $q \equiv M_\rmP/M_\rmG$ is the mass ratio of the in-falling perturber and the host galaxy, while $q_{\rm enc}(R) \equiv M_\rmP/M_\rmG(R)$ is the mass ratio of the perturber and the galaxy enclosed within $R$. The distances between the COM and the galactic center and between the COM and the perturber are given by $q_\rmG R$ and $q_\rmP R$, respectively, where
\begin{align}
q_\rmG &= \frac{M_\rmP}{M_\rmP + M_\rmG(R)} = \frac{q_{\rm enc}(R)}{1+q_{\rm enc}(R)}\,, \nonumber \\
q_\rmP &= \frac{M_\rmG(R)}{M_\rmP + M_\rmG(R)} = \frac{1}{1+q_{\rm enc}(R)}\,.
\label{qgqp}
\end{align}

Throughout this chapter, we adopt dimensionless units to describe our dynamical system. All length scales are expressed in units of $r_\rms$, the scale radius of the galaxy, masses are expressed in units of the mass of the galaxy, $M_\rmG$, and velocities are expressed in units of $\sigma = (G M_\rmG / r_\rms)^{1/2}$. The corresponding, characteristic time-scale is $r_\rms/\sigma$.

For convenience, we consider the perturber to be a point mass, but we emphasize that the analysis that follows can be easily extended to accommodate any other (spherically symmetric) perturber potential. In our dimensionless units, we then have that the perturber potential,
\begin{equation}
\Phi_\rmP = -q /r_\rmP.
\label{subjectpotential}
\end{equation}
Throughout we adopt $q = 0.004$ (i.e., the mass of the perturber is only 0.4 percent of that of the galaxy). Unlike the perturbative treatments in TW84 and KS18, though, which require $q$ to be small, our analysis is also valid for more massive perturbers.

\begin{figure}
\centering
\includegraphics[width=1\textwidth]{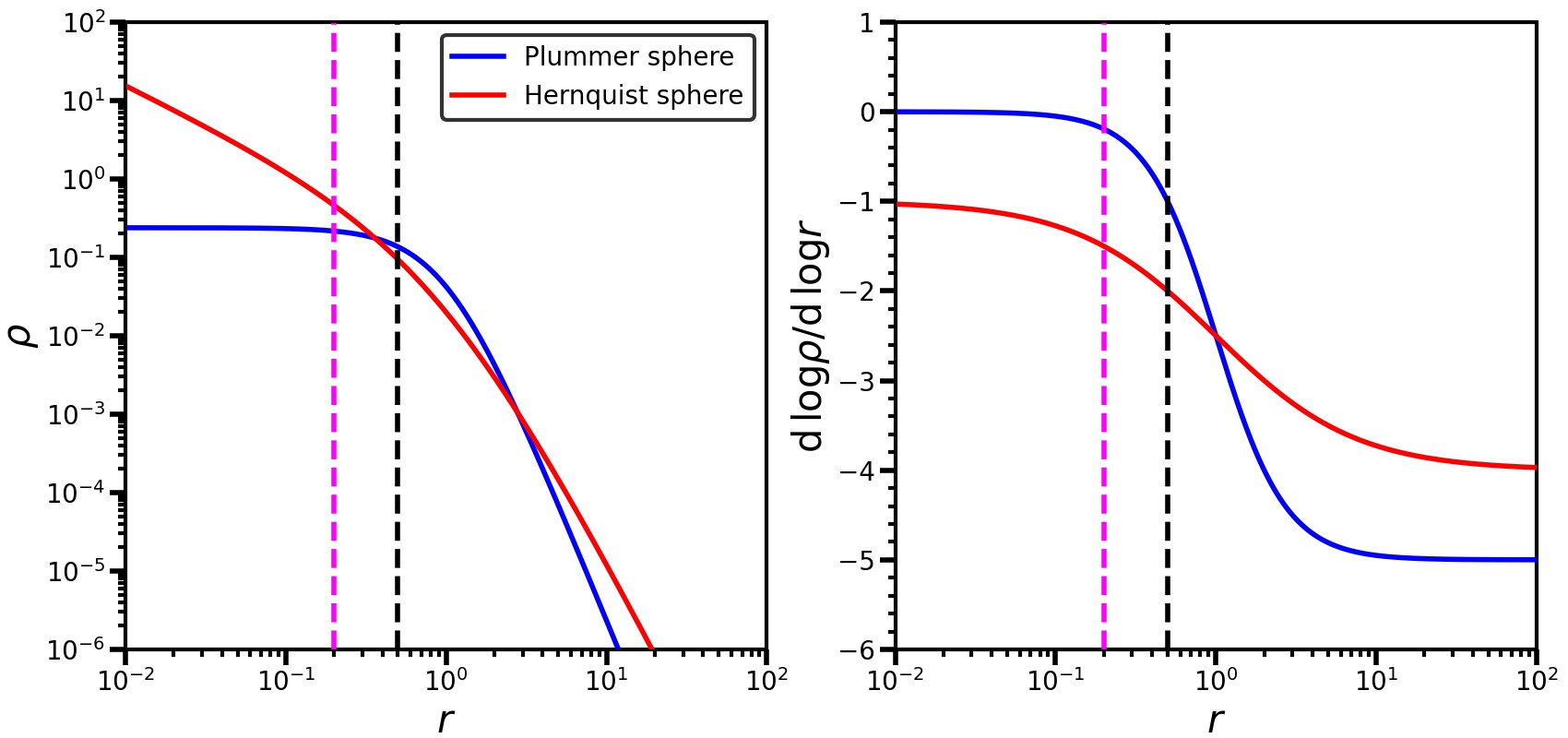}
\caption{\small Density (left-hand panel) and logarithmic slope       $\rmd\log\rho/\rmd\log r$ (right-hand panel) as functions of galacto-centric radius $r$ for the Plummer (blue) and Hernquist (red) spheres used in this chapter. The dashed magenta and black lines indicate the orbital radii, $R=0.2$ and $0.5$, considered in this chapter. These two radii bracket the bifurcation radius for the Plummer sphere and a $q=0.004$ perturber, at which the torque exerted on the perturber changes from being retarding to enhancing (see sections~\ref{sec:orbits} and \ref{sec:resonances} for details). No such transition occurs for the Hernquist sphere.}
\label{fig:densities}
\end{figure}

In order to contrast dynamical friction in cored and cuspy density profiles, we consider two different density profiles for the galaxy: a Plummer sphere, which has a central constant density core with central logarithmic density gradient, $\gamma \equiv \lim_{r \to 0}\rmd\log\rho/\rmd\log r=0$ \citep{Plummer.11}, and a Hernquist sphere, which has a central $\gamma=-1$ cusp \citep{Hernquist.90}. Both have the advantage that the density and potential are given by simple, analytical expressions. For the Plummer sphere, the density and potential (in our dimensionless units) are given by
\begin{equation}
\rho_\rmG(r) = \frac{3}{4\pi} \frac{1}{\left(1+r^2\right)^{5/2}}\,,
\,\,\,\,\,\,\,\,\,\,\,\,\,
\Phi_\rmG(r) = -\frac{1}{\sqrt{1+r^2}}\,,
\label{plummer}
\end{equation}
while for the Hernquist sphere we have that
\begin{equation}
\rho_\rmG(r) = \frac{1}{2\pi} \, \frac{1}{r \, (1+r)^3}\,,
\,\,\,\,\,\,\,\,\,\,
\Phi_\rmG(r) = -\frac{1}{1+r}\,.
\label{hernquist}
\end{equation}
Figure~\ref{fig:densities} plots these density profiles (left-hand panel) and corresponding logarithmic density gradients, $\rmd\log\rho/\rmd\log r$ (right-hand panel), as functions of radius. The magenta and black vertical dashed lines indicate $R=0.2$ and $0.5$, respectively. These are the orbital radii of the perturber considered in this chapter. As we demonstrate below, in the case of the Plummer host these radii bracket the bifurcation radius, $\Rbif$ ($\approx 0.39$ for our fiducial case), at which the orbital make-up of the Plummer sphere undergoes a drastic change due to a bifurcation of some of the Lagrange points, which in turn impacts the nature (retarding vs. enhancing) of the torque on the perturber. In the case of the Hernquist sphere, no such bifurcation occurs.

Throughout, we assume that the galaxies have isotropic velocity distributions, such that their distribution functions are ergodic (i.e., depend only on energy). In the case of the Plummer sphere we have
\begin{align}
f_0(\varepsilon) &= \frac{3}{7\pi^3} {\left(2\varepsilon\right)}^{7/2}\,,
\label{fdistP}
\end{align}
while for the Hernquist sphere 
\begin{align}
f_0(\varepsilon) &= \frac{1}{8 \sqrt{2} \pi^3}\nonumber \\
&\times \frac{3 \sin^{-1}\sqrt{\varepsilon} + \sqrt{\varepsilon (1 - \varepsilon)} (1 - 2 \varepsilon) (8\varepsilon^2 - 8 \varepsilon - 3)}{(1-\varepsilon)^{5/2}}\,.
\label{fdistH}
\end{align}
Here $\varepsilon=-E_{0\rmG}$ ($E_{0\rmG}$ is the unperturbed galactocentric energy), and the subscript `0' indicates that these distribution functions correspond to the unperturbed galaxies. Both distribution functions have been normalized such that 
\begin{align}
\rho_\rmG(r) = 4 \pi \int_0^{\Psi_\rmG} \sqrt{2(\Psi_\rmG-\varepsilon)} \, f_0(\varepsilon) \, \rmd\varepsilon\,,
\end{align}
with $\Psi_\rmG = -\Phi_\rmG$.
\\

\subsection{Hamiltonian dynamics in the co-rotating frame}
\label{sec:Hamilton_5}

Since the gravitational potential, and hence the Hamiltonian, in the restricted three body problem is time-variable, energy is not a conserved quantity. And due to the lack of spherical symmetry, neither is angular momentum. However, as is well known \citep[see e.g.,][]{Binney.Tremaine.08}, the Jacobi integral,
\begin{equation}
E_{\rm J} = E -{\bf \Omega_{\rm{\bP}}} \cdot \bL = \frac{1}{2}{\dot{\br}}^2 + \Phi_{\rm eff}\left(\br\right),
\label{Ej_5}
\end{equation}
is a conserved quantity. Here $\br$ is the position vector of the field particle with respect to the COM (see Fig.~\ref{fig:schematic}), and ${\bf \Omega_{\rm{\bP}}} = (0,0,\Omega_\rmP)$ with
\begin{equation}
\Omega_\rmP = \sqrt{\frac{G\,[M_\rmG(R) + M_\rmP]}{R^3}}\,,
\end{equation}
the angular frequency of the perturber with respect to the COM, which, in our dimensionless units, is given by 
\begin{equation}
\Omega_\rmP = \left( \frac{1}{R} \, \left.\frac{\partial \Phi_\rmG}{\partial r_\rmG}\right|_{r_\rmG=R}+\frac{q}{R^3}\right)^{1/2}\,.
\label{Omegap}
\end{equation}
$E$ and $\bL$ are, respectively, the perturbed energy and angular momentum (per unit mass) of the field particle in the non-rotating, inertial frame, given by
\begin{align}\label{Eperturbed_5}
E& =E_0+\Phi_\rmP=\frac{1}{2}{\vert{\dot{\br}}+\bf{\Omega_{\rm{\bP}}}\times \br \vert}^2 + \Phi_\rmG\left(\br\right) + \Phi_\rmP\left(\br\right),\\
\bL &= \br \times \left(\bf{\dot{r}}+\bf{\Omega_{\rm{\bP}}}\times \br\right)\,.
\end{align}
Here $E_0$ is the unperturbed energy, i.e., the part of the Hamiltonian without the perturber potential, and $\Phi_\rmG$ and $\Phi_\rmP$ are the gravitational potentials due to the galaxy and the perturber, respectively. The effective potential in equation~(\ref{Ej_5}) is defined as
\begin{equation}
\Phi_{\rm eff}\left(\br\right) = \Phi_\rmG(r_\rmG) + \Phi_\rmP(r_\rmP) - \frac{1}{2} \vert{\bf \Omega_{\rm{\bP}}} \times \br \vert^2,
\label{phieff}
\end{equation}
where $r_\rmG$ and $r_\rmP$ are the distances to the field particle from the galactic center and the perturber respectively, and are given by
\begin{align}
r_\rmG^2 &= r^2 + q_\rmG^2 R^2 + 2\, q_\rmG R\, r \cos{\phi},\nonumber \\
r_\rmP^2 &= r^2 + q_\rmP^2 R^2 - 2\, q_\rmP R\, r \cos{\phi}.
\label{r1r2}
\end{align}
Here $r = \vert \br \vert$, $\phi$ is the counter-clockwise angle between $\br$ and the line connecting the COM and the perturber positioned along the positive $x-$axis (see Fig.~\ref{fig:schematic}), and $q_\rmG$ and $q_\rmP$ are the mass ratios given by equation~(\ref{qgqp}). The third term in equation~(\ref{phieff}) is the potential due to the centrifugal force. Plugging in the expression for $\Omega_\rmP$, and using the fact that $\partial \Phi_\rmG/\partial r = G M_\rmG(r)/r^2$ and $q_{\rm enc}(R)=M_\rmP/M_\rmG(R)$, the effective potential reduces to
\begin{align}
\Phi_{\rm eff}(\br) = \Phi_\rmG(r_\rmG) - \frac{q}{r_\rmP} - \frac{1+q_{\rm enc}(R)}{2}\, \frac{r^2}{R} \, \left.\frac{\partial \Phi_\rmG}{\partial r_\rmG}\right|_{r_\rmG=R}\,.
\end{align}

\renewcommand{\arraystretch}{0.6}
\begin{sidewaystable}
\centering

\tabcolsep=0.03 cm
\scalebox{0.9}{\begin{tabular}{|c|ccc|c|cc|}
 \hline
 Orbit type & $E_{\rm{Jc}}^{(4)}$ & $E_{\rm{Jc}}^{(0)}$ & $E_{\rm{Jc}}^\rmP$ & $L$ & COC & Friction (F)/\\
 & & & & & & Buoyancy (B)/ \\
 & & & & & & Negligible (N) \\
 (1) & (2) & (3) & (4) & (5) & (6) & (7) \\
 \hline 
 & & & & & &\\
 \horseshoe $(\gamma=0)$ & $E_{\rm{Jc}}^{(4)}<E_\rmJ^{(3)}$ & $E_{\rm{Jc}}^{(0)}>\max{\left[E_\rmJ^{(1)},E_\rmJ^{(2)}\right]}$  & -- & -- & L3  & F \\ & & & & & & \\
 \horseshoe $(\gamma<0)$ & $E_{\rm{Jc}}^{(4)}<E_\rmJ^{(3)}$ & --  & $E_{\rm Jc}^\rmP>\max{\left[E_\rmJ^{(1)},E_\rmJ^{(2)}\right]}$ & -- & L3  & F \\ & & & & & & \\
 \pacman $(\gamma=0)$ & -- & $E_{\rm{Jc}}^{(0)}<E_\rmJ^{(1)}$ & $E_{\rm{Jc}}^\rmP>E_\rmJ^{(2)}$ & $L^{(1)}<L<L^{(2)}$ & L0 & F/B \\ & & & & & & \\
 \pacman $(\gamma<0)$ & -- & -- & $E_\rmJ^{(2)}<E_{\rm{Jc}}^\rmP<E_\rmJ^{(1)}$ & $L^{(1)}<L<L^{(2)}$ & L0 & F/B \\ & & & & & & \\
 tadpole ($R>\Rbif$) & $E_\rmJ^{(3)}<E_{\rm{Jc}}^{(4)}<E_\rmJ^{(4)}$ & -- & -- & -- & L4/L5 & F/B \\ & & & & & & \\
 tadpole ($R\leq\Rbif$) & $E_\rmJ^{(0)}<E_{\rm{Jc}}^{(4)}<E_\rmJ^{(4)}$ & -- & -- & -- & L4/L5 & F/B \\ & & & & & & \\
\hline
& & & & & & \\
 center-phylic $(\gamma=0)$    & -- &   $E_\rmJ^{(0)}<E_{\rm{Jc}}^{(0)}<E_\rmJ^{(1)}$ & -- & $L<L^{(1)}$ & L0 & N\\ & & & & & & \\
 center-phylic $(\gamma<0)$    & -- &   -- & $E_{\rm Jc}^\rmP<E_\rmJ^{(1)}$ & $L<L^{(1)}$ & L0/cusp & N\\ & & & & & & \\
 perturber-phylic & -- & -- & $E_\rmJ^\rmP<E_{\rm{Jc}}^\rmP<\min{\left[E_\rmJ^{(1)},E_\rmJ^{(2)}\right]}$ & $L^{(1)}<L<L^{(2)}$ & P & N\\ & & & & & & \\
 COM-phylic & -- & -- & $E_{\rm{Jc}}^\rmP<E_\rmJ^{(2)}$ & $L>L^{(2)}$ & COM & N\\ & & & & & & \\
 \hline
\end{tabular}}

\caption{\small Different orbital families in the co-rotating frame of our restricted three-body framework. Column (1) indicates the name of the orbital family used throughout this chapter. Columns (2), (3) and (4) indicate the bounds on the circular part of the Jacobi energy, $E_{\rm Jc}\approx E_\rmJ-\kappa_0 J_r$ ($\kappa_0$ is the value of the radial epicyclic frequency evaluated at the center of perturbation and $J_r$ is the radial action; see Appendix~\ref{App:orb_class} for details), evaluated in the neighborhood of L4/L5, L0 and the perturber, i.e., $E_{\rm Jc}^{(4)}$, $E_{\rm Jc}^{(0)}$ and $E_{\rm Jc}^\rmP$, respectively. Column (5) indicates the angular momentum, $L$. Column (6) indicates the center-of-circulation (COC), where `P' refers to the perturber, and column (7) indicates whether these orbits contribute significantly to dynamical friction (F) or buoyancy (B) or negligibly to either of the two (N). $E_\rmJ^{(k)}$, with $k=0,1,..,5$, approximately denotes the value of $\Phi_{\rm eff}$ at the $k^{\rm th}$ Lagrange point (see Appendix~\ref{App:orb_class} for details), while $E_\rmJ^\rmP$ denotes that at the location of the perturber ($E_\rmJ^\rmP=-\infty$ for a point mass). $L^{(k)}$, with $k=1,2$, denotes the value of the angular momentum at the $k^{\rm th}$ Lagrange point. Note that \pacman orbits are absent when $E_\rmJ^{(2)}>E_\rmJ^{(1)}$, which is always the case if the galaxy has a central cusp or the perturber is at large $R$ in a cored galaxy. Orbits that are further away from co-rotation resonance can cross the separatrix corresponding to L1, L2 or L3 due to changes in $J_r$, thereby taking on the morphology of a different orbital family, constituting what we call `Chimera orbits' (see section~\ref{sec:orbfam} and Appendix~\ref{App:Chimera} for details).}

\label{tab:Ej}
\end{sidewaystable}

\subsection{Lagrange points}
\label{sec:lagrangepoints}

The fixed points of the system are known as the Lagrange points, where the effective force in the co-rotating frame vanishes. These are given by the roots of
\begin{equation}
\nabla \Phi_{\mathrm{eff}} = 0\,,
\end{equation}
and are therefore solutions to the following set of equations:
\begin{align}
&\frac{\partial \Phi_{\rm eff}}{\partial x} = \frac{\partial \Phi_{\rmG}}{\partial r_\rmG}\frac{x+q_\rmG R}{r_\rmG} + \frac{q\left(x-q_\rmP R\right)}{r_\rmP^3}\nonumber \\
&-\left(\frac{1+q_{\rm enc}(R)}{R}\frac{\partial \Phi_{\rmG}}{\partial R}\right)x = 0\,, \nonumber \\
&\frac{\partial \Phi_{\rm eff}}{\partial y} = \left(\frac{1}{r_\rmG}\frac{\partial \Phi_{\rmG}}{\partial r_\rmG}+\frac{q}{r_\rmP^3}-\frac{1+q_{\rm enc}(R)}{R}\frac{\partial \Phi_{\rmG}}{\partial R}\right) y =0\,.
\label{fixedpt}
\end{align}

For $y=0$, $r_\rmG = \vert x + q_\rmG R \vert$ and $r_\rmP = \vert x - q_\rmP R \vert$, and equations~(\ref{fixedpt}) reduce to
\begin{align}
&\frac{\partial \Phi_{\rmG}}{\partial r_\rmG}\sgn\left(x+q_\rmG R\right) + \frac{q}{r_\rmP^2}\sgn\left(x-q_\rmP R\right) \nonumber \\
&-\left(\frac{1+q_{\rm enc}(R)}{R}\frac{\partial \Phi_{\rmG}}{\partial R}\right)x=0\,,
\label{L0123}
\end{align}
with $\sgn(x)$ the sign function. This equation can be solved to obtain the Lagrange points along the $x$-axis. The number of such fixed points depends on the galactocentric distance of the subject, $R$, and on the radial gradient of the density profile. In a Hernquist sphere, there are always three Lagrange points along the $x$-axis; L1, L2 and L3. This situation is similar to the well-known restricted three-body treatment of the dynamics of a body of negligible mass in the Earth-Sun system. The picture is however very different when a central core (here defined as having $\gamma > -1$) is present, such as in the case of the Plummer sphere. In this case there is an additional Lagrange point, which we call L0, at the galactic centre (one can easily check that $x=-q_\rmG R$ is a solution to equation~[\ref{L0123}]). This is expected since the gravitational force tends to zero towards the centre if $\gamma > -1$ and the force due to the subject is exactly balanced by the centrifugal force. A stability analysis (see Appendix~\ref{App:stability}) shows that L1, L2 and L3 are saddle points, and thus unstable under small perturbation, while L0 is stable.

When $y\neq 0$, we can simultaneously solve equations~(\ref{fixedpt}) to obtain
\begin{align}
&\frac{1}{r_\rmG} \frac{\partial \Phi_\rmG}{\partial r_\rmG} = \frac{1}{R} \frac{\partial \Phi_\rmG}{\partial R} \,\,\,\,\,\,\,\,\Rightarrow \,\,\,\,\,\,\,r_\rmG=R \nonumber, \\
&\frac{q}{r_\rmP^3} = \frac{q_{\rm enc}(R)}{R}\frac{\partial \Phi_\rmG}{\partial R}.
\end{align}
Using the expressions for $r_\rmG$ and $r_\rmP$ (equation~[\ref{r1r2}]), this reduces to
\begin{align}
x = \frac{R}{2}\left[\frac{1-q_{\rm enc}(R)}{1+q_{\rm enc}(R)}\right]\,,\;\;\;\;\;\;\;\;\;
&y = \pm \frac{\sqrt{3}}{2}R\,.
\label{xyL4L5}
\end{align}
These are the $x$ and $y$ coordinates of the Lagrange points L4 and L5. Note that both L4 and L5 form equilateral triangles with the galactic centre and the subject. A stability analysis (see Appendix~\ref{App:stability}) shows that these two Lagrange points are stable under small perturbations. 

As we discuss in more detail in Section~\ref{sec:bifurcation}, the number of Lagrange points present in the co-rotating frame of a perturbed potential depends on both the detailed potential of the galaxy (in particular, on the central, logarithmic slope $\gamma$) and the galacto-centric distance $R$ of the subject. All six Lagrange points (L0, L1,..., L5) are present in a galaxy with a shallow density profile, but only when the subject is sufficiently far away from the galactic centre, i.e., when the Roche lobes surrounding the galactic centre and the subject remain separated by the inner saddle point L1. As the subject approaches the galactic centre, the two Roche lobes coalesce to form a single lobe surrounding the subject. This coincides with the merging (bifurcation) of several of the Lagrange points, after which only L0, L2, L4 and L5 remain. In a cuspy galaxy, though, there is no L0, and all five Lagrange points (L1, L2,...,L5) survive throughout, for any $R$. As we demonstrate in the subsequent sections, the number and nature of Lagrange points dictates the orbital families available for the field particles, which is an important factor in how dynamical friction operates in galaxies with different density profiles.

\section{Survey of Orbits}
\label{sec:orbits}

\subsection{Equations of motion}
\label{sec:eom}

As already mentioned above, in the perturbed potential, energy and angular momentum are no longer constants of motion. Instead, the only conserved quantity in the restricted three body case considered here is the Jacobi energy,  $E_{\rm J}$. A field particle therefore gains and loses energy and angular momentum (which is exchanged with the perturber) as it traverses its orbit. In order to compute the rates at which the energy and angular momentum of a field particle change as function of time, we integrate its orbit using the equation of motion in the co-rotating frame \citep[][]{Binney.Tremaine.87}, which is given by
\begin{align}
&\ddot{\bf r} = -\nabla \Phi_\rmG - \nabla \Phi_\rmP - 2\left(\bf \Omega_{\rm{\bP}} \times \dot{\bf r}\right) - \bf \Omega_{\rm{\bP}} \times \left(\bf \Omega_{\rm{\bP}} \times \bf r\right)\,.
\label{geneqmotion}
\end{align}
Here the first and second terms on the RHS denote the gravitational accelerations due to the galaxy and the perturber, respectively, while the third and the fourth terms correspond to accelerations due to the Coriolis and centrifugal forces, respectively. In cylindrical coordinates, the above reduces to the following radial and azimuthal equations of motion:
\begin{align}
&\ddot{r} - r\,\dot{\phi}^2 = -\frac{\partial \Phi_\rmG}{\partial r} - \frac{\partial \Phi_\rmP}{\partial r} + 2\,\Omega_\rmP \, r \, \dot{\phi} + \Omega^2_\rmP \, r\,, \nonumber \\
&r \, \ddot{\phi} + 2 \dot{r}\,\dot{\phi} = -\frac{1}{r}\,\frac{\partial \Phi_\rmG}{\partial \phi}-\frac{1}{r}\frac{\partial \Phi_\rmP}{\partial \phi} - 2\,\Omega_\rmP\,\dot{r}\,.
\label{eqmotion}
\end{align}
The latter can be combined with equations~(\ref{r1r2}) to yield an expression for the torque,
\begin{align}
\calT=\frac{\rmd L}{\rmd t}&=-\frac{\partial \Phi_\rmG}{\partial \phi}-\frac{\partial \Phi_\rmP}{\partial \phi}\nonumber \\
&=\frac{r R\, \mathrm{sin}\, \phi}{1+q_{\rm enc}(R)}\left[\frac{q_{\rm enc}(R)}{r_\rmG}\frac{\partial \Phi_\rmG}{\partial r_\rmG}-\frac{1}{r_\rmP}\frac{\partial \Phi_\rmP}{\partial r_\rmP}\right],
\label{singletorque1}
\end{align}
where $L=r^2(\dot{\phi}+\Omega_\rmP)$ is the total angular momentum of the field particle in the inertial frame. Equation~(\ref{singletorque1}) is an expression for the combined torque, exerted by both the perturber and the galaxy on the field particle. For a slowly evolving circular orbit of the perturber, i.e., nearly constant $\Omega_\rmP$, as considered in this chapter, $E_\rmJ = E - {\bf \Omega_{\rm{\bP}}} \cdot \bL$ is a conserved quantity. Hence, the corresponding rate of energy change of the field particle is simply given by
\begin{equation}
\frac{\rmd E}{\rmd t} = {\bf \Omega_{\rm{\bP}}} \cdot \frac{\rmd \bL}{\rmd t} \,.
\label{dEdt}
\end{equation}
Because of this equality, throughout this chapter we will talk about $\Delta E$ and $\Delta L$ interchangeably. Note that, depending on the sign of the torque $\calT = \rmd L/\rmd t$, the perturber can either lose (dynamical friction) or gain energy (dynamical buoyancy). Also note that dynamical friction or buoyancy results in a non-zero  time-derivative of ${\bf \Omega_{\rm{\bP}}}$, which, following TW84 and KS18, has been ignored in the above equations. Since we are mainly interested in examining dynamical friction near the core-stalling radius, where $\vert \rmd{\bf \Omega_{\rm{\bP}}}/\rmd t \vert$ vanishes, this is justified. In fact, it is justified as long as the time scale for dynamical friction is sufficiently long, i.e., we are in what TW84 refer to as the `slow' regime.

Throughout this chapter, all orbit integrations are performed using an exactly Hamiltonian conserving algorithm proposed by \cite{Kotovych_2002} for simulating general $N-$body systems. It ensures that the Jacobi Hamiltonian is conserved up to machine precision for all the orbits we have integrated. 

\begin{figure*}[t!]
    \centering
    \subfloat[Plummer sphere: outside core ($R=0.5$)]{\includegraphics[width=0.45\textwidth,height=0.45\textwidth]{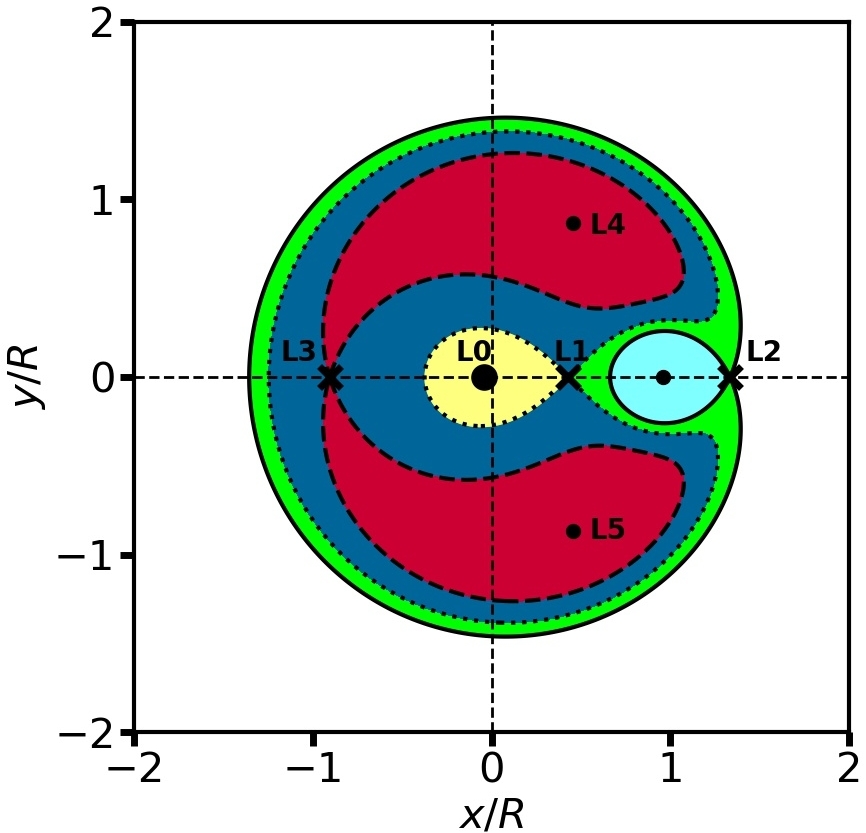}
    \label{Plu_out}}
    \subfloat[Plummer sphere: inside core ($R=0.2$)]{\includegraphics[width=0.466\textwidth,height=0.45\textwidth]{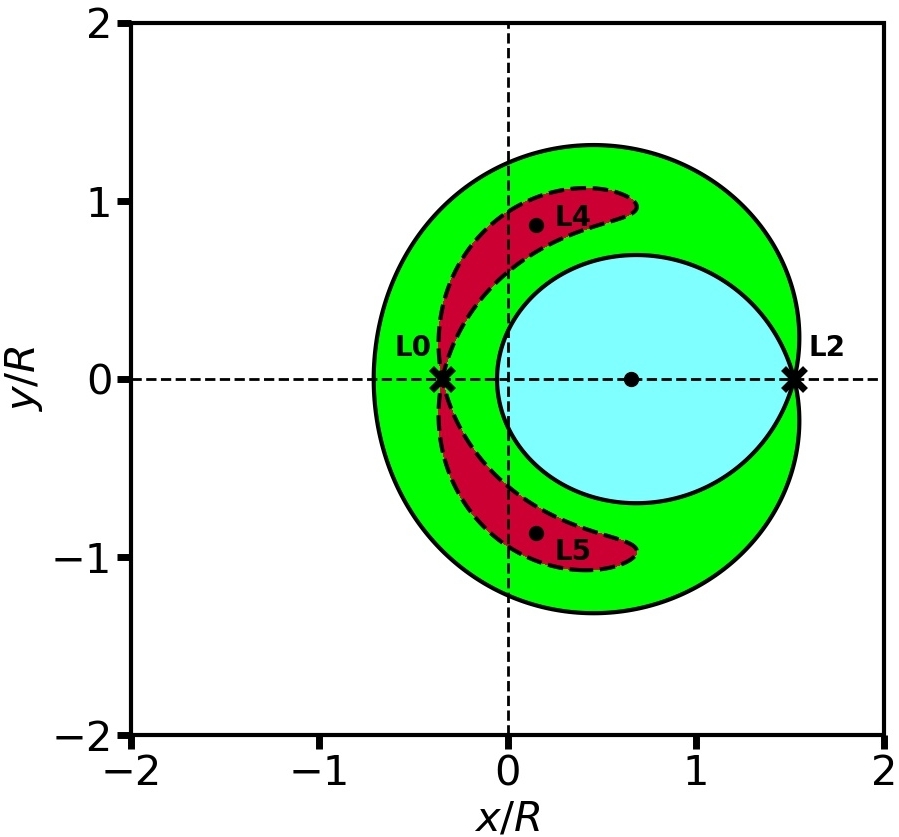}
    \label{Plu_in}}
    \\
    \subfloat[Hernquist sphere ($R=0.5$)]{\includegraphics[width=0.45\textwidth,height=0.45\textwidth]{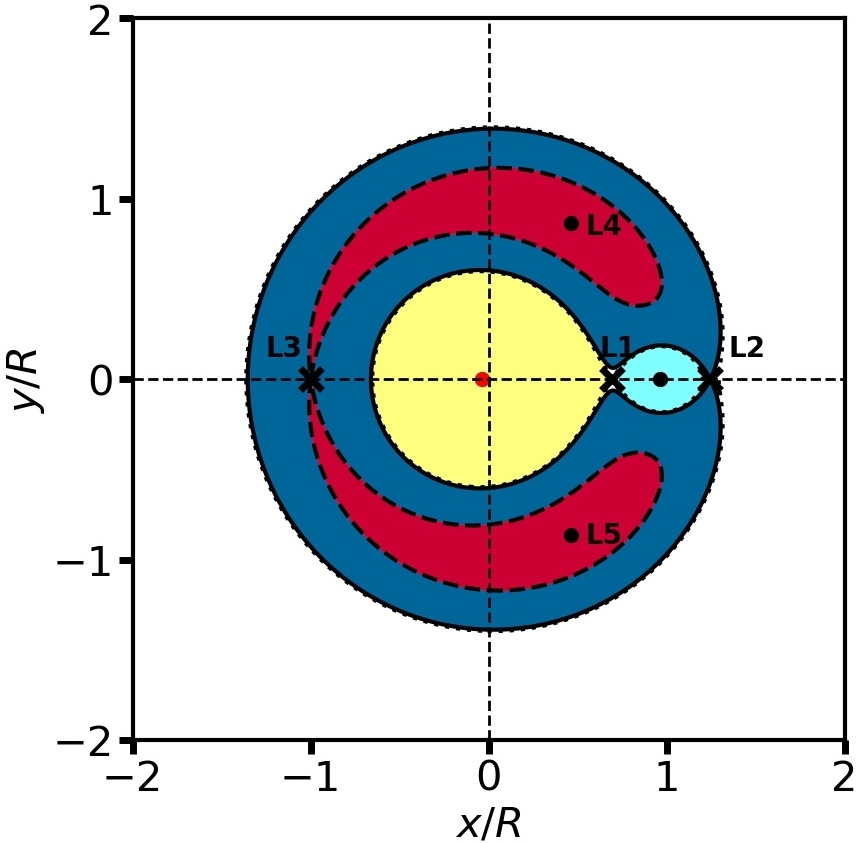}
    \label{Her_out}}
    \subfloat[Hernquist sphere ($R=0.2$)]{\includegraphics[width=0.466\textwidth,height=0.45\textwidth]{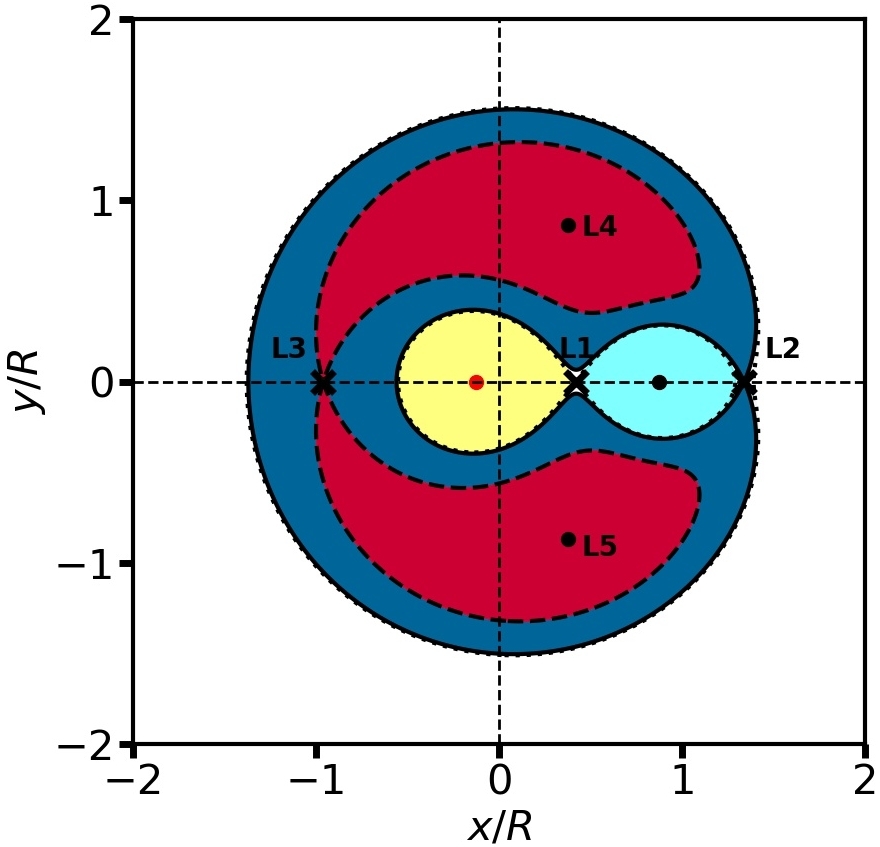}
    \label{Her_in}}
    
    \caption{\small Effective potential of the galaxy plus perturber with $(x,y)=(0,0)$ corresponding to the COM (see Fig.~\ref{fig:schematic}). The various Lagrange points (fixed points in the co-rotating frame) are indicated, and the different colored regions mark the intervals in Jacobi energy for the zero-velocity curves (ZVCs) of the various near-circular orbital families: \horseshoe (dark blue), \pacman (green), tadpole (red), perturber-phylic (cyan), center-phylic (yellow), and COM-phylic (white). Note that there are no \pacman orbits in a Hernquist galaxy (lower two panels), and that the \horseshoe and center-phylic orbits disappear when the perturber approaches a core (cf. upper two panels). Be aware that the color coding only indicates the locations of the ZVCs: the invariance of the Jacobi energy only limits accessible phase-space from one direction; particles with Jacobi energy $E_\rmJ$ cannot access areas where $\Phi_{\rm eff}(\br) > E_\rmJ$, but given sufficient kinetic energy they can in principle reach any location where  $\Phi_{\rm eff}(\br) < E_\rmJ$. For example, \horseshoe orbits can never enter the red regions, but they can make excursions into the regions that are shaded green, cyan, yellow or white.}
  \label{fig:Phi_eff}
\end{figure*}

\subsection{Orbital Families}
\label{sec:orbfam}

To get a better understanding of dynamical friction, it is instructive to study the different kinds of stellar orbits that arise in presence of the perturber. Using equation~(\ref{geneqmotion}), we numerically integrate stellar orbits in the co-rotating frame under the combined gravitational potential of the perturber plus galaxy. Along each orbit we then register the time-evolution of the orbital energy and angular momentum. We emphasize that in doing so, the perturber is fully accounted for (i.e., is not treated as a small perturbation). For the sake of simplicity, though, we restrict ourselves to 2D, and only study the dynamics in the orbital plane of the perturber. 

One can gain valuable insight regarding the orbital families by examining the system's equipotential contours, which can be parametrized by
\begin{equation}
\Phi_{\rm eff}\left(\bf r\right) = E_{\rmJ}\,.
\end{equation}
These contours are zero-velocity curves (ZVCs) since they map out the locations in the co-rotating frame where the field particles of a given Jacobi energy $E_
\rmJ$ have zero velocity (in the co-rotating frame). Therefore, field particles along an orbit can only occasionally touch its ZVC and can only access regions on the side of its ZVC where its Jacobi energy $E_\rmJ>\Phi_{\rm eff}(\br)$.

Of particular relevance are the fixed points, also known as the Lagrange points, where the effective force in the co-rotating frame vanishes. These are given by the roots of
\begin{equation}
\nabla \Phi_{\mathrm{eff}} = 0\,.
\end{equation}
As we discuss below, the number of Lagrange points depends on the inner logarithmic slope $\gamma$ of the galaxy density profile and the galacto-centric distance $R$ of the perturber.

All orbits in the restricted three-body problem have some sense of circulation, either around the galactic center, around the perturber, around the COM, or around a specific Lagrange point.\footnote{The only exceptions are orbits associated with the (stable) Lagrange points, L4 and L5, which are stationary in the co-rotating frame and perfectly circular in the inertial frame.} We can discriminate between these different cases by considering the circular part of their Jacobi energy, $E_{\rm Jc}$, evaluated in the neighborhood of a center of perturbation (COP) (either the location of the perturber or a stable Lagrange point such as L4, L5 or L0),
\begin{align}
E_{\rm Jc}&=E_\rmJ - \frac{{\left(\Delta\Omega_0\right)}^2}{4\left|c_0\right|} - \left(\kappa_0+\frac{b_0}{\left|c_0\right|}\Delta \Omega_0\right)J_r - \left(a_0+\frac{b^2_0}{\left|c_0\right|}\right)J^2_r.
\label{Ejcirc}
\end{align}
Here $J_r$ is the radial action, $\Delta\Omega_0=\Omega_0-\Omega_\rmP$, $\Omega_0$ and $\kappa_0$ are the azimuthal and radial epicyclic frequencies, respectively, and $a_0$, $b_0$ and $c_0$ are constants that depend on the galaxy potential, evaluated at the COP (see Appendix~\ref{App:orb_class} for details). The family of an orbit is dictated by the values of $E_{\rm Jc}$ computed in the neighborhood of L4/L5 ($E_{\rm Jc}^{(4)}$), L0 ($E_{\rm Jc}^{(0)}$) and the perturber ($E_{\rm Jc}^\rmP$) respectively, relative to the values of the effective potential, $\Phi_{\rm eff}$, at the various Lagrange points and the location of the perturber. In what follows, we use $E_\rmJ^{(k)}$, with $k=0,1,..,5$ to (approximately) indicate the value of $\Phi_{\rm eff}$ at the $k^{\rm th}$ Lagrange point (e.g., $E_\rmJ^{(3)}$ indicates the $\Phi_{\rm eff}$ value corresponding to the equipotential/zero-velocity contour that passes through L3), and $E_\rmJ^\rmP$ to indicate the value of the effective potential at the location of the perturber (see Appendix~\ref{App:orb_class} for details). For nearly circular orbits with $J_r\approx 0$, $E_{\rm Jc}\approx E_\rmJ$ and the orbital families are roughly dictated by the equipotential contours.

\begin{figure*}[t!]
  \centering
  \subfloat{\includegraphics[width=1\textwidth]{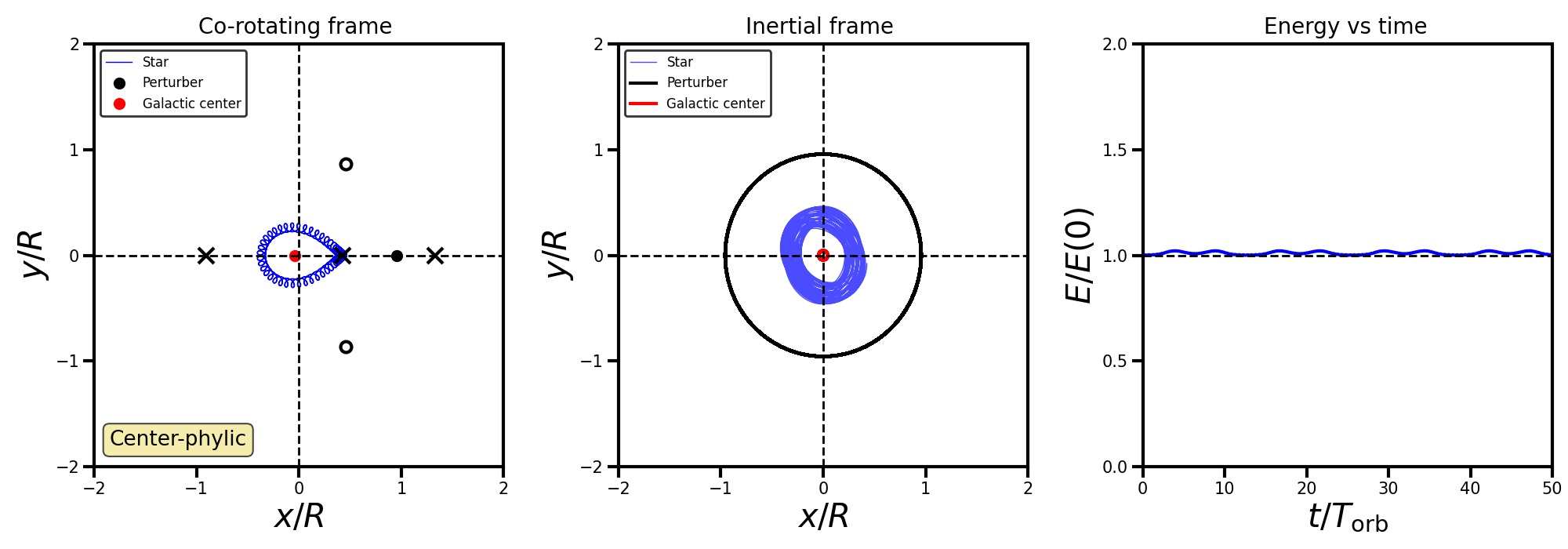}\label{orbitphyla}}
  \\
  \subfloat{\includegraphics[width=1\textwidth]{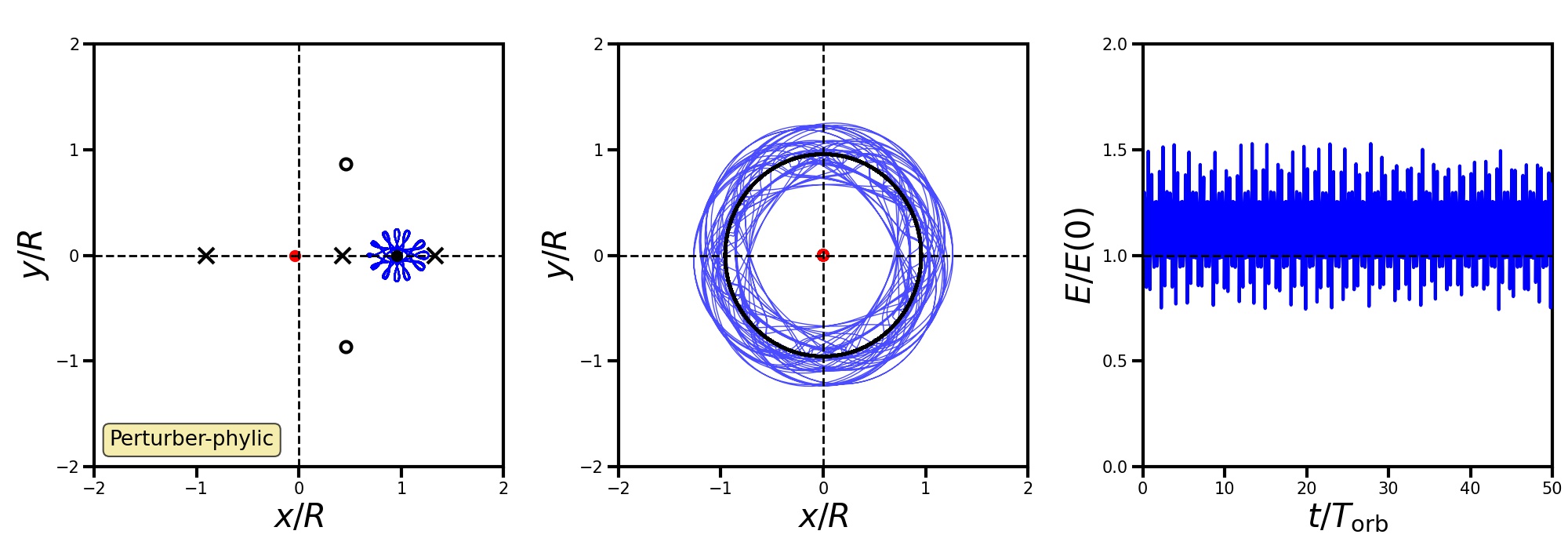}\label{orbitphylb}}
  \\
  \subfloat{\includegraphics[width=1\textwidth]{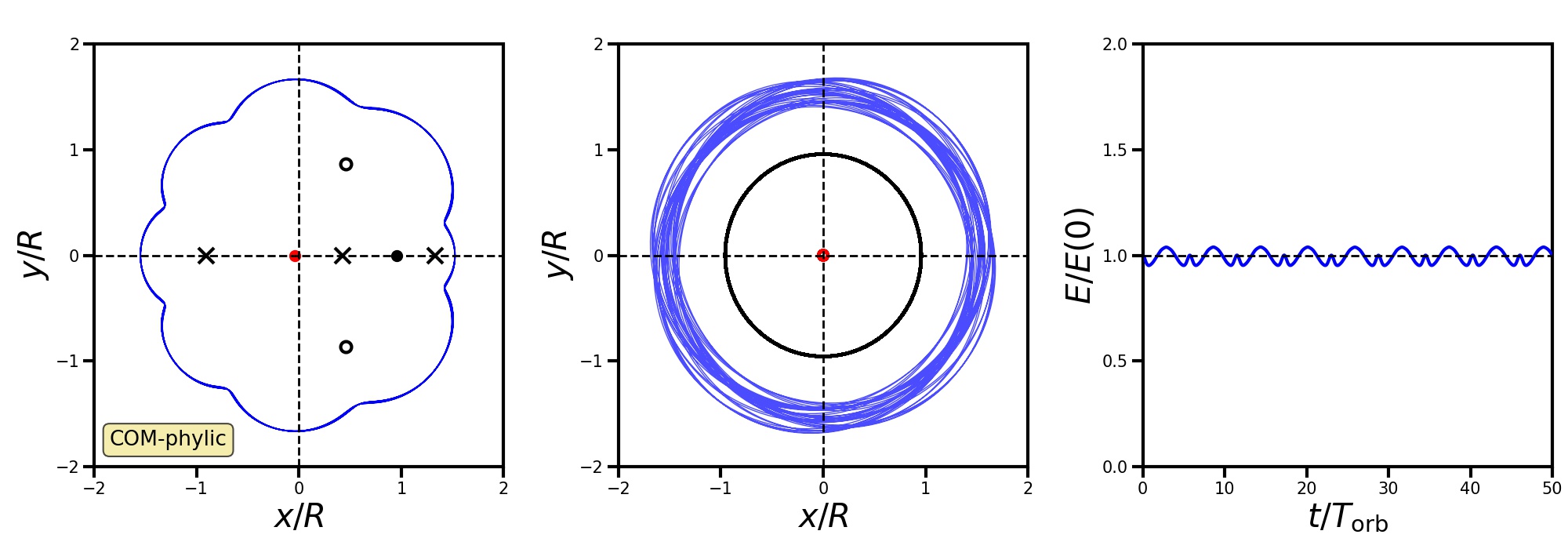}\label{orbitphylc}}
  
  \caption{\small Three orbital families (from top to bottom, center-phylic, perturber-phylic and COM-phylic) in a Plummer sphere with a perturber ($q=0.004$) on a circular orbit outside the core ($R=0.5$). As always, (x,y)=(0,0) corresponds to the COM (see Fig.~\ref{fig:schematic}). In each row, the left-hand panel shows the orbit in the co-rotating frame. The black dot indicates the perturber, the red dot marks the galactic center, and the open circles and crosses mark the stable and unstable Lagrange points, respectively. The middle panels show the orbits in the inertial frame, and the right-hand panels show the evolution in energy (as a function of time in units of $T_{\rm orb}$, the orbital time of the perturber) for a particle moving along the orbit. As discussed in the text, none of these orbital families significantly contribute to dynamical friction.}
  \label{fig:orbit1}
\end{figure*}

We start our census of the orbital families by considering a Plummer galaxy with a massive perturber (as always, assumed to be a point mass with $q=0.004$) orbiting at $R=0.5$, which is outside of the bifurcation radius (see Fig.~\ref{fig:densities}). The equipotential contours ($\Phi_{\rm eff} = E_\rmJ$) in this case are as depicted in Fig.~\ref{Plu_out}. The system has 6 Lagrange points 
(L0, L1, L2 , L3, L4 and L5) as indicated. Of these, L0 (which coincides with the galactic center), L1, L2 and L3 all lie along the $x$-axis, while L4 and L5 are located symmetrically on both sides of it, each forming an equilateral triangle with L0 and the perturber. As discussed in detail in Appendix~\ref{App:stability}, the Lagrange points L0, L4 and L5 are stable fixed points (centers), while L1, L2 and L3 are unstable fixed points (saddles).

We identify different orbital families based on the circular part of the Jacobi energy, $E_{\rm Jc}$, as specified in Table~\ref{tab:Ej}; for nearly circular orbits, this amounts to considering only the Jacobi energy since they lie close to their ZVCs. All such near-circular orbits with ZVCs inside the same shaded region in Fig.~\ref{Plu_out} have similar morphology and are taken to belong to the same orbital family. These families are separated by the ZVCs passing through the saddle points, known as separatrices. Note though that since $J_r$ can vary along an orbit, certain orbits (especially those with higher eccentricities in the inertial frame) can transition between different orbital families by undergoing separatrix-crossings. We shall address these special kinds of orbits separately towards the end of this section and proceed with the delineation of orbital families using $E_{\rm Jc}$ for now.

Let's start with the yellow-shaded region in Fig.~\ref{Plu_out}. These are orbits that circulate the galactic center (which coincides with the stable Lagrange point L0 for $\gamma>-1$ and the central cusp for $\gamma\leq -1$). These are characterized by $E_\rmJ^{(0)} < E_{\rm Jc}^{(0)} < E_\rmJ^{(1)}$ for central cores ($\gamma=0$) and $E_{\rm Jc}^\rmP<E_\rmJ^{(1)}$ for steeper profiles ($\gamma<0$). Additionally, they have lower angular momentum than that at L1, $L^{(1)}$, i.e., have $L<L^{(1)}$. Their orbital frequency is typically much larger than that of the perturber, and particles on these orbits are thus far from co-rotation resonance. In what follows we shall refer to such orbits as `center-phylic'. An example is shown in the top row of Fig~\ref{fig:orbit1}. As is evident from the right-hand panel, the orbital energy varies very little with orbital phase. As a consequence, field particles on these center-phylic orbits exchange very little energy with the perturber, and thus do not contribute significantly to dynamical friction. 

There is a similar family of non-resonant orbits, with $E_\rmJ^\rmP < E_{\rm Jc}^\rmP < \min[E_\rmJ^{(1)},E_\rmJ^{(2)}]$, that, in the co-rotating frame, only circulate the perturber. These orbits, which we call `perturber-phylic', are restricted to the Roche-lobe centered on the perturber (shaded light-blue in Fig.~\ref{Plu_out}). Their angular momentum is higher than that at L1, $L^{(1)}$, but smaller than that at L2, $L^{(2)}$, i.e., they have $L^{(1)}<L<L^{(2)}$. An example is shown in the middle row of Fig~\ref{fig:orbit1}. Note that, due to the proximity to the perturber, the orbital energy along this orbit changes drastically, and rapidly. Because of the rapid oscillations of orbital energy, the {\it net} energy exchange from {\it all} field particles on these perturber-phylic orbits is negligible, and this orbital family therefore is also not a significant contributor to dynamical friction.

Next, there is a family of low-$E_\rmJ$ orbits with $E_{\rm Jc}^\rmP < E_\rmJ^{(2)}$, that circulate the COM of the combined galaxy$+$perturber system. Their ZVCs (for near-circular orbits) fall in the unshaded region of Fig.~\ref{Plu_out} (outside of the equipotential contour that passes through L2), as their angular momentum prevents them from entering the `central' (shaded) regions, i.e., they have $L>L^{(2)}$. An example of such a `COM-phylic' orbit is shown in the bottom row of Fig~\ref{fig:orbit1}. It reveals small fluctuations in orbital energy on a relatively short timescale. Since there are roughly equal numbers of field particles along each phase of these COM-phylic orbits, they also have a negligible, net contribution to dynamical friction (i.e., at each point in time, these orbits contribute roughly equal numbers of energy gainers as energy losers).

\begin{figure*}[t!]
  \centering
  \subfloat{\includegraphics[width=1\textwidth]{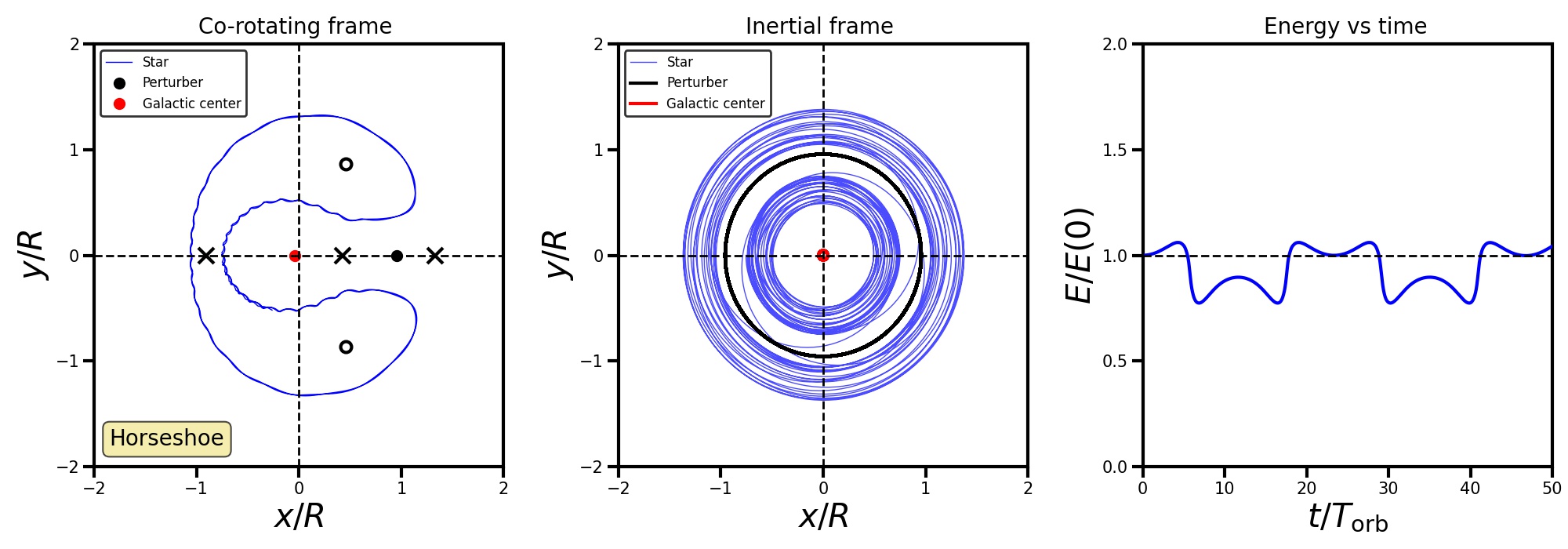}\label{orbitouta}}
  \\
  \subfloat{\includegraphics[width=1\textwidth]{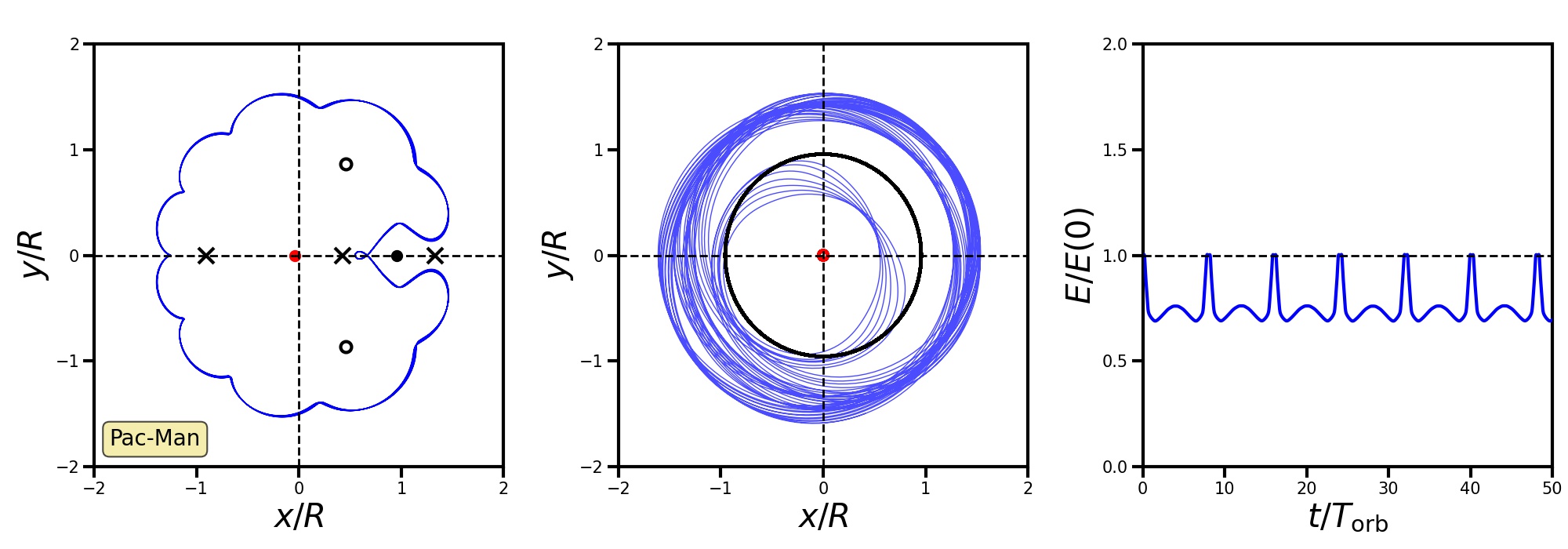}\label{orbitoutb}}
  \\
  \subfloat{\includegraphics[width=1\textwidth]{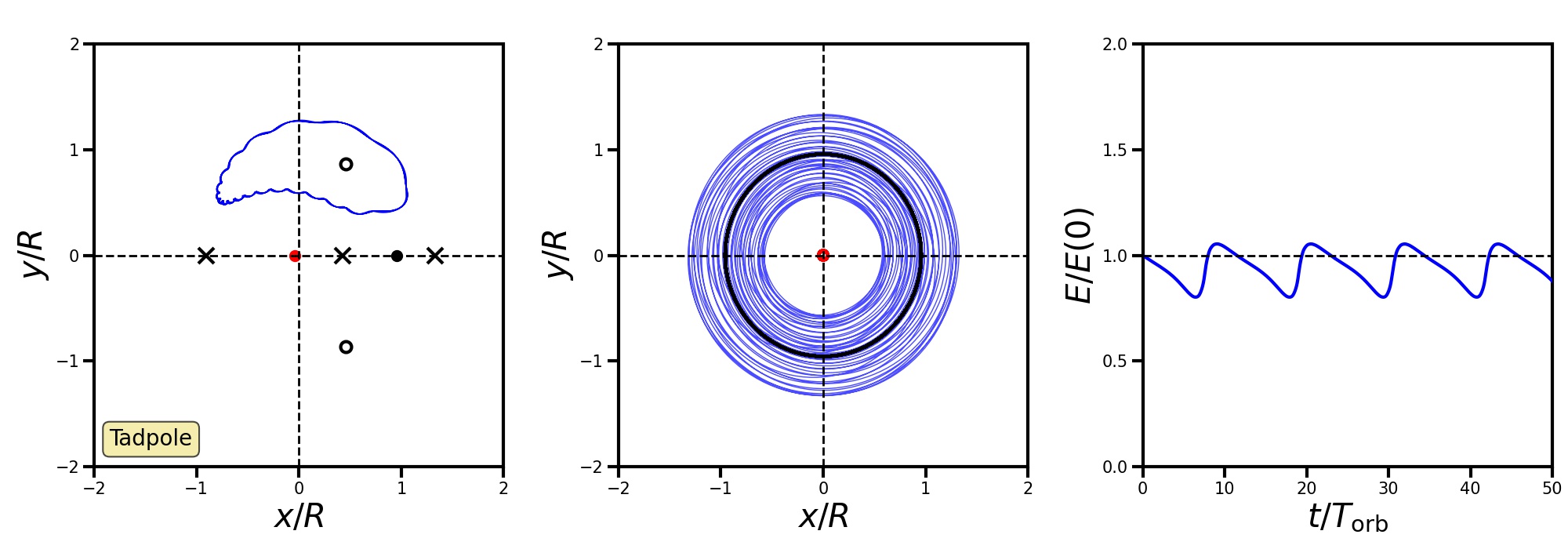}\label{orbitoutc}}

  \caption{\small Same as Fig~\ref{fig:orbit1}, but for the three NCRR families (from top to bottom, \horseshoen, \pacman and tadpole) that make significant contribution to dynamical friction.}
  \label{fig:orbit2}
\end{figure*}

Next, we discuss the three families that are the dominant contributors to dynamical friction. They all have azimuthal frequencies that are comparable to that of the perturber, i.e., $\Omega_\phi\approx \Omega_\rmP$, such that their libration time in the co-rotating frame is long. In fact, along these orbits, $\Omega_\phi-\Omega_\rmP$ oscillates back and forth about $\pm\Omega_r/N$, where $\Omega_r$ is the radial frequency and the integer $N$ is the number of radial excursions or epicycles for every libration. The typical range of $N$ is $[2,\infty)$ for realistic galaxy profiles, with $N\to \infty$ marking the co-rotation resonance, i.e., $N$ is larger the closer the orbit is to co-rotation. Therefore they are `near-co-rotation-resonant' (NCRR), i.e., they librate about the near-co-rotation resonances, $\Omega_\phi-\Omega_\rmP=\pm\Omega_r/N$. When the perturber is farther out, $M_\rmG(R)\gg M_\rmP$, implying $\Omega_\rmP\approx \sqrt{GM_\rmG(R)/R^3}\approx \Omega_\phi$ in the vicinity of L4 and L5 (since these two Lagrange points are both at a distance, $R$, from the galactic center). Therefore, $N$ is large, i.e., $N\gg 1$, and the orbits librating about L4 and L5 are close to co-rotation resonance. As the perturber penetrates deeper into the core region, $M_\rmP$ becomes comparable to $M_\rmG(R)$, and $\Omega_\rmP$ significantly exceeds $\Omega_\phi$ near L4 and L5, thereby pushing the orbits farther away from co-rotation resonance (smaller $N$), as pointed out by KS18.

The first, and probably most well-known, among the NCRR orbits is the family of so-called `\horseshoe orbits', which we already encountered in Section~\ref{sec:concept}. These have $E_{\rm Jc}^{(4)} < E_\rmJ^{(3)}$ and $E_{\rm Jc}^{(0)} > \max[E_\rmJ^{(1)},E_\rmJ^{(2)}]$ for central cores ($\gamma=0$), while $E_{\rm Jc}^{(4)} < E_\rmJ^{(3)}$ and $E_{\rm Jc}^\rmP>\max[E_\rmJ^{(1)},E_\rmJ^{(2)}]$ for steeper profiles ($\gamma<0$). The ZVCs of near-circular \horseshoe orbits fall within the dark blue-shaded region in Fig.~\ref{Plu_out} and can only cross the $x$-axis at the side of L0 opposite to the perturber; the Lagrange point L1 acts as a barrier, forcing the particle to take a long `detour' around the center of the galaxy. They have a net sense of circulation around L3, with a libration frequency $\vert \Omega_{\rm lib} \vert \ll \Omega_\rmP$. As is evident from the top row of Fig.~\ref{fig:orbit2} (see also Fig.~\ref{fig:horseshoe}), the orbital energy can vary drastically along the orbit, undergoing rapid changes when close to the perturber, where the perturber's force pulls the field particle either inward or outward. 

Somewhat similar to the \horseshoe orbits is a family of orbits that we call `\pacmann' orbits. These are characterized by $E_{\rm Jc}^{(0)} < E_\rmJ^{(1)}$ and $E_{\rm Jc}^\rmP>E_\rmJ^{(2)}$ for central cores ($\gamma=0$), while $E_\rmJ^{(2)}<E_{\rm Jc}^\rmP<E_\rmJ^{(1)}$ for steeper profiles ($\gamma<0$). Additionally, they have $L^{(1)}<L<L^{(2)}$. They differ from the \horseshoe orbits in that they have a net sense of circulation around L0. The Jacobi energy of the near-circular \pacman orbits is less than that of L1, which allows their ZVCs to cross the $x$-axis at the side of L0 that coincides with the perturber, and fall within the green-shaded region of Fig.~\ref{Plu_out}. Rather than taking a `detour', these orbits can therefore take a `short-cut', which changes their characteristic shape such that they resemble the iconic flashing-dots eating character of the popular 1980's computer-game Pac-Man (see middle row of Fig.~\ref{fig:orbit2}). We emphasize that \pacman orbits are only present when $E_\rmJ^{(1)} > E_\rmJ^{(2)}$. For a given galaxy potential and mass of the perturber, this puts a constraint on the galacto-centric distance of the perturber, $R$; for the Plummer potential and our fiducial mass ratio $q=0.004$, \pacman orbits are only present when the perturber is located at $R \lta 1.23$. When further out, \pacman orbits are absent such that the equipotential contours and orbital families are similar for both cored and cuspy galaxy profiles. 

The final family of NCRR orbits are known as `tadpole' orbits, a name that again relates to their characteristic shape in the co-rotating frame (see bottom row of Fig.~\ref{fig:orbit2}). These are characterized by $E_\rmJ^{(3)} (E_\rmJ^{(0)}) < E_{\rm Jc}^{(4)} < E_\rmJ^{(4)}=E_\rmJ^{(5)}$ for $R>\Rbif$ ($R\leq \Rbif$), and have a net sense of circulation around either L4 or L5. Their ZVCs fall within the red-shaded region of Fig.~\ref{Plu_out}.

\subsection{Slow versus fast actions}
\label{sec:slowfast}

Along all NCRR orbits (\horseshoen, \pacman and tadpole), the energy and angular momentum oscillate with a large amplitude and long period, and the star is up/down-scattered through near-co-rotation resonances by interactions with the perturber. This can be understood in terms of slow and fast action-angle variables, which exist in the neighborhood of a resonance and are related to the radial and azimuthal action-angle variables by a canonical transformation \citep[e.g.,][]{Tremaine.Weinberg.84,Lichtenberg.Lieberman.92,Chiba.Schonrich.22}. The NCRR orbits librate about the commensurability condition $\Omega_\phi -\Omega_\rmP \mp \Omega_r/N = 0$. The corresponding angle, $\theta_s = \theta_\phi \mp \theta_r/N - \Omega_\rmP t$, is called the {\it slow angle}, and the action conjugate to it is called the slow action, $J_s$, which is proportional to the angular momentum. Note that close to the commensurability condition $\rmd\theta_s/\rmd t = \Omega_\phi - \Omega_\rmP \mp \Omega_r/N \simeq 0$, indicating that $\theta_s$ indeed varies slowly. And while it does, the corresponding slow action undergoes large changes.  Both $J_s$ and $\theta_s$ librate about the near-co-rotation resonances with a time period, $T_{\rm lib}$, which is much larger than the orbital time of the perturber \citep[see][for detailed derivations using perturbative expansions of the Hamiltonian around resonances]{Contopoulos.73, Chiba.Schonrich.22}. In fact, for orbits that come arbitrarily close to the separatrices, $T_{\rm lib}$ approaches infinity.

Contrary to the slow angle, the {\it fast angle}, which is nothing but the radial angle, $\theta_r$, varies rapidly along an orbit, while its conjugate action, the {\it fast action}, $J_f=J_r\pm L/N$, is nearly invariant. In general, the faster the angle changes, the closer its corresponding fast action is to an adiabatic invariant. Therefore, the NCRR orbits have two integrals of motion, the Jacobi Hamiltonian, $E_\rmJ$ (which is exactly conserved), and the fast action, $J_f$ (which is {\it very nearly} conserved), and are {\it nearly integrable}\footnote{In 3D, the near-resonant orbits possess a second pair of fast action-angle variables, where the fast angle corresponds to the azimuthal angle along the orbital plane of the field particle, which can be inclined wrt the perturber's plane of orbit.}. For the very nearly co-rotation resonant orbits, $N\gg 1$, and therefore $J_f\approx J_r$, i.e., the orbital eccentricity (in the inertial frame) remains nearly constant. This is however not the case for orbits farther away from co-rotation resonance, which can show very interesting dynamics, as we shall see shortly.

\subsection{Orbital make-up}
\label{sec:makeup}

The relative abundances of the different orbital families depend on the orbital radius $R$ of the perturber. For example, Fig.~\ref{Plu_in} shows the equipotential contours of the same Plummer galaxy as in Fig.~\ref{Plu_out}, but with the perturber orbiting inside the central core, at $R = 0.2$. Now only four Lagrange points are present; both L1 and L3 have disappeared. As the perturber approaches the galactic center, the Roche lobes around the galactic center and the perturber coalesce to form a single lobe surrounding the perturber. As we show in section~\ref{sec:bifurcation}, this is associated with the merging, or `bifurcation' of L3, L0 and L1 at a critical bifurcation radius, $\Rbif$, which leaves only L0, L2, L4 and L5, and changes the stability of L0 from being a center to a saddle.  As a consequence, neither \horseshoe nor center-phylic orbits survive. In addition, the contribution of the tadpole orbits is also significantly diminished. Instead, the dominant orbital families in the central core region are the perturber-phylic orbits and the \pacman orbits. As we will see, this has profound implications for dynamical friction. 

The orbital configuration is particularly sensitive to the density profile of the galaxy. The lower two panels of Fig.~\ref{fig:Phi_eff} show the equipotential contours of a Hernquist galaxy with a perturber at $R=0.5$ (Fig.~\ref{Her_out}) and $R=0.2$ (Fig.~\ref{Her_in}). In such a cuspy galaxy, there is no L0 (L0 is replaced by the cusp), and the five Lagrange points (L1, L2, L3, L4 and L5) survive throughout, for any value of the orbital radius of the perturber, $R$, without the occurrence of any bifurcation. As a consequence, in this galaxy potential, there are never any \pacman orbits and the relative abundances of different orbital families show a much weaker dependence on $R$ than in the case of the Plummer sphere. How all of this relates to dynamical friction will be discussed in more detail in sections~\ref{sec:resonances}-\ref{sec:core}.

\subsection{Separatrix crossing and Chimera orbits}
\label{sec:crossing}

Before proceeding with the computation of the dynamical friction torque from the various orbits, we first discuss a potential complication. We have defined orbital families on the basis of $E_{\rm Jc}$, but family is not an invariant property for all orbits. In fact, an orbit can change its family in course of its evolution. This is because the orbit-determinant, $E_{\rm Jc}$, as expressed in equation~(\ref{Ejcirc}), is not an invariant quantity. It not only involves $E_\rmJ$, which is an integral of motion and thus conserved, but also the radial action, $J_r$, which is typically not constant along an orbit. In particular, $J_r$ can undergo significant changes along orbits that are farther away from co-rotation resonance, since only a linear combination of $J_r$ and $L$, and not $J_r$ alone, is the fast action in this case. Therefore the value of $E_{\rm Jc}$ can potentially cross over from that corresponding to one orbital family to another, which corresponds to the orbit undergoing separatrix-crossing due to a change in the radial action enabled by the perturber, altering its morphological appearance. We call such orbits `Chimera orbits'\footnote{The Chimera orbits are named after the hybrid creature in Greek mythology that is composed of parts of more than one animal.}. These Chimera-like transitions occur between trapped regions of neighboring resonances on either side of a separatrix (see Appendix~\ref{App:orb_class}) or a chaotic island formed by the {\it overlap} of resonances \citep[see][for a detailed discussion in the context of bar-like perturbations]{Chiba.Schonrich.22}. For example, the metamorphosis between \horseshoes and tadpoles occurs near L3, while that between \horseshoens, \pacmans and center-phylic orbits happens near L1. And finally the transition between \pacmann, COM-phylic and perturber-phylic orbits occurs in the neighborhood of L2. We show several examples of such Chimera orbits in Appendix~\ref{App:Chimera}. Not all orbits show this Chimera behavior. The very nearly co-rotation resonant orbits are nearly circular and thus have small $J_r$. Since $J_r$ is a fast action along such orbits, it remains almost constant, i.e., the orbits remain nearly circular and do not exhibit Chimera characteristics. 

When the separatrix crossing along a Chimera orbit results in a perturber-phylic phase, we speak of resonant capture \citep[][]{Henrard.82}, which as pointed out in \cite{Tremaine.Weinberg.84}, can `dress' the perturber with a cloud of captured stars. Note, though, that in the `slow' regime considered here, in which the orbital radius of the perturber is taken to be invariant, these stars can undergo separatrix crossing again, transitioning back to a \pacman or a COM-phylic orbit. Similarly, when a separatrix-crossing results in a transition from a `trapped' NCRR state to an `untrapped' COM-phylic state, the transition is sometimes called `scattering', e.g., \cite{Daniel.Wyse.15}.

Chimera orbits are difficult to account for in our treatment because they do not have a clear periodic behaviour, i.e., do not have a well-defined libration time. However, we find that most of them typically behave as an archetypal orbit of their family for many orbital periods before revealing their Chimera nature, i.e., they are `semi-ergodic' (similar to the semi-ergodic orbits identified by \cite{Athanassoula.etal.83} in their study of barred galaxies). This is akin to how Arnold diffusion in KAM theory can cause chaotic orbits to behave quasi-regularly for extended periods  \citep[e.g.][]{Lichtenberg.Lieberman.92}. Hence, we conjecture that their relevance to dynamical friction is captured, at least to leading order, by our following treatment of the NCRR orbital families.

\section{The origin of dynamical friction in the\\ non-perturbative case} 
\label{sec:resonances}

As described in Section~\ref{sec:concept}, in our non-perturbative framework the net torque on the perturber arises from an {\it imbalance} between field particles {\it along the same orbit} that are {\it up-scattered} vs. {\it down-scattered} in energy. We now proceed to compute the torque on the perturber due to individual orbits. Using the results from a large ensemble of such orbits, we then highlight the transition from a net retarding to a net enhancing torque when approaching the core of a Plummer sphere.

\subsection{The net torque from individual orbits}
\label{sec:integrated_energy}

In order to compute the torque on the perturber due to a single orbit, we proceed as follows. We numerically integrate the orbit of a massless field particle in the presence of the perturber, registering its position $\br$, velocity $\dot{\br}$, energy $E$, and angular momentum $\bL$, as a function of time $t'$. We use $t'$ to indicate the phase of a particle along this orbit. We have seen in sections~\ref{sec:concept} and~\ref{sec:orbfam} that as a particle moves along the perturbed orbit, it undergoes changes in energy and angular momentum due to exchanges with the perturber. Hence, after some time $\Delta t$, a particle starting from phase $t'$ has transferred a net amount of energy $\Delta E(\Delta t) = E(t' + \Delta t) - E(t')$ to the perturber. Here $E(t')$ is the perturbed energy of a particle at phase $t'$, given by equation~(\ref{Eperturbed_5}). To work out the total energy exchanged with the perturber by all stars associated with the orbit in question, we need to integrate $\Delta E(\Delta t)$ along the orbit, weighted by the relative number of stars at each point along the orbit. This weight is given by $f_0(E_{0\rmG}(t'))$, with $f_0$ the unperturbed DF, and

\begin{align}
E_{0\rmG}(t')=\frac{1}{2}{\left|\dot{\br}+\bf\Omega_{\rm{\bP}}\times \br-\bv_{\rm\bG}\right|}^2+\Phi_\rmG
\end{align}
the galactocentric energy of the star at phase $t'$ in absence of the perturber, where $\bv_{\rm\bG}=-\Omega_\rmP\, q_\rmG R\, \hat{y}$ is the circular velocity of the galactic center about the COM. If we use $s(t')$ to parameterize the path-length along the phase-space trajectory traced out by the orbit, then the total energy exchanged with the perturber along this orbit, some time $\Delta t$ after the perturber was introduced, is given by the following line-integral
\begin{equation}
\Delta E(\Delta t) = \frac{1}{\calA} \int_s \rmd s(t') \, \left[E(t'+\Delta t)-E(t')\right] \, f_0(E_{0\rmG}(t'))\,.
\label{deltaElineintegral}
\end{equation}
with $\calA$ a normalization factor (see below).

Typically, an orbit in the co-rotating frame will not be exactly closed and the integration limit therefore will have no boundaries. However, for the NCRR orbits discussed in Section~\ref{sec:orbfam}, the orbit is {\it approximately} periodic in the co-rotating frame, with a period $T_{\rm lib}$ set by the time it takes the particle to librate about its Lagrange point (the COC in column~6 of Table~\ref{tab:Ej}), which we compute by a Fourier analysis of the orbit in the co-rotating frame. In the vicinity of the stable Lagrange points, L4 and L5, $T_{\rm lib}$ can be analytically computed using a perturbative method, as discussed in Appendix~\ref{App:stability}.

\begin{figure}
\centering
\hspace{-1mm}
\includegraphics[width=0.9\textwidth]{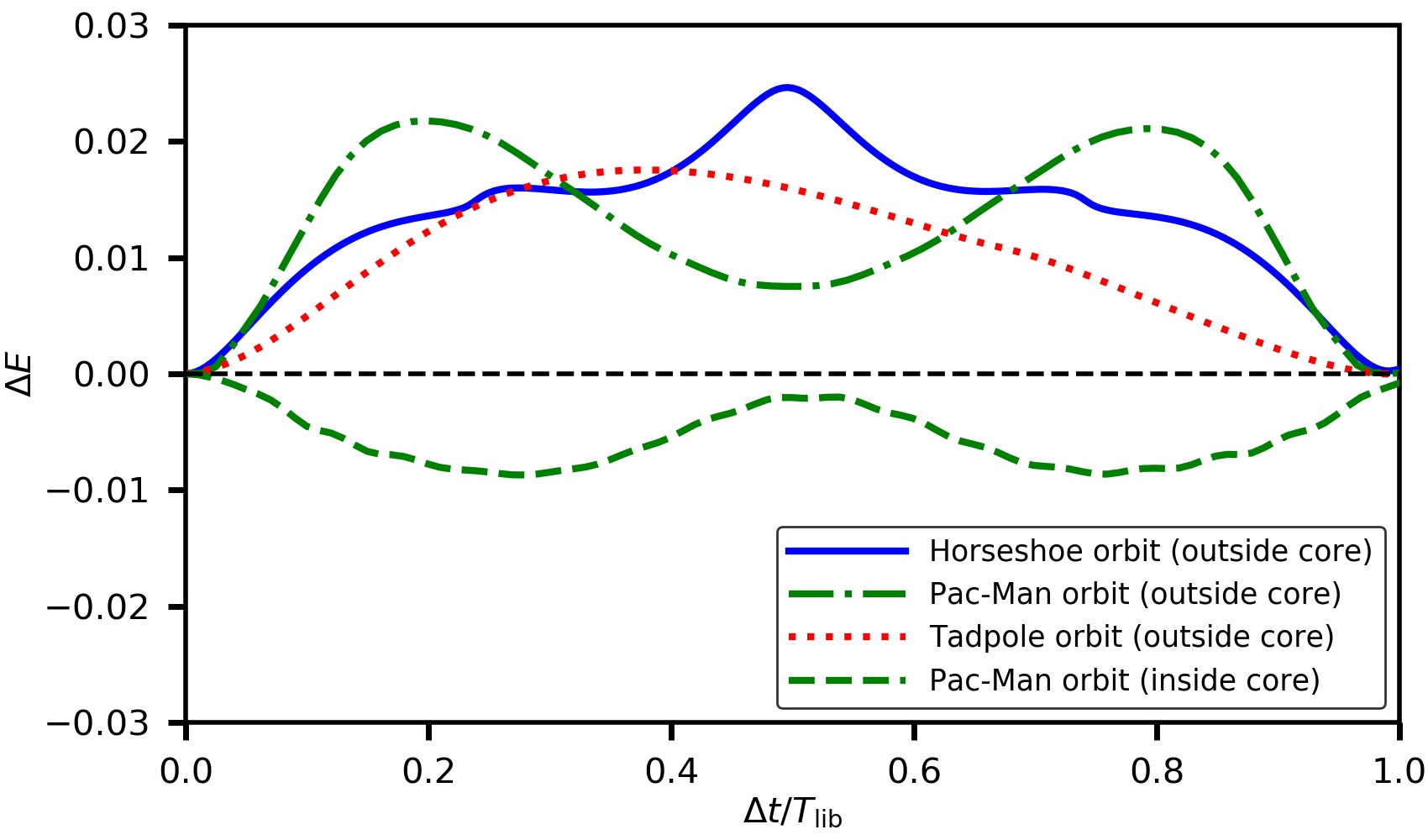}
\caption{\small The solid blue, dot-dashed green and dotted red curves respectively show the average energy change per star (equation~[\ref{deltaE}]) along individual NCRR \horseshoen, \pacman and tadpole orbits shown in Fig.~\ref{fig:orbit2} as a function of time (in units of the libration time, $T_{\rm lib}$). All these are examples of orbits in the case where the perturber is orbiting outside of the core of a Plummer sphere, at $R=0.5$. For comparison, the green dashed curve shows the integrated energy change for a \pacman orbit when the perturber is orbiting inside the core, at $R=0.2$. See text for details.}
\label{fig:integrated_energy}
\end{figure}

The line integral in Eq.~(\ref{deltaElineintegral}) has to be performed along the phase-space trajectory and therefore the differential line element $\rmd s(t')$ is given by $\rmd s = \sqrt{{\left|\rmd \br_i\right|}^2+{\left|\rmd \dot{\br}_i\right|}^2}$. Using that the Jacobian for the transformation from $t'$ to the arc-length $s(t')$ is given by  
\begin{align}
\frac{\rmd s}{\rmd t'} &= \sqrt{{\left|{\dot{\br}}_{\mathrm{\bi}}\right|}^2+{\left|{\ddot{\br}}_{\mathrm{\bi}}\right|}^2}\,,
\end{align}
with ${\dot{\br}}_{\mathrm{\bi}}$ and ${\ddot{\br}}_{\mathrm{\bi}}$ the velocity and acceleration in the inertial frame, respectively, we can approximate the line integral as
\begin{align}
\Delta E(\Delta t) &\approx \frac{1}{\calA} \int_0^{T_{\rm{lib}}} \rmd t' \, \sqrt{{\left|{\dot{\br}}_{\mathrm{\bi}}\right|}^2+{\left|{\ddot{\br}}_{\mathrm{\bi}}\right|}^2}\nonumber \\
&\times \left[E(t'+\Delta t)-E(t')\right] \, f_0(E_{0\rmG}(t'))\,,
\label{deltaE}
\end{align}
with
\begin{align}
\calA &= \int_s \rmd s(t')\,f_0(E_{0\rmG}(t')) \nonumber \\
&= \int_0^{T_{\rm{lib}}} \rmd t' \, \sqrt{{\left|{\dot{\br}}_{\mathrm{\bi}}\right|}^2+{\left|{\ddot{\br}}_{\mathrm{\bi}}\right|}^2} \, f_0(E_{0\rmG}(t'))\,.
\label{norm}
\end{align}
Note that, with this normalization, $\Delta E(\Delta t)$ is the average energy {\it per star} exchanged with the perturber in a time $\Delta t$ along the orbit in question.

The inertial acceleration vector is given by
\begin{align}
{\ddot{\br}}_{\mathrm{\bi}} = -\nabla \Phi\,,
\end{align}
where $\Phi = \Phi_\rmP + \Phi_\rmG$ is the total potential, while the velocity vector in the inertial frame is related to that in the co-rotating frame, 
$\dot{\br}$, by 
\begin{align}
{\dot{\br}}_{\mathrm{\bi}}=\dot{\br}+\bf \Omega_{\rm{\bP}}\times \br\,.
\end{align}

We perform this line integral for the three NCRR orbits (\horseshoen, \pacman and tadpole) shown in Fig.~\ref{fig:orbit2}. All three orbits correspond to our fiducial $q=0.004$ point-mass perturber in a Plummer potential at $R=0.5$. The solid blue, dot-dashed green and dotted red lines in Fig.~\ref{fig:integrated_energy} show the resulting $\Delta E$ for the \horseshoen, \pacman and tadpole orbits respectively as function of $\Delta t$. Note that $\Delta E(\Delta t = T_{\rm lib}) = 0$; as discussed in Section~\ref{sec:concept}, along each NCRR orbit particles both gain and loose energy, and the net effect for a single particle over a full libration period is zero. However, due to the non-uniform phase distribution along each orbit, which arises from the unperturbed phase-space distribution, $f_0(E_{0\rmG})$, we see that $\Delta E$ is positive for all $0 < \Delta t < T_{\rm lib}$. A positive $\Delta E$ indicates that the field particles along these orbits {\it gain} net energy from the perturber, and thus that the perturber experiences dynamical friction. As the field particles gain energy, their $\Omega_\phi$ decreases. The perturber in turn loses energy and falls in, with increasing $\Omega_\rmP$. This puts the original NCRR orbits out of near-co-rotation resonance. Therefore, $\Delta E(\Delta t)$ is only relevant for the dynamics of the system for relatively small $\Delta t$. The exact choice of $\Delta t$ to consider is somewhat ambiguous; it should be indicative of the time scale over which the perturber moves through the resonances, which in turn depends on the strength of dynamical friction. In what follows, we take $\Delta t = T_{\rm orb}$, the orbital time of the perturber. None of our qualitative conclusions are sensitive to this particular choice.

\begin{figure*}[t!]
  \centering
  \subfloat{\includegraphics[width=0.48\textwidth]{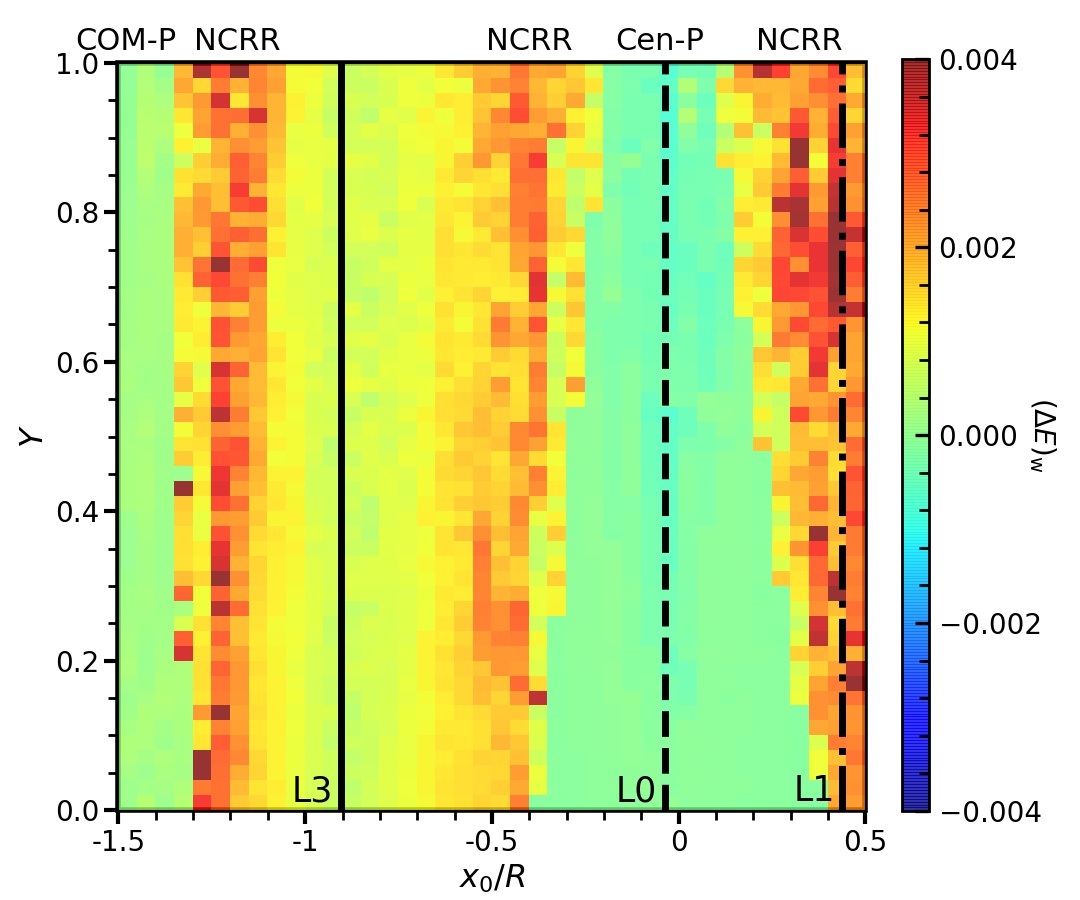}}
  \hspace{2mm}
  \subfloat{\includegraphics[width=0.49\textwidth]{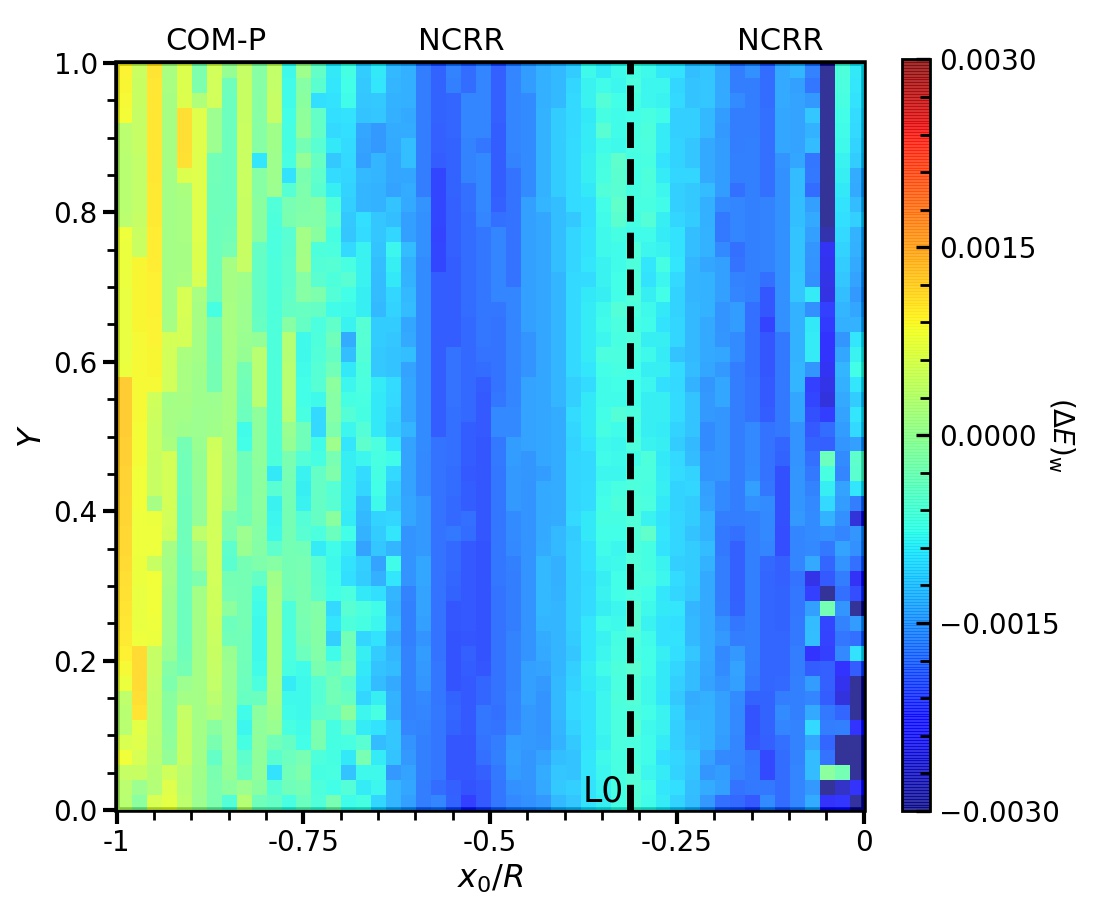}}

  \caption{\small Energy change per unit phase-space, $\dE$, of field particles moving along orbits in a cored Plummer potential with a perturber $(q=0.004)$ on a circular orbit at $R=0.5$ (left-hand panel) and $R=0.2$ (right-hand panel). The initial conditions for the orbits are sampled uniformly in $x_0$ and $Y \equiv [E_\rmJ  -\Phi_{\rm eff} (x_0,0)] / [E_\rmJ^{(4)} - \Phi_{\rm eff}(x_0,0)]$ (for every $x_0$), with $y_0=0$ and $\left|v_{\rmy,0}\right| = \frac{1}{3}v_0$, where $v_0 = \sqrt{2[E_\rmJ-\Phi_{\rm eff}(x_0,0)]}$.  Solid, dashed, and dot-dashed vertical lines indicate the positions of L3, L0 (the galactic center) and L1, respectively. Note that when the perturber is located outside the core, at $R=0.5$, $\dE$ is predominantly positive (red) suggesting ongoing dynamical friction. Inside the core, though, at $R=0.2$, $\dE$ is predominantly negative (blue) indicating dynamical buoyancy. The red and blue bands are due to NCRR orbits (causing a larger $|\dE|$), while bands of greenish color (small $|\dE|$) generally indicate non-resonant orbits. In particular, the wide green band in the left panel centered on $x_0=0$ corresponds to the non-resonant center-phylic (Cen-P) orbits, while the green band in the extreme left of both panels indicates COM-phylic (COM-P) orbits. As discussed in the text, due to a bifurcation of Lagrange points there are no center-phylic orbits when the perturber is inside $R \sim 0.39$.}
  \label{fig:delE}
\end{figure*}

The solid blue, dot-dashed green and dotted red curves in Fig.~\ref{fig:integrated_energy} correspond to NCRR orbits in the case where the perturber is orbiting at $R=0.5$, just outside the core of the Plummer sphere. For comparison, the dashed, green curve in Fig.~\ref{fig:integrated_energy} indicates the $\Delta E(\Delta t)$ for a \pacman orbit in the case where the perturber is at $R=0.2$, well inside the core of the Plummer galaxy. In this case $\Delta E$ is negative, indicating that this orbit contributes a positive, enhancing torque. Note that, since the torque on the field particle is given by $\rmd L/\rmd t =  \Omega_\rmP^{-1} (\rmd E/\rmd t)$ (cf. equation~[\ref{dEdt}]), the average torque {\it on the perturber} due to an orbit between $t=0$ and $t=T_{\rm orb}$ is equal to $-\Delta E/\left(\Omega_\rmP T_{\rm orb}\right)=-\Delta E/(2\pi)$, i.e., ${\rm sign}(\calT)=-{\rm sign}(\Delta E)$. We thus see that some of the NCRR orbits can give rise to dynamical buoyancy, rather than friction. An explanation of the latter is discussed in Section~\ref{sec:core}.

\subsection{Scanning Orbital Parameter Space}
\label{sec:scan}

Having demonstrated how to compute the contribution to dynamical friction  from individual orbits, in the form of $\Delta E(T_{\rm orb})$, one can in principle obtain the total torque by summing over all orbits, properly weighted by their relative contribution to the distribution function. In practice, though, this is far from trivial. First of all, sampling all orbits numerically is tedious to the point that one is better off just running an $N$-body simulation. Secondly, some orbits are difficult to integrate accurately, especially some Chimera orbits which reveal semi-ergodic behavior, and the perturber-phylic orbits along which the energy varies rapidly with time. Hence, the non-perturbative method adopted in this chapter is not well suited to accurately compute the total dynamical friction torque. Notwithstanding, it gives valuable insight as to the inner workings, in an orbit-based sense, of dynamical friction and buoyancy.

As an example, we now proceed to investigate the contribution to the torque, in terms of $\Delta E(T_{\rm orb})$, from a modest sub-sample of orbits. In what follows we continue to treat the dynamics in 2D (i.e., we only consider orbits in the $x$-$y$ plane depicted in Fig.~\ref{fig:schematic}). We densely sample the part of the orbital parameter space corresponding to the NCRR orbits, which is most relevant for dynamical friction. We first sample the starting point $(x_0,y_0)$ by setting $y_0 = 0$ and sampling $x_0$ uniformly over the range dominated by the NCRR \horseshoe and \pacman orbits (roughly the region inside the $E_\rmJ^{(2)}$ separatrix marked by the solid line in Fig.~\ref{fig:Phi_eff}). Note that by sampling orbits that intersect the $x$-axis, we exclude tadpole orbits with large $E_\rmJ$ that librate in small regions around L4 and L5. After sampling $x_0$, we uniformly sample $E_\rmJ$ over the range $[\Phi_{\rm eff}(x_0,0), E_\rmJ^{(4)}]$. Although orbits with $E_\rmJ\gg E_\rmJ^{(4)}$ are far from co-rotation resonance, thereby contributing less to dynamical friction, those with small, positive values of $E_\rmJ-E^{(4)}_\rmJ$ are NCRR and have similar contribution to the torque as those with $E_\rmJ \lesssim E^{(4)}_\rmJ$. Therefore we consider $E^{(4)}_\rmJ$ to be only an approximate rather than a hard cut-off for the NCRR orbits. Finally, we sample the initial velocities, $v_{\rm x,0}$ and $v_{\rm y,0}$, under the constraint that
\begin{align}
 v_0 = \sqrt{v^2_{\rm x,0} + v^2_{\rm y,0}} = \sqrt{2[E_\rmJ - \Phi_{\rm eff}(x_0,0)]}\,.
\end{align}
Note that both $\bx_0$ and $\bv_0$ are defined in the co-rotating frame. We numerically integrate the orbits for $100\, T_{\rm orb}$, with $T_{\rm orb}$ the orbital time of the perturber, after which we estimate the libration time, $T_{\rm lib}$, by noting the consecutive time-stamps at which each orbit crosses the abscissa of its center-of-circulation (see Table~\ref{tab:Ej}) after making a $2\pi$ circulation about it. Finally, we compute $\Delta E \equiv \Delta E(T_{\rm orb}|x_0, E_\rmJ, v_{\rm x,0}, v_{\rm y,0})$ using equation~(\ref{deltaE}). 

In order to allow for a meaningful comparison of the torque contribution from each of these orbits, we weight the $\Delta E$ {\it per star}, given by equations~(\ref{deltaE})-(\ref{norm}), by the average phase-space density associated with that orbit. This yields the total energy exchange per unit phase-space from an orbit, given by
\begin{align}
\dE & \equiv \frac{\int_s \rmd s(t') f_0(E_{0\rmG}(t'))}{\int_s \rmd s(t')}\,\Delta E \,.
\end{align}

Using that the time-averaged torque (per unit phase-space) on the perturber contributed by an individual orbit is given by
\begin{align}
 \calT_\rmw = -\frac{1}{\Omega_\rmP} \, \frac{\dE}{\Delta t}
\end{align}
(cf. equation~[\ref{dEdt}]), we have that the torque per unit phase-space contributed by the orbit can be expressed as
\begin{align}
&\calT_\rmw = -\frac{1}{2\pi} \frac{\int_0^{T_{\rm{lib}}} \rmd t' \, \sqrt{{\left|{\dot{\br}}_{\mathrm{\bi}}\right|}^2+{\left|{\ddot{\br}}_{\mathrm{\bi}}\right|}^2} \left[E(t'+\Delta t)-E(t')\right] f_0(E_{0\rmG}(t'))}{\int_0^{T_{\rm{lib}}} \rmd t' \, \sqrt{{\left|{\dot{\br}}_{\mathrm{\bi}}\right|}^2+{\left|{\ddot{\br}}_{\mathrm{\bi}}\right|}^2}},
\end{align}
where we have used the fact that we adopt $\Delta t = T_{\rm orb} = 2 \pi/\Omega_\rmP$, and we have rewritten $\dE$ using equations~(\ref{deltaE}) and (\ref{norm}).

Fig.~\ref{fig:delE} plots $\dE$ for the Plummer sphere as a function of $x_0$ and $E_\rmJ$ for $\left|v_{\rm y,0}\right| = \frac{1}{3}v_0$. Results for other values of $\left|v_{\rm y,0}\right|$ are very similar, but with the overall amplitudes in $\dE$ decreasing as $\left|v_{\rm y,0}\right| \to v_0$. For each $\left(x_0,E_\rmJ,\left|v_{\rm y,0}\right|\right)$, there are four combinations of $(v_{\rm x,0},v_{\rm y,0})$, given by $(\pm \sqrt{v^2_0 - v^2_{\rm y,0}}, \pm \left|v_{\rm y,0}\right|)$. The values of $\dE$ shown are the sums of these four cases combined.

\begin{figure*}[t!]
  \centering
  \subfloat{\includegraphics[width=0.485\textwidth]{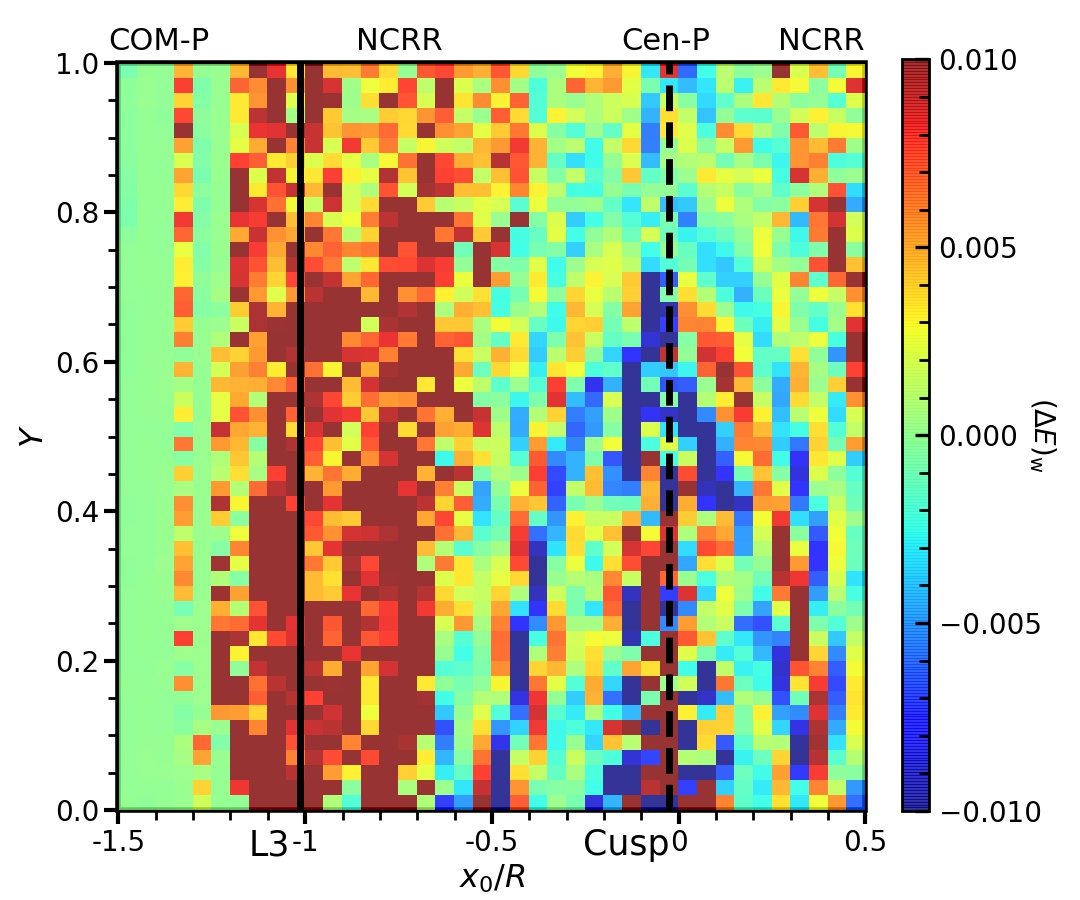}}
  \hspace{2mm}
  \subfloat{\includegraphics[width=0.463\textwidth]{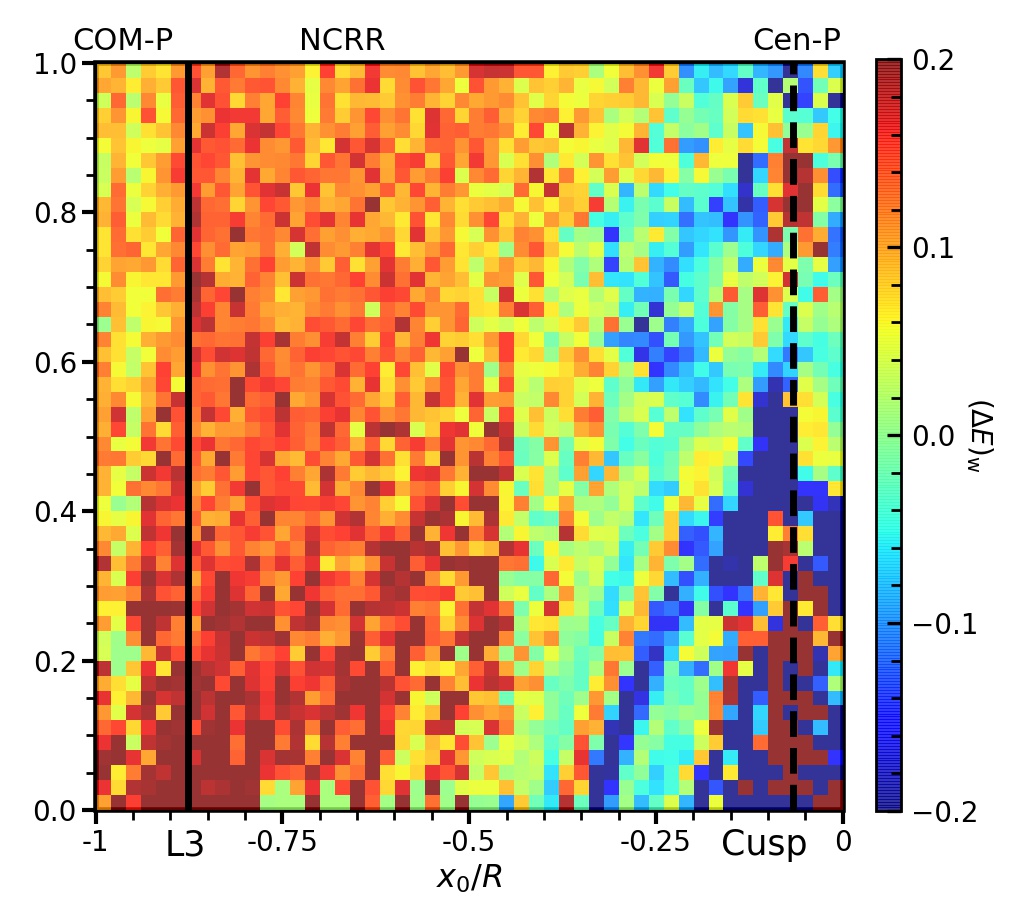}}
  \caption{Same as \ref{fig:delE} but for the cuspy Hernquist potential. Note that $\dE$ is predominantly positive, indicative of a negative (retarding) torque on the perturber. See text for discussion.}
  \label{fig:delE_Hern}
\end{figure*}

Left- and right-hand panels correspond to $R=0.5$ and $R=0.2$, respectively. They show the results for a total of $4 \times 2,500$ different orbits. Redder colors denote more positive values of $\dE$ (i.e., stronger dynamical friction), while bluer colors indicate more negative values (i.e., more pronounced dynamical buoyancy). Note that for $R=0.5$, i.e., when the perturber is outside the core, $\dE$ is predominantly positive, indicating that nearly all the NCRR orbits (\horseshoens, \pacmans and some tadpoles, with $x_0$ on either side of L3 and L1) exert a retarding torque (i.e., dynamical friction). However, when $R=0.2$ and the perturber is orbiting inside the core, almost the entire orbital parameter space (dominated by the NCRR \pacmans and tadpoles) contributes to dynamical buoyancy (i.e., $\dE < 0$). Clearly, there is a profound transition in the total torque once the perturber enters the core.

When the perturber is outside the core (left-hand panel), the contribution from the center-phylic orbits, which occupy the range of $x_0$ on either side of the galactic center (L0, marked by the vertical, dashed line) is completely negligible. The same holds for the COM-phylic orbits near the left-most edge of the plot. When the perturber is inside the core (right-hand panel), one again sees that orbits with starting positions close to L0 contribute a negligible torque. Unlike in the left-hand panel, though, these are not center-phylic orbits. After all, those vanish when the perturber crosses the bifurcation radius. Rather, these are predominantly \pacman and tadpole orbits, but unlike their counterparts with starting positions a bit further away from the (unstable) L0, they happen to exert negligible torque. Note that some of the COM-phylic orbits with $x_0/R \lta -0.75$ also contribute a (positive) torque. Their net contribution, though, is significantly smaller than that from the NCRR \pacman orbits, and rapidly weakens when $x_0/R$ becomes smaller (i.e., further away from the galactic center).

Fig.~\ref{fig:delE_Hern} is the same as Fig.~\ref{fig:delE}, but for our Hernquist galaxy. For both $R=0.5$ (left-hand panel) and $R=0.2$ (right-hand panel), it is clear that the total torque is negative (retarding) and dominated by the NCRR orbits. Most importantly, there is no transition in the sign of the total torque as one approaches the center, consistent with the notion that buoyancy and core-stalling are absent if the central density profile is cuspy.
Another difference with respect to the Plummer sphere is that while there is no significant contribution to the torque from the COM-phylic orbits, neither for $R=0.5$, nor for $R=0.2$, the  center-phylic orbits now make a significant contribution to the total torque. Although each of these orbits has a very small $\Delta E(T_{\rm orb})$, the steepness of the distribution function towards the galactic center means that they are abundant, thus receiving a large weight. When the perturber is at $R=0.5$, there are roughly equal numbers of center-phylic orbits with positive and negative $\dE$ (note the alternating red and blue stripes on either side of the galactic center). As a consequence, the net torque contribution from the entire population of center-phylic orbits is small.

Finally, we emphasize that the above inventory of the torque from individual orbits is incomplete. First of all, we have restricted the range of $x_0$ such that it does not include any perturber-phylic orbits. The reason is that they are difficult to integrate, while their contribution to the torque is negligible for reasons discussed in Section~\ref{sec:orbfam}. Secondly, by only picking starting points along the $x$-axis, we have selected against tadpole orbits with large $E_\rmJ$, which are typically confined to small regions centered on L4 or L5. We have examined several of such orbits and found their behavior to be very similar to that of the \horseshoe and \pacman orbits in terms of their contribution to the torque. Thirdly, we have restricted the $E_\rmJ$ values of the orbits up to $E_\rmJ^{(4)}=E_\rmJ^{(5)}$. This is because orbits with $E_\rmJ\gg E_\rmJ^{(4)}$ are far from co-rotation resonance (with drift time steeply falling with increasing $E_\rmJ$) and consequently less important for dynamical friction. However, orbits with small, positive values of $E_\rmJ-E^{(4)}_\rmJ$ have similar contribution to the torque as those with $E_\rmJ\lesssim E^{(4)}_\rmJ$. Thus we use $E^{(4)}_\rmJ$ only as an approximate cut-off for the NCRR orbits. Finally, and most significantly, we have only considered orbits of field particles confined to the orbital plane of the perturber, i.e., those with $z=0$ and $v_\rmz=0$. We presume that this doesn't significantly impact any of our conclusions regarding the contributions of the NCRR \horseshoe and \pacman orbits, as the third dimension merely allows for an additional vertical oscillation not accounted for in our 2D planar treatment (in particular, no new orbital families are introduced by allowing motion in the $z$-direction since there exist no Lagrange points off the orbital plane). However, the relative contributions of the different NCRR orbits to the total torque may be significantly different from what emerges from the 2D analysis presented here. In particular, the tadpole orbits would dominate the phase-space and therefore might contribute more significantly to the overall torque in 3D. This is a caveat of our approach that we leave for future work.

\begin{figure*}[t!]
  \centering  \includegraphics[width=1\textwidth]{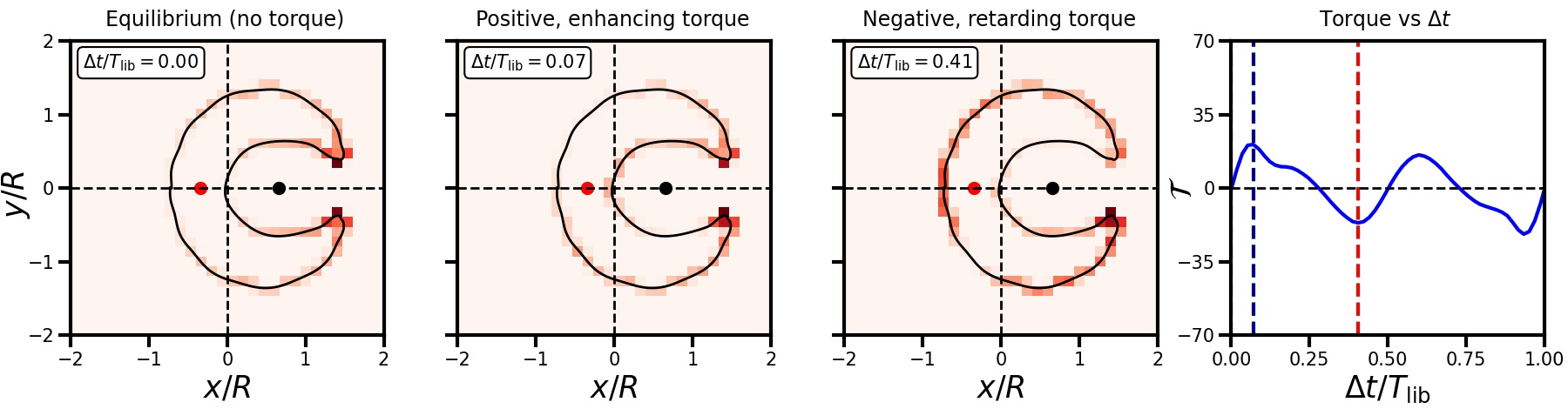}
  \caption{\small Same as Fig.~\ref{fig:wake}, but for a \pacman orbit when the perturber is inside of the core region ($R=0.2$). The first panel from the left shows the unperturbed phase distribution that exerts no torque. In the second panel (corresponding to $\Delta t$ marked by the blue dashed line in the rightmost panel showing the time evolution of the torque) one can note overdensities along the orbit in quadrants I and  III that are responsible for a positive, enhancing torque on the perturber + galactic center. In the third panel (corresponding to $\Delta t$ marked by the red dashed line in the rightmost panel) similar overdensities can be noted in quadrants II and IV, resulting in a negative, retarding torque. Note that the initial torque from this orbit is positive/enhancing, indicating that it will contribute to dynamical buoyancy on the perturber.}
  \label{fig:dynamical_buoyancy}
\end{figure*}

\subsection{Dynamical buoyancy and core-stalling} 
\label{sec:core}

When the perturber approaches the core region, a bifurcation of some of the Lagrange points causes a drastic change in the orbital structure. As we discuss in detail in section~\ref{sec:bifurcation}, the L3, L0 and L1 points undergo bifurcation at a certain radius $\Rbif$ ($\approx 0.39$ for our fiducial Plummer galaxy plus $q=0.004$ perturber), in which L1 and L3 are annihilated and L0 changes its stability from a center to a saddle. This is associated with the disappearance of the NCRR \horseshoe orbits. The torque from the remaining NCRR \pacman orbits changes from being retarding and contributing to dynamical friction, to being enhancing and contributing to dynamical buoyancy. In this section we discuss why it is that the \pacman orbits (and to some extent also the tadpole orbits) suddenly change the sign of their torque.

When the perturber is well beyond the core radius, the \pacman and tadpole orbits drain energy and angular momentum from the perturber in the same way as the \horseshoe orbits. As described in Section~\ref{sec:concept}, due to the large, radial gradient in the density profile outside of the core, the number density of field particles along the inner section (part of the orbit inside the perturber's radius) of these orbits, which is closer to the galactic center, is larger than that along the outer section (part of the orbit outside the perturber's radius). Due to the clockwise drift motion (in the co-rotating frame), the overdensity along the inner section shifts to the region behind the perturber, creating a `wake' that exerts a retarding torque. This in turn causes the perturber to experience dynamical friction, and thus to move radially inwards (see Fig.~\ref{fig:wake}).

When the perturber is inside the core radius, this picture changes profoundly. The unperturbed galaxy density profile is now very shallow and therefore there no longer is a sharp density contrast of field particles between the inner and outer sections. The equilibrium distribution of particles along the orbit is now dominated by the Jacobian $\sqrt{{\left|{\dot{\br}}_{\mathrm{\bi}}\right|}^2+{\left|{\ddot{\br}}_{\mathrm{\bi}}\right|}^2}$ rather than by the unperturbed distribution function $f_0(E_{0\rmG})$ (cf., equation~[\ref{deltaE}]). And since the particles speed up while approaching the perturber and slow down while receding from it, an overcrowding of particles develops around the inter-section junctions above and below the perturber, as shown in the leftmost panel of Fig.~\ref{fig:dynamical_buoyancy}. As the particles drift along the orbit in a clockwise direction, the overdensity ahead of the perturber approaches it and spreads over the inner section while that behind the perturber moves further away onto the outer section (see the second panel from the left in Fig.~\ref{fig:dynamical_buoyancy}). Hence, contrary to the \horseshoe orbit shown in Fig.~\ref{fig:wake}, here an overdensity of particles first forms {\it ahead} of the perturber, exerting a positive, enhancing torque (marked by the blue dashed line in the rightmost panel that shows the time evolution of the torque) which implies that this orbit will give rise to dynamical buoyancy.

When the perturber is inside the core, some of the tadpole orbits exhibit a similar behavior as the \pacman orbits, thereby contributing to dynamical buoyancy due to orbital-phase-crowding that gives rise to an enhancing torque. The perturber ultimately stalls at a radius where the buoyancy from these orbits is balanced out by friction from the others. As we show in the next section, core-stalling occurs near the bifurcation radius.

\section{Bifurcation of Lagrange points} 
\label{sec:bifurcation}

The relative abundance of the orbits is governed by the configuration of the equipotential surfaces of the effective potential (see Fig.~\ref{fig:Phi_eff}), which in turn strongly depends on the topology of the Lagrange points. The L0, L4 and L5 Lagrange points are `centers', which are stable under small perturbations along any direction, while L1, L2 and L3 are `saddles', which are unstable under small perturbations along certain directions.

\begin{figure*}
\captionsetup[subfigure]{labelformat=empty}
\hspace{-30pt}\subfloat[\hspace{30pt}(a) Plummer sphere]{\includegraphics[width=0.79\textwidth]{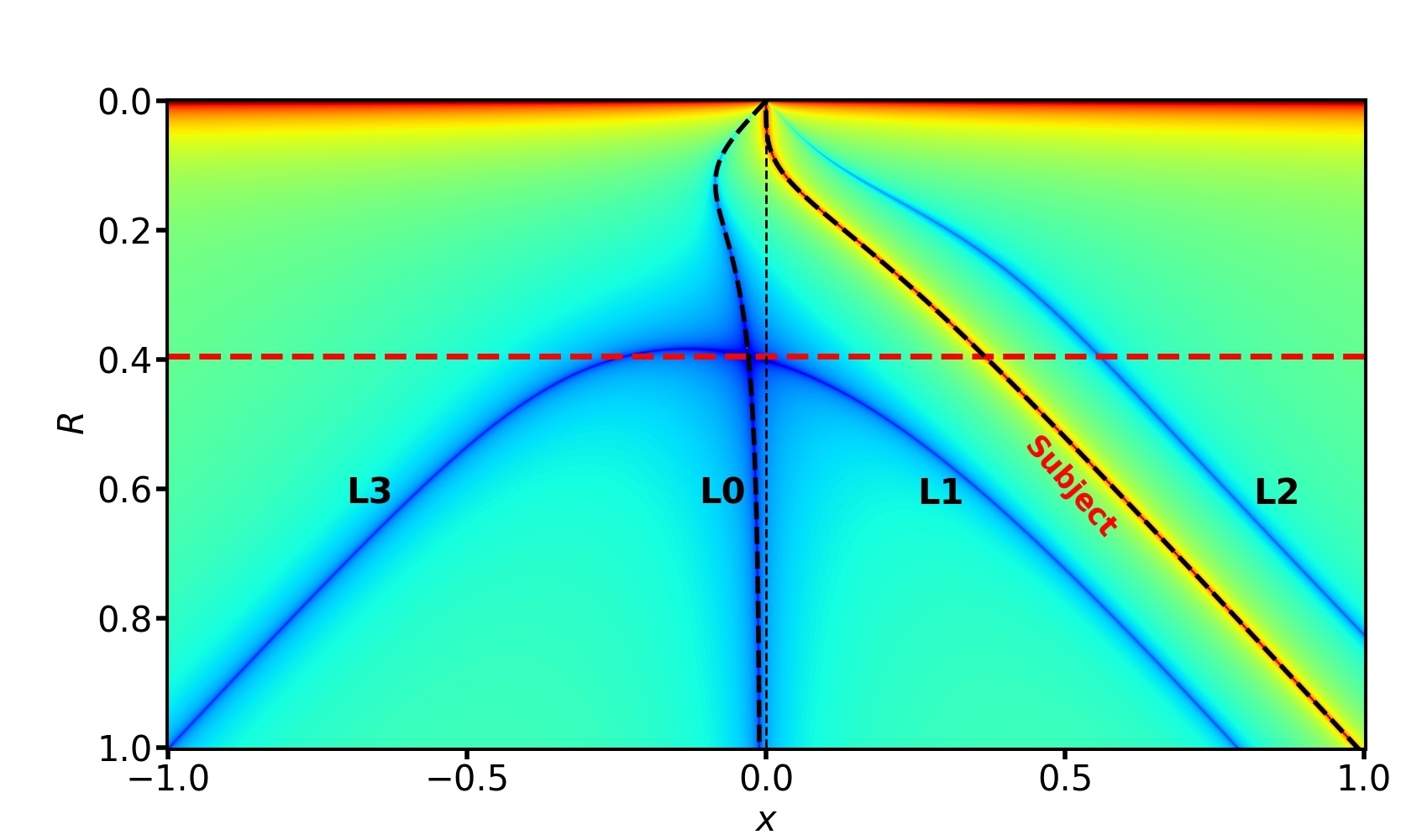}\label{core_bif_diag}}\\
\centering
\subfloat[\hspace{-40pt}(b) Hernquist sphere]{\includegraphics[width=0.95\textwidth]{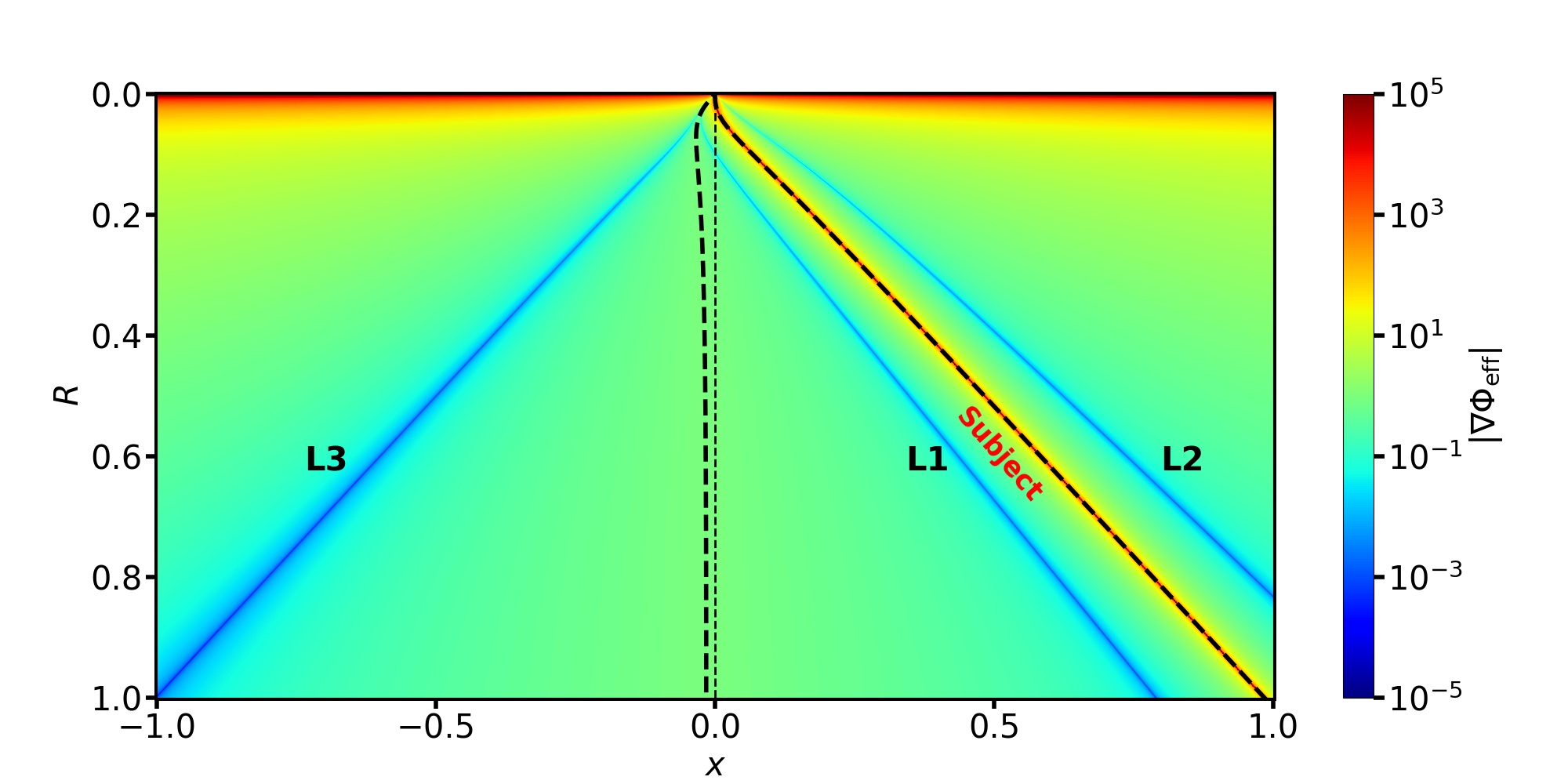}\label{cusp_bif_diag}}
\caption{Bifurcation diagram showing the position of the Lagrange points (left panel: L3, L0, L1 and L2 from the left for Plummer sphere, right panel: L3, L1 and L2 for Hernquist sphere) as a function of the galacto-centric radius $R$, color-mapped in terms of $\left|\nabla \Phi_{\rm eff}\right|$ from red (high) to blue (low). The blue lines mark the positions of the fixed points ($\nabla \Phi_{\rm eff}=0$) as a function of R. The black dashed lines mark the positions of the galactic center (which coincides with L0 for Plummer sphere) and the subject. The horizontal brown and yellow dashed lines in the left panel respectively indicate our estimate for the stalling radius and the estimate for the same from KS18. Note that in the left panel (Plummer sphere), L0, L1 and L3 merge together in a pitch-fork bifurcation and in the process L0 changes its stability from center to saddle. Under close examination, this bifurcation can be found to be a combination of two separate ones that happen right after one another- 1. trans-critical bifurcation between L0 and L1 where they exchange their stability, and 2. saddle-node bifurcation between L1 (center) and L3 (saddle). However in the right panel (Hernquist sphere), no such bifurcation happens and L3, L1 and L2 maintain their stability and relative positions throughout the in-fall history of the subject.}
\label{fig:bifurcation}
\end{figure*}

As $R$ changes, the location, stability and even the existence of these Lagrange points changes. This is shown in the heat map of Fig.~\ref{fig:bifurcation}, which shows the gradient of the effective potential (color shading) along the $x$-axis ($y=0$) as a function of $x$  and $R$. Let us first focus on the left-hand panel, which shows the results for a Plummer sphere (see Section~\ref{sec:models}). The locations of the Lagrange points (L0, L1, L2 and L3) correspond to blue colors, as indicated, while that of the perturber is indicated by the black dashed line on top of the yellow shading. For large $R$, the system has 4 Lagrange points along the $x$-axis: L0, L1, L2, and L3, with the perturber located roughly midway between L1 and L2. As the perturber approaches the center, the Lagrange points start to converge. When $R \sim 0.4$, L0, L1 and L3 `merge' together to form a single new Lagrange point. This is an example of bifurcation, which is a property of non-linear dynamical systems with multiple fixed points in which the smooth change of a parameter can cause a sudden qualitative or topological change in the dynamical nature of the system. In this case, a smooth change in $R$ causes what appears to be a pitch-fork bifurcation in which three fixed points (two saddles and a center) merge together to form a single fixed point (saddle). In fact, upon closer examination, it is apparent that the bifurcation happens in two stages. First L0 and L1 cross-over in a transcritical bifurcation, exchanging stability along the way (i.e., after the bifurcation, L1, which is now located in between L3 and L0, is stable, whereas L0, which still coincides with the center of the galaxy has become an unstable saddle. At a slightly smaller $R$, L3 and L1 annihilate each other in what is called a saddle-center bifurcation (also known as a tangential bifurcation). At $R \lta 0.39$ the only surviving Lagrange points along the $x$-axis are the saddle points L0 and L2. L0 is now a saddle that connects the stable regions surrounding L4 and L5. Hence, it has some of the characteristic hall-marks of L3 prior to bifurcation. However, from the bifurcation trajectories and the fact that it is located at the center of the galaxy, it is clear that this fixed point is L0.

For a cuspy galaxy, e.g., the Hernquist galaxy shown in the right-hand panel of Fig.~\ref{fig:bifurcation}, the central cusp always separates the L1 and L3 saddle points, preventing the occurrence of a bifurcation.

The orbital radius, $R_{\rm bif}$, at which bifurcation happens, depends on the mass-ratio $q$. In fact, $R_{\rm bif}$ increases with $q$, i.e., bifurcation occurs farther out for more massive perturbers. In the following section, we provide a generic bifurcation criterion (obtained using a stability analysis of Lagrange points given in Appendix~\ref{App:stability}) that provides an approximate power law scaling of $R_{\rm bif}$ with $q$, where the exponent depends on the specific galaxy density profile.

\subsection{Bifurcation condition}

As discussed in Appendix~\ref{App:stability}, the general bifurcation condition is characterized by a vanishing eigenfrequency $\omega$, and is given by
\begin{equation}
\omega^2=0 \;\;\;\;\; \Rightarrow \;\;\;\;\; \Phi_{\rm eff,xx} \Phi_{\rm eff,yy} = \Phi_{\rm eff,xy}^2\,,
\label{genbifcon}
\end{equation}
where $\Phi_{\rm eff,ij} = \partial^2\Phi_{\rm eff}/\partial x_i\partial x_j$ at the point of bifurcation. Since the bifurcation of interest involves Lagrange points that all lie on the $x$-axis, this reduces to $\Phi_{\rm eff,xx} \Phi_{\rm eff,yy} = 0$. Furthermore, for our Plummer galaxy, $\Phi_{\rm eff,yy} = 1 - (1+R^2)^{-3/2} > 0$ at L0, where $x = -q_\rmG R$ (and thus $r_\rmG=0$ and $r_\rmP=R$). Therefore the bifurcation condition reduces to 
\begin{equation}
\Phi_{\rm eff,xx}=0. 
\end{equation}
This along with $\Phi_{\rm eff,x}=0$ (since we are looking at fixed points) amounts to solving the following equations for $R$ and $x$:
\begin{align}
&\frac{\partial^2 \Phi_\rmG}{\partial r^2_\rmG}-\frac{2q}{r_\rmP^3}-\frac{1+q_{\rm enc}(R)}{R}\frac{\partial \Phi_\rmG}{\partial R}=0\,, \nonumber \\
&\frac{\partial \Phi_\rmG}{\partial r_\rmG}-\frac{q}{r_\rmP^2}-\left(\frac{1+q_{\rm enc} (R)}{R}\frac{\partial \Phi_\rmG}{\partial R}\right)x=0
\label{bifcondition1}
\end{align}
Since the bifurcation involves L0, $x=-q_\rmG R$ is a solution to the second of these equations. Substituting $x=-q_\rmG R$ therefore in the first equation yields the following bifurcation condition for $R$:
\begin{equation}\label{Rbif}
\frac{1}{R}\frac{\partial \Phi_\rmG}{\partial R} = \left.\frac{\partial^2 \Phi_\rmG}{\partial r^2_\rmG}\right\vert_{r_\rmG=0} - \frac{3q}{R^3}\,.
\end{equation}
For the Plummer sphere (equation~[\ref{plummer}]), this reduces to
\begin{equation}
\left(1+R^2\right)^{-3/2} = 1 - \frac{3q}{R^3}\,.
\label{bifcondition2}
\end{equation}
Bifurcation typically happens for $R<1$ and so $R^2$ is sufficiently smaller than $1$. Therefore, with the approximation that $R^2 \ll 1$, we can binomially expand the LHS of equation~(\ref{bifcondition2}) to obtain the following expression for the bifurcation radius
\begin{equation}\label{bifrad}
R_{\rm bif} \approx (2q)^{1/5}.
\end{equation}
For our fiducial value of $q=0.004$ this yields $R_{\rm bif} \simeq 0.38$, in excellent agreement with the bifurcation radius inferred from the heat-map in Fig.~\ref{core_bif_diag} (horizontal, dashed line). Bifurcation happens when the mass of the subject is of order the enclosed mass of the galaxy within $R$ and is associated with the disappearance of the Roche-lobe centred on the galaxy. Post-bifurcation, only four of the six Lagrange points remain (L0, L2, L4 and L5), which causes a sudden change in the equipotential contours, and hence the orbital make-up of the system. 

In a cuspy profile ($\gamma \leq -1$), though, the situation is very different. First of all, as already mentioned in Section~\ref{sec:lagrangepoints}, there is no L0 Lagrange point. In addition, bifurcation does not happen and the inner saddle point L1 survives throughout the in-fall history of the perturber.

\begin{figure*}

\subfloat[Burkert profile]{\includegraphics[width=0.33\textwidth]{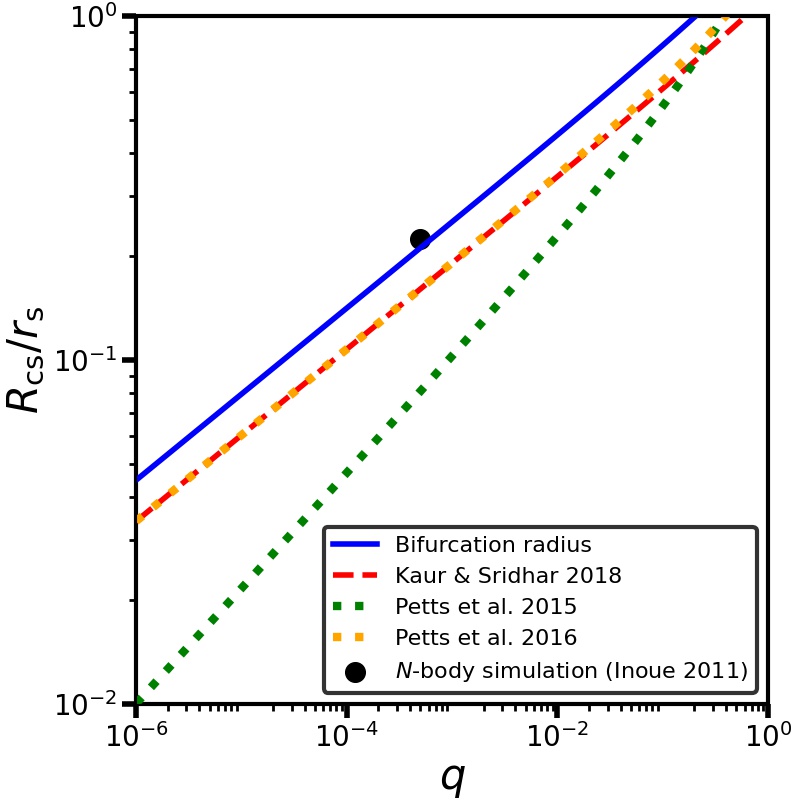}\label{bifurcation_b}}
\subfloat[Henon's isochrone profile]{\includegraphics[width=0.33\textwidth]{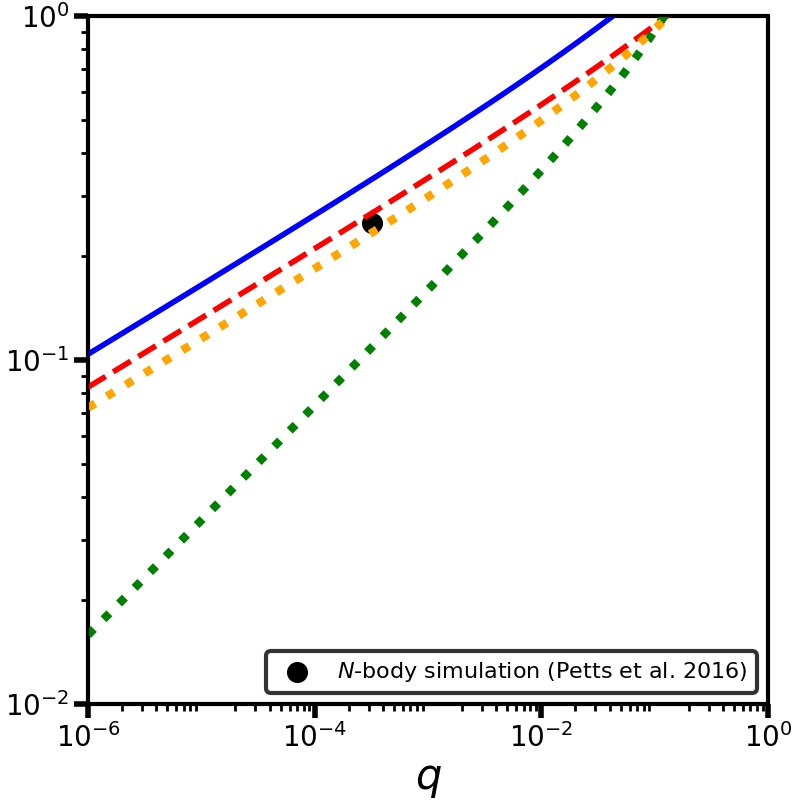}\label{bifurcation_i}}
\subfloat[Dehnen sphere ($\gamma=0$)]{\includegraphics[width=0.33\textwidth]{figures/stalling_radius_burkert.jpg}\label{bifurcation_d}}

\caption{Estimates of the core-stalling radius, $R_{\rm cs}$, in units of the galaxy's scale radius $r_\rms$, as a function of the mass ratio $q = M_\rmp/M_\rmG$,  for a cored Burkert profile (left-hand panel), Henon's isochrone profile (middle panel), and a cored ($\gamma=0$) Dehnen profile (right-hand panel). The solid blue lines correspond to the bifurcation radius (the root of equation~[\ref{bifcondition2}]), advocated in this chapter as a rough estimate of the core-stalling radius, and is compared to other estimates in the literature: KS18 (dashed red-lines), \citet[dotted orange]{Petts.etal.16}, and \citet[dotted green]{Petts.etal.15}. Solid, black dots denote the results from $N$-body simulations, as indicated.}
\label{fig:stalling_radius}
\end{figure*}

\subsection{Bifurcation radius as core stalling radius}
\label{sec:rcs}

Core-stalling is the cessation of dynamical friction driven inspiral of a massive perturber in a cored galaxy. We inferred in BB21 and section~\ref{sec:resonances} that core-stalling occurs due to a subtle balance between friction and buoyancy. We shall now discuss how this transition from friction to buoyancy and the associated stalling of the perturber's in-fall relates to the bifurcation of the Lagrange points in a cored galaxy. 

When the perturber is outside the bifurcation radius $R_{\rm bif}$, the near-resonant orbits comprise of the NCRR \horseshoens, \pacmans and tadpoles that librate about the co-rotation resonances and are the dominant drivers of dynamical friction. As the perturber penetrates deeper into the core, the Lagrange points L0, L1 and L3 undergo bifurcation. As a result, several interesting things happen. Firstly, the phase-space available for the \horseshoe orbits becomes vanishingly small, which is synonymous to the suppression of the lower order co-rotation resonances in the core as pointed out by KS18. Secondly, the eigenfrequency of perturbations about L0, i.e., the libration frequency, $\Omega_{\rm lib}$, of the \pacman orbits goes to zero at bifurcation, which is why it takes infinitely long for the response density to develop from these slowly librating NCRR orbits, thereby weakening the dynamical friction torque and causing the perturber to stall. This stagnation of the orbital dynamics near bifurcation is like a `bottleneck effect'. Thirdly, within the bifurcation radius, the torque from the remaining NCRR \pacman orbits becomes enhancing (positive) rather than retarding (negative), as shown in section~\ref{sec:resonances}. This gives rise to dynamical buoyancy, which counteracts friction from the other NCRR (tadpole) orbits, causing the perturber to stall where friction and buoyancy balance each other.

In a cuspy galaxy ($\gamma \leq -1$), the force is infinite at the cusp, i.e., the L0 Lagrange point is absent. The L1 saddle point always separates the Roche lobe of the galaxy from that of the subject. Therefore bifurcation does not occur, ensuring the ongoing presence of the Roche lobe centred on the galaxy and the NCRR \horseshoe orbits. Moreover, the \pacman orbits which are the dominant contributors to dynamical buoyancy in the core region, are absent in the cusp. The steep gradient in the distribution function ensures that the NCRR \horseshoe and tadpole orbits continue to drain energy from the subject, causing dynamical friction. 

Dynamical friction in cored galaxies has been explored in detail by the analytical studies of \citet[]{Petts.etal.15,Petts.etal.16,Kaur.Sridhar.18} and so on. Our explanation for core-stalling based on the bifurcation of the Lagrange points is radically different from any of the previous studies. In order to make the comparison of the different explanations for core-stalling more quantitative, we now compare the predicted core-stalling radii, $R_{\rm cs}$, to existing simulation results. In the interpretation of core-stalling advocated in this chapter, it is a consequence of two effects: (i) slowing down of the libration of the NCRR \pacman orbits near bifurcation (bottleneck effect), and (ii) a balancing act between buoyancy, which dominates within the bifurcation radius, and friction, which dominates further out. We therefore associate the onset of buoyancy with the bifurcation of the L0, L1, and L3 Lagrange points. Hence, we postulate that core-stalling happens close to the bifurcation radius, $R_{\rm bif}$, introduced in Section~\ref{sec:bifurcation}. The solid blue line in Fig.~\ref{fig:stalling_radius} plots $R_{\rm bif}$ as a function of the mass ratio $q = M_\rmp/M_\rmG$ for three different cored galaxy profiles: \cite{Burkert.95} profile (left-hand panel), \cite{Henon.59} isochrone profile (middle panel), and a $\gamma=0$ \cite{Dehnen.93} sphere (right-hand panel). For comparison, we also plot the $R_{\rm cs}$ predictions for the models of \cite{Petts.etal.15},  \cite{Petts.etal.16} and KS18, as indicated. Note that the latter two are almost indistinguishable in most cases. The solid dot in the left-hand panel marks the core-stalling radius measured by \cite{Inoue.11} in his high-resolution simulation of dynamical friction in a spherical galaxy with a Burkert profile. In the middle and right-hand panels, the solid dots indicate the simulation results of \cite{Petts.etal.16} for spherical galaxies with an isochrone and a $\gamma=0$ Dehnen profile, respectively. Note that the \cite{Petts.etal.15} estimate of the stalling radius falls well below all the other estimates (including ours) as well as the simulation results, for a wide range of $q$. This suggests that in most cases core-stalling happens well before the subject mass equals the enclosed galaxy mass, which is the criterion for core-stalling advocated by \cite{Petts.etal.15}. As is evident, the bifurcation radius scales with the mass ratio $q$ in exactly the same way as the core-stalling radius suggested by KS18, and the tidal-stalling radius advocated by \cite{Petts.etal.16}. The scaling of $R_{\rm bif}$ with $q$ is given by
\begin{align}
R_{\rm bif} \approx 
    \begin{cases}
    q^{1/4}, & \textnormal{$\gamma=0$ Dehnen sphere}\\
    {\left(4q\right)}^{1/4}, & \textnormal{Burkert profile}\\
    {\left(12q\right)}^{1/5}, & \textnormal{Henon's isochrone}.
    \end{cases}
\end{align}
Here $q$ is equal to $M_\rmP/M_\rmG$ for the Dehnen sphere and Isochrone profiles, but $M_\rmP/M_\rmc$ for the Burkert profile, where $M_\rmc=4\pi \rho_\rmG(0)r^3_\rms/3$ is defined as the core mass (the total mass for the Burkert profile is infinite). Note that $R_{\rm bif}$ is slightly larger, in better agreement with the simulation results of \cite{Inoue.11} for the Burkert profile, while being slightly higher than the stalling radii found in the simulation results of \citep{Petts.etal.16} for Henon's isochrone and the cored Dehnen sphere. Although more simulations, spanning a wider range in $q$ and galaxy profile, are needed before any meaningful conclusions can be drawn from such a comparison, at the very least, this comparison is in good agreement with our suggested explanation for core-stalling.

\section{Conclusion}
\label{sec:concl_5}

Numerical simulations have shown that dynamical friction becomes inefficient inside constant density cores, causing the inward motion of a massive object (the perturber) to stall near the core radius. Objects placed inside this stalling radius are furthermore found to experience dynamical buoyancy that pushes them out towards the stalling radius. This phenomenology is neither predicted by Chandrasekhar's treatment of dynamical friction, nor by the more sophisticated linear, perturbative treatments of TW84 and KS18. The latter infer that dynamical friction arises from the LBK torque due to purely resonant orbits that is exclusively retarding. In BB21, we demonstrated that the LBK torque provides an incomplete description of dynamical friction, which is especially acute in cored galaxies. In particular, we derived an expression for the `self-consistent torque', which includes a memory term that depends on the entire in-fall history of the perturber. As the perturber approaches a core, this memory term causes the net torque to flip sign, i.e., become enhancing, inside a critical radius, $R_{\rm crit}$. Although this formalism thus seems to offer a natural explanation for core stalling, in terms of a balance between friction and buoyancy, it is still based on linear perturbation theory which, as discussed in Section~\ref{sec:intro_5}, is not justified in the core region of a galaxy, where the perturber can no longer be treated as a weak perturbation, or whenever the perturber does not sweep through the resonances fast enough to prevent non-linearities from building up, i.e., when core-stalling takes effect. 

In this chapter, in an attempt to overcome these conceptual problems, we have examined dynamical friction using an alternative, non-perturbative, orbit-based approach. This paints a view of dynamical friction that is subtly different from the standard resonant picture developed in TW84 and KS18. Interactions between the perturber and field particles cause the frequencies (and actions) of the field particles to evolve with time, to the extent that one can no longer talk about particles that obey a commensurability condition (i.e., are in resonance with the perturber) throughout their orbital evolution; rather they are trapped/librating about resonances. As such, dynamical friction does not arise from resonances per se, but rather from an imbalance between the number of particles that are `up-scattered' in angular momentum (or energy) versus those that are `down-scattered' along {\it near-co-rotation-resonant orbits}. This imbalance owes its origin to a non-zero gradient in the distribution function.

We have investigated the inner workings of dynamical friction, on an orbit-by-orbit basis, for the case of a point mass perturber on a circular orbit at radius $R$ in a spherical host galaxy. By assuming that $|\rmd R/\rmd t|$ of the perturber is small, i.e., we are in the `slow' regime, appropriate for studying core stalling, the motion of the field particles can be treated as a restricted three-body problem in which the Jacobi energy of the field particles is conserved.  Each individual orbit in the fully perturbed, time-dependent potential of the galaxy$+$perturber system has phases at which the orbital angular momentum and energy increase (the perturber experiences friction) and decrease (the perturber experiences buoyancy). Since the Jacobi energy is conserved, the net effect of these energy changes, when integrated over a full libration period in the frame co-rotating with the perturber, is nullified. For dynamical friction to emerge, then, two conditions need to be satisfied: (i) there need to be orbits with a non-uniform phase distribution along the orbit for which the time-lag between the two phases corresponding to retarding (friction) and enhancing (buoyancy) torques is sufficiently long, and (ii) there needs to be sufficient phase-coherence among different orbits. In that case, if all these phase-coherent orbits first exert friction on the perturber, the latter can sink in, modifying its frequency significantly, before the orbits would enter their buoyancy-exerting phase.

We have shown that dynamical friction is dominated by orbits that have (unperturbed) azimuthal frequencies similar to the circular frequency of the perturber. These near-co-rotation-resonant (NCRR) orbits all have a long libration time in the frame co-rotating with the perturber, assuring a long time-lag between the orbit's contribution to a retarding torque (friction) and an enhancing torque (buoyancy). And since all NCRR orbits have the same sense of rotation in the co-rotating frame, phase-coherence is guaranteed. Other orbits, such as the center- or perturber-phylic ones have a time-lag between the friction and buoyancy phases that is too short for a net, coherent torque to emerge. In other cases, especially the COM-phylic orbits, the phase-density along the orbit is almost uniform, such that again no net torque arises.

We have identified three different families of NCRR orbits that dominate the contribution to dynamical friction: \horseshoe orbits that circulate the Lagrange point L3, tadpole orbits that librate around either L4 or L5, and \pacman orbits which circulate the galactic center and pass through the region between the equipotential contours corresponding to L1 and L2. \Horseshoe and tadpole orbits are relatively well-known in planetary dynamics \citep[e.g.,][]{Dermott.Murray.81, Goldreich.Tremaine.82}. For example, objects on tadpole orbits are known as trojans, which includes large swarms of trojan asteroids associated with Jupiter, as well as several trojan moons in the Saturn system. There are several asteroids known to be on \horseshoe orbits in the Earth-Sun system \citep[][]{Connors.etal.02, Brasser.etal.04, Christou.Asher.11}, while the Saturnian moons Janus and Epimetheus are known to be horse-shoeing each other \citep[][]{Dermott.Murray.81b}. The \horseshoe and tadpole orbits are also key players in galactic dynamics, where they are often referred to as `trapped' orbits. They play a key role in phenomena such as the radial migration of stars in disk galaxies induced by perturbations due to a bar or spiral arm \citep[e.g.,][]{Barbanis.76, Carlberg.Sellwood.85, Sellwood.Binney.02, Daniel.Wyse.15}. However, to our knowledge, the orbits that we have called \pacman orbits, because of their characteristic shape, have hitherto not been identified as a separate orbital class. Pacman orbits are only present if the galaxy has a cored density profile ($\rmd\log\rho/\rmd\log r>-1$), in which case, the center of the galaxy is a stationary Lagrange point, which we have dubbed L0.

In a cusp, \pacman orbits are absent, and dynamical friction is mainly caused by field particles moving along the NCRR \horseshoe and tadpole orbits. Due to a large, negative gradient in the distribution function, the vast majority of these orbits yield a net retarding torque, draining orbital energy and angular momentum from the perturber.

In a cored profile, the behavior is very different. Well outside the core, where the density gradient is steep, the NCRR \horseshoen, tadpole and \pacman orbits exert a retarding torque, just as in the case of a cuspy density profile.  However, as the perturber enters the core region, the orbital configuration changes drastically. First of all, as we show in section~\ref{sec:bifurcation}, the  Lagrange points L3, L1 and L0 undergo a bifurcation in which L3 and L1 are annihilated, while L0 changes from a stable center to an unstable saddle. As a consequence, the \horseshoe orbits disappear, making the tadpoles and \pacmans the only surviving NCRR orbits. This disappearance of the \horseshoe orbits is equivalent to the suppression of low-order resonances in the core region, advocated by KS18 as the main cause of core stalling. However, we have demonstrated that a large number of NCRR orbits (\pacmans and tadpoles) remain, which continue to exchange energy and angular momentum with the perturber. The pre-eminent cause of core stalling, therefore, is not the disappearance of resonances, but the fact that most of the remaining \pacman and tadpole orbits now give rise to a net enhancing torque, thereby effectuating `dynamical buoyancy'. The main reason for this reversal in the sign of the net torque is the dramatic change in the radial gradient of the density distribution as described in Section~\ref{sec:core}. 

With dynamical buoyancy dominating over dynamical friction in the central region of a cored density profile, the perturber will ultimately settle at a core-stalling radius where the outward buoyant force balances friction. This notion of buoyancy counteracting friction in the core region is supported by numerical simulations \citep[][]{Inoue.11, Cole.etal.12, Petts.etal.16}, and provides a natural explanation for core stalling. In section~\ref{sec:bifurcation} we show that core-stalling happens close to a critical `bifurcation radius', $R_{\rm bif}$.

Dynamical buoyancy has a number of important astrophysical implications. It can prevent massive objects like black holes, globular clusters, and satellite galaxies from sinking all the way to the center of their host system, if the latter has a central constant density core. Hence, buoyancy acts as a natural barrier for, among others, the merging of supermassive black holes (SMBHs), with implications for the expected rates of such events to be detected by future gravitational wave detectors such as LISA \citep[e.g.,][]{Rhook.Wyithe.05, Tremmel.etal.18, Ricarte.Natarajan.18}, and for the creation of nuclear star clusters through the merging of globular clusters \citep[e.g.,][]{Tremaine.etal.75, ArcaSedda.CapuzzoDolcetta.17, Boldrini.etal.19}. Put differently, if their formation mechanism is merger driven, then the presence of central SMBHs and/or nuclear star clusters would favor cuspy density profiles for their hosts, which could help to constrain the particle nature of dark matter \citep[e.g.,][and references therein]{Brooks.14}.

However, many outstanding issues remain. For example, the analysis presented here has largely focused on orbits in 2D, and needs to be extended to 3D. It is also important to examine how friction and buoyancy act on perturbers along non-circular orbits and/or in non-spherical potentials, both of which are expected to result in a much richer dynamics \citep[e.g.,][]{Capuzzo-Dolcetta.Vicari.05}. In our analysis we also neglected the radial motion of the perturber due to friction/buoyancy itself. Although a reasonable approximation to make for studying dynamical friction in the `slow' regime, especially core-stalling, BB21 have shown that the memory effect of dynamical friction, i.e., the dependence on the perturber's past in-fall-history, can play an important role, something that warrants further investigation within the non-perturbative framework presented here. And finally, more work is needed to assess if and how core-stalling depends on the central, logarithmic slope, $\gamma$, of the host galaxy.  In this chapter we have focused exclusively on two special cases; a constant density core with $\gamma = 0$, and a steep NFW-like cusp with $\gamma = -1$. Numerical simulations suggest that core stalling might be present as long as $\gamma > -1$ \citep[][]{Goerdt.etal.10}. In future work we intend to examine dynamical friction and core stalling in host-galaxies with a variety of different central density slopes in the range $-1 \leq \gamma \leq 0$, using a combination of numerical simulations and the orbit-based formalism presented here.


    \begin{subappendices}


\chapter*{Appendix}

\section{Orbit classification}
\label{App:orb_class}

As shown in \cite{Daniel.Wyse.15}, the Jacobi energy of an NCRR orbit can be obtained by a perturbative expansion around the co-rotation L4/L5 points in terms of angular momentum and radial action, using the third order epicyclic theory of \cite{Contopoulos.75}. We extend this analysis to include orbits in the vicinity of not only L4 and L5, but also L0 and the perturber. About any such stable fixed point, which we shall refer to as the center of perturbation (COP) hereon, the Jacobi energy (defined wrt the galactic center) can be perturbatively expanded up to second order in actions as the following series
\citep[equation~A29 of][]{Contopoulos.75}:
\begin{align}
E'_\rmJ = h'_0 + \Delta\Omega_0 J_\varphi + \kappa_0 J_r + a_0 J^2_r + 2 b_0 J_r J_\varphi + c_0 J^2_\varphi + \Phi_1(r_\rmG,\varphi),
\label{Ej_perturb}
\end{align}
where $r_\rmG$ is the distance from the galactic center and $\varphi$ is the anticlockwise angle measured from the x-axis wrt the galactic center. $h'_0$ is the unperturbed Jacobi energy (wrt the galactic center) evaluated at the COP. The Jacobi energy, $E_\rmJ$, wrt the COM of the galaxy-perturber system is related to that wrt the galactic center, $E'_\rmJ$, as follows

\begin{align}
E_\rmJ = E'_\rmJ + \frac{1}{2}\Omega^2_\rmP q^2_\rmG R^2.
\end{align}
$J_r$ is the radial action (wrt the galactic center), $J_\varphi=L-L_0$ is the angular momentum relative to the COP (with $L=r^2_\rmG\dot{\varphi}$ and $L_0=r^2_{\rmG 0}\Omega_0$, where $r_{\rmG0}$ is the distance of the COP from the galactic center), and $\Delta \Omega_0=\Omega_0-\Omega_\rmP$, with $\Omega_0$ the azimuthal frequency evaluated at the COP. The radial epicyclic frequency, $\kappa_0$, and the constants $a_0$, $b_0$ and $c_0$ are evaluated at the COP in terms of the galaxy potential as follows
\begin{align}
\kappa^2_0 &= \Phi^{''}_\rmG+3\Omega^2_0,\nonumber \\
a_0 &= \frac{1}{16}\kappa^2_0\left[\Phi^{''''}_\rmG+\frac{60\Phi^{'}_\rmG}{r^3_0}-\frac{5}{3\kappa^2_0}{\left(\Phi^{'''}_\rmG-\frac{12\Phi^{'}_\rmG}{r^2_0}\right)}^2\right],\nonumber \\
b_0 &= \frac{\Omega_0\kappa^{'}_0}{r_0\kappa^2_0},\nonumber \\
c_0 &= \frac{\Omega_0\Omega^{'}_0}{r_0\kappa^2_0},
\end{align}
where each prime denotes a derivative with respect to $r_\rmG$. $\Phi_1$ is the disturbing potential that includes the perturber potential $\Phi_\rmP$ and the tidal potential due to the orbital motion of the galactic center about the COM, i.e.,
\begin{align}
\Phi_1(r_\rmG,\varphi) &= \Phi_\rmP + \frac{q\,r_\rmG\cos{\varphi}}{R^2} \nonumber\\
&= q\left[-\frac{1}{\sqrt{r^2_\rmG+R^2-2r_\rmG R\cos{\varphi}}}+\frac{r_\rmG\cos{\varphi}}{R^2}\right].
\label{Phip_tidal}
\end{align}

Following \citet{Daniel.Wyse.15}, equation~(\ref{Ej_perturb}) can be written in the following
quadratic form:
\begin{align}
A J^2_\varphi+B J_\varphi + C = 0,
\end{align}
where $A$, $B$ and $C$ are given by
\begin{align}
A &= c_0,\nonumber \\
B &= \Delta \Omega_0 + 2 b_0 J_r,\nonumber \\
C &= a_0 J^2_r + \kappa_0 J_r + h'_0 + \Phi_1(r,\varphi) - E'_\rmJ.
\end{align}
The particle can only venture into those regions of $(r_\rmG,\varphi)$ where $J_\varphi$ has real solutions, which occurs when
\begin{align}
B^2-4 A C \geq 0,
\end{align}
i.e.,
\begin{align}
E'_{\rm Jc}=E'_\rmJ - \frac{{\left(\Delta\Omega_0\right)}^2}{4\left|c_0\right|} - \left(\kappa_0+\frac{b_0}{\left|c_0\right|}\Delta \Omega_0\right)J_r - \left(a_0+\frac{b^2_0}{\left|c_0\right|}\right)J^2_r \leq h'_0+\Phi_1(r,\varphi).
\label{Ejcirc_app}
\end{align}
Here $E'_{\rm Jc}$ is the circular part of the Jacobi energy (we have used the fact that $c_0=-\left|c_0\right|<0$ for realistic galaxy profiles). The region accessible to the particle, i.e., the range of $\varphi$ for which $J_\varphi$ has real roots, depends on the value of $E'_{\rm Jc}$ relative to that of $\Phi_1$ at the separatrix, the zero-velocity curve (ZVC) passing through the saddle point $(r_{\rm sep},\varphi_{\rm sep})$ nearest to the COP. If the accessible region is `inside' (`outside') the separatrix, i.e., $E'_{\rm Jc}>h'_0+\Phi_1(r_{\rm sep},\varphi_{\rm sep})$ ($E'_{\rm Jc}<h'_0+\Phi_1(r_{\rm sep},\varphi_{\rm sep})$), then the orbit is said to be `trapped' (`untrapped'). However, since $J_r$ can oscillate along an orbit, especially for orbits that are  further away from co-rotation resonance (see \S\ref{sec:slowfast}), the trapping criterion is not guaranteed to be satisfied forever; the field particle can oscillate in and out of the trapped region by crossing the separatrix and transitioning between different orbit-families (see \S\ref{sec:crossing}).

\subsection{Perturbation about L4 and L5}

L4 and L5 are located at a distance $r_\rmG=R$ from the galactic center, and at an angle $\varphi=\pm\pi/3$. The nearest saddle point to L4/L5 is L3, at a distance of $r_\rmG=r_{\rm sep}=r_3$ from the galactic center, and with $\varphi=\varphi_{\rm sep}=\pi$. In the region centred on L4/L5 and bounded by the L3 separatrix, the disturbing potential $\Phi_1$, given by equation~(\ref{Phip_tidal}), is bounded by
\begin{align}
-q\left[\frac{1}{r_3 + R} + \frac{r_3}{R^2}\right] = \Phi_1(r_3,\pi) < \Phi_1 < \Phi_1(R,\pm\pi/3) = -\frac{q}{2 R}.
\end{align}
Therefore, in the vicinity of L4, L5 and L3, $J_\phi$ has real roots for $\varphi$ that are restricted to $0<\varphi<\pi$ (or $-\pi<\varphi<0$) when $h_0^{'(4)}+\Phi_1(r_3,\pi) < E_{\rm Jc}^{'(4)} < h_0^{'(4)} + \Phi_1(R,\pm\pi/3)$, i.e., when the `trapping criterion',
\begin{align}
h^{'(4)}_0-q\left[\frac{1}{r_3 + R}+\frac{r_3}{R^2}\right]=E^{'(3)}_\rmJ < E_{\rm Jc}^{'(4)} < E^{'(4)}_\rmJ=h^{'(4)}_0 - \frac{q}{2 R},
\end{align}
is satisfied. Here $E_{\rm Jc}^{'(4)}$ and $h_0^{'(4)}$ are, respectively, the circular part of the Jacobi energy given by equation~(\ref{Ejcirc_app}) and the unperturbed Jacobi energy, both evaluated at the COP, L4/L5. These `trapped' orbits that lie inside the L3 separatrix are the tadpole orbits. The `untrapped' orbits that lie beyond the L3 separatrix ($\varphi>\varphi_{\rm sep}=\pi$), i.e., that satisfy the condition,
\begin{align}
E_{\rm Jc}^{'(4)} < E^{'(3)}_\rmJ,
\end{align}
are the horse-shoe orbits. The tadpole orbits are trapped, librating around L4/L5, while the \horseshoe orbits are untrapped wrt L4/L5. However, as we shall see shortly, the \horseshoe orbits are trapped inside the L1/L2 separatrix and are therefore still librating about near-co-rotation resonances. Since $J_r$ can vary along an orbit, certain orbits with high $J_r$ can cross the L3 separatrix, and the trapped tadpoles and untrapped \horseshoes can metamorphose into each other, showing Chimera behavior (see Appendix~\ref{App:Chimera}).

\subsection{Perturbation about L0}

A perturbative analysis around L0 can only be performed for a perfect central core, i.e., inner log scope $\gamma \equiv \lim_{r \to 0}\rmd\log\rho/\rmd\log r=0$. For density profiles with $\gamma<0$, $\Omega_0$ and $\kappa_0$ diverge like $r^{\gamma/2}_\rmG$ towards the galactic center, around which the perturbative expansion of $E'_\rmJ$ given in equation~(\ref{Ej_perturb}) is thus not defined. 

For a cored galaxy, L0 is a stable Lagrange point located at the galactic centre, i.e., $r_\rmG=0$. The nearest saddle point to L0 is L1, located along the x-axis ($\varphi=\varphi_{\rm sep}=0$) at a distance $r_\rmG=r_{\rm sep}=r_1$ from the galactic center.  In the region centred on L0 and bounded by the L1-sparatrix, $\Phi_1$ is bounded by
\begin{align}
q\left[-\frac{1}{\left|r_1 - R\right|}+\frac{r_1}{R^2}\right] < \Phi_1 < q\left[-\frac{1}{R}+\frac{r_1}{2 R^2}\right].
\end{align}
Hence, in the vicinity of L1, the roots of $J_\varphi$ are real for restricted values of $\varphi$ when $E_{\rm Jc}^{'(0)} > h_0^{'(0)} + \Phi_1(r_1,0)$, i.e., when
\begin{align}
E_{\rm Jc}^{'(0)} > E^{'(1)}_\rmJ = h^{'(0)}_0 + q\left[-\frac{1}{\left|r_1 - R\right|} + \frac{r_1}{R^2}\right].
\end{align}
Here $E_{\rm Jc}^{'(0)}$ and $h_0^{'(0)}$ are, respectively, the circular part of the Jacobi energy given by equation~(\ref{Ejcirc_app}) and the unperturbed Jacobi energy, both evaluated at the COP, L0. These trapped orbits that lie inside the L1 separatrix are the \horseshoe orbits. These orbits are therefore in a librating state around near-co-rotation resonances (despite being untrapped wrt L4/L5). The untrapped orbits that lie outside the L1 separatrix, and that satisfy the condition,
\begin{align}
E_{\rm Jc}^{'(0)} < E^{'(1)}_\rmJ,
\end{align}
are the \pacman and the center-phylic orbits. The \pacmans have higher angular momentum ($L^{(1)}<L<L^{(2)}$) than the center-phylic orbits ($L<L^{(1)}$), where $L^{(k)}$, with $k=1,2$, denotes the value of $L$ at the $k^{\rm th}$ Lagrange point. Although the \pacmans are beyond the L1 separatrix, they are still trapped inside the L2 separatrix (as we shall see shortly) and therefore librating about near-co-rotation resonances. The center-phylic orbits on the other hand rotate about the galactic center. Since $J_r$ can vary along an orbit, certain orbits with high $J_r$ can cross the L1 separatrix, resulting in Chimera-like metamorphosis between the trapped \horseshoe and untrapped \pacman and center-phylic orbital families (see Appendix~\ref{App:Chimera}).

\subsection{Perturbation about the perturber}

In the vicinity of the perturber (i.e., the region centred on the perturber and bounded by the L2-separatrix), for a given $r_\rmG$, $\Phi_1$ varies in the range, 
\begin{align}
q\left[-\frac{1}{\left|r_\rmG - R\right|}+\frac{r_\rmG}{R^2}\right]<\Phi_1<q\left[-\frac{1}{R}+\frac{r_\rmG}{2 R^2}\right].
\end{align}
The nearest saddle point to the perturber is L2, located along the x-axis ($\varphi=\varphi_{\rm sep}=0$) at a distance $r_\rmG=r_{\rm sep}=r_2$ from the galactic center. Hence, in the neighborhood of L2, $J_\varphi$ has real roots for restricted values of $\varphi$ when $E_{\rm Jc}^{'\rmP} > h_0^{'\rmP} + \Phi_1(r_2,0)$, i.e., when
\begin{align}
E_{\rm Jc}^{'\rmP} > E^{'(2)}_\rmJ = h^{'\rmP}_0 + q\left[-\frac{1}{\left|r_2 - R\right|}+\frac{r_2}{R^2}\right].
\end{align}
Here $E_{\rm Jc}^{'\rmP}$ and $h_0^{'\rmP}$ are, respectively, the circular part of the Jacobi energy given by equation~(\ref{Ejcirc_app} and the unperturbed Jacobi energy, both evaluated at the COP, which in this case is the perturber. These trapped orbits that lie inside the L2 separatrix are the \pacmans when $E^{'(1)}_\rmJ>E^{'(2)}_\rmJ$ and \horseshoes otherwise. The \pacmans are therefore in a librating state about near-co-rotation resonances, even if they are untrapped wrt L0. The untrapped orbits that lie outside the L2 separatrix, and that satisfy the condition,
\begin{align}
E_{\rm Jc}^{'\rmP} < E^{'(2)}_\rmJ,
\end{align}
are the COM-phylic, perturber-phylic and center-phylic (for $\gamma<0$ profiles) orbits. The COM-phylic orbits have higher angular momentum ($L>L^{(2)}$) than the perturber-phylic orbits ($L^{(1)}<L<L^{(2)}$). While the COM-phylic orbits rotate about the COM of the galaxy-perturber system, the perturber-phylic orbits rotate about the perturber. Due to variation of $J_r$ along an orbit, Chimera-like transitions can occur between the trapped \pacman and untrapped COM-phylic and perturber-phylic orbital families (see Appendix~\ref{App:Chimera}).

\section{Chimera Orbits}
\label{App:Chimera}

\begin{figure}[H]
  \centering
  \subfloat{\includegraphics[width=0.91\textwidth]{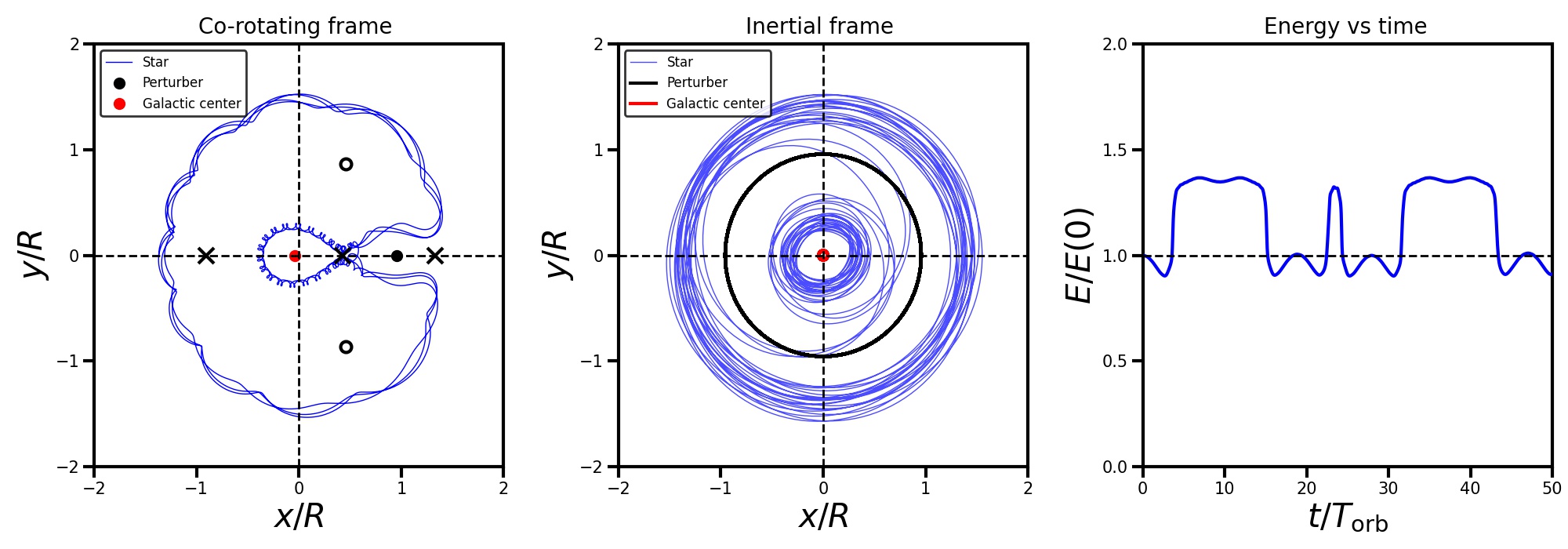}}
  \\
  \subfloat{\includegraphics[width=0.91\textwidth]{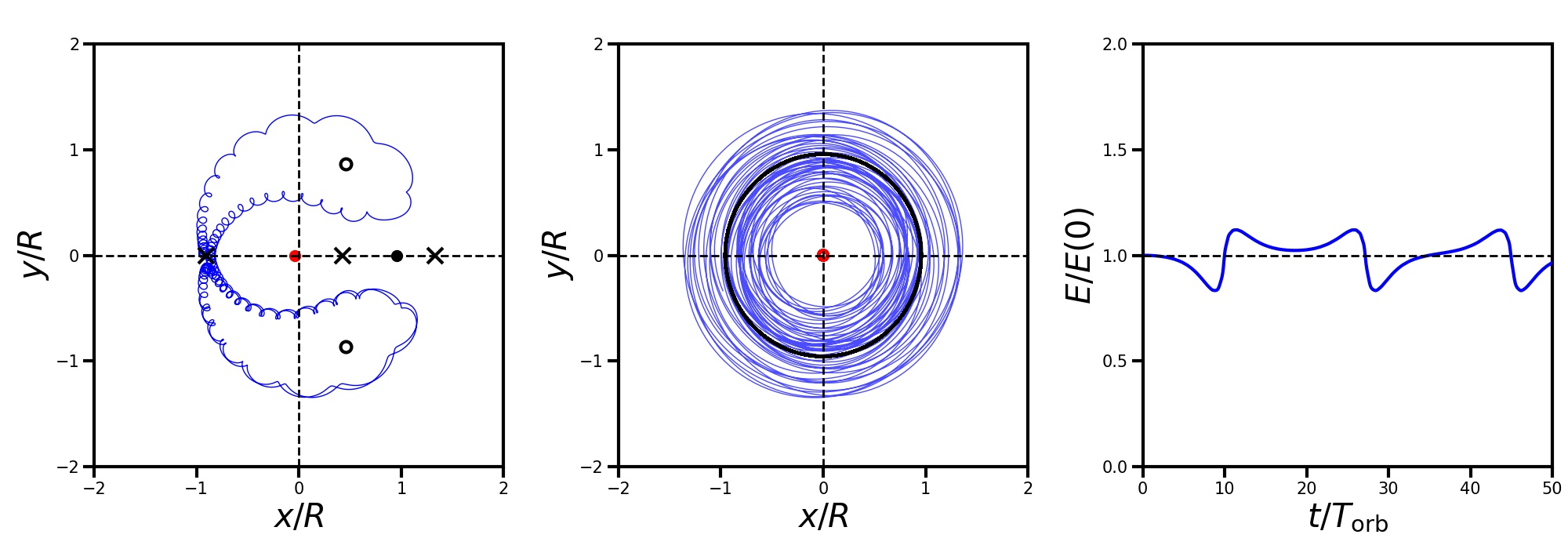}}
  \\
  \subfloat{\includegraphics[width=0.91\textwidth]{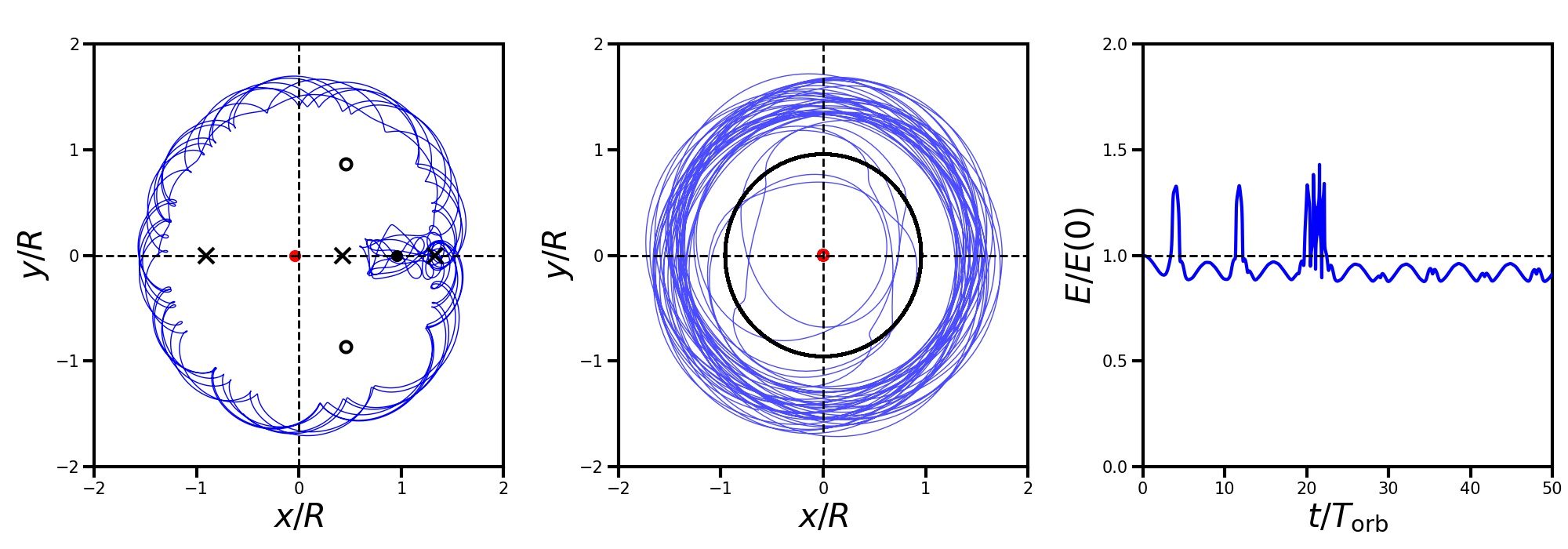}}
  \\
  \subfloat{\includegraphics[width=0.91\textwidth]{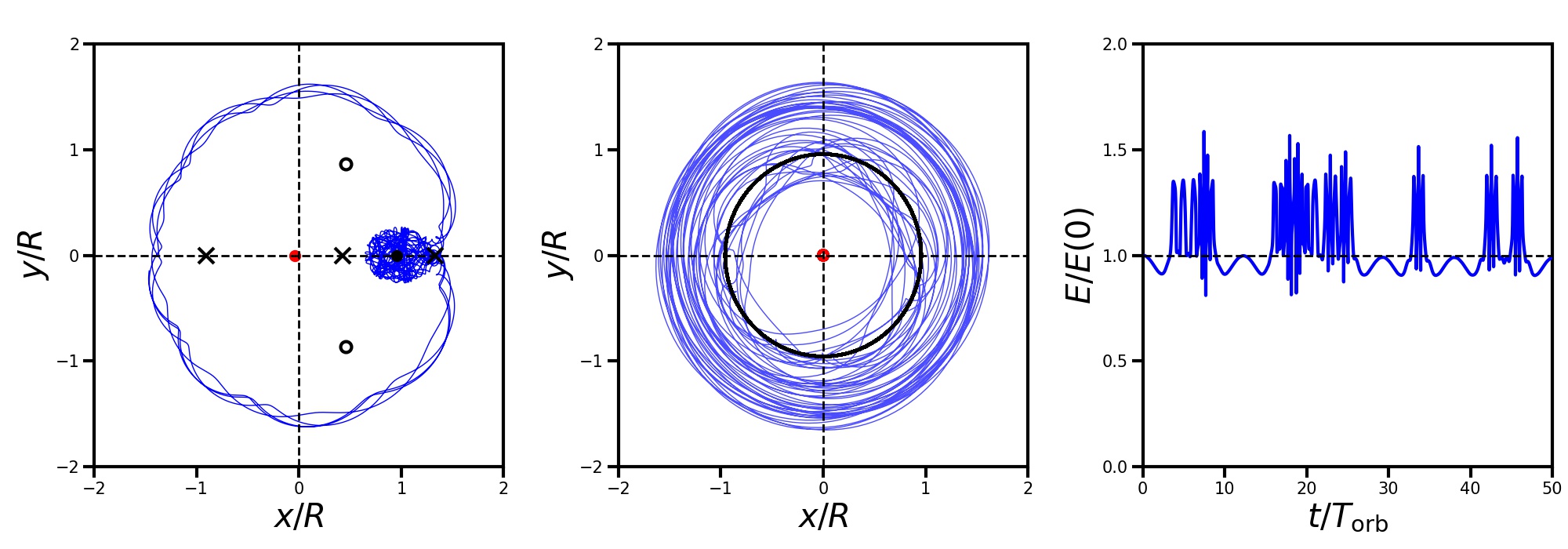}}
  
  \caption{Examples of Chimera orbits. From top to bottom the panels depict (i) a Chimera orbit initially classified as a \horseshoen, which occasionally undergoes separatrix crossing to transform into a \pacmann, (ii) an initial \horseshoe that transforms into a tadpole, (iii) an initial \pacman that transforms into perturber-phylic and COM-phylic orbits, and (iv) an initial \pacman that occasionally transforms into a perturber-phylic orbit. See text for details.}
  \label{fig:orbit_chimera}
\end{figure}

A significant subset of the orbits that we classify, based on the circular part of their Jacobi energy (see Appendix~\ref{App:orb_class}), in different orbital families (see Table.~\ref{tab:Ej}), occasionally undergo an inter-family metamorphosis triggered by a separatrix-crossing due to a change in the radial action enabled by the perturber. Fig.~\ref{fig:orbit_chimera} shows a few examples of such Chimera orbits. 

The first row of Fig.~\ref{fig:orbit_chimera} shows a Chimera orbit that is initially classified as a \horseshoe based on the criteria given in Table~\ref{tab:Ej}. However, after taking a detour around L0 like a typical \horseshoen, trapped between the L3 and L1 separatrices, during its first passage along the inner section (part of the orbit inside the perturber's radius), the field particle comes arbitrarily close to L1 during its second passage. Since its ZVC lies very close to the equipotential contour passing through L1, the particle undergoes a separatrix crossing (L1 separatrix) after which it takes a shortcut in between L0 and the perturber and becomes a \pacman orbit (based on the criteria given in Table~\ref{tab:Ej}), trapped between the L1 and L2 separatrices. After behaving like a \pacman during its second passage, the particle crosses the L1 separatrix again during its third passage to re-enter the \horseshoe phase. These \horseshoe $\rightarrow$ \pacman $\rightarrow$ \horseshoe transformations of the Chimera orbit are evident from the energy curve (right-hand panel), where a short-period oscillation corresponding to the \pacman phase is sandwiched between two long-period oscillations corresponding to the \horseshoe phases (cf. top and middle rows of Fig.~\ref{fig:orbit2}).

The second row depicts a Chimera orbit that is initially classified as a \horseshoe (trapped between the L3 and L1 separatrices), but which transforms into a tadpole (trapped within the L3 separatrix). In its \horseshoe phase, the particle makes a full circulation around L4 and L5 and its energy undergoes a long period oscillation (see right-hand panel). Then it enters its tadpole phase where it circulates only L5 and its energy undergoes a short period oscillation with a period exactly half of that of its \horseshoe phase. The separatrix-crossing in this case is triggered when the particle comes arbitrarily close to L3. This particular orbit has a ZVC that lies close to the L3 separatrix, which is why both the \horseshoe and tadpole phases have very long libration periods ($T_{\rm lib}$ asymptotes to infinity as the particle approaches the separatrix).

The third row shows a Chimera orbit that is initially classified as a \pacman orbit (trapped between the L1 and L2 separatrices), but which transforms into a perturber-phylic and a COM-phylic orbit, both of which lie beyond the L2 separatrix. In its initial \pacman phase, the particle undergoes regular, long period oscillations in energy (see right-hand panel). Then it comes arbitrarily close to L2 and undergoes a separatrix-crossing (L2 separatrix) to enter the perturber-phylic phase, which is reminiscent of resonant capture. In this phase the particle rotates around the perturber, associated with rapid oscillations in energy. At some point the particle approaches L2 again and enters a COM-phylic phase associated with energy oscillations that have much smaller amplitude than those during the \pacman and perturber-phylic phases.

Finally, the fourth row depicts a Chimera orbit that is initially classified as a \pacman orbit, trapped between the L1 and L2 separatrices, but which undergoes frequent L2 separatrix-crossings to become perturber-phylic. Note how the regular, long-period oscillations in energy corresponding to its \pacman phase are interspersed with rapid oscillations corresponding to its perturber-phylic phase. 

\section{The Stability of Lagrange Points}
\label{App:stability}

Lagrange points are fixed points in the co-rotating frame characterized by $\nabla \Phi_{\rm eff}=0$. Some Lagrange points are `centres', which are stable under small perturbations along any direction, while others are `saddles' with a stable and an unstable eigen-direction. In this appendix we analyze the stability of the Lagrange points using a linear perturbation analysis. The effective potential $\Phi_{\rm eff}$ can be Taylor-expanded about a Lagrange point $\bx_\rmL = (x_\rmL,y_\rmL)$ as follows
\begin{align}
\Phi_{\rm eff}(x,y) & = \Phi_{\rm eff}(x_\rmL,y_\rmL) + \frac{1}{2} \Phi_{\rm eff,xx} (\delta x)^2 + \frac{1}{2} \Phi_{\rm eff,yy} (\delta y)^2 + \Phi_{\rm eff,xy} \delta x\,\delta y\,,
\label{phiefftaylor}
\end{align}
where $\delta x = x - x_\rmL$, $\delta y = y - y_\rmL$, 
\begin{align}
\Phi_{\rm eff,xx} = \frac{\partial^2 \Phi_{\rm eff}}{\partial x^2}\bigg\rvert_{\bx_\rmL}\,,\;\;\;\;\;\; 
\Phi_{\rm eff,xy} = \frac{\partial^2\Phi_{\rm eff}}{\partial x \partial y}\bigg\rvert_{\bx_\rmL}\,,\;\;\;\;\;{\rm etc.,} 
\label{PhieffderivsA}
\end{align}
and we have used that $\nabla \Phi_{\rm eff}(x_\rmL,y_\rmL)=0$. Here and throughout we adopt the Cartesian $(x,y)$-coordinate system depicted in Fig.~\ref{fig:schematic}, with the origin at the COM of the galaxy$+$subject system, and the $x$-axis in the direction towards the subject.

The double derivatives of $\Phi_{\rm eff}$ can be written in terms of the derivatives of the galactic and subject potentials as follows
\begin{align}
\frac{\partial^2\Phi_{\rm eff}}{\partial x^2}&=\left(\frac{\partial^2\Phi_\rmG}{\partial r_\rmG^2}-\frac{1}{r_\rmG}\frac{\partial \Phi_\rmG}{\partial r_\rmG}\right){\left(\frac{x+q_\rmG R}{r_\rmG}\right)}^2 + \frac{1}{r_\rmG}\frac{\partial \Phi_\rmG}{\partial r_\rmG}+ \frac{q}{r_\rmP^3}\left[1-3{\left(\frac{x-q_\rmP R}{r_\rmP}\right)}^2\right] \nonumber \\
&- \frac{1+q_{\rm enc}(R)}{R}\frac{\partial \Phi_\rmG}{\partial R},\nonumber \\
\nonumber \\
\frac{\partial^2\Phi_{\rm eff}}{\partial y^2}&=\left(\frac{\partial^2\Phi_\rmG}{\partial r_\rmG^2}-\frac{1}{r_\rmG}\frac{\partial \Phi_\rmG}{\partial r_\rmG}\right){\left(\frac{y}{r_\rmG}\right)}^2 + \frac{1}{r_\rmG}\frac{\partial \Phi_\rmG}{\partial r_\rmG}+ \frac{q}{r_\rmP^3}\left[1-3{\left(\frac{y}{r_\rmP}\right)}^2\right] \nonumber\\
&- \frac{1+q_{\rm enc}(R)}{R}\frac{\partial \Phi_\rmG}{\partial R},\nonumber \\ \nonumber \\
\frac{\partial^2\Phi_{\rm eff}}{\partial x \partial y}&=y\left[\left(\frac{\partial^2\Phi_\rmG}{\partial r_\rmG^2}-\frac{1}{r_\rmG}\frac{\partial \Phi_\rmG}{\partial r_\rmG}\right)\frac{x+q_\rmG R}{r_\rmG^2}\nonumber -\frac{3q\left(x-q_\rmP R\right)}{r_\rmP^5}\right] \nonumber \,. \\
\label{phieffdderiv}
\end{align}

Using the expansion for $\Phi_{\rm eff}(x,y)$ given in equation~(\ref{phiefftaylor}), one can write down the equations of motion in the vicinity of $(x_\rmL,y_\rmL)$ as
\begin{align}
&\delta \ddot{x} - 2\,\Omega_\rmP\,\delta \dot{y} = -\Phi_{\rm eff,xx}\,\delta x - \Phi_{\rm eff,xy}\,\delta y, \nonumber \\
&\delta \ddot{y} + 2\,\Omega_\rmP\,\delta \dot{x} = -\Phi_{\rm eff,xy}\,\delta x - \Phi_{\rm eff,yy}\,\delta y\,,
\label{perturbedeqmotion}
\end{align}
The second term on the LHS corresponds to the Coriolis force. The centrifugal force and the gravitational fields due to the galaxy and the subject are included in the terms on the RHS. The equations of motion are linear and hence can be solved using the standard exponential ansatz
\begin{align}
&\delta x(t)=\delta x_0 \,e^{\omega t} \nonumber, \\
&\delta y(t)=\delta y_0 \,e^{\omega t}.
\end{align}

Substituting the above expressions for $\delta x(t)$ and $\delta y(t)$ in equations~(\ref{perturbedeqmotion}), we arrive at the following eigenvalue problem
\begin{equation}
\begin{bmatrix} 
{\omega}^2+\Phi_{\rm eff,xx} & -2\Omega_\rmP \omega+\Phi_{\rm eff,xy} \\ 
2\Omega_\rmP \omega+\Phi_{\rm eff,xy} & {\omega}^2+\Phi_{\rm eff,yy} 
\end{bmatrix}
\begin{bmatrix}
\delta x_0\\ 
\delta y_0
\end{bmatrix}
=\begin{bmatrix}
0\\ 
0
\end{bmatrix},
\end{equation}
which only has a non-trivial solution if the determinant of the $2\times 2$-matrix on the LHS vanishes. The resulting `characteristic' equation can be re-formulated as a quadratic equation in $\omega^2$:
\begin{equation}
\omega^4+B \omega^2+C=0,
\label{gammaquad}
\end{equation}
where 
\begin{equation}\label{omegaB}
B = 4\Omega^2_\rmP + \Phi_{\rm eff,xx} + \Phi_{\rm eff,yy}\,,
\end{equation}
and
\begin{equation}\label{omegaC}
C = \Phi_{\rm eff,xx}\Phi_{\rm eff,yy} - \Phi_{\rm eff,xy}^2\,.
\end{equation}
The roots of equation~(\ref{gammaquad}) are
\begin{equation}
\omega^2 = -\frac{B}{2}\pm\sqrt{{\left(\frac{B}{2}\right)}^2-C}.
\label{omegaroot}
\end{equation}

The stability of the fixed point depends on the sign of $\omega^2$. When $\omega^2<0$, $\omega$ is imaginary, which implies that $\delta x$ and $\delta y$ are sinusoidal functions of time, i.e. the fixed point is a center and therefore stable under small perturbations. The higher of the two values of $\Omega=\sqrt{-\omega^2}$ is the epicycle frequency of the field particle in the rotating frame while the smaller root is the libration frequency about the fixed point, i.e.,
\begin{align}\label{Omlib}
\Omega_{\rm epi}=\sqrt{\frac{B}{2}+\sqrt{{\left(\frac{B}{2}\right)}^2-C}}, \nonumber \\
\Omega_{\rm lib}=\sqrt{\frac{B}{2}-\sqrt{{\left(\frac{B}{2}\right)}^2-C}}.
\end{align}

Since in general ${\left(B/2\right)}^2-C>0$ \citep[see][]{Binney.Tremaine.08}, we have that all four roots of $\omega$ are imaginary (corresponding to a stable solution) when $C>0$. Using equation~(\ref{omegaC}) this translates into the following stability condition
\begin{equation}
\Phi_{\rm eff,xx} \Phi_{\rm eff,yy} > \Phi_{\rm eff,xy}^2\,,
\label{genstabilitycondition}
\end{equation}
which is found to be satisfied for L4 and L5. Hence these two Lagrange points are stable. For the Lagrange points that lie on the $x$-axis (i.e., L0, L1, L2, and L3), we have that $\Phi_{\rm eff,xy} = 0$ (this is immediately evident from the fact that  $\Phi_{\rm eff,xy} \propto y$, see equation~[\ref{phieffdderiv}]), and the stability condition reduces to
\begin{equation}
\Phi_{\rm eff,xx} \Phi_{\rm eff,yy} > 0\,.
\label{stabilitycondition}
\end{equation}
This condition is satisfied (violated) for L0 before (after) bifurcation. The Lagrange points L1, L2 and L3, on the other hand, always violate this condition, implying that they are unstable Lagrange points. They have positive and negative eigenvalues, corresponding to the unstable and stable eigen-directions, respectively, indicating that they are saddle points.

Since the effective potential, and hence its derivatives, changes when the distance $R$ between the subject and the galactic centre changes, whether or not a particular Lagrange point is stable, according to the general stability criterion of equation~(\ref{genstabilitycondition}), may depend on $R$. Such a change in the stability (or existence) of a fixed point under the change of $R$ is called a bifurcation. As is evident from the general stability criterion, bifurcation can happen whenever a Lagrange point is neither a centre nor a saddle, but is undergoing a transition from one to the other. This requires
\begin{equation}
\omega^2=0 \;\;\;\;\;\;\;\;\;\;\Rightarrow \;\;\;\;\;\;\;\;\; \Phi_{\rm eff,xx} \Phi_{\rm eff,yy} = \Phi_{\rm eff,xy}^2.
\label{genbifcondition}
\end{equation}
which thus represents the general bifurcation condition.
    \end{subappendices}


    \chapter{Summary and Future Work}\label{chapter: thesis_summary}

\section{Summary}

This dissertation contributes to the fields of galactic dynamics and non-equilibrium statistical mechanics of self-gravitating collisionless systems like galaxies and cold dark matter halos. We show how perturbed galaxies and halos relax to equilibrium and how this relaxation process in turn drives the secular evolution of the perturber's orbit (dynamical friction). We discuss how collisionless relaxation and secular evolution play out in different regimes of gravitational encounters: impulsive, resonant, and adiabatic. Our theories for relaxation and dynamical friction explain hitherto unexplained dynamical phenomena, thereby pushing the frontiers of galactic dynamics and galaxy formation and evolution research. The conclusions of our investigation are summarized as follows.

Chapter~\ref{chapter: paper1} investigates gravitational encounters between astrophysical objects like galaxies and dark matter halos in the impulsive limit. Impulsive encounters are usually treated using the distant tide approximation (DTA) for which the impact parameter, $b$, is assumed to be significantly larger than the characteristic radii of the subject, $r_\rmS$, and the perturber, $r_\rmP$. The perturber potential is then expanded as a multipole series and truncated at the quadrupole term. When the perturber is more extended than the subject, this standard approach can be extended to the case where $r_\rmS \ll b < r_\rmP$. However, for encounters with $b$ of order $r_\rmS$ or smaller, the DTA typically overpredicts the impulse, $\Delta \bv$, and hence the internal energy change of the subject, $\Delta E_{\rm int}$. This is unfortunate, as these close encounters are the most interesting, potentially leading to tidal capture, mass stripping, or tidal disruption. Another drawback of the DTA is that $\Delta E_{\rm int}$ is proportional to the moment of inertia, which diverges unless the subject is truncated or has a density profile that falls off faster than $r^{-5}$. To overcome these shortcomings, this chapter presents a fully general, non-perturbative treatment of impulsive encounters which is valid for any impact parameter, and not hampered by divergence issues, thereby negating the necessity to truncate the subject. We present analytical expressions for $\Delta \bv$ for a variety of perturber profiles, and apply our formalism to both straight-path encounters and eccentric orbits. We show that our non-perturbative treatment of impulsive encounters adequately describes the mass loss due to tidal shocks in gravitational encounters between equal mass galaxies.

In chapters~\ref{chapter: paper2} and \ref{chapter: paper3} we study the impact of both impulsive and non-impulsive perturbations on the dynamics of axisymmetric systems (disks). Galactic disks are highly responsive systems that often undergo external perturbations and subsequent collisionless equilibration, predominantly via phase-mixing. In Chapter~\ref{chapter: paper2}, we use linear perturbation theory to study the response of infinite isothermal slab analogues of disks to perturbations with diverse spatio-temporal characteristics. Without self-gravity of the response, the dominant Fourier modes that get excited in a disk are the bending and breathing modes, which, due to vertical phase-mixing, trigger local phase-space spirals that are one- and two-armed, respectively. We demonstrate how the lateral streaming motion of slab stars causes phase spirals to damp out over time (in a coarse-grained sense). The ratio of the perturbation timescale ($\tau_\rmP$) to the local, vertical oscillation time ($\tau_z$) ultimately decides which of the two modes is excited. Faster, more impulsive ($\tau_\rmP < \tau_z$) and slower, more adiabatic ($\tau_\rmP > \tau_z$) perturbations excite stronger breathing and bending modes, respectively, although the response to very slow perturbations is exponentially suppressed. For encounters with satellite galaxies, this translates to more distant and more perpendicular encounters triggering stronger bending modes. We compute the direct response of the Milky Way disk to several of its satellite galaxies, and find that recent encounters with all of them excite bending modes in the Solar neighborhood. The encounter with Sagittarius triggers a response that is at least $1-2$ orders of magnitude larger than that due to any other satellite, including the Large Magellanic Cloud. We briefly discuss how ignoring the presence of a dark matter halo and the self-gravity of the response might impact our conclusions.

In chapter~\ref{chapter: paper3}, we develop a linear perturbative formalism to compute the response of an inhomogeneous stellar disk embedded in a non-responsive dark matter (DM) halo to various perturbations such as bars, spiral arms and encounters with satellite galaxies. Without self-gravity to reinforce it, the response of a Fourier mode phase mixes away due to an intrinsic spread in the vertical ($\Omega_z$), radial ($\Omega_r$) and azimuthal ($\Omega_\phi$) frequencies, giving rise to local phase-space spirals. Collisional diffusion due to scattering of stars by structures like giant molecular clouds causes a fine-grained super-exponential damping of the phase spiral amplitude. The $z-v_z$ phase spiral turns out to be one-armed (two-armed) for vertically anti-symmetric (symmetric) bending (breathing) modes. Among bar and spiral arm perturbations, only transient ones that vary over timescales ($\tau_\rmP$) comparable to the vertical oscillation period ($\tau_z=2\pi/\Omega_z$) can trigger vertical phase spirals. Each $(n,l,m)$ mode of the response to impulsive ($\tau_\rmP<\tau=1/(n\Omega_z+l\Omega_r+m\Omega_\phi)$) perturbations is power law ($\sim \tau_\rmP/\tau$) suppressed. On the other hand, for adiabatic ($\tau_\rmP>\tau$) perturbations, the response for each mode is exponentially weak ($\sim \exp{\left[-\left(\tau_\rmP/\tau\right)^\alpha\right]}$) except resonant ($\tau\to \infty$) modes, where $\alpha$ is dictated by the exact time-dependence of the perturbing potential. Slower ($\tau_\rmP>\tau_z$) perturbations, which for satellite galaxies correspond to more distant encounters, induce stronger bending modes. Sagittarius dominates the Solar neighborhood response of the Milky Way (MW) disk to satellite encounters. Thus, if the Gaia phase spiral was triggered by a MW satellite, Sagittarius is the leading contender. However, the impact triggering this phase spiral must have occurred $\sim 0.6-0.7\Gyr$ ago in order for it to have survived collisional damping. We discuss the impact of the detailed galactic potential on the shape of phase spirals: phase-mixing occurs more slowly and thus phase spirals are more loosely wound in the outer disk and in presence of an ambient DM halo.

Chapters~\ref{chapter: paper1}, \ref{chapter: paper2} and \ref{chapter: paper3} describe the perturbation and relaxation of the subject during gravitational interactions. In chapters~\ref{chapter: paper4} and \ref{chapter: paper5}, we discuss how the back reaction of the response of the subject/host exerts dynamical friction on the perturber and drives the secular evolution of its orbit. In the standard resonance picture, dynamical friction is regarded as a secular process, in which the perturber evolves very slowly (secular approximation), and has been introduced to the host over a long time (adiabatic approximation). These assumptions imply that dynamical friction arises from the LBK torque with non-zero contribution only from purely resonant orbits. However, dynamical friction is only of astrophysical interest if its timescale is shorter than the age of the universe. In this thesis, we therefore relax the adiabatic and secular approximations. We first derive a generalized LBK torque, which reduces to the LBK torque in the adiabatic limit, and show that it gives rise to transient oscillations due to non-resonant orbits that slowly damp out due to phase-mixing, giving way to the LBK torque. This is analogous to how a forced, damped oscillator undergoes transients before settling to a steady state, except that here the damping is due to phase-mixing rather than dissipation. Next, we present a self-consistent treatment, that properly accounts for the time-dependence of the perturber potential and circular frequency (memory effect), which we use to examine orbital decay in a cored galaxy. We find that the memory effect results in a phase of accelerated, super-Chandrasekhar friction before the perturber stalls at a critical radius, $\Rcrit$, in the core (core-stalling). Inside of $\Rcrit$ the torque flips sign, giving rise to dynamical buoyancy, which counteracts friction and causes the perturber to stall. This phenomenology is consistent with $N$-body simulations, but has thus far eluded a proper explanation.

In Chapter~\ref{chapter: paper5}, we examine the origin of dynamical friction using a non-perturbative, orbit-based approach. In the standard perturbative approach, dynamical friction arises from the LBK torque due to pure resonances, whereas in the self-consistent perturbative approach discussed in chapter~\ref{chapter: paper4}, it arises from near-resonant orbits. This chapter provides an alternative, complementary view of dynamical friction that nicely illustrates how a massive perturber significantly (non-perturbatively) changes the energies and angular momenta of field particles on near-resonant orbits, with friction arising from an imbalance between particles that gain energy and those that lose energy. We treat dynamical friction in a spherical host system as a restricted three-body problem. This treatment is applicable in the `slow' regime, in which the perturber sinks slowly and the standard perturbative framework fails due to the onset of non-linearities. Hence it is especially suited to investigate the origin of core-stalling: the cessation of dynamical friction in central constant-density cores. We identify three different families of near-co-rotation-resonant (NCRR) orbits that dominate the contribution to dynamical friction. Their relative contribution is governed by the Lagrange points (fixed points in the co-rotating frame). In particular, one of the three families, which we call \pacman orbits because of their appearance in the co-rotating frame, is unique to cored density distributions. When the perturber reaches a central core, a bifurcation of the inner Lagrange points, L3, L0 and L1, and the associated tidal disruption of the core, drastically change the orbital make-up, with \pacman orbits becoming dominant. In addition, due to relatively small gradients in the distribution function inside a core, the net torque from these \pacman orbits becomes positive (enhancing), thereby effectuating a dynamical buoyancy. Core stalling occurs at the radius of transition from friction to buoyancy. The pre-eminent causes of stalling are (i) the choking away of the phase-space available to the NCRR \horseshoe orbits, which is equivalent to the suppression of co-rotation resonances as noted by \cite{Kaur.Sridhar.18}, and (ii) the shooting up of the libration timescale of the remaining NCRR \pacman orbits and the consequent delay in the development of the near-resonant response density along these orbits that drives dynamical friction (bottleneck effect). Both these effects are the outcomes of the bifurcation of the inner Lagrange points, which occurs when the perturber enters the core region of a host galaxy with a central constant density core.

\section{Implications and future work}

This dissertation presents novel techniques to study gravitational encounters and relaxation phenomena. Our non-perturbative treatment of impulsive heating can be applied to the study of tidal shocks in penetrating gravitational encounters, unlike the standard approach which only works for distant encounters. The perturbative formalism we developed for disk response and phase-mixing can be wielded in combination with observations of non-equilibrium phase-space features like phase spirals to constrain the dynamical history and gravitational potential of our Milky Way galaxy. Our self-consistent perturbative and non-perturbative orbit-based treatments of dynamical friction explain the origin of hitherto unexplained dynamical phenomena observed in the $N$-body simulations of cored galaxies: core-stalling and dynamical buoyancy, which counteract the dynamical friction-driven orbital inspiral of massive perturbers in the core region. 

The occurrence of core-stalling and buoyancy in cored galaxies have far reaching implications for various astrophysical processes. For example, the in-fall and merger of supermassive black holes (SMBHs) can be choked in the core region of galaxies due to core-stalling. We obtained constraints on the distribution of these wandering SMBHs using the distortions of gravitational lensing arcs caused by such perturbers \citep[][]{Banik.etal.19}. The presence of cores in the stellar and dark matter components of galaxies thus arrests SMBH mergers, posing a challenge for the merger-driven formation of SMBHs if the merging host galaxies that harbor the SMBHs possess cores or a central core forms in the post-merger galaxy. Even if the host galaxy initially has a central density cusp, the dynamical friction-driven inspiral of a massive perturber can form a core \citep[][]{Goerdt.etal.10}, which can in turn cause the perturber to stall. If somehow two SMBHs can inspiral all the way to the center (e.g., in a cuspy host) and then merge, the gravitational wave recoil kick experienced by the merger remnant, which is a heavier SMBH, can push it quite far from the center. And since the merging SMBHs might have scoured a core out of the initial host profile, the occurrence of stalling and buoyancy may prevent the final SMBH from falling back to the center. We intend to address the detailed implications of stalling and buoyancy in SMBH mergers and the resultant gravitational wave observations to be performed by LISA in future work. Because of the above phenomena of stalling and buoyancy, a significant population of SMBHs in the universe may not lie in the centers of galaxies but rather be wandering, i.e., offset from galactic centers \citep[][]{Tremmel.etal.18,Mezcua.Dominguez.20,Bellovary.etal.21}. This dissertation describes how dynamical friction, buoyancy and core-stalling play out in collisionless systems. Dynamical friction in collisional systems like gas can behave differently. Gas dynamical friction can cause a massive perturber to continually inspiral as long as the perturber lies in the supersonic regime \citep[][]{Ostriker.99}, i.e., the local temperature and/or turbulent velocity dispersion of gas is low enough. In the subsonic regime, however, the perturber experiences much weaker dynamical friction. The strength of dynamical friction due to gas depends of course on the amount of gas in the host system as well as the sound speed profile of the gas. The relative importance of dark matter, stars and gas in dynamical friction and SMBH mergers deserves detailed future investigation using $N$-body simulations and analytical techniques.

The phenomena of core-stalling and buoyancy in dark matter halos are sensitive to the nature of dark matter. In cold dark matter (CDM) halos with a central constant density core (which can form due to baryonic feedback or dark matter self-interactions), a more massive perturber stalls further out since it tidally disrupts the core earlier than a less massive one does. Therefore, the observed offset radius of massive perturbers like SMBHs, globular clusters, nuclear star clusters, etc. with respect to the galactic center should show an increasing trend with the perturber-to-host mass ratio if dark matter is cold. If dark matter is `fuzzy', i.e., composed of ultralight axions \citep[][]{Hui.etal.17} that are bosonic in nature and possess an astronomically large de Broglie wavelength, the halo consists of a central soliton core (a Bose-Einstein condensate) and quantum interference patterns called `quasiparticles'. A massive perturber orbiting in a fuzzy dark matter (FDM) halo experiences dynamical friction from less massive quasiparticles and diffusive heating from more massive ones, and stalls when its mass matches the local quasiparticle mass \citep[][]{Bar-Or.etal.19,DuttaChowdhury.etal.20}. Since, more massive quasiparticles reside further in, a more massive perturber stalls further in as well. This behaviour is remarkably different from that of perturbers orbiting in CDM halos, where stalling occurs due to a subtle balance between friction and buoyancy (rather than friction and diffusive heating in FDM), and more massive perturbers stall farther out. Unlike CDM, the offset radius of massive perturbers in an FDM halo should therefore show a decreasing trend with the perturber-to-host mass ratio. Hence, a statistical analysis of the offset distributions of massive objects in galaxies can potentially put constraints on the inner density profiles of galaxies and the nature of dark matter. 

We have investigated the operating mechanism of dynamical friction in a central constant density core $(\gamma=\rmd\log\rho/\rmd\log r=0)$ and compared it to a $\gamma=-1$ NFW-like cusp. We found that core-stalling and dynamical buoyancy should occur in Plummer and Isochrone cores but not in $\gamma=-1$ cusps. However, preliminary analysis using the orbit-based approach we developed shows that a shallower core (shallower transition from the outer to the inner log slope of the density profile) than the Plummer and Isochrone cores, e.g., the $\gamma=0$ \cite{Dehnen.93} sphere, exhibits a bifurcation of Lagrange points and core-stalling but not buoyancy. It appears that bifurcation and therefore stalling occurs in any constant density ($\gamma=0$) core, but buoyancy only occurs in profiles with a steep enough transition from the outer to the inner slope. These are profiles that possess a shallow enough distribution function (small enough $\left|\partial f_0/\partial E_0\right|$ as $E_0$ goes to the central potential) in the central region. In future we intend to investigate how dynamical friction, buoyancy and stalling operate in host systems with diverse density profiles, i.e., for $-1\leq \gamma \leq 0$ as well as varying steepness of transition from the outer to the inner slope.

This dissertation investigates collisionless relaxation phenomena and secular evolution mainly with the help of linear perturbation theory (except the non-perturbative orbit-based treatment of dynamical friction). In all perturbative treatments, we have made the simplifying assumption of negligible self-gravity of the response. For computing the perturbative response of an inhomogeneous disk, this assumption is justified since we were mostly interested in studying the phase-mixing of the response that spawns phase spirals. The dominant effect of self-gravity is the occurrence of coherent point mode oscillations of the disk as a whole, which are decoupled from the phase-mixing component of the response in the linear regime. \cite{Mathur.90} and \cite{Weinberg.91} analyzed the self-gravitating response of a perturbed slab. \cite{Widrow.23} investigated the impact of self-gravity on the response of a shearing box to impulsive perturbations while \cite{Dootson.Magorrian.22} studied the self-gravitating response of an inhomogeneous but razor-thin disk to bar perturbations. However, computing the self-gravitating response of an inhomogeneous thick disk to generic perturbations is still an unsolved problem. We intend to address this in future work using the \cite{Kalnajs.77} matrix method. The impact of self-gravity on the response of a spherical host to a massive perturber and the resultant secular evolution of the perturber's orbit have been studied by \cite{Weinberg.89}, who found that self-gravity typically reduces the lag of the response behind the perturber and thus weakens the dynamical friction inspiral rate by a factor of $\sim 2-3$ compared to the non self-gravitating case. The inclusion of self-gravity is however a complicated exercise even without a self-consistent computation (taking into account the dependence of the torque on the radial motion of the perturber), which has been the standard practice \citep[e.g.,][etc]{Tremaine.Weinberg.84,Weinberg.89,Kaur.Sridhar.18,Kaur.Stone.22} as of now. Since we compute the self-consistent torque in \cite{Banik.vdBosch.21a}, we neglect the self-gravity of the response for the sake of simplicity. We hope to address the impact of self-gravity on the self-consistent treatment of dynamical friction in future work.

The perturbative techniques developed as part of this dissertation are reliable as long as the perturbation series converges, i.e., $f_{i+1}/f_i < 1$, where $f_i$ is the $i^{\rm th}$ order perturbation in the distribution function of a system in response to an external perturbing potential, $\Phi_\rmP$. This is the case when $\Phi_\rmP$ is weak, i.e., $\Phi_\rmP < \Phi_0$, with $\Phi_0$ the potential of the unperturbed system, or in other words the perturbation is in the linear regime. Although a useful tool to analyze the response of a system to weak perturbations, perturbation theory becomes questionable in the non-linear regime of strong perturbations, e.g., the rapid virialization or violent relaxation of perturbed galaxies and CDM halos, the non-linear saturation of gravitationally unstable modes (bars and spiral arms growing out of disk galaxies), etc. Such cases have only been studied using $N$-body simulations as of date since a general (either analytical or numerical) solution of the Boltzmann-Poisson system of equations has been beyond our reach. Even in the collisionless limit, it is a system of highly non-linear, coupled, integro-differential equations in $2d$ dimensions (for $d$ spatial dimensions). This makes it extremely challenging to solve. However, using the method of integral transforms, these equations can be significantly simplified. This simplification is manifest in the $(\bx,\bv)$ space, but not in the $(\bw,\bI)$ space, since the action-angle coordinates are natural coordinates only for the Vlasov equation but not for the Poisson equation. Some of the challenges involved in solving these equations are (i) the high dimensionality of the problem (equal to $7$ in 3D, including $3$ components of $\bx$, $3$ of $\bv$ and $1$ of time) and (ii) the attainment of the resolution required to obtain the fine structures that a Boltzmann equation seems to predict, just like the turbulent eddies predicted by the Navier-Stokes equations. In future work, we intend to come up with a computationally efficient Boltzmann-Poisson solver that can act as a competitive tool alongside $N$-body simulations. Through this general treatment of the Boltzmann-Poisson equations, we hope to enhance our understanding of non-linear relaxation phenomena such as violent relaxation and non-linear structure formation, and elucidate how equilibrium or quasi-equilibrium end-states of collisionless relaxation are attained.

    \addcontentsline{toc}{chapter}{Bibliography} 
    \setstretch{1.0}
    

    \Urlmuskip=0mu plus 2mu
    \printbibliography

\end{document}